%% file: main.tex
\documentclass[a4paper,11pt]{article}
\usepackage{jheppub} 

\makeatletter
\def\@fpheader{}
\makeatother

\usepackage{units}
\usepackage{xcolor}
\usepackage{array}
\usepackage{graphicx}
\usepackage{subcaption}
\usepackage{longtable}
\usepackage{multicol}
\usepackage{setspace}
\usepackage{comment}
\usepackage{pdfpages}
\usepackage{graphicx}
\usepackage[normalem]{ulem}
\usepackage{notoccite}

\usepackage{breqn}
\usepackage{amsmath}

\usepackage{tabularx}
\usepackage{ragged2e}

\def \L {\mathcal{L}} 

\def \vec#1{{\boldsymbol{#1}}}
\newcommand{\hc}{\ensuremath{\text{h.c.}}}

\newcolumntype{L}{>{$}l<{$}} 
\newcolumntype{C}{>{$}c<{$}}
\newcolumntype{P}{>{$}p<{$}}

\newcommand{\website}{\href{https://uva.theopenscholar.com/heeck/BNV}{uva.theopenscholar.com/heeck/BNV}}

\usepackage[most]{tcolorbox}
\usepackage{xparse}

\newlength{\SplitTableNameWidth}
\newcommand{\SplitTableOverlay}[1]{
  \draw[line width=.4pt]
    ([xshift=\SplitTableNameWidth]frame.north west) --
    ([xshift=\SplitTableNameWidth]frame.south west);
  \node[anchor=west, inner sep=2pt, align=left,
        text width=\SplitTableNameWidth]
    at (frame.west) {#1};
}

\NewDocumentCommand{\SplitRow}{m +m}{
  \begin{tcolorbox}[
    enhanced,
    breakable,
    sharp corners,
    colback=white,
    colframe=black,
    boxrule=.4pt,
    left=\dimexpr\SplitTableNameWidth+2pt\relax,
    right=2pt, top=2pt, bottom=2pt,
    before skip=0pt, after skip=0pt,
    width=\linewidth,
    overlay unbroken={\SplitTableOverlay{#1}},
    overlay first={\SplitTableOverlay{#1}},
    overlay middle={\SplitTableOverlay{#1}},
    overlay last={\SplitTableOverlay{#1}},
  ]
  #2
  \end{tcolorbox}
}

\NewDocumentCommand{\SplitHeader}{m m}{
  \begin{tcolorbox}[
    enhanced,
    sharp corners,
    colback=white,
    colframe=black,
    boxrule=.4pt,
    left=\dimexpr\SplitTableNameWidth+2pt\relax,
    right=2pt, top=2pt, bottom=2pt,
    before skip=0pt, after skip=0pt,
    width=\linewidth,
    overlay={\SplitTableOverlay{#1}},
  ]
  #2
  \end{tcolorbox}
}

\NewDocumentEnvironment{SplitTable}{O{4.0cm}}{
  \par\setlength{\SplitTableNameWidth}{#1}\noindent
}{\par}

\allowdisplaybreaks

\title{Opening up baryon-number-violating operators}

\author[a,1]{Julian Heeck,%
\note{ORCID: \href{https://orcid.org/0000-0003-2653-5962}{0000-0003-2653-5962}}}
\emailAdd{heeck@virginia.edu}

\author[a,2]{Diana Sokhashvili,%
\note{ORCID: \href{https://orcid.org/0009-0002-0428-6264}{0009-0002-0428-6264}}}
\emailAdd{bkq5jz@virginia.edu}

\author[a,b,3]{Anil Thapa%
\note{ORCID: \href{https://orcid.org/0000-0003-4471-2336}{0000-0003-4471-2336}}}
\emailAdd{a.thapa@colostate.edu}

\affiliation[a]{Department of Physics, University of Virginia,
Charlottesville, Virginia 22904-4714, USA}
\affiliation[b]{Physics Department, Colorado State University, Fort Collins, CO 80523, USA}

\abstract{
Baryon number violation is our most sensitive probe of physics beyond the Standard Model. Its realization through heavy new particles can be conveniently encoded in higher-dimensional operators that allow for model-agnostic analyses. The unparalleled sensitivity of nuclear decays to baryon number violation makes it possible to probe effective operators of very high mass dimension, far beyond the commonly discussed dimension-six operators. To facilitate studies of this ginormous and scarcely explored testable operator landscape we provide the exhaustive set of tree-level UV completions consisting of scalars, fermions, and vectors for non-derivative baryon-number-violating operators in this Standard Model effective field theory up to mass dimension 15, which corresponds roughly to the border of sensitivity. In addition to the known Standard Model fields we also include right-handed neutrinos in our operators. Our public code can be used to UV-complete any non-derivative operator and match it onto an operator basis. 
}

\begin{document}

\maketitle
\flushbottom
\clearpage
\addcontentsline{toc}{section}{\listtablename}
\listoftables
\flushbottom
\clearpage
\addcontentsline{toc}{section}{\listfigurename}
\listoffigures
\flushbottom
\clearpage

\pagebreak

\section{Introduction}
\label{sec:intro}

The fantastically successful Standard Model (SM) of particle physics predicts, among other things, that baryon and lepton number, $B$ and $L$, are conserved in all perturbative interactions. This by itself would be sufficient motivation to look for baryon number violation (BNV) as a test of the SM~\cite{FileviezPerez:2022ypk,Broussard:2025opd}. As it happens, many extensions of the SM actually \textit{do} break $B$ and/or $L$ at potentially testable rates, promoting BNV searches to discovery probes. The same holds for searches for lepton number violation and lepton \emph{flavor} violation~\cite{Davidson:2022jai}, of course, but BNV is special since it can be tested by observing SM-\emph{stable} particles such as protons or nuclei, which are relatively easy to collect in large numbers and observe for long times, even several decades. This renders BNV one of the most sensitive probes of physics beyond the SM at our disposal~\cite{Heeck:2019kgr}. To wit, some well-motivated BNV proton decay channels such as $p\to e^+\pi^0$ are constrained by Super-Kamiokande to lifetimes above $\unit[10^{34}]{yr}$~\cite{Super-Kamiokande:2018apg,Super-Kamiokande:2020tor,Super-Kamiokande:2020wjk,Super-Kamiokande:2024qbv}, exceeding the age of our universe by 24 orders of magnitude. The detectors JUNO~\cite{JUNO:2015zny} (taking data since 2025), Hyper-Kamiokande~\cite{Hyper-Kamiokande:2018ofw} (to start 2028), DUNE~\cite{DUNE:2020ypp} (to start 2029), and potentially THEIA~\cite{Theia:2019non}, will further advance these BNV searches~\cite{Takhistov:2026aln}. 
The stringent lifetime limits in turn put constraints on particles beyond the SM as heavy as $\unit[10^{15}]{GeV}$, a mass scale that is impossible to probe by any other means~\cite{FileviezPerez:2022ypk}.
BNV searches thus provide a unique window to new heavy particles. Indeed, it is \emph{so} sensitive that we have yet to map out all kinds of new particles that are probed through BNV~\cite{Heeck:2019kgr}. The article at hand is intended to start this mapping program.

Since we have not yet discovered any particles beyond the SM in our careful exploration of physics up to the TeV scale, it is reasonable to imagine that all new particles are much heavier, with some loopholes discussed below. Although frustrating from the direct-discovery side, this would allow us to describe the effect of any new particles using higher-dimensional effective operators, essentially by integrating out the heavy degrees of freedom. This Standard Model Effective Field Theory (SMEFT) paradigm has become very popular in recent years, see Refs.~\cite{Isidori:2023pyp,Aebischer:2025qhh} for recent reviews. The approach is fairly straightforward: write down all gauge- and Lorentz-invariant local operators $\mathcal{O}$ involving only SM fields and order them according to their mass dimension~$d$. The $d\leq 4$ terms just give the renormalizable SM Lagrangian $\mathcal{L}_{\text{SM}}$; the $d \geq 5$ ones can be interpreted as the effects of heavy new particles, leading to the full SMEFT Lagrangian
\begin{align}
\mathcal{L} &= \mathcal{L}_{\text{SM}} + 
\left[\sum_{d \geq 5}\sum_{j} \frac{C^j_{d}}{\Lambda^{d-4}} \mathcal{O}_{d}^j +\hc\right],
\end{align}
where $j$ sums over the different operators with same $d$, and $C^j_{d}$ are their dimensionless coefficients. 
For this to be useful, we want to truncate this sum at some $d_{\text{max}}$, arguing that the prefactor $1/\Lambda^{d-4}$ with $\Lambda \gg \unit{TeV}$ suppresses the remaining $d>d_{\text{max}}$ operators to negligible magnitudes. 
By this logic, BNV phenomenology should be dominated by the BNV operators with the lowest mass dimension, which is $d=6$~\cite{Weinberg:1979sa,Wilczek:1979hc}. Those $d=6$ BNV operators, which all happen to conserve the linear combination $B-L$, are indeed the focus of most theoretical and experimental work~\cite{FileviezPerez:2022ypk}.
  
However, as already pointed out by Weinberg~\cite{Weinberg:1980bf}, BNV operators with $d>6$ can conserve \emph{different} linear combinations of $B$ and $L$~\cite{Weldon:1980gi,Rao:1983sd,Degrande:2012wf, deGouvea:2014lva, Henning:2015alf,Heeck:2019kgr, Kobach:2016ami,Helset:2019eyc, Heeck:2025btc}, so the relative importance of BNV operators does not only depend on the mass dimension $d$, but also on the details of how the global SM symmetry $U(1)_B\times U(1)_L$ is broken, as this can lead to vanishing Wilson coefficients $C^j_{d}$ that overthrow the naive $1/\Lambda^{d-4}$ power-counting. Together with the fact that proton decay could be sensitive to BNV operators up to $d\sim 15$ if $\Lambda\sim\unit{TeV}$~\cite{Heeck:2019kgr}, this opens up a vast field of potentially interesting BNV operators with different $d$, $B$, and $L$.
In addition to these $B + a L$ symmetries pointed out by Weinberg, with $a\in \mathbb{Q}$, many other ways have since been found to suppress the naively-dominant $d=6$ BNV operators, from other continuous or discrete (flavor) symmetries~\cite{Zee:1981vr,Sarkar:1983tm,Ibanez:1991pr,Carone:1995pu,Babu:2003qh,Hambye:2017qix,Heeck:2019kgr,Fonseca:2018ehk,Helset:2019eyc,Heeck:2024jei} to extra dimensions~\cite{Appelquist:2001mj,Nussinov:2001rb,Lee:2002mn,Mohapatra:2002ug,Girmohanta:2019fsx}.

BNV is then one of the few areas of particle physics that is sensitive to SMEFT operators with $d\gg 6$. In most other areas, one can simply employ the established SMEFT results, notably a non-redundant basis of operators, which has been constructed for 
$d=6$~\cite{Buchmuller:1985jz,Grzadkowski:2010es}, $d=7$~\cite{Lehman:2014jma,Liao:2016hru}, $d=8$~\cite{Li:2020gnx,Murphy:2020rsh}, and $d=9$~\cite{Li:2020xlh,Liao:2020jmn}. Ref.~\cite{Harlander:2023psl} has pushed this up to $d=12$ through a code in computer-readable form. To \emph{fully} explore BNV, we have to go further still and will do so by using Fonseca's \texttt{Mathematica} package \texttt{Sym2Int}~\cite{Fonseca:2017lem,Fonseca:2019yya}, which can generate SMEFT operators even up to $d\sim 20$ in reasonably short time~\cite{Heeck:2025btc}.

Powerful as the SMEFT might be, there are benefits in considering renormalizable ultraviolet (UV) complete realizations of the underlying operators. For $\Lambda$ near the TeV scale, this allows us to compare SMEFT-operator limits to on-shell particle searches at the LHC, and to scrutinize the validity of the approximations that underlie the SMEFT. But even for larger $\Lambda$ there are benefits of knowing UV completions, as they i) provide a more physical basis for SMEFT operators, ii) can imbue operators with extra symmetries that would seem unmotivated in the SMEFT, iii) provide a catalog of elementary particles that could induce an experimentally observed SMEFT operator. 
Tree-level UV-completions of the $d=5$ SMEFT operator that leads to Majorana neutrino masses~\cite{Weinberg:1979sa} are well known and go by the names of type-I~\cite{Minkowski:1977sc,Yanagida:1979as,Gell-Mann:1979vob,Mohapatra:1979ia}, type-II~\cite{Magg:1980ut,Schechter:1980gr,Wetterich:1981bx,Lazarides:1980nt,Mohapatra:1980yp,Cheng:1980qt}, and type-III~\cite{Foot:1988aq} seesaw mechanism~\cite{Ma:1998dn}.
(Loop-level realization have also been constructed~\cite{Farzan:2012ev}.)
UV completions of all $d=6$ SMEFT operators can be found in Ref.~\cite{deBlas:2017xtg}, those of $d=7$ in Refs.~\cite{Li:2022abx,Li:2023cwy}, and those of $d=8$ in Ref.~\cite{Li:2023pfw}. 
The SMEFT assumes no additional light degrees of freedom beyond the SM fields, listed in Tab.~\ref{tab:fields}. Given that neutrino masses could well be generated by light new particles, notably right-handed/sterile neutrinos, it is not uncommon to also include three $\nu$ in the SMEFT, then calling it $\nu$SMEFT (or $\nu_R$SMEFT or $N_R$SMEFT).
Aside from more operators there is nothing complicated about this SMEFT generalization, and it can have interesting phenomenological consequences for BNV~\cite{Helo:2018bgb}. A basis of operators up to $d=9$ has been constructed here as well, see Ref.~\cite{Li:2021tsq}, and UV completions of $\nu$SMEFT operators up to $d=7$ can be found in Refs.~\cite{Bischer:2019ttk,Beltran:2023ymm}.

\begin{table}[tb]
\centering
{\renewcommand{\arraystretch}{1.2}
\begin{tabular}{c c c c} 
 \hline
 field & chirality & generations & $SU(3)_C\times SU(2)_L\times U(1)_Y$ representation  \\ [0.5ex] 
 \hline\hline
 $Q$ & left & 3 & $\left(\vec{3},\vec{2},\tfrac16\right)$\\
 $u$ & right & 3 & $\left(\vec{3},\vec{1},\tfrac23\right)$\\
 $d$ & right & 3 & $\left(\vec{3},\vec{1},-\tfrac13\right)$\\
 $L$ & left & 3 & $\left(\vec{1},\vec{2},-\tfrac12\right)$\\
 $e$ & right & 3 & $\left(\vec{1},\vec{1},-1\right)$\\
 $\nu$ & right & 3 & $\left(\vec{1},\vec{1},0\right)$\\
 $H$ & scalar & 1 & $\left(\vec{1},\vec{2},\tfrac12\right)$\\
 \hline
\end{tabular}
}
\caption{SM fields (plus right-handed neutrinos $\nu$) and quantum numbers; hypercharge $Y$ is related to electric charge $Q_e$ via $Q_e = Y + T_3$, with diagonal $SU(2)_L$ generator $T_3$.}
\label{tab:fields}
\end{table}

Only partial results are known for $d>8$, mostly for operators that break $L$ by two units and thus lead to Majorana neutrino masses; UV completions for those $\Delta L =2$ operators have been constructed up to $d=11$ by now~\cite{Babu:2001ex,deGouvea:2007qla,Angel:2012ug,Gargalionis:2020xvt}. Operators of such high mass dimension can still be relevant due to the sensitivity of neutrino masses and lepton-number-violating neutrinoless double-beta decay~\cite{Rodejohann:2011mu}.
Of course \emph{baryon number violation} is apparently even more suppressed and even more sensitive than lepton number violation, which motivates us to repeat these efforts for BNV operators and push for even higher mass dimension.
Some work in this direction exists already, notably Weinberg's seminal paper~\cite{Weinberg:1980bf}; UV completions of many BNV operators involving just (leptoquark) scalars were given in Refs.~\cite{Bowes:1996xy,Kovalenko:2002eh,Klapdor-Kleingrothaus:2002rvk,Arnold:2012sd,Dorsner:2022twk}. 
We aim to be exhaustive in this effort to produce a valuable resource for the community. Along the way, we also uncover some errors in the literature.

The rest of our article is organized as follows: in Sec.~\ref{sec:d6}, we discuss the well-known $d=6$ BNV $\nu$SMEFT operators and their tree-level UV completions. This example serves to illustrate our method for UV-completing operators and to establish our notation.
We explain our general procedure to obtain UV completions in Sec.~\ref{sec:general_procedure}, which also contains the tables that define the names of all new heavy scalars, fermions, and vectors.
In this article, we restrict ourselves to \emph{non-derivative} BNV operators, which we justify in Sec.~\ref{sec:derivative_operators}.
Some UV completions contain gauge-singlet particles, which could lead to novel effects if they are light, as discussed in Sec.~\ref{sec:light_particles}. 
We conclude in Sec.~\ref{sec:conclusions}.
The actual UV completions for $d>6$ are deferred to appendices in the interest of readability. App.~\ref{sec:d7} lists the UV completions and relevant Feynman-diagram topologies for $d=7$ BNV operators; App.~\ref{sec:d8} does the same for $d=8$, App.~\ref{sec:d9} for $d=9$, split into the different $(\Delta B,\Delta L)$ cases that arise at that order.
App.~\ref{sec:d10} is for $d=10$, App.~\ref{sec:d11} for $d=11$, App.~\ref{sec:d12} for $d=12$, App.~\ref{sec:d13} for $d=13$, App.~\ref{sec:d14} for $d=14$, and App.~\ref{sec:d15} for $d=15$.
We only provide a relatively small subset of UV completions in this article to keep the number of pages in the triple digits; the complete data tables of topologies and UV completions in multiple formats as well as the \texttt{Mathematica} notebooks used to create and manipulate them can be found online at \website. Our code is in no way specific to BNV operators and can be used to generate tree-level UV completions for any non-derivative operator at practically any mass dimension.

\section{Dimension 6: \texorpdfstring{$\Delta B=1, \Delta L=1$}{Delta B=1,Delta L=1}}
\label{sec:d6}

In the ($\nu$)SMEFT, operators that violate baryon number start to appear at operator mass dimension $d=6$~\cite{Weinberg:1979sa,Wilczek:1979hc}. These will serve as a useful example to explain our general procedure. Keeping the flavor structure, these operators all violate $\Delta B=\Delta L =1$ and can be written as
\begin{align}
\mathcal{L}_{d=6} &= 
\frac{1}{\Lambda^2}\sum_{j,a,b,c,d} C_{abcd}^j \mathcal{O}_{6\, abcd}^j+\text{h.c.} \,,
\label{eq:dequal6}
\end{align}
with basis operators
\begin{align}
\mathcal{O}_{6\,abcd}^1 &\equiv \epsilon^{\alpha\beta\gamma} \epsilon_{ij} (\overline{d}^C_{a,\alpha} u_{b,\beta})(\overline{Q}^C_{i,c,\gamma}L_{j,d}) \,, \\
\mathcal{O}_{6\,abcd}^2 &\equiv  \epsilon^{\alpha\beta\gamma} \epsilon_{ij} (\overline{Q}^C_{i,a,\alpha}Q_{j,b,\beta})(\overline{u}^C_{c,\gamma} e_{d})\,, \\
\mathcal{O}_{6\,abcd}^3 &\equiv  \epsilon^{\alpha\beta\gamma} \epsilon_{il}\epsilon_{jk} (\overline{Q}^C_{i,a,\alpha} Q_{j,b,\beta})(\overline{Q}^C_{k,c,\gamma}  L_{l,d}) \,,\\
\mathcal{O}_{6\,abcd}^4 &\equiv  \epsilon^{\alpha\beta\gamma} (\overline{d}^C_{a,\alpha} u_{b,\beta})(\overline{u}^C_{c,\gamma} e_{d}) \,,\\
\mathcal{O}_{6\,abcd}^5 &\equiv \epsilon^{\alpha\beta\gamma} \epsilon_{ij}(\overline{Q}^C_{i, a,\alpha} Q_{j,b,\beta})(\overline{d}^C_{c,\gamma} \nu_{d}) \,, \\
\mathcal{O}_{6\,abcd}^6 &\equiv \epsilon^{\alpha\beta\gamma}(\overline{d}^C_{a,\alpha} u_{b,\beta})(\overline{d}^C_{c,\gamma} \nu_{d}) \,,
\label{eq:dequal6basis}
\end{align}
where  $\alpha,\beta,\gamma$ denote the $SU(3)_C$ color, $i,j,k,l$ the $SU(2)_L$ isospin, and $a,b,c,d$ the family indices~\cite{Weinberg:1979sa,Wilczek:1979hc,Weinberg:1980bf,Weldon:1980gi,Abbott:1980zj,delAguila:2008ir,Beltran:2023ymm}.
$u$, $d$, $e$, and $\nu$ are the right-handed up-quark, down-quark, charged lepton, and neutrino fields, while $Q$ and $L$ are the left-handed quark and lepton doublets, respectively, see Tab.~\ref{tab:fields}.
The Wilson coefficients $C^j$ are dimensionless and the first-generation entries are constrained to be $<\Lambda^2/ (\unit[\mathcal{O}(10^{15})]{GeV})^{2}$ due to the induced two-body nucleon decays such as $p\to e^+\pi^0$ that have not been observed yet~\cite{Heeck:2019kgr}.

Since effective field theories by construction hide the details of the underlying interactions, infinitely many new physics models could generate the same effective operators. It then seems that even in an ideal world, where we experimentally pin down the coefficients of these operators, we cannot know their fundamental origin.
We can, however, appeal to Occam's razor and look for the \emph{simplest} realization of these operators, which in field-theory language corresponds to their tree-level generation.\footnote{One-loop realizations of the $d=6$ SMEFT BNV operators can be found in Ref.~\cite{Helo:2019yqp}, see also~\cite{Dorsner:2022twk,Gargalionis:2024nij}.}
It is a straightforward exercise to write down tree-level Feynman diagrams involving only renormalizable interactions that reduce to the above operators upon integrating out the -- by assumption heavy -- non-SM fields. At $d=6$, all tree-level realizations involve scalar or vector leptoquarks (LQs)~\cite{Buchmuller:1986zs,Dorsner:2016wpm,deBlas:2017xtg}; we collect their quantum numbers in Tab.~\ref{tab:LQs} in the naming scheme of Refs.~\cite{Buchmuller:1986zs,Dorsner:2016wpm}.

\begin{table}[tb]
\centering
{\renewcommand{\arraystretch}{1.3}
\begin{tabular}{c c c|| c c c} 
  \hline
  leptoquark & spin & representation & leptoquark & spin & representation\\
  \hline
  \hline
$\mathcal S_1$ ($S_2$) & 0 & ($\bar{\vec{3}},\vec{1},1/3$)  &   $\mathcal S_3$ ($S_4$) & 0 & ($\bar{\vec{3}},\vec{3},1/3$)\\
    \hline
$\widetilde{\mathcal S}_1$  ($S_1$) & 0 & ($\bar{\vec{3}},\vec{1},4/3$) & $\mathcal V_2$ ($V_1$) & 1 & ($\bar{\vec{3}},\vec{2},5/6$)\\
    \hline
 $\bar{\mathcal S}_1$ ($S_3$) & 0 &  ($\bar{\vec{3}},\vec{1},-2/3$)  &  $\Tilde{\mathcal V}_2$ ($V_2$) & 1& ($\bar{\vec{3}},\vec{2},-1/6$) \\
    \hline
\end{tabular}
}
\caption{Scalar and vector leptoquark representations under $SU(3)_C\times SU(2)_L\times U(1)_Y$ that generate the $d=6$ $\Delta B = 1$ operators, in the notation of Refs.~\cite{Buchmuller:1986zs,Dorsner:2016wpm} in calligraphic font and in our notation as $S_j$ and $V_j$.}
\label{tab:LQs}
\end{table}

Integrating out these LQs individually generates, to leading order in $1/m_{\text{LQ}}^2$, linear combinations of the six basis operators of Eq.~\eqref{eq:dequal6}; explicitly:
\begin{align}
\L_{\mathcal{S}_1} &\supset - m_{\mathcal{S}_1}^2 |\mathcal{S}_1|^2 + \left(
    y^{LL}_{1\, ab} \bar{Q}^C_{i, a} \mathcal{S}_1 \epsilon^{ij}  L_{j,b} +
  y^{RR}_{1\, ab} \bar{u} ^C_{a} \mathcal{S}_1 e_{b} +
  y^{\overline{RR}}_{1\, ab} \bar{d} ^C_{a} \mathcal{S}_1 \nu_{b} \right. \nonumber\\
  &\left. ~~+
  z^{LL}_{1\, ab} \bar{Q} ^C_{ i, a} \mathcal{S}_1^{*} \epsilon ^{ij}  Q_{j, b} +
  z^{RR}_{1\, ab} \bar{u} ^C_{a} \mathcal{S}_1^{*} d_{b} 
    + \hc\right) \\
    &\to \frac{y^{LL}_{1\, cd}\ z^{RR}_{1\, ba}}{m_{\mathcal{S}_1}^2} \mathcal{O}^1_{6\,abcd} - \frac{y^{RR}_{1\, cd}\ z^{LL}_{1\, ab}}{m_{\mathcal{S}_1}^2} \mathcal{O}^2_{6\,abcd} - \frac{y^{LL}_{1\, cd}\ z^{LL}_{1\, ab}}{m_{\mathcal{S}_1}^2} \mathcal{O}^3_{6\,abcd}+ \frac{y^{RR}_{1\, cd}\ z^{RR}_{1\, ba}}{m_{\mathcal{S}_1}^2} \mathcal{O}^4_{6\,abcd}  \nonumber \\ &
    \qquad - \frac{y^{\overline{RR}}_{1\, cd}\ z^{LL}_{1\, ab}}{m_{\mathcal{S}_1}^2} \mathcal{O}^5_{6\,abcd} + \frac{y^{\overline{RR}}_{1\, cd}\ z^{RR}_{1\, ba}}{m_{\mathcal{S}_1}^2} \mathcal{O}^6_{6\,abcd} + \hc\,, \\
\L_{\tilde{\mathcal{S}}_1} &\supset - m_{\tilde{\mathcal{S}}_1}^2 |\tilde{\mathcal{S}}_1|^2 + \left(
    \tilde{y}^{RR}_{1\, ab}\, \bar{d}^C_{a} \tilde{\mathcal{S}}_1 e_{b}
    + \tilde{z}^{RR}_{1\, ab}\, \bar{u}^C_{a} \tilde{\mathcal{S}}_1^* u_{b}
    + \hc\right) \\
    &\to \frac{2\ \tilde{y}^{RR}_{1\, ad}\ \tilde{z}^{RR}_{1\, bc}}{m_{\tilde{\mathcal{S}}_1}^2}\mathcal{O}^4_{6\,abcd} + \hc\,, \\
\L_{\bar{\mathcal{S}}_1} &\supset - m_{\bar{\mathcal{S}}_1}^2 |\bar{\mathcal{S}}_1|^2 + \left(
    \bar{y} ^{\overline{RR}}_{1\, ab} \bar{u}^C_{a} \bar{\mathcal{S}}_1 \nu_{b} +
  \bar{z} ^{RR}_{1\, ab} \bar{d}^C_{a} \bar{\mathcal{S}}_1^{* } d_{b} + \hc\right)\\
    &\to -\frac{2\ \bar{y} ^{\overline{RR}}_{1\, bd}\ \bar{z} ^{RR}_{1\, ac}}{m^{2}_{\bar{\mathcal{S}}_1}} \mathcal{O}^6_{6\,abcd}  + \hc\,, \\
\L_{\mathcal{S}_3} &\supset - m_{\mathcal{S}_3}^2 |\mathcal{S}_3|^2 + \left(
    \tilde{y}^{LL}_{3\, ab}\, \bar{Q}^C_{i, a}  \epsilon^{ij} (\tau^p \mathcal{S}_3^p)^{jk} L_{k, b}
    + \tilde{z}^{LL}_{3\, ab}\, \bar{Q}^C_{i, a}  \epsilon^{ij} ((\tau^p \mathcal{S}_3^p)^\dagger) ^{jk} Q_{k, b}
    + \hc\right)\\
    &\to \frac{\tilde{y}^{LL}_{3\, cd}\ \tilde{z}^{LL}_{3\, ab}}{ m_{\mathcal{S}_3}^2}\mathcal{O}^3_{6\,abcd} + \hc \,,\\
\L_{\mathcal{V}_2} &\supset - m_{\mathcal{V}_2}^2 |\mathcal{V}_2|^2 + \left(
    x ^{RL}_{2\, ab} \bar{d}^C_{i, a} \gamma ^{\mu} \epsilon ^{ij}  \mathcal{V}_{2, \mu} L_{j, b} +
  x ^{LR}_{2\, ab} \bar{Q}^C_{i, a} \gamma ^{\mu} \epsilon ^{ij}  \mathcal{V}_{2, \mu, j} e_{b} \right.\nonumber\\
  &\left.~~+ \omega ^{LR}_{2\, ab} \bar{Q} ^C_{i, a} \gamma ^{\mu} \mathcal{V}_{2, \mu, i}^{*} u_{b} + \hc\right)\\
    &\to \frac{2x^{RL}_{2\, ad}\ \omega^{LR}_{2\, cb}}{m^{2}_{\mathcal{V}_2}} \mathcal{O}^1_{6\,abcd}  + \frac{2\ x^{LR}_{2\, ad}\ \omega^{LR}_{2\, bc}}{m^{2}_{\mathcal{V}_2}} \mathcal{O}^2_{6\,abcd} +\hc\,, \\
\L_{\tilde{\mathcal{V}}_2} &\supset - m_{\tilde{\mathcal{V}}_2}^2 |\tilde{\mathcal{V}}_2|^2 + \left(
    \tilde{x} ^{RL}_{2\, ab} \bar{u}^C_{a} \gamma ^{\mu} \tilde{\mathcal{V}}_{2, \mu, i} \epsilon ^{ij}  L _{j, b} +
  \tilde{x} ^{\overline{LR}}_{2\, ab} \bar{Q}^C_{i, a} \gamma ^{\mu} \epsilon ^{ij} \tilde{\mathcal{V}}_{2, \mu, j} \nu_{b} \right.\nonumber\\
  &\left. ~~ + \tilde{\omega}^{RL}_{2\, ab}\ \bar{d} ^C_{a} \gamma ^{\mu} \tilde{\mathcal{V}}_{2, \mu, i}^{*} Q_{i, b} + \hc\right)\\
    &\to -\frac{2\ \tilde{x} ^{RL}_{2\, bd}\  \tilde{\omega} ^{RL}_{2\, ac}}{m^{2}_{\tilde{\mathcal{V}}_2}} \mathcal{O}^1_{6\,abcd}  - \frac{2\ \tilde{x}^{\overline{LR}}_{2\, bd}\  \tilde{\omega} ^{RL}_{2\, ca}}{m^{2}_{\tilde{\mathcal{V}}_2}} \mathcal{O}^5_{6\,abcd} +\hc 
\end{align}
We are only interested in the $\Delta B =1$ effects, the $\Delta B = 0$ operators for all cases can be found in Ref.~\cite{Dorsner:2016wpm}.
Here and below, we allow for vector bosons in our UV completions, even though it is far from obvious that their interactions are renormalizable.
In the example above, the vector LQs $\mathcal{V}_{2}$ and $\tilde{\mathcal{V}}_{2}$ can indeed be realized as gauge bosons of a renormalizable theory, namely the grand unified theories of $SU(5)$~\cite{Georgi:1974sy} and $SO(10)$~\cite{Fritzsch:1974nn}, respectively.\footnote{For many other vector bosons, the corresponding gauge groups can found in Ref.~\cite{Fonseca:2016jbm}.} However, their simple gauge origin forbids the arbitrary Yukawa couplings assumed here, illustrating the restrictions that arise for UV completions involving vector bosons. Alternatively, one can view the vector bosons themselves as effective and imagine them as bound states of elementary particles, which allows more freedom in their couplings at the prize of not having a full UV completion. 
We shall remain agnostic about the origin of vector bosons and include them mainly to enable comparisons with other studies in the literature.

UV-completing the $d=6$ BNV operators reveals some interesting aspects that are impossible to see by looking at $\nu$SMEFT operators:
\begin{enumerate}
\item There are six different LQs generating the six $d=6$ operators, allowing us to realize arbitrary Wilson coefficients $C^j$, including the often studied case of single-operator dominance. In some cases, this is easy, e.g.~$\mathcal{S}_3$ only generates $\mathcal{O}^3_6$. Exclusively generating $\mathcal{O}^1_6$, on the other hand, requires \emph{all} LQs and a careful cancellations of all other operators. Without \emph{vector} LQs -- a reasonable restriction given their own difficult renormalization -- the operators $\mathcal{O}^1_6$, $\mathcal{O}^2_6$ and $\mathcal{O}^5_6$ \emph{inevitably} arise together, even if we use the other scalar leptoquarks to cancel $\mathcal{O}^{3,4,6}_6$.
\item The UV completions provide an alternative operator basis. Rather than the typically considered single-operator dominance, we can consider the more physically-motivated single-\emph{particle} dominance, which can lead to novel predictions. For example, introducing only $\mathcal{S}_1$ generates \emph{all} six $d=6$ operators, but since there are only five independent Yukawa couplings we end up with a restricted set of Wilson coefficients: $C^1_{abcd} C^2_{abcd} = C^3_{abcd} C^4_{abcd}$ and $C^3_{abcd} C^6_{abcd} = C^1_{abcd} C^5_{abcd}$. Such relations would be difficult to motivate within the $\nu$SMEFT but are easy to establish given the UV completions.
Similarly, extending the SM only by $\tilde{\mathcal{S}}_1$ generates the single operator $\mathcal{O}^4_6$, seemingly bringing us back to single-operator dominance. However, the $\tilde{\mathcal{S}}_1 u u$ coupling is \emph{antisymmetric} in flavor space and thus enforces BNV operators involving charm~\cite{Dorsner:2012nq} or top quarks~\cite{Dong:2011rh}. Once again, the $\nu$SMEFT would not have led us to this interesting scenario.
\end{enumerate}

\begin{figure}
\centering
 \begin{subfigure}[]{0.4\textwidth}
 \centering
    \includegraphics[scale=1.0]{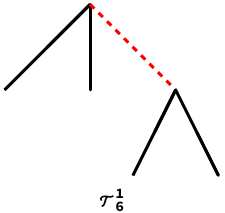}
      \end{subfigure}
       \caption{Feynman-diagram topology for the $d=6$ operators from Eq.~\eqref{eq:dequal6basis_simple}, involving four external SM fermions (solid black lines) and one new heavy boson (dashed red line).}
    \label{fig:d=6_topology}
\end{figure}

Let the above suffice as motivation to find UV completions for effective BNV operators. 
Explicitly integrating out the new particles and matching them to a known minimal operator basis is easy in this case, but cumbersome for $d\gg 6$ operators, both due to the large number of new particles, diagrams, and operators, and because convenient compact basis operators have been constructed only up to $d=8$~\cite{Murphy:2020rsh}.
Because of this, we will switch to a simpler notation that suppresses all Lorentz and gauge group contractions and only keeps the particles. In Ref.~\cite{Gargalionis:2024nij}, these were denoted as \emph{field strings}, similar to \emph{operator types} in Ref.~\cite{Fonseca:2019yya}. For $d=6$, they read
\begin{align}
    \mathcal{O}_{6}^1 \equiv d u Q L, &&   
    \mathcal{O}_{6}^2 \equiv  QQ u e,  &&  
    \mathcal{O}_{6}^3 \equiv  QQQL, && 
    \mathcal{O}_{6}^4 \equiv  duue,  && 
    \mathcal{O}_{6}^5 \equiv QQd\nu , &&
    \mathcal{O}_{6}^6 \equiv dud\nu \,,
    \label{eq:dequal6basis_simple}
\end{align}
with no particular ordering of the fields inside each operator, although we always list right-handed neutrinos $\nu$ at the end to make a restriction to the SMEFT easier. 
The existence of at least one total singlet contraction is fairly easy to ascertain using group theory codes such as \texttt{GroupMath}~\cite{Fonseca:2020vke} or \texttt{Sym2Int}~\cite{Fonseca:2017lem}, whereas the explicit construction of a minimal number of effective terms is more difficult for $d\gg 6$ and will be left for future work.
Since all these $d=6$ operators involve four fermions, the tree-level UV completions can be described through the simple Feynman-diagram topology shown in Fig.~\ref{fig:d=6_topology}, where the red internal line describes  a heavy boson. For each operator, we distribute the fields over the external lines in every possible permutation and then read off the quantum numbers of the internal line using \texttt{GroupMath}'s products of representations~\cite{Fonseca:2020vke}, assuming for concreteness that all external particle momenta are incoming. 
The results are given in table~\ref{tab:T6Top1B1L1}, which simply lists which new particles can generate each operator; notice that we add extra subscripts indicating $(\Delta B,\Delta L)$ for each operator for clarity: $\mathcal{O}_{d,(\Delta B,\Delta L)}^j$. The actual quantum numbers of the new particles in our naming scheme are collected in Tabs.~\ref{tab:newScalars},~\ref{tab:newFermions}, and~\ref{tab:newVectors}.  
For $d=6$, this, of course, matches the leptoquark representations listed above and agrees with results in the literature~\cite{Bischer:2019ttk,deBlas:2017xtg,Beltran:2023ymm}.

\input{Tables_dim6/Table_dim6_Top1_B1_L1.tex}

Having obtained the new-particle representations in this simplified approach, we could then study the Lagrangian, integrate out the heavy particles, and obtain \emph{actual} operators with well-defined group contractions. This is a fairly straightforward task, with many codes and packages on the market that could be used in principle~\cite{Fuentes-Martin:2022jrf,terHoeve:2023pvs,Harlander:2023ozs}. Another helpful step here is the definition of a complete irreducible basis of operators at a given mass dimension onto which the so-generated operators can be matched.  
In the interest of article length, and since a compact basis of operators has only been constructed up to $d=8$, we will not perform this step and rather restrict ourselves to the more difficult part of finding the UV-completing particles.

\section{General procedure}
\label{sec:general_procedure}

The $d=6$ example above illustrates the general procedure we will use below to obtain the tree-level UV completions for non-derivative $\Delta B$ operators up to high mass dimension:\footnote{Similar UV-completion algorithms can be found in the literature~\cite{Gargalionis:2020xvt} (even at loop level~\cite{Cepedello:2022pyx}).}
\begin{enumerate}
\item We obtain simplified effective $\Delta B$ operators including right-handed neutrinos up to $d=15$ using \texttt{Sym2Int}~\cite{Fonseca:2017lem}, which is sufficiently fast even on a personal computer. Whenever possible, we compare the output to the literature. The number of $\Delta B$ operators grows exponentially with mass dimension $d$~\cite{Heeck:2025btc}, see Fig.~\ref{fig:Op VS d}. The right-handed neutrinos lead to a significant increase in operators at large $d$. For $d=6$, this just gives the $\Delta B = \Delta L = 1$ operators of Eq.~\eqref{eq:dequal6basis_simple}; in general, we split operators into groups of fixed $\Delta B$ and $\Delta L$.

\begin{figure}[tb]
    \centering
    \includegraphics[width=0.85\textwidth]{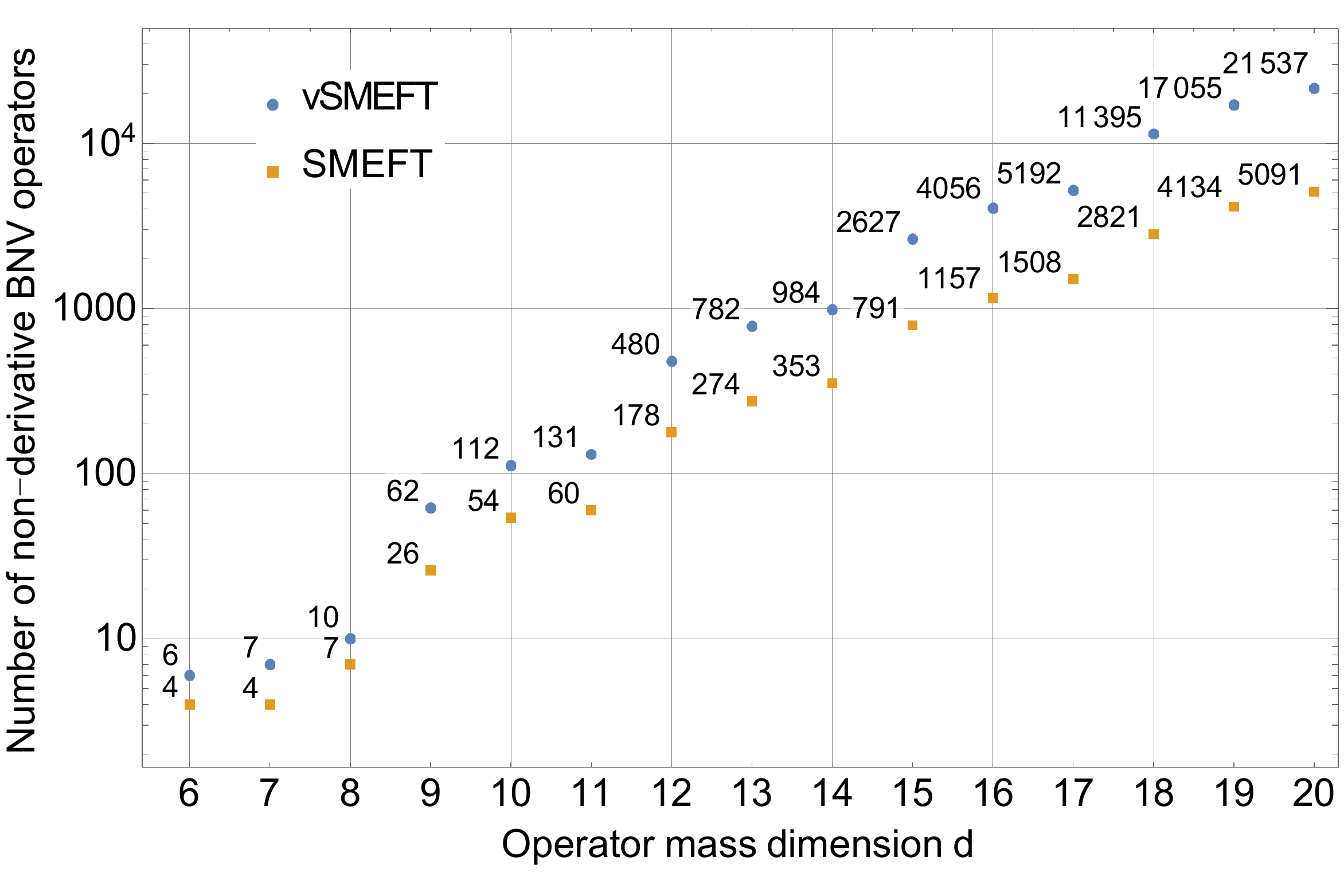}
    \caption{Number of non-derivative $\Delta B$ operators per mass dimension $d$. Circular blue dots correspond to $\nu$SMEFT operators and square orange dots to SMEFT operators, i.e.~operators without right-handed neutrinos. See Ref.~\cite{Heeck:2025btc} for a more detailed graph.
    }
    \label{fig:Op VS d}
\end{figure}

\item Next, we require tree-level Feynman diagrams with the same number of external fermions and bosons as the operators. Since we go through all permutations of how the operator-fields are distributed over the diagram, we only need to keep the topologically different diagrams, i.e.~we do not need to distinguish $s$, $t$, and $u$ channels in the $d=6$ case but just need the one diagram of Fig.~\ref{fig:d=6_topology}. For up to about 10 external particles, \texttt{FeynArts}~\cite{Kublbeck:1990xc} is sufficiently fast to accomplish the task of generating all tree-level diagrams with a fixed number of external fermions and scalars; these results are then translated into graphs with colored edges and topologically isomorphic graphs are eliminated using the \texttt{Mathematica} package \texttt{IGraph/M}~\cite{Horv_t_2023}, specifically the command \\
\centerline{{\tt IGVF2IsomorphicQ[graph1, graph2] }}
\\
that determines if \texttt{graph1} and \texttt{graph2} are isomorphic. 
\texttt{FeynArts}' algorithm, which generates all possible permutations of external lines, becomes too slow and memory-intensive for more than 10 external particles, so we devised our own \texttt{Mathematica} code using \texttt{IGraph/M}: we generate nested lists of graphs by iteratively adding vertices and edges to the \textit{initial} graphs, which just consist of a cubic\\
\phantom{xx}{ \small{\texttt{Graph[\{UndirectedEdge[2, 1], UndirectedEdge[3, 1], UndirectedEdge[4, 1]\}]}}} \\
  or quartic\\
\phantom{xx}{\small{\texttt{Graph[\{\allowbreak UndirectedEdge[2, 1], \allowbreak UndirectedEdge[3, 1], \allowbreak UndirectedEdge[4, 1], \\ \allowbreak \phantom{xxxxxxxx} UndirectedEdge[5, 1]\}]}}}
\\
 vertex. 
These bare base topologies are then fed into \texttt{FeynArts}, which threads the fermion and scalar lines through it based on a custom \texttt{FeynArts} model file that defines allowed 3- and 4-point interaction vertices between the particles. 
Each graph is then outfitted with an edge style (colored depending on incoming particle or internal propagator) and labels (boson or fermion). Duplicate and topologically equivalent graphs are removed using \texttt{IGraph/M}. Finally, we apply a sorting algorithm to label the incoming particles and propagators such that two known fields in the list of three or four particles always appear first. 

One example of a $d=9$ topology with four external fermions and three external scalars is shown below to illustrate the sorting procedure and the format used in our topology files:\\
\includegraphics[width=0.93\textwidth,trim={1.2cm 0 0 0},clip]{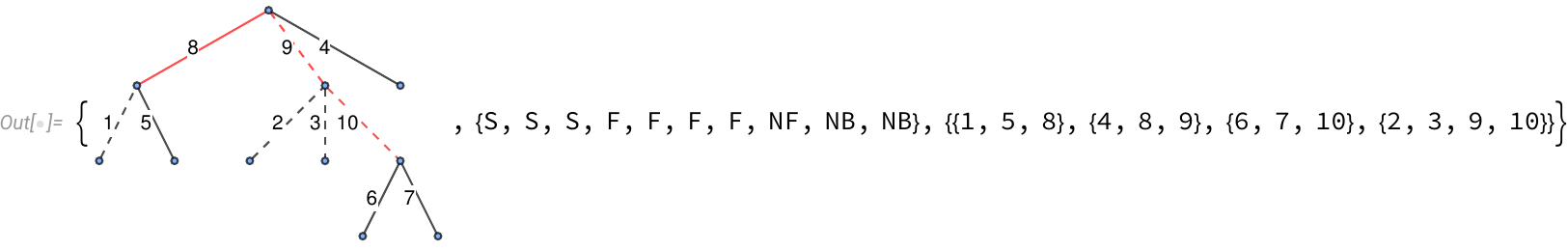}\\
The first entry is the actual graph object, with red lines corresponding to heavy new particles, and each line is numbered. Dashed lines denote bosons and solid lines fermions. The second entry simply collects which particles are scalars (S), fermions (F), internal New Bosons (NB), and internal New Fermions (NF), with positions in this list corresponding to the numbers in the graph. The last entry lists the vertices, i.e.~which particles meet at which vertex. 
This topology format is redundant in that the graph itself is equivalent to the second plus third entries, but proves convenient. 
The particle numbers are sorted so that the quantum numbers of the new particles can be obtained by sequentially going through the vertex list: particles 1 and 5 are known, since they are external particles, so the $\{1,5,8\}$ vertex can be used to calculate the quantum numbers of particle 8 by requiring the vertex to be a total singlet. The second vertex, $\{4,8,9\}$, now uses the known quantum numbers of particles 4 and 8 to determine 9, the third vertex determines particle 10, and the last vertex is the final cross check and always needs to be a complete singlet. This numbering/ordering, albeit in no way unique, is highly convenient for a fully automated calculation such as ours since it turns the potentially complicated set of equations that the new-particle quantum numbers need to fulfill into a straightforward sequential task.

\begin{figure}[tb]
    \centering
    \includegraphics[width=0.8\textwidth]{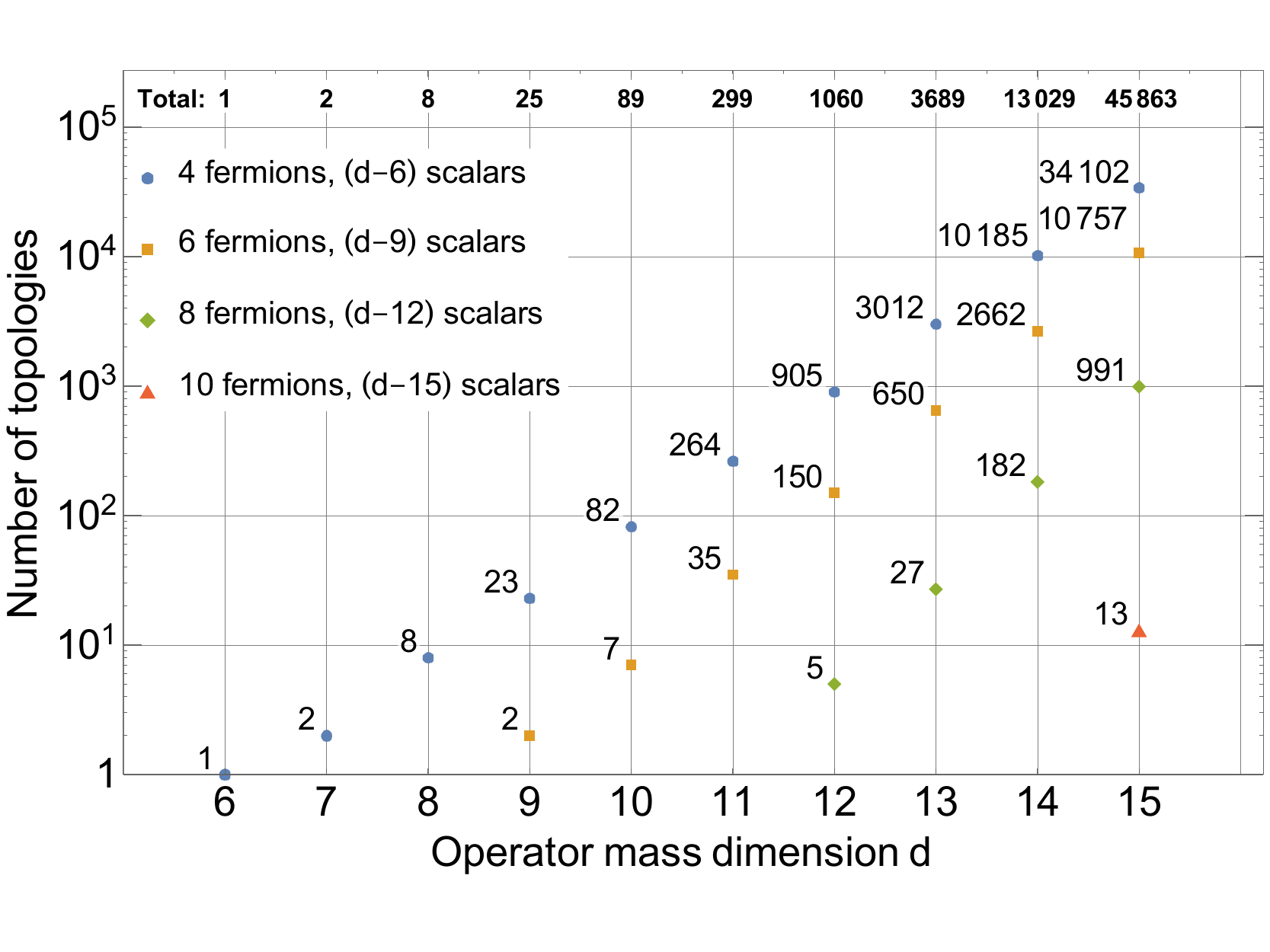}
    \caption{The number of distinct topologies categorized by the associated operator mass dimension~$d$ and the number of external fermions and scalars. The numbers at the very top give the \emph{total} number of topologies for each mass dimension.}
    \label{fig:Top VS d}
\end{figure}

The number of base topologies grows exponentially with the number of propagators or mass dimension,\footnote{The number of tree graphs with just cubic or just quartic vertices are listed as A000672 and A000602 in the On-Line Encyclopedia of Integer Sequences~\cite{oeis}, respectively.} as illustrated in Fig.~\ref{fig:Top VS d}. 
We ran the code on a high-performance computing cluster to achieve fast results. 
Notice that these topology files only need to be generated once and can then be used for any operator with the appropriate number of external fermions and bosons. We make our topology files -- including many topologies not used in this study --  available at \website.

\item 
For each effective operator at dimension $d$, we take a topology with the correct number of external fermions and bosons, distribute the operator fields over the external graph lines, and then obtain the quantum numbers of the internal fields by successively multiplying group representations using \texttt{GroupMath}. The quantum numbers correspond to the particle representation under the SM gauge group $SU(3)\times SU(2)\times U(1)$, as well as the global Lorentz group $SU(2)\times SU(2)$. Including the Lorentz group allows us to determine the spin and chirality of the internal particles. Notice that internal particles are massive by assumption so we take all internal fermions to be vector-like or Dirac, and we ensure that each fermion line comes with a mass flip to eliminate derivative operators, see Sec.~\ref{sec:derivative_operators}.
By successively multiplying particle representations and going through the vertex list of our topology we eventually end up with a small number of viable internal quantum numbers, i.e.~UV completions of the underlying operator.
We then distribute the operator fields over the topology in all possible permutations and repeat the procedure, leading to all possible UV completions of that operator using that topology. Some of these UV completions can be equivalent, however, due to symmetries in the topology; to eliminate equivalent cases, we turn our Feynman diagrams into directed graphs that have arrows indicating the charge flow, flip arrows as needed so that all internal particles are indeed particles rather than antiparticles, as defined below, then convert the particle labels to colors and let \texttt{IGraph/M} filter out isomorphic colored directed graphs. 
We repeat this procedure for all viable topologies, which then gives us all tree-level UV completions of this operator. 

We follow this algorithm for all operators and collect the UV completions for these operator--topology pairs. For infinite generations of fermions and scalars, this would already be the end of the story, but the finite number of SM fields imposes additional restrictions explained below.

\item 
The finite number of SM fermions and scalars restricts the number of non-vanishing operators and topologies; \texttt{Sym2Int} automatically takes this into account, eliminating  operators that vanish for three generations such as $\nu^8$~\cite{Heeck:2025btc}. UV completions for the surviving operators can nevertheless end up with vanishing contractions on account of exchange symmetries. \\
Most importantly, care must be taken for vertices involving two or three Higgs doublets $H$. Whenever $HH$ ($HHH$) is contracted to an $SU(2)_L$ singlet (doublet), the vertex vanishes on account of being antisymmetric. (For similar fermion vertices, the generation indices ensure non-vanishing terms.)
Our code automatically identifies such vertices and eliminates the corresponding UV completion. For $d \leq 12$ operators, we cross-checked this procedure by explicitly constructing the Lagrangian involving all fields in the diagram using \texttt{Sym2Int}, which is computationally slow but confirmed the results of our automated method.\\
The above procedure leads us to Feynman diagrams populated with particles in which each vertex is a non-vanishing gauge and Lorentz singlet. Alas, this alone does not yet guarantee a UV completion of a particular operator. The non-derivative operators under investigation here are obtained by dropping the momentum dependence of the internal propagators, effectively turning them into constants.  This leading term in the heavy-mass expansion is of course a total singlet, but might end up \textit{vanishing}, unlike the total amplitude including the internal momenta. 
For example, take the UV completion of the $\mathcal{O}_{8, \,(1 , 1)}^{1} \equiv HHddQL$ operator shown in Fig.~\ref{fig:Higgs_antisym}; every vertex and thus the entire diagram is a total gauge and Lorentz singlet, inducing $B-L$ conserving nucleon $\to$ meson $+$ anti-lepton processes.
However, in this example, $QL$ clearly forms an $SU(2)_L$ singlet, so $HH$ is also forced to contract into an $SU(2)_L$ singlet. Such a contraction is antisymmetric and, therefore, vanishes in the single-Higgs SM setup.
Expanding the $S_5$ propagator to $p^2/m_{S_5}^4$ order instead gives a non-vanishing contribution, since $H D^2 H$ can be a non-vanishing $SU(2)_L$ singlet, which overall corresponds to the $d=10$ derivative operator $D^2 HHddQL$. This point was already noted in Ref.~\cite{Gargalionis:2020xvt} for $\Delta L=2$ operators and will be discussed again in Sec.~\ref{sec:derivative_operators}. 
To weed out such cases, we need to ensure that UV completions of operators involving multiple $H$ are not antisymmetric under $H\leftrightarrow H$ exchanges, and the same for $\bar{H}$. To implement this, we explicitly calculate the vertex invariants in each diagram using \texttt{GroupMath}, treat propagators as constants and contract the internal multiplet components, giving us the amplitude for this diagram to lowest order in the heavy-mass expansion. Then we check if the resulting expression is antisymmetric under any Higgs exchanges. While reasonably fast for a given diagram, this extra filtering of UV completions significantly increases the computational time for higher mass dimensions.

\begin{figure}[tb]
    \centering
    \includegraphics[scale=0.8]{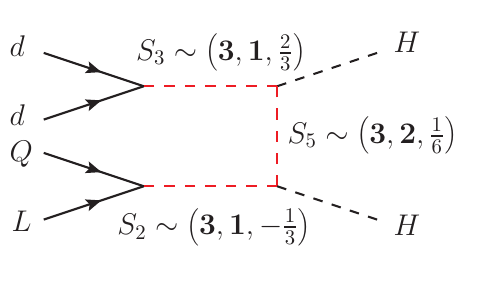}
  \caption{Feynman diagram that seemingly UV-completes the operator $\mathcal{O}_{8, \,(1 , 1)}^{1} \equiv HHddQL$, with all momenta incoming, but actually generates the $d=10$ derivative operator $HddD^2( HQL)$ at leading order in large $S_j$ masses.}
    \label{fig:Higgs_antisym}
\end{figure}

The same effect can occur for the \textit{fermions} inside an operator, as they can also have exchange symmetries that are not respected by a given constant-propagator diagram; the multitude of fermion generations only makes this relevant at much higher $d$ though and is considerably more tedious to check for. We only need to check operators with at least four identical fermions that contain a fully-antisymmetric singlet contraction. For example, the $d=9$ operator $uu dddd$ for $n$ generations contains a contraction that is fully antisymmetric under $d$ exchange according to \texttt{Sym2Int}; for $n=3$ generations, this contraction is then trivially zero. Overall, there are 363 operators of this type in our dataset, and all of their UV completions need to be checked for an antisymmetric amplitude that would vanish for three generations. Similar to the Higgs case, this filtering is computationally expensive due to the large number of cases to check. Ultimately, none of the UV completions induce exclusively fully antisymmetric amplitudes. In some cases, the amplitude can be decomposed into terms that \textit{contain} an antisymmetric piece, and sometimes a UV completion gives an antisymmetric amplitude but \textit{also} generates a non-vanishing amplitude using different couplings; either way, at least within our dataset, three vs. infinite fermion generations do not change the types of UV completions.

Finally, we assume all \textit{internal} particles to be distinguishable, even if they share the same quantum numbers. In many cases, one copy of the same field could well suffice to UV-complete an operator, but in other cases antisymmetric vertices require more than one copy. To avoid checking the symmetry of all internal vertices, we simply assume a sufficient number of new-particle generations to end up with non-vanishing vertices.

\end{enumerate}

Once the desired list of operators and topologies is generated, the UV-completion procedure itself is relatively fast. However, exponential growth in both the number of operators and the corresponding topologies quickly results in extremely large data sets.  The most computationally intensive cases are operators with a large number of Higgs fields. 
Note that our code is not specific to $\Delta B$ operators and can therefore be used to find tree-level UV completions for \emph{any} non-derivative effective operator. We provide all necessary \texttt{Mathematica} notebooks and data files at \website.

The quantum numbers of the new particles are saved in \texttt{Mathematica} notebooks together with their corresponding Feynman diagrams.
We extract the quantum numbers of all scalars, fermions, and vectors, order them by representation size, and name them consecutively. We assign names so that all particles have an un-barred $SU(3)_C$ representation (in the naming convention of \texttt{GroupMath}); for real $SU(3)_C$ representations, we define particles to have positive hypercharge. Our particle definitions are given in Tab.~\ref{tab:newScalars} for scalars, Tab.~\ref{tab:newFermions} for fermions, and Tab.~\ref{tab:newVectors} for vectors. There, we also indicate the mass dimension where they first appear in our UV completions.

\begingroup
\renewcommand{\arraystretch}{1.2}
\newcommand{\br}{,\linebreak[1]}
\newcolumntype{C}{>{\centering\arraybackslash\setstretch{1.25}}X}

\begin{table}[h!]
\small
\centering
\begin{tabularx}{\textwidth}{|c|C|}
\hline
      $d$   & {\rm Scalars}   \\
         \hline \hline

    6& $S_{1}=(\bold{3},\bold{1},-\frac{4}{3})$, $S_{2}=(\bold{3},\bold{1},-\frac{1}{3})$,
    $S_{3}=(\bold{3},\bold{1},\frac{2}{3})$, $S_{4}=(\bold{3},\bold{3},-\frac{1}{3})$.\\
    \hline
    7& $S_{5}=(\bold{3},\bold{2},\frac{1}{6})$, $S_{6}=(\bold{3},\bold{2},\frac{7}{6})$.\\
    \hline
    8& $S_{7}=(\bold{1},\bold{1},0)$, $S_{8}=(\bold{1},\bold{3},0)$, $S_{9}=(\bold{1},\bold{3},1)$,
    $S_{10}=(\bold{3},\bold{2},-\frac{11}{6})$, $S_{11}=(\bold{3},\bold{2},-\frac{5}{6})$,
    $S_{12}=(\bold{3},\bold{4},-\frac{5}{6})$, $S_{13}=(\bold{3},\bold{4},\frac{1}{6})$.\\
    \hline
    9& $S_{14}=(\bold{1},\bold{1},1)$, $S_{15}=(\bold{1},\bold{1},2)$,
    $S_{16}=(\bold{1},\bold{2},\frac{1}{2})$, $S_{17}=(\bold{1},\bold{4},\frac{1}{2})$,
    $S_{18}=(\bold{1},\bold{4},\frac{3}{2})$, $S_{19}=(\bold{3},\bold{3},\frac{2}{3})$,
    $S_{20}=(\bold{3},\bold{3},\frac{5}{3})$, $S_{21}=(\bold{6},\bold{1},-\frac{4}{3})$,
    $S_{22}=(\bold{6},\bold{1},-\frac{1}{3})$, $S_{23}=(\bold{6},\bold{1},\frac{2}{3})$,
    $S_{24}=(\bold{6},\bold{3},-\frac{1}{3})$, $S_{25}=(\bold{8},\bold{2},\frac{1}{2})$.\\
    \hline
    10& $S_{26}=(\bold{1},\bold{2},\frac{3}{2})$, $S_{27}=(\bold{1},\bold{2},\frac{5}{2})$,
    $S_{28}=(\bold{1},\bold{5},0)$, $S_{29}=(\bold{3},\bold{1},\frac{5}{3})$,
    $S_{30}=(\bold{3},\bold{3},-\frac{7}{3})$, $S_{31}=(\bold{3},\bold{3},-\frac{4}{3})$,
    $S_{32}=(\bold{3},\bold{5},-\frac{4}{3})$, $S_{33}=(\bold{3},\bold{5},-\frac{1}{3})$,
    $S_{34}=(\bold{3},\bold{5},\frac{2}{3})$, $S_{35}=(\bold{6},\bold{2},-\frac{11}{6})$,
    $S_{36}=(\bold{6},\bold{2},-\frac{5}{6})$, $S_{37}=(\bold{6},\bold{2},\frac{1}{6})$,
    $S_{38}=(\bold{6},\bold{2},\frac{7}{6})$, $S_{39}=(\bold{6},\bold{4},-\frac{5}{6})$,
    $S_{40}=(\bold{6},\bold{4},\frac{1}{6})$, $S_{41}=(\bold{8},\bold{1},0)$,
    $S_{42}=(\bold{8},\bold{1},1)$, $S_{43}=(\bold{8},\bold{3},0)$, $S_{44}=(\bold{8},\bold{3},1)$.\\
    \hline
    11& $S_{45}=(\bold{1},\bold{5},1)$, $S_{46}=(\bold{1},\bold{5},2)$,
    $S_{47}=(\bold{3},\bold{4},\frac{7}{6})$, $S_{48}=(\bold{3},\bold{4},\frac{13}{6})$,
    $S_{49}=(\bold{6},\bold{3},\frac{2}{3})$, $S_{50}=(\bold{6},\bold{3},\frac{5}{3})$,
    $S_{51}=(\bold{6},\bold{5},\frac{2}{3})$, $S_{52}=(\bold{8},\bold{2},\frac{3}{2})$,
    $S_{53}=(\bold{8},\bold{4},\frac{1}{2})$, $S_{54}=(\bold{8},\bold{4},\frac{3}{2})$.\\
    \hline
    12& $S_{55}=(\bold{1},\bold{3},2)$, $S_{56}=(\bold{1},\bold{3},3)$,
    $S_{57}=(\bold{1},\bold{6},\frac{1}{2})$, $S_{58}=(\bold{3},\bold{1},-\frac{10}{3})$,
    $S_{59}=(\bold{3},\bold{1},-\frac{7}{3})$, $S_{60}=(\bold{3},\bold{1},\frac{8}{3})$,
    $S_{61}=(\bold{3},\bold{2},\frac{13}{6})$, $S_{62}=(\bold{3},\bold{4},-\frac{17}{6})$,
    $S_{63}=(\bold{3},\bold{4},-\frac{11}{6})$, $S_{64}=(\bold{3},\bold{6},-\frac{11}{6})$,
    $S_{65}=(\bold{3},\bold{6},-\frac{5}{6})$, $S_{66}=(\bold{3},\bold{6},\frac{1}{6})$,
    $S_{67}=(\bold{3},\bold{6},\frac{7}{6})$, $S_{68}=(\bold{6},\bold{1},\frac{5}{3})$,
    $S_{69}=(\bold{6},\bold{1},\frac{8}{3})$, $S_{70}=(\bold{6},\bold{3},-\frac{7}{3})$,
    $S_{71}=(\bold{6},\bold{3},-\frac{4}{3})$, $S_{72}=(\bold{6},\bold{5},-\frac{4}{3})$,
    $S_{73}=(\bold{6},\bold{5},-\frac{1}{3})$, $S_{74}=(\bold{8},\bold{1},2)$,
    $S_{75}=(\bold{8},\bold{2},\frac{5}{2})$, $S_{76}=(\bold{8},\bold{5},0)$,
    $S_{77}=(\bold{10},\bold{1},-2)$, $S_{78}=(\bold{10},\bold{1},-1)$,
    $S_{79}=(\bold{10},\bold{1},0)$, $S_{80}=(\bold{10},\bold{1},1)$,
    $S_{81}=(\bold{10},\bold{1},2)$, $S_{82}=(\bold{10},\bold{3},-1)$,
    $S_{83}=(\bold{10},\bold{3},0)$, $S_{84}=(\bold{10},\bold{3},1)$,
    $S_{85}=(\bold{10},\bold{5},0)$, $S_{86}=(\bold{15},\bold{1},-\frac{4}{3})$,
    $S_{87}=(\bold{15},\bold{1},-\frac{1}{3})$, $S_{88}=(\bold{15},\bold{1},\frac{2}{3})$,
    $S_{89}=(\bold{15},\bold{1},\frac{5}{3})$, $S_{90}=(\bold{15},\bold{1},\frac{8}{3})$,
    $S_{91}=(\bold{15},\bold{2},-\frac{11}{6})$, $S_{92}=(\bold{15},\bold{2},-\frac{5}{6})$,
    $S_{93}=(\bold{15},\bold{2},\frac{1}{6})$, $S_{94}=(\bold{15},\bold{2},\frac{7}{6})$,
    $S_{95}=(\bold{15},\bold{3},-\frac{1}{3})$, $S_{96}=(\bold{15},\bold{3},\frac{2}{3})$,
    $S_{97}=(\bold{15},\bold{3},\frac{5}{3})$, $S_{98}=(\bold{15},\bold{4},-\frac{5}{6})$,
    $S_{99}=(\bold{15},\bold{4},\frac{1}{6})$, $S_{100}=(\bold{15},\bold{5},\frac{2}{3})$.\\
    \hline
    13& $S_{101}=(\bold{1},\bold{1},3)$, $S_{102}=(\bold{1},\bold{6},\frac{3}{2})$,
    $S_{103}=(\bold{1},\bold{6},\frac{5}{2})$, $S_{104}=(\bold{3},\bold{2},\frac{19}{6})$,
    $S_{105}=(\bold{3},\bold{5},\frac{5}{3})$, $S_{106}=(\bold{3},\bold{5},\frac{8}{3})$,
    $S_{107}=(\bold{6},\bold{1},-\frac{7}{3})$, $S_{108}=(\bold{6},\bold{4},\frac{7}{6})$,
    $S_{109}=(\bold{6},\bold{4},\frac{13}{6})$, $S_{110}=(\bold{6},\bold{6},\frac{1}{6})$,
    $S_{111}=(\bold{6},\bold{6},\frac{7}{6})$, $S_{112}=(\bold{8},\bold{3},2)$,
    $S_{113}=(\bold{8},\bold{5},1)$, $S_{114}=(\bold{8},\bold{5},2)$,
    $S_{115}=(\bold{10},\bold{2},-\frac{1}{2})$, $S_{116}=(\bold{10},\bold{2},\frac{1}{2})$,
    $S_{117}=(\bold{10},\bold{2},\frac{3}{2})$, $S_{118}=(\bold{10},\bold{2},\frac{5}{2})$,
    $S_{119}=(\bold{10},\bold{4},\frac{1}{2})$, $S_{120}=(\bold{10},\bold{4},\frac{3}{2})$.\\
    \hline
    14& $S_{121}=(\bold{1},\bold{2},\frac{7}{2})$, $S_{122}=(\bold{1},\bold{4},\frac{5}{2})$,
    $S_{123}=(\bold{1},\bold{4},\frac{7}{2})$, $S_{124}=(\bold{1},\bold{7},0)$,
    $S_{125}=(\bold{1},\bold{7},1)$, $S_{126}=(\bold{3},\bold{2},-\frac{23}{6})$,
    $S_{127}=(\bold{3},\bold{3},-\frac{10}{3})$, $S_{128}=(\bold{3},\bold{3},\frac{8}{3})$,
    $S_{129}=(\bold{3},\bold{5},-\frac{10}{3})$, $S_{130}=(\bold{3},\bold{5},-\frac{7}{3})$,
    $S_{131}=(\bold{3},\bold{7},-\frac{7}{3})$, $S_{132}=(\bold{3},\bold{7},-\frac{4}{3})$,
    $S_{133}=(\bold{3},\bold{7},-\frac{1}{3})$, $S_{134}=(\bold{3},\bold{7},\frac{2}{3})$,
    $S_{135}=(\bold{3},\bold{7},\frac{5}{3})$, $S_{136}=(\bold{6},\bold{2},-\frac{17}{6})$,
    $S_{137}=(\bold{6},\bold{2},\frac{13}{6})$, $S_{138}=(\bold{6},\bold{2},\frac{19}{6})$,
    $S_{139}=(\bold{6},\bold{4},-\frac{17}{6})$, $S_{140}=(\bold{6},\bold{4},-\frac{11}{6})$,
    $S_{141}=(\bold{6},\bold{6},-\frac{11}{6})$, $S_{142}=(\bold{6},\bold{6},-\frac{5}{6})$,
    $S_{143}=(\bold{8},\bold{1},3)$, $S_{144}=(\bold{8},\bold{3},3)$,
    $S_{145}=(\bold{8},\bold{6},\frac{1}{2})$, $S_{146}=(\bold{10},\bold{2},-\frac{5}{2})$,
    $S_{147}=(\bold{10},\bold{2},-\frac{3}{2})$, $S_{148}=(\bold{10},\bold{4},-\frac{3}{2})$,
    $S_{149}=(\bold{10},\bold{4},-\frac{1}{2})$, $S_{150}=(\bold{10},\bold{6},-\frac{1}{2})$,
    $S_{151}=(\bold{10},\bold{6},\frac{1}{2})$, $S_{152}=(\bold{15},\bold{1},-\frac{7}{3})$,
    $S_{153}=(\bold{15},\bold{2},\frac{13}{6})$, $S_{154}=(\bold{15},\bold{2},\frac{19}{6})$,
    $S_{155}=(\bold{15},\bold{3},-\frac{7}{3})$, $S_{156}=(\bold{15},\bold{3},-\frac{4}{3})$,
    $S_{157}=(\bold{15},\bold{4},\frac{7}{6})$, $S_{158}=(\bold{15},\bold{4},\frac{13}{6})$,
    $S_{159}=(\bold{15},\bold{5},-\frac{4}{3})$, $S_{160}=(\bold{15},\bold{5},-\frac{1}{3})$,
    $S_{161}=(\bold{15},\bold{6},\frac{1}{6})$, $S_{162}=(\bold{15},\bold{6},\frac{7}{6})$. \\
    \hline
    15& $S_{163}=(\bold{1},\bold{1},4)$. \\
    \hline
    \end{tabularx}
    \caption{List of new scalars and mass dimension $d$ at which they first appear. }
    \label{tab:newScalars}
\end{table}

\begin{table}[h!]
\small
\centering
\begin{tabularx}{\textwidth}{|c|C|}
\hline
      $d$   & {\rm Fermions}   \\
         \hline \hline
         
    6& \\
    \hline
    7& $F_{1}=(\bold{1},\bold{1},0)$,
    $F_{2}=(\bold{1},\bold{1},1)$, $F_{3}=(\bold{1},\bold{2},\frac{1}{2})$,
    $F_{4}=(\bold{1},\bold{3},0)$, $F_{5}=(\bold{3},\bold{1},-\frac{1}{3})$,
    $F_{6}=(\bold{3},\bold{1},\frac{2}{3})$, $F_{7}=(\bold{3},\bold{2},-\frac{5}{6})$,
    $F_{8}=(\bold{3},\bold{2},\frac{1}{6})$, $F_{9}=(\bold{3},\bold{3},-\frac{1}{3})$.\\
    \hline
    8& $F_{10}=(\bold{1},\bold{2},\frac{3}{2})$, $F_{11}=(\bold{1},\bold{3},1)$,
    $F_{12}=(\bold{1},\bold{4},\frac{1}{2})$, $F_{13}=(\bold{3},\bold{2},\frac{7}{6})$,
    $F_{14}=(\bold{3},\bold{3},\frac{2}{3})$, $F_{15}=(\bold{3},\bold{4},\frac{1}{6})$.\\
    \hline
        9& $F_{16}=(\bold{1},\bold{1},2)$,
    $F_{17}=(\bold{3},\bold{1},-\frac{4}{3})$, $F_{18}=(\bold{3},\bold{1},\frac{5}{3})$,
    $F_{19}=(\bold{3},\bold{3},-\frac{4}{3})$, $F_{20}=(\bold{3},\bold{4},-\frac{5}{6})$,
    $F_{21}=(\bold{6},\bold{1},-\frac{4}{3})$, $F_{22}=(\bold{6},\bold{1},-\frac{1}{3})$,
    $F_{23}=(\bold{6},\bold{1},\frac{2}{3})$, $F_{24}=(\bold{6},\bold{2},-\frac{5}{6})$,
    $F_{25}=(\bold{6},\bold{2},\frac{1}{6})$, $F_{26}=(\bold{6},\bold{3},-\frac{1}{3})$,
    $F_{27}=(\bold{8},\bold{1},0)$, $F_{28}=(\bold{8},\bold{1},1)$,
    $F_{29}=(\bold{8},\bold{2},\frac{1}{2})$, $F_{30}=(\bold{8},\bold{3},0)$.\\
    \hline
    10& $F_{31}=(\bold{1},\bold{3},2)$, $F_{32}=(\bold{1},\bold{4},\frac{3}{2})$,
    $F_{33}=(\bold{1},\bold{5},0)$, $F_{34}=(\bold{1},\bold{5},1)$,
    $F_{35}=(\bold{3},\bold{1},-\frac{7}{3})$, $F_{36}=(\bold{3},\bold{2},-\frac{11}{6})$,
    $F_{37}=(\bold{3},\bold{3},\frac{5}{3})$, $F_{38}=(\bold{3},\bold{4},\frac{7}{6})$,
    $F_{39}=(\bold{3},\bold{5},-\frac{1}{3})$, $F_{40}=(\bold{3},\bold{5},\frac{2}{3})$,
    $F_{41}=(\bold{6},\bold{1},\frac{5}{3})$, $F_{42}=(\bold{6},\bold{2},\frac{7}{6})$,
    $F_{43}=(\bold{6},\bold{3},\frac{2}{3})$, $F_{44}=(\bold{6},\bold{4},\frac{1}{6})$,
    $F_{45}=(\bold{8},\bold{1},2)$, $F_{46}=(\bold{8},\bold{2},\frac{3}{2})$,
    $F_{47}=(\bold{8},\bold{3},1)$, $F_{48}=(\bold{8},\bold{4},\frac{1}{2})$.\\
    \hline
    11& $F_{49}=(\bold{1},\bold{2},\frac{5}{2})$,
    $F_{50}=(\bold{3},\bold{2},\frac{13}{6})$, $F_{51}=(\bold{3},\bold{4},-\frac{11}{6})$,
    $F_{52}=(\bold{3},\bold{5},-\frac{4}{3})$, $F_{53}=(\bold{6},\bold{2},-\frac{11}{6})$,
    $F_{54}=(\bold{6},\bold{3},-\frac{4}{3})$, $F_{55}=(\bold{6},\bold{4},-\frac{5}{6})$,
    $F_{56}=(\bold{6},\bold{5},-\frac{1}{3})$, $F_{57}=(\bold{8},\bold{5},0)$.\\
    \hline
        12& $F_{58}=(\bold{1},\bold{1},3)$, $F_{59}=(\bold{1},\bold{4},\frac{5}{2})$,
    $F_{60}=(\bold{1},\bold{5},2)$, $F_{61}=(\bold{1},\bold{6},\frac{1}{2})$,
    $F_{62}=(\bold{1},\bold{6},\frac{3}{2})$, $F_{63}=(\bold{3},\bold{1},\frac{8}{3})$,
    $F_{64}=(\bold{3},\bold{2},-\frac{17}{6})$, $F_{65}=(\bold{3},\bold{3},-\frac{7}{3})$,
    $F_{66}=(\bold{3},\bold{4},\frac{13}{6})$, $F_{67}=(\bold{3},\bold{5},\frac{5}{3})$,
    $F_{68}=(\bold{3},\bold{6},-\frac{5}{6})$, $F_{69}=(\bold{3},\bold{6},\frac{1}{6})$,
    $F_{70}=(\bold{3},\bold{6},\frac{7}{6})$, $F_{71}=(\bold{6},\bold{1},-\frac{7}{3})$,
    $F_{72}=(\bold{6},\bold{2},\frac{13}{6})$, $F_{73}=(\bold{6},\bold{3},\frac{5}{3})$,
    $F_{74}=(\bold{6},\bold{4},\frac{7}{6})$, $F_{75}=(\bold{6},\bold{5},\frac{2}{3})$,
    $F_{76}=(\bold{8},\bold{2},\frac{5}{2})$, $F_{77}=(\bold{8},\bold{3},2)$,
    $F_{78}=(\bold{8},\bold{4},\frac{3}{2})$, $F_{79}=(\bold{8},\bold{5},1)$,
    $F_{80}=(\bold{10},\bold{1},-1)$, $F_{81}=(\bold{10},\bold{1},0)$,
    $F_{82}=(\bold{10},\bold{1},1)$, $F_{83}=(\bold{10},\bold{1},2)$,
    $F_{84}=(\bold{10},\bold{2},-\frac{1}{2})$, $F_{85}=(\bold{10},\bold{2},\frac{1}{2})$,
    $F_{86}=(\bold{10},\bold{2},\frac{3}{2})$, $F_{87}=(\bold{10},\bold{3},0)$,
    $F_{88}=(\bold{10},\bold{3},1)$, $F_{89}=(\bold{10},\bold{4},\frac{1}{2})$,
    $F_{90}=(\bold{15},\bold{1},-\frac{4}{3})$, $F_{91}=(\bold{15},\bold{1},-\frac{1}{3})$,
    $F_{92}=(\bold{15},\bold{1},\frac{2}{3})$, $F_{93}=(\bold{15},\bold{1},\frac{5}{3})$,
    $F_{94}=(\bold{15},\bold{2},-\frac{5}{6})$, $F_{95}=(\bold{15},\bold{2},\frac{1}{6})$,
    $F_{96}=(\bold{15},\bold{2},\frac{7}{6})$, $F_{97}=(\bold{15},\bold{3},-\frac{1}{3})$,
    $F_{98}=(\bold{15},\bold{3},\frac{2}{3})$, $F_{99}=(\bold{15},\bold{4},\frac{1}{6})$.\\
    \hline
    13& $F_{100}=(\bold{1},\bold{3},3)$, $F_{101}=(\bold{3},\bold{3},\frac{8}{3})$,
    $F_{102}=(\bold{3},\bold{5},-\frac{7}{3})$, $F_{103}=(\bold{3},\bold{6},-\frac{11}{6})$,
    $F_{104}=(\bold{6},\bold{3},-\frac{7}{3})$, $F_{105}=(\bold{6},\bold{4},-\frac{11}{6})$,
    $F_{106}=(\bold{6},\bold{5},-\frac{4}{3})$, $F_{107}=(\bold{6},\bold{6},-\frac{5}{6})$,
    $F_{108}=(\bold{6},\bold{6},\frac{1}{6})$, $F_{109}=(\bold{8},\bold{6},\frac{1}{2})$,
    $F_{110}=(\bold{10},\bold{2},-\frac{3}{2})$, $F_{111}=(\bold{10},\bold{3},-1)$,
    $F_{112}=(\bold{10},\bold{4},-\frac{1}{2})$, $F_{113}=(\bold{10},\bold{5},0)$,
    $F_{114}=(\bold{15},\bold{2},-\frac{11}{6})$, $F_{115}=(\bold{15},\bold{2},\frac{13}{6})$,
    $F_{116}=(\bold{15},\bold{3},-\frac{4}{3})$, $F_{117}=(\bold{15},\bold{3},\frac{5}{3})$,
    $F_{118}=(\bold{15},\bold{4},-\frac{5}{6})$, $F_{119}=(\bold{15},\bold{4},\frac{7}{6})$,
    $F_{120}=(\bold{15},\bold{5},-\frac{1}{3})$, $F_{121}=(\bold{15},\bold{5},\frac{2}{3})$.\\
    \hline
    14& $F_{122}=(\bold{1},\bold{2},\frac{7}{2})$,
    $F_{123}=(\bold{1},\bold{5},3)$, $F_{124}=(\bold{1},\bold{6},\frac{5}{2})$,
    $F_{125}=(\bold{1},\bold{7},0)$, $F_{126}=(\bold{1},\bold{7},1)$,
    $F_{127}=(\bold{1},\bold{7},2)$, $F_{128}=(\bold{3},\bold{1},-\frac{10}{3})$,
    $F_{129}=(\bold{3},\bold{2},\frac{19}{6})$, $F_{130}=(\bold{3},\bold{3},-\frac{10}{3})$,
    $F_{131}=(\bold{3},\bold{4},-\frac{17}{6})$, $F_{132}=(\bold{3},\bold{5},\frac{8}{3})$,
    $F_{133}=(\bold{3},\bold{6},\frac{13}{6})$, $F_{134}=(\bold{3},\bold{7},-\frac{4}{3})$,
    $F_{135}=(\bold{3},\bold{7},-\frac{1}{3})$, $F_{136}=(\bold{3},\bold{7},\frac{2}{3})$,
    $F_{137}=(\bold{3},\bold{7},\frac{5}{3})$, $F_{138}=(\bold{6},\bold{1},\frac{8}{3})$,
    $F_{139}=(\bold{6},\bold{2},-\frac{17}{6})$, $F_{140}=(\bold{6},\bold{3},\frac{8}{3})$,
    $F_{141}=(\bold{6},\bold{4},\frac{13}{6})$, $F_{142}=(\bold{6},\bold{5},\frac{5}{3})$,
    $F_{143}=(\bold{6},\bold{6},\frac{7}{6})$, $F_{144}=(\bold{8},\bold{1},3)$,
    $F_{145}=(\bold{8},\bold{3},3)$, $F_{146}=(\bold{8},\bold{4},\frac{5}{2})$,
    $F_{147}=(\bold{8},\bold{5},2)$, $F_{148}=(\bold{8},\bold{6},\frac{3}{2})$,
    $F_{149}=(\bold{10},\bold{2},\frac{5}{2})$, $F_{150}=(\bold{10},\bold{3},2)$,
    $F_{151}=(\bold{10},\bold{4},\frac{3}{2})$, $F_{152}=(\bold{10},\bold{5},1)$,
    $F_{153}=(\bold{10},\bold{6},\frac{1}{2})$, $F_{154}=(\bold{15},\bold{6},\frac{1}{6})$.\\
    \hline
    15& $F_{155}=(\bold{15},\bold{1},-\frac{7}{3})$.\\
    \hline
    \end{tabularx}
    \caption{List of new fermions and mass dimension $d$ at which they first appear.}
    \label{tab:newFermions}
\end{table}

\begin{table}[h!]
\small
\centering
\begin{tabularx}{\textwidth}{|c|C|}
\hline
      $d$   & {\rm Vectors}   \\
         \hline \hline
         
    6& $V_{1}=(\bold{3},\bold{2},-\frac{5}{6})$, $V_{2}=(\bold{3},\bold{2},\frac{1}{6})$.\\
    \hline
    7& $V_{3}=(\bold{3},\bold{1},-\frac{1}{3})$,
    $V_{4}=(\bold{3},\bold{1},\frac{2}{3})$, $V_{5}=(\bold{3},\bold{3},\frac{2}{3})$.\\
    \hline
    8& $V_{6}=(\bold{3},\bold{1},-\frac{4}{3})$,
    $V_{7}=(\bold{3},\bold{3},-\frac{4}{3})$, $V_{8}=(\bold{3},\bold{3},-\frac{1}{3})$.\\
    \hline
    9& $V_{9}=(\bold{1},\bold{1},0)$, $V_{10}=(\bold{1},\bold{1},1)$,
    $V_{11}=(\bold{1},\bold{2},\frac{1}{2})$, $V_{12}=(\bold{1},\bold{2},\frac{3}{2})$,
    $V_{13}=(\bold{1},\bold{3},0)$, $V_{14}=(\bold{3},\bold{1},\frac{5}{3})$,
    $V_{15}=(\bold{3},\bold{2},\frac{7}{6})$, $V_{16}=(\bold{3},\bold{4},\frac{1}{6})$,
    $V_{17}=(\bold{3},\bold{4},\frac{7}{6})$, $V_{18}=(\bold{6},\bold{2},-\frac{5}{6})$,
    $V_{19}=(\bold{6},\bold{2},\frac{1}{6})$, $V_{20}=(\bold{8},\bold{1},0)$,
    $V_{21}=(\bold{8},\bold{1},1)$, $V_{22}=(\bold{8},\bold{3},0)$.\\
    \hline
       10& $V_{23}=(\bold{1},\bold{1},2)$, $V_{24}=(\bold{1},\bold{3},1)$, $V_{25}=(\bold{1},\bold{3},2)$,
    $V_{26}=(\bold{1},\bold{4},\frac{1}{2})$, $V_{27}=(\bold{3},\bold{2},-\frac{11}{6})$,
    $V_{28}=(\bold{3},\bold{2},\frac{13}{6})$, $V_{29}=(\bold{3},\bold{4},-\frac{11}{6})$,
    $V_{30}=(\bold{3},\bold{4},-\frac{5}{6})$, $V_{31}=(\bold{6},\bold{1},-\frac{4}{3})$,
    $V_{32}=(\bold{6},\bold{1},-\frac{1}{3})$, $V_{33}=(\bold{6},\bold{1},\frac{2}{3})$,
    $V_{34}=(\bold{6},\bold{3},-\frac{4}{3})$, $V_{35}=(\bold{6},\bold{3},-\frac{1}{3})$,
    $V_{36}=(\bold{6},\bold{3},\frac{2}{3})$, $V_{37}=(\bold{8},\bold{2},\frac{1}{2})$,
    $V_{38}=(\bold{8},\bold{2},\frac{3}{2})$, $V_{39}=(\bold{8},\bold{4},\frac{1}{2})$.\\
    \hline
    11& $V_{40}=(\bold{1},\bold{5},1)$, $V_{41}=(\bold{3},\bold{3},\frac{5}{3})$,
    $V_{42}=(\bold{3},\bold{5},-\frac{1}{3})$, $V_{43}=(\bold{3},\bold{5},\frac{2}{3})$,
    $V_{44}=(\bold{3},\bold{5},\frac{5}{3})$, $V_{45}=(\bold{6},\bold{2},\frac{7}{6})$,
    $V_{46}=(\bold{6},\bold{4},\frac{1}{6})$, $V_{47}=(\bold{6},\bold{4},\frac{7}{6})$,
    $V_{48}=(\bold{8},\bold{3},1)$, $V_{49}=(\bold{8},\bold{3},2)$, $V_{50}=(\bold{8},\bold{5},1)$.\\
    \hline
    12& $V_{51}=(\bold{1},\bold{1},3)$, $V_{52}=(\bold{1},\bold{2},\frac{5}{2})$,
    $V_{53}=(\bold{1},\bold{4},\frac{3}{2})$, $V_{54}=(\bold{1},\bold{4},\frac{5}{2})$,
    $V_{55}=(\bold{1},\bold{5},0)$, $V_{56}=(\bold{3},\bold{1},-\frac{7}{3})$,
    $V_{57}=(\bold{3},\bold{2},-\frac{17}{6})$, $V_{58}=(\bold{3},\bold{3},-\frac{7}{3})$,
    $V_{59}=(\bold{3},\bold{3},\frac{8}{3})$, $V_{60}=(\bold{3},\bold{5},-\frac{7}{3})$,
    $V_{61}=(\bold{3},\bold{5},-\frac{4}{3})$, $V_{62}=(\bold{6},\bold{1},-\frac{7}{3})$,
    $V_{63}=(\bold{6},\bold{1},\frac{5}{3})$, $V_{64}=(\bold{6},\bold{2},-\frac{11}{6})$,
    $V_{65}=(\bold{6},\bold{2},\frac{13}{6})$, $V_{66}=(\bold{6},\bold{4},-\frac{11}{6})$,
    $V_{67}=(\bold{6},\bold{4},-\frac{5}{6})$, $V_{68}=(\bold{6},\bold{5},-\frac{1}{3})$,
    $V_{69}=(\bold{8},\bold{1},2)$, $V_{70}=(\bold{8},\bold{1},3)$,
    $V_{71}=(\bold{8},\bold{5},0)$, $V_{72}=(\bold{10},\bold{2},-\frac{3}{2})$,
    $V_{73}=(\bold{10},\bold{2},-\frac{1}{2})$, $V_{74}=(\bold{10},\bold{2},\frac{1}{2})$,
    $V_{75}=(\bold{10},\bold{2},\frac{3}{2})$, $V_{76}=(\bold{10},\bold{4},-\frac{1}{2})$,
    $V_{77}=(\bold{10},\bold{4},\frac{1}{2})$, $V_{78}=(\bold{15},\bold{1},-\frac{7}{3})$,
    $V_{79}=(\bold{15},\bold{1},-\frac{4}{3})$, $V_{80}=(\bold{15},\bold{1},-\frac{1}{3})$,
    $V_{81}=(\bold{15},\bold{1},\frac{2}{3})$, $V_{82}=(\bold{15},\bold{1},\frac{5}{3})$,
    $V_{83}=(\bold{15},\bold{2},-\frac{5}{6})$, $V_{84}=(\bold{15},\bold{2},\frac{1}{6})$,
    $V_{85}=(\bold{15},\bold{2},\frac{7}{6})$, $V_{86}=(\bold{15},\bold{2},\frac{13}{6})$,
    $V_{87}=(\bold{15},\bold{3},-\frac{4}{3})$, $V_{88}=(\bold{15},\bold{3},-\frac{1}{3})$,
    $V_{89}=(\bold{15},\bold{3},\frac{2}{3})$, $V_{90}=(\bold{15},\bold{4},\frac{1}{6})$,
    $V_{91}=(\bold{15},\bold{4},\frac{7}{6})$, $V_{92}=(\bold{15},\bold{5},-\frac{1}{3})$.\\
    \hline
    13& $V_{93}=(\bold{1},\bold{2},\frac{7}{2})$,
    $V_{94}=(\bold{1},\bold{6},\frac{1}{2})$, $V_{95}=(\bold{1},\bold{6},\frac{3}{2})$,
    $V_{96}=(\bold{3},\bold{1},\frac{8}{3})$, $V_{97}=(\bold{3},\bold{4},\frac{13}{6})$,
    $V_{98}=(\bold{3},\bold{6},-\frac{5}{6})$, $V_{99}=(\bold{3},\bold{6},\frac{1}{6})$,
    $V_{100}=(\bold{3},\bold{6},\frac{7}{6})$, $V_{101}=(\bold{3},\bold{6},\frac{13}{6})$,
    $V_{102}=(\bold{6},\bold{2},-\frac{17}{6})$, $V_{103}=(\bold{6},\bold{3},\frac{5}{3})$,
    $V_{104}=(\bold{6},\bold{5},\frac{2}{3})$, $V_{105}=(\bold{6},\bold{5},\frac{5}{3})$,
    $V_{106}=(\bold{8},\bold{4},\frac{3}{2})$, $V_{107}=(\bold{8},\bold{4},\frac{5}{2})$,
    $V_{108}=(\bold{8},\bold{6},\frac{1}{2})$, $V_{109}=(\bold{8},\bold{6},\frac{3}{2})$,
    $V_{110}=(\bold{10},\bold{1},-1)$, $V_{111}=(\bold{10},\bold{1},0)$,
    $V_{112}=(\bold{10},\bold{1},1)$, $V_{113}=(\bold{10},\bold{1},2)$,
    $V_{114}=(\bold{10},\bold{3},0)$, $V_{115}=(\bold{10},\bold{3},1)$,
    $V_{116}=(\bold{10},\bold{3},2)$, $V_{117}=(\bold{10},\bold{5},1)$.\\
    \hline
    14&  $V_{118}=(\bold{1},\bold{3},3)$,
    $V_{119}=(\bold{1},\bold{5},2)$, $V_{120}=(\bold{1},\bold{5},3)$,
    $V_{121}=(\bold{3},\bold{1},-\frac{10}{3})$, $V_{122}=(\bold{3},\bold{2},\frac{19}{6})$,
    $V_{123}=(\bold{3},\bold{3},-\frac{10}{3})$, $V_{124}=(\bold{3},\bold{4},-\frac{17}{6})$,
    $V_{125}=(\bold{3},\bold{4},\frac{19}{6})$, $V_{126}=(\bold{3},\bold{6},-\frac{17}{6})$,
    $V_{127}=(\bold{3},\bold{6},-\frac{11}{6})$, $V_{128}=(\bold{6},\bold{1},\frac{8}{3})$,
    $V_{129}=(\bold{6},\bold{3},-\frac{7}{3})$, $V_{130}=(\bold{6},\bold{3},\frac{8}{3})$,
    $V_{131}=(\bold{6},\bold{5},-\frac{7}{3})$, $V_{132}=(\bold{6},\bold{5},-\frac{4}{3})$,
    $V_{133}=(\bold{6},\bold{6},-\frac{5}{6})$, $V_{134}=(\bold{6},\bold{6},\frac{1}{6})$,
    $V_{135}=(\bold{8},\bold{2},\frac{5}{2})$, $V_{136}=(\bold{8},\bold{2},\frac{7}{2})$,
    $V_{137}=(\bold{10},\bold{1},-2)$, $V_{138}=(\bold{10},\bold{3},-2)$,
    $V_{139}=(\bold{10},\bold{3},-1)$, $V_{140}=(\bold{10},\bold{5},-1)$,
    $V_{141}=(\bold{10},\bold{5},0)$, $V_{142}=(\bold{15},\bold{1},\frac{8}{3})$,
    $V_{143}=(\bold{15},\bold{2},-\frac{17}{6})$, $V_{144}=(\bold{15},\bold{2},-\frac{11}{6})$,
    $V_{145}=(\bold{15},\bold{3},\frac{5}{3})$, $V_{146}=(\bold{15},\bold{3},\frac{8}{3})$,
    $V_{147}=(\bold{15},\bold{4},-\frac{11}{6})$, $V_{148}=(\bold{15},\bold{4},-\frac{5}{6})$,
    $V_{149}=(\bold{15},\bold{5},\frac{2}{3})$, $V_{150}=(\bold{15},\bold{5},\frac{5}{3})$,
    $V_{151}=(\bold{15},\bold{6},-\frac{5}{6})$, $V_{152}=(\bold{15},\bold{6},\frac{1}{6})$.\\
    \hline
    15& \\
    \hline
    \end{tabularx}
    \caption{List of new vectors and mass dimension $d$ at which they first appear. }
    \label{tab:newVectors}
\end{table}

\endgroup
\clearpage

The \textit{full} results, including Feynman diagrams and chirality information, are provided online at  \website. 
In the appendices of this article, we provide a significantly shorter version that only shows the new particles that UV complete each operator per topology, without retaining how exactly the particles are distributed over the diagram. 
Furthermore, for $d > 9$, we only display UV completions for \textit{symmetry-protected} operators, i.e.~those operators with a $(\Delta B,\Delta L)$ not found at lower $d$, which can thus be easily singled-out by imposing a $B+a L$ symmetry~\cite{Weinberg:1980bf,Heeck:2019kgr}.
For $d\geq 12$, even that is impossible given the large number of UV completions, so we refer to \website, although the Latex tables for $d=12$ and $d=13$ are also included as ancillary files in the arXiv version.
We stress again that our procedure is in principle fast enough to be used for even higher mass dimensions, but the vast output size makes the results increasingly difficult to use.

 \section{Derivative operators}
\label{sec:derivative_operators}

We only study non-derivative operators in this article, in part because derivative operators are technically more complicated to UV-complete, sometimes lacking tree-level completions altogether.
More importantly, derivatives inside operators generically arise from the momentum of an internal particle,\footnote{In the presence of vector bosons, one could also have derivative vertices such as $V^\mu S_1 \partial_\mu S_2$. Just like any other vector-boson interactions it is far from clear that these should be considered renormalizable.} essentially from higher-order terms in the large-mass expansion. For example, an internal scalar line comes with a propagator
\begin{align}
\frac{1}{M^2 - p^2} = \frac{1}{M^2}+\frac{p^2}{M^4} + \mathcal{O}\left(\frac{p^4}{M^6}\right) .
\end{align}
The first term on the right-hand side generates a non-derivative operator of some mass dimension $d$, while the second term gives rise to a two-derivative operator of mass dimension $d+2$, being suppressed by two more powers of $1/M$. Since $p$ corresponds to the momentum of the SM particles, e.g.~of order of the proton mass in proton-decay processes, this additional $p^2/M^2$ is a devastating suppression. Generically, derivative operators are thus heavily suppressed compared to non-derivative operators. Furthermore, the above argument illustrates that the UV completions of derivative operators will involve the same particles that UV-completed non-derivative operators of lower mass dimension.

There are loopholes to these arguments. For one, as already pointed out in Ref.~\cite{Gargalionis:2020xvt}, non-derivative operators involving multiple Higgs doublets could end up with a vanishing UV completion to order $p^0$ but with a non-vanishing term at order $p^1$ or $p^2$, see the example in Fig.~\ref{fig:Higgs_antisym}. In those special cases, one could find UV completions for derivative operators that did not already show up as UV completions for non-derivative operators at lower mass dimension.
As far as phenomenological importance goes, one can also imagine several UV completions for some non-derivative operator that interfere destructively; the higher-order momentum-dependent terms in the amplitudes have no reason to cancel in the same way, leaving dominant derivative operators. 
We postpone the discussion of derivative operators, their UV completions, and phenomenological consequences to a separate article~\cite{Heeck2025derivative}.

\section{Light new particles}
\label{sec:light_particles}

By construction, $(\nu)$SMEFT operators are generated by integrating out \emph{heavy} new particles. Our UV completions need not be restricted to this region of parameter space though, as nucleon-decay limits could equally well be satisfied with tiny Yukawa couplings instead of heavy masses. Stringent LHC limits nevertheless force almost all new particles to $\mathcal{O}(\unit{TeV})$ masses on account of their large gauge-boson-mediated production rates. 
The obvious exception are gauge \textit{singlets}, which are far less constrained and could well be at or below the GeV mass scale. This opens up the possibility of emitting such light particles in BNV nucleon decays. In general, this setup requires the construction of a new effective field theory that includes the light new particle, constructed with focus on BNV for light fermions and (pseudo)scalars in Refs.~\cite{Helo:2018bgb,Li:2025slp,Helo:2025kgx} and~\cite{Li:2024liy,Ma:2025mjy,Helo:2025kgx}, respectively. (Technically, the $\nu$SMEFT already covers light new singlet fermions.)

\begin{figure}[tb]
    \centering
    \includegraphics[scale=0.77]{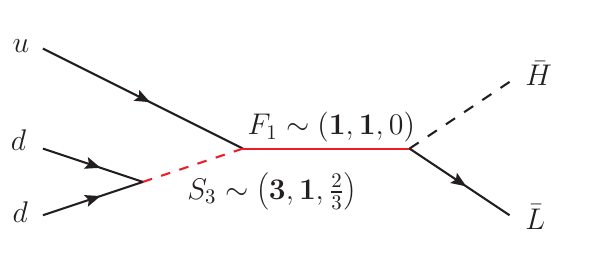}\hspace{4ex}
    \includegraphics[scale=0.77]{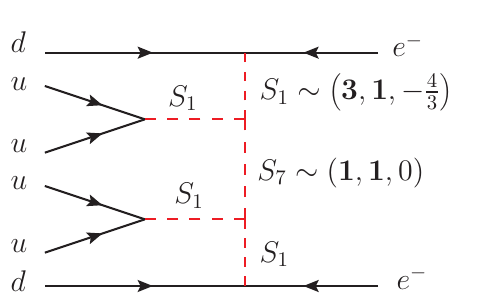}
  \caption{Feynman diagrams involving gauge-singlet particles that UV-complete the operator $\mathcal{O}^{3}_{7,(1,-1)}=\bar{H}ddu\bar{L}$ (left) and $\mathcal{O}^{10}_{12,(2,2)} = uuuuddee$ (right), with all momenta incoming.}
    \label{fig:light_particles}
\end{figure}

Our results automatically contain an interesting  \textit{subset} of those results, typically in the form of light particles with baryon and lepton numbers~\cite{Heeck:2020nbq}. Let us focus on UV completions that contain a singlet \textit{fermion}, $F_1$ in our notation, and make the mass small, say below the proton mass. Depending on the operator and diagram, this can lead to a BNV process in which $F_1$ is emitted \textit{on shell} and then decays. For example, the $F_1 S_3$ UV completion of the $\mathcal{O}^{3}_{7,(1,-1)}=\bar{H}ddu\bar{L}$ operator, Fig.~\ref{fig:light_particles} (left), can then induce $p\to K^+ F_1$ followed by $F_1\to \pi^+ e^- $, 
recently studied in Refs.~\cite{Fajfer:2020tqf,Domingo:2024qoj,Heeck:2025uwh}, even without imposing any $B+ a L$ symmetry to forbid $d=6$ operators. Decay channels of $F_1$ into $L=3$ ($L=5$) final states arise at $d=9$ ($d=12$) in complete analogy, but typically require $F_1$ to carry appropriate lepton number to forbid lower-$d$ processes.

As an example for a light gauge-singlet \textit{scalar}, $S_7$ in our notation, we take the UV completion $S_7 S_1 S_1 S_1 S_1$ of $\mathcal{O}^{10}_{12,(2,2)} = uuuuddee$, see Fig.~\ref{fig:light_particles} (right). To ensure absence of $d<12$ BNV operators, two copies of $S_1$ are necessary, one of them odd under a $\mathbb{Z}_2^L$, allowing for the LQ coupling $\bar{S}_1 d e$, and the other one even, so it couples to $S_1 uu$. $S_7$ and leptons are also odd under this $\mathbb{Z}_2^L$, and $B-L$ is automatically conserved as well.
For heavy $S_7$, this generates $pp\to e^+ e^+$~\cite{Super-Kamiokande:2018apg}, but for light $S_7$ it gives $p\to e^+ S_7$~\cite{Super-Kamiokande:2015pys,McKeen:2020zni,Heeck:2025uwh}.
For $m_{S_7}\sim m_n$, nucleon decays are kinematically forbidden, but the di-nucleon decay $pp\to e^+ e^+$ involves a fairly light $t$-channel $S_7$, likely changing the calculation somewhat compared to the SMEFT case. 
Notice that the $S_1 uu$ coupling is antisymmetric and thus involves a charm or top quark; for $m_n\lesssim m_{S_7}\lesssim m_{\Lambda_c^+}$, there is then a decay chain $\Lambda_c^+ \to e^+ S_7$, followed by $S_7\to \bar{p} e^+$ -- a displaced version of the decay $\Lambda_c^+ \to \bar{p} e^+e^+$ searched for in BaBar~\cite{BaBar:2011ouc} --  although the better limits on this model likely still come from the di-nucleon decay $pp\to e^+ e^+$, despite the loop and CKM suppression required to get first-generation quarks.
Other cases involving light $S_7$ can be found using our UV completions, particularly fruitful in the $\nu$SMEFT because $S_7$ can couple to $\nu\nu$.

In principle one can also study light gauge-singlet \textit{vector} bosons, $V_9$ in our notation, but since renormalizability becomes an even bigger issue for light vectors, we refrain from an analogous discussion.

\begin{figure}[tb]
    \centering
    \includegraphics[scale=0.8]{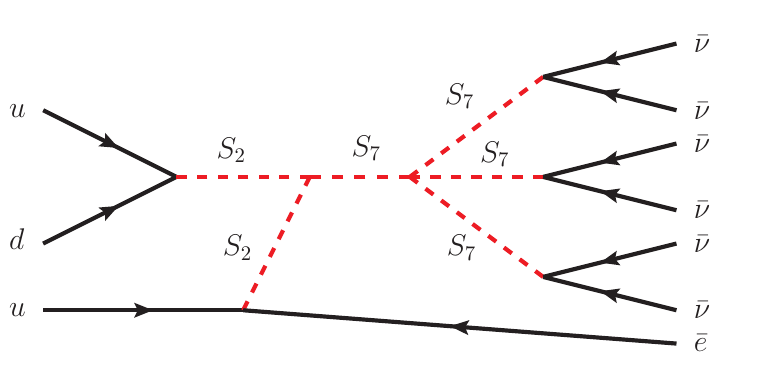}
  \caption{Example UV completion for the $d=15$ operator $ \mathcal{O}_{15, \,(1 , 7)}^{11} \equiv  u u d e \nu \nu \nu \nu \nu \nu$ that leads to $p\to e^+\bar{\nu}\bar{\nu}\bar{\nu}\bar{\nu}\bar{\nu}\bar{\nu}$ with heavy scalars $S_2 \sim (\vec{3},\vec{1},-1/3)$ and $S_7 \sim (\vec{1},\vec{1},0)$.}
    \label{fig:d15_proton_decay}
\end{figure}

For another example that illustrates the benefit of studying UV completions over effective operators, we turn to operators with $(\Delta B,\Delta L) = (1,7)$, which first appear at $d=15$ (see App.~\ref{sec:d15}). These all involve between three and six right-handed neutrinos and are hence part of $\nu$SMEFT rather than SMEFT. 
The Pauli exclusion principle forbids operators such as $\nu\nu\nu\nu$ in which all neutrinos have the same flavor~\cite{Fonseca:2018aav}, so operators such as $\mathcal{O}_{15, \,(1 , 7)}^{11} \equiv  u u d e \nu \nu \nu \nu \nu \nu $ involve right-handed neutrinos of all three generations. 
The high mass dimension of these operators together with the phase space suppression from at least seven final-state particles renders these operators near impossible to observe. For example, $ \mathcal{O}_{15, \,(1 , 7)}^{11} \equiv  u u d e \nu \nu \nu \nu \nu \nu$ gives
\begin{align}
        \Gamma (p\to e^+\bar{\nu}\bar{\nu}\bar{\nu}\bar{\nu}\bar{\nu}\bar{\nu})\simeq \frac{\alpha^2 m_p^{17}}{2^{30}3^3 5^2\pi^{11}\Lambda^{22}}\simeq \left( \unit[10^{34}]{yr}\right)^{-1} \left(\frac{\unit[100]{GeV}}{\Lambda} \right)^{22} ,
\end{align}
where $\alpha \simeq \unit[0.01]{GeV^3}$ is a QCD matrix element~\cite{Yoo:2021gql} and the phase space factor is taken from Ref.~\cite{Kleiss:1985gy}. Even for $\Lambda$ at the electroweak scale, the proton lifetime is too long to be measured with current technology, given the rather unspectacular signature.
However, our UV completions are more flexible than SMEFT operators. Take the UV completion from Fig.~\ref{fig:d15_proton_decay}, involving four singlet scalars $S_7$, which have no direct couplings to SM particles and can thus easily have masses below the electroweak scale. Even if $S_2$ has multi-TeV mass, lowering $m_{S_7}$ to a few GeV can generate fast proton decay. Assuming all dimensionless couplings to be of order one and the cubic coupling $y_{S_2 S_2 S_7}$ to be of order $m_{S_7}$, we estimate
\begin{align}
        \Gamma (p\to e^+\bar{\nu}\bar{\nu}\bar{\nu}\bar{\nu}\bar{\nu}\bar{\nu})\simeq  \left( \unit[5\times 10^{28}]{yr}\right)^{-1} \left(\frac{\unit[10]{TeV}}{m_{S_2}} \right)^{8}\left(\frac{\unit[3]{GeV}}{m_{S_7}} \right)^{14} .
\end{align}
By properly assigning global charges of the conserved $U(1)_{B- L/7}$ to the new scalars in Fig.~\ref{fig:d15_proton_decay} -- which requires two different $S_2$ and two different $S_7$ -- we ensure that there is no BNV from operators $d<15$, leaving us with $p\to e^+\bar{\nu}\bar{\nu}\bar{\nu}\bar{\nu}\bar{\nu}\bar{\nu}$ as the only signature.
This illustrates once more that BNV is sensitive to extremely high operator dimensions which could well be the dominant source of BNV.
At such high mass dimensions it is crucial to study UV-complete models to properly take into account constraints on particle masses and couplings, rather than simply demanding $\Lambda > \unit{TeV}$. 
    
\section{Conclusions}
\label{sec:conclusions}

The conservation of baryon number is a prediction of the SM that appears to be either exact or the best approximate symmetry in physics. Realistic models that break baryon number have been around for half a century, yet no experimental observation has been made despite serious efforts. This is all the more surprising since BNV is generically \textit{required} to explain the matter--antimatter asymmetry of our universe~\cite{Sakharov:1967dj}. 
It might just be that we are not looking for BNV in the right places, thwarted by theoretical bias towards the ``simplest'' models. Exploring the \textit{full} BNV landscape is a daunting task given the ridiculous sensitivity of large underground detectors to nucleon decays: whereas most areas of particle physics are sensitive to SMEFT operators of mass dimension 6 at most, BNV operators up to mass dimension $\sim 15$ could conceivably give testable nuclear decay rates. What is more, the different units of $B$ and $L$ carried by $d> 6$ operators make it possible to suppress the naively-dominant lower-dimensional operators, giving rise to oodles of potentially dominant BNV operators, few of which have been studied carefully so far.
In this article, we provide an exhaustive list of non-derivative BNV operators, or at least types of BNV operators, up to $d=15$ in the $\nu$SMEFT, and furthermore provide their tree-level UV completions through new scalars, fermions, and vectors. This goes far beyond existing results and corrects many mistakes in the literature. 

The full list of UV completions can be obtained at \website, with only a selection of results presented here given the sheer size of the tables. We also provide the underlying \texttt{Mathematica} code, which can be used to UV-complete any non-derivative operator that could possibly be of interest in particle physics, and also map it onto a linear combination of \texttt{GroupMath} invariants. Our work should provide a useful resource for the BNV community and hopefully facilitate broader exploration of the BNV landscape.

In follow-up work, we will 
\begin{itemize}
    \item use these results to explore a qualitatively different way to make $d\gg 6$ BNV operators dominant over lower-$d$ one's: rather than imposing $B+a L$ or other symmetries, many UV completions provide an accidental protection in that they simply do not generate lower-$d$ BNV operators at tree level. Preliminary results indicate that this is not a rare case but common enough that the majority of BNV operators could be singled-out by this procedure, including operators that have the same $(B,L)$ as lower-$d$ one's. This vastly enlarges the number of potentially interesting BNV operators;

    \item study derivative BNV operators and identify cases in which they are not subdominant to non-derivative operators. Once again, it appears that this is not as rare as naively expected, again enlarging the number of interesting BNV operators;

    \item construct a non-redundant BNV operator basis for $d\gg 6$, which would allow for proper matching of our UV completions onto the $\nu$SMEFT. A significantly more complicated step is the translation of these oodles of SMEFT operators -- including derivatives and a potentially large number of quark fields -- into hadronic operators that describe the induced nucleon decays;

    \item explore the connection of these UV completions to baryogenesis.
\end{itemize}

Given the unparalleled sensitivity of BNV to new physics and the upcoming new generation of large underground detectors it is about time to systematically explore the BNV landscape, which might just hold a surprise for us in a previously overlooked corner.

\section*{Acknowledgements}

We thank Ian Shoemaker, Christopher Murphy, and John Gargalionis for related discussions, and Mikheil Sokhashvili for spotting bugs in our code and general coding advice.  We have made excessive use of Renato Fonseca's fantastic \texttt{Mathematica} packages \texttt{Sym2Int}~\cite{Fonseca:2017lem} and \texttt{GroupMath}~\cite{Fonseca:2020vke}.
This work was supported in part by the U.S. Department of Energy under Grant No. DE-SC0007974  and a 4-VA at UVA Collaborative Research Grant.
We acknowledge Research Computing at The University of Virginia for providing computational resources that have contributed to the results reported within this publication.

\clearpage

\appendix


\section{Dimension 7: \texorpdfstring{$\Delta B=1, \Delta L=-1$}{Delta B=1,Delta L=-1}}
\label{sec:d7}

\begin{figure}[tb]
    \centering
    \includegraphics{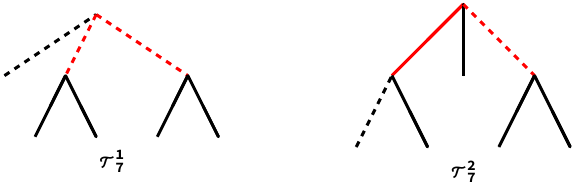}
  \caption{Feynman-diagram topologies for the $d=7$, $\Delta B = 1$ operators, involving four external SM fermions (solid black lines), one external scalar (dashed black) and new heavy fermions (red) or bosons (dashed red). }
    \label{fig:d=7_topologies}
\end{figure}

At $d=7$, there are 7 simplified non-derivative BNV operators in the $\nu$SMEFT, all with $\Delta B=1$, $\Delta L=-1$:
 \begin{multicols}{3}
 \begin{spacing}{1.4}
 \noindent
 \begin{math}
\mathcal{O}_{7, \,(1 , -1)}^{1} \equiv \bar{H}ddQ\bar{e} \,, \\
\mathcal{O}_{7, \,(1 , -1)}^{2} \equiv \bar{H}dQQ\bar{L} \,, \\
\mathcal{O}_{7, \,(1 , -1)}^{3} \equiv \bar{H}ddu\bar{L} \,, \\
\mathcal{O}_{7, \,(1 , -1)}^{4} \equiv Hddd\bar{L} \,, \\
\mathcal{O}_{7, \,(1 , -1)}^{5} \equiv HddQ\bar{\nu} \,, \\
\mathcal{O}_{7, \,(1 , -1)}^{6} \equiv \bar{H}QQQ\bar{\nu} \,, \\
\mathcal{O}_{7, \,(1 , -1)}^{7} \equiv \bar{H}duQ\bar{\nu} \,. 
 \label{eq:dequal7basisB1L-1}
 \end{math}
 \end{spacing}
 \end{multicols}
 Since $d=7$ is the lowest mass dimension for $(\Delta B,\Delta L)=(1,-1)$ operators, these can be symmetry protected, e.g.~by imposing $U(1)_{B+L}$. 
 The SMEFT operators (and a few UV completions) were already listed in Ref.~\cite{Weinberg:1980bf}.
All operators involve four fermions and one scalar and have tree-level UV completions through the two topologies shown in Fig.~\ref{fig:d=7_topologies}. The UV completions for topology $\mathcal{T}_ 7 ^{ 1 }$, involving two bosons, are given in Tab.~\ref{tab:T7Top1B1L-1}, the ones for $\mathcal{T}_ 7 ^{ 2 }$, involving one boson and one fermion, in Tab.~\ref{tab:T7Top2B1L-1}.
Our UV completions for the SMEFT operators $\mathcal{O}_{7}^{1,2,3,4}$ agree with Ref.~\cite{Li:2023cwy}. 
Our UV completions for the operators involving right-handed neutrinos, $\mathcal{O}_{7}^{5,6,7}$, contain the solutions given in Ref.~\cite{Beltran:2023ymm}, but we found three additional solutions: $V_2 V_3$ for $\mathcal{O}_{7}^{5}$, as well as  $V_{2} V_4$ and $V_{1} V_3$ for $\mathcal{O}_{7}^{7}$.

Ref.~\cite{Babu:2012iv} also provides some UV completions for $d=7$ BNV SMEFT operators in grand unified theories, in our notation corresponding to the $S_2 S_5$ UV completions of $\mathcal{O}_{7, \,(1 , -1)}^{2} $ and $\mathcal{O}_{7, \,(1 , -1)}^{3}$.

\input{Tables_dim7/Table_dim7_Top1_B1_L-1.tex}
\input{Tables_dim7/Table_dim7_Top2_B1_L-1.tex}

\clearpage


\section{Dimension 8: 
\texorpdfstring{$\Delta B=1, \Delta L=1$}{Delta B=1,Delta L=1}}
\label{sec:d8}

The $d=8$ BNV operators all obey $\Delta B=1$, $\Delta L =1$ and can be written as
 \begin{multicols}{3}
 \begin{spacing}{1.4}
 \noindent
 \begin{math}
\mathcal{O}_{8, \,(1 , 1)}^{1} \equiv HHddQL \,, \\
\mathcal{O}_{8, \,(1 , 1)}^{2} \equiv HHdQQe \,, \\
\mathcal{O}_{8, \,(1 , 1)}^{3} \equiv \bar{H}\bar{H}uuQL \,, \\
\mathcal{O}_{8, \,(1 , 1)}^{4} \equiv H\bar{H}QQQL \,, \\
\mathcal{O}_{8, \,(1 , 1)}^{5} \equiv H\bar{H}duQL \,, \\
\mathcal{O}_{8, \,(1 , 1)}^{6} \equiv H\bar{H}uQQe \,, \\
\mathcal{O}_{8, \,(1 , 1)}^{7} \equiv H\bar{H}duue \,, \\
\mathcal{O}_{8, \,(1 , 1)}^{8} \equiv H\bar{H}dQQ\nu \,, \\
\mathcal{O}_{8, \,(1 , 1)}^{9} \equiv \bar{H}\bar{H}uQQ\nu \,, \\
\mathcal{O}_{8, \,(1 , 1)}^{10} \equiv H\bar{H}ddu\nu \,. 
 \label{eq:dequal8basisB1L1}
 \end{math}
 \end{spacing}
 \end{multicols}
\begin{figure}
    \centering
\includegraphics[scale=0.8]{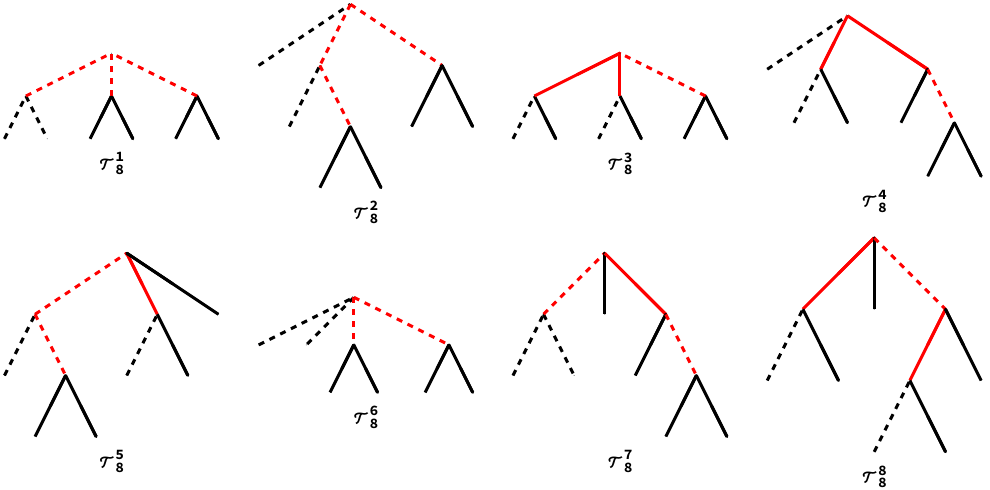}
    \caption{Feynman-diagram topologies for $d=8$ BNV operators.}
    \label{fig:d=8_topologies}
\end{figure}
The relevant four-fermion-two-scalar topologies are given in Fig.~\ref{fig:d=8_topologies} and the resulting UV completions in Tabs.~\ref{tab:T8Top1B1L1}--\ref{tab:T8Top8B1L1}. 
By and large our results agree with the UV completions of Ref.~\cite{Li:2023pfw}. However, we find almost 200 solutions that are not listed in Ref.~\cite{Li:2023pfw}, an example being our UV completion $V_1 V_2 V_8$ of operator $\mathcal{O}_{8, \,(1 , 1)}^{1}$ with topology $\mathcal{T}_8^2$.
Ref.~\cite{Li:2023pfw} also lists almost 70 UV completions that we disagree with, as they do not involve a chirality flip on a fermion line and thus lead to \textit{derivative} operators instead. For example, Ref.~\cite{Li:2023pfw} states that $\mathcal{O}_{8, \,(1 , 1)}^{4}$ can be UV-completed with the particles (in our notation) $F_4\sim (\vec{1},\vec{3},0)$, $F_9\sim (\vec{3},\vec{3},-\frac{1}{3})$, and $V_{16}\sim (\vec{3},\vec{4},\frac{1}{6})$ via topology $\mathcal{T}_8^8$; one can indeed draw the diagram, but since it only involves \emph{one} chirality of $F_9$, it does not contain an $F_9$ mass flip and thus gives a $d=9$ derivative operator.
The other operators listed in Ref.~\cite{Li:2023pfw} but not in our tables share the same chirality-flip problem and apparently only arise for operators $\mathcal{O}_{8, \,(1 , 1)}^{4}$ and $\mathcal{O}_{8, \,(1 , 1)}^{7}$, i.e.~the operators that have four same-chirality SM particles.

A few UV completions were also discussed in Ref.~\cite{Gargalionis:2024nij}, for example the $S_3 F_4 F_{14}$ UV completion of $\mathcal{O}_{8, \,(1 , 1)}^{1}$.

\input{Tables_dim8/Table_dim8_Top1_B1_L1.tex} \input{Tables_dim8/Table_dim8_Top2_B1_L1.tex}
\input{Tables_dim8/Table_dim8_Top3_B1_L1.tex} \input{Tables_dim8/Table_dim8_Top4_B1_L1.tex}
\input{Tables_dim8/Table_dim8_Top5_B1_L1.tex} \input{Tables_dim8/Table_dim8_Top6_B1_L1.tex}
\input{Tables_dim8/Table_dim8_Top7_B1_L1.tex} \input{Tables_dim8/Table_dim8_Top8_B1_L1.tex}

\clearpage


\section{Dimension 9} 
\label{sec:d9}
 
\begin{figure}
    \centering
    \includegraphics[scale=0.7]{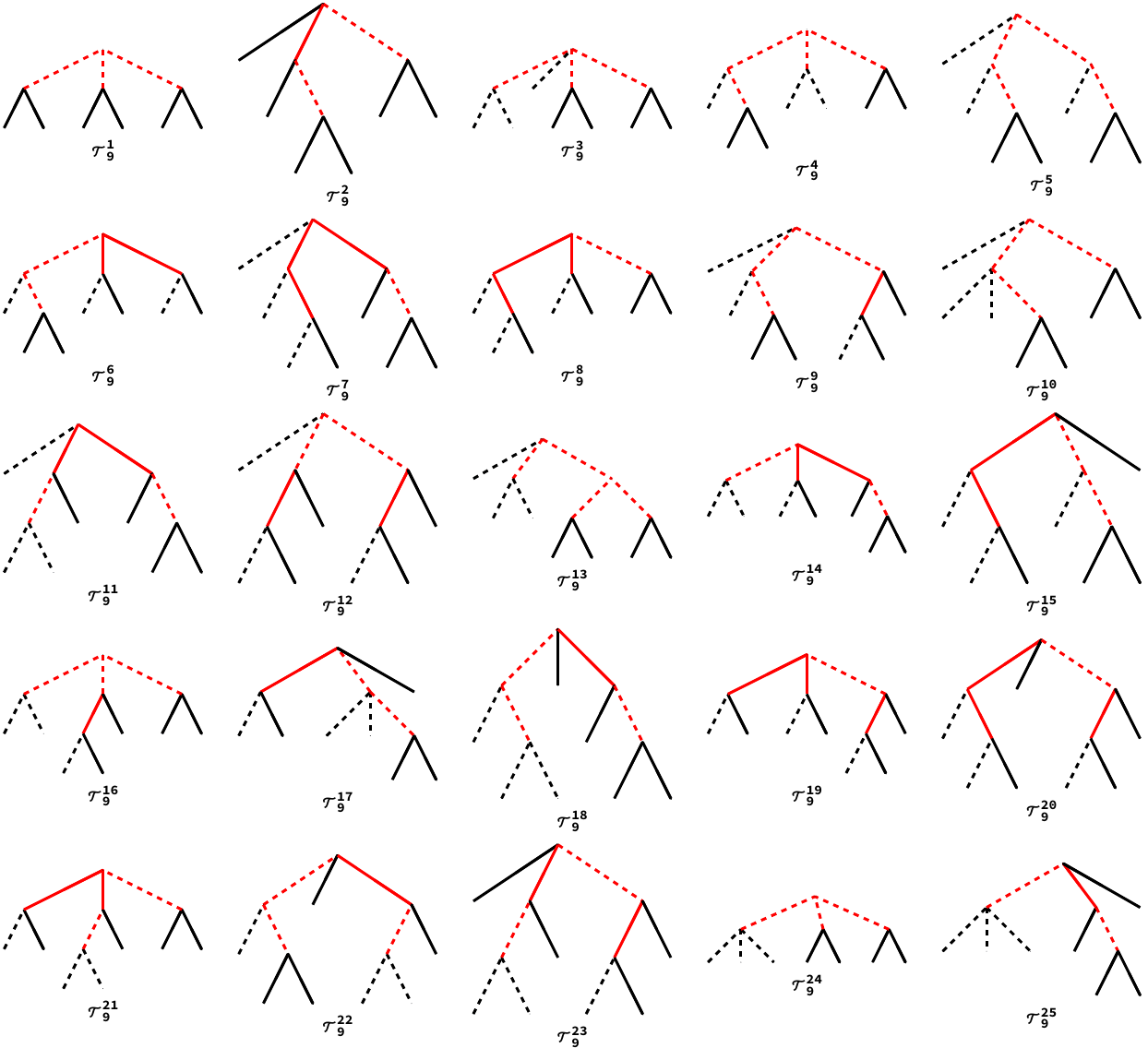}
    \caption{Feynman-diagram topologies for $d=9$ BNV operators.}
    \label{fig:d=9_topologies}
\end{figure}

At $d=9$, we find operators with different $\Delta B$ and $\Delta L$, which we discuss separately. The different topologies for all $d=9$ cases are shown in Fig.~\ref{fig:d=9_topologies}.

 \subsection{\texorpdfstring{$\Delta B=2, \Delta L=0$}{Delta B=2,Delta L=0}}

$d=9$ is the lowest mass dimension for $(\Delta B,\Delta L)=(2,0)$ operators, so these can be symmetry protected:
 \begin{multicols}{3}
 \begin{spacing}{1.4}
 \noindent
 \begin{math}
\mathcal{O}_{9, \,(2 , 0)}^{1}  \equiv ddQQQQ \,, \\
 \mathcal{O}_{9, \,(2 , 0)}^{2}  \equiv ddduQQ \,, \\
 \mathcal{O}_{9, \,(2 , 0)}^{3}  \equiv dddduu \,.
 \label{eq:dequal9basisB2L0}
 \end{math}
 \end{spacing}
 \end{multicols}
 We list the UV completions in Tabs.~\ref{tab:T9Top1B2L0}--\ref{tab:T9Top2B2L0}.  
The first two operators were mentioned in Ref.~\cite{Weinberg:1980bf}; some scalar UV completions were given in Ref.~\cite{Arnold:2012sd}, a few other UV completions in Ref.~\cite{Dorsner:2025pwe}.

Ref.~\cite{Chen:2022gjd} provides those UV completions (without vectors) for all three operators that lead to $n$-$\bar{n}$ oscillations, i.e.~contain first-generation quarks. All of their UV completions can be found in our lists as well. However, Ref.~\cite{Chen:2022gjd} determines the quark exchange symmetries that enforce second- or third-generation quarks purely through the individual vertices in the diagram, not the diagram as a whole in the constant-propagator limit. As a result, some UV completions listed in Ref.~\cite{Chen:2022gjd} actually do not satisfy their desired criterion. For example, they claim that one fermion in the SM gauge-group representation $(\vec{8},\vec{1},0)$ and two scalars in $(\bar{\vec{3}},\vec{1},1/3)$ and $(\bar{\vec{6}},\vec{1},1/3)$ (in our convention regarding $\vec{6}$ vs.~$\bar{\vec{6}}$) would generate the operator $\mathcal{O}_{9, \,(2 , 0)}^{3}$ with first-generation quarks. According to our calculation, the constant-propagator amplitude for this UV completion is antisymmetric under quark exchange and thus vanishes for first-generation quarks. Operator $\mathcal{O}_{9, \,(2 , 0)}^{3}$ is hence induced with some heavier quarks with this UV completion, or with first-generation quarks but additional derivatives and accompanying heavy-mass suppression. Some other UV completions in Ref.~\cite{Chen:2022gjd} suffer from the same problem of not checking exchange symmetries of the full amplitude.

\input{Tables_dim9/Table_dim9_Top1_B2_L0.tex}
\input{Tables_dim9/Table_dim9_Top2_B2_L0.tex}

 \subsection{\texorpdfstring{$\Delta B=1, \Delta L=3$}{Delta B=1,Delta L=3}}
 
$d=9$ is the lowest mass dimension for $(\Delta B,\Delta L)=(1,3)$ operators, so these can be symmetry protected:
 \begin{multicols}{3}
 \begin{spacing}{1.5}
 \noindent
 \begin{math}
\mathcal{O}_{9, \,(1 , 3)}^{1}  \equiv uuQLLL \,, \\
 \mathcal{O}_{9, \,(1 , 3)}^{2}  \equiv uuueLL \,, \\
 \mathcal{O}_{9, \,(1 , 3)}^{3}  \equiv uQQLL\nu \,, \\
 \mathcal{O}_{9, \,(1 , 3)}^{4}  \equiv duuLL\nu \,, \\
 \mathcal{O}_{9, \,(1 , 3)}^{5}  \equiv uuQeL\nu \,, \\
 \mathcal{O}_{9, \,(1 , 3)}^{6}  \equiv uuuee\nu \,, \\
 \mathcal{O}_{9, \,(1 , 3)}^{7}  \equiv QQQL\nu\nu \,, \\
 \mathcal{O}_{9, \,(1 , 3)}^{8}  \equiv duQL\nu\nu \,, \\
 \mathcal{O}_{9, \,(1 , 3)}^{9}  \equiv uQQe\nu\nu \,, \\
 \mathcal{O}_{9, \,(1 , 3)}^{10}  \equiv duue\nu\nu \,, \\
 \mathcal{O}_{9, \,(1 , 3)}^{11}  \equiv dQQ\nu\nu\nu \,, \\
 \mathcal{O}_{9, \,(1 , 3)}^{12}  \equiv ddu\nu\nu\nu \,. 
 \label{eq:dequal9basisB1L3}
 \end{math}
 \end{spacing}
 \end{multicols}
We list the UV completions in Tabs.~\ref{tab:T9Top1B1L3}--\ref{tab:T9Top2B1L3}. 
The two SMEFT operators have been found and discussed in Refs.~\cite{Weinberg:1980bf,Fonseca:2018ehk}. Ref.~\cite{ODonnell:1993kdg} discusses $\mathcal{O}_{9, \,(1 , 3)}^{1}$, and Ref.~\cite{Bhoonah:2025qzd} discusses one UV completion of that operator. Ref.~\cite{Durieux:2012gj} also provides one UV completion.

\input{Tables_dim9/Table_dim9_Top1_B1_L3.tex}
\input{Tables_dim9/Table_dim9_Top2_B1_L3.tex}

 \subsection{\texorpdfstring{$\Delta B=1, \Delta L=-1$}{Delta B=1,Delta L=-1}}
\label{sec:d9_B1_L-1}

These operators, listed below, have the same quantum numbers as the $d=7$ BNV operators, so cannot be protected using $B+aL$ symmetries. We list our UV completions in Tabs.~\ref{tab:T9Top1B1L-1}--\ref{tab:T9Top11B1L-1}.

Ref.~\cite{ODonnell:1993kdg} discusses $\mathcal{O}_{9, \,(1 , -1)}^{10}  \equiv dQQe\bar{L}\bar{L}$; 
Ref.~\cite{Hambye:2017qix} discusses the SMEFT operators that contain three leptons, and Ref.~\cite{Faroughy:2014tfa} provides some UV completions of these operators in a supersymmetric SM extension. Ref.~\cite{Heeck:2019kgr} provides one UV completion for $ \mathcal{O}_{9, \,(1 , -1)}^{2}  \equiv ddde\bar{e}\bar{e}$.
Ref.~\cite{Kovalenko:2002eh} also shows one UV completion that agrees with ours (operator $\mathcal{O}_{9, \,(1 , -1)}^{11}  \equiv ddue\bar{L}\bar{L}$ with topology $\mathcal{T}^1_9$ and new scalars $S_2S_5S_5$).

 \begin{multicols}{3}
 \begin{spacing}{1.5}
 \noindent
 \begin{math}
\mathcal{O}_{9, \,(1 , -1)}^{1}  \equiv dddd\bar{d}\bar{e} \,, \\
 \mathcal{O}_{9, \,(1 , -1)}^{2}  \equiv ddde\bar{e}\bar{e} \,, \\
 \mathcal{O}_{9, \,(1 , -1)}^{3}  \equiv dddQ\bar{Q}\bar{e} \,, \\
 \mathcal{O}_{9, \,(1 , -1)}^{4}  \equiv dd\bar{u}QQ\bar{e} \,, \\
 \mathcal{O}_{9, \,(1 , -1)}^{5}  \equiv dddu\bar{u}\bar{e} \,, \\
 \mathcal{O}_{9, \,(1 , -1)}^{6}  \equiv ddd\bar{d}Q\bar{L} \,, \\
 \mathcal{O}_{9, \,(1 , -1)}^{7}  \equiv ddd\bar{e}L\bar{L} \,, \\
 \mathcal{O}_{9, \,(1 , -1)}^{8}  \equiv ddQe\bar{e}\bar{L} \,, \\
 \mathcal{O}_{9, \,(1 , -1)}^{9}  \equiv ddQL\bar{L}\bar{L} \,, \\
 \mathcal{O}_{9, \,(1 , -1)}^{10}  \equiv dQQe\bar{L}\bar{L} \,, \\
 \mathcal{O}_{9, \,(1 , -1)}^{11}  \equiv ddue\bar{L}\bar{L} \,, \\
 \mathcal{O}_{9, \,(1 , -1)}^{12}  \equiv ddQQ\bar{Q}\bar{L} \,, \\
 \mathcal{O}_{9, \,(1 , -1)}^{13}  \equiv dddu\bar{Q}\bar{L} \,, \\
 \mathcal{O}_{9, \,(1 , -1)}^{14}  \equiv d\bar{u}QQQ\bar{L} \,, \\
 \mathcal{O}_{9, \,(1 , -1)}^{15}  \equiv ddu\bar{u}Q\bar{L} \,, \\
 \mathcal{O}_{9, \,(1 , -1)}^{16}  \equiv ddd\bar{L}\bar{L}\nu \,, \\
 \mathcal{O}_{9, \,(1 , -1)}^{17}  \equiv dd\bar{d}QQ\bar{\nu} \,, \\
 \mathcal{O}_{9, \,(1 , -1)}^{18}  \equiv ddd\bar{d}u\bar{\nu} \,, \\
 \mathcal{O}_{9, \,(1 , -1)}^{19}  \equiv ddQ\bar{e}L\bar{\nu} \,, \\
 \mathcal{O}_{9, \,(1 , -1)}^{20}  \equiv dQQe\bar{e}\bar{\nu} \,, \\
 \mathcal{O}_{9, \,(1 , -1)}^{21}  \equiv ddue\bar{e}\bar{\nu} \,, \\
 \mathcal{O}_{9, \,(1 , -1)}^{22}  \equiv dQQL\bar{L}\bar{\nu} \,, \\
 \mathcal{O}_{9, \,(1 , -1)}^{23}  \equiv QQQe\bar{L}\bar{\nu} \,, \\
 \mathcal{O}_{9, \,(1 , -1)}^{24}  \equiv dduL\bar{L}\bar{\nu} \,, \\
 \mathcal{O}_{9, \,(1 , -1)}^{25}  \equiv duQe\bar{L}\bar{\nu} \,, \\
 \mathcal{O}_{9, \,(1 , -1)}^{26}  \equiv dQQQ\bar{Q}\bar{\nu} \,, \\
 \mathcal{O}_{9, \,(1 , -1)}^{27}  \equiv dduQ\bar{Q}\bar{\nu} \,, \\
 \mathcal{O}_{9, \,(1 , -1)}^{28}  \equiv \bar{u}QQQQ\bar{\nu} \,, \\
 \mathcal{O}_{9, \,(1 , -1)}^{29}  \equiv du\bar{u}QQ\bar{\nu} \,, \\
 \mathcal{O}_{9, \,(1 , -1)}^{30}  \equiv dduu\bar{u}\bar{\nu} \,, \\
 \mathcal{O}_{9, \,(1 , -1)}^{31}  \equiv ddd\bar{e}\nu\bar{\nu} \,, \\
 \mathcal{O}_{9, \,(1 , -1)}^{32}  \equiv ddQ\bar{L}\nu\bar{\nu} \,, \\
 \mathcal{O}_{9, \,(1 , -1)}^{33}  \equiv QQQL\bar{\nu}\bar{\nu} \,, \\
 \mathcal{O}_{9, \,(1 , -1)}^{34}  \equiv duQL\bar{\nu}\bar{\nu} \,, \\
 \mathcal{O}_{9, \,(1 , -1)}^{35}  \equiv uQQe\bar{\nu}\bar{\nu} \,, \\
 \mathcal{O}_{9, \,(1 , -1)}^{36}  \equiv duue\bar{\nu}\bar{\nu} \,, \\
 \mathcal{O}_{9, \,(1 , -1)}^{37}  \equiv dQQ\nu\bar{\nu}\bar{\nu} \,, \\
 \mathcal{O}_{9, \,(1 , -1)}^{38}  \equiv ddu\nu\bar{\nu}\bar{\nu} \,, \\
 \mathcal{O}_{9, \,(1 , -1)}^{39}  \equiv H\bar{H}\bar{H}ddQ\bar{e} \,, \\
 \mathcal{O}_{9, \,(1 , -1)}^{40}  \equiv H\bar{H}\bar{H}dQQ\bar{L} \,, \\
 \mathcal{O}_{9, \,(1 , -1)}^{41}  \equiv H\bar{H}\bar{H}ddu\bar{L} \,, \\
 \mathcal{O}_{9, \,(1 , -1)}^{42}  \equiv HH\bar{H}ddd\bar{L} \,, \\
 \mathcal{O}_{9, \,(1 , -1)}^{43}  \equiv \bar{H}\bar{H}\bar{H}QQQ\bar{e} \,, \\
 \mathcal{O}_{9, \,(1 , -1)}^{44}  \equiv \bar{H}\bar{H}\bar{H}uQQ\bar{L} \,, \\
 \mathcal{O}_{9, \,(1 , -1)}^{45}  \equiv HH\bar{H}ddQ\bar{\nu} \,, \\
 \mathcal{O}_{9, \,(1 , -1)}^{46}  \equiv H\bar{H}\bar{H}QQQ\bar{\nu} \,, \\
 \mathcal{O}_{9, \,(1 , -1)}^{47}  \equiv H\bar{H}\bar{H}duQ\bar{\nu} \,.
 \label{eq:dequal9basisB1L-1}
 \end{math}
 \end{spacing}
 \end{multicols}

\input{Tables_dim9/Table_dim9_Top1_B1_L-1.tex} \input{Tables_dim9/Table_dim9_Top2_B1_L-1.tex}

\input{Tables_dim9/Table_dim9_Top3_B1_L-1.tex} \input{Tables_dim9/Table_dim9_Top4_B1_L-1.tex}
\input{Tables_dim9/Table_dim9_Top5_B1_L-1.tex} \input{Tables_dim9/Table_dim9_Top6_B1_L-1.tex}
\input{Tables_dim9/Table_dim9_Top7_B1_L-1.tex} \input{Tables_dim9/Table_dim9_Top8_B1_L-1.tex}
\input{Tables_dim9/Table_dim9_Top9_B1_L-1.tex}
\input{Tables_dim9/Table_dim9_Top10_B1_L-1.tex} \input{Tables_dim9/Table_dim9_Top11_B1_L-1.tex}
\input{Tables_dim9/Table_dim9_Top12_B1_L-1.tex} \input{Tables_dim9/Table_dim9_Top13_B1_L-1.tex}
\input{Tables_dim9/Table_dim9_Top14_B1_L-1.tex} \input{Tables_dim9/Table_dim9_Top15_B1_L-1.tex}
\input{Tables_dim9/Table_dim9_Top16_B1_L-1.tex} \input{Tables_dim9/Table_dim9_Top17_B1_L-1.tex}
\input{Tables_dim9/Table_dim9_Top18_B1_L-1.tex} \input{Tables_dim9/Table_dim9_Top19_B1_L-1.tex}
\input{Tables_dim9/Table_dim9_Top20_B1_L-1.tex} \input{Tables_dim9/Table_dim9_Top21_B1_L-1.tex}
\input{Tables_dim9/Table_dim9_Top22_B1_L-1.tex} \input{Tables_dim9/Table_dim9_Top23_B1_L-1.tex}
\input{Tables_dim9/Table_dim9_Top24_B1_L-1.tex} \input{Tables_dim9/Table_dim9_Top25_B1_L-1.tex}

\clearpage
 
 
\section{Dimension 10}
\label{sec:d10}

At $d=10$, we find operators with different $\Delta B$ and $\Delta L$, which we discuss separately. The different topologies for all cases are shown in Fig.~\ref{fig:d=10_topologies}.

\begin{figure}[tbh]
    \centering
    \includegraphics[scale=0.53]{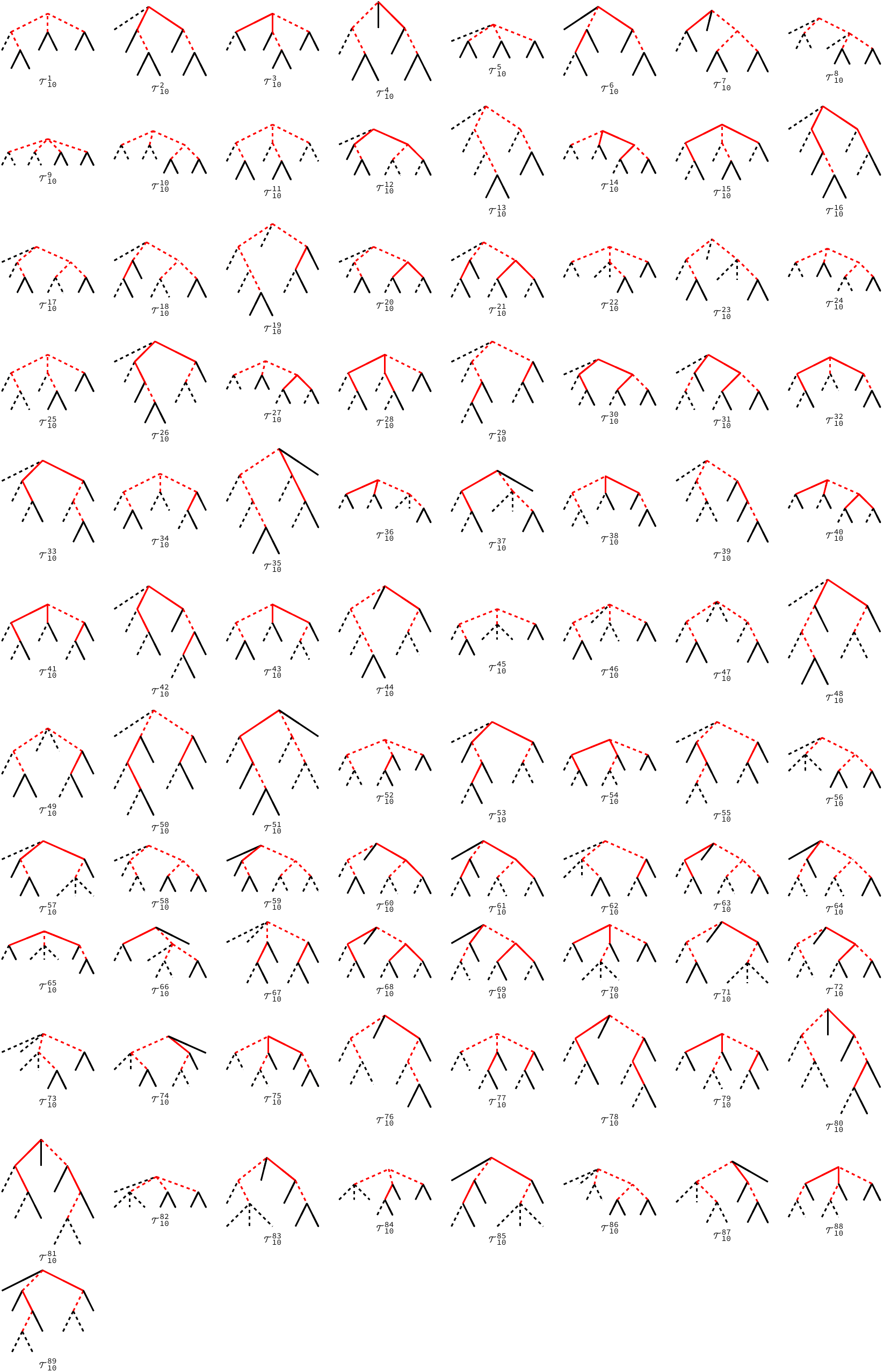} 
    \caption{Feynman-diagram topologies for $d=10$ BNV operators.}
     \label{fig:d=10_topologies}
\end{figure}

\clearpage
 
 \subsection{\texorpdfstring{$\Delta B=1, \Delta L=-3$}{Delta B=1,Delta L=-3}}
 
$d=10$ is the lowest mass dimension for $(\Delta B,\Delta L)=(1,-3)$ operators, so these can be symmetry protected:
 \begin{multicols}{3}
 \begin{spacing}{1.5}
 \noindent
 \begin{math}
\mathcal{O}_{10, \,(1 , -3)}^{1} \equiv \bar{H}ddd\bar{L}\bar{L}\bar{L} \,, \\
 \mathcal{O}_{10, \,(1 , -3)}^{2} \equiv \bar{H}ddd\bar{e}\bar{L}\bar{\nu} \,, \\
 \mathcal{O}_{10, \,(1 , -3)}^{3} \equiv \bar{H}ddQ\bar{L}\bar{L}\bar{\nu} \,, \\
 \mathcal{O}_{10, \,(1 , -3)}^{4} \equiv \bar{H}ddu\bar{L}\bar{\nu}\bar{\nu} \,, \\
 \mathcal{O}_{10, \,(1 , -3)}^{5} \equiv Hddd\bar{L}\bar{\nu}\bar{\nu} \,, \\
 \mathcal{O}_{10, \,(1 , -3)}^{6} \equiv \bar{H}ddQ\bar{e}\bar{\nu}\bar{\nu} \,, \\
 \mathcal{O}_{10, \,(1 , -3)}^{7} \equiv \bar{H}dQQ\bar{L}\bar{\nu}\bar{\nu} \,, \\
 \mathcal{O}_{10, \,(1 , -3)}^{8} \equiv HddQ\bar{\nu}\bar{\nu}\bar{\nu} \,, \\
 \mathcal{O}_{10, \,(1 , -3)}^{9} \equiv \bar{H}duQ\bar{\nu}\bar{\nu}\bar{\nu} \,, \\
 \mathcal{O}_{10, \,(1 , -3)}^{10} \equiv \bar{H}QQQ\bar{\nu}\bar{\nu}\bar{\nu} \, .
 \label{eq:dequal10basisB1L-3}
 \end{math}
 \end{spacing}
 \end{multicols}
We list the UV completions for these operators in Tabs.~\ref{tab:T10Top1B1L-3}--\ref{tab:T10Top7B1L-3}.  
The single SMEFT operator has been found and discussed in Refs.~\cite{Weinberg:1980bf,Fonseca:2018ehk}, with one UV completion given in Refs.~\cite{Weinberg:1980bf,Arnold:2012sd}.

\input{Tables_dim10/Table_dim10_Top1_B1_L-3.tex} \input{Tables_dim10/Table_dim10_Top2_B1_L-3.tex}
\input{Tables_dim10/Table_dim10_Top3_B1_L-3.tex} \input{Tables_dim10/Table_dim10_Top4_B1_L-3.tex}
\input{Tables_dim10/Table_dim10_Top5_B1_L-3.tex} \input{Tables_dim10/Table_dim10_Top6_B1_L-3.tex}
\input{Tables_dim10/Table_dim10_Top7_B1_L-3.tex}

\subsection{\texorpdfstring{$\Delta B=1, \Delta L=1$}{Delta B=1,Delta L=1}}

The $\Delta B=\Delta L=1$ operators have the same quantum numbers as the BNV operators at $d=6$ and $d=8$ and cannot be protected by $B+a L$ symmetries.
 \begin{multicols}{3}
 \begin{spacing}{1.5}
 \noindent
 \begin{math}
\mathcal{O}_{10, \,(1 , 1)}^{1} \equiv Hdd\bar{d}QQL \,, \\
 \mathcal{O}_{10, \,(1 , 1)}^{2} \equiv Hd\bar{d}QQQe \,, \\
 \mathcal{O}_{10, \,(1 , 1)}^{3} \equiv Hddd\bar{d}uL \,, \\
 \mathcal{O}_{10, \,(1 , 1)}^{4} \equiv Hdd\bar{d}uQe \,, \\
 \mathcal{O}_{10, \,(1 , 1)}^{5} \equiv HddQ\bar{e}LL \,, \\
 \mathcal{O}_{10, \,(1 , 1)}^{6} \equiv HdQQe\bar{e}L \,, \\
 \mathcal{O}_{10, \,(1 , 1)}^{7} \equiv HQQQee\bar{e} \,, \\
 \mathcal{O}_{10, \,(1 , 1)}^{8} \equiv Hddue\bar{e}L \,, \\
 \mathcal{O}_{10, \,(1 , 1)}^{9} \equiv HduQee\bar{e} \,, \\
 \mathcal{O}_{10, \,(1 , 1)}^{10} \equiv \bar{H}\bar{d}QQQQL \,, \\
 \mathcal{O}_{10, \,(1 , 1)}^{11} \equiv \bar{H}d\bar{d}uQQL \,, \\
 \mathcal{O}_{10, \,(1 , 1)}^{12} \equiv \bar{H}\bar{d}uQQQe \,, \\
 \mathcal{O}_{10, \,(1 , 1)}^{13} \equiv \bar{H}dd\bar{d}uuL \,, \\
 \mathcal{O}_{10, \,(1 , 1)}^{14} \equiv \bar{H}d\bar{d}uuQe \,, \\
 \mathcal{O}_{10, \,(1 , 1)}^{15} \equiv \bar{H}QQQ\bar{e}LL \,, \\
 \mathcal{O}_{10, \,(1 , 1)}^{16} \equiv \bar{H}duQ\bar{e}LL \,, \\
 \mathcal{O}_{10, \,(1 , 1)}^{17} \equiv \bar{H}uQQe\bar{e}L \,, \\
 \mathcal{O}_{10, \,(1 , 1)}^{18} \equiv \bar{H}duue\bar{e}L \,, \\
 \mathcal{O}_{10, \,(1 , 1)}^{19} \equiv \bar{H}uuQee\bar{e} \,, \\
 \mathcal{O}_{10, \,(1 , 1)}^{20} \equiv \bar{H}uQQLL\bar{L} \,, \\
 \mathcal{O}_{10, \,(1 , 1)}^{21} \equiv \bar{H}duuLL\bar{L} \,, \\
 \mathcal{O}_{10, \,(1 , 1)}^{22} \equiv \bar{H}uuQeL\bar{L} \,, \\
 \mathcal{O}_{10, \,(1 , 1)}^{23} \equiv \bar{H}uuuee\bar{L} \,, \\
 \mathcal{O}_{10, \,(1 , 1)}^{24} \equiv \bar{H}uQQQ\bar{Q}L \,, \\
 \mathcal{O}_{10, \,(1 , 1)}^{25} \equiv \bar{H}duuQ\bar{Q}L \,, \\
 \mathcal{O}_{10, \,(1 , 1)}^{26} \equiv \bar{H}uuQQ\bar{Q}e \,, \\
 \mathcal{O}_{10, \,(1 , 1)}^{27} \equiv \bar{H}duuu\bar{Q}e \,, \\
 \mathcal{O}_{10, \,(1 , 1)}^{28} \equiv \bar{H}uu\bar{u}QQL \,, \\
 \mathcal{O}_{10, \,(1 , 1)}^{29} \equiv \bar{H}duuu\bar{u}L \,, \\
 \mathcal{O}_{10, \,(1 , 1)}^{30} \equiv \bar{H}uuu\bar{u}Qe \,, \\
 \mathcal{O}_{10, \,(1 , 1)}^{31} \equiv HdQQLL\bar{L} \,, \\
 \mathcal{O}_{10, \,(1 , 1)}^{32} \equiv HQQQeL\bar{L} \,, \\
 \mathcal{O}_{10, \,(1 , 1)}^{33} \equiv HdduLL\bar{L} \,, \\
 \mathcal{O}_{10, \,(1 , 1)}^{34} \equiv HduQeL\bar{L} \,, \\
 \mathcal{O}_{10, \,(1 , 1)}^{35} \equiv HuQQee\bar{L} \,, \\
 \mathcal{O}_{10, \,(1 , 1)}^{36} \equiv Hduuee\bar{L} \,, \\
 \mathcal{O}_{10, \,(1 , 1)}^{37} \equiv HdQQQ\bar{Q}L \,, \\
 \mathcal{O}_{10, \,(1 , 1)}^{38} \equiv HQQQQ\bar{Q}e \,, \\
 \mathcal{O}_{10, \,(1 , 1)}^{39} \equiv HdduQ\bar{Q}L \,, \\
 \mathcal{O}_{10, \,(1 , 1)}^{40} \equiv HduQQ\bar{Q}e \,, \\
 \mathcal{O}_{10, \,(1 , 1)}^{41} \equiv Hdduu\bar{Q}e \,, \\
 \mathcal{O}_{10, \,(1 , 1)}^{42} \equiv H\bar{u}QQQQL \,, \\
 \mathcal{O}_{10, \,(1 , 1)}^{43} \equiv Hdu\bar{u}QQL \,, \\
 \mathcal{O}_{10, \,(1 , 1)}^{44} \equiv Hu\bar{u}QQQe \,, \\
 \mathcal{O}_{10, \,(1 , 1)}^{45} \equiv Hdduu\bar{u}L \,, \\
 \mathcal{O}_{10, \,(1 , 1)}^{46} \equiv Hduu\bar{u}Qe \,, \\
 \mathcal{O}_{10, \,(1 , 1)}^{47} \equiv Hddd\bar{d}Q\nu \,, \\
 \mathcal{O}_{10, \,(1 , 1)}^{48} \equiv Hddd\bar{e}L\nu \,, \\
 \mathcal{O}_{10, \,(1 , 1)}^{49} \equiv HddQe\bar{e}\nu \,, \\
 \mathcal{O}_{10, \,(1 , 1)}^{50} \equiv \bar{H}d\bar{d}QQQ\nu \,, \\
 \mathcal{O}_{10, \,(1 , 1)}^{51} \equiv \bar{H}dd\bar{d}uQ\nu \,, \\
 \mathcal{O}_{10, \,(1 , 1)}^{52} \equiv \bar{H}dQQ\bar{e}L\nu \,, \\
 \mathcal{O}_{10, \,(1 , 1)}^{53} \equiv \bar{H}QQQe\bar{e}\nu \,, \\
 \mathcal{O}_{10, \,(1 , 1)}^{54} \equiv \bar{H}ddu\bar{e}L\nu \,, \\
 \mathcal{O}_{10, \,(1 , 1)}^{55} \equiv \bar{H}duQe\bar{e}\nu \,, \\
 \mathcal{O}_{10, \,(1 , 1)}^{56} \equiv \bar{H}QQQL\bar{L}\nu \,, \\
 \mathcal{O}_{10, \,(1 , 1)}^{57} \equiv \bar{H}duQL\bar{L}\nu \,, \\
 \mathcal{O}_{10, \,(1 , 1)}^{58} \equiv \bar{H}uQQe\bar{L}\nu \,, \\
 \mathcal{O}_{10, \,(1 , 1)}^{59} \equiv \bar{H}duue\bar{L}\nu \,, \\
 \mathcal{O}_{10, \,(1 , 1)}^{60} \equiv \bar{H}QQQQ\bar{Q}\nu \,, \\
 \mathcal{O}_{10, \,(1 , 1)}^{61} \equiv \bar{H}duQQ\bar{Q}\nu \,, \\
 \mathcal{O}_{10, \,(1 , 1)}^{62} \equiv \bar{H}dduu\bar{Q}\nu \,, \\
 \mathcal{O}_{10, \,(1 , 1)}^{63} \equiv \bar{H}u\bar{u}QQQ\nu \,, \\
 \mathcal{O}_{10, \,(1 , 1)}^{64} \equiv \bar{H}duu\bar{u}Q\nu \,, \\
 \mathcal{O}_{10, \,(1 , 1)}^{65} \equiv HddQL\bar{L}\nu \,, \\
 \mathcal{O}_{10, \,(1 , 1)}^{66} \equiv HdQQe\bar{L}\nu \,, \\
 \mathcal{O}_{10, \,(1 , 1)}^{67} \equiv Hddue\bar{L}\nu \,, \\
 \mathcal{O}_{10, \,(1 , 1)}^{68} \equiv \bar{H}uuQLL\bar{\nu} \,, \\
 \mathcal{O}_{10, \,(1 , 1)}^{69} \equiv \bar{H}uuueL\bar{\nu} \,, \\
 \mathcal{O}_{10, \,(1 , 1)}^{70} \equiv HQQQLL\bar{\nu} \,, \\
 \mathcal{O}_{10, \,(1 , 1)}^{71} \equiv HduQLL\bar{\nu} \,, \\
 \mathcal{O}_{10, \,(1 , 1)}^{72} \equiv HuQQeL\bar{\nu} \,, \\
 \mathcal{O}_{10, \,(1 , 1)}^{73} \equiv HduueL\bar{\nu} \,, \\
 \mathcal{O}_{10, \,(1 , 1)}^{74} \equiv HuuQee\bar{\nu} \,, \\
 \mathcal{O}_{10, \,(1 , 1)}^{75} \equiv HddQQ\bar{Q}\nu \,, \\
 \mathcal{O}_{10, \,(1 , 1)}^{76} \equiv Hdddu\bar{Q}\nu \,, \\
 \mathcal{O}_{10, \,(1 , 1)}^{77} \equiv Hd\bar{u}QQQ\nu \,, \\
 \mathcal{O}_{10, \,(1 , 1)}^{78} \equiv Hddu\bar{u}Q\nu \,, \\
 \mathcal{O}_{10, \,(1 , 1)}^{79} \equiv \bar{H}ddQ\bar{e}\nu\nu \,, \\
 \mathcal{O}_{10, \,(1 , 1)}^{80} \equiv \bar{H}dQQ\bar{L}\nu\nu \,, \\
 \mathcal{O}_{10, \,(1 , 1)}^{81} \equiv \bar{H}ddu\bar{L}\nu\nu \,, \\
 \mathcal{O}_{10, \,(1 , 1)}^{82} \equiv Hddd\bar{L}\nu\nu \,, \\
 \mathcal{O}_{10, \,(1 , 1)}^{83} \equiv HdQQL\nu\bar{\nu} \,, \\
 \mathcal{O}_{10, \,(1 , 1)}^{84} \equiv \bar{H}uQQL\nu\bar{\nu} \,, \\
 \mathcal{O}_{10, \,(1 , 1)}^{85} \equiv \bar{H}duuL\nu\bar{\nu} \,, \\
 \mathcal{O}_{10, \,(1 , 1)}^{86} \equiv \bar{H}uuQe\nu\bar{\nu} \,, \\
 \mathcal{O}_{10, \,(1 , 1)}^{87} \equiv HQQQe\nu\bar{\nu} \,, \\
 \mathcal{O}_{10, \,(1 , 1)}^{88} \equiv HdduL\nu\bar{\nu} \,, \\
 \mathcal{O}_{10, \,(1 , 1)}^{89} \equiv HduQe\nu\bar{\nu} \,, \\
 \mathcal{O}_{10, \,(1 , 1)}^{90} \equiv HddQ\nu\nu\bar{\nu} \,, \\
 \mathcal{O}_{10, \,(1 , 1)}^{91} \equiv \bar{H}QQQ\nu\nu\bar{\nu} \,, \\
 \mathcal{O}_{10, \,(1 , 1)}^{92} \equiv \bar{H}duQ\nu\nu\bar{\nu} \,, \\
 \mathcal{O}_{10, \,(1 , 1)}^{93} \equiv HHH\bar{H}ddQL \,, \\
 \mathcal{O}_{10, \,(1 , 1)}^{94} \equiv HHH\bar{H}dQQe \,, \\
 \mathcal{O}_{10, \,(1 , 1)}^{95} \equiv H\bar{H}\bar{H}\bar{H}uuQL \,, \\
 \mathcal{O}_{10, \,(1 , 1)}^{96} \equiv HH\bar{H}\bar{H}QQQL \,, \\
 \mathcal{O}_{10, \,(1 , 1)}^{97} \equiv HH\bar{H}\bar{H}duQL \,, \\
 \mathcal{O}_{10, \,(1 , 1)}^{98} \equiv HH\bar{H}\bar{H}uQQe \,, \\
 \mathcal{O}_{10, \,(1 , 1)}^{99} \equiv HH\bar{H}\bar{H}duue \,, \\
 \mathcal{O}_{10, \,(1 , 1)}^{100} \equiv HH\bar{H}\bar{H}dQQ\nu \,, \\
 \mathcal{O}_{10, \,(1 , 1)}^{101} \equiv H\bar{H}\bar{H}\bar{H}uQQ\nu \,, \\
 \mathcal{O}_{10, \,(1 , 1)}^{102} \equiv HH\bar{H}\bar{H}ddu\nu \,.
 \label{eq:dequal10basisB1L1}
 \end{math}
 \end{spacing}
 \end{multicols}

Ref.~\cite{ODonnell:1993kdg} discusses four of these operators;
Ref.~\cite{Hambye:2017qix} discusses the SMEFT operators that contain three leptons and provides a few UV completions.
Ref.~\cite{Liao:2025vlj} also gives one UV completion (see their Fig.~1).

\clearpage
 

\section{Dimension 11}
\label{sec:d11}

At $d=11$, we find operators with different $\Delta B$ and $\Delta L$, which we discuss separately. None of them can be protected using $B+a L$ symmetries. The different topologies for all cases are shown in Fig.~\ref{fig:d=11_topologies}.
 
 \begin{figure}
    \centering
    \includegraphics[scale=0.5]{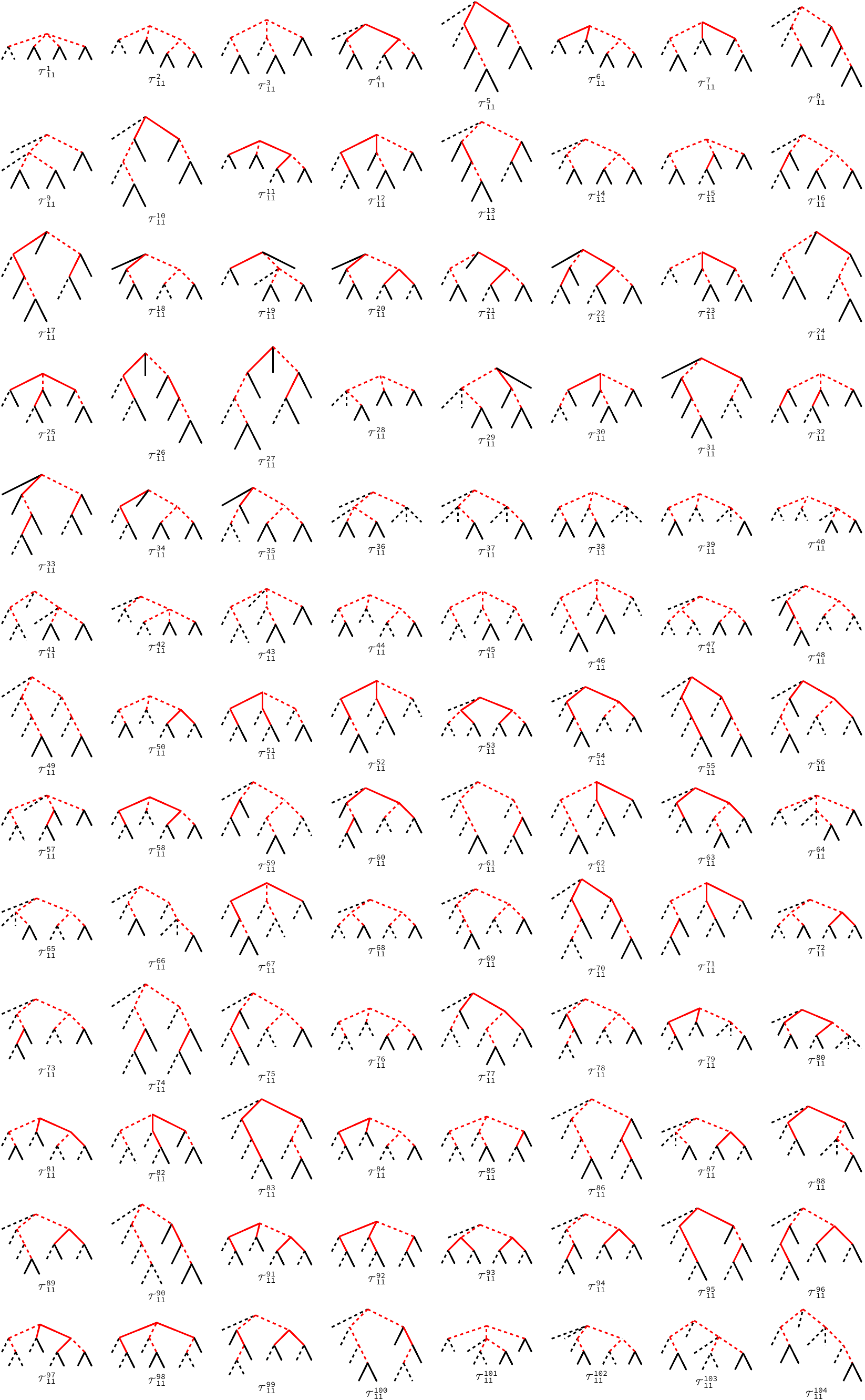}
    \caption*{\hfill Contd.}
\end{figure}

\begin{figure}
    \centering
    \includegraphics[scale=0.5]{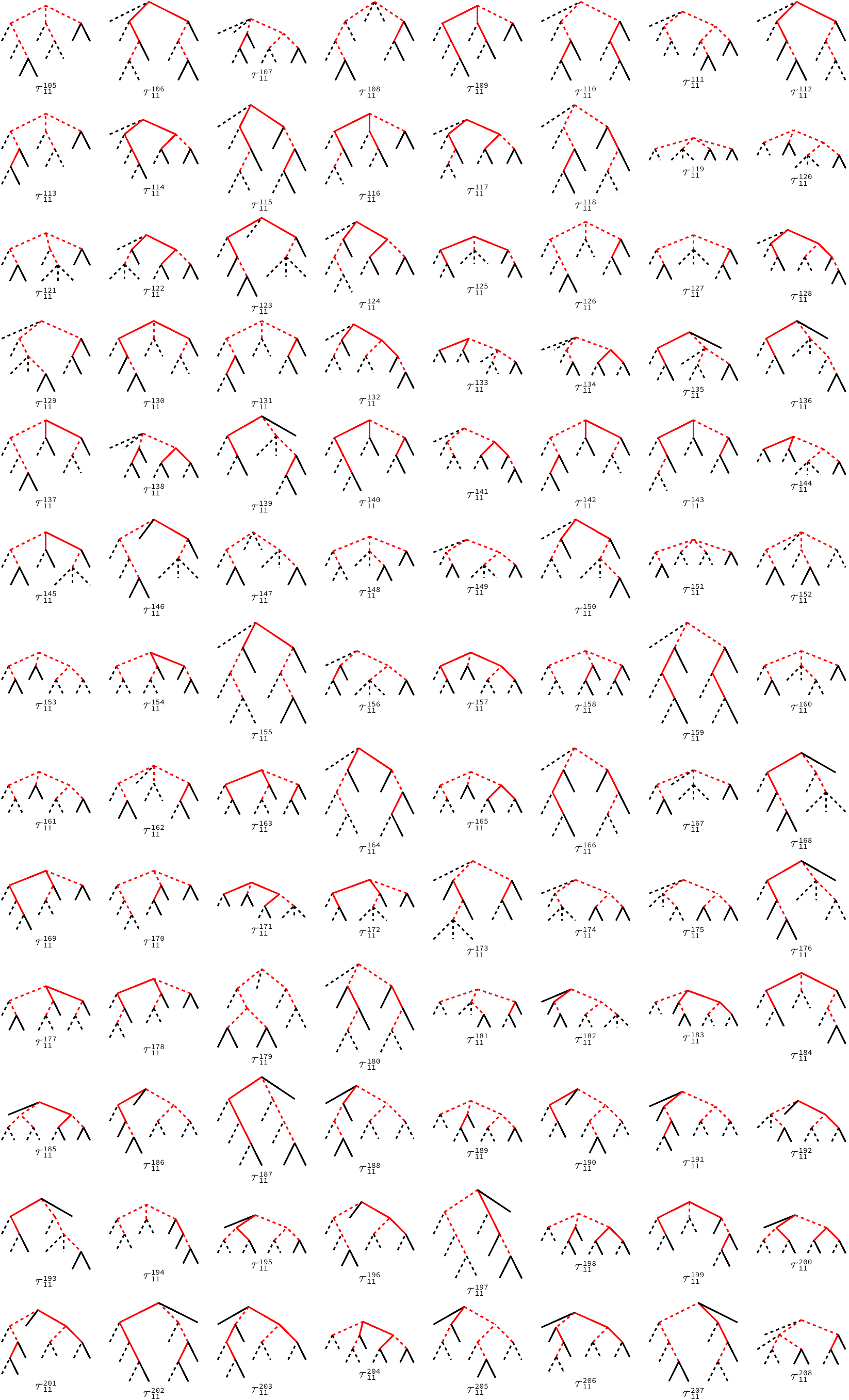}
    \caption*{\hfill Contd.}
\end{figure}

\begin{figure}
    \centering
    \includegraphics[scale=0.5]{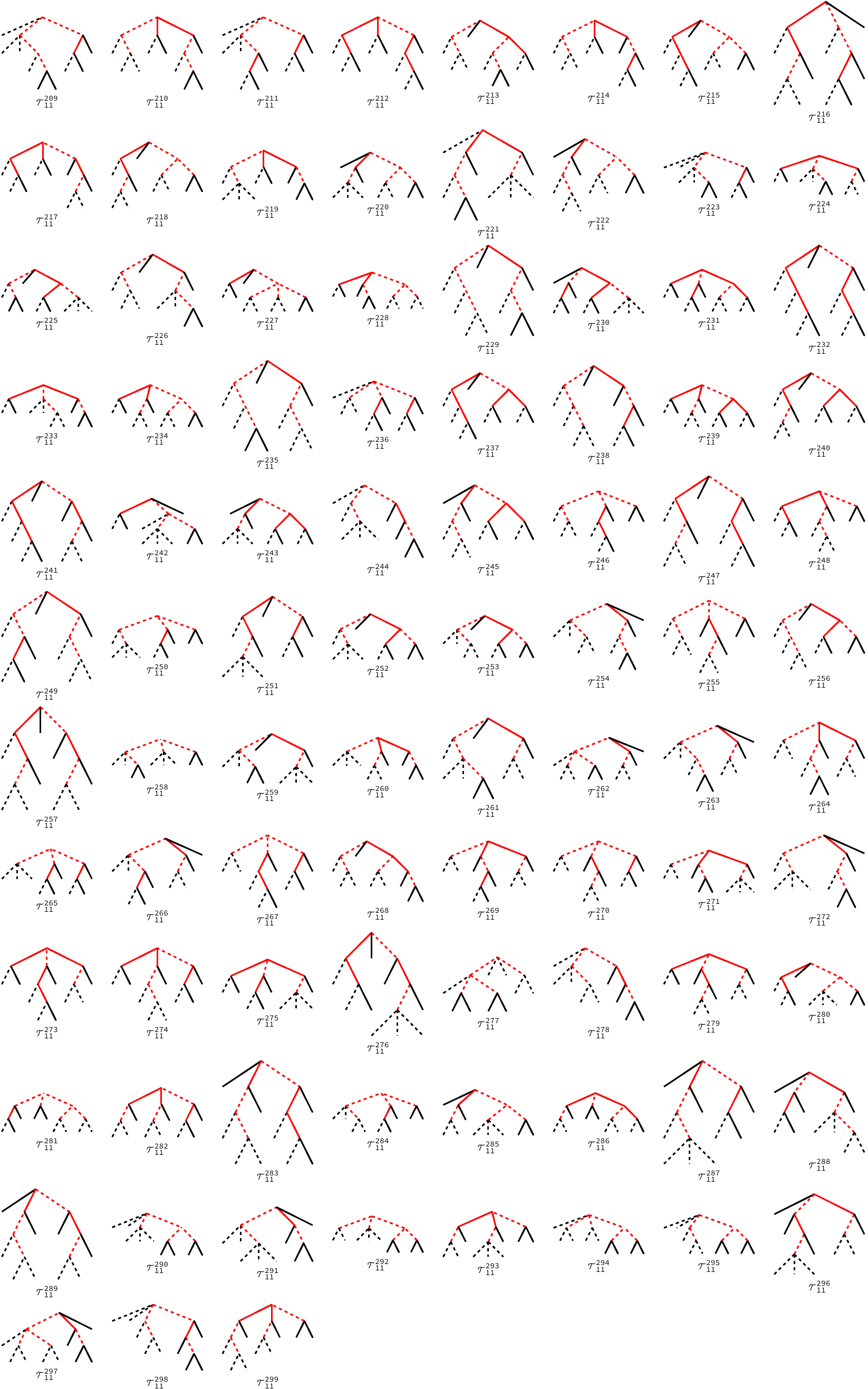}
    \caption{Feynman-diagram topologies for $d=11$ BNV operators.}
    \label{fig:d=11_topologies}
\end{figure}

  \subsection{\texorpdfstring{$\Delta B=1, \Delta L=-1$}{Delta B=1,Delta L=-1}}

 \begin{multicols}{3}
 \begin{spacing}{1.5}
 \noindent
 \begin{math}
 \mathcal{O}_{11, \,(1 , -1)}^{1} \equiv H\bar{H}dddd\bar{d}\bar{e} \,, \\
 \mathcal{O}_{11, \,(1 , -1)}^{2} \equiv H\bar{H}ddde\bar{e}\bar{e} \,, \\
 \mathcal{O}_{11, \,(1 , -1)}^{3} \equiv H\bar{H}dddQ\bar{Q}\bar{e} \,, \\
 \mathcal{O}_{11, \,(1 , -1)}^{4} \equiv H\bar{H}dd\bar{u}QQ\bar{e} \,, \\
 \mathcal{O}_{11, \,(1 , -1)}^{5} \equiv H\bar{H}dddu\bar{u}\bar{e} \,, \\
 \mathcal{O}_{11, \,(1 , -1)}^{6} \equiv \bar{H}\bar{H}dd\bar{d}QQ\bar{e} \,, \\
 \mathcal{O}_{11, \,(1 , -1)}^{7} \equiv \bar{H}\bar{H}ddQ\bar{e}\bar{e}L \,, \\
 \mathcal{O}_{11, \,(1 , -1)}^{8} \equiv \bar{H}\bar{H}dQQe\bar{e}\bar{e} \,, \\
 \mathcal{O}_{11, \,(1 , -1)}^{9} \equiv \bar{H}\bar{H}dQQQ\bar{Q}\bar{e} \,, \\
 \mathcal{O}_{11, \,(1 , -1)}^{10} \equiv \bar{H}\bar{H}dduQ\bar{Q}\bar{e} \,, \\
 \mathcal{O}_{11, \,(1 , -1)}^{11} \equiv \bar{H}\bar{H}\bar{u}QQQQ\bar{e} \,, \\
 \mathcal{O}_{11, \,(1 , -1)}^{12} \equiv \bar{H}\bar{H}du\bar{u}QQ\bar{e} \,, \\
 \mathcal{O}_{11, \,(1 , -1)}^{13} \equiv \bar{H}\bar{H}d\bar{d}QQQ\bar{L} \,, \\
 \mathcal{O}_{11, \,(1 , -1)}^{14} \equiv \bar{H}\bar{H}dd\bar{d}uQ\bar{L} \,, \\
 \mathcal{O}_{11, \,(1 , -1)}^{15} \equiv \bar{H}\bar{H}dQQ\bar{e}L\bar{L} \,, \\
 \mathcal{O}_{11, \,(1 , -1)}^{16} \equiv \bar{H}\bar{H}QQQe\bar{e}\bar{L} \,, \\
 \mathcal{O}_{11, \,(1 , -1)}^{17} \equiv \bar{H}\bar{H}ddu\bar{e}L\bar{L} \,, \\
 \mathcal{O}_{11, \,(1 , -1)}^{18} \equiv \bar{H}\bar{H}duQe\bar{e}\bar{L} \,, \\
 \mathcal{O}_{11, \,(1 , -1)}^{19} \equiv \bar{H}\bar{H}QQQL\bar{L}\bar{L} \,, \\
 \mathcal{O}_{11, \,(1 , -1)}^{20} \equiv \bar{H}\bar{H}duQL\bar{L}\bar{L} \,, \\
 \mathcal{O}_{11, \,(1 , -1)}^{21} \equiv \bar{H}\bar{H}uQQe\bar{L}\bar{L} \,, \\
 \mathcal{O}_{11, \,(1 , -1)}^{22} \equiv \bar{H}\bar{H}duue\bar{L}\bar{L} \,, \\
 \mathcal{O}_{11, \,(1 , -1)}^{23} \equiv \bar{H}\bar{H}QQQQ\bar{Q}\bar{L} \,, \\
 \mathcal{O}_{11, \,(1 , -1)}^{24} \equiv \bar{H}\bar{H}duQQ\bar{Q}\bar{L} \,, \\
 \mathcal{O}_{11, \,(1 , -1)}^{25} \equiv \bar{H}\bar{H}dduu\bar{Q}\bar{L} \,, \\
 \mathcal{O}_{11, \,(1 , -1)}^{26} \equiv \bar{H}\bar{H}u\bar{u}QQQ\bar{L} \,, \\
 \mathcal{O}_{11, \,(1 , -1)}^{27} \equiv \bar{H}\bar{H}duu\bar{u}Q\bar{L} \,, \\
 \mathcal{O}_{11, \,(1 , -1)}^{28} \equiv H\bar{H}ddd\bar{d}Q\bar{L} \,, \\
 \mathcal{O}_{11, \,(1 , -1)}^{29} \equiv H\bar{H}ddd\bar{e}L\bar{L} \,, \\
 \mathcal{O}_{11, \,(1 , -1)}^{30} \equiv H\bar{H}ddQe\bar{e}\bar{L} \,, \\
 \mathcal{O}_{11, \,(1 , -1)}^{31} \equiv H\bar{H}ddQL\bar{L}\bar{L} \,, \\
 \mathcal{O}_{11, \,(1 , -1)}^{32} \equiv H\bar{H}dQQe\bar{L}\bar{L} \,, \\
 \mathcal{O}_{11, \,(1 , -1)}^{33} \equiv H\bar{H}ddue\bar{L}\bar{L} \,, \\
 \mathcal{O}_{11, \,(1 , -1)}^{34} \equiv H\bar{H}ddQQ\bar{Q}\bar{L} \,, \\
 \mathcal{O}_{11, \,(1 , -1)}^{35} \equiv H\bar{H}dddu\bar{Q}\bar{L} \,, \\
 \mathcal{O}_{11, \,(1 , -1)}^{36} \equiv H\bar{H}d\bar{u}QQQ\bar{L} \,, \\
 \mathcal{O}_{11, \,(1 , -1)}^{37} \equiv H\bar{H}ddu\bar{u}Q\bar{L} \,, \\
 \mathcal{O}_{11, \,(1 , -1)}^{38} \equiv HHddde\bar{L}\bar{L} \,, \\
 \mathcal{O}_{11, \,(1 , -1)}^{39} \equiv HHdddd\bar{Q}\bar{L} \,, \\
 \mathcal{O}_{11, \,(1 , -1)}^{40} \equiv HHddd\bar{u}Q\bar{L} \,, \\
 \mathcal{O}_{11, \,(1 , -1)}^{41} \equiv H\bar{H}ddd\bar{L}\bar{L}\nu \,, \\
 \mathcal{O}_{11, \,(1 , -1)}^{42} \equiv \bar{H}\bar{H}ddu\bar{L}\bar{L}\nu \,, \\
 \mathcal{O}_{11, \,(1 , -1)}^{43} \equiv \bar{H}\bar{H}ddQ\bar{e}\bar{L}\nu \,, \\
 \mathcal{O}_{11, \,(1 , -1)}^{44} \equiv \bar{H}\bar{H}dQQ\bar{L}\bar{L}\nu \,, \\
 \mathcal{O}_{11, \,(1 , -1)}^{45} \equiv H\bar{H}dd\bar{d}QQ\bar{\nu} \,, \\
 \mathcal{O}_{11, \,(1 , -1)}^{46} \equiv H\bar{H}ddd\bar{d}u\bar{\nu} \,, \\
 \mathcal{O}_{11, \,(1 , -1)}^{47} \equiv H\bar{H}ddQ\bar{e}L\bar{\nu} \,, \\
 \mathcal{O}_{11, \,(1 , -1)}^{48} \equiv H\bar{H}dQQe\bar{e}\bar{\nu} \,, \\
 \mathcal{O}_{11, \,(1 , -1)}^{49} \equiv H\bar{H}ddue\bar{e}\bar{\nu} \,, \\
 \mathcal{O}_{11, \,(1 , -1)}^{50} \equiv \bar{H}\bar{H}\bar{d}QQQQ\bar{\nu} \,, \\
 \mathcal{O}_{11, \,(1 , -1)}^{51} \equiv \bar{H}\bar{H}d\bar{d}uQQ\bar{\nu} \,, \\
 \mathcal{O}_{11, \,(1 , -1)}^{52} \equiv \bar{H}\bar{H}QQQ\bar{e}L\bar{\nu} \,, \\
 \mathcal{O}_{11, \,(1 , -1)}^{53} \equiv \bar{H}\bar{H}duQ\bar{e}L\bar{\nu} \,, \\
 \mathcal{O}_{11, \,(1 , -1)}^{54} \equiv \bar{H}\bar{H}uQQe\bar{e}\bar{\nu} \,, \\
 \mathcal{O}_{11, \,(1 , -1)}^{55} \equiv \bar{H}\bar{H}uQQL\bar{L}\bar{\nu} \,, \\
 \mathcal{O}_{11, \,(1 , -1)}^{56} \equiv \bar{H}\bar{H}duuL\bar{L}\bar{\nu} \,, \\
 \mathcal{O}_{11, \,(1 , -1)}^{57} \equiv \bar{H}\bar{H}uuQe\bar{L}\bar{\nu} \,, \\
 \mathcal{O}_{11, \,(1 , -1)}^{58} \equiv \bar{H}\bar{H}uQQQ\bar{Q}\bar{\nu} \,, \\
 \mathcal{O}_{11, \,(1 , -1)}^{59} \equiv \bar{H}\bar{H}duuQ\bar{Q}\bar{\nu} \,, \\
 \mathcal{O}_{11, \,(1 , -1)}^{60} \equiv \bar{H}\bar{H}uu\bar{u}QQ\bar{\nu} \,, \\
 \mathcal{O}_{11, \,(1 , -1)}^{61} \equiv H\bar{H}dQQL\bar{L}\bar{\nu} \,, \\
 \mathcal{O}_{11, \,(1 , -1)}^{62} \equiv H\bar{H}QQQe\bar{L}\bar{\nu} \,, \\
 \mathcal{O}_{11, \,(1 , -1)}^{63} \equiv H\bar{H}dduL\bar{L}\bar{\nu} \,, \\
 \mathcal{O}_{11, \,(1 , -1)}^{64} \equiv H\bar{H}duQe\bar{L}\bar{\nu} \,, \\
 \mathcal{O}_{11, \,(1 , -1)}^{65} \equiv H\bar{H}dQQQ\bar{Q}\bar{\nu} \,, \\
 \mathcal{O}_{11, \,(1 , -1)}^{66} \equiv H\bar{H}dduQ\bar{Q}\bar{\nu} \,, \\
 \mathcal{O}_{11, \,(1 , -1)}^{67} \equiv H\bar{H}\bar{u}QQQQ\bar{\nu} \,, \\
 \mathcal{O}_{11, \,(1 , -1)}^{68} \equiv H\bar{H}du\bar{u}QQ\bar{\nu} \,, \\
 \mathcal{O}_{11, \,(1 , -1)}^{69} \equiv H\bar{H}dduu\bar{u}\bar{\nu} \,, \\
 \mathcal{O}_{11, \,(1 , -1)}^{70} \equiv HHdddL\bar{L}\bar{\nu} \,, \\
 \mathcal{O}_{11, \,(1 , -1)}^{71} \equiv HHddQe\bar{L}\bar{\nu} \,, \\
 \mathcal{O}_{11, \,(1 , -1)}^{72} \equiv HHdddQ\bar{Q}\bar{\nu} \,, \\
 \mathcal{O}_{11, \,(1 , -1)}^{73} \equiv HHdd\bar{u}QQ\bar{\nu} \,, \\
 \mathcal{O}_{11, \,(1 , -1)}^{74} \equiv H\bar{H}ddd\bar{e}\nu\bar{\nu} \,, \\
 \mathcal{O}_{11, \,(1 , -1)}^{75} \equiv \bar{H}\bar{H}dQQ\bar{e}\nu\bar{\nu} \,, \\
 \mathcal{O}_{11, \,(1 , -1)}^{76} \equiv \bar{H}\bar{H}QQQ\bar{L}\nu\bar{\nu} \,, \\
 \mathcal{O}_{11, \,(1 , -1)}^{77} \equiv \bar{H}\bar{H}duQ\bar{L}\nu\bar{\nu} \,, \\
 \mathcal{O}_{11, \,(1 , -1)}^{78} \equiv H\bar{H}ddQ\bar{L}\nu\bar{\nu} \,, \\
 \mathcal{O}_{11, \,(1 , -1)}^{79} \equiv HHddQL\bar{\nu}\bar{\nu} \,, \\
 \mathcal{O}_{11, \,(1 , -1)}^{80} \equiv HHdQQe\bar{\nu}\bar{\nu} \,, \\
 \mathcal{O}_{11, \,(1 , -1)}^{81} \equiv \bar{H}\bar{H}uuQL\bar{\nu}\bar{\nu} \,, \\
 \mathcal{O}_{11, \,(1 , -1)}^{82} \equiv H\bar{H}QQQL\bar{\nu}\bar{\nu} \,, \\
 \mathcal{O}_{11, \,(1 , -1)}^{83} \equiv H\bar{H}duQL\bar{\nu}\bar{\nu} \,, \\
 \mathcal{O}_{11, \,(1 , -1)}^{84} \equiv H\bar{H}uQQe\bar{\nu}\bar{\nu} \,, \\
 \mathcal{O}_{11, \,(1 , -1)}^{85} \equiv H\bar{H}duue\bar{\nu}\bar{\nu} \,, \\
 \mathcal{O}_{11, \,(1 , -1)}^{86} \equiv H\bar{H}dQQ\nu\bar{\nu}\bar{\nu} \,, \\
 \mathcal{O}_{11, \,(1 , -1)}^{87} \equiv \bar{H}\bar{H}uQQ\nu\bar{\nu}\bar{\nu} \,, \\
 \mathcal{O}_{11, \,(1 , -1)}^{88} \equiv H\bar{H}ddu\nu\bar{\nu}\bar{\nu} \,, \\
 \mathcal{O}_{11, \,(1 , -1)}^{89} \equiv H\bar{H}\bar{H}\bar{H}\bar{H}QQQ\bar{e} \,, \\
 \mathcal{O}_{11, \,(1 , -1)}^{90} \equiv H\bar{H}\bar{H}\bar{H}\bar{H}uQQ\bar{L} \,, \\
 \mathcal{O}_{11, \,(1 , -1)}^{91} \equiv HH\bar{H}\bar{H}\bar{H}ddQ\bar{e} \,, \\
 \mathcal{O}_{11, \,(1 , -1)}^{92} \equiv HH\bar{H}\bar{H}\bar{H}dQQ\bar{L} \,, \\
 \mathcal{O}_{11, \,(1 , -1)}^{93} \equiv HH\bar{H}\bar{H}\bar{H}ddu\bar{L} \,, \\
 \mathcal{O}_{11, \,(1 , -1)}^{94} \equiv HHH\bar{H}\bar{H}ddd\bar{L} \,, \\
 \mathcal{O}_{11, \,(1 , -1)}^{95} \equiv HHH\bar{H}\bar{H}ddQ\bar{\nu} \,, \\
 \mathcal{O}_{11, \,(1 , -1)}^{96} \equiv HH\bar{H}\bar{H}\bar{H}QQQ\bar{\nu} \,, \\
 \mathcal{O}_{11, \,(1 , -1)}^{97} \equiv HH\bar{H}\bar{H}\bar{H}duQ\bar{\nu} \,.
 \label{eq:dequal11basisB1L-1}
 \end{math}
 \end{spacing}
 \end{multicols}

 \subsection{\texorpdfstring{$\Delta B=1, \Delta L=3$}{Delta B=1,Delta L=3}}
 

 \begin{multicols}{3}
 \begin{spacing}{1.5}
 \noindent
 \begin{math}
 \mathcal{O}_{11, \,(1 , 3)}^{1} \equiv H\bar{H}uuQLLL \,, \\
 \mathcal{O}_{11, \,(1 , 3)}^{2} \equiv H\bar{H}uuueLL \,, \\
 \mathcal{O}_{11, \,(1 , 3)}^{3} \equiv HHQQQLLL \,, \\
 \mathcal{O}_{11, \,(1 , 3)}^{4} \equiv HHduQLLL \,, \\
 \mathcal{O}_{11, \,(1 , 3)}^{5} \equiv HHuQQeLL \,, \\
 \mathcal{O}_{11, \,(1 , 3)}^{6} \equiv HHduueLL \,, \\
 \mathcal{O}_{11, \,(1 , 3)}^{7} \equiv HHuuQeeL \,, \\
 \mathcal{O}_{11, \,(1 , 3)}^{8} \equiv HHdQQLL\nu \,, \\
 \mathcal{O}_{11, \,(1 , 3)}^{9} \equiv \bar{H}\bar{H}uuuLL\nu \,, \\
 \mathcal{O}_{11, \,(1 , 3)}^{10} \equiv H\bar{H}uQQLL\nu \,, \\
 \mathcal{O}_{11, \,(1 , 3)}^{11} \equiv H\bar{H}duuLL\nu \,, \\
 \mathcal{O}_{11, \,(1 , 3)}^{12} \equiv H\bar{H}uuQeL\nu \,, \\
 \mathcal{O}_{11, \,(1 , 3)}^{13} \equiv H\bar{H}uuuee\nu \,, \\
 \mathcal{O}_{11, \,(1 , 3)}^{14} \equiv HHQQQeL\nu \,, \\
 \mathcal{O}_{11, \,(1 , 3)}^{15} \equiv HHdduLL\nu \,, \\
 \mathcal{O}_{11, \,(1 , 3)}^{16} \equiv HHduQeL\nu \,, \\
 \mathcal{O}_{11, \,(1 , 3)}^{17} \equiv HHuQQee\nu \,, \\
 \mathcal{O}_{11, \,(1 , 3)}^{18} \equiv HHddQL\nu\nu \,, \\
 \mathcal{O}_{11, \,(1 , 3)}^{19} \equiv HHdQQe\nu\nu \,, \\
 \mathcal{O}_{11, \,(1 , 3)}^{20} \equiv \bar{H}\bar{H}uuQL\nu\nu \,, \\
 \mathcal{O}_{11, \,(1 , 3)}^{21} \equiv H\bar{H}QQQL\nu\nu \,, \\
 \mathcal{O}_{11, \,(1 , 3)}^{22} \equiv H\bar{H}duQL\nu\nu \,, \\
 \mathcal{O}_{11, \,(1 , 3)}^{23} \equiv H\bar{H}uQQe\nu\nu \,, \\
 \mathcal{O}_{11, \,(1 , 3)}^{24} \equiv H\bar{H}duue\nu\nu \,, \\
 \mathcal{O}_{11, \,(1 , 3)}^{25} \equiv H\bar{H}dQQ\nu\nu\nu \,, \\
 \mathcal{O}_{11, \,(1 , 3)}^{26} \equiv \bar{H}\bar{H}uQQ\nu\nu\nu \,, \\
 \mathcal{O}_{11, \,(1 , 3)}^{27} \equiv H\bar{H}ddu\nu\nu\nu \, .
 \label{eq:dequal11basisB1L3}
 \end{math}
 \end{spacing}
 \end{multicols}

 \subsection{\texorpdfstring{$\Delta B=2, \Delta L=0$}{Delta B=2,Delta L=0}}
 \begin{multicols}{3}
 \begin{spacing}{1.5}
 \noindent
 \begin{math}
 \mathcal{O}_{11, \,(2 , 0)}^{1} \equiv HHddddQQ \,, \\
 \mathcal{O}_{11, \,(2 , 0)}^{2} \equiv H\bar{H}ddQQQQ \,, \\
 \mathcal{O}_{11, \,(2 , 0)}^{3} \equiv \bar{H}\bar{H}QQQQQQ \,, \\
 \mathcal{O}_{11, \,(2 , 0)}^{4} \equiv \bar{H}\bar{H}duQQQQ \,, \\
 \mathcal{O}_{11, \,(2 , 0)}^{5} \equiv \bar{H}\bar{H}dduuQQ \,, \\
 \mathcal{O}_{11, \,(2 , 0)}^{6} \equiv H\bar{H}ddduQQ \,, \\
 \mathcal{O}_{11, \,(2 , 0)}^{7} \equiv H\bar{H}dddduu \, .
 \label{eq:dequal11basisB2L0}
 \end{math}
 \end{spacing}
 \end{multicols}

\clearpage
 

\section{Dimension 12}
\label{sec:d12}
 
\begin{figure}
     \centering
     \includegraphics[width=1\linewidth]{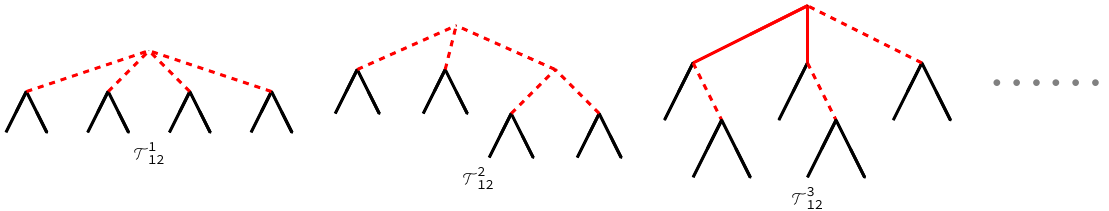}
     \caption{A few representative Feynman-diagram topologies for mass dimension~12. All 1060 topologies are provided at \website. }
     \label{fig:d=12_topologies}
 \end{figure}

At $d=12$, we find operators with different $\Delta B$ and $\Delta L$, which we discuss separately. The number of topologies is too large to include here, see Fig.~\ref{fig:d=12_topologies}.

 \subsection{\texorpdfstring{$\Delta B=2, \Delta L=-2$}{Delta B=2,Delta L=-2}}
 
$d=12$ is the lowest mass dimension for $(\Delta B,\Delta L)=(2,-2)$ operators, so these can be symmetry protected.  
 \begin{multicols}{3}
 \begin{spacing}{1.5}
 \noindent
 \begin{math}
\mathcal{O}_{12, \ (2,-2)}^{1}\equiv dddddd\bar{e}\bar{e}\,,\\
\mathcal{O}_{12, \ (2,-2)}^{2}\equiv dddddQ\bar{L}\bar{e}\,,\\
\mathcal{O}_{12, \ (2,-2)}^{3}\equiv ddddQQ\bar{L}\bar{L}\,,\\
\mathcal{O}_{12, \ (2,-2)}^{4}\equiv dddddu\bar{L}\bar{L}\,,\\
\mathcal{O}_{12, \ (2,-2)}^{5}\equiv ddddQQ\bar{e}\bar{\nu}\,,\\
\mathcal{O}_{12, \ (2,-2)}^{6}\equiv dddddu\bar{e}\bar{\nu}\,,\\
\mathcal{O}_{12, \ (2,-2)}^{7}\equiv dddQQQ\bar{L}\bar{\nu}\,,\\
\mathcal{O}_{12, \ (2,-2)}^{8}\equiv dddduQ\bar{L}\bar{\nu}\,,\\
\mathcal{O}_{12, \ (2,-2)}^{9}\equiv ddQQQQ\bar{\nu}\bar{\nu}\,,\\
\mathcal{O}_{12, \ (2,-2)}^{10}\equiv ddduQQ\bar{\nu}\bar{\nu}\,,\\
\mathcal{O}_{12, \ (2,-2)}^{11}\equiv dddduu\bar{\nu}\bar{\nu}\,.
 \label{eq:dequal12basisB2L-2}
 \end{math}
 \end{spacing}
 \end{multicols}
 Some scalar UV completions were given in Ref.~\cite{Arnold:2012sd}.



 \subsection{\texorpdfstring{$\Delta B=2, \Delta L=2$}{Delta B=2,Delta L=2}}
 
 $d=12$ is the lowest mass dimension for $(\Delta B,\Delta L)=(2,2)$ operators, so these can be symmetry protected. 
 \begin{multicols}{3}
 \begin{spacing}{1.5}
 \noindent
 \begin{math}
\mathcal{O}_{12, \ (2,2)}^{1}\equiv QQQQQQLL\,,\\
\mathcal{O}_{12, \ (2,2)}^{2}\equiv duQQQQLL\,,\\
\mathcal{O}_{12, \ (2,2)}^{3}\equiv uQQQQQLe\,,\\
\mathcal{O}_{12, \ (2,2)}^{4}\equiv dduuQQLL\,,\\
\mathcal{O}_{12, \ (2,2)}^{5}\equiv duuQQQLe\,,\\
\mathcal{O}_{12, \ (2,2)}^{6}\equiv uuQQQQee\,,\\
\mathcal{O}_{12, \ (2,2)}^{7}\equiv ddduuuLL\,,\\
\mathcal{O}_{12, \ (2,2)}^{8}\equiv dduuuQLe\,,\\
\mathcal{O}_{12, \ (2,2)}^{9}\equiv duuuQQee\,,\\
\mathcal{O}_{12, \ (2,2)}^{10}\equiv dduuuuee\,,\\
\mathcal{O}_{12, \ (2,2)}^{11}\equiv dQQQQQL\nu\,,\\
\mathcal{O}_{12, \ (2,2)}^{12}\equiv QQQQQQe\nu\,,\\
\mathcal{O}_{12, \ (2,2)}^{13}\equiv dduQQQL\nu\,,\\
\mathcal{O}_{12, \ (2,2)}^{14}\equiv duQQQQe\nu\,,\\
\mathcal{O}_{12, \ (2,2)}^{15}\equiv ddduuQL\nu\,,\\
\mathcal{O}_{12, \ (2,2)}^{16}\equiv dduuQQe\nu\,,\\
\mathcal{O}_{12, \ (2,2)}^{17}\equiv ddduuue\nu\,,\\
\mathcal{O}_{12, \ (2,2)}^{18}\equiv ddQQQQ\nu\nu\,,\\
\mathcal{O}_{12, \ (2,2)}^{19}\equiv ddduQQ\nu\nu\,,\\
\mathcal{O}_{12, \ (2,2)}^{20}\equiv dddduu\nu\nu\,.
 \label{eq:dequal12basisB2L2}
 \end{math}
 \end{spacing}
 \end{multicols}
 Some scalar UV completions were given in Refs.~\cite{Arnellos:1982nt,Arnold:2012sd}.
 

 
 \subsection{\texorpdfstring{$\Delta B=1, \Delta L=5$}{Delta B=1,Delta L=5}}
 
$d=12$ is the lowest mass dimension for $(\Delta B,\Delta L)=(1,5)$ operators, so these can be symmetry protected. Notice that they all involve right-handed neutrinos and thus only arise in the $\nu$SMEFT.
\begin{multicols}{3}
 \begin{spacing}{1.5}
 \noindent
 \begin{math}
\mathcal{O}_{12, \ (1,5)}^{1}\equiv uuuLLLL\nu\,,\\
\mathcal{O}_{12, \ (1,5)}^{2}\equiv uuQLLL\nu\nu\,,\\
\mathcal{O}_{12, \ (1,5)}^{3}\equiv uuuLLe\nu\nu\,,\\
\mathcal{O}_{12, \ (1,5)}^{4}\equiv uQQLL\nu\nu\nu\,,\\
\mathcal{O}_{12, \ (1,5)}^{5}\equiv duuLL\nu\nu\nu\,,\\
\mathcal{O}_{12, \ (1,5)}^{6}\equiv uuQLe\nu\nu\nu\,,\\
\mathcal{O}_{12, \ (1,5)}^{7}\equiv uuuee\nu\nu\nu\,,\\
\mathcal{O}_{12, \ (1,5)}^{8}\equiv QQQL\nu\nu\nu\nu\,,\\
\mathcal{O}_{12, \ (1,5)}^{9}\equiv duQL\nu\nu\nu\nu\,,\\
\mathcal{O}_{12, \ (1,5)}^{10}\equiv uQQe\nu\nu\nu\nu\,,\\
\mathcal{O}_{12, \ (1,5)}^{11}\equiv duue\nu\nu\nu\nu\,,\\
\mathcal{O}_{12, \ (1,5)}^{12}\equiv dQQ\nu\nu\nu\nu\nu\,,\\
\mathcal{O}_{12, \ (1,5)}^{13}\equiv ddu\nu\nu\nu\nu\nu\,.
 \label{eq:dequal12basisB1L5}
 \end{math}
 \end{spacing}
 \end{multicols}

  \subsection{\texorpdfstring{$\Delta B=1, \Delta L=-3$}{Delta B=1,Delta L=-3}}
  
 The operators $\mathcal{O}_{12, \,(1 , -3)}^{1}\equiv\bar{L} \bar{L} \bar{e} \bar{u} d d d d$ and $ \mathcal{O}_{12, \,(1 , -3)}^{53} \equiv  d d u \bar{L} \bar{L} \bar{L} \bar{H} \bar{H} \bar{H}$  have been mentioned in Ref.~\cite{Weinberg:1980bf}.
 \begin{multicols}{3}
 \begin{spacing}{1.5}
 \noindent
 \begin{math}
 \mathcal{O}_{12, \ (1,-3)}^{1}\equiv dddd\bar{u}\bar{L}\bar{L}\bar{e}\,,\\
\mathcal{O}_{12, \ (1,-3)}^{2}\equiv ddd\bar{L}\bar{L}\bar{L}\bar{L}e\,,\\
\mathcal{O}_{12, \ (1,-3)}^{3}\equiv dddd\bar{Q}\bar{L}\bar{L}\bar{L}\,,\\
\mathcal{O}_{12, \ (1,-3)}^{4}\equiv ddd\bar{u}Q\bar{L}\bar{L}\bar{L}\,,\\
\mathcal{O}_{12, \ (1,-3)}^{5}\equiv dddd\bar{u}\bar{e}\bar{e}\bar{\nu}\,,\\
\mathcal{O}_{12, \ (1,-3)}^{6}\equiv dddd\bar{Q}\bar{L}\bar{e}\bar{\nu}\,,\\
\mathcal{O}_{12, \ (1,-3)}^{7}\equiv ddd\bar{u}Q\bar{L}\bar{e}\bar{\nu}\,,\\
\mathcal{O}_{12, \ (1,-3)}^{8}\equiv dddd\bar{d}\bar{L}\bar{L}\bar{\nu}\,,\\
\mathcal{O}_{12, \ (1,-3)}^{9}\equiv ddd\bar{L}\bar{L}e\bar{e}\bar{\nu}\,,\\
\mathcal{O}_{12, \ (1,-3)}^{10}\equiv dddL\bar{L}\bar{L}\bar{L}\bar{\nu}\,,\\
\mathcal{O}_{12, \ (1,-3)}^{11}\equiv ddQ\bar{L}\bar{L}\bar{L}e\bar{\nu}\,,\\
\mathcal{O}_{12, \ (1,-3)}^{12}\equiv dddQ\bar{Q}\bar{L}\bar{L}\bar{\nu}\,,\\
\mathcal{O}_{12, \ (1,-3)}^{13}\equiv dd\bar{u}QQ\bar{L}\bar{L}\bar{\nu}\,,\\
\mathcal{O}_{12, \ (1,-3)}^{14}\equiv dddu\bar{u}\bar{L}\bar{L}\bar{\nu}\,,\\
\mathcal{O}_{12, \ (1,-3)}^{15}\equiv dddd\bar{d}\bar{e}\bar{\nu}\bar{\nu}\,,\\
\mathcal{O}_{12, \ (1,-3)}^{16}\equiv ddde\bar{e}\bar{e}\bar{\nu}\bar{\nu}\,,\\
\mathcal{O}_{12, \ (1,-3)}^{17}\equiv dddQ\bar{Q}\bar{e}\bar{\nu}\bar{\nu}\,,\\
\mathcal{O}_{12, \ (1,-3)}^{18}\equiv dd\bar{u}QQ\bar{e}\bar{\nu}\bar{\nu}\,,\\
\mathcal{O}_{12, \ (1,-3)}^{19}\equiv dddu\bar{u}\bar{e}\bar{\nu}\bar{\nu}\,,\\
\mathcal{O}_{12, \ (1,-3)}^{20}\equiv ddd\bar{d}Q\bar{L}\bar{\nu}\bar{\nu}\,,\\
\mathcal{O}_{12, \ (1,-3)}^{21}\equiv dddL\bar{L}\bar{e}\bar{\nu}\bar{\nu}\,,\\
\mathcal{O}_{12, \ (1,-3)}^{22}\equiv ddQ\bar{L}e\bar{e}\bar{\nu}\bar{\nu}\,,\\
\mathcal{O}_{12, \ (1,-3)}^{23}\equiv ddQL\bar{L}\bar{L}\bar{\nu}\bar{\nu}\,,\\
\mathcal{O}_{12, \ (1,-3)}^{24}\equiv dQQ\bar{L}\bar{L}e\bar{\nu}\bar{\nu}\,,\\
\mathcal{O}_{12, \ (1,-3)}^{25}\equiv ddu\bar{L}\bar{L}e\bar{\nu}\bar{\nu}\,,\\
\mathcal{O}_{12, \ (1,-3)}^{26}\equiv ddQQ\bar{Q}\bar{L}\bar{\nu}\bar{\nu}\,,\\
\mathcal{O}_{12, \ (1,-3)}^{27}\equiv dddu\bar{Q}\bar{L}\bar{\nu}\bar{\nu}\,,\\
\mathcal{O}_{12, \ (1,-3)}^{28}\equiv d\bar{u}QQQ\bar{L}\bar{\nu}\bar{\nu}\,,\\
\mathcal{O}_{12, \ (1,-3)}^{29}\equiv ddu\bar{u}Q\bar{L}\bar{\nu}\bar{\nu}\,,\\
\mathcal{O}_{12, \ (1,-3)}^{30}\equiv ddd\bar{L}\bar{L}\nu\bar{\nu}\bar{\nu}\,,\\
\mathcal{O}_{12, \ (1,-3)}^{31}\equiv dd\bar{d}QQ\bar{\nu}\bar{\nu}\bar{\nu}\,,\\
\mathcal{O}_{12, \ (1,-3)}^{32}\equiv ddd\bar{d}u\bar{\nu}\bar{\nu}\bar{\nu}\,,\\
\mathcal{O}_{12, \ (1,-3)}^{33}\equiv ddQL\bar{e}\bar{\nu}\bar{\nu}\bar{\nu}\,,\\
\mathcal{O}_{12, \ (1,-3)}^{34}\equiv dQQe\bar{e}\bar{\nu}\bar{\nu}\bar{\nu}\,,\\
\mathcal{O}_{12, \ (1,-3)}^{35}\equiv ddue\bar{e}\bar{\nu}\bar{\nu}\bar{\nu}\,,\\
\mathcal{O}_{12, \ (1,-3)}^{36}\equiv dQQL\bar{L}\bar{\nu}\bar{\nu}\bar{\nu}\,,\\
\mathcal{O}_{12, \ (1,-3)}^{37}\equiv QQQ\bar{L}e\bar{\nu}\bar{\nu}\bar{\nu}\,,\\
\mathcal{O}_{12, \ (1,-3)}^{38}\equiv dduL\bar{L}\bar{\nu}\bar{\nu}\bar{\nu}\,,\\
\mathcal{O}_{12, \ (1,-3)}^{39}\equiv duQ\bar{L}e\bar{\nu}\bar{\nu}\bar{\nu}\,,\\
\mathcal{O}_{12, \ (1,-3)}^{40}\equiv dQQQ\bar{Q}\bar{\nu}\bar{\nu}\bar{\nu}\,,\\
\mathcal{O}_{12, \ (1,-3)}^{41}\equiv dduQ\bar{Q}\bar{\nu}\bar{\nu}\bar{\nu}\,,\\
\mathcal{O}_{12, \ (1,-3)}^{42}\equiv \bar{u}QQQQ\bar{\nu}\bar{\nu}\bar{\nu}\,,\\
\mathcal{O}_{12, \ (1,-3)}^{43}\equiv du\bar{u}QQ\bar{\nu}\bar{\nu}\bar{\nu}\,,\\
\mathcal{O}_{12, \ (1,-3)}^{44}\equiv dduu\bar{u}\bar{\nu}\bar{\nu}\bar{\nu}\,,\\
\mathcal{O}_{12, \ (1,-3)}^{45}\equiv ddd\bar{e}\nu\bar{\nu}\bar{\nu}\bar{\nu}\,,\\
\mathcal{O}_{12, \ (1,-3)}^{46}\equiv ddQ\bar{L}\nu\bar{\nu}\bar{\nu}\bar{\nu}\,,\\
\mathcal{O}_{12, \ (1,-3)}^{47}\equiv QQQL\bar{\nu}\bar{\nu}\bar{\nu}\bar{\nu}\,,\\
\mathcal{O}_{12, \ (1,-3)}^{48}\equiv duQL\bar{\nu}\bar{\nu}\bar{\nu}\bar{\nu}\,,\\
\mathcal{O}_{12, \ (1,-3)}^{49}\equiv uQQe\bar{\nu}\bar{\nu}\bar{\nu}\bar{\nu}\,,\\
\mathcal{O}_{12, \ (1,-3)}^{50}\equiv duue\bar{\nu}\bar{\nu}\bar{\nu}\bar{\nu}\,,\\
\mathcal{O}_{12, \ (1,-3)}^{51}\equiv dQQ\nu\bar{\nu}\bar{\nu}\bar{\nu}\bar{\nu}\,,\\
\mathcal{O}_{12, \ (1,-3)}^{52}\equiv ddu\nu\bar{\nu}\bar{\nu}\bar{\nu}\bar{\nu}\,,\\
\mathcal{O}_{12, \ (1,-3)}^{53}\equiv \bar{H}\bar{H}\bar{H}ddu\bar{L}\bar{L}\bar{L}\,,\\
\mathcal{O}_{12, \ (1,-3)}^{54}\equiv H\bar{H}\bar{H}ddd\bar{L}\bar{L}\bar{L}\,,\\
\mathcal{O}_{12, \ (1,-3)}^{55}\equiv \bar{H}\bar{H}\bar{H}ddQ\bar{L}\bar{L}\bar{e}\,,\\
\mathcal{O}_{12, \ (1,-3)}^{56}\equiv \bar{H}\bar{H}\bar{H}dQQ\bar{L}\bar{L}\bar{L}\,,\\
\mathcal{O}_{12, \ (1,-3)}^{57}\equiv H\bar{H}\bar{H}ddd\bar{L}\bar{e}\bar{\nu}\,,\\
\mathcal{O}_{12, \ (1,-3)}^{58}\equiv H\bar{H}\bar{H}ddQ\bar{L}\bar{L}\bar{\nu}\,,\\
\mathcal{O}_{12, \ (1,-3)}^{59}\equiv \bar{H}\bar{H}\bar{H}duQ\bar{L}\bar{L}\bar{\nu}\,,\\
\mathcal{O}_{12, \ (1,-3)}^{60}\equiv \bar{H}\bar{H}\bar{H}dQQ\bar{L}\bar{e}\bar{\nu}\,,\\
\mathcal{O}_{12, \ (1,-3)}^{61}\equiv \bar{H}\bar{H}\bar{H}QQQ\bar{L}\bar{L}\bar{\nu}\,,\\
\mathcal{O}_{12, \ (1,-3)}^{62}\equiv H\bar{H}\bar{H}ddu\bar{L}\bar{\nu}\bar{\nu}\,,\\
\mathcal{O}_{12, \ (1,-3)}^{63}\equiv H\bar{H}\bar{H}ddQ\bar{e}\bar{\nu}\bar{\nu}\,,\\
\mathcal{O}_{12, \ (1,-3)}^{64}\equiv H\bar{H}\bar{H}dQQ\bar{L}\bar{\nu}\bar{\nu}\,,\\
\mathcal{O}_{12, \ (1,-3)}^{65}\equiv HH\bar{H}ddd\bar{L}\bar{\nu}\bar{\nu}\,,\\
\mathcal{O}_{12, \ (1,-3)}^{66}\equiv \bar{H}\bar{H}\bar{H}QQQ\bar{e}\bar{\nu}\bar{\nu}\,,\\
\mathcal{O}_{12, \ (1,-3)}^{67}\equiv \bar{H}\bar{H}\bar{H}uQQ\bar{L}\bar{\nu}\bar{\nu}\,,\\
\mathcal{O}_{12, \ (1,-3)}^{68}\equiv H\bar{H}\bar{H}duQ\bar{\nu}\bar{\nu}\bar{\nu}\,,\\
\mathcal{O}_{12, \ (1,-3)}^{69}\equiv H\bar{H}\bar{H}QQQ\bar{\nu}\bar{\nu}\bar{\nu}\,,\\
\mathcal{O}_{12, \ (1,-3)}^{70}\equiv HH\bar{H}ddQ\bar{\nu}\bar{\nu}\bar{\nu}\,.
 \label{eq:dequal12basisB1L-3}
 \end{math}
 \end{spacing}
 \end{multicols}

 \subsection{\texorpdfstring{$\Delta B=1, \Delta L=1$}{Delta B=1,Delta L=1}}
 
Ref.~\cite{Hambye:2017qix} mentioned that $ \mathcal{O}_{12, \,(1 , 1)}^{25} \equiv  \bar{e} \bar{e} u u d e e e$, and other operators involving multiple leptons, can be made dominant by imposing lepton-flavor symmetries.
 \begin{multicols}{3}
 \begin{spacing}{1.7}
 \noindent
 \begin{math}
\mathcal{O}_{12, \ (1,1)}^{1}\equiv dd\bar{d}\bar{d}QQQL\,,\\
\mathcal{O}_{12, \ (1,1)}^{2}\equiv d\bar{d}\bar{d}QQQQe\,,\\
\mathcal{O}_{12, \ (1,1)}^{3}\equiv ddd\bar{d}\bar{d}uQL\,,\\
\mathcal{O}_{12, \ (1,1)}^{4}\equiv dd\bar{d}\bar{d}uQQe\,,\\
\mathcal{O}_{12, \ (1,1)}^{5}\equiv ddd\bar{d}\bar{d}uue\,,\\
\mathcal{O}_{12, \ (1,1)}^{6}\equiv \bar{d}\bar{u}QQQQQL\,,\\
\mathcal{O}_{12, \ (1,1)}^{7}\equiv d\bar{d}u\bar{u}QQQL\,,\\
\mathcal{O}_{12, \ (1,1)}^{8}\equiv \bar{d}u\bar{u}QQQQe\,,\\
\mathcal{O}_{12, \ (1,1)}^{9}\equiv dd\bar{d}uu\bar{u}QL\,,\\
\mathcal{O}_{12, \ (1,1)}^{10}\equiv d\bar{d}uu\bar{u}QQe\,,\\
\mathcal{O}_{12, \ (1,1)}^{11}\equiv dd\bar{d}uuu\bar{u}e\,,\\
\mathcal{O}_{12, \ (1,1)}^{12}\equiv dd\bar{d}QQLL\bar{e}\,,\\
\mathcal{O}_{12, \ (1,1)}^{13}\equiv d\bar{d}QQQLe\bar{e}\,,\\
\mathcal{O}_{12, \ (1,1)}^{14}\equiv \bar{d}QQQQee\bar{e}\,,\\
\mathcal{O}_{12, \ (1,1)}^{15}\equiv ddd\bar{d}uLL\bar{e}\,,\\
\mathcal{O}_{12, \ (1,1)}^{16}\equiv dd\bar{d}uQLe\bar{e}\,,\\
\mathcal{O}_{12, \ (1,1)}^{17}\equiv d\bar{d}uQQee\bar{e}\,,\\
\mathcal{O}_{12, \ (1,1)}^{18}\equiv dd\bar{d}uuee\bar{e}\,,\\
\mathcal{O}_{12, \ (1,1)}^{19}\equiv ddQLLL\bar{e}\bar{e}\,,\\
\mathcal{O}_{12, \ (1,1)}^{20}\equiv dQQLLe\bar{e}\bar{e}\,,\\
\mathcal{O}_{12, \ (1,1)}^{21}\equiv QQQLee\bar{e}\bar{e}\,,\\
\mathcal{O}_{12, \ (1,1)}^{22}\equiv dduLLe\bar{e}\bar{e}\,,\\
\mathcal{O}_{12, \ (1,1)}^{23}\equiv duQLee\bar{e}\bar{e}\,,\\
\mathcal{O}_{12, \ (1,1)}^{24}\equiv uQQeee\bar{e}\bar{e}\,,\\
\mathcal{O}_{12, \ (1,1)}^{25}\equiv duueee\bar{e}\bar{e}\,,\\
\mathcal{O}_{12, \ (1,1)}^{26}\equiv dQQQ\bar{Q}LL\bar{e}\,,\\
\mathcal{O}_{12, \ (1,1)}^{27}\equiv QQQQ\bar{Q}Le\bar{e}\,,\\
\mathcal{O}_{12, \ (1,1)}^{28}\equiv dduQ\bar{Q}LL\bar{e}\,,\\
\mathcal{O}_{12, \ (1,1)}^{29}\equiv duQQ\bar{Q}Le\bar{e}\,,\\
\mathcal{O}_{12, \ (1,1)}^{30}\equiv uQQQ\bar{Q}ee\bar{e}\,,\\
\mathcal{O}_{12, \ (1,1)}^{31}\equiv dduu\bar{Q}Le\bar{e}\,,\\
\mathcal{O}_{12, \ (1,1)}^{32}\equiv duuQ\bar{Q}ee\bar{e}\,,\\
\mathcal{O}_{12, \ (1,1)}^{33}\equiv \bar{u}QQQQLL\bar{e}\,,\\
\mathcal{O}_{12, \ (1,1)}^{34}\equiv du\bar{u}QQLL\bar{e}\,,\\
\mathcal{O}_{12, \ (1,1)}^{35}\equiv u\bar{u}QQQLe\bar{e}\,,\\
\mathcal{O}_{12, \ (1,1)}^{36}\equiv dduu\bar{u}LL\bar{e}\,,\\
\mathcal{O}_{12, \ (1,1)}^{37}\equiv duu\bar{u}QLe\bar{e}\,,\\
\mathcal{O}_{12, \ (1,1)}^{38}\equiv uu\bar{u}QQee\bar{e}\,,\\
\mathcal{O}_{12, \ (1,1)}^{39}\equiv duuu\bar{u}ee\bar{e}\,,\\
\mathcal{O}_{12, \ (1,1)}^{40}\equiv d\bar{d}QQQLL\bar{L}\,,\\
\mathcal{O}_{12, \ (1,1)}^{41}\equiv \bar{d}QQQQL\bar{L}e\,,\\
\mathcal{O}_{12, \ (1,1)}^{42}\equiv dd\bar{d}uQLL\bar{L}\,,\\
\mathcal{O}_{12, \ (1,1)}^{43}\equiv d\bar{d}uQQL\bar{L}e\,,\\
\mathcal{O}_{12, \ (1,1)}^{44}\equiv \bar{d}uQQQ\bar{L}ee\,,\\
\mathcal{O}_{12, \ (1,1)}^{45}\equiv dd\bar{d}uuL\bar{L}e\,,\\
\mathcal{O}_{12, \ (1,1)}^{46}\equiv d\bar{d}uuQ\bar{L}ee\,,\\
\mathcal{O}_{12, \ (1,1)}^{47}\equiv dQQLLL\bar{L}\bar{e}\,,\\
\mathcal{O}_{12, \ (1,1)}^{48}\equiv QQQLL\bar{L}e\bar{e}\,,\\
\mathcal{O}_{12, \ (1,1)}^{49}\equiv dduLLL\bar{L}\bar{e}\,,\\
\mathcal{O}_{12, \ (1,1)}^{50}\equiv duQLL\bar{L}e\bar{e}\,,\\
\mathcal{O}_{12, \ (1,1)}^{51}\equiv uQQL\bar{L}ee\bar{e}\,,\\
\mathcal{O}_{12, \ (1,1)}^{52}\equiv duuL\bar{L}ee\bar{e}\,,\\
\mathcal{O}_{12, \ (1,1)}^{53}\equiv uuQ\bar{L}eee\bar{e}\,,\\
\mathcal{O}_{12, \ (1,1)}^{54}\equiv QQQLLL\bar{L}\bar{L}\,,\\
\mathcal{O}_{12, \ (1,1)}^{55}\equiv duQLLL\bar{L}\bar{L}\,,\\
\mathcal{O}_{12, \ (1,1)}^{56}\equiv uQQLL\bar{L}\bar{L}e\,,\\
\mathcal{O}_{12, \ (1,1)}^{57}\equiv duuLL\bar{L}\bar{L}e\,,\\
\mathcal{O}_{12, \ (1,1)}^{58}\equiv uuQL\bar{L}\bar{L}ee\,,\\
\mathcal{O}_{12, \ (1,1)}^{59}\equiv uuu\bar{L}\bar{L}eee\,,\\
\mathcal{O}_{12, \ (1,1)}^{60}\equiv QQQQ\bar{Q}LL\bar{L}\,,\\
\mathcal{O}_{12, \ (1,1)}^{61}\equiv duQQ\bar{Q}LL\bar{L}\,,\\
\mathcal{O}_{12, \ (1,1)}^{62}\equiv uQQQ\bar{Q}L\bar{L}e\,,\\
\mathcal{O}_{12, \ (1,1)}^{63}\equiv dduu\bar{Q}LL\bar{L}\,,\\
\mathcal{O}_{12, \ (1,1)}^{64}\equiv duuQ\bar{Q}L\bar{L}e\,,\\
\mathcal{O}_{12, \ (1,1)}^{65}\equiv uuQQ\bar{Q}\bar{L}ee\,,\\
\mathcal{O}_{12, \ (1,1)}^{66}\equiv duuu\bar{Q}\bar{L}ee\,,\\
\mathcal{O}_{12, \ (1,1)}^{67}\equiv u\bar{u}QQQLL\bar{L}\,,\\
\mathcal{O}_{12, \ (1,1)}^{68}\equiv duu\bar{u}QLL\bar{L}\,,\\
\mathcal{O}_{12, \ (1,1)}^{69}\equiv uu\bar{u}QQL\bar{L}e\,,\\
\mathcal{O}_{12, \ (1,1)}^{70}\equiv duuu\bar{u}L\bar{L}e\,,\\
\mathcal{O}_{12, \ (1,1)}^{71}\equiv uuu\bar{u}Q\bar{L}ee\,,\\
\mathcal{O}_{12, \ (1,1)}^{72}\equiv d\bar{d}QQQQ\bar{Q}L\,,\\
\mathcal{O}_{12, \ (1,1)}^{73}\equiv \bar{d}QQQQQ\bar{Q}e\,,\\
\mathcal{O}_{12, \ (1,1)}^{74}\equiv dd\bar{d}uQQ\bar{Q}L\,,\\
\mathcal{O}_{12, \ (1,1)}^{75}\equiv d\bar{d}uQQQ\bar{Q}e\,,\\
\mathcal{O}_{12, \ (1,1)}^{76}\equiv ddd\bar{d}uu\bar{Q}L\,,\\
\mathcal{O}_{12, \ (1,1)}^{77}\equiv dd\bar{d}uuQ\bar{Q}e\,,\\
\mathcal{O}_{12, \ (1,1)}^{78}\equiv QQQQQ\bar{Q}\bar{Q}L\,,\\
\mathcal{O}_{12, \ (1,1)}^{79}\equiv duQQQ\bar{Q}\bar{Q}L\,,\\
\mathcal{O}_{12, \ (1,1)}^{80}\equiv uQQQQ\bar{Q}\bar{Q}e\,,\\
\mathcal{O}_{12, \ (1,1)}^{81}\equiv dduuQ\bar{Q}\bar{Q}L\,,\\
\mathcal{O}_{12, \ (1,1)}^{82}\equiv duuQQ\bar{Q}\bar{Q}e\,,\\
\mathcal{O}_{12, \ (1,1)}^{83}\equiv dduuu\bar{Q}\bar{Q}e\,,\\
\mathcal{O}_{12, \ (1,1)}^{84}\equiv u\bar{u}QQQQ\bar{Q}L\,,\\
\mathcal{O}_{12, \ (1,1)}^{85}\equiv duu\bar{u}QQ\bar{Q}L\,,\\
\mathcal{O}_{12, \ (1,1)}^{86}\equiv uu\bar{u}QQQ\bar{Q}e\,,\\
\mathcal{O}_{12, \ (1,1)}^{87}\equiv dduuu\bar{u}\bar{Q}L\,,\\
\mathcal{O}_{12, \ (1,1)}^{88}\equiv duuu\bar{u}Q\bar{Q}e\,,\\
\mathcal{O}_{12, \ (1,1)}^{89}\equiv uu\bar{u}\bar{u}QQQL\,,\\
\mathcal{O}_{12, \ (1,1)}^{90}\equiv duuu\bar{u}\bar{u}QL\,,\\
\mathcal{O}_{12, \ (1,1)}^{91}\equiv uuu\bar{u}\bar{u}QQe\,,\\
\mathcal{O}_{12, \ (1,1)}^{92}\equiv duuuu\bar{u}\bar{u}e\,,\\
\mathcal{O}_{12, \ (1,1)}^{93}\equiv ddd\bar{d}\bar{d}QQ\nu\,,\\
\mathcal{O}_{12, \ (1,1)}^{94}\equiv dddd\bar{d}\bar{d}u\nu\,,\\
\mathcal{O}_{12, \ (1,1)}^{95}\equiv d\bar{d}\bar{u}QQQQ\nu\,,\\
\mathcal{O}_{12, \ (1,1)}^{96}\equiv dd\bar{d}u\bar{u}QQ\nu\,,\\
\mathcal{O}_{12, \ (1,1)}^{97}\equiv ddd\bar{d}uu\bar{u}\nu\,,\\
\mathcal{O}_{12, \ (1,1)}^{98}\equiv ddd\bar{d}QL\bar{e}\nu\,,\\
\mathcal{O}_{12, \ (1,1)}^{99}\equiv dd\bar{d}QQe\bar{e}\nu\,,\\
\mathcal{O}_{12, \ (1,1)}^{100}\equiv ddd\bar{d}ue\bar{e}\nu\,,\\
\mathcal{O}_{12, \ (1,1)}^{101}\equiv dddLL\bar{e}\bar{e}\nu\,,\\
\mathcal{O}_{12, \ (1,1)}^{102}\equiv ddQLe\bar{e}\bar{e}\nu\,,\\
\mathcal{O}_{12, \ (1,1)}^{103}\equiv dQQee\bar{e}\bar{e}\nu\,,\\
\mathcal{O}_{12, \ (1,1)}^{104}\equiv dduee\bar{e}\bar{e}\nu\,,\\
\mathcal{O}_{12, \ (1,1)}^{105}\equiv ddQQ\bar{Q}L\bar{e}\nu\,,\\
\mathcal{O}_{12, \ (1,1)}^{106}\equiv dQQQ\bar{Q}e\bar{e}\nu\,,\\
\mathcal{O}_{12, \ (1,1)}^{107}\equiv dddu\bar{Q}L\bar{e}\nu\,,\\
\mathcal{O}_{12, \ (1,1)}^{108}\equiv dduQ\bar{Q}e\bar{e}\nu\,,\\
\mathcal{O}_{12, \ (1,1)}^{109}\equiv d\bar{u}QQQL\bar{e}\nu\,,\\
\mathcal{O}_{12, \ (1,1)}^{110}\equiv \bar{u}QQQQe\bar{e}\nu\,,\\
\mathcal{O}_{12, \ (1,1)}^{111}\equiv ddu\bar{u}QL\bar{e}\nu\,,\\
\mathcal{O}_{12, \ (1,1)}^{112}\equiv du\bar{u}QQe\bar{e}\nu\,,\\
\mathcal{O}_{12, \ (1,1)}^{113}\equiv dduu\bar{u}e\bar{e}\nu\,,\\
\mathcal{O}_{12, \ (1,1)}^{114}\equiv dd\bar{d}QQL\bar{L}\nu\,,\\
\mathcal{O}_{12, \ (1,1)}^{115}\equiv d\bar{d}QQQ\bar{L}e\nu\,,\\
\mathcal{O}_{12, \ (1,1)}^{116}\equiv ddd\bar{d}uL\bar{L}\nu\,,\\
\mathcal{O}_{12, \ (1,1)}^{117}\equiv dd\bar{d}uQ\bar{L}e\nu\,,\\
\mathcal{O}_{12, \ (1,1)}^{118}\equiv ddQLL\bar{L}\bar{e}\nu\,,\\
\mathcal{O}_{12, \ (1,1)}^{119}\equiv dQQL\bar{L}e\bar{e}\nu\,,\\
\mathcal{O}_{12, \ (1,1)}^{120}\equiv QQQ\bar{L}ee\bar{e}\nu\,,\\
\mathcal{O}_{12, \ (1,1)}^{121}\equiv dduL\bar{L}e\bar{e}\nu\,,\\
\mathcal{O}_{12, \ (1,1)}^{122}\equiv duQ\bar{L}ee\bar{e}\nu\,,\\
\mathcal{O}_{12, \ (1,1)}^{123}\equiv dQQLL\bar{L}\bar{L}\nu\,,\\
\mathcal{O}_{12, \ (1,1)}^{124}\equiv QQQL\bar{L}\bar{L}e\nu\,,\\
\mathcal{O}_{12, \ (1,1)}^{125}\equiv dduLL\bar{L}\bar{L}\nu\,,\\
\mathcal{O}_{12, \ (1,1)}^{126}\equiv duQL\bar{L}\bar{L}e\nu\,,\\
\mathcal{O}_{12, \ (1,1)}^{127}\equiv uQQ\bar{L}\bar{L}ee\nu\,,\\
\mathcal{O}_{12, \ (1,1)}^{128}\equiv duu\bar{L}\bar{L}ee\nu\,,\\
\mathcal{O}_{12, \ (1,1)}^{129}\equiv dQQQ\bar{Q}L\bar{L}\nu\,,\\
\mathcal{O}_{12, \ (1,1)}^{130}\equiv QQQQ\bar{Q}\bar{L}e\nu\,,\\
\mathcal{O}_{12, \ (1,1)}^{131}\equiv dduQ\bar{Q}L\bar{L}\nu\,,\\
\mathcal{O}_{12, \ (1,1)}^{132}\equiv duQQ\bar{Q}\bar{L}e\nu\,,\\
\mathcal{O}_{12, \ (1,1)}^{133}\equiv dduu\bar{Q}\bar{L}e\nu\,,\\
\mathcal{O}_{12, \ (1,1)}^{134}\equiv \bar{u}QQQQL\bar{L}\nu\,,\\
\mathcal{O}_{12, \ (1,1)}^{135}\equiv du\bar{u}QQL\bar{L}\nu\,,\\
\mathcal{O}_{12, \ (1,1)}^{136}\equiv u\bar{u}QQQ\bar{L}e\nu\,,\\
\mathcal{O}_{12, \ (1,1)}^{137}\equiv dduu\bar{u}L\bar{L}\nu\,,\\
\mathcal{O}_{12, \ (1,1)}^{138}\equiv duu\bar{u}Q\bar{L}e\nu\,,\\
\mathcal{O}_{12, \ (1,1)}^{139}\equiv \bar{d}QQQQLL\bar{\nu}\,,\\
\mathcal{O}_{12, \ (1,1)}^{140}\equiv d\bar{d}uQQLL\bar{\nu}\,,\\
\mathcal{O}_{12, \ (1,1)}^{141}\equiv \bar{d}uQQQLe\bar{\nu}\,,\\
\mathcal{O}_{12, \ (1,1)}^{142}\equiv dd\bar{d}uuLL\bar{\nu}\,,\\
\mathcal{O}_{12, \ (1,1)}^{143}\equiv d\bar{d}uuQLe\bar{\nu}\,,\\
\mathcal{O}_{12, \ (1,1)}^{144}\equiv \bar{d}uuQQee\bar{\nu}\,,\\
\mathcal{O}_{12, \ (1,1)}^{145}\equiv d\bar{d}uuuee\bar{\nu}\,,\\
\mathcal{O}_{12, \ (1,1)}^{146}\equiv QQQLLL\bar{e}\bar{\nu}\,,\\
\mathcal{O}_{12, \ (1,1)}^{147}\equiv duQLLL\bar{e}\bar{\nu}\,,\\
\mathcal{O}_{12, \ (1,1)}^{148}\equiv uQQLLe\bar{e}\bar{\nu}\,,\\
\mathcal{O}_{12, \ (1,1)}^{149}\equiv duuLLe\bar{e}\bar{\nu}\,,\\
\mathcal{O}_{12, \ (1,1)}^{150}\equiv uuQLee\bar{e}\bar{\nu}\,,\\
\mathcal{O}_{12, \ (1,1)}^{151}\equiv uuueee\bar{e}\bar{\nu}\,,\\
\mathcal{O}_{12, \ (1,1)}^{152}\equiv uQQLLL\bar{L}\bar{\nu}\,,\\
\mathcal{O}_{12, \ (1,1)}^{153}\equiv duuLLL\bar{L}\bar{\nu}\,,\\
\mathcal{O}_{12, \ (1,1)}^{154}\equiv uuQLL\bar{L}e\bar{\nu}\,,\\
\mathcal{O}_{12, \ (1,1)}^{155}\equiv uuuL\bar{L}ee\bar{\nu}\,,\\
\mathcal{O}_{12, \ (1,1)}^{156}\equiv uQQQ\bar{Q}LL\bar{\nu}\,,\\
\mathcal{O}_{12, \ (1,1)}^{157}\equiv duuQ\bar{Q}LL\bar{\nu}\,,\\
\mathcal{O}_{12, \ (1,1)}^{158}\equiv uuQQ\bar{Q}Le\bar{\nu}\,,\\
\mathcal{O}_{12, \ (1,1)}^{159}\equiv duuu\bar{Q}Le\bar{\nu}\,,\\
\mathcal{O}_{12, \ (1,1)}^{160}\equiv uuuQ\bar{Q}ee\bar{\nu}\,,\\
\mathcal{O}_{12, \ (1,1)}^{161}\equiv uu\bar{u}QQLL\bar{\nu}\,,\\
\mathcal{O}_{12, \ (1,1)}^{162}\equiv duuu\bar{u}LL\bar{\nu}\,,\\
\mathcal{O}_{12, \ (1,1)}^{163}\equiv uuu\bar{u}QLe\bar{\nu}\,,\\
\mathcal{O}_{12, \ (1,1)}^{164}\equiv uuuu\bar{u}ee\bar{\nu}\,,\\
\mathcal{O}_{12, \ (1,1)}^{165}\equiv dd\bar{d}QQQ\bar{Q}\nu\,,\\
\mathcal{O}_{12, \ (1,1)}^{166}\equiv ddd\bar{d}uQ\bar{Q}\nu\,,\\
\mathcal{O}_{12, \ (1,1)}^{167}\equiv dQQQQ\bar{Q}\bar{Q}\nu\,,\\
\mathcal{O}_{12, \ (1,1)}^{168}\equiv dduQQ\bar{Q}\bar{Q}\nu\,,\\
\mathcal{O}_{12, \ (1,1)}^{169}\equiv ddduu\bar{Q}\bar{Q}\nu\,,\\
\mathcal{O}_{12, \ (1,1)}^{170}\equiv \bar{u}QQQQQ\bar{Q}\nu\,,\\
\mathcal{O}_{12, \ (1,1)}^{171}\equiv du\bar{u}QQQ\bar{Q}\nu\,,\\
\mathcal{O}_{12, \ (1,1)}^{172}\equiv dduu\bar{u}Q\bar{Q}\nu\,,\\
\mathcal{O}_{12, \ (1,1)}^{173}\equiv u\bar{u}\bar{u}QQQQ\nu\,,\\
\mathcal{O}_{12, \ (1,1)}^{174}\equiv duu\bar{u}\bar{u}QQ\nu\,,\\
\mathcal{O}_{12, \ (1,1)}^{175}\equiv dduuu\bar{u}\bar{u}\nu\,,\\
\mathcal{O}_{12, \ (1,1)}^{176}\equiv dddd\bar{d}\bar{e}\nu\nu\,,\\
\mathcal{O}_{12, \ (1,1)}^{177}\equiv ddde\bar{e}\bar{e}\nu\nu\,,\\
\mathcal{O}_{12, \ (1,1)}^{178}\equiv dddQ\bar{Q}\bar{e}\nu\nu\,,\\
\mathcal{O}_{12, \ (1,1)}^{179}\equiv dd\bar{u}QQ\bar{e}\nu\nu\,,\\
\mathcal{O}_{12, \ (1,1)}^{180}\equiv dddu\bar{u}\bar{e}\nu\nu\,,\\
\mathcal{O}_{12, \ (1,1)}^{181}\equiv ddd\bar{d}Q\bar{L}\nu\nu\,,\\
\mathcal{O}_{12, \ (1,1)}^{182}\equiv dddL\bar{L}\bar{e}\nu\nu\,,\\
\mathcal{O}_{12, \ (1,1)}^{183}\equiv ddQ\bar{L}e\bar{e}\nu\nu\,,\\
\mathcal{O}_{12, \ (1,1)}^{184}\equiv ddQL\bar{L}\bar{L}\nu\nu\,,\\
\mathcal{O}_{12, \ (1,1)}^{185}\equiv dQQ\bar{L}\bar{L}e\nu\nu\,,\\
\mathcal{O}_{12, \ (1,1)}^{186}\equiv ddu\bar{L}\bar{L}e\nu\nu\,,\\
\mathcal{O}_{12, \ (1,1)}^{187}\equiv ddQQ\bar{Q}\bar{L}\nu\nu\,,\\
\mathcal{O}_{12, \ (1,1)}^{188}\equiv dddu\bar{Q}\bar{L}\nu\nu\,,\\
\mathcal{O}_{12, \ (1,1)}^{189}\equiv d\bar{u}QQQ\bar{L}\nu\nu\,,\\
\mathcal{O}_{12, \ (1,1)}^{190}\equiv ddu\bar{u}Q\bar{L}\nu\nu\,,\\
\mathcal{O}_{12, \ (1,1)}^{191}\equiv d\bar{d}QQQL\nu\bar{\nu}\,,\\
\mathcal{O}_{12, \ (1,1)}^{192}\equiv \bar{d}QQQQe\nu\bar{\nu}\,,\\
\mathcal{O}_{12, \ (1,1)}^{193}\equiv dd\bar{d}uQL\nu\bar{\nu}\,,\\
\mathcal{O}_{12, \ (1,1)}^{194}\equiv d\bar{d}uQQe\nu\bar{\nu}\,,\\
\mathcal{O}_{12, \ (1,1)}^{195}\equiv dd\bar{d}uue\nu\bar{\nu}\,,\\
\mathcal{O}_{12, \ (1,1)}^{196}\equiv dQQLL\bar{e}\nu\bar{\nu}\,,\\
\mathcal{O}_{12, \ (1,1)}^{197}\equiv QQQLe\bar{e}\nu\bar{\nu}\,,\\
\mathcal{O}_{12, \ (1,1)}^{198}\equiv dduLL\bar{e}\nu\bar{\nu}\,,\\
\mathcal{O}_{12, \ (1,1)}^{199}\equiv duQLe\bar{e}\nu\bar{\nu}\,,\\
\mathcal{O}_{12, \ (1,1)}^{200}\equiv uQQee\bar{e}\nu\bar{\nu}\,,\\
\mathcal{O}_{12, \ (1,1)}^{201}\equiv duuee\bar{e}\nu\bar{\nu}\,,\\
\mathcal{O}_{12, \ (1,1)}^{202}\equiv QQQLL\bar{L}\nu\bar{\nu}\,,\\
\mathcal{O}_{12, \ (1,1)}^{203}\equiv duQLL\bar{L}\nu\bar{\nu}\,,\\
\mathcal{O}_{12, \ (1,1)}^{204}\equiv uQQL\bar{L}e\nu\bar{\nu}\,,\\
\mathcal{O}_{12, \ (1,1)}^{205}\equiv duuL\bar{L}e\nu\bar{\nu}\,,\\
\mathcal{O}_{12, \ (1,1)}^{206}\equiv uuQ\bar{L}ee\nu\bar{\nu}\,,\\
\mathcal{O}_{12, \ (1,1)}^{207}\equiv uuQLLL\bar{\nu}\bar{\nu}\,,\\
\mathcal{O}_{12, \ (1,1)}^{208}\equiv uuuLLe\bar{\nu}\bar{\nu}\,,\\
\mathcal{O}_{12, \ (1,1)}^{209}\equiv QQQQ\bar{Q}L\nu\bar{\nu}\,,\\
\mathcal{O}_{12, \ (1,1)}^{210}\equiv duQQ\bar{Q}L\nu\bar{\nu}\,,\\
\mathcal{O}_{12, \ (1,1)}^{211}\equiv uQQQ\bar{Q}e\nu\bar{\nu}\,,\\
\mathcal{O}_{12, \ (1,1)}^{212}\equiv dduu\bar{Q}L\nu\bar{\nu}\,,\\
\mathcal{O}_{12, \ (1,1)}^{213}\equiv duuQ\bar{Q}e\nu\bar{\nu}\,,\\
\mathcal{O}_{12, \ (1,1)}^{214}\equiv u\bar{u}QQQL\nu\bar{\nu}\,,\\
\mathcal{O}_{12, \ (1,1)}^{215}\equiv duu\bar{u}QL\nu\bar{\nu}\,,\\
\mathcal{O}_{12, \ (1,1)}^{216}\equiv uu\bar{u}QQe\nu\bar{\nu}\,,\\
\mathcal{O}_{12, \ (1,1)}^{217}\equiv duuu\bar{u}e\nu\bar{\nu}\,,\\
\mathcal{O}_{12, \ (1,1)}^{218}\equiv ddd\bar{L}\bar{L}\nu\nu\nu\,,\\
\mathcal{O}_{12, \ (1,1)}^{219}\equiv dd\bar{d}QQ\nu\nu\bar{\nu}\,,\\
\mathcal{O}_{12, \ (1,1)}^{220}\equiv ddd\bar{d}u\nu\nu\bar{\nu}\,,\\
\mathcal{O}_{12, \ (1,1)}^{221}\equiv ddQL\bar{e}\nu\nu\bar{\nu}\,,\\
\mathcal{O}_{12, \ (1,1)}^{222}\equiv dQQe\bar{e}\nu\nu\bar{\nu}\,,\\
\mathcal{O}_{12, \ (1,1)}^{223}\equiv ddue\bar{e}\nu\nu\bar{\nu}\,,\\
\mathcal{O}_{12, \ (1,1)}^{224}\equiv dQQL\bar{L}\nu\nu\bar{\nu}\,,\\
\mathcal{O}_{12, \ (1,1)}^{225}\equiv QQQ\bar{L}e\nu\nu\bar{\nu}\,,\\
\mathcal{O}_{12, \ (1,1)}^{226}\equiv dduL\bar{L}\nu\nu\bar{\nu}\,,\\
\mathcal{O}_{12, \ (1,1)}^{227}\equiv duQ\bar{L}e\nu\nu\bar{\nu}\,,\\
\mathcal{O}_{12, \ (1,1)}^{228}\equiv uQQLL\nu\bar{\nu}\bar{\nu}\,,\\
\mathcal{O}_{12, \ (1,1)}^{229}\equiv duuLL\nu\bar{\nu}\bar{\nu}\,,\\
\mathcal{O}_{12, \ (1,1)}^{230}\equiv uuQLe\nu\bar{\nu}\bar{\nu}\,,\\
\mathcal{O}_{12, \ (1,1)}^{231}\equiv uuuee\nu\bar{\nu}\bar{\nu}\,,\\
\mathcal{O}_{12, \ (1,1)}^{232}\equiv dQQQ\bar{Q}\nu\nu\bar{\nu}\,,\\
\mathcal{O}_{12, \ (1,1)}^{233}\equiv dduQ\bar{Q}\nu\nu\bar{\nu}\,,\\
\mathcal{O}_{12, \ (1,1)}^{234}\equiv \bar{u}QQQQ\nu\nu\bar{\nu}\,,\\
\mathcal{O}_{12, \ (1,1)}^{235}\equiv du\bar{u}QQ\nu\nu\bar{\nu}\,,\\
\mathcal{O}_{12, \ (1,1)}^{236}\equiv dduu\bar{u}\nu\nu\bar{\nu}\,,\\
\mathcal{O}_{12, \ (1,1)}^{237}\equiv ddd\bar{e}\nu\nu\nu\bar{\nu}\,,\\
\mathcal{O}_{12, \ (1,1)}^{238}\equiv ddQ\bar{L}\nu\nu\nu\bar{\nu}\,,\\
\mathcal{O}_{12, \ (1,1)}^{239}\equiv QQQL\nu\nu\bar{\nu}\bar{\nu}\,,\\
\mathcal{O}_{12, \ (1,1)}^{240}\equiv duQL\nu\nu\bar{\nu}\bar{\nu}\,,\\
\mathcal{O}_{12, \ (1,1)}^{241}\equiv uQQe\nu\nu\bar{\nu}\bar{\nu}\,,\\
\mathcal{O}_{12, \ (1,1)}^{242}\equiv duue\nu\nu\bar{\nu}\bar{\nu}\,,\\
\mathcal{O}_{12, \ (1,1)}^{243}\equiv dQQ\nu\nu\nu\bar{\nu}\bar{\nu}\,,\\
\mathcal{O}_{12, \ (1,1)}^{244}\equiv ddu\nu\nu\nu\bar{\nu}\bar{\nu}\,,\\
\mathcal{O}_{12, \ (1,1)}^{245}\equiv HH\bar{H}dd\bar{d}QQL\,,\\
\mathcal{O}_{12, \ (1,1)}^{246}\equiv HH\bar{H}d\bar{d}QQQe\,,\\
\mathcal{O}_{12, \ (1,1)}^{247}\equiv HH\bar{H}ddd\bar{d}uL\,,\\
\mathcal{O}_{12, \ (1,1)}^{248}\equiv HH\bar{H}dd\bar{d}uQe\,,\\
\mathcal{O}_{12, \ (1,1)}^{249}\equiv HH\bar{H}ddQLL\bar{e}\,,\\
\mathcal{O}_{12, \ (1,1)}^{250}\equiv HH\bar{H}dQQLe\bar{e}\,,\\
\mathcal{O}_{12, \ (1,1)}^{251}\equiv HH\bar{H}QQQee\bar{e}\,,\\
\mathcal{O}_{12, \ (1,1)}^{252}\equiv HH\bar{H}dduLe\bar{e}\,,\\
\mathcal{O}_{12, \ (1,1)}^{253}\equiv HH\bar{H}duQee\bar{e}\,,\\
\mathcal{O}_{12, \ (1,1)}^{254}\equiv H\bar{H}\bar{H}\bar{d}QQQQL\,,\\
\mathcal{O}_{12, \ (1,1)}^{255}\equiv H\bar{H}\bar{H}d\bar{d}uQQL\,,\\
\mathcal{O}_{12, \ (1,1)}^{256}\equiv H\bar{H}\bar{H}\bar{d}uQQQe\,,\\
\mathcal{O}_{12, \ (1,1)}^{257}\equiv H\bar{H}\bar{H}dd\bar{d}uuL\,,\\
\mathcal{O}_{12, \ (1,1)}^{258}\equiv H\bar{H}\bar{H}d\bar{d}uuQe\,,\\
\mathcal{O}_{12, \ (1,1)}^{259}\equiv H\bar{H}\bar{H}QQQLL\bar{e}\,,\\
\mathcal{O}_{12, \ (1,1)}^{260}\equiv H\bar{H}\bar{H}duQLL\bar{e}\,,\\
\mathcal{O}_{12, \ (1,1)}^{261}\equiv H\bar{H}\bar{H}uQQLe\bar{e}\,,\\
\mathcal{O}_{12, \ (1,1)}^{262}\equiv H\bar{H}\bar{H}duuLe\bar{e}\,,\\
\mathcal{O}_{12, \ (1,1)}^{263}\equiv H\bar{H}\bar{H}uuQee\bar{e}\,,\\
\mathcal{O}_{12, \ (1,1)}^{264}\equiv \bar{H}\bar{H}\bar{H}\bar{d}uuQQL\,,\\
\mathcal{O}_{12, \ (1,1)}^{265}\equiv \bar{H}\bar{H}\bar{H}uuQLL\bar{e}\,,\\
\mathcal{O}_{12, \ (1,1)}^{266}\equiv \bar{H}\bar{H}\bar{H}uuuLL\bar{L}\,,\\
\mathcal{O}_{12, \ (1,1)}^{267}\equiv \bar{H}\bar{H}\bar{H}uuuQ\bar{Q}L\,,\\
\mathcal{O}_{12, \ (1,1)}^{268}\equiv H\bar{H}\bar{H}uQQLL\bar{L}\,,\\
\mathcal{O}_{12, \ (1,1)}^{269}\equiv H\bar{H}\bar{H}duuLL\bar{L}\,,\\
\mathcal{O}_{12, \ (1,1)}^{270}\equiv H\bar{H}\bar{H}uuQL\bar{L}e\,,\\
\mathcal{O}_{12, \ (1,1)}^{271}\equiv H\bar{H}\bar{H}uuu\bar{L}ee\,,\\
\mathcal{O}_{12, \ (1,1)}^{272}\equiv H\bar{H}\bar{H}uQQQ\bar{Q}L\,,\\
\mathcal{O}_{12, \ (1,1)}^{273}\equiv H\bar{H}\bar{H}duuQ\bar{Q}L\,,\\
\mathcal{O}_{12, \ (1,1)}^{274}\equiv H\bar{H}\bar{H}uuQQ\bar{Q}e\,,\\
\mathcal{O}_{12, \ (1,1)}^{275}\equiv H\bar{H}\bar{H}duuu\bar{Q}e\,,\\
\mathcal{O}_{12, \ (1,1)}^{276}\equiv H\bar{H}\bar{H}uu\bar{u}QQL\,,\\
\mathcal{O}_{12, \ (1,1)}^{277}\equiv H\bar{H}\bar{H}duuu\bar{u}L\,,\\
\mathcal{O}_{12, \ (1,1)}^{278}\equiv H\bar{H}\bar{H}uuu\bar{u}Qe\,,\\
\mathcal{O}_{12, \ (1,1)}^{279}\equiv HH\bar{H}dQQLL\bar{L}\,,\\
\mathcal{O}_{12, \ (1,1)}^{280}\equiv HH\bar{H}QQQL\bar{L}e\,,\\
\mathcal{O}_{12, \ (1,1)}^{281}\equiv HH\bar{H}dduLL\bar{L}\,,\\
\mathcal{O}_{12, \ (1,1)}^{282}\equiv HH\bar{H}duQL\bar{L}e\,,\\
\mathcal{O}_{12, \ (1,1)}^{283}\equiv HH\bar{H}uQQ\bar{L}ee\,,\\
\mathcal{O}_{12, \ (1,1)}^{284}\equiv HH\bar{H}duu\bar{L}ee\,,\\
\mathcal{O}_{12, \ (1,1)}^{285}\equiv HH\bar{H}dQQQ\bar{Q}L\,,\\
\mathcal{O}_{12, \ (1,1)}^{286}\equiv HH\bar{H}QQQQ\bar{Q}e\,,\\
\mathcal{O}_{12, \ (1,1)}^{287}\equiv HH\bar{H}dduQ\bar{Q}L\,,\\
\mathcal{O}_{12, \ (1,1)}^{288}\equiv HH\bar{H}duQQ\bar{Q}e\,,\\
\mathcal{O}_{12, \ (1,1)}^{289}\equiv HH\bar{H}dduu\bar{Q}e\,,\\
\mathcal{O}_{12, \ (1,1)}^{290}\equiv HH\bar{H}\bar{u}QQQQL\,,\\
\mathcal{O}_{12, \ (1,1)}^{291}\equiv HH\bar{H}du\bar{u}QQL\,,\\
\mathcal{O}_{12, \ (1,1)}^{292}\equiv HH\bar{H}u\bar{u}QQQe\,,\\
\mathcal{O}_{12, \ (1,1)}^{293}\equiv HH\bar{H}dduu\bar{u}L\,,\\
\mathcal{O}_{12, \ (1,1)}^{294}\equiv HH\bar{H}duu\bar{u}Qe\,,\\
\mathcal{O}_{12, \ (1,1)}^{295}\equiv HHHdddLL\bar{L}\,,\\
\mathcal{O}_{12, \ (1,1)}^{296}\equiv HHHddQL\bar{L}e\,,\\
\mathcal{O}_{12, \ (1,1)}^{297}\equiv HHHdQQ\bar{L}ee\,,\\
\mathcal{O}_{12, \ (1,1)}^{298}\equiv HHHdddQ\bar{Q}L\,,\\
\mathcal{O}_{12, \ (1,1)}^{299}\equiv HHHddQQ\bar{Q}e\,,\\
\mathcal{O}_{12, \ (1,1)}^{300}\equiv HHHdd\bar{u}QQL\,,\\
\mathcal{O}_{12, \ (1,1)}^{301}\equiv HHHd\bar{u}QQQe\,,\\
\mathcal{O}_{12, \ (1,1)}^{302}\equiv HH\bar{H}ddd\bar{d}Q\nu\,,\\
\mathcal{O}_{12, \ (1,1)}^{303}\equiv HH\bar{H}dddL\bar{e}\nu\,,\\
\mathcal{O}_{12, \ (1,1)}^{304}\equiv HH\bar{H}ddQe\bar{e}\nu\,,\\
\mathcal{O}_{12, \ (1,1)}^{305}\equiv H\bar{H}\bar{H}d\bar{d}QQQ\nu\,,\\
\mathcal{O}_{12, \ (1,1)}^{306}\equiv H\bar{H}\bar{H}dd\bar{d}uQ\nu\,,\\
\mathcal{O}_{12, \ (1,1)}^{307}\equiv H\bar{H}\bar{H}dQQL\bar{e}\nu\,,\\
\mathcal{O}_{12, \ (1,1)}^{308}\equiv H\bar{H}\bar{H}QQQe\bar{e}\nu\,,\\
\mathcal{O}_{12, \ (1,1)}^{309}\equiv H\bar{H}\bar{H}dduL\bar{e}\nu\,,\\
\mathcal{O}_{12, \ (1,1)}^{310}\equiv H\bar{H}\bar{H}duQe\bar{e}\nu\,,\\
\mathcal{O}_{12, \ (1,1)}^{311}\equiv \bar{H}\bar{H}\bar{H}\bar{d}uQQQ\nu\,,\\
\mathcal{O}_{12, \ (1,1)}^{312}\equiv \bar{H}\bar{H}\bar{H}uQQL\bar{e}\nu\,,\\
\mathcal{O}_{12, \ (1,1)}^{313}\equiv \bar{H}\bar{H}\bar{H}uuQL\bar{L}\nu\,,\\
\mathcal{O}_{12, \ (1,1)}^{314}\equiv \bar{H}\bar{H}\bar{H}uuQQ\bar{Q}\nu\,,\\
\mathcal{O}_{12, \ (1,1)}^{315}\equiv H\bar{H}\bar{H}QQQL\bar{L}\nu\,,\\
\mathcal{O}_{12, \ (1,1)}^{316}\equiv H\bar{H}\bar{H}duQL\bar{L}\nu\,,\\
\mathcal{O}_{12, \ (1,1)}^{317}\equiv H\bar{H}\bar{H}uQQ\bar{L}e\nu\,,\\
\mathcal{O}_{12, \ (1,1)}^{318}\equiv H\bar{H}\bar{H}duu\bar{L}e\nu\,,\\
\mathcal{O}_{12, \ (1,1)}^{319}\equiv H\bar{H}\bar{H}QQQQ\bar{Q}\nu\,,\\
\mathcal{O}_{12, \ (1,1)}^{320}\equiv H\bar{H}\bar{H}duQQ\bar{Q}\nu\,,\\
\mathcal{O}_{12, \ (1,1)}^{321}\equiv H\bar{H}\bar{H}dduu\bar{Q}\nu\,,\\
\mathcal{O}_{12, \ (1,1)}^{322}\equiv H\bar{H}\bar{H}u\bar{u}QQQ\nu\,,\\
\mathcal{O}_{12, \ (1,1)}^{323}\equiv H\bar{H}\bar{H}duu\bar{u}Q\nu\,,\\
\mathcal{O}_{12, \ (1,1)}^{324}\equiv HH\bar{H}ddQL\bar{L}\nu\,,\\
\mathcal{O}_{12, \ (1,1)}^{325}\equiv HH\bar{H}dQQ\bar{L}e\nu\,,\\
\mathcal{O}_{12, \ (1,1)}^{326}\equiv HH\bar{H}ddu\bar{L}e\nu\,,\\
\mathcal{O}_{12, \ (1,1)}^{327}\equiv HH\bar{H}ddQQ\bar{Q}\nu\,,\\
\mathcal{O}_{12, \ (1,1)}^{328}\equiv HH\bar{H}dddu\bar{Q}\nu\,,\\
\mathcal{O}_{12, \ (1,1)}^{329}\equiv HH\bar{H}d\bar{u}QQQ\nu\,,\\
\mathcal{O}_{12, \ (1,1)}^{330}\equiv HH\bar{H}ddu\bar{u}Q\nu\,,\\
\mathcal{O}_{12, \ (1,1)}^{331}\equiv HHHddQLL\bar{\nu}\,,\\
\mathcal{O}_{12, \ (1,1)}^{332}\equiv HHHdQQLe\bar{\nu}\,,\\
\mathcal{O}_{12, \ (1,1)}^{333}\equiv H\bar{H}\bar{H}uuQLL\bar{\nu}\,,\\
\mathcal{O}_{12, \ (1,1)}^{334}\equiv H\bar{H}\bar{H}uuuLe\bar{\nu}\,,\\
\mathcal{O}_{12, \ (1,1)}^{335}\equiv HH\bar{H}QQQLL\bar{\nu}\,,\\
\mathcal{O}_{12, \ (1,1)}^{336}\equiv HH\bar{H}duQLL\bar{\nu}\,,\\
\mathcal{O}_{12, \ (1,1)}^{337}\equiv HH\bar{H}uQQLe\bar{\nu}\,,\\
\mathcal{O}_{12, \ (1,1)}^{338}\equiv HH\bar{H}duuLe\bar{\nu}\,,\\
\mathcal{O}_{12, \ (1,1)}^{339}\equiv HH\bar{H}uuQee\bar{\nu}\,,\\
\mathcal{O}_{12, \ (1,1)}^{340}\equiv HHHQQQee\bar{\nu}\,,\\
\mathcal{O}_{12, \ (1,1)}^{341}\equiv H\bar{H}\bar{H}ddQ\bar{e}\nu\nu\,,\\
\mathcal{O}_{12, \ (1,1)}^{342}\equiv \bar{H}\bar{H}\bar{H}QQQ\bar{e}\nu\nu\,,\\
\mathcal{O}_{12, \ (1,1)}^{343}\equiv \bar{H}\bar{H}\bar{H}uQQ\bar{L}\nu\nu\,,\\
\mathcal{O}_{12, \ (1,1)}^{344}\equiv H\bar{H}\bar{H}dQQ\bar{L}\nu\nu\,,\\
\mathcal{O}_{12, \ (1,1)}^{345}\equiv H\bar{H}\bar{H}ddu\bar{L}\nu\nu\,,\\
\mathcal{O}_{12, \ (1,1)}^{346}\equiv HH\bar{H}ddd\bar{L}\nu\nu\,,\\
\mathcal{O}_{12, \ (1,1)}^{347}\equiv HH\bar{H}dQQL\nu\bar{\nu}\,,\\
\mathcal{O}_{12, \ (1,1)}^{348}\equiv H\bar{H}\bar{H}uQQL\nu\bar{\nu}\,,\\
\mathcal{O}_{12, \ (1,1)}^{349}\equiv H\bar{H}\bar{H}duuL\nu\bar{\nu}\,,\\
\mathcal{O}_{12, \ (1,1)}^{350}\equiv H\bar{H}\bar{H}uuQe\nu\bar{\nu}\,,\\
\mathcal{O}_{12, \ (1,1)}^{351}\equiv HH\bar{H}QQQe\nu\bar{\nu}\,,\\
\mathcal{O}_{12, \ (1,1)}^{352}\equiv HH\bar{H}dduL\nu\bar{\nu}\,,\\
\mathcal{O}_{12, \ (1,1)}^{353}\equiv HH\bar{H}duQe\nu\bar{\nu}\,,\\
\mathcal{O}_{12, \ (1,1)}^{354}\equiv HH\bar{H}ddQ\nu\nu\bar{\nu}\,,\\
\mathcal{O}_{12, \ (1,1)}^{355}\equiv H\bar{H}\bar{H}QQQ\nu\nu\bar{\nu}\,,\\
\mathcal{O}_{12, \ (1,1)}^{356}\equiv H\bar{H}\bar{H}duQ\nu\nu\bar{\nu}\,,\\
\mathcal{O}_{12, \ (1,1)}^{357}\equiv HHHH\bar{H}\bar{H}ddQL\,,\\
\mathcal{O}_{12, \ (1,1)}^{358}\equiv HHHH\bar{H}\bar{H}dQQe\,,\\
\mathcal{O}_{12, \ (1,1)}^{359}\equiv HH\bar{H}\bar{H}\bar{H}\bar{H}uuQL\,,\\
\mathcal{O}_{12, \ (1,1)}^{360}\equiv HHH\bar{H}\bar{H}\bar{H}QQQL\,,\\
\mathcal{O}_{12, \ (1,1)}^{361}\equiv HHH\bar{H}\bar{H}\bar{H}duQL\,,\\
\mathcal{O}_{12, \ (1,1)}^{362}\equiv HHH\bar{H}\bar{H}\bar{H}uQQe\,,\\
\mathcal{O}_{12, \ (1,1)}^{363}\equiv HHH\bar{H}\bar{H}\bar{H}duue\,,\\
\mathcal{O}_{12, \ (1,1)}^{364}\equiv HHH\bar{H}\bar{H}\bar{H}dQQ\nu\,,\\
\mathcal{O}_{12, \ (1,1)}^{365}\equiv HH\bar{H}\bar{H}\bar{H}\bar{H}uQQ\nu\,,\\
\mathcal{O}_{12, \ (1,1)}^{366}\equiv HHH\bar{H}\bar{H}\bar{H}ddu\nu\,.
 \label{eq:dequal12basisB1L1}
 \end{math}
 \end{spacing}
 \end{multicols}

\clearpage
 

\section{Dimension 13}
\label{sec:d13}
 
\begin{figure}
     \centering
     \includegraphics[width=1\linewidth]{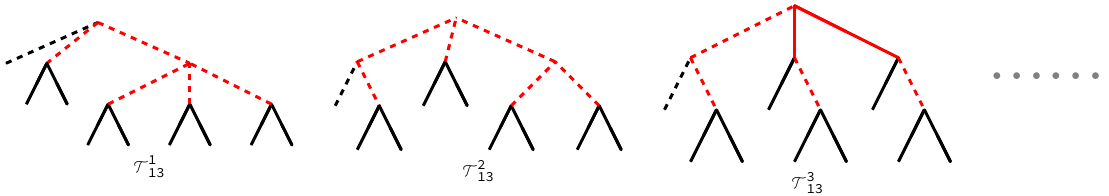}
     \caption{A few representative Feynman-diagram topologies for mass dimension~13. All 3689 topologies are provided at \website. }
     \label{fig:d=13_topologies}
 \end{figure}

At $d=13$, we find operators with different $\Delta B$ and $\Delta L$, which we discuss separately. The number of topologies is too large to include here, see Fig.~\ref{fig:d=13_topologies}.

 \subsection{\texorpdfstring{$\Delta B=1, \Delta L=-5$}{Delta B=1,Delta L=-5}}

$d=13$ is the lowest mass dimension for $(\Delta B,\Delta L)=(1,-5)$ operators, so these can be symmetry protected. 
   \begin{multicols}{3}
 \begin{spacing}{1.5}
 \noindent
 \begin{math}
 \mathcal{O}_{13, \ (1,-5)}^{1}\equiv \bar{H}ddd\bar{L}\bar{L}\bar{L}\bar{\nu}\bar{\nu}\,,\\
\mathcal{O}_{13, \ (1,-5)}^{2}\equiv \bar{H}ddd\bar{L}\bar{e}\bar{\nu}\bar{\nu}\bar{\nu}\,,\\
\mathcal{O}_{13, \ (1,-5)}^{3}\equiv \bar{H}ddQ\bar{L}\bar{L}\bar{\nu}\bar{\nu}\bar{\nu}\,,\\
\mathcal{O}_{13, \ (1,-5)}^{4}\equiv \bar{H}ddu\bar{L}\bar{\nu}\bar{\nu}\bar{\nu}\bar{\nu}\,,\\
\mathcal{O}_{13, \ (1,-5)}^{5}\equiv Hddd\bar{L}\bar{\nu}\bar{\nu}\bar{\nu}\bar{\nu}\,,\\
\mathcal{O}_{13, \ (1,-5)}^{6}\equiv \bar{H}ddQ\bar{e}\bar{\nu}\bar{\nu}\bar{\nu}\bar{\nu}\,,\\
\mathcal{O}_{13, \ (1,-5)}^{7}\equiv \bar{H}dQQ\bar{L}\bar{\nu}\bar{\nu}\bar{\nu}\bar{\nu}\,,\\
\mathcal{O}_{13, \ (1,-5)}^{8}\equiv HddQ\bar{\nu}\bar{\nu}\bar{\nu}\bar{\nu}\bar{\nu}\,,\\
\mathcal{O}_{13, \ (1,-5)}^{9}\equiv \bar{H}duQ\bar{\nu}\bar{\nu}\bar{\nu}\bar{\nu}\bar{\nu}\,,\\
\mathcal{O}_{13, \ (1,-5)}^{10}\equiv \bar{H}QQQ\bar{\nu}\bar{\nu}\bar{\nu}\bar{\nu}\bar{\nu}\,.
 \label{eq:dequal13basisB1L-5}
 \end{math}
 \end{spacing}
 \end{multicols}
 These $(\Delta B ,\Delta L) =(1,-5)$ operators require between two and five right-handed neutrinos and thus do not show up in the SMEFT, or at least not at $d=13$. 
 The high mass dimension of these operators together with the phase-space suppression from at least five final-state particles renders these operators very suppressed. For example, $ \mathcal{O}_{13, \,(1 , -5)}^{9} \equiv  \bar{H}duQ\bar{\nu}\bar{\nu}\bar{\nu}\bar{\nu}\bar{\nu} $ gives 
 \begin{align}
        \Gamma (n\to \nu\nu\nu\nu\nu )\simeq \frac{\alpha^2 m_n^{11}\langle H\rangle^2}{2^{19} 3^2 \pi^7 \Lambda^{18}}\simeq \left( \unit[10^{32}]{yr}\right)^{-1} \left(\frac{\unit[1]{TeV}}{\Lambda} \right)^{18} ,
\end{align}
where $\alpha \simeq \unit[0.01]{GeV^3}$ is a QCD matrix element~\cite{Yoo:2021gql} and the phase space factor is taken from Ref.~\cite{Kleiss:1985gy}. The signature is just invisible neutron decay in this case, currently limited to lifetimes exceeding $\unit[9\times 10^{29}]{yr}$~\cite{SNO:2022trz}, to be improved by a factor of 50  with JUNO~\cite{JUNO:2024pur}. Even these doubly suppressed $d=13$ five-body decays could be observable if all the new particles are near the TeV scale. The decay rate can be larger for UV completions involving SM singlets, which could have masses below TeV.

  \subsection{\texorpdfstring{$\Delta B=1, \Delta L=-1$}{Delta B=1,Delta L=-1}}
 \begin{multicols}{2}
 \begin{spacing}{1.5}
 \noindent
 \begin{math}
 \mathcal{O}_{13, \ (1,-1)}^{1}\equiv Hdddd\bar{d}\bar{u}Q\bar{e}\,,\\
\mathcal{O}_{13, \ (1,-1)}^{2}\equiv Hdddd\bar{Q}e\bar{e}\bar{e}\,,\\
\mathcal{O}_{13, \ (1,-1)}^{3}\equiv Hdddd\bar{u}L\bar{e}\bar{e}\,,\\
\mathcal{O}_{13, \ (1,-1)}^{4}\equiv Hddd\bar{u}Qe\bar{e}\bar{e}\,,\\
\mathcal{O}_{13, \ (1,-1)}^{5}\equiv Hddddd\bar{d}\bar{Q}\bar{e}\,,\\
\mathcal{O}_{13, \ (1,-1)}^{6}\equiv HddddQ\bar{Q}\bar{Q}\bar{e}\,,\\
\mathcal{O}_{13, \ (1,-1)}^{7}\equiv Hddd\bar{u}QQ\bar{Q}\bar{e}\,,\\
\mathcal{O}_{13, \ (1,-1)}^{8}\equiv Hddddu\bar{u}\bar{Q}\bar{e}\,,\\
\mathcal{O}_{13, \ (1,-1)}^{9}\equiv Hdd\bar{u}\bar{u}QQQ\bar{e}\,,\\
\mathcal{O}_{13, \ (1,-1)}^{10}\equiv Hdddu\bar{u}\bar{u}Q\bar{e}\,,\\
\mathcal{O}_{13, \ (1,-1)}^{11}\equiv \bar{H}dddd\bar{d}\bar{d}Q\bar{e}\,,\\
\mathcal{O}_{13, \ (1,-1)}^{12}\equiv \bar{H}dd\bar{d}\bar{u}QQQ\bar{e}\,,\\
\mathcal{O}_{13, \ (1,-1)}^{13}\equiv \bar{H}ddd\bar{d}u\bar{u}Q\bar{e}\,,\\
\mathcal{O}_{13, \ (1,-1)}^{14}\equiv \bar{H}dddd\bar{d}L\bar{e}\bar{e}\,,\\
\mathcal{O}_{13, \ (1,-1)}^{15}\equiv \bar{H}ddd\bar{d}Qe\bar{e}\bar{e}\,,\\
\mathcal{O}_{13, \ (1,-1)}^{16}\equiv \bar{H}dddLe\bar{e}\bar{e}\bar{e}\,,\\
\mathcal{O}_{13, \ (1,-1)}^{17}\equiv \bar{H}ddQee\bar{e}\bar{e}\bar{e}\,,\\
\mathcal{O}_{13, \ (1,-1)}^{18}\equiv \bar{H}dddQ\bar{Q}L\bar{e}\bar{e}\,,\\
\mathcal{O}_{13, \ (1,-1)}^{19}\equiv \bar{H}ddQQ\bar{Q}e\bar{e}\bar{e}\,,\\
\mathcal{O}_{13, \ (1,-1)}^{20}\equiv \bar{H}dddu\bar{Q}e\bar{e}\bar{e}\,,\\
\mathcal{O}_{13, \ (1,-1)}^{21}\equiv \bar{H}dd\bar{u}QQL\bar{e}\bar{e}\,,\\
\mathcal{O}_{13, \ (1,-1)}^{22}\equiv \bar{H}d\bar{u}QQQe\bar{e}\bar{e}\,,\\
\mathcal{O}_{13, \ (1,-1)}^{23}\equiv \bar{H}dddu\bar{u}L\bar{e}\bar{e}\,,\\
\mathcal{O}_{13, \ (1,-1)}^{24}\equiv \bar{H}ddu\bar{u}Qe\bar{e}\bar{e}\,,\\
\mathcal{O}_{13, \ (1,-1)}^{25}\equiv \bar{H}ddd\bar{d}QQ\bar{Q}\bar{e}\,,\\
\mathcal{O}_{13, \ (1,-1)}^{26}\equiv \bar{H}dddd\bar{d}u\bar{Q}\bar{e}\,,\\
\mathcal{O}_{13, \ (1,-1)}^{27}\equiv \bar{H}ddQQQ\bar{Q}\bar{Q}\bar{e}\,,\\
\mathcal{O}_{13, \ (1,-1)}^{28}\equiv \bar{H}ddduQ\bar{Q}\bar{Q}\bar{e}\,,\\
\mathcal{O}_{13, \ (1,-1)}^{29}\equiv \bar{H}d\bar{u}QQQQ\bar{Q}\bar{e}\,,\\
\mathcal{O}_{13, \ (1,-1)}^{30}\equiv \bar{H}ddu\bar{u}QQ\bar{Q}\bar{e}\,,\\
\mathcal{O}_{13, \ (1,-1)}^{31}\equiv \bar{H}ddduu\bar{u}\bar{Q}\bar{e}\,,\\
\mathcal{O}_{13, \ (1,-1)}^{32}\equiv \bar{H}\bar{u}\bar{u}QQQQQ\bar{e}\,,\\
\mathcal{O}_{13, \ (1,-1)}^{33}\equiv \bar{H}du\bar{u}\bar{u}QQQ\bar{e}\,,\\
\mathcal{O}_{13, \ (1,-1)}^{34}\equiv \bar{H}dduu\bar{u}\bar{u}Q\bar{e}\,,\\
\mathcal{O}_{13, \ (1,-1)}^{35}\equiv \bar{H}ddd\bar{d}\bar{d}QQ\bar{L}\,,\\
\mathcal{O}_{13, \ (1,-1)}^{36}\equiv \bar{H}dddd\bar{d}\bar{d}u\bar{L}\,,\\
\mathcal{O}_{13, \ (1,-1)}^{37}\equiv \bar{H}d\bar{d}\bar{u}QQQQ\bar{L}\,,\\
\mathcal{O}_{13, \ (1,-1)}^{38}\equiv \bar{H}dd\bar{d}u\bar{u}QQ\bar{L}\,,\\
\mathcal{O}_{13, \ (1,-1)}^{39}\equiv \bar{H}ddd\bar{d}uu\bar{u}\bar{L}\,,\\
\mathcal{O}_{13, \ (1,-1)}^{40}\equiv \bar{H}ddd\bar{d}QL\bar{L}\bar{e}\,,\\
\mathcal{O}_{13, \ (1,-1)}^{41}\equiv \bar{H}dd\bar{d}QQ\bar{L}e\bar{e}\,,\\
\mathcal{O}_{13, \ (1,-1)}^{42}\equiv \bar{H}ddd\bar{d}u\bar{L}e\bar{e}\,,\\
\mathcal{O}_{13, \ (1,-1)}^{43}\equiv \bar{H}dddLL\bar{L}\bar{e}\bar{e}\,,\\
\mathcal{O}_{13, \ (1,-1)}^{44}\equiv \bar{H}ddQL\bar{L}e\bar{e}\bar{e}\,,\\
\mathcal{O}_{13, \ (1,-1)}^{45}\equiv \bar{H}dQQ\bar{L}ee\bar{e}\bar{e}\,,\\
\mathcal{O}_{13, \ (1,-1)}^{46}\equiv \bar{H}ddu\bar{L}ee\bar{e}\bar{e}\,,\\
\mathcal{O}_{13, \ (1,-1)}^{47}\equiv \bar{H}ddQQ\bar{Q}L\bar{L}\bar{e}\,,\\
\mathcal{O}_{13, \ (1,-1)}^{48}\equiv \bar{H}dQQQ\bar{Q}\bar{L}e\bar{e}\,,\\
\mathcal{O}_{13, \ (1,-1)}^{49}\equiv \bar{H}dddu\bar{Q}L\bar{L}\bar{e}\,,\\
\mathcal{O}_{13, \ (1,-1)}^{50}\equiv \bar{H}dduQ\bar{Q}\bar{L}e\bar{e}\,,\\
\mathcal{O}_{13, \ (1,-1)}^{51}\equiv \bar{H}d\bar{u}QQQL\bar{L}\bar{e}\,,\\
\mathcal{O}_{13, \ (1,-1)}^{52}\equiv \bar{H}\bar{u}QQQQ\bar{L}e\bar{e}\,,\\
\mathcal{O}_{13, \ (1,-1)}^{53}\equiv \bar{H}ddu\bar{u}QL\bar{L}\bar{e}\,,\\
\mathcal{O}_{13, \ (1,-1)}^{54}\equiv \bar{H}du\bar{u}QQ\bar{L}e\bar{e}\,,\\
\mathcal{O}_{13, \ (1,-1)}^{55}\equiv \bar{H}dduu\bar{u}\bar{L}e\bar{e}\,,\\
\mathcal{O}_{13, \ (1,-1)}^{56}\equiv \bar{H}dd\bar{d}QQL\bar{L}\bar{L}\,,\\
\mathcal{O}_{13, \ (1,-1)}^{57}\equiv \bar{H}d\bar{d}QQQ\bar{L}\bar{L}e\,,\\
\mathcal{O}_{13, \ (1,-1)}^{58}\equiv \bar{H}ddd\bar{d}uL\bar{L}\bar{L}\,,\\
\mathcal{O}_{13, \ (1,-1)}^{59}\equiv \bar{H}dd\bar{d}uQ\bar{L}\bar{L}e\,,\\
\mathcal{O}_{13, \ (1,-1)}^{60}\equiv \bar{H}ddQLL\bar{L}\bar{L}\bar{e}\,,\\
\mathcal{O}_{13, \ (1,-1)}^{61}\equiv \bar{H}dQQL\bar{L}\bar{L}e\bar{e}\,,\\
\mathcal{O}_{13, \ (1,-1)}^{62}\equiv \bar{H}QQQ\bar{L}\bar{L}ee\bar{e}\,,\\
\mathcal{O}_{13, \ (1,-1)}^{63}\equiv \bar{H}dduL\bar{L}\bar{L}e\bar{e}\,,\\
\mathcal{O}_{13, \ (1,-1)}^{64}\equiv \bar{H}duQ\bar{L}\bar{L}ee\bar{e}\,,\\
\mathcal{O}_{13, \ (1,-1)}^{65}\equiv \bar{H}dQQLL\bar{L}\bar{L}\bar{L}\,,\\
\mathcal{O}_{13, \ (1,-1)}^{66}\equiv \bar{H}QQQL\bar{L}\bar{L}\bar{L}e\,,\\
\mathcal{O}_{13, \ (1,-1)}^{67}\equiv \bar{H}dduLL\bar{L}\bar{L}\bar{L}\,,\\
\mathcal{O}_{13, \ (1,-1)}^{68}\equiv \bar{H}duQL\bar{L}\bar{L}\bar{L}e\,,\\
\mathcal{O}_{13, \ (1,-1)}^{69}\equiv \bar{H}uQQ\bar{L}\bar{L}\bar{L}ee\,,\\
\mathcal{O}_{13, \ (1,-1)}^{70}\equiv \bar{H}duu\bar{L}\bar{L}\bar{L}ee\,,\\
\mathcal{O}_{13, \ (1,-1)}^{71}\equiv \bar{H}dQQQ\bar{Q}L\bar{L}\bar{L}\,,\\
\mathcal{O}_{13, \ (1,-1)}^{72}\equiv \bar{H}QQQQ\bar{Q}\bar{L}\bar{L}e\,,\\
\mathcal{O}_{13, \ (1,-1)}^{73}\equiv \bar{H}dduQ\bar{Q}L\bar{L}\bar{L}\,,\\
\mathcal{O}_{13, \ (1,-1)}^{74}\equiv \bar{H}duQQ\bar{Q}\bar{L}\bar{L}e\,,\\
\mathcal{O}_{13, \ (1,-1)}^{75}\equiv \bar{H}dduu\bar{Q}\bar{L}\bar{L}e\,,\\
\mathcal{O}_{13, \ (1,-1)}^{76}\equiv \bar{H}\bar{u}QQQQL\bar{L}\bar{L}\,,\\
\mathcal{O}_{13, \ (1,-1)}^{77}\equiv \bar{H}du\bar{u}QQL\bar{L}\bar{L}\,,\\
\mathcal{O}_{13, \ (1,-1)}^{78}\equiv \bar{H}u\bar{u}QQQ\bar{L}\bar{L}e\,,\\
\mathcal{O}_{13, \ (1,-1)}^{79}\equiv \bar{H}dduu\bar{u}L\bar{L}\bar{L}\,,\\
\mathcal{O}_{13, \ (1,-1)}^{80}\equiv \bar{H}duu\bar{u}Q\bar{L}\bar{L}e\,,\\
\mathcal{O}_{13, \ (1,-1)}^{81}\equiv \bar{H}dd\bar{d}QQQ\bar{Q}\bar{L}\,,\\
\mathcal{O}_{13, \ (1,-1)}^{82}\equiv \bar{H}ddd\bar{d}uQ\bar{Q}\bar{L}\,,\\
\mathcal{O}_{13, \ (1,-1)}^{83}\equiv \bar{H}dQQQQ\bar{Q}\bar{Q}\bar{L}\,,\\
\mathcal{O}_{13, \ (1,-1)}^{84}\equiv \bar{H}dduQQ\bar{Q}\bar{Q}\bar{L}\,,\\
\mathcal{O}_{13, \ (1,-1)}^{85}\equiv \bar{H}ddduu\bar{Q}\bar{Q}\bar{L}\,,\\
\mathcal{O}_{13, \ (1,-1)}^{86}\equiv \bar{H}\bar{u}QQQQQ\bar{Q}\bar{L}\,,\\
\mathcal{O}_{13, \ (1,-1)}^{87}\equiv \bar{H}du\bar{u}QQQ\bar{Q}\bar{L}\,,\\
\mathcal{O}_{13, \ (1,-1)}^{88}\equiv \bar{H}dduu\bar{u}Q\bar{Q}\bar{L}\,,\\
\mathcal{O}_{13, \ (1,-1)}^{89}\equiv \bar{H}u\bar{u}\bar{u}QQQQ\bar{L}\,,\\
\mathcal{O}_{13, \ (1,-1)}^{90}\equiv \bar{H}duu\bar{u}\bar{u}QQ\bar{L}\,,\\
\mathcal{O}_{13, \ (1,-1)}^{91}\equiv \bar{H}dduuu\bar{u}\bar{u}\bar{L}\,,\\
\mathcal{O}_{13, \ (1,-1)}^{92}\equiv Hddddd\bar{d}\bar{d}\bar{L}\,,\\
\mathcal{O}_{13, \ (1,-1)}^{93}\equiv Hddd\bar{d}\bar{u}QQ\bar{L}\,,\\
\mathcal{O}_{13, \ (1,-1)}^{94}\equiv Hdddd\bar{d}u\bar{u}\bar{L}\,,\\
\mathcal{O}_{13, \ (1,-1)}^{95}\equiv Hdddd\bar{d}\bar{L}e\bar{e}\,,\\
\mathcal{O}_{13, \ (1,-1)}^{96}\equiv Hddd\bar{L}ee\bar{e}\bar{e}\,,\\
\mathcal{O}_{13, \ (1,-1)}^{97}\equiv Hdddd\bar{Q}L\bar{L}\bar{e}\,,\\
\mathcal{O}_{13, \ (1,-1)}^{98}\equiv HdddQ\bar{Q}\bar{L}e\bar{e}\,,\\
\mathcal{O}_{13, \ (1,-1)}^{99}\equiv Hddd\bar{u}QL\bar{L}\bar{e}\,,\\
\mathcal{O}_{13, \ (1,-1)}^{100}\equiv Hdd\bar{u}QQ\bar{L}e\bar{e}\,,\\
\mathcal{O}_{13, \ (1,-1)}^{101}\equiv Hdddu\bar{u}\bar{L}e\bar{e}\,,\\
\mathcal{O}_{13, \ (1,-1)}^{102}\equiv Hdddd\bar{d}L\bar{L}\bar{L}\,,\\
\mathcal{O}_{13, \ (1,-1)}^{103}\equiv Hddd\bar{d}Q\bar{L}\bar{L}e\,,\\
\mathcal{O}_{13, \ (1,-1)}^{104}\equiv HdddL\bar{L}\bar{L}e\bar{e}\,,\\
\mathcal{O}_{13, \ (1,-1)}^{105}\equiv HddQ\bar{L}\bar{L}ee\bar{e}\,,\\
\mathcal{O}_{13, \ (1,-1)}^{106}\equiv HdddLL\bar{L}\bar{L}\bar{L}\,,\\
\mathcal{O}_{13, \ (1,-1)}^{107}\equiv HddQL\bar{L}\bar{L}\bar{L}e\,,\\
\mathcal{O}_{13, \ (1,-1)}^{108}\equiv HdQQ\bar{L}\bar{L}\bar{L}ee\,,\\
\mathcal{O}_{13, \ (1,-1)}^{109}\equiv Hddu\bar{L}\bar{L}\bar{L}ee\,,\\
\mathcal{O}_{13, \ (1,-1)}^{110}\equiv HdddQ\bar{Q}L\bar{L}\bar{L}\,,\\
\mathcal{O}_{13, \ (1,-1)}^{111}\equiv HddQQ\bar{Q}\bar{L}\bar{L}e\,,\\
\mathcal{O}_{13, \ (1,-1)}^{112}\equiv Hdddu\bar{Q}\bar{L}\bar{L}e\,,\\
\mathcal{O}_{13, \ (1,-1)}^{113}\equiv Hdd\bar{u}QQL\bar{L}\bar{L}\,,\\
\mathcal{O}_{13, \ (1,-1)}^{114}\equiv Hd\bar{u}QQQ\bar{L}\bar{L}e\,,\\
\mathcal{O}_{13, \ (1,-1)}^{115}\equiv Hdddu\bar{u}L\bar{L}\bar{L}\,,\\
\mathcal{O}_{13, \ (1,-1)}^{116}\equiv Hddu\bar{u}Q\bar{L}\bar{L}e\,,\\
\mathcal{O}_{13, \ (1,-1)}^{117}\equiv Hdddd\bar{d}Q\bar{Q}\bar{L}\,,\\
\mathcal{O}_{13, \ (1,-1)}^{118}\equiv HdddQQ\bar{Q}\bar{Q}\bar{L}\,,\\
\mathcal{O}_{13, \ (1,-1)}^{119}\equiv Hddddu\bar{Q}\bar{Q}\bar{L}\,,\\
\mathcal{O}_{13, \ (1,-1)}^{120}\equiv Hdd\bar{u}QQQ\bar{Q}\bar{L}\,,\\
\mathcal{O}_{13, \ (1,-1)}^{121}\equiv Hdddu\bar{u}Q\bar{Q}\bar{L}\,,\\
\mathcal{O}_{13, \ (1,-1)}^{122}\equiv Hd\bar{u}\bar{u}QQQQ\bar{L}\,,\\
\mathcal{O}_{13, \ (1,-1)}^{123}\equiv Hddu\bar{u}\bar{u}QQ\bar{L}\,,\\
\mathcal{O}_{13, \ (1,-1)}^{124}\equiv Hddduu\bar{u}\bar{u}\bar{L}\,,\\
\mathcal{O}_{13, \ (1,-1)}^{125}\equiv \bar{H}dddd\bar{Q}\bar{e}\bar{e}\nu\,,\\
\mathcal{O}_{13, \ (1,-1)}^{126}\equiv \bar{H}ddd\bar{u}Q\bar{e}\bar{e}\nu\,,\\
\mathcal{O}_{13, \ (1,-1)}^{127}\equiv \bar{H}dddd\bar{d}\bar{L}\bar{e}\nu\,,\\
\mathcal{O}_{13, \ (1,-1)}^{128}\equiv \bar{H}ddd\bar{L}e\bar{e}\bar{e}\nu\,,\\
\mathcal{O}_{13, \ (1,-1)}^{129}\equiv \bar{H}dddQ\bar{Q}\bar{L}\bar{e}\nu\,,\\
\mathcal{O}_{13, \ (1,-1)}^{130}\equiv \bar{H}dd\bar{u}QQ\bar{L}\bar{e}\nu\,,\\
\mathcal{O}_{13, \ (1,-1)}^{131}\equiv \bar{H}dddu\bar{u}\bar{L}\bar{e}\nu\,,\\
\mathcal{O}_{13, \ (1,-1)}^{132}\equiv \bar{H}ddd\bar{d}Q\bar{L}\bar{L}\nu\,,\\
\mathcal{O}_{13, \ (1,-1)}^{133}\equiv \bar{H}dddL\bar{L}\bar{L}\bar{e}\nu\,,\\
\mathcal{O}_{13, \ (1,-1)}^{134}\equiv \bar{H}ddQ\bar{L}\bar{L}e\bar{e}\nu\,,\\
\mathcal{O}_{13, \ (1,-1)}^{135}\equiv \bar{H}ddQL\bar{L}\bar{L}\bar{L}\nu\,,\\
\mathcal{O}_{13, \ (1,-1)}^{136}\equiv \bar{H}dQQ\bar{L}\bar{L}\bar{L}e\nu\,,\\
\mathcal{O}_{13, \ (1,-1)}^{137}\equiv \bar{H}ddu\bar{L}\bar{L}\bar{L}e\nu\,,\\
\mathcal{O}_{13, \ (1,-1)}^{138}\equiv \bar{H}ddQQ\bar{Q}\bar{L}\bar{L}\nu\,,\\
\mathcal{O}_{13, \ (1,-1)}^{139}\equiv \bar{H}dddu\bar{Q}\bar{L}\bar{L}\nu\,,\\
\mathcal{O}_{13, \ (1,-1)}^{140}\equiv \bar{H}d\bar{u}QQQ\bar{L}\bar{L}\nu\,,\\
\mathcal{O}_{13, \ (1,-1)}^{141}\equiv \bar{H}ddu\bar{u}Q\bar{L}\bar{L}\nu\,,\\
\mathcal{O}_{13, \ (1,-1)}^{142}\equiv Hdddd\bar{u}\bar{L}\bar{e}\nu\,,\\
\mathcal{O}_{13, \ (1,-1)}^{143}\equiv Hddd\bar{L}\bar{L}\bar{L}e\nu\,,\\
\mathcal{O}_{13, \ (1,-1)}^{144}\equiv Hdddd\bar{Q}\bar{L}\bar{L}\nu\,,\\
\mathcal{O}_{13, \ (1,-1)}^{145}\equiv Hddd\bar{u}Q\bar{L}\bar{L}\nu\,,\\
\mathcal{O}_{13, \ (1,-1)}^{146}\equiv Hdddd\bar{d}\bar{d}Q\bar{\nu}\,,\\
\mathcal{O}_{13, \ (1,-1)}^{147}\equiv Hdd\bar{d}\bar{u}QQQ\bar{\nu}\,,\\
\mathcal{O}_{13, \ (1,-1)}^{148}\equiv Hddd\bar{d}u\bar{u}Q\bar{\nu}\,,\\
\mathcal{O}_{13, \ (1,-1)}^{149}\equiv Hdddd\bar{d}L\bar{e}\bar{\nu}\,,\\
\mathcal{O}_{13, \ (1,-1)}^{150}\equiv Hddd\bar{d}Qe\bar{e}\bar{\nu}\,,\\
\mathcal{O}_{13, \ (1,-1)}^{151}\equiv HdddLe\bar{e}\bar{e}\bar{\nu}\,,\\
\mathcal{O}_{13, \ (1,-1)}^{152}\equiv HddQee\bar{e}\bar{e}\bar{\nu}\,,\\
\mathcal{O}_{13, \ (1,-1)}^{153}\equiv HdddQ\bar{Q}L\bar{e}\bar{\nu}\,,\\
\mathcal{O}_{13, \ (1,-1)}^{154}\equiv HddQQ\bar{Q}e\bar{e}\bar{\nu}\,,\\
\mathcal{O}_{13, \ (1,-1)}^{155}\equiv Hdddu\bar{Q}e\bar{e}\bar{\nu}\,,\\
\mathcal{O}_{13, \ (1,-1)}^{156}\equiv Hdd\bar{u}QQL\bar{e}\bar{\nu}\,,\\
\mathcal{O}_{13, \ (1,-1)}^{157}\equiv Hd\bar{u}QQQe\bar{e}\bar{\nu}\,,\\
\mathcal{O}_{13, \ (1,-1)}^{158}\equiv Hdddu\bar{u}L\bar{e}\bar{\nu}\,,\\
\mathcal{O}_{13, \ (1,-1)}^{159}\equiv Hddu\bar{u}Qe\bar{e}\bar{\nu}\,,\\
\mathcal{O}_{13, \ (1,-1)}^{160}\equiv \bar{H}dd\bar{d}\bar{d}QQQ\bar{\nu}\,,\\
\mathcal{O}_{13, \ (1,-1)}^{161}\equiv \bar{H}ddd\bar{d}\bar{d}uQ\bar{\nu}\,,\\
\mathcal{O}_{13, \ (1,-1)}^{162}\equiv \bar{H}\bar{d}\bar{u}QQQQQ\bar{\nu}\,,\\
\mathcal{O}_{13, \ (1,-1)}^{163}\equiv \bar{H}d\bar{d}u\bar{u}QQQ\bar{\nu}\,,\\
\mathcal{O}_{13, \ (1,-1)}^{164}\equiv \bar{H}dd\bar{d}uu\bar{u}Q\bar{\nu}\,,\\
\mathcal{O}_{13, \ (1,-1)}^{165}\equiv \bar{H}dd\bar{d}QQL\bar{e}\bar{\nu}\,,\\
\mathcal{O}_{13, \ (1,-1)}^{166}\equiv \bar{H}d\bar{d}QQQe\bar{e}\bar{\nu}\,,\\
\mathcal{O}_{13, \ (1,-1)}^{167}\equiv \bar{H}ddd\bar{d}uL\bar{e}\bar{\nu}\,,\\
\mathcal{O}_{13, \ (1,-1)}^{168}\equiv \bar{H}dd\bar{d}uQe\bar{e}\bar{\nu}\,,\\
\mathcal{O}_{13, \ (1,-1)}^{169}\equiv \bar{H}ddQLL\bar{e}\bar{e}\bar{\nu}\,,\\
\mathcal{O}_{13, \ (1,-1)}^{170}\equiv \bar{H}dQQLe\bar{e}\bar{e}\bar{\nu}\,,\\
\mathcal{O}_{13, \ (1,-1)}^{171}\equiv \bar{H}QQQee\bar{e}\bar{e}\bar{\nu}\,,\\
\mathcal{O}_{13, \ (1,-1)}^{172}\equiv \bar{H}dduLe\bar{e}\bar{e}\bar{\nu}\,,\\
\mathcal{O}_{13, \ (1,-1)}^{173}\equiv \bar{H}duQee\bar{e}\bar{e}\bar{\nu}\,,\\
\mathcal{O}_{13, \ (1,-1)}^{174}\equiv \bar{H}dQQQ\bar{Q}L\bar{e}\bar{\nu}\,,\\
\mathcal{O}_{13, \ (1,-1)}^{175}\equiv \bar{H}QQQQ\bar{Q}e\bar{e}\bar{\nu}\,,\\
\mathcal{O}_{13, \ (1,-1)}^{176}\equiv \bar{H}dduQ\bar{Q}L\bar{e}\bar{\nu}\,,\\
\mathcal{O}_{13, \ (1,-1)}^{177}\equiv \bar{H}duQQ\bar{Q}e\bar{e}\bar{\nu}\,,\\
\mathcal{O}_{13, \ (1,-1)}^{178}\equiv \bar{H}dduu\bar{Q}e\bar{e}\bar{\nu}\,,\\
\mathcal{O}_{13, \ (1,-1)}^{179}\equiv \bar{H}\bar{u}QQQQL\bar{e}\bar{\nu}\,,\\
\mathcal{O}_{13, \ (1,-1)}^{180}\equiv \bar{H}du\bar{u}QQL\bar{e}\bar{\nu}\,,\\
\mathcal{O}_{13, \ (1,-1)}^{181}\equiv \bar{H}u\bar{u}QQQe\bar{e}\bar{\nu}\,,\\
\mathcal{O}_{13, \ (1,-1)}^{182}\equiv \bar{H}dduu\bar{u}L\bar{e}\bar{\nu}\,,\\
\mathcal{O}_{13, \ (1,-1)}^{183}\equiv \bar{H}duu\bar{u}Qe\bar{e}\bar{\nu}\,,\\
\mathcal{O}_{13, \ (1,-1)}^{184}\equiv \bar{H}d\bar{d}QQQL\bar{L}\bar{\nu}\,,\\
\mathcal{O}_{13, \ (1,-1)}^{185}\equiv \bar{H}\bar{d}QQQQ\bar{L}e\bar{\nu}\,,\\
\mathcal{O}_{13, \ (1,-1)}^{186}\equiv \bar{H}dd\bar{d}uQL\bar{L}\bar{\nu}\,,\\
\mathcal{O}_{13, \ (1,-1)}^{187}\equiv \bar{H}d\bar{d}uQQ\bar{L}e\bar{\nu}\,,\\
\mathcal{O}_{13, \ (1,-1)}^{188}\equiv \bar{H}dd\bar{d}uu\bar{L}e\bar{\nu}\,,\\
\mathcal{O}_{13, \ (1,-1)}^{189}\equiv \bar{H}dQQLL\bar{L}\bar{e}\bar{\nu}\,,\\
\mathcal{O}_{13, \ (1,-1)}^{190}\equiv \bar{H}QQQL\bar{L}e\bar{e}\bar{\nu}\,,\\
\mathcal{O}_{13, \ (1,-1)}^{191}\equiv \bar{H}dduLL\bar{L}\bar{e}\bar{\nu}\,,\\
\mathcal{O}_{13, \ (1,-1)}^{192}\equiv \bar{H}duQL\bar{L}e\bar{e}\bar{\nu}\,,\\
\mathcal{O}_{13, \ (1,-1)}^{193}\equiv \bar{H}uQQ\bar{L}ee\bar{e}\bar{\nu}\,,\\
\mathcal{O}_{13, \ (1,-1)}^{194}\equiv \bar{H}duu\bar{L}ee\bar{e}\bar{\nu}\,,\\
\mathcal{O}_{13, \ (1,-1)}^{195}\equiv \bar{H}QQQLL\bar{L}\bar{L}\bar{\nu}\,,\\
\mathcal{O}_{13, \ (1,-1)}^{196}\equiv \bar{H}duQLL\bar{L}\bar{L}\bar{\nu}\,,\\
\mathcal{O}_{13, \ (1,-1)}^{197}\equiv \bar{H}uQQL\bar{L}\bar{L}e\bar{\nu}\,,\\
\mathcal{O}_{13, \ (1,-1)}^{198}\equiv \bar{H}duuL\bar{L}\bar{L}e\bar{\nu}\,,\\
\mathcal{O}_{13, \ (1,-1)}^{199}\equiv \bar{H}uuQ\bar{L}\bar{L}ee\bar{\nu}\,,\\
\mathcal{O}_{13, \ (1,-1)}^{200}\equiv \bar{H}QQQQ\bar{Q}L\bar{L}\bar{\nu}\,,\\
\mathcal{O}_{13, \ (1,-1)}^{201}\equiv \bar{H}duQQ\bar{Q}L\bar{L}\bar{\nu}\,,\\
\mathcal{O}_{13, \ (1,-1)}^{202}\equiv \bar{H}uQQQ\bar{Q}\bar{L}e\bar{\nu}\,,\\
\mathcal{O}_{13, \ (1,-1)}^{203}\equiv \bar{H}dduu\bar{Q}L\bar{L}\bar{\nu}\,,\\
\mathcal{O}_{13, \ (1,-1)}^{204}\equiv \bar{H}duuQ\bar{Q}\bar{L}e\bar{\nu}\,,\\
\mathcal{O}_{13, \ (1,-1)}^{205}\equiv \bar{H}u\bar{u}QQQL\bar{L}\bar{\nu}\,,\\
\mathcal{O}_{13, \ (1,-1)}^{206}\equiv \bar{H}duu\bar{u}QL\bar{L}\bar{\nu}\,,\\
\mathcal{O}_{13, \ (1,-1)}^{207}\equiv \bar{H}uu\bar{u}QQ\bar{L}e\bar{\nu}\,,\\
\mathcal{O}_{13, \ (1,-1)}^{208}\equiv \bar{H}duuu\bar{u}\bar{L}e\bar{\nu}\,,\\
\mathcal{O}_{13, \ (1,-1)}^{209}\equiv \bar{H}d\bar{d}QQQQ\bar{Q}\bar{\nu}\,,\\
\mathcal{O}_{13, \ (1,-1)}^{210}\equiv \bar{H}dd\bar{d}uQQ\bar{Q}\bar{\nu}\,,\\
\mathcal{O}_{13, \ (1,-1)}^{211}\equiv \bar{H}ddd\bar{d}uu\bar{Q}\bar{\nu}\,,\\
\mathcal{O}_{13, \ (1,-1)}^{212}\equiv \bar{H}QQQQQ\bar{Q}\bar{Q}\bar{\nu}\,,\\
\mathcal{O}_{13, \ (1,-1)}^{213}\equiv \bar{H}duQQQ\bar{Q}\bar{Q}\bar{\nu}\,,\\
\mathcal{O}_{13, \ (1,-1)}^{214}\equiv \bar{H}dduuQ\bar{Q}\bar{Q}\bar{\nu}\,,\\
\mathcal{O}_{13, \ (1,-1)}^{215}\equiv \bar{H}u\bar{u}QQQQ\bar{Q}\bar{\nu}\,,\\
\mathcal{O}_{13, \ (1,-1)}^{216}\equiv \bar{H}duu\bar{u}QQ\bar{Q}\bar{\nu}\,,\\
\mathcal{O}_{13, \ (1,-1)}^{217}\equiv \bar{H}dduuu\bar{u}\bar{Q}\bar{\nu}\,,\\
\mathcal{O}_{13, \ (1,-1)}^{218}\equiv \bar{H}uu\bar{u}\bar{u}QQQ\bar{\nu}\,,\\
\mathcal{O}_{13, \ (1,-1)}^{219}\equiv \bar{H}duuu\bar{u}\bar{u}Q\bar{\nu}\,,\\
\mathcal{O}_{13, \ (1,-1)}^{220}\equiv Hddd\bar{d}QL\bar{L}\bar{\nu}\,,\\
\mathcal{O}_{13, \ (1,-1)}^{221}\equiv Hdd\bar{d}QQ\bar{L}e\bar{\nu}\,,\\
\mathcal{O}_{13, \ (1,-1)}^{222}\equiv Hddd\bar{d}u\bar{L}e\bar{\nu}\,,\\
\mathcal{O}_{13, \ (1,-1)}^{223}\equiv HdddLL\bar{L}\bar{e}\bar{\nu}\,,\\
\mathcal{O}_{13, \ (1,-1)}^{224}\equiv HddQL\bar{L}e\bar{e}\bar{\nu}\,,\\
\mathcal{O}_{13, \ (1,-1)}^{225}\equiv HdQQ\bar{L}ee\bar{e}\bar{\nu}\,,\\
\mathcal{O}_{13, \ (1,-1)}^{226}\equiv Hddu\bar{L}ee\bar{e}\bar{\nu}\,,\\
\mathcal{O}_{13, \ (1,-1)}^{227}\equiv HddQLL\bar{L}\bar{L}\bar{\nu}\,,\\
\mathcal{O}_{13, \ (1,-1)}^{228}\equiv HdQQL\bar{L}\bar{L}e\bar{\nu}\,,\\
\mathcal{O}_{13, \ (1,-1)}^{229}\equiv HQQQ\bar{L}\bar{L}ee\bar{\nu}\,,\\
\mathcal{O}_{13, \ (1,-1)}^{230}\equiv HdduL\bar{L}\bar{L}e\bar{\nu}\,,\\
\mathcal{O}_{13, \ (1,-1)}^{231}\equiv HduQ\bar{L}\bar{L}ee\bar{\nu}\,,\\
\mathcal{O}_{13, \ (1,-1)}^{232}\equiv HddQQ\bar{Q}L\bar{L}\bar{\nu}\,,\\
\mathcal{O}_{13, \ (1,-1)}^{233}\equiv HdQQQ\bar{Q}\bar{L}e\bar{\nu}\,,\\
\mathcal{O}_{13, \ (1,-1)}^{234}\equiv Hdddu\bar{Q}L\bar{L}\bar{\nu}\,,\\
\mathcal{O}_{13, \ (1,-1)}^{235}\equiv HdduQ\bar{Q}\bar{L}e\bar{\nu}\,,\\
\mathcal{O}_{13, \ (1,-1)}^{236}\equiv Hd\bar{u}QQQL\bar{L}\bar{\nu}\,,\\
\mathcal{O}_{13, \ (1,-1)}^{237}\equiv H\bar{u}QQQQ\bar{L}e\bar{\nu}\,,\\
\mathcal{O}_{13, \ (1,-1)}^{238}\equiv Hddu\bar{u}QL\bar{L}\bar{\nu}\,,\\
\mathcal{O}_{13, \ (1,-1)}^{239}\equiv Hdu\bar{u}QQ\bar{L}e\bar{\nu}\,,\\
\mathcal{O}_{13, \ (1,-1)}^{240}\equiv Hdduu\bar{u}\bar{L}e\bar{\nu}\,,\\
\mathcal{O}_{13, \ (1,-1)}^{241}\equiv Hddd\bar{d}QQ\bar{Q}\bar{\nu}\,,\\
\mathcal{O}_{13, \ (1,-1)}^{242}\equiv Hdddd\bar{d}u\bar{Q}\bar{\nu}\,,\\
\mathcal{O}_{13, \ (1,-1)}^{243}\equiv HddQQQ\bar{Q}\bar{Q}\bar{\nu}\,,\\
\mathcal{O}_{13, \ (1,-1)}^{244}\equiv HddduQ\bar{Q}\bar{Q}\bar{\nu}\,,\\
\mathcal{O}_{13, \ (1,-1)}^{245}\equiv Hd\bar{u}QQQQ\bar{Q}\bar{\nu}\,,\\
\mathcal{O}_{13, \ (1,-1)}^{246}\equiv Hddu\bar{u}QQ\bar{Q}\bar{\nu}\,,\\
\mathcal{O}_{13, \ (1,-1)}^{247}\equiv Hddduu\bar{u}\bar{Q}\bar{\nu}\,,\\
\mathcal{O}_{13, \ (1,-1)}^{248}\equiv H\bar{u}\bar{u}QQQQQ\bar{\nu}\,,\\
\mathcal{O}_{13, \ (1,-1)}^{249}\equiv Hdu\bar{u}\bar{u}QQQ\bar{\nu}\,,\\
\mathcal{O}_{13, \ (1,-1)}^{250}\equiv Hdduu\bar{u}\bar{u}Q\bar{\nu}\,,\\
\mathcal{O}_{13, \ (1,-1)}^{251}\equiv \bar{H}ddd\bar{L}\bar{L}\bar{L}\nu\nu\,,\\
\mathcal{O}_{13, \ (1,-1)}^{252}\equiv Hdddd\bar{Q}\bar{e}\nu\bar{\nu}\,,\\
\mathcal{O}_{13, \ (1,-1)}^{253}\equiv Hddd\bar{u}Q\bar{e}\nu\bar{\nu}\,,\\
\mathcal{O}_{13, \ (1,-1)}^{254}\equiv \bar{H}ddd\bar{d}Q\bar{e}\nu\bar{\nu}\,,\\
\mathcal{O}_{13, \ (1,-1)}^{255}\equiv \bar{H}dddL\bar{e}\bar{e}\nu\bar{\nu}\,,\\
\mathcal{O}_{13, \ (1,-1)}^{256}\equiv \bar{H}ddQe\bar{e}\bar{e}\nu\bar{\nu}\,,\\
\mathcal{O}_{13, \ (1,-1)}^{257}\equiv \bar{H}ddQQ\bar{Q}\bar{e}\nu\bar{\nu}\,,\\
\mathcal{O}_{13, \ (1,-1)}^{258}\equiv \bar{H}dddu\bar{Q}\bar{e}\nu\bar{\nu}\,,\\
\mathcal{O}_{13, \ (1,-1)}^{259}\equiv \bar{H}d\bar{u}QQQ\bar{e}\nu\bar{\nu}\,,\\
\mathcal{O}_{13, \ (1,-1)}^{260}\equiv \bar{H}ddu\bar{u}Q\bar{e}\nu\bar{\nu}\,,\\
\mathcal{O}_{13, \ (1,-1)}^{261}\equiv \bar{H}dd\bar{d}QQ\bar{L}\nu\bar{\nu}\,,\\
\mathcal{O}_{13, \ (1,-1)}^{262}\equiv \bar{H}ddd\bar{d}u\bar{L}\nu\bar{\nu}\,,\\
\mathcal{O}_{13, \ (1,-1)}^{263}\equiv \bar{H}ddQL\bar{L}\bar{e}\nu\bar{\nu}\,,\\
\mathcal{O}_{13, \ (1,-1)}^{264}\equiv \bar{H}dQQ\bar{L}e\bar{e}\nu\bar{\nu}\,,\\
\mathcal{O}_{13, \ (1,-1)}^{265}\equiv \bar{H}ddu\bar{L}e\bar{e}\nu\bar{\nu}\,,\\
\mathcal{O}_{13, \ (1,-1)}^{266}\equiv \bar{H}dQQL\bar{L}\bar{L}\nu\bar{\nu}\,,\\
\mathcal{O}_{13, \ (1,-1)}^{267}\equiv \bar{H}QQQ\bar{L}\bar{L}e\nu\bar{\nu}\,,\\
\mathcal{O}_{13, \ (1,-1)}^{268}\equiv \bar{H}dduL\bar{L}\bar{L}\nu\bar{\nu}\,,\\
\mathcal{O}_{13, \ (1,-1)}^{269}\equiv \bar{H}duQ\bar{L}\bar{L}e\nu\bar{\nu}\,,\\
\mathcal{O}_{13, \ (1,-1)}^{270}\equiv \bar{H}dQQQ\bar{Q}\bar{L}\nu\bar{\nu}\,,\\
\mathcal{O}_{13, \ (1,-1)}^{271}\equiv \bar{H}dduQ\bar{Q}\bar{L}\nu\bar{\nu}\,,\\
\mathcal{O}_{13, \ (1,-1)}^{272}\equiv \bar{H}\bar{u}QQQQ\bar{L}\nu\bar{\nu}\,,\\
\mathcal{O}_{13, \ (1,-1)}^{273}\equiv \bar{H}du\bar{u}QQ\bar{L}\nu\bar{\nu}\,,\\
\mathcal{O}_{13, \ (1,-1)}^{274}\equiv \bar{H}dduu\bar{u}\bar{L}\nu\bar{\nu}\,,\\
\mathcal{O}_{13, \ (1,-1)}^{275}\equiv Hdddd\bar{d}\bar{L}\nu\bar{\nu}\,,\\
\mathcal{O}_{13, \ (1,-1)}^{276}\equiv Hddd\bar{L}e\bar{e}\nu\bar{\nu}\,,\\
\mathcal{O}_{13, \ (1,-1)}^{277}\equiv HdddL\bar{L}\bar{L}\nu\bar{\nu}\,,\\
\mathcal{O}_{13, \ (1,-1)}^{278}\equiv HddQ\bar{L}\bar{L}e\nu\bar{\nu}\,,\\
\mathcal{O}_{13, \ (1,-1)}^{279}\equiv HdddQ\bar{Q}\bar{L}\nu\bar{\nu}\,,\\
\mathcal{O}_{13, \ (1,-1)}^{280}\equiv Hdd\bar{u}QQ\bar{L}\nu\bar{\nu}\,,\\
\mathcal{O}_{13, \ (1,-1)}^{281}\equiv Hdddu\bar{u}\bar{L}\nu\bar{\nu}\,,\\
\mathcal{O}_{13, \ (1,-1)}^{282}\equiv Hdd\bar{d}QQL\bar{\nu}\bar{\nu}\,,\\
\mathcal{O}_{13, \ (1,-1)}^{283}\equiv Hd\bar{d}QQQe\bar{\nu}\bar{\nu}\,,\\
\mathcal{O}_{13, \ (1,-1)}^{284}\equiv Hddd\bar{d}uL\bar{\nu}\bar{\nu}\,,\\
\mathcal{O}_{13, \ (1,-1)}^{285}\equiv Hdd\bar{d}uQe\bar{\nu}\bar{\nu}\,,\\
\mathcal{O}_{13, \ (1,-1)}^{286}\equiv HddQLL\bar{e}\bar{\nu}\bar{\nu}\,,\\
\mathcal{O}_{13, \ (1,-1)}^{287}\equiv HdQQLe\bar{e}\bar{\nu}\bar{\nu}\,,\\
\mathcal{O}_{13, \ (1,-1)}^{288}\equiv HQQQee\bar{e}\bar{\nu}\bar{\nu}\,,\\
\mathcal{O}_{13, \ (1,-1)}^{289}\equiv HdduLe\bar{e}\bar{\nu}\bar{\nu}\,,\\
\mathcal{O}_{13, \ (1,-1)}^{290}\equiv HduQee\bar{e}\bar{\nu}\bar{\nu}\,,\\
\mathcal{O}_{13, \ (1,-1)}^{291}\equiv \bar{H}\bar{d}QQQQL\bar{\nu}\bar{\nu}\,,\\
\mathcal{O}_{13, \ (1,-1)}^{292}\equiv \bar{H}d\bar{d}uQQL\bar{\nu}\bar{\nu}\,,\\
\mathcal{O}_{13, \ (1,-1)}^{293}\equiv \bar{H}\bar{d}uQQQe\bar{\nu}\bar{\nu}\,,\\
\mathcal{O}_{13, \ (1,-1)}^{294}\equiv \bar{H}dd\bar{d}uuL\bar{\nu}\bar{\nu}\,,\\
\mathcal{O}_{13, \ (1,-1)}^{295}\equiv \bar{H}d\bar{d}uuQe\bar{\nu}\bar{\nu}\,,\\
\mathcal{O}_{13, \ (1,-1)}^{296}\equiv \bar{H}QQQLL\bar{e}\bar{\nu}\bar{\nu}\,,\\
\mathcal{O}_{13, \ (1,-1)}^{297}\equiv \bar{H}duQLL\bar{e}\bar{\nu}\bar{\nu}\,,\\
\mathcal{O}_{13, \ (1,-1)}^{298}\equiv \bar{H}uQQLe\bar{e}\bar{\nu}\bar{\nu}\,,\\
\mathcal{O}_{13, \ (1,-1)}^{299}\equiv \bar{H}duuLe\bar{e}\bar{\nu}\bar{\nu}\,,\\
\mathcal{O}_{13, \ (1,-1)}^{300}\equiv \bar{H}uuQee\bar{e}\bar{\nu}\bar{\nu}\,,\\
\mathcal{O}_{13, \ (1,-1)}^{301}\equiv \bar{H}uQQLL\bar{L}\bar{\nu}\bar{\nu}\,,\\
\mathcal{O}_{13, \ (1,-1)}^{302}\equiv \bar{H}duuLL\bar{L}\bar{\nu}\bar{\nu}\,,\\
\mathcal{O}_{13, \ (1,-1)}^{303}\equiv \bar{H}uuQL\bar{L}e\bar{\nu}\bar{\nu}\,,\\
\mathcal{O}_{13, \ (1,-1)}^{304}\equiv \bar{H}uuu\bar{L}ee\bar{\nu}\bar{\nu}\,,\\
\mathcal{O}_{13, \ (1,-1)}^{305}\equiv \bar{H}uQQQ\bar{Q}L\bar{\nu}\bar{\nu}\,,\\
\mathcal{O}_{13, \ (1,-1)}^{306}\equiv \bar{H}duuQ\bar{Q}L\bar{\nu}\bar{\nu}\,,\\
\mathcal{O}_{13, \ (1,-1)}^{307}\equiv \bar{H}uuQQ\bar{Q}e\bar{\nu}\bar{\nu}\,,\\
\mathcal{O}_{13, \ (1,-1)}^{308}\equiv \bar{H}duuu\bar{Q}e\bar{\nu}\bar{\nu}\,,\\
\mathcal{O}_{13, \ (1,-1)}^{309}\equiv \bar{H}uu\bar{u}QQL\bar{\nu}\bar{\nu}\,,\\
\mathcal{O}_{13, \ (1,-1)}^{310}\equiv \bar{H}duuu\bar{u}L\bar{\nu}\bar{\nu}\,,\\
\mathcal{O}_{13, \ (1,-1)}^{311}\equiv \bar{H}uuu\bar{u}Qe\bar{\nu}\bar{\nu}\,,\\
\mathcal{O}_{13, \ (1,-1)}^{312}\equiv HdQQLL\bar{L}\bar{\nu}\bar{\nu}\,,\\
\mathcal{O}_{13, \ (1,-1)}^{313}\equiv HQQQL\bar{L}e\bar{\nu}\bar{\nu}\,,\\
\mathcal{O}_{13, \ (1,-1)}^{314}\equiv HdduLL\bar{L}\bar{\nu}\bar{\nu}\,,\\
\mathcal{O}_{13, \ (1,-1)}^{315}\equiv HduQL\bar{L}e\bar{\nu}\bar{\nu}\,,\\
\mathcal{O}_{13, \ (1,-1)}^{316}\equiv HuQQ\bar{L}ee\bar{\nu}\bar{\nu}\,,\\
\mathcal{O}_{13, \ (1,-1)}^{317}\equiv Hduu\bar{L}ee\bar{\nu}\bar{\nu}\,,\\
\mathcal{O}_{13, \ (1,-1)}^{318}\equiv HdQQQ\bar{Q}L\bar{\nu}\bar{\nu}\,,\\
\mathcal{O}_{13, \ (1,-1)}^{319}\equiv HQQQQ\bar{Q}e\bar{\nu}\bar{\nu}\,,\\
\mathcal{O}_{13, \ (1,-1)}^{320}\equiv HdduQ\bar{Q}L\bar{\nu}\bar{\nu}\,,\\
\mathcal{O}_{13, \ (1,-1)}^{321}\equiv HduQQ\bar{Q}e\bar{\nu}\bar{\nu}\,,\\
\mathcal{O}_{13, \ (1,-1)}^{322}\equiv Hdduu\bar{Q}e\bar{\nu}\bar{\nu}\,,\\
\mathcal{O}_{13, \ (1,-1)}^{323}\equiv H\bar{u}QQQQL\bar{\nu}\bar{\nu}\,,\\
\mathcal{O}_{13, \ (1,-1)}^{324}\equiv Hdu\bar{u}QQL\bar{\nu}\bar{\nu}\,,\\
\mathcal{O}_{13, \ (1,-1)}^{325}\equiv Hu\bar{u}QQQe\bar{\nu}\bar{\nu}\,,\\
\mathcal{O}_{13, \ (1,-1)}^{326}\equiv Hdduu\bar{u}L\bar{\nu}\bar{\nu}\,,\\
\mathcal{O}_{13, \ (1,-1)}^{327}\equiv Hduu\bar{u}Qe\bar{\nu}\bar{\nu}\,,\\
\mathcal{O}_{13, \ (1,-1)}^{328}\equiv \bar{H}ddd\bar{L}\bar{e}\nu\nu\bar{\nu}\,,\\
\mathcal{O}_{13, \ (1,-1)}^{329}\equiv \bar{H}ddQ\bar{L}\bar{L}\nu\nu\bar{\nu}\,,\\
\mathcal{O}_{13, \ (1,-1)}^{330}\equiv Hddd\bar{d}Q\nu\bar{\nu}\bar{\nu}\,,\\
\mathcal{O}_{13, \ (1,-1)}^{331}\equiv HdddL\bar{e}\nu\bar{\nu}\bar{\nu}\,,\\
\mathcal{O}_{13, \ (1,-1)}^{332}\equiv HddQe\bar{e}\nu\bar{\nu}\bar{\nu}\,,\\
\mathcal{O}_{13, \ (1,-1)}^{333}\equiv \bar{H}d\bar{d}QQQ\nu\bar{\nu}\bar{\nu}\,,\\
\mathcal{O}_{13, \ (1,-1)}^{334}\equiv \bar{H}dd\bar{d}uQ\nu\bar{\nu}\bar{\nu}\,,\\
\mathcal{O}_{13, \ (1,-1)}^{335}\equiv \bar{H}dQQL\bar{e}\nu\bar{\nu}\bar{\nu}\,,\\
\mathcal{O}_{13, \ (1,-1)}^{336}\equiv \bar{H}QQQe\bar{e}\nu\bar{\nu}\bar{\nu}\,,\\
\mathcal{O}_{13, \ (1,-1)}^{337}\equiv \bar{H}dduL\bar{e}\nu\bar{\nu}\bar{\nu}\,,\\
\mathcal{O}_{13, \ (1,-1)}^{338}\equiv \bar{H}duQe\bar{e}\nu\bar{\nu}\bar{\nu}\,,\\
\mathcal{O}_{13, \ (1,-1)}^{339}\equiv \bar{H}QQQL\bar{L}\nu\bar{\nu}\bar{\nu}\,,\\
\mathcal{O}_{13, \ (1,-1)}^{340}\equiv \bar{H}duQL\bar{L}\nu\bar{\nu}\bar{\nu}\,,\\
\mathcal{O}_{13, \ (1,-1)}^{341}\equiv \bar{H}uQQ\bar{L}e\nu\bar{\nu}\bar{\nu}\,,\\
\mathcal{O}_{13, \ (1,-1)}^{342}\equiv \bar{H}duu\bar{L}e\nu\bar{\nu}\bar{\nu}\,,\\
\mathcal{O}_{13, \ (1,-1)}^{343}\equiv \bar{H}QQQQ\bar{Q}\nu\bar{\nu}\bar{\nu}\,,\\
\mathcal{O}_{13, \ (1,-1)}^{344}\equiv \bar{H}duQQ\bar{Q}\nu\bar{\nu}\bar{\nu}\,,\\
\mathcal{O}_{13, \ (1,-1)}^{345}\equiv \bar{H}dduu\bar{Q}\nu\bar{\nu}\bar{\nu}\,,\\
\mathcal{O}_{13, \ (1,-1)}^{346}\equiv \bar{H}u\bar{u}QQQ\nu\bar{\nu}\bar{\nu}\,,\\
\mathcal{O}_{13, \ (1,-1)}^{347}\equiv \bar{H}duu\bar{u}Q\nu\bar{\nu}\bar{\nu}\,,\\
\mathcal{O}_{13, \ (1,-1)}^{348}\equiv HddQL\bar{L}\nu\bar{\nu}\bar{\nu}\,,\\
\mathcal{O}_{13, \ (1,-1)}^{349}\equiv HdQQ\bar{L}e\nu\bar{\nu}\bar{\nu}\,,\\
\mathcal{O}_{13, \ (1,-1)}^{350}\equiv Hddu\bar{L}e\nu\bar{\nu}\bar{\nu}\,,\\
\mathcal{O}_{13, \ (1,-1)}^{351}\equiv \bar{H}uuQLL\bar{\nu}\bar{\nu}\bar{\nu}\,,\\
\mathcal{O}_{13, \ (1,-1)}^{352}\equiv \bar{H}uuuLe\bar{\nu}\bar{\nu}\bar{\nu}\,,\\
\mathcal{O}_{13, \ (1,-1)}^{353}\equiv HQQQLL\bar{\nu}\bar{\nu}\bar{\nu}\,,\\
\mathcal{O}_{13, \ (1,-1)}^{354}\equiv HduQLL\bar{\nu}\bar{\nu}\bar{\nu}\,,\\
\mathcal{O}_{13, \ (1,-1)}^{355}\equiv HuQQLe\bar{\nu}\bar{\nu}\bar{\nu}\,,\\
\mathcal{O}_{13, \ (1,-1)}^{356}\equiv HduuLe\bar{\nu}\bar{\nu}\bar{\nu}\,,\\
\mathcal{O}_{13, \ (1,-1)}^{357}\equiv HuuQee\bar{\nu}\bar{\nu}\bar{\nu}\,,\\
\mathcal{O}_{13, \ (1,-1)}^{358}\equiv HddQQ\bar{Q}\nu\bar{\nu}\bar{\nu}\,,\\
\mathcal{O}_{13, \ (1,-1)}^{359}\equiv Hdddu\bar{Q}\nu\bar{\nu}\bar{\nu}\,,\\
\mathcal{O}_{13, \ (1,-1)}^{360}\equiv Hd\bar{u}QQQ\nu\bar{\nu}\bar{\nu}\,,\\
\mathcal{O}_{13, \ (1,-1)}^{361}\equiv Hddu\bar{u}Q\nu\bar{\nu}\bar{\nu}\,,\\
\mathcal{O}_{13, \ (1,-1)}^{362}\equiv \bar{H}ddQ\bar{e}\nu\nu\bar{\nu}\bar{\nu}\,,\\
\mathcal{O}_{13, \ (1,-1)}^{363}\equiv \bar{H}dQQ\bar{L}\nu\nu\bar{\nu}\bar{\nu}\,,\\
\mathcal{O}_{13, \ (1,-1)}^{364}\equiv \bar{H}ddu\bar{L}\nu\nu\bar{\nu}\bar{\nu}\,,\\
\mathcal{O}_{13, \ (1,-1)}^{365}\equiv Hddd\bar{L}\nu\nu\bar{\nu}\bar{\nu}\,,\\
\mathcal{O}_{13, \ (1,-1)}^{366}\equiv HdQQL\nu\bar{\nu}\bar{\nu}\bar{\nu}\,,\\
\mathcal{O}_{13, \ (1,-1)}^{367}\equiv \bar{H}uQQL\nu\bar{\nu}\bar{\nu}\bar{\nu}\,,\\
\mathcal{O}_{13, \ (1,-1)}^{368}\equiv \bar{H}duuL\nu\bar{\nu}\bar{\nu}\bar{\nu}\,,\\
\mathcal{O}_{13, \ (1,-1)}^{369}\equiv \bar{H}uuQe\nu\bar{\nu}\bar{\nu}\bar{\nu}\,,\\
\mathcal{O}_{13, \ (1,-1)}^{370}\equiv HQQQe\nu\bar{\nu}\bar{\nu}\bar{\nu}\,,\\
\mathcal{O}_{13, \ (1,-1)}^{371}\equiv HdduL\nu\bar{\nu}\bar{\nu}\bar{\nu}\,,\\
\mathcal{O}_{13, \ (1,-1)}^{372}\equiv HduQe\nu\bar{\nu}\bar{\nu}\bar{\nu}\,,\\
\mathcal{O}_{13, \ (1,-1)}^{373}\equiv HddQ\nu\nu\bar{\nu}\bar{\nu}\bar{\nu}\,,\\
\mathcal{O}_{13, \ (1,-1)}^{374}\equiv \bar{H}QQQ\nu\nu\bar{\nu}\bar{\nu}\bar{\nu}\,,\\
\mathcal{O}_{13, \ (1,-1)}^{375}\equiv \bar{H}duQ\nu\nu\bar{\nu}\bar{\nu}\bar{\nu}\,,\\
\mathcal{O}_{13, \ (1,-1)}^{376}\equiv HH\bar{H}\bar{H}dddd\bar{d}\bar{e}\,,\\
\mathcal{O}_{13, \ (1,-1)}^{377}\equiv HH\bar{H}\bar{H}ddde\bar{e}\bar{e}\,,\\
\mathcal{O}_{13, \ (1,-1)}^{378}\equiv HH\bar{H}\bar{H}dddQ\bar{Q}\bar{e}\,,\\
\mathcal{O}_{13, \ (1,-1)}^{379}\equiv HH\bar{H}\bar{H}dd\bar{u}QQ\bar{e}\,,\\
\mathcal{O}_{13, \ (1,-1)}^{380}\equiv HH\bar{H}\bar{H}dddu\bar{u}\bar{e}\,,\\
\mathcal{O}_{13, \ (1,-1)}^{381}\equiv H\bar{H}\bar{H}\bar{H}dd\bar{d}QQ\bar{e}\,,\\
\mathcal{O}_{13, \ (1,-1)}^{382}\equiv H\bar{H}\bar{H}\bar{H}ddQL\bar{e}\bar{e}\,,\\
\mathcal{O}_{13, \ (1,-1)}^{383}\equiv H\bar{H}\bar{H}\bar{H}dQQe\bar{e}\bar{e}\,,\\
\mathcal{O}_{13, \ (1,-1)}^{384}\equiv H\bar{H}\bar{H}\bar{H}dQQQ\bar{Q}\bar{e}\,,\\
\mathcal{O}_{13, \ (1,-1)}^{385}\equiv H\bar{H}\bar{H}\bar{H}dduQ\bar{Q}\bar{e}\,,\\
\mathcal{O}_{13, \ (1,-1)}^{386}\equiv H\bar{H}\bar{H}\bar{H}\bar{u}QQQQ\bar{e}\,,\\
\mathcal{O}_{13, \ (1,-1)}^{387}\equiv H\bar{H}\bar{H}\bar{H}du\bar{u}QQ\bar{e}\,,\\
\mathcal{O}_{13, \ (1,-1)}^{388}\equiv H\bar{H}\bar{H}\bar{H}d\bar{d}QQQ\bar{L}\,,\\
\mathcal{O}_{13, \ (1,-1)}^{389}\equiv H\bar{H}\bar{H}\bar{H}dd\bar{d}uQ\bar{L}\,,\\
\mathcal{O}_{13, \ (1,-1)}^{390}\equiv H\bar{H}\bar{H}\bar{H}dQQL\bar{L}\bar{e}\,,\\
\mathcal{O}_{13, \ (1,-1)}^{391}\equiv H\bar{H}\bar{H}\bar{H}QQQ\bar{L}e\bar{e}\,,\\
\mathcal{O}_{13, \ (1,-1)}^{392}\equiv H\bar{H}\bar{H}\bar{H}dduL\bar{L}\bar{e}\,,\\
\mathcal{O}_{13, \ (1,-1)}^{393}\equiv H\bar{H}\bar{H}\bar{H}duQ\bar{L}e\bar{e}\,,\\
\mathcal{O}_{13, \ (1,-1)}^{394}\equiv H\bar{H}\bar{H}\bar{H}QQQL\bar{L}\bar{L}\,,\\
\mathcal{O}_{13, \ (1,-1)}^{395}\equiv H\bar{H}\bar{H}\bar{H}duQL\bar{L}\bar{L}\,,\\
\mathcal{O}_{13, \ (1,-1)}^{396}\equiv H\bar{H}\bar{H}\bar{H}uQQ\bar{L}\bar{L}e\,,\\
\mathcal{O}_{13, \ (1,-1)}^{397}\equiv H\bar{H}\bar{H}\bar{H}duu\bar{L}\bar{L}e\,,\\
\mathcal{O}_{13, \ (1,-1)}^{398}\equiv H\bar{H}\bar{H}\bar{H}QQQQ\bar{Q}\bar{L}\,,\\
\mathcal{O}_{13, \ (1,-1)}^{399}\equiv H\bar{H}\bar{H}\bar{H}duQQ\bar{Q}\bar{L}\,,\\
\mathcal{O}_{13, \ (1,-1)}^{400}\equiv H\bar{H}\bar{H}\bar{H}dduu\bar{Q}\bar{L}\,,\\
\mathcal{O}_{13, \ (1,-1)}^{401}\equiv H\bar{H}\bar{H}\bar{H}u\bar{u}QQQ\bar{L}\,,\\
\mathcal{O}_{13, \ (1,-1)}^{402}\equiv H\bar{H}\bar{H}\bar{H}duu\bar{u}Q\bar{L}\,,\\
\mathcal{O}_{13, \ (1,-1)}^{403}\equiv HH\bar{H}\bar{H}ddd\bar{d}Q\bar{L}\,,\\
\mathcal{O}_{13, \ (1,-1)}^{404}\equiv HH\bar{H}\bar{H}dddL\bar{L}\bar{e}\,,\\
\mathcal{O}_{13, \ (1,-1)}^{405}\equiv HH\bar{H}\bar{H}ddQ\bar{L}e\bar{e}\,,\\
\mathcal{O}_{13, \ (1,-1)}^{406}\equiv HH\bar{H}\bar{H}ddQL\bar{L}\bar{L}\,,\\
\mathcal{O}_{13, \ (1,-1)}^{407}\equiv HH\bar{H}\bar{H}dQQ\bar{L}\bar{L}e\,,\\
\mathcal{O}_{13, \ (1,-1)}^{408}\equiv HH\bar{H}\bar{H}ddu\bar{L}\bar{L}e\,,\\
\mathcal{O}_{13, \ (1,-1)}^{409}\equiv HH\bar{H}\bar{H}ddQQ\bar{Q}\bar{L}\,,\\
\mathcal{O}_{13, \ (1,-1)}^{410}\equiv HH\bar{H}\bar{H}dddu\bar{Q}\bar{L}\,,\\
\mathcal{O}_{13, \ (1,-1)}^{411}\equiv HH\bar{H}\bar{H}d\bar{u}QQQ\bar{L}\,,\\
\mathcal{O}_{13, \ (1,-1)}^{412}\equiv HH\bar{H}\bar{H}ddu\bar{u}Q\bar{L}\,,\\
\mathcal{O}_{13, \ (1,-1)}^{413}\equiv HHH\bar{H}ddd\bar{L}\bar{L}e\,,\\
\mathcal{O}_{13, \ (1,-1)}^{414}\equiv HHH\bar{H}dddd\bar{Q}\bar{L}\,,\\
\mathcal{O}_{13, \ (1,-1)}^{415}\equiv HHH\bar{H}ddd\bar{u}Q\bar{L}\,,\\
\mathcal{O}_{13, \ (1,-1)}^{416}\equiv \bar{H}\bar{H}\bar{H}\bar{H}QQQL\bar{e}\bar{e}\,,\\
\mathcal{O}_{13, \ (1,-1)}^{417}\equiv \bar{H}\bar{H}\bar{H}\bar{H}uQQL\bar{L}\bar{e}\,,\\
\mathcal{O}_{13, \ (1,-1)}^{418}\equiv \bar{H}\bar{H}\bar{H}\bar{H}uuQL\bar{L}\bar{L}\,,\\
\mathcal{O}_{13, \ (1,-1)}^{419}\equiv \bar{H}\bar{H}\bar{H}\bar{H}\bar{d}QQQQ\bar{e}\,,\\
\mathcal{O}_{13, \ (1,-1)}^{420}\equiv \bar{H}\bar{H}\bar{H}\bar{H}\bar{d}uQQQ\bar{L}\,,\\
\mathcal{O}_{13, \ (1,-1)}^{421}\equiv \bar{H}\bar{H}\bar{H}\bar{H}uQQQ\bar{Q}\bar{e}\,,\\
\mathcal{O}_{13, \ (1,-1)}^{422}\equiv \bar{H}\bar{H}\bar{H}\bar{H}uuQQ\bar{Q}\bar{L}\,,\\
\mathcal{O}_{13, \ (1,-1)}^{423}\equiv HH\bar{H}\bar{H}ddd\bar{L}\bar{L}\nu\,,\\
\mathcal{O}_{13, \ (1,-1)}^{424}\equiv H\bar{H}\bar{H}\bar{H}ddu\bar{L}\bar{L}\nu\,,\\
\mathcal{O}_{13, \ (1,-1)}^{425}\equiv H\bar{H}\bar{H}\bar{H}ddQ\bar{L}\bar{e}\nu\,,\\
\mathcal{O}_{13, \ (1,-1)}^{426}\equiv H\bar{H}\bar{H}\bar{H}dQQ\bar{L}\bar{L}\nu\,,\\
\mathcal{O}_{13, \ (1,-1)}^{427}\equiv \bar{H}\bar{H}\bar{H}\bar{H}QQQ\bar{L}\bar{e}\nu\,,\\
\mathcal{O}_{13, \ (1,-1)}^{428}\equiv \bar{H}\bar{H}\bar{H}\bar{H}uQQ\bar{L}\bar{L}\nu\,,\\
\mathcal{O}_{13, \ (1,-1)}^{429}\equiv HH\bar{H}\bar{H}dd\bar{d}QQ\bar{\nu}\,,\\
\mathcal{O}_{13, \ (1,-1)}^{430}\equiv HH\bar{H}\bar{H}ddd\bar{d}u\bar{\nu}\,,\\
\mathcal{O}_{13, \ (1,-1)}^{431}\equiv HH\bar{H}\bar{H}ddQL\bar{e}\bar{\nu}\,,\\
\mathcal{O}_{13, \ (1,-1)}^{432}\equiv HH\bar{H}\bar{H}dQQe\bar{e}\bar{\nu}\,,\\
\mathcal{O}_{13, \ (1,-1)}^{433}\equiv HH\bar{H}\bar{H}ddue\bar{e}\bar{\nu}\,,\\
\mathcal{O}_{13, \ (1,-1)}^{434}\equiv H\bar{H}\bar{H}\bar{H}\bar{d}QQQQ\bar{\nu}\,,\\
\mathcal{O}_{13, \ (1,-1)}^{435}\equiv H\bar{H}\bar{H}\bar{H}d\bar{d}uQQ\bar{\nu}\,,\\
\mathcal{O}_{13, \ (1,-1)}^{436}\equiv H\bar{H}\bar{H}\bar{H}QQQL\bar{e}\bar{\nu}\,,\\
\mathcal{O}_{13, \ (1,-1)}^{437}\equiv H\bar{H}\bar{H}\bar{H}duQL\bar{e}\bar{\nu}\,,\\
\mathcal{O}_{13, \ (1,-1)}^{438}\equiv H\bar{H}\bar{H}\bar{H}uQQe\bar{e}\bar{\nu}\,,\\
\mathcal{O}_{13, \ (1,-1)}^{439}\equiv H\bar{H}\bar{H}\bar{H}uQQL\bar{L}\bar{\nu}\,,\\
\mathcal{O}_{13, \ (1,-1)}^{440}\equiv H\bar{H}\bar{H}\bar{H}duuL\bar{L}\bar{\nu}\,,\\
\mathcal{O}_{13, \ (1,-1)}^{441}\equiv H\bar{H}\bar{H}\bar{H}uuQ\bar{L}e\bar{\nu}\,,\\
\mathcal{O}_{13, \ (1,-1)}^{442}\equiv H\bar{H}\bar{H}\bar{H}uQQQ\bar{Q}\bar{\nu}\,,\\
\mathcal{O}_{13, \ (1,-1)}^{443}\equiv H\bar{H}\bar{H}\bar{H}duuQ\bar{Q}\bar{\nu}\,,\\
\mathcal{O}_{13, \ (1,-1)}^{444}\equiv H\bar{H}\bar{H}\bar{H}uu\bar{u}QQ\bar{\nu}\,,\\
\mathcal{O}_{13, \ (1,-1)}^{445}\equiv HH\bar{H}\bar{H}dQQL\bar{L}\bar{\nu}\,,\\
\mathcal{O}_{13, \ (1,-1)}^{446}\equiv HH\bar{H}\bar{H}QQQ\bar{L}e\bar{\nu}\,,\\
\mathcal{O}_{13, \ (1,-1)}^{447}\equiv HH\bar{H}\bar{H}dduL\bar{L}\bar{\nu}\,,\\
\mathcal{O}_{13, \ (1,-1)}^{448}\equiv HH\bar{H}\bar{H}duQ\bar{L}e\bar{\nu}\,,\\
\mathcal{O}_{13, \ (1,-1)}^{449}\equiv HH\bar{H}\bar{H}dQQQ\bar{Q}\bar{\nu}\,,\\
\mathcal{O}_{13, \ (1,-1)}^{450}\equiv HH\bar{H}\bar{H}dduQ\bar{Q}\bar{\nu}\,,\\
\mathcal{O}_{13, \ (1,-1)}^{451}\equiv HH\bar{H}\bar{H}\bar{u}QQQQ\bar{\nu}\,,\\
\mathcal{O}_{13, \ (1,-1)}^{452}\equiv HH\bar{H}\bar{H}du\bar{u}QQ\bar{\nu}\,,\\
\mathcal{O}_{13, \ (1,-1)}^{453}\equiv HH\bar{H}\bar{H}dduu\bar{u}\bar{\nu}\,,\\
\mathcal{O}_{13, \ (1,-1)}^{454}\equiv HHH\bar{H}dddL\bar{L}\bar{\nu}\,,\\
\mathcal{O}_{13, \ (1,-1)}^{455}\equiv HHH\bar{H}ddQ\bar{L}e\bar{\nu}\,,\\
\mathcal{O}_{13, \ (1,-1)}^{456}\equiv HHH\bar{H}dddQ\bar{Q}\bar{\nu}\,,\\
\mathcal{O}_{13, \ (1,-1)}^{457}\equiv HHH\bar{H}dd\bar{u}QQ\bar{\nu}\,,\\
\mathcal{O}_{13, \ (1,-1)}^{458}\equiv HH\bar{H}\bar{H}ddd\bar{e}\nu\bar{\nu}\,,\\
\mathcal{O}_{13, \ (1,-1)}^{459}\equiv H\bar{H}\bar{H}\bar{H}dQQ\bar{e}\nu\bar{\nu}\,,\\
\mathcal{O}_{13, \ (1,-1)}^{460}\equiv H\bar{H}\bar{H}\bar{H}QQQ\bar{L}\nu\bar{\nu}\,,\\
\mathcal{O}_{13, \ (1,-1)}^{461}\equiv H\bar{H}\bar{H}\bar{H}duQ\bar{L}\nu\bar{\nu}\,,\\
\mathcal{O}_{13, \ (1,-1)}^{462}\equiv HH\bar{H}\bar{H}ddQ\bar{L}\nu\bar{\nu}\,,\\
\mathcal{O}_{13, \ (1,-1)}^{463}\equiv HHH\bar{H}ddQL\bar{\nu}\bar{\nu}\,,\\
\mathcal{O}_{13, \ (1,-1)}^{464}\equiv HHH\bar{H}dQQe\bar{\nu}\bar{\nu}\,,\\
\mathcal{O}_{13, \ (1,-1)}^{465}\equiv H\bar{H}\bar{H}\bar{H}uuQL\bar{\nu}\bar{\nu}\,,\\
\mathcal{O}_{13, \ (1,-1)}^{466}\equiv HH\bar{H}\bar{H}QQQL\bar{\nu}\bar{\nu}\,,\\
\mathcal{O}_{13, \ (1,-1)}^{467}\equiv HH\bar{H}\bar{H}duQL\bar{\nu}\bar{\nu}\,,\\
\mathcal{O}_{13, \ (1,-1)}^{468}\equiv HH\bar{H}\bar{H}uQQe\bar{\nu}\bar{\nu}\,,\\
\mathcal{O}_{13, \ (1,-1)}^{469}\equiv HH\bar{H}\bar{H}duue\bar{\nu}\bar{\nu}\,,\\
\mathcal{O}_{13, \ (1,-1)}^{470}\equiv HH\bar{H}\bar{H}dQQ\nu\bar{\nu}\bar{\nu}\,,\\
\mathcal{O}_{13, \ (1,-1)}^{471}\equiv H\bar{H}\bar{H}\bar{H}uQQ\nu\bar{\nu}\bar{\nu}\,,\\
\mathcal{O}_{13, \ (1,-1)}^{472}\equiv HH\bar{H}\bar{H}ddu\nu\bar{\nu}\bar{\nu}\,,\\
\mathcal{O}_{13, \ (1,-1)}^{473}\equiv HH\bar{H}\bar{H}\bar{H}\bar{H}\bar{H}QQQ\bar{e}\,,\\
\mathcal{O}_{13, \ (1,-1)}^{474}\equiv HH\bar{H}\bar{H}\bar{H}\bar{H}\bar{H}uQQ\bar{L}\,,\\
\mathcal{O}_{13, \ (1,-1)}^{475}\equiv HHH\bar{H}\bar{H}\bar{H}\bar{H}ddQ\bar{e}\,,\\
\mathcal{O}_{13, \ (1,-1)}^{476}\equiv HHH\bar{H}\bar{H}\bar{H}\bar{H}dQQ\bar{L}\,,\\
\mathcal{O}_{13, \ (1,-1)}^{477}\equiv HHH\bar{H}\bar{H}\bar{H}\bar{H}ddu\bar{L}\,,\\
\mathcal{O}_{13, \ (1,-1)}^{478}\equiv HHHH\bar{H}\bar{H}\bar{H}ddd\bar{L}\,,\\
\mathcal{O}_{13, \ (1,-1)}^{479}\equiv HHHH\bar{H}\bar{H}\bar{H}ddQ\bar{\nu}\,,\\
\mathcal{O}_{13, \ (1,-1)}^{480}\equiv HHH\bar{H}\bar{H}\bar{H}\bar{H}QQQ\bar{\nu}\,,\\
\mathcal{O}_{13, \ (1,-1)}^{481}\equiv HHH\bar{H}\bar{H}\bar{H}\bar{H}duQ\bar{\nu}\,.
 \label{eq:dequal13basisB1L-1}
 \end{math}
 \end{spacing}
 \end{multicols}

 \subsection{\texorpdfstring{$\Delta B=1, \Delta L=3$}{Delta B=1,Delta L=3}}
 \begin{multicols}{3}
 \begin{spacing}{1.5}
 \noindent
 \begin{math}
 \mathcal{O}_{13, \ (1,3)}^{1}\equiv H\bar{d}QQQQLLL\,,\\
\mathcal{O}_{13, \ (1,3)}^{2}\equiv Hd\bar{d}uQQLLL\,,\\
\mathcal{O}_{13, \ (1,3)}^{3}\equiv H\bar{d}uQQQLLe\,,\\
\mathcal{O}_{13, \ (1,3)}^{4}\equiv Hdd\bar{d}uuLLL\,,\\
\mathcal{O}_{13, \ (1,3)}^{5}\equiv Hd\bar{d}uuQLLe\,,\\
\mathcal{O}_{13, \ (1,3)}^{6}\equiv H\bar{d}uuQQLee\,,\\
\mathcal{O}_{13, \ (1,3)}^{7}\equiv Hd\bar{d}uuuLee\,,\\
\mathcal{O}_{13, \ (1,3)}^{8}\equiv H\bar{d}uuuQeee\,,\\
\mathcal{O}_{13, \ (1,3)}^{9}\equiv HQQQLLLL\bar{e}\,,\\
\mathcal{O}_{13, \ (1,3)}^{10}\equiv HduQLLLL\bar{e}\,,\\
\mathcal{O}_{13, \ (1,3)}^{11}\equiv HuQQLLLe\bar{e}\,,\\
\mathcal{O}_{13, \ (1,3)}^{12}\equiv HduuLLLe\bar{e}\,,\\
\mathcal{O}_{13, \ (1,3)}^{13}\equiv HuuQLLee\bar{e}\,,\\
\mathcal{O}_{13, \ (1,3)}^{14}\equiv HuuuLeee\bar{e}\,,\\
\mathcal{O}_{13, \ (1,3)}^{15}\equiv \bar{H}\bar{d}uuQQLLL\,,\\
\mathcal{O}_{13, \ (1,3)}^{16}\equiv \bar{H}d\bar{d}uuuLLL\,,\\
\mathcal{O}_{13, \ (1,3)}^{17}\equiv \bar{H}\bar{d}uuuQLLe\,,\\
\mathcal{O}_{13, \ (1,3)}^{18}\equiv \bar{H}\bar{d}uuuuLee\,,\\
\mathcal{O}_{13, \ (1,3)}^{19}\equiv \bar{H}uuQLLLL\bar{e}\,,\\
\mathcal{O}_{13, \ (1,3)}^{20}\equiv \bar{H}uuuLLLe\bar{e}\,,\\
\mathcal{O}_{13, \ (1,3)}^{21}\equiv \bar{H}uuuLLLL\bar{L}\,,\\
\mathcal{O}_{13, \ (1,3)}^{22}\equiv \bar{H}uuuQ\bar{Q}LLL\,,\\
\mathcal{O}_{13, \ (1,3)}^{23}\equiv \bar{H}uuuu\bar{Q}LLe\,,\\
\mathcal{O}_{13, \ (1,3)}^{24}\equiv \bar{H}uuuu\bar{u}LLL\,,\\
\mathcal{O}_{13, \ (1,3)}^{25}\equiv HuQQLLLL\bar{L}\,,\\
\mathcal{O}_{13, \ (1,3)}^{26}\equiv HduuLLLL\bar{L}\,,\\
\mathcal{O}_{13, \ (1,3)}^{27}\equiv HuuQLLL\bar{L}e\,,\\
\mathcal{O}_{13, \ (1,3)}^{28}\equiv HuuuLL\bar{L}ee\,,\\
\mathcal{O}_{13, \ (1,3)}^{29}\equiv HuQQQ\bar{Q}LLL\,,\\
\mathcal{O}_{13, \ (1,3)}^{30}\equiv HduuQ\bar{Q}LLL\,,\\
\mathcal{O}_{13, \ (1,3)}^{31}\equiv HuuQQ\bar{Q}LLe\,,\\
\mathcal{O}_{13, \ (1,3)}^{32}\equiv Hduuu\bar{Q}LLe\,,\\
\mathcal{O}_{13, \ (1,3)}^{33}\equiv HuuuQ\bar{Q}Lee\,,\\
\mathcal{O}_{13, \ (1,3)}^{34}\equiv Huuuu\bar{Q}eee\,,\\
\mathcal{O}_{13, \ (1,3)}^{35}\equiv Huu\bar{u}QQLLL\,,\\
\mathcal{O}_{13, \ (1,3)}^{36}\equiv Hduuu\bar{u}LLL\,,\\
\mathcal{O}_{13, \ (1,3)}^{37}\equiv Huuu\bar{u}QLLe\,,\\
\mathcal{O}_{13, \ (1,3)}^{38}\equiv Huuuu\bar{u}Lee\,,\\
\mathcal{O}_{13, \ (1,3)}^{39}\equiv Hd\bar{d}QQQLL\nu\,,\\
\mathcal{O}_{13, \ (1,3)}^{40}\equiv H\bar{d}QQQQLe\nu\,,\\
\mathcal{O}_{13, \ (1,3)}^{41}\equiv Hdd\bar{d}uQLL\nu\,,\\
\mathcal{O}_{13, \ (1,3)}^{42}\equiv Hd\bar{d}uQQLe\nu\,,\\
\mathcal{O}_{13, \ (1,3)}^{43}\equiv H\bar{d}uQQQee\nu\,,\\
\mathcal{O}_{13, \ (1,3)}^{44}\equiv Hdd\bar{d}uuLe\nu\,,\\
\mathcal{O}_{13, \ (1,3)}^{45}\equiv Hd\bar{d}uuQee\nu\,,\\
\mathcal{O}_{13, \ (1,3)}^{46}\equiv HdQQLLL\bar{e}\nu\,,\\
\mathcal{O}_{13, \ (1,3)}^{47}\equiv HQQQLLe\bar{e}\nu\,,\\
\mathcal{O}_{13, \ (1,3)}^{48}\equiv HdduLLL\bar{e}\nu\,,\\
\mathcal{O}_{13, \ (1,3)}^{49}\equiv HduQLLe\bar{e}\nu\,,\\
\mathcal{O}_{13, \ (1,3)}^{50}\equiv HuQQLee\bar{e}\nu\,,\\
\mathcal{O}_{13, \ (1,3)}^{51}\equiv HduuLee\bar{e}\nu\,,\\
\mathcal{O}_{13, \ (1,3)}^{52}\equiv HuuQeee\bar{e}\nu\,,\\
\mathcal{O}_{13, \ (1,3)}^{53}\equiv \bar{H}\bar{d}uQQQLL\nu\,,\\
\mathcal{O}_{13, \ (1,3)}^{54}\equiv \bar{H}d\bar{d}uuQLL\nu\,,\\
\mathcal{O}_{13, \ (1,3)}^{55}\equiv \bar{H}\bar{d}uuQQLe\nu\,,\\
\mathcal{O}_{13, \ (1,3)}^{56}\equiv \bar{H}d\bar{d}uuuLe\nu\,,\\
\mathcal{O}_{13, \ (1,3)}^{57}\equiv \bar{H}\bar{d}uuuQee\nu\,,\\
\mathcal{O}_{13, \ (1,3)}^{58}\equiv \bar{H}uQQLLL\bar{e}\nu\,,\\
\mathcal{O}_{13, \ (1,3)}^{59}\equiv \bar{H}duuLLL\bar{e}\nu\,,\\
\mathcal{O}_{13, \ (1,3)}^{60}\equiv \bar{H}uuQLLe\bar{e}\nu\,,\\
\mathcal{O}_{13, \ (1,3)}^{61}\equiv \bar{H}uuuLee\bar{e}\nu\,,\\
\mathcal{O}_{13, \ (1,3)}^{62}\equiv \bar{H}uuQLLL\bar{L}\nu\,,\\
\mathcal{O}_{13, \ (1,3)}^{63}\equiv \bar{H}uuuLL\bar{L}e\nu\,,\\
\mathcal{O}_{13, \ (1,3)}^{64}\equiv \bar{H}uuQQ\bar{Q}LL\nu\,,\\
\mathcal{O}_{13, \ (1,3)}^{65}\equiv \bar{H}duuu\bar{Q}LL\nu\,,\\
\mathcal{O}_{13, \ (1,3)}^{66}\equiv \bar{H}uuuQ\bar{Q}Le\nu\,,\\
\mathcal{O}_{13, \ (1,3)}^{67}\equiv \bar{H}uuuu\bar{Q}ee\nu\,,\\
\mathcal{O}_{13, \ (1,3)}^{68}\equiv \bar{H}uuu\bar{u}QLL\nu\,,\\
\mathcal{O}_{13, \ (1,3)}^{69}\equiv \bar{H}uuuu\bar{u}Le\nu\,,\\
\mathcal{O}_{13, \ (1,3)}^{70}\equiv HQQQLLL\bar{L}\nu\,,\\
\mathcal{O}_{13, \ (1,3)}^{71}\equiv HduQLLL\bar{L}\nu\,,\\
\mathcal{O}_{13, \ (1,3)}^{72}\equiv HuQQLL\bar{L}e\nu\,,\\
\mathcal{O}_{13, \ (1,3)}^{73}\equiv HduuLL\bar{L}e\nu\,,\\
\mathcal{O}_{13, \ (1,3)}^{74}\equiv HuuQL\bar{L}ee\nu\,,\\
\mathcal{O}_{13, \ (1,3)}^{75}\equiv Huuu\bar{L}eee\nu\,,\\
\mathcal{O}_{13, \ (1,3)}^{76}\equiv HuuQLLLL\bar{\nu}\,,\\
\mathcal{O}_{13, \ (1,3)}^{77}\equiv HuuuLLLe\bar{\nu}\,,\\
\mathcal{O}_{13, \ (1,3)}^{78}\equiv HQQQQ\bar{Q}LL\nu\,,\\
\mathcal{O}_{13, \ (1,3)}^{79}\equiv HduQQ\bar{Q}LL\nu\,,\\
\mathcal{O}_{13, \ (1,3)}^{80}\equiv HuQQQ\bar{Q}Le\nu\,,\\
\mathcal{O}_{13, \ (1,3)}^{81}\equiv Hdduu\bar{Q}LL\nu\,,\\
\mathcal{O}_{13, \ (1,3)}^{82}\equiv HduuQ\bar{Q}Le\nu\,,\\
\mathcal{O}_{13, \ (1,3)}^{83}\equiv HuuQQ\bar{Q}ee\nu\,,\\
\mathcal{O}_{13, \ (1,3)}^{84}\equiv Hduuu\bar{Q}ee\nu\,,\\
\mathcal{O}_{13, \ (1,3)}^{85}\equiv Hu\bar{u}QQQLL\nu\,,\\
\mathcal{O}_{13, \ (1,3)}^{86}\equiv Hduu\bar{u}QLL\nu\,,\\
\mathcal{O}_{13, \ (1,3)}^{87}\equiv Huu\bar{u}QQLe\nu\,,\\
\mathcal{O}_{13, \ (1,3)}^{88}\equiv Hduuu\bar{u}Le\nu\,,\\
\mathcal{O}_{13, \ (1,3)}^{89}\equiv Huuu\bar{u}Qee\nu\,,\\
\mathcal{O}_{13, \ (1,3)}^{90}\equiv Hdd\bar{d}QQL\nu\nu\,,\\
\mathcal{O}_{13, \ (1,3)}^{91}\equiv Hd\bar{d}QQQe\nu\nu\,,\\
\mathcal{O}_{13, \ (1,3)}^{92}\equiv Hddd\bar{d}uL\nu\nu\,,\\
\mathcal{O}_{13, \ (1,3)}^{93}\equiv Hdd\bar{d}uQe\nu\nu\,,\\
\mathcal{O}_{13, \ (1,3)}^{94}\equiv HddQLL\bar{e}\nu\nu\,,\\
\mathcal{O}_{13, \ (1,3)}^{95}\equiv HdQQLe\bar{e}\nu\nu\,,\\
\mathcal{O}_{13, \ (1,3)}^{96}\equiv HQQQee\bar{e}\nu\nu\,,\\
\mathcal{O}_{13, \ (1,3)}^{97}\equiv HdduLe\bar{e}\nu\nu\,,\\
\mathcal{O}_{13, \ (1,3)}^{98}\equiv HduQee\bar{e}\nu\nu\,,\\
\mathcal{O}_{13, \ (1,3)}^{99}\equiv \bar{H}\bar{d}QQQQL\nu\nu\,,\\
\mathcal{O}_{13, \ (1,3)}^{100}\equiv \bar{H}d\bar{d}uQQL\nu\nu\,,\\
\mathcal{O}_{13, \ (1,3)}^{101}\equiv \bar{H}\bar{d}uQQQe\nu\nu\,,\\
\mathcal{O}_{13, \ (1,3)}^{102}\equiv \bar{H}dd\bar{d}uuL\nu\nu\,,\\
\mathcal{O}_{13, \ (1,3)}^{103}\equiv \bar{H}d\bar{d}uuQe\nu\nu\,,\\
\mathcal{O}_{13, \ (1,3)}^{104}\equiv \bar{H}QQQLL\bar{e}\nu\nu\,,\\
\mathcal{O}_{13, \ (1,3)}^{105}\equiv \bar{H}duQLL\bar{e}\nu\nu\,,\\
\mathcal{O}_{13, \ (1,3)}^{106}\equiv \bar{H}uQQLe\bar{e}\nu\nu\,,\\
\mathcal{O}_{13, \ (1,3)}^{107}\equiv \bar{H}duuLe\bar{e}\nu\nu\,,\\
\mathcal{O}_{13, \ (1,3)}^{108}\equiv \bar{H}uuQee\bar{e}\nu\nu\,,\\
\mathcal{O}_{13, \ (1,3)}^{109}\equiv \bar{H}uQQLL\bar{L}\nu\nu\,,\\
\mathcal{O}_{13, \ (1,3)}^{110}\equiv \bar{H}duuLL\bar{L}\nu\nu\,,\\
\mathcal{O}_{13, \ (1,3)}^{111}\equiv \bar{H}uuQL\bar{L}e\nu\nu\,,\\
\mathcal{O}_{13, \ (1,3)}^{112}\equiv \bar{H}uuu\bar{L}ee\nu\nu\,,\\
\mathcal{O}_{13, \ (1,3)}^{113}\equiv \bar{H}uQQQ\bar{Q}L\nu\nu\,,\\
\mathcal{O}_{13, \ (1,3)}^{114}\equiv \bar{H}duuQ\bar{Q}L\nu\nu\,,\\
\mathcal{O}_{13, \ (1,3)}^{115}\equiv \bar{H}uuQQ\bar{Q}e\nu\nu\,,\\
\mathcal{O}_{13, \ (1,3)}^{116}\equiv \bar{H}duuu\bar{Q}e\nu\nu\,,\\
\mathcal{O}_{13, \ (1,3)}^{117}\equiv \bar{H}uu\bar{u}QQL\nu\nu\,,\\
\mathcal{O}_{13, \ (1,3)}^{118}\equiv \bar{H}duuu\bar{u}L\nu\nu\,,\\
\mathcal{O}_{13, \ (1,3)}^{119}\equiv \bar{H}uuu\bar{u}Qe\nu\nu\,,\\
\mathcal{O}_{13, \ (1,3)}^{120}\equiv HdQQLL\bar{L}\nu\nu\,,\\
\mathcal{O}_{13, \ (1,3)}^{121}\equiv HQQQL\bar{L}e\nu\nu\,,\\
\mathcal{O}_{13, \ (1,3)}^{122}\equiv HdduLL\bar{L}\nu\nu\,,\\
\mathcal{O}_{13, \ (1,3)}^{123}\equiv HduQL\bar{L}e\nu\nu\,,\\
\mathcal{O}_{13, \ (1,3)}^{124}\equiv HuQQ\bar{L}ee\nu\nu\,,\\
\mathcal{O}_{13, \ (1,3)}^{125}\equiv Hduu\bar{L}ee\nu\nu\,,\\
\mathcal{O}_{13, \ (1,3)}^{126}\equiv \bar{H}uuuLLL\nu\bar{\nu}\,,\\
\mathcal{O}_{13, \ (1,3)}^{127}\equiv HuQQLLL\nu\bar{\nu}\,,\\
\mathcal{O}_{13, \ (1,3)}^{128}\equiv HduuLLL\nu\bar{\nu}\,,\\
\mathcal{O}_{13, \ (1,3)}^{129}\equiv HuuQLLe\nu\bar{\nu}\,,\\
\mathcal{O}_{13, \ (1,3)}^{130}\equiv HuuuLee\nu\bar{\nu}\,,\\
\mathcal{O}_{13, \ (1,3)}^{131}\equiv HdQQQ\bar{Q}L\nu\nu\,,\\
\mathcal{O}_{13, \ (1,3)}^{132}\equiv HQQQQ\bar{Q}e\nu\nu\,,\\
\mathcal{O}_{13, \ (1,3)}^{133}\equiv HdduQ\bar{Q}L\nu\nu\,,\\
\mathcal{O}_{13, \ (1,3)}^{134}\equiv HduQQ\bar{Q}e\nu\nu\,,\\
\mathcal{O}_{13, \ (1,3)}^{135}\equiv Hdduu\bar{Q}e\nu\nu\,,\\
\mathcal{O}_{13, \ (1,3)}^{136}\equiv H\bar{u}QQQQL\nu\nu\,,\\
\mathcal{O}_{13, \ (1,3)}^{137}\equiv Hdu\bar{u}QQL\nu\nu\,,\\
\mathcal{O}_{13, \ (1,3)}^{138}\equiv Hu\bar{u}QQQe\nu\nu\,,\\
\mathcal{O}_{13, \ (1,3)}^{139}\equiv Hdduu\bar{u}L\nu\nu\,,\\
\mathcal{O}_{13, \ (1,3)}^{140}\equiv Hduu\bar{u}Qe\nu\nu\,,\\
\mathcal{O}_{13, \ (1,3)}^{141}\equiv Hddd\bar{d}Q\nu\nu\nu\,,\\
\mathcal{O}_{13, \ (1,3)}^{142}\equiv HdddL\bar{e}\nu\nu\nu\,,\\
\mathcal{O}_{13, \ (1,3)}^{143}\equiv HddQe\bar{e}\nu\nu\nu\,,\\
\mathcal{O}_{13, \ (1,3)}^{144}\equiv \bar{H}d\bar{d}QQQ\nu\nu\nu\,,\\
\mathcal{O}_{13, \ (1,3)}^{145}\equiv \bar{H}dd\bar{d}uQ\nu\nu\nu\,,\\
\mathcal{O}_{13, \ (1,3)}^{146}\equiv \bar{H}dQQL\bar{e}\nu\nu\nu\,,\\
\mathcal{O}_{13, \ (1,3)}^{147}\equiv \bar{H}QQQe\bar{e}\nu\nu\nu\,,\\
\mathcal{O}_{13, \ (1,3)}^{148}\equiv \bar{H}dduL\bar{e}\nu\nu\nu\,,\\
\mathcal{O}_{13, \ (1,3)}^{149}\equiv \bar{H}duQe\bar{e}\nu\nu\nu\,,\\
\mathcal{O}_{13, \ (1,3)}^{150}\equiv \bar{H}QQQL\bar{L}\nu\nu\nu\,,\\
\mathcal{O}_{13, \ (1,3)}^{151}\equiv \bar{H}duQL\bar{L}\nu\nu\nu\,,\\
\mathcal{O}_{13, \ (1,3)}^{152}\equiv \bar{H}uQQ\bar{L}e\nu\nu\nu\,,\\
\mathcal{O}_{13, \ (1,3)}^{153}\equiv \bar{H}duu\bar{L}e\nu\nu\nu\,,\\
\mathcal{O}_{13, \ (1,3)}^{154}\equiv \bar{H}QQQQ\bar{Q}\nu\nu\nu\,,\\
\mathcal{O}_{13, \ (1,3)}^{155}\equiv \bar{H}duQQ\bar{Q}\nu\nu\nu\,,\\
\mathcal{O}_{13, \ (1,3)}^{156}\equiv \bar{H}dduu\bar{Q}\nu\nu\nu\,,\\
\mathcal{O}_{13, \ (1,3)}^{157}\equiv \bar{H}u\bar{u}QQQ\nu\nu\nu\,,\\
\mathcal{O}_{13, \ (1,3)}^{158}\equiv \bar{H}duu\bar{u}Q\nu\nu\nu\,,\\
\mathcal{O}_{13, \ (1,3)}^{159}\equiv HddQL\bar{L}\nu\nu\nu\,,\\
\mathcal{O}_{13, \ (1,3)}^{160}\equiv HdQQ\bar{L}e\nu\nu\nu\,,\\
\mathcal{O}_{13, \ (1,3)}^{161}\equiv Hddu\bar{L}e\nu\nu\nu\,,\\
\mathcal{O}_{13, \ (1,3)}^{162}\equiv \bar{H}uuQLL\nu\nu\bar{\nu}\,,\\
\mathcal{O}_{13, \ (1,3)}^{163}\equiv \bar{H}uuuLe\nu\nu\bar{\nu}\,,\\
\mathcal{O}_{13, \ (1,3)}^{164}\equiv HQQQLL\nu\nu\bar{\nu}\,,\\
\mathcal{O}_{13, \ (1,3)}^{165}\equiv HduQLL\nu\nu\bar{\nu}\,,\\
\mathcal{O}_{13, \ (1,3)}^{166}\equiv HuQQLe\nu\nu\bar{\nu}\,,\\
\mathcal{O}_{13, \ (1,3)}^{167}\equiv HduuLe\nu\nu\bar{\nu}\,,\\
\mathcal{O}_{13, \ (1,3)}^{168}\equiv HuuQee\nu\nu\bar{\nu}\,,\\
\mathcal{O}_{13, \ (1,3)}^{169}\equiv HddQQ\bar{Q}\nu\nu\nu\,,\\
\mathcal{O}_{13, \ (1,3)}^{170}\equiv Hdddu\bar{Q}\nu\nu\nu\,,\\
\mathcal{O}_{13, \ (1,3)}^{171}\equiv Hd\bar{u}QQQ\nu\nu\nu\,,\\
\mathcal{O}_{13, \ (1,3)}^{172}\equiv Hddu\bar{u}Q\nu\nu\nu\,,\\
\mathcal{O}_{13, \ (1,3)}^{173}\equiv \bar{H}ddQ\bar{e}\nu\nu\nu\nu\,,\\
\mathcal{O}_{13, \ (1,3)}^{174}\equiv \bar{H}dQQ\bar{L}\nu\nu\nu\nu\,,\\
\mathcal{O}_{13, \ (1,3)}^{175}\equiv \bar{H}ddu\bar{L}\nu\nu\nu\nu\,,\\
\mathcal{O}_{13, \ (1,3)}^{176}\equiv Hddd\bar{L}\nu\nu\nu\nu\,,\\
\mathcal{O}_{13, \ (1,3)}^{177}\equiv HdQQL\nu\nu\nu\bar{\nu}\,,\\
\mathcal{O}_{13, \ (1,3)}^{178}\equiv \bar{H}uQQL\nu\nu\nu\bar{\nu}\,,\\
\mathcal{O}_{13, \ (1,3)}^{179}\equiv \bar{H}duuL\nu\nu\nu\bar{\nu}\,,\\
\mathcal{O}_{13, \ (1,3)}^{180}\equiv \bar{H}uuQe\nu\nu\nu\bar{\nu}\,,\\
\mathcal{O}_{13, \ (1,3)}^{181}\equiv HQQQe\nu\nu\nu\bar{\nu}\,,\\
\mathcal{O}_{13, \ (1,3)}^{182}\equiv HdduL\nu\nu\nu\bar{\nu}\,,\\
\mathcal{O}_{13, \ (1,3)}^{183}\equiv HduQe\nu\nu\nu\bar{\nu}\,,\\
\mathcal{O}_{13, \ (1,3)}^{184}\equiv HddQ\nu\nu\nu\nu\bar{\nu}\,,\\
\mathcal{O}_{13, \ (1,3)}^{185}\equiv \bar{H}QQQ\nu\nu\nu\nu\bar{\nu}\,,\\
\mathcal{O}_{13, \ (1,3)}^{186}\equiv \bar{H}duQ\nu\nu\nu\nu\bar{\nu}\,,\\
\mathcal{O}_{13, \ (1,3)}^{187}\equiv HHHHddQLLL\,,\\
\mathcal{O}_{13, \ (1,3)}^{188}\equiv HHHHdQQLLe\,,\\
\mathcal{O}_{13, \ (1,3)}^{189}\equiv HH\bar{H}\bar{H}uuQLLL\,,\\
\mathcal{O}_{13, \ (1,3)}^{190}\equiv HH\bar{H}\bar{H}uuuLLe\,,\\
\mathcal{O}_{13, \ (1,3)}^{191}\equiv HHH\bar{H}QQQLLL\,,\\
\mathcal{O}_{13, \ (1,3)}^{192}\equiv HHH\bar{H}duQLLL\,,\\
\mathcal{O}_{13, \ (1,3)}^{193}\equiv HHH\bar{H}uQQLLe\,,\\
\mathcal{O}_{13, \ (1,3)}^{194}\equiv HHH\bar{H}duuLLe\,,\\
\mathcal{O}_{13, \ (1,3)}^{195}\equiv HHH\bar{H}uuQLee\,,\\
\mathcal{O}_{13, \ (1,3)}^{196}\equiv HHHHQQQLee\,,\\
\mathcal{O}_{13, \ (1,3)}^{197}\equiv HHH\bar{H}dQQLL\nu\,,\\
\mathcal{O}_{13, \ (1,3)}^{198}\equiv H\bar{H}\bar{H}\bar{H}uuuLL\nu\,,\\
\mathcal{O}_{13, \ (1,3)}^{199}\equiv HH\bar{H}\bar{H}uQQLL\nu\,,\\
\mathcal{O}_{13, \ (1,3)}^{200}\equiv HH\bar{H}\bar{H}duuLL\nu\,,\\
\mathcal{O}_{13, \ (1,3)}^{201}\equiv HH\bar{H}\bar{H}uuQLe\nu\,,\\
\mathcal{O}_{13, \ (1,3)}^{202}\equiv HH\bar{H}\bar{H}uuuee\nu\,,\\
\mathcal{O}_{13, \ (1,3)}^{203}\equiv HHH\bar{H}QQQLe\nu\,,\\
\mathcal{O}_{13, \ (1,3)}^{204}\equiv HHH\bar{H}dduLL\nu\,,\\
\mathcal{O}_{13, \ (1,3)}^{205}\equiv HHH\bar{H}duQLe\nu\,,\\
\mathcal{O}_{13, \ (1,3)}^{206}\equiv HHH\bar{H}uQQee\nu\,,\\
\mathcal{O}_{13, \ (1,3)}^{207}\equiv HHH\bar{H}ddQL\nu\nu\,,\\
\mathcal{O}_{13, \ (1,3)}^{208}\equiv HHH\bar{H}dQQe\nu\nu\,,\\
\mathcal{O}_{13, \ (1,3)}^{209}\equiv H\bar{H}\bar{H}\bar{H}uuQL\nu\nu\,,\\
\mathcal{O}_{13, \ (1,3)}^{210}\equiv HH\bar{H}\bar{H}QQQL\nu\nu\,,\\
\mathcal{O}_{13, \ (1,3)}^{211}\equiv HH\bar{H}\bar{H}duQL\nu\nu\,,\\
\mathcal{O}_{13, \ (1,3)}^{212}\equiv HH\bar{H}\bar{H}uQQe\nu\nu\,,\\
\mathcal{O}_{13, \ (1,3)}^{213}\equiv HH\bar{H}\bar{H}duue\nu\nu\,,\\
\mathcal{O}_{13, \ (1,3)}^{214}\equiv HH\bar{H}\bar{H}dQQ\nu\nu\nu\,,\\
\mathcal{O}_{13, \ (1,3)}^{215}\equiv H\bar{H}\bar{H}\bar{H}uQQ\nu\nu\nu\,,\\
\mathcal{O}_{13, \ (1,3)}^{216}\equiv HH\bar{H}\bar{H}ddu\nu\nu\nu\,.
 \label{eq:dequal13basisB1L3}
 \end{math}
 \end{spacing}
 \end{multicols}

 \subsection{\texorpdfstring{$\Delta B=2, \Delta L=0$}{Delta B=2,Delta L=0}}
 \begin{multicols}{3}
 \begin{spacing}{1.5}
 \noindent
 \begin{math}
\mathcal{O}_{13, \ (2,0)}^{1}\equiv Hdddd\bar{d}QQQ\,,\\
\mathcal{O}_{13, \ (2,0)}^{2}\equiv Hddddd\bar{d}uQ\,,\\
\mathcal{O}_{13, \ (2,0)}^{3}\equiv HddddQQL\bar{e}\,,\\
\mathcal{O}_{13, \ (2,0)}^{4}\equiv HdddQQQe\bar{e}\,,\\
\mathcal{O}_{13, \ (2,0)}^{5}\equiv HddddduL\bar{e}\,,\\
\mathcal{O}_{13, \ (2,0)}^{6}\equiv HdddduQe\bar{e}\,,\\
\mathcal{O}_{13, \ (2,0)}^{7}\equiv \bar{H}dd\bar{d}QQQQQ\,,\\
\mathcal{O}_{13, \ (2,0)}^{8}\equiv \bar{H}ddd\bar{d}uQQQ\,,\\
\mathcal{O}_{13, \ (2,0)}^{9}\equiv \bar{H}dddd\bar{d}uuQ\,,\\
\mathcal{O}_{13, \ (2,0)}^{10}\equiv \bar{H}ddQQQQL\bar{e}\,,\\
\mathcal{O}_{13, \ (2,0)}^{11}\equiv \bar{H}dQQQQQe\bar{e}\,,\\
\mathcal{O}_{13, \ (2,0)}^{12}\equiv \bar{H}ddduQQL\bar{e}\,,\\
\mathcal{O}_{13, \ (2,0)}^{13}\equiv \bar{H}dduQQQe\bar{e}\,,\\
\mathcal{O}_{13, \ (2,0)}^{14}\equiv \bar{H}dddduuL\bar{e}\,,\\
\mathcal{O}_{13, \ (2,0)}^{15}\equiv \bar{H}ddduuQe\bar{e}\,,\\
\mathcal{O}_{13, \ (2,0)}^{16}\equiv \bar{H}dQQQQQL\bar{L}\,,\\
\mathcal{O}_{13, \ (2,0)}^{17}\equiv \bar{H}QQQQQQ\bar{L}e\,,\\
\mathcal{O}_{13, \ (2,0)}^{18}\equiv \bar{H}dduQQQL\bar{L}\,,\\
\mathcal{O}_{13, \ (2,0)}^{19}\equiv \bar{H}duQQQQ\bar{L}e\,,\\
\mathcal{O}_{13, \ (2,0)}^{20}\equiv \bar{H}ddduuQL\bar{L}\,,\\
\mathcal{O}_{13, \ (2,0)}^{21}\equiv \bar{H}dduuQQ\bar{L}e\,,\\
\mathcal{O}_{13, \ (2,0)}^{22}\equiv \bar{H}ddduuu\bar{L}e\,,\\
\mathcal{O}_{13, \ (2,0)}^{23}\equiv \bar{H}dQQQQQQ\bar{Q}\,,\\
\mathcal{O}_{13, \ (2,0)}^{24}\equiv \bar{H}dduQQQQ\bar{Q}\,,\\
\mathcal{O}_{13, \ (2,0)}^{25}\equiv \bar{H}ddduuQQ\bar{Q}\,,\\
\mathcal{O}_{13, \ (2,0)}^{26}\equiv \bar{H}dddduuu\bar{Q}\,,\\
\mathcal{O}_{13, \ (2,0)}^{27}\equiv \bar{H}\bar{u}QQQQQQQ\,,\\
\mathcal{O}_{13, \ (2,0)}^{28}\equiv \bar{H}du\bar{u}QQQQQ\,,\\
\mathcal{O}_{13, \ (2,0)}^{29}\equiv \bar{H}dduu\bar{u}QQQ\,,\\
\mathcal{O}_{13, \ (2,0)}^{30}\equiv \bar{H}ddduuu\bar{u}Q\,,\\
\mathcal{O}_{13, \ (2,0)}^{31}\equiv HdddQQQL\bar{L}\,,\\
\mathcal{O}_{13, \ (2,0)}^{32}\equiv HddQQQQ\bar{L}e\,,\\
\mathcal{O}_{13, \ (2,0)}^{33}\equiv HdddduQL\bar{L}\,,\\
\mathcal{O}_{13, \ (2,0)}^{34}\equiv HddduQQ\bar{L}e\,,\\
\mathcal{O}_{13, \ (2,0)}^{35}\equiv Hdddduu\bar{L}e\,,\\
\mathcal{O}_{13, \ (2,0)}^{36}\equiv HdddQQQQ\bar{Q}\,,\\
\mathcal{O}_{13, \ (2,0)}^{37}\equiv HdddduQQ\bar{Q}\,,\\
\mathcal{O}_{13, \ (2,0)}^{38}\equiv Hddddduu\bar{Q}\,,\\
\mathcal{O}_{13, \ (2,0)}^{39}\equiv Hdd\bar{u}QQQQQ\,,\\
\mathcal{O}_{13, \ (2,0)}^{40}\equiv Hdddu\bar{u}QQQ\,,\\
\mathcal{O}_{13, \ (2,0)}^{41}\equiv Hdddduu\bar{u}Q\,,\\
\mathcal{O}_{13, \ (2,0)}^{42}\equiv HdddddQ\bar{e}\nu\,,\\
\mathcal{O}_{13, \ (2,0)}^{43}\equiv \bar{H}dddQQQ\bar{e}\nu\,,\\
\mathcal{O}_{13, \ (2,0)}^{44}\equiv \bar{H}dddduQ\bar{e}\nu\,,\\
\mathcal{O}_{13, \ (2,0)}^{45}\equiv \bar{H}ddQQQQ\bar{L}\nu\,,\\
\mathcal{O}_{13, \ (2,0)}^{46}\equiv \bar{H}ddduQQ\bar{L}\nu\,,\\
\mathcal{O}_{13, \ (2,0)}^{47}\equiv \bar{H}dddduu\bar{L}\nu\,,\\
\mathcal{O}_{13, \ (2,0)}^{48}\equiv HddddQQ\bar{L}\nu\,,\\
\mathcal{O}_{13, \ (2,0)}^{49}\equiv Hdddddu\bar{L}\nu\,,\\
\mathcal{O}_{13, \ (2,0)}^{50}\equiv HddQQQQL\bar{\nu}\,,\\
\mathcal{O}_{13, \ (2,0)}^{51}\equiv HdQQQQQe\bar{\nu}\,,\\
\mathcal{O}_{13, \ (2,0)}^{52}\equiv \bar{H}QQQQQQL\bar{\nu}\,,\\
\mathcal{O}_{13, \ (2,0)}^{53}\equiv \bar{H}duQQQQL\bar{\nu}\,,\\
\mathcal{O}_{13, \ (2,0)}^{54}\equiv \bar{H}uQQQQQe\bar{\nu}\,,\\
\mathcal{O}_{13, \ (2,0)}^{55}\equiv \bar{H}dduuQQL\bar{\nu}\,,\\
\mathcal{O}_{13, \ (2,0)}^{56}\equiv \bar{H}duuQQQe\bar{\nu}\,,\\
\mathcal{O}_{13, \ (2,0)}^{57}\equiv \bar{H}ddduuuL\bar{\nu}\,,\\
\mathcal{O}_{13, \ (2,0)}^{58}\equiv \bar{H}dduuuQe\bar{\nu}\,,\\
\mathcal{O}_{13, \ (2,0)}^{59}\equiv HddduQQL\bar{\nu}\,,\\
\mathcal{O}_{13, \ (2,0)}^{60}\equiv HdduQQQe\bar{\nu}\,,\\
\mathcal{O}_{13, \ (2,0)}^{61}\equiv HdddduuL\bar{\nu}\,,\\
\mathcal{O}_{13, \ (2,0)}^{62}\equiv HddduuQe\bar{\nu}\,,\\
\mathcal{O}_{13, \ (2,0)}^{63}\equiv HdddQQQ\nu\bar{\nu}\,,\\
\mathcal{O}_{13, \ (2,0)}^{64}\equiv \bar{H}dQQQQQ\nu\bar{\nu}\,,\\
\mathcal{O}_{13, \ (2,0)}^{65}\equiv \bar{H}dduQQQ\nu\bar{\nu}\,,\\
\mathcal{O}_{13, \ (2,0)}^{66}\equiv \bar{H}ddduuQ\nu\bar{\nu}\,,\\
\mathcal{O}_{13, \ (2,0)}^{67}\equiv HdddduQ\nu\bar{\nu}\,,\\
\mathcal{O}_{13, \ (2,0)}^{68}\equiv HHH\bar{H}ddddQQ\,,\\
\mathcal{O}_{13, \ (2,0)}^{69}\equiv HH\bar{H}\bar{H}ddQQQQ\,,\\
\mathcal{O}_{13, \ (2,0)}^{70}\equiv \bar{H}\bar{H}\bar{H}\bar{H}uuQQQQ\,,\\
\mathcal{O}_{13, \ (2,0)}^{71}\equiv H\bar{H}\bar{H}\bar{H}QQQQQQ\,,\\
\mathcal{O}_{13, \ (2,0)}^{72}\equiv H\bar{H}\bar{H}\bar{H}duQQQQ\,,\\
\mathcal{O}_{13, \ (2,0)}^{73}\equiv H\bar{H}\bar{H}\bar{H}dduuQQ\,,\\
\mathcal{O}_{13, \ (2,0)}^{74}\equiv HH\bar{H}\bar{H}ddduQQ\,,\\
\mathcal{O}_{13, \ (2,0)}^{75}\equiv HH\bar{H}\bar{H}dddduu\,.
 \label{eq:dequal13basisB2L0}
 \end{math}
 \end{spacing}
 \end{multicols}

\clearpage
 
 
\section{Dimension 14}
\label{sec:d14}

\begin{figure}
     \centering
  \includegraphics[width=1\linewidth]{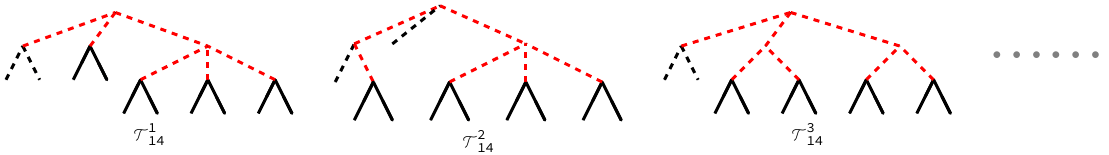}
     \caption{A few representative Feynman-diagram topologies for mass dimension~14. All 13029 topologies are provided at \website. }
     \label{fig:d=14_topologies}
 \end{figure}

At $d=14$, we find operators with different $\Delta B$ and $\Delta L$, which we list separately. None of them can be protected using $B+a L$ symmetries and the number of topologies is too large to include here, see Fig.~\ref{fig:d=14_topologies}.

Ignoring the question of how to make any of these $d=14$ operators dominant for now, the decay rates could be large enough to be observable. Take for example the operator $\mathcal{O}_{14, \ (1,1)}^{702}\equiv HHHH\bar{H}\bar{H}\bar{H}\bar{H}QQQL$, which, after electroweak symmetry breaking, induces
\begin{align}
        \Gamma (p\to e^+\pi^0)\simeq \frac{\alpha^2 \langle H\rangle^{16}}{8 \pi m_p\Lambda^{20}}\simeq \left( \unit[10^{34}]{yr}\right)^{-1} \left(\frac{\unit[64]{TeV}}{\Lambda} \right)^{20} ,
\end{align}
where $\alpha \simeq \unit[0.01]{GeV^3}$ is a QCD matrix element~\cite{Yoo:2021gql}. The large vacuum expectation value $\langle H\rangle \simeq\unit[174]{GeV}$ allows for testable rates for $\Lambda \lesssim \unit[64]{TeV}$, despite the suppression by 20 orders of $\Lambda$!

  \subsection{\texorpdfstring{$\Delta B=1, \Delta L=-3$}{Delta B=1,Delta L=-3}}
 \begin{multicols}{2}
 \begin{spacing}{1.5}
 \noindent
 \begin{math}
\mathcal{O}_{14, \ (1,-3)}^{1}\equiv \bar{H}\bar{H}dddd\bar{d}\bar{L}\bar{L}\bar{e}\,,\\
\mathcal{O}_{14, \ (1,-3)}^{2}\equiv \bar{H}\bar{H}dddd\bar{Q}\bar{L}\bar{e}\bar{e}\,,\\
\mathcal{O}_{14, \ (1,-3)}^{3}\equiv \bar{H}\bar{H}dddu\bar{Q}\bar{L}\bar{L}\bar{L}\,,\\
\mathcal{O}_{14, \ (1,-3)}^{4}\equiv \bar{H}\bar{H}dddu\bar{u}\bar{L}\bar{L}\bar{e}\,,\\
\mathcal{O}_{14, \ (1,-3)}^{5}\equiv \bar{H}\bar{H}ddd\bar{L}\bar{L}e\bar{e}\bar{e}\,,\\
\mathcal{O}_{14, \ (1,-3)}^{6}\equiv \bar{H}\bar{H}ddu\bar{L}\bar{L}\bar{L}\bar{L}e\,,\\
\mathcal{O}_{14, \ (1,-3)}^{7}\equiv \bar{H}\bar{H}ddQ\bar{L}\bar{L}\bar{L}e\bar{e}\,,\\
\mathcal{O}_{14, \ (1,-3)}^{8}\equiv \bar{H}\bar{H}dQQ\bar{L}\bar{L}\bar{L}\bar{L}e\,,\\
\mathcal{O}_{14, \ (1,-3)}^{9}\equiv H\bar{H}dddd\bar{u}\bar{L}\bar{L}\bar{e}\,,\\
\mathcal{O}_{14, \ (1,-3)}^{10}\equiv H\bar{H}ddd\bar{L}\bar{L}\bar{L}\bar{L}e\,,\\
\mathcal{O}_{14, \ (1,-3)}^{11}\equiv H\bar{H}dddd\bar{Q}\bar{L}\bar{L}\bar{L}\,,\\
\mathcal{O}_{14, \ (1,-3)}^{12}\equiv H\bar{H}ddd\bar{u}Q\bar{L}\bar{L}\bar{L}\,,\\
\mathcal{O}_{14, \ (1,-3)}^{13}\equiv \bar{H}\bar{H}dddL\bar{L}\bar{L}\bar{L}\bar{e}\,,\\
\mathcal{O}_{14, \ (1,-3)}^{14}\equiv \bar{H}\bar{H}ddQL\bar{L}\bar{L}\bar{L}\bar{L}\,,\\
\mathcal{O}_{14, \ (1,-3)}^{15}\equiv \bar{H}\bar{H}ddd\bar{d}Q\bar{L}\bar{L}\bar{L}\,,\\
\mathcal{O}_{14, \ (1,-3)}^{16}\equiv \bar{H}\bar{H}dddQ\bar{Q}\bar{L}\bar{L}\bar{e}\,,\\
\mathcal{O}_{14, \ (1,-3)}^{17}\equiv \bar{H}\bar{H}ddd\bar{u}Q\bar{L}\bar{e}\bar{e}\,,\\
\mathcal{O}_{14, \ (1,-3)}^{18}\equiv \bar{H}\bar{H}ddu\bar{u}Q\bar{L}\bar{L}\bar{L}\,,\\
\mathcal{O}_{14, \ (1,-3)}^{19}\equiv \bar{H}\bar{H}ddQQ\bar{Q}\bar{L}\bar{L}\bar{L}\,,\\
\mathcal{O}_{14, \ (1,-3)}^{20}\equiv \bar{H}\bar{H}dd\bar{u}QQ\bar{L}\bar{L}\bar{e}\,,\\
\mathcal{O}_{14, \ (1,-3)}^{21}\equiv \bar{H}\bar{H}d\bar{u}QQQ\bar{L}\bar{L}\bar{L}\,,\\
\mathcal{O}_{14, \ (1,-3)}^{22}\equiv \bar{H}\bar{H}ddd\bar{d}u\bar{L}\bar{L}\bar{\nu}\,,\\
\mathcal{O}_{14, \ (1,-3)}^{23}\equiv \bar{H}\bar{H}dddu\bar{Q}\bar{L}\bar{e}\bar{\nu}\,,\\
\mathcal{O}_{14, \ (1,-3)}^{24}\equiv \bar{H}\bar{H}dduu\bar{u}\bar{L}\bar{L}\bar{\nu}\,,\\
\mathcal{O}_{14, \ (1,-3)}^{25}\equiv \bar{H}\bar{H}ddu\bar{L}\bar{L}e\bar{e}\bar{\nu}\,,\\
\mathcal{O}_{14, \ (1,-3)}^{26}\equiv \bar{H}\bar{H}ddQ\bar{L}e\bar{e}\bar{e}\bar{\nu}\,,\\
\mathcal{O}_{14, \ (1,-3)}^{27}\equiv \bar{H}\bar{H}duQ\bar{L}\bar{L}\bar{L}e\bar{\nu}\,,\\
\mathcal{O}_{14, \ (1,-3)}^{28}\equiv \bar{H}\bar{H}dQQ\bar{L}\bar{L}e\bar{e}\bar{\nu}\,,\\
\mathcal{O}_{14, \ (1,-3)}^{29}\equiv \bar{H}\bar{H}QQQ\bar{L}\bar{L}\bar{L}e\bar{\nu}\,,\\
\mathcal{O}_{14, \ (1,-3)}^{30}\equiv \bar{H}\bar{H}dddL\bar{L}\bar{e}\bar{e}\bar{\nu}\,,\\
\mathcal{O}_{14, \ (1,-3)}^{31}\equiv \bar{H}\bar{H}dduL\bar{L}\bar{L}\bar{L}\bar{\nu}\,,\\
\mathcal{O}_{14, \ (1,-3)}^{32}\equiv \bar{H}\bar{H}ddQL\bar{L}\bar{L}\bar{e}\bar{\nu}\,,\\
\mathcal{O}_{14, \ (1,-3)}^{33}\equiv \bar{H}\bar{H}dQQL\bar{L}\bar{L}\bar{L}\bar{\nu}\,,\\
\mathcal{O}_{14, \ (1,-3)}^{34}\equiv \bar{H}\bar{H}ddd\bar{L}\bar{L}\bar{L}\bar{L}\nu\,,\\
\mathcal{O}_{14, \ (1,-3)}^{35}\equiv H\bar{H}dddd\bar{u}\bar{e}\bar{e}\bar{\nu}\,,\\
\mathcal{O}_{14, \ (1,-3)}^{36}\equiv H\bar{H}dddd\bar{Q}\bar{L}\bar{e}\bar{\nu}\,,\\
\mathcal{O}_{14, \ (1,-3)}^{37}\equiv H\bar{H}ddd\bar{u}Q\bar{L}\bar{e}\bar{\nu}\,,\\
\mathcal{O}_{14, \ (1,-3)}^{38}\equiv H\bar{H}dddd\bar{d}\bar{L}\bar{L}\bar{\nu}\,,\\
\mathcal{O}_{14, \ (1,-3)}^{39}\equiv H\bar{H}ddd\bar{L}\bar{L}e\bar{e}\bar{\nu}\,,\\
\mathcal{O}_{14, \ (1,-3)}^{40}\equiv H\bar{H}dddL\bar{L}\bar{L}\bar{L}\bar{\nu}\,,\\
\mathcal{O}_{14, \ (1,-3)}^{41}\equiv H\bar{H}ddQ\bar{L}\bar{L}\bar{L}e\bar{\nu}\,,\\
\mathcal{O}_{14, \ (1,-3)}^{42}\equiv H\bar{H}dddQ\bar{Q}\bar{L}\bar{L}\bar{\nu}\,,\\
\mathcal{O}_{14, \ (1,-3)}^{43}\equiv H\bar{H}dd\bar{u}QQ\bar{L}\bar{L}\bar{\nu}\,,\\
\mathcal{O}_{14, \ (1,-3)}^{44}\equiv H\bar{H}dddu\bar{u}\bar{L}\bar{L}\bar{\nu}\,,\\
\mathcal{O}_{14, \ (1,-3)}^{45}\equiv HHdddd\bar{u}\bar{L}\bar{L}\bar{\nu}\,,\\
\mathcal{O}_{14, \ (1,-3)}^{46}\equiv \bar{H}\bar{H}ddd\bar{d}Q\bar{L}\bar{e}\bar{\nu}\,,\\
\mathcal{O}_{14, \ (1,-3)}^{47}\equiv \bar{H}\bar{H}dddQ\bar{Q}\bar{e}\bar{e}\bar{\nu}\,,\\
\mathcal{O}_{14, \ (1,-3)}^{48}\equiv \bar{H}\bar{H}dduQ\bar{Q}\bar{L}\bar{L}\bar{\nu}\,,\\
\mathcal{O}_{14, \ (1,-3)}^{49}\equiv \bar{H}\bar{H}ddu\bar{u}Q\bar{L}\bar{e}\bar{\nu}\,,\\
\mathcal{O}_{14, \ (1,-3)}^{50}\equiv \bar{H}\bar{H}dd\bar{d}QQ\bar{L}\bar{L}\bar{\nu}\,,\\
\mathcal{O}_{14, \ (1,-3)}^{51}\equiv \bar{H}\bar{H}ddQQ\bar{Q}\bar{L}\bar{e}\bar{\nu}\,,\\
\mathcal{O}_{14, \ (1,-3)}^{52}\equiv \bar{H}\bar{H}dd\bar{u}QQ\bar{e}\bar{e}\bar{\nu}\,,\\
\mathcal{O}_{14, \ (1,-3)}^{53}\equiv \bar{H}\bar{H}du\bar{u}QQ\bar{L}\bar{L}\bar{\nu}\,,\\
\mathcal{O}_{14, \ (1,-3)}^{54}\equiv \bar{H}\bar{H}dQQQ\bar{Q}\bar{L}\bar{L}\bar{\nu}\,,\\
\mathcal{O}_{14, \ (1,-3)}^{55}\equiv \bar{H}\bar{H}d\bar{u}QQQ\bar{L}\bar{e}\bar{\nu}\,,\\
\mathcal{O}_{14, \ (1,-3)}^{56}\equiv \bar{H}\bar{H}\bar{u}QQQQ\bar{L}\bar{L}\bar{\nu}\,,\\
\mathcal{O}_{14, \ (1,-3)}^{57}\equiv \bar{H}\bar{H}dduu\bar{Q}\bar{L}\bar{\nu}\bar{\nu}\,,\\
\mathcal{O}_{14, \ (1,-3)}^{58}\equiv \bar{H}\bar{H}duu\bar{L}\bar{L}e\bar{\nu}\bar{\nu}\,,\\
\mathcal{O}_{14, \ (1,-3)}^{59}\equiv \bar{H}\bar{H}duQ\bar{L}e\bar{e}\bar{\nu}\bar{\nu}\,,\\
\mathcal{O}_{14, \ (1,-3)}^{60}\equiv \bar{H}\bar{H}dQQe\bar{e}\bar{e}\bar{\nu}\bar{\nu}\,,\\
\mathcal{O}_{14, \ (1,-3)}^{61}\equiv \bar{H}\bar{H}QQQ\bar{L}e\bar{e}\bar{\nu}\bar{\nu}\,,\\
\mathcal{O}_{14, \ (1,-3)}^{62}\equiv \bar{H}\bar{H}uQQ\bar{L}\bar{L}e\bar{\nu}\bar{\nu}\,,\\
\mathcal{O}_{14, \ (1,-3)}^{63}\equiv \bar{H}\bar{H}dduL\bar{L}\bar{e}\bar{\nu}\bar{\nu}\,,\\
\mathcal{O}_{14, \ (1,-3)}^{64}\equiv \bar{H}\bar{H}ddQL\bar{e}\bar{e}\bar{\nu}\bar{\nu}\,,\\
\mathcal{O}_{14, \ (1,-3)}^{65}\equiv \bar{H}\bar{H}duQL\bar{L}\bar{L}\bar{\nu}\bar{\nu}\,,\\
\mathcal{O}_{14, \ (1,-3)}^{66}\equiv \bar{H}\bar{H}dQQL\bar{L}\bar{e}\bar{\nu}\bar{\nu}\,,\\
\mathcal{O}_{14, \ (1,-3)}^{67}\equiv \bar{H}\bar{H}QQQL\bar{L}\bar{L}\bar{\nu}\bar{\nu}\,,\\
\mathcal{O}_{14, \ (1,-3)}^{68}\equiv \bar{H}\bar{H}ddd\bar{L}\bar{L}\bar{e}\nu\bar{\nu}\,,\\
\mathcal{O}_{14, \ (1,-3)}^{69}\equiv \bar{H}\bar{H}ddQ\bar{L}\bar{L}\bar{L}\nu\bar{\nu}\,,\\
\mathcal{O}_{14, \ (1,-3)}^{70}\equiv H\bar{H}dddd\bar{d}\bar{e}\bar{\nu}\bar{\nu}\,,\\
\mathcal{O}_{14, \ (1,-3)}^{71}\equiv H\bar{H}ddde\bar{e}\bar{e}\bar{\nu}\bar{\nu}\,,\\
\mathcal{O}_{14, \ (1,-3)}^{72}\equiv H\bar{H}dddQ\bar{Q}\bar{e}\bar{\nu}\bar{\nu}\,,\\
\mathcal{O}_{14, \ (1,-3)}^{73}\equiv H\bar{H}dd\bar{u}QQ\bar{e}\bar{\nu}\bar{\nu}\,,\\
\mathcal{O}_{14, \ (1,-3)}^{74}\equiv H\bar{H}dddu\bar{u}\bar{e}\bar{\nu}\bar{\nu}\,,\\
\mathcal{O}_{14, \ (1,-3)}^{75}\equiv H\bar{H}ddd\bar{d}Q\bar{L}\bar{\nu}\bar{\nu}\,,\\
\mathcal{O}_{14, \ (1,-3)}^{76}\equiv H\bar{H}dddL\bar{L}\bar{e}\bar{\nu}\bar{\nu}\,,\\
\mathcal{O}_{14, \ (1,-3)}^{77}\equiv H\bar{H}ddQ\bar{L}e\bar{e}\bar{\nu}\bar{\nu}\,,\\
\mathcal{O}_{14, \ (1,-3)}^{78}\equiv H\bar{H}ddQL\bar{L}\bar{L}\bar{\nu}\bar{\nu}\,,\\
\mathcal{O}_{14, \ (1,-3)}^{79}\equiv H\bar{H}dQQ\bar{L}\bar{L}e\bar{\nu}\bar{\nu}\,,\\
\mathcal{O}_{14, \ (1,-3)}^{80}\equiv H\bar{H}ddu\bar{L}\bar{L}e\bar{\nu}\bar{\nu}\,,\\
\mathcal{O}_{14, \ (1,-3)}^{81}\equiv H\bar{H}ddQQ\bar{Q}\bar{L}\bar{\nu}\bar{\nu}\,,\\
\mathcal{O}_{14, \ (1,-3)}^{82}\equiv H\bar{H}dddu\bar{Q}\bar{L}\bar{\nu}\bar{\nu}\,,\\
\mathcal{O}_{14, \ (1,-3)}^{83}\equiv H\bar{H}d\bar{u}QQQ\bar{L}\bar{\nu}\bar{\nu}\,,\\
\mathcal{O}_{14, \ (1,-3)}^{84}\equiv H\bar{H}ddu\bar{u}Q\bar{L}\bar{\nu}\bar{\nu}\,,\\
\mathcal{O}_{14, \ (1,-3)}^{85}\equiv HHddd\bar{L}\bar{L}e\bar{\nu}\bar{\nu}\,,\\
\mathcal{O}_{14, \ (1,-3)}^{86}\equiv HHdddd\bar{Q}\bar{L}\bar{\nu}\bar{\nu}\,,\\
\mathcal{O}_{14, \ (1,-3)}^{87}\equiv HHddd\bar{u}Q\bar{L}\bar{\nu}\bar{\nu}\,,\\
\mathcal{O}_{14, \ (1,-3)}^{88}\equiv \bar{H}\bar{H}dd\bar{d}uQ\bar{L}\bar{\nu}\bar{\nu}\,,\\
\mathcal{O}_{14, \ (1,-3)}^{89}\equiv \bar{H}\bar{H}dduQ\bar{Q}\bar{e}\bar{\nu}\bar{\nu}\,,\\
\mathcal{O}_{14, \ (1,-3)}^{90}\equiv \bar{H}\bar{H}duu\bar{u}Q\bar{L}\bar{\nu}\bar{\nu}\,,\\
\mathcal{O}_{14, \ (1,-3)}^{91}\equiv \bar{H}\bar{H}dd\bar{d}QQ\bar{e}\bar{\nu}\bar{\nu}\,,\\
\mathcal{O}_{14, \ (1,-3)}^{92}\equiv \bar{H}\bar{H}duQQ\bar{Q}\bar{L}\bar{\nu}\bar{\nu}\,,\\
\mathcal{O}_{14, \ (1,-3)}^{93}\equiv \bar{H}\bar{H}du\bar{u}QQ\bar{e}\bar{\nu}\bar{\nu}\,,\\
\mathcal{O}_{14, \ (1,-3)}^{94}\equiv \bar{H}\bar{H}d\bar{d}QQQ\bar{L}\bar{\nu}\bar{\nu}\,,\\
\mathcal{O}_{14, \ (1,-3)}^{95}\equiv \bar{H}\bar{H}dQQQ\bar{Q}\bar{e}\bar{\nu}\bar{\nu}\,,\\
\mathcal{O}_{14, \ (1,-3)}^{96}\equiv \bar{H}\bar{H}QQQQ\bar{Q}\bar{L}\bar{\nu}\bar{\nu}\,,\\
\mathcal{O}_{14, \ (1,-3)}^{97}\equiv \bar{H}\bar{H}\bar{u}QQQQ\bar{e}\bar{\nu}\bar{\nu}\,,\\
\mathcal{O}_{14, \ (1,-3)}^{98}\equiv \bar{H}\bar{H}u\bar{u}QQQ\bar{L}\bar{\nu}\bar{\nu}\,,\\
\mathcal{O}_{14, \ (1,-3)}^{99}\equiv \bar{H}\bar{H}uQQe\bar{e}\bar{\nu}\bar{\nu}\bar{\nu}\,,\\
\mathcal{O}_{14, \ (1,-3)}^{100}\equiv \bar{H}\bar{H}uuQ\bar{L}e\bar{\nu}\bar{\nu}\bar{\nu}\,,\\
\mathcal{O}_{14, \ (1,-3)}^{101}\equiv \bar{H}\bar{H}duuL\bar{L}\bar{\nu}\bar{\nu}\bar{\nu}\,,\\
\mathcal{O}_{14, \ (1,-3)}^{102}\equiv \bar{H}\bar{H}duQL\bar{e}\bar{\nu}\bar{\nu}\bar{\nu}\,,\\
\mathcal{O}_{14, \ (1,-3)}^{103}\equiv \bar{H}\bar{H}QQQL\bar{e}\bar{\nu}\bar{\nu}\bar{\nu}\,,\\
\mathcal{O}_{14, \ (1,-3)}^{104}\equiv \bar{H}\bar{H}uQQL\bar{L}\bar{\nu}\bar{\nu}\bar{\nu}\,,\\
\mathcal{O}_{14, \ (1,-3)}^{105}\equiv \bar{H}\bar{H}ddu\bar{L}\bar{L}\nu\bar{\nu}\bar{\nu}\,,\\
\mathcal{O}_{14, \ (1,-3)}^{106}\equiv \bar{H}\bar{H}ddQ\bar{L}\bar{e}\nu\bar{\nu}\bar{\nu}\,,\\
\mathcal{O}_{14, \ (1,-3)}^{107}\equiv \bar{H}\bar{H}dQQ\bar{L}\bar{L}\nu\bar{\nu}\bar{\nu}\,,\\
\mathcal{O}_{14, \ (1,-3)}^{108}\equiv H\bar{H}ddd\bar{L}\bar{L}\nu\bar{\nu}\bar{\nu}\,,\\
\mathcal{O}_{14, \ (1,-3)}^{109}\equiv H\bar{H}dd\bar{d}QQ\bar{\nu}\bar{\nu}\bar{\nu}\,,\\
\mathcal{O}_{14, \ (1,-3)}^{110}\equiv H\bar{H}ddd\bar{d}u\bar{\nu}\bar{\nu}\bar{\nu}\,,\\
\mathcal{O}_{14, \ (1,-3)}^{111}\equiv H\bar{H}ddQL\bar{e}\bar{\nu}\bar{\nu}\bar{\nu}\,,\\
\mathcal{O}_{14, \ (1,-3)}^{112}\equiv H\bar{H}dQQe\bar{e}\bar{\nu}\bar{\nu}\bar{\nu}\,,\\
\mathcal{O}_{14, \ (1,-3)}^{113}\equiv H\bar{H}ddue\bar{e}\bar{\nu}\bar{\nu}\bar{\nu}\,,\\
\mathcal{O}_{14, \ (1,-3)}^{114}\equiv H\bar{H}dQQL\bar{L}\bar{\nu}\bar{\nu}\bar{\nu}\,,\\
\mathcal{O}_{14, \ (1,-3)}^{115}\equiv H\bar{H}QQQ\bar{L}e\bar{\nu}\bar{\nu}\bar{\nu}\,,\\
\mathcal{O}_{14, \ (1,-3)}^{116}\equiv H\bar{H}dduL\bar{L}\bar{\nu}\bar{\nu}\bar{\nu}\,,\\
\mathcal{O}_{14, \ (1,-3)}^{117}\equiv H\bar{H}duQ\bar{L}e\bar{\nu}\bar{\nu}\bar{\nu}\,,\\
\mathcal{O}_{14, \ (1,-3)}^{118}\equiv H\bar{H}dQQQ\bar{Q}\bar{\nu}\bar{\nu}\bar{\nu}\,,\\
\mathcal{O}_{14, \ (1,-3)}^{119}\equiv H\bar{H}dduQ\bar{Q}\bar{\nu}\bar{\nu}\bar{\nu}\,,\\
\mathcal{O}_{14, \ (1,-3)}^{120}\equiv H\bar{H}\bar{u}QQQQ\bar{\nu}\bar{\nu}\bar{\nu}\,,\\
\mathcal{O}_{14, \ (1,-3)}^{121}\equiv H\bar{H}du\bar{u}QQ\bar{\nu}\bar{\nu}\bar{\nu}\,,\\
\mathcal{O}_{14, \ (1,-3)}^{122}\equiv H\bar{H}dduu\bar{u}\bar{\nu}\bar{\nu}\bar{\nu}\,,\\
\mathcal{O}_{14, \ (1,-3)}^{123}\equiv HHdddL\bar{L}\bar{\nu}\bar{\nu}\bar{\nu}\,,\\
\mathcal{O}_{14, \ (1,-3)}^{124}\equiv HHddQ\bar{L}e\bar{\nu}\bar{\nu}\bar{\nu}\,,\\
\mathcal{O}_{14, \ (1,-3)}^{125}\equiv HHdddQ\bar{Q}\bar{\nu}\bar{\nu}\bar{\nu}\,,\\
\mathcal{O}_{14, \ (1,-3)}^{126}\equiv HHdd\bar{u}QQ\bar{\nu}\bar{\nu}\bar{\nu}\,,\\
\mathcal{O}_{14, \ (1,-3)}^{127}\equiv \bar{H}\bar{H}duuQ\bar{Q}\bar{\nu}\bar{\nu}\bar{\nu}\,,\\
\mathcal{O}_{14, \ (1,-3)}^{128}\equiv \bar{H}\bar{H}d\bar{d}uQQ\bar{\nu}\bar{\nu}\bar{\nu}\,,\\
\mathcal{O}_{14, \ (1,-3)}^{129}\equiv \bar{H}\bar{H}\bar{d}QQQQ\bar{\nu}\bar{\nu}\bar{\nu}\,,\\
\mathcal{O}_{14, \ (1,-3)}^{130}\equiv \bar{H}\bar{H}uQQQ\bar{Q}\bar{\nu}\bar{\nu}\bar{\nu}\,,\\
\mathcal{O}_{14, \ (1,-3)}^{131}\equiv \bar{H}\bar{H}uu\bar{u}QQ\bar{\nu}\bar{\nu}\bar{\nu}\,,\\
\mathcal{O}_{14, \ (1,-3)}^{132}\equiv \bar{H}\bar{H}uuQL\bar{\nu}\bar{\nu}\bar{\nu}\bar{\nu}\,,\\
\mathcal{O}_{14, \ (1,-3)}^{133}\equiv \bar{H}\bar{H}duQ\bar{L}\nu\bar{\nu}\bar{\nu}\bar{\nu}\,,\\
\mathcal{O}_{14, \ (1,-3)}^{134}\equiv \bar{H}\bar{H}dQQ\bar{e}\nu\bar{\nu}\bar{\nu}\bar{\nu}\,,\\
\mathcal{O}_{14, \ (1,-3)}^{135}\equiv \bar{H}\bar{H}QQQ\bar{L}\nu\bar{\nu}\bar{\nu}\bar{\nu}\,,\\
\mathcal{O}_{14, \ (1,-3)}^{136}\equiv H\bar{H}ddd\bar{e}\nu\bar{\nu}\bar{\nu}\bar{\nu}\,,\\
\mathcal{O}_{14, \ (1,-3)}^{137}\equiv H\bar{H}ddQ\bar{L}\nu\bar{\nu}\bar{\nu}\bar{\nu}\,,\\
\mathcal{O}_{14, \ (1,-3)}^{138}\equiv HHddQL\bar{\nu}\bar{\nu}\bar{\nu}\bar{\nu}\,,\\
\mathcal{O}_{14, \ (1,-3)}^{139}\equiv HHdQQe\bar{\nu}\bar{\nu}\bar{\nu}\bar{\nu}\,,\\
\mathcal{O}_{14, \ (1,-3)}^{140}\equiv H\bar{H}QQQL\bar{\nu}\bar{\nu}\bar{\nu}\bar{\nu}\,,\\
\mathcal{O}_{14, \ (1,-3)}^{141}\equiv H\bar{H}duQL\bar{\nu}\bar{\nu}\bar{\nu}\bar{\nu}\,,\\
\mathcal{O}_{14, \ (1,-3)}^{142}\equiv H\bar{H}uQQe\bar{\nu}\bar{\nu}\bar{\nu}\bar{\nu}\,,\\
\mathcal{O}_{14, \ (1,-3)}^{143}\equiv H\bar{H}duue\bar{\nu}\bar{\nu}\bar{\nu}\bar{\nu}\,,\\
\mathcal{O}_{14, \ (1,-3)}^{144}\equiv \bar{H}\bar{H}uQQ\nu\bar{\nu}\bar{\nu}\bar{\nu}\bar{\nu}\,,\\
\mathcal{O}_{14, \ (1,-3)}^{145}\equiv H\bar{H}dQQ\nu\bar{\nu}\bar{\nu}\bar{\nu}\bar{\nu}\,,\\
\mathcal{O}_{14, \ (1,-3)}^{146}\equiv H\bar{H}ddu\nu\bar{\nu}\bar{\nu}\bar{\nu}\bar{\nu}\,,\\
\mathcal{O}_{14, \ (1,-3)}^{147}\equiv H\bar{H}\bar{H}\bar{H}\bar{H}ddu\bar{L}\bar{L}\bar{L}\,,\\
\mathcal{O}_{14, \ (1,-3)}^{148}\equiv HH\bar{H}\bar{H}\bar{H}ddd\bar{L}\bar{L}\bar{L}\,,\\
\mathcal{O}_{14, \ (1,-3)}^{149}\equiv H\bar{H}\bar{H}\bar{H}\bar{H}ddQ\bar{L}\bar{L}\bar{e}\,,\\
\mathcal{O}_{14, \ (1,-3)}^{150}\equiv H\bar{H}\bar{H}\bar{H}\bar{H}dQQ\bar{L}\bar{L}\bar{L}\,,\\
\mathcal{O}_{14, \ (1,-3)}^{151}\equiv \bar{H}\bar{H}\bar{H}\bar{H}\bar{H}QQQ\bar{L}\bar{L}\bar{e}\,,\\
\mathcal{O}_{14, \ (1,-3)}^{152}\equiv \bar{H}\bar{H}\bar{H}\bar{H}\bar{H}uQQ\bar{L}\bar{L}\bar{L}\,,\\
\mathcal{O}_{14, \ (1,-3)}^{153}\equiv HH\bar{H}\bar{H}\bar{H}ddd\bar{L}\bar{e}\bar{\nu}\,,\\
\mathcal{O}_{14, \ (1,-3)}^{154}\equiv HH\bar{H}\bar{H}\bar{H}ddQ\bar{L}\bar{L}\bar{\nu}\,,\\
\mathcal{O}_{14, \ (1,-3)}^{155}\equiv H\bar{H}\bar{H}\bar{H}\bar{H}duQ\bar{L}\bar{L}\bar{\nu}\,,\\
\mathcal{O}_{14, \ (1,-3)}^{156}\equiv H\bar{H}\bar{H}\bar{H}\bar{H}dQQ\bar{L}\bar{e}\bar{\nu}\,,\\
\mathcal{O}_{14, \ (1,-3)}^{157}\equiv H\bar{H}\bar{H}\bar{H}\bar{H}QQQ\bar{L}\bar{L}\bar{\nu}\,,\\
\mathcal{O}_{14, \ (1,-3)}^{158}\equiv HH\bar{H}\bar{H}\bar{H}ddu\bar{L}\bar{\nu}\bar{\nu}\,,\\
\mathcal{O}_{14, \ (1,-3)}^{159}\equiv HH\bar{H}\bar{H}\bar{H}ddQ\bar{e}\bar{\nu}\bar{\nu}\,,\\
\mathcal{O}_{14, \ (1,-3)}^{160}\equiv HH\bar{H}\bar{H}\bar{H}dQQ\bar{L}\bar{\nu}\bar{\nu}\,,\\
\mathcal{O}_{14, \ (1,-3)}^{161}\equiv H\bar{H}\bar{H}\bar{H}\bar{H}QQQ\bar{e}\bar{\nu}\bar{\nu}\,,\\
\mathcal{O}_{14, \ (1,-3)}^{162}\equiv H\bar{H}\bar{H}\bar{H}\bar{H}uQQ\bar{L}\bar{\nu}\bar{\nu}\,,\\
\mathcal{O}_{14, \ (1,-3)}^{163}\equiv HHH\bar{H}\bar{H}ddd\bar{L}\bar{\nu}\bar{\nu}\,,\\
\mathcal{O}_{14, \ (1,-3)}^{164}\equiv HH\bar{H}\bar{H}\bar{H}duQ\bar{\nu}\bar{\nu}\bar{\nu}\,,\\
\mathcal{O}_{14, \ (1,-3)}^{165}\equiv HH\bar{H}\bar{H}\bar{H}QQQ\bar{\nu}\bar{\nu}\bar{\nu}\,,\\
\mathcal{O}_{14, \ (1,-3)}^{166}\equiv HHH\bar{H}\bar{H}ddQ\bar{\nu}\bar{\nu}\bar{\nu}\,.
 \label{eq:dequal14basisB1L-3}
 \end{math}
 \end{spacing}
 \end{multicols}

 \subsection{\texorpdfstring{$\Delta B=1, \Delta L=1$}{Delta B=1,Delta L=1}}
 \begin{multicols}{2}
 \begin{spacing}{1.5}
 \noindent
 \begin{math}
\mathcal{O}_{14, \ (1,1)}^{1}\equiv HHdddd\bar{d}\bar{d}QL\,,\\
\mathcal{O}_{14, \ (1,1)}^{2}\equiv HHddd\bar{d}\bar{d}QQe\,,\\
\mathcal{O}_{14, \ (1,1)}^{3}\equiv HHdd\bar{d}\bar{u}QQQL\,,\\
\mathcal{O}_{14, \ (1,1)}^{4}\equiv HHd\bar{d}\bar{u}QQQQe\,,\\
\mathcal{O}_{14, \ (1,1)}^{5}\equiv HHddd\bar{d}u\bar{u}QL\,,\\
\mathcal{O}_{14, \ (1,1)}^{6}\equiv HHdd\bar{d}u\bar{u}QQe\,,\\
\mathcal{O}_{14, \ (1,1)}^{7}\equiv HHdddd\bar{d}LL\bar{e}\,,\\
\mathcal{O}_{14, \ (1,1)}^{8}\equiv HHddd\bar{d}QLe\bar{e}\,,\\
\mathcal{O}_{14, \ (1,1)}^{9}\equiv HHdd\bar{d}QQee\bar{e}\,,\\
\mathcal{O}_{14, \ (1,1)}^{10}\equiv HHdddLLe\bar{e}\bar{e}\,,\\
\mathcal{O}_{14, \ (1,1)}^{11}\equiv HHddQLee\bar{e}\bar{e}\,,\\
\mathcal{O}_{14, \ (1,1)}^{12}\equiv HHdQQeee\bar{e}\bar{e}\,,\\
\mathcal{O}_{14, \ (1,1)}^{13}\equiv HHdddQ\bar{Q}LL\bar{e}\,,\\
\mathcal{O}_{14, \ (1,1)}^{14}\equiv HHddQQ\bar{Q}Le\bar{e}\,,\\
\mathcal{O}_{14, \ (1,1)}^{15}\equiv HHdQQQ\bar{Q}ee\bar{e}\,,\\
\mathcal{O}_{14, \ (1,1)}^{16}\equiv HHdddu\bar{Q}Le\bar{e}\,,\\
\mathcal{O}_{14, \ (1,1)}^{17}\equiv HHdduQ\bar{Q}ee\bar{e}\,,\\
\mathcal{O}_{14, \ (1,1)}^{18}\equiv HHdd\bar{u}QQLL\bar{e}\,,\\
\mathcal{O}_{14, \ (1,1)}^{19}\equiv HHd\bar{u}QQQLe\bar{e}\,,\\
\mathcal{O}_{14, \ (1,1)}^{20}\equiv HH\bar{u}QQQQee\bar{e}\,,\\
\mathcal{O}_{14, \ (1,1)}^{21}\equiv HHdddu\bar{u}LL\bar{e}\,,\\
\mathcal{O}_{14, \ (1,1)}^{22}\equiv HHddu\bar{u}QLe\bar{e}\,,\\
\mathcal{O}_{14, \ (1,1)}^{23}\equiv HHdu\bar{u}QQee\bar{e}\,,\\
\mathcal{O}_{14, \ (1,1)}^{24}\equiv H\bar{H}dd\bar{d}\bar{d}QQQL\,,\\
\mathcal{O}_{14, \ (1,1)}^{25}\equiv H\bar{H}d\bar{d}\bar{d}QQQQe\,,\\
\mathcal{O}_{14, \ (1,1)}^{26}\equiv H\bar{H}ddd\bar{d}\bar{d}uQL\,,\\
\mathcal{O}_{14, \ (1,1)}^{27}\equiv H\bar{H}dd\bar{d}\bar{d}uQQe\,,\\
\mathcal{O}_{14, \ (1,1)}^{28}\equiv H\bar{H}ddd\bar{d}\bar{d}uue\,,\\
\mathcal{O}_{14, \ (1,1)}^{29}\equiv H\bar{H}\bar{d}\bar{u}QQQQQL\,,\\
\mathcal{O}_{14, \ (1,1)}^{30}\equiv H\bar{H}d\bar{d}u\bar{u}QQQL\,,\\
\mathcal{O}_{14, \ (1,1)}^{31}\equiv H\bar{H}\bar{d}u\bar{u}QQQQe\,,\\
\mathcal{O}_{14, \ (1,1)}^{32}\equiv H\bar{H}dd\bar{d}uu\bar{u}QL\,,\\
\mathcal{O}_{14, \ (1,1)}^{33}\equiv H\bar{H}d\bar{d}uu\bar{u}QQe\,,\\
\mathcal{O}_{14, \ (1,1)}^{34}\equiv H\bar{H}dd\bar{d}uuu\bar{u}e\,,\\
\mathcal{O}_{14, \ (1,1)}^{35}\equiv H\bar{H}dd\bar{d}QQLL\bar{e}\,,\\
\mathcal{O}_{14, \ (1,1)}^{36}\equiv H\bar{H}d\bar{d}QQQLe\bar{e}\,,\\
\mathcal{O}_{14, \ (1,1)}^{37}\equiv H\bar{H}\bar{d}QQQQee\bar{e}\,,\\
\mathcal{O}_{14, \ (1,1)}^{38}\equiv H\bar{H}ddd\bar{d}uLL\bar{e}\,,\\
\mathcal{O}_{14, \ (1,1)}^{39}\equiv H\bar{H}dd\bar{d}uQLe\bar{e}\,,\\
\mathcal{O}_{14, \ (1,1)}^{40}\equiv H\bar{H}d\bar{d}uQQee\bar{e}\,,\\
\mathcal{O}_{14, \ (1,1)}^{41}\equiv H\bar{H}dd\bar{d}uuee\bar{e}\,,\\
\mathcal{O}_{14, \ (1,1)}^{42}\equiv H\bar{H}ddQLLL\bar{e}\bar{e}\,,\\
\mathcal{O}_{14, \ (1,1)}^{43}\equiv H\bar{H}dQQLLe\bar{e}\bar{e}\,,\\
\mathcal{O}_{14, \ (1,1)}^{44}\equiv H\bar{H}QQQLee\bar{e}\bar{e}\,,\\
\mathcal{O}_{14, \ (1,1)}^{45}\equiv H\bar{H}dduLLe\bar{e}\bar{e}\,,\\
\mathcal{O}_{14, \ (1,1)}^{46}\equiv H\bar{H}duQLee\bar{e}\bar{e}\,,\\
\mathcal{O}_{14, \ (1,1)}^{47}\equiv H\bar{H}uQQeee\bar{e}\bar{e}\,,\\
\mathcal{O}_{14, \ (1,1)}^{48}\equiv H\bar{H}duueee\bar{e}\bar{e}\,,\\
\mathcal{O}_{14, \ (1,1)}^{49}\equiv H\bar{H}dQQQ\bar{Q}LL\bar{e}\,,\\
\mathcal{O}_{14, \ (1,1)}^{50}\equiv H\bar{H}QQQQ\bar{Q}Le\bar{e}\,,\\
\mathcal{O}_{14, \ (1,1)}^{51}\equiv H\bar{H}dduQ\bar{Q}LL\bar{e}\,,\\
\mathcal{O}_{14, \ (1,1)}^{52}\equiv H\bar{H}duQQ\bar{Q}Le\bar{e}\,,\\
\mathcal{O}_{14, \ (1,1)}^{53}\equiv H\bar{H}uQQQ\bar{Q}ee\bar{e}\,,\\
\mathcal{O}_{14, \ (1,1)}^{54}\equiv H\bar{H}dduu\bar{Q}Le\bar{e}\,,\\
\mathcal{O}_{14, \ (1,1)}^{55}\equiv H\bar{H}duuQ\bar{Q}ee\bar{e}\,,\\
\mathcal{O}_{14, \ (1,1)}^{56}\equiv H\bar{H}\bar{u}QQQQLL\bar{e}\,,\\
\mathcal{O}_{14, \ (1,1)}^{57}\equiv H\bar{H}du\bar{u}QQLL\bar{e}\,,\\
\mathcal{O}_{14, \ (1,1)}^{58}\equiv H\bar{H}u\bar{u}QQQLe\bar{e}\,,\\
\mathcal{O}_{14, \ (1,1)}^{59}\equiv H\bar{H}dduu\bar{u}LL\bar{e}\,,\\
\mathcal{O}_{14, \ (1,1)}^{60}\equiv H\bar{H}duu\bar{u}QLe\bar{e}\,,\\
\mathcal{O}_{14, \ (1,1)}^{61}\equiv H\bar{H}uu\bar{u}QQee\bar{e}\,,\\
\mathcal{O}_{14, \ (1,1)}^{62}\equiv H\bar{H}duuu\bar{u}ee\bar{e}\,,\\
\mathcal{O}_{14, \ (1,1)}^{63}\equiv \bar{H}\bar{H}\bar{d}\bar{d}QQQQQL\,,\\
\mathcal{O}_{14, \ (1,1)}^{64}\equiv \bar{H}\bar{H}d\bar{d}\bar{d}uQQQL\,,\\
\mathcal{O}_{14, \ (1,1)}^{65}\equiv \bar{H}\bar{H}\bar{d}\bar{d}uQQQQe\,,\\
\mathcal{O}_{14, \ (1,1)}^{66}\equiv \bar{H}\bar{H}dd\bar{d}\bar{d}uuQL\,,\\
\mathcal{O}_{14, \ (1,1)}^{67}\equiv \bar{H}\bar{H}d\bar{d}\bar{d}uuQQe\,,\\
\mathcal{O}_{14, \ (1,1)}^{68}\equiv \bar{H}\bar{H}\bar{d}uu\bar{u}QQQL\,,\\
\mathcal{O}_{14, \ (1,1)}^{69}\equiv \bar{H}\bar{H}d\bar{d}uuu\bar{u}QL\,,\\
\mathcal{O}_{14, \ (1,1)}^{70}\equiv \bar{H}\bar{H}\bar{d}uuu\bar{u}QQe\,,\\
\mathcal{O}_{14, \ (1,1)}^{71}\equiv \bar{H}\bar{H}\bar{d}QQQQLL\bar{e}\,,\\
\mathcal{O}_{14, \ (1,1)}^{72}\equiv \bar{H}\bar{H}d\bar{d}uQQLL\bar{e}\,,\\
\mathcal{O}_{14, \ (1,1)}^{73}\equiv \bar{H}\bar{H}\bar{d}uQQQLe\bar{e}\,,\\
\mathcal{O}_{14, \ (1,1)}^{74}\equiv \bar{H}\bar{H}dd\bar{d}uuLL\bar{e}\,,\\
\mathcal{O}_{14, \ (1,1)}^{75}\equiv \bar{H}\bar{H}d\bar{d}uuQLe\bar{e}\,,\\
\mathcal{O}_{14, \ (1,1)}^{76}\equiv \bar{H}\bar{H}\bar{d}uuQQee\bar{e}\,,\\
\mathcal{O}_{14, \ (1,1)}^{77}\equiv \bar{H}\bar{H}QQQLLL\bar{e}\bar{e}\,,\\
\mathcal{O}_{14, \ (1,1)}^{78}\equiv \bar{H}\bar{H}duQLLL\bar{e}\bar{e}\,,\\
\mathcal{O}_{14, \ (1,1)}^{79}\equiv \bar{H}\bar{H}uQQLLe\bar{e}\bar{e}\,,\\
\mathcal{O}_{14, \ (1,1)}^{80}\equiv \bar{H}\bar{H}duuLLe\bar{e}\bar{e}\,,\\
\mathcal{O}_{14, \ (1,1)}^{81}\equiv \bar{H}\bar{H}uuQLee\bar{e}\bar{e}\,,\\
\mathcal{O}_{14, \ (1,1)}^{82}\equiv \bar{H}\bar{H}uQQQ\bar{Q}LL\bar{e}\,,\\
\mathcal{O}_{14, \ (1,1)}^{83}\equiv \bar{H}\bar{H}duuQ\bar{Q}LL\bar{e}\,,\\
\mathcal{O}_{14, \ (1,1)}^{84}\equiv \bar{H}\bar{H}uuQQ\bar{Q}Le\bar{e}\,,\\
\mathcal{O}_{14, \ (1,1)}^{85}\equiv \bar{H}\bar{H}duuu\bar{Q}Le\bar{e}\,,\\
\mathcal{O}_{14, \ (1,1)}^{86}\equiv \bar{H}\bar{H}uuuQ\bar{Q}ee\bar{e}\,,\\
\mathcal{O}_{14, \ (1,1)}^{87}\equiv \bar{H}\bar{H}uu\bar{u}QQLL\bar{e}\,,\\
\mathcal{O}_{14, \ (1,1)}^{88}\equiv \bar{H}\bar{H}duuu\bar{u}LL\bar{e}\,,\\
\mathcal{O}_{14, \ (1,1)}^{89}\equiv \bar{H}\bar{H}uuu\bar{u}QLe\bar{e}\,,\\
\mathcal{O}_{14, \ (1,1)}^{90}\equiv \bar{H}\bar{H}\bar{d}uQQQLL\bar{L}\,,\\
\mathcal{O}_{14, \ (1,1)}^{91}\equiv \bar{H}\bar{H}d\bar{d}uuQLL\bar{L}\,,\\
\mathcal{O}_{14, \ (1,1)}^{92}\equiv \bar{H}\bar{H}\bar{d}uuQQL\bar{L}e\,,\\
\mathcal{O}_{14, \ (1,1)}^{93}\equiv \bar{H}\bar{H}d\bar{d}uuuL\bar{L}e\,,\\
\mathcal{O}_{14, \ (1,1)}^{94}\equiv \bar{H}\bar{H}\bar{d}uuuQ\bar{L}ee\,,\\
\mathcal{O}_{14, \ (1,1)}^{95}\equiv \bar{H}\bar{H}uQQLLL\bar{L}\bar{e}\,,\\
\mathcal{O}_{14, \ (1,1)}^{96}\equiv \bar{H}\bar{H}duuLLL\bar{L}\bar{e}\,,\\
\mathcal{O}_{14, \ (1,1)}^{97}\equiv \bar{H}\bar{H}uuQLL\bar{L}e\bar{e}\,,\\
\mathcal{O}_{14, \ (1,1)}^{98}\equiv \bar{H}\bar{H}uuuL\bar{L}ee\bar{e}\,,\\
\mathcal{O}_{14, \ (1,1)}^{99}\equiv \bar{H}\bar{H}uuQLLL\bar{L}\bar{L}\,,\\
\mathcal{O}_{14, \ (1,1)}^{100}\equiv \bar{H}\bar{H}uuuLL\bar{L}\bar{L}e\,,\\
\mathcal{O}_{14, \ (1,1)}^{101}\equiv \bar{H}\bar{H}uuQQ\bar{Q}LL\bar{L}\,,\\
\mathcal{O}_{14, \ (1,1)}^{102}\equiv \bar{H}\bar{H}duuu\bar{Q}LL\bar{L}\,,\\
\mathcal{O}_{14, \ (1,1)}^{103}\equiv \bar{H}\bar{H}uuuQ\bar{Q}L\bar{L}e\,,\\
\mathcal{O}_{14, \ (1,1)}^{104}\equiv \bar{H}\bar{H}uuuu\bar{Q}\bar{L}ee\,,\\
\mathcal{O}_{14, \ (1,1)}^{105}\equiv \bar{H}\bar{H}uuu\bar{u}QLL\bar{L}\,,\\
\mathcal{O}_{14, \ (1,1)}^{106}\equiv \bar{H}\bar{H}uuuu\bar{u}L\bar{L}e\,,\\
\mathcal{O}_{14, \ (1,1)}^{107}\equiv \bar{H}\bar{H}\bar{d}uQQQQ\bar{Q}L\,,\\
\mathcal{O}_{14, \ (1,1)}^{108}\equiv \bar{H}\bar{H}d\bar{d}uuQQ\bar{Q}L\,,\\
\mathcal{O}_{14, \ (1,1)}^{109}\equiv \bar{H}\bar{H}\bar{d}uuQQQ\bar{Q}e\,,\\
\mathcal{O}_{14, \ (1,1)}^{110}\equiv \bar{H}\bar{H}dd\bar{d}uuu\bar{Q}L\,,\\
\mathcal{O}_{14, \ (1,1)}^{111}\equiv \bar{H}\bar{H}d\bar{d}uuuQ\bar{Q}e\,,\\
\mathcal{O}_{14, \ (1,1)}^{112}\equiv \bar{H}\bar{H}uuQQQ\bar{Q}\bar{Q}L\,,\\
\mathcal{O}_{14, \ (1,1)}^{113}\equiv \bar{H}\bar{H}duuuQ\bar{Q}\bar{Q}L\,,\\
\mathcal{O}_{14, \ (1,1)}^{114}\equiv \bar{H}\bar{H}uuuQQ\bar{Q}\bar{Q}e\,,\\
\mathcal{O}_{14, \ (1,1)}^{115}\equiv \bar{H}\bar{H}duuuu\bar{Q}\bar{Q}e\,,\\
\mathcal{O}_{14, \ (1,1)}^{116}\equiv \bar{H}\bar{H}uuu\bar{u}QQ\bar{Q}L\,,\\
\mathcal{O}_{14, \ (1,1)}^{117}\equiv \bar{H}\bar{H}duuuu\bar{u}\bar{Q}L\,,\\
\mathcal{O}_{14, \ (1,1)}^{118}\equiv \bar{H}\bar{H}uuuu\bar{u}Q\bar{Q}e\,,\\
\mathcal{O}_{14, \ (1,1)}^{119}\equiv \bar{H}\bar{H}uuuu\bar{u}\bar{u}QL\,,\\
\mathcal{O}_{14, \ (1,1)}^{120}\equiv H\bar{H}d\bar{d}QQQLL\bar{L}\,,\\
\mathcal{O}_{14, \ (1,1)}^{121}\equiv H\bar{H}\bar{d}QQQQL\bar{L}e\,,\\
\mathcal{O}_{14, \ (1,1)}^{122}\equiv H\bar{H}dd\bar{d}uQLL\bar{L}\,,\\
\mathcal{O}_{14, \ (1,1)}^{123}\equiv H\bar{H}d\bar{d}uQQL\bar{L}e\,,\\
\mathcal{O}_{14, \ (1,1)}^{124}\equiv H\bar{H}\bar{d}uQQQ\bar{L}ee\,,\\
\mathcal{O}_{14, \ (1,1)}^{125}\equiv H\bar{H}dd\bar{d}uuL\bar{L}e\,,\\
\mathcal{O}_{14, \ (1,1)}^{126}\equiv H\bar{H}d\bar{d}uuQ\bar{L}ee\,,\\
\mathcal{O}_{14, \ (1,1)}^{127}\equiv H\bar{H}dQQLLL\bar{L}\bar{e}\,,\\
\mathcal{O}_{14, \ (1,1)}^{128}\equiv H\bar{H}QQQLL\bar{L}e\bar{e}\,,\\
\mathcal{O}_{14, \ (1,1)}^{129}\equiv H\bar{H}dduLLL\bar{L}\bar{e}\,,\\
\mathcal{O}_{14, \ (1,1)}^{130}\equiv H\bar{H}duQLL\bar{L}e\bar{e}\,,\\
\mathcal{O}_{14, \ (1,1)}^{131}\equiv H\bar{H}uQQL\bar{L}ee\bar{e}\,,\\
\mathcal{O}_{14, \ (1,1)}^{132}\equiv H\bar{H}duuL\bar{L}ee\bar{e}\,,\\
\mathcal{O}_{14, \ (1,1)}^{133}\equiv H\bar{H}uuQ\bar{L}eee\bar{e}\,,\\
\mathcal{O}_{14, \ (1,1)}^{134}\equiv H\bar{H}QQQLLL\bar{L}\bar{L}\,,\\
\mathcal{O}_{14, \ (1,1)}^{135}\equiv H\bar{H}duQLLL\bar{L}\bar{L}\,,\\
\mathcal{O}_{14, \ (1,1)}^{136}\equiv H\bar{H}uQQLL\bar{L}\bar{L}e\,,\\
\mathcal{O}_{14, \ (1,1)}^{137}\equiv H\bar{H}duuLL\bar{L}\bar{L}e\,,\\
\mathcal{O}_{14, \ (1,1)}^{138}\equiv H\bar{H}uuQL\bar{L}\bar{L}ee\,,\\
\mathcal{O}_{14, \ (1,1)}^{139}\equiv H\bar{H}uuu\bar{L}\bar{L}eee\,,\\
\mathcal{O}_{14, \ (1,1)}^{140}\equiv H\bar{H}QQQQ\bar{Q}LL\bar{L}\,,\\
\mathcal{O}_{14, \ (1,1)}^{141}\equiv H\bar{H}duQQ\bar{Q}LL\bar{L}\,,\\
\mathcal{O}_{14, \ (1,1)}^{142}\equiv H\bar{H}uQQQ\bar{Q}L\bar{L}e\,,\\
\mathcal{O}_{14, \ (1,1)}^{143}\equiv H\bar{H}dduu\bar{Q}LL\bar{L}\,,\\
\mathcal{O}_{14, \ (1,1)}^{144}\equiv H\bar{H}duuQ\bar{Q}L\bar{L}e\,,\\
\mathcal{O}_{14, \ (1,1)}^{145}\equiv H\bar{H}uuQQ\bar{Q}\bar{L}ee\,,\\
\mathcal{O}_{14, \ (1,1)}^{146}\equiv H\bar{H}duuu\bar{Q}\bar{L}ee\,,\\
\mathcal{O}_{14, \ (1,1)}^{147}\equiv H\bar{H}u\bar{u}QQQLL\bar{L}\,,\\
\mathcal{O}_{14, \ (1,1)}^{148}\equiv H\bar{H}duu\bar{u}QLL\bar{L}\,,\\
\mathcal{O}_{14, \ (1,1)}^{149}\equiv H\bar{H}uu\bar{u}QQL\bar{L}e\,,\\
\mathcal{O}_{14, \ (1,1)}^{150}\equiv H\bar{H}duuu\bar{u}L\bar{L}e\,,\\
\mathcal{O}_{14, \ (1,1)}^{151}\equiv H\bar{H}uuu\bar{u}Q\bar{L}ee\,,\\
\mathcal{O}_{14, \ (1,1)}^{152}\equiv H\bar{H}d\bar{d}QQQQ\bar{Q}L\,,\\
\mathcal{O}_{14, \ (1,1)}^{153}\equiv H\bar{H}\bar{d}QQQQQ\bar{Q}e\,,\\
\mathcal{O}_{14, \ (1,1)}^{154}\equiv H\bar{H}dd\bar{d}uQQ\bar{Q}L\,,\\
\mathcal{O}_{14, \ (1,1)}^{155}\equiv H\bar{H}d\bar{d}uQQQ\bar{Q}e\,,\\
\mathcal{O}_{14, \ (1,1)}^{156}\equiv H\bar{H}ddd\bar{d}uu\bar{Q}L\,,\\
\mathcal{O}_{14, \ (1,1)}^{157}\equiv H\bar{H}dd\bar{d}uuQ\bar{Q}e\,,\\
\mathcal{O}_{14, \ (1,1)}^{158}\equiv H\bar{H}QQQQQ\bar{Q}\bar{Q}L\,,\\
\mathcal{O}_{14, \ (1,1)}^{159}\equiv H\bar{H}duQQQ\bar{Q}\bar{Q}L\,,\\
\mathcal{O}_{14, \ (1,1)}^{160}\equiv H\bar{H}uQQQQ\bar{Q}\bar{Q}e\,,\\
\mathcal{O}_{14, \ (1,1)}^{161}\equiv H\bar{H}dduuQ\bar{Q}\bar{Q}L\,,\\
\mathcal{O}_{14, \ (1,1)}^{162}\equiv H\bar{H}duuQQ\bar{Q}\bar{Q}e\,,\\
\mathcal{O}_{14, \ (1,1)}^{163}\equiv H\bar{H}dduuu\bar{Q}\bar{Q}e\,,\\
\mathcal{O}_{14, \ (1,1)}^{164}\equiv H\bar{H}u\bar{u}QQQQ\bar{Q}L\,,\\
\mathcal{O}_{14, \ (1,1)}^{165}\equiv H\bar{H}duu\bar{u}QQ\bar{Q}L\,,\\
\mathcal{O}_{14, \ (1,1)}^{166}\equiv H\bar{H}uu\bar{u}QQQ\bar{Q}e\,,\\
\mathcal{O}_{14, \ (1,1)}^{167}\equiv H\bar{H}dduuu\bar{u}\bar{Q}L\,,\\
\mathcal{O}_{14, \ (1,1)}^{168}\equiv H\bar{H}duuu\bar{u}Q\bar{Q}e\,,\\
\mathcal{O}_{14, \ (1,1)}^{169}\equiv H\bar{H}uu\bar{u}\bar{u}QQQL\,,\\
\mathcal{O}_{14, \ (1,1)}^{170}\equiv H\bar{H}duuu\bar{u}\bar{u}QL\,,\\
\mathcal{O}_{14, \ (1,1)}^{171}\equiv H\bar{H}uuu\bar{u}\bar{u}QQe\,,\\
\mathcal{O}_{14, \ (1,1)}^{172}\equiv H\bar{H}duuuu\bar{u}\bar{u}e\,,\\
\mathcal{O}_{14, \ (1,1)}^{173}\equiv HHddd\bar{d}QLL\bar{L}\,,\\
\mathcal{O}_{14, \ (1,1)}^{174}\equiv HHdd\bar{d}QQL\bar{L}e\,,\\
\mathcal{O}_{14, \ (1,1)}^{175}\equiv HHd\bar{d}QQQ\bar{L}ee\,,\\
\mathcal{O}_{14, \ (1,1)}^{176}\equiv HHddd\bar{d}uL\bar{L}e\,,\\
\mathcal{O}_{14, \ (1,1)}^{177}\equiv HHdd\bar{d}uQ\bar{L}ee\,,\\
\mathcal{O}_{14, \ (1,1)}^{178}\equiv HHdddLLL\bar{L}\bar{e}\,,\\
\mathcal{O}_{14, \ (1,1)}^{179}\equiv HHddQLL\bar{L}e\bar{e}\,,\\
\mathcal{O}_{14, \ (1,1)}^{180}\equiv HHdQQL\bar{L}ee\bar{e}\,,\\
\mathcal{O}_{14, \ (1,1)}^{181}\equiv HHQQQ\bar{L}eee\bar{e}\,,\\
\mathcal{O}_{14, \ (1,1)}^{182}\equiv HHdduL\bar{L}ee\bar{e}\,,\\
\mathcal{O}_{14, \ (1,1)}^{183}\equiv HHduQ\bar{L}eee\bar{e}\,,\\
\mathcal{O}_{14, \ (1,1)}^{184}\equiv HHddQLLL\bar{L}\bar{L}\,,\\
\mathcal{O}_{14, \ (1,1)}^{185}\equiv HHdQQLL\bar{L}\bar{L}e\,,\\
\mathcal{O}_{14, \ (1,1)}^{186}\equiv HHQQQL\bar{L}\bar{L}ee\,,\\
\mathcal{O}_{14, \ (1,1)}^{187}\equiv HHdduLL\bar{L}\bar{L}e\,,\\
\mathcal{O}_{14, \ (1,1)}^{188}\equiv HHduQL\bar{L}\bar{L}ee\,,\\
\mathcal{O}_{14, \ (1,1)}^{189}\equiv HHuQQ\bar{L}\bar{L}eee\,,\\
\mathcal{O}_{14, \ (1,1)}^{190}\equiv HHduu\bar{L}\bar{L}eee\,,\\
\mathcal{O}_{14, \ (1,1)}^{191}\equiv HHddQQ\bar{Q}LL\bar{L}\,,\\
\mathcal{O}_{14, \ (1,1)}^{192}\equiv HHdQQQ\bar{Q}L\bar{L}e\,,\\
\mathcal{O}_{14, \ (1,1)}^{193}\equiv HHQQQQ\bar{Q}\bar{L}ee\,,\\
\mathcal{O}_{14, \ (1,1)}^{194}\equiv HHdddu\bar{Q}LL\bar{L}\,,\\
\mathcal{O}_{14, \ (1,1)}^{195}\equiv HHdduQ\bar{Q}L\bar{L}e\,,\\
\mathcal{O}_{14, \ (1,1)}^{196}\equiv HHduQQ\bar{Q}\bar{L}ee\,,\\
\mathcal{O}_{14, \ (1,1)}^{197}\equiv HHdduu\bar{Q}\bar{L}ee\,,\\
\mathcal{O}_{14, \ (1,1)}^{198}\equiv HHd\bar{u}QQQLL\bar{L}\,,\\
\mathcal{O}_{14, \ (1,1)}^{199}\equiv HH\bar{u}QQQQL\bar{L}e\,,\\
\mathcal{O}_{14, \ (1,1)}^{200}\equiv HHddu\bar{u}QLL\bar{L}\,,\\
\mathcal{O}_{14, \ (1,1)}^{201}\equiv HHdu\bar{u}QQL\bar{L}e\,,\\
\mathcal{O}_{14, \ (1,1)}^{202}\equiv HHu\bar{u}QQQ\bar{L}ee\,,\\
\mathcal{O}_{14, \ (1,1)}^{203}\equiv HHdduu\bar{u}L\bar{L}e\,,\\
\mathcal{O}_{14, \ (1,1)}^{204}\equiv HHduu\bar{u}Q\bar{L}ee\,,\\
\mathcal{O}_{14, \ (1,1)}^{205}\equiv HHddd\bar{d}QQ\bar{Q}L\,,\\
\mathcal{O}_{14, \ (1,1)}^{206}\equiv HHdd\bar{d}QQQ\bar{Q}e\,,\\
\mathcal{O}_{14, \ (1,1)}^{207}\equiv HHdddd\bar{d}u\bar{Q}L\,,\\
\mathcal{O}_{14, \ (1,1)}^{208}\equiv HHddd\bar{d}uQ\bar{Q}e\,,\\
\mathcal{O}_{14, \ (1,1)}^{209}\equiv HHddQQQ\bar{Q}\bar{Q}L\,,\\
\mathcal{O}_{14, \ (1,1)}^{210}\equiv HHdQQQQ\bar{Q}\bar{Q}e\,,\\
\mathcal{O}_{14, \ (1,1)}^{211}\equiv HHddduQ\bar{Q}\bar{Q}L\,,\\
\mathcal{O}_{14, \ (1,1)}^{212}\equiv HHdduQQ\bar{Q}\bar{Q}e\,,\\
\mathcal{O}_{14, \ (1,1)}^{213}\equiv HHddduu\bar{Q}\bar{Q}e\,,\\
\mathcal{O}_{14, \ (1,1)}^{214}\equiv HHd\bar{u}QQQQ\bar{Q}L\,,\\
\mathcal{O}_{14, \ (1,1)}^{215}\equiv HH\bar{u}QQQQQ\bar{Q}e\,,\\
\mathcal{O}_{14, \ (1,1)}^{216}\equiv HHddu\bar{u}QQ\bar{Q}L\,,\\
\mathcal{O}_{14, \ (1,1)}^{217}\equiv HHdu\bar{u}QQQ\bar{Q}e\,,\\
\mathcal{O}_{14, \ (1,1)}^{218}\equiv HHddduu\bar{u}\bar{Q}L\,,\\
\mathcal{O}_{14, \ (1,1)}^{219}\equiv HHdduu\bar{u}Q\bar{Q}e\,,\\
\mathcal{O}_{14, \ (1,1)}^{220}\equiv HH\bar{u}\bar{u}QQQQQL\,,\\
\mathcal{O}_{14, \ (1,1)}^{221}\equiv HHdu\bar{u}\bar{u}QQQL\,,\\
\mathcal{O}_{14, \ (1,1)}^{222}\equiv HHu\bar{u}\bar{u}QQQQe\,,\\
\mathcal{O}_{14, \ (1,1)}^{223}\equiv HHdduu\bar{u}\bar{u}QL\,,\\
\mathcal{O}_{14, \ (1,1)}^{224}\equiv HHduu\bar{u}\bar{u}QQe\,,\\
\mathcal{O}_{14, \ (1,1)}^{225}\equiv HHddd\bar{d}\bar{u}QQ\nu\,,\\
\mathcal{O}_{14, \ (1,1)}^{226}\equiv HHdddd\bar{Q}L\bar{e}\nu\,,\\
\mathcal{O}_{14, \ (1,1)}^{227}\equiv HHdddQ\bar{Q}e\bar{e}\nu\,,\\
\mathcal{O}_{14, \ (1,1)}^{228}\equiv HHddd\bar{u}QL\bar{e}\nu\,,\\
\mathcal{O}_{14, \ (1,1)}^{229}\equiv HHdd\bar{u}QQe\bar{e}\nu\,,\\
\mathcal{O}_{14, \ (1,1)}^{230}\equiv H\bar{H}ddd\bar{d}\bar{d}QQ\nu\,,\\
\mathcal{O}_{14, \ (1,1)}^{231}\equiv H\bar{H}dddd\bar{d}\bar{d}u\nu\,,\\
\mathcal{O}_{14, \ (1,1)}^{232}\equiv H\bar{H}d\bar{d}\bar{u}QQQQ\nu\,,\\
\mathcal{O}_{14, \ (1,1)}^{233}\equiv H\bar{H}dd\bar{d}u\bar{u}QQ\nu\,,\\
\mathcal{O}_{14, \ (1,1)}^{234}\equiv H\bar{H}ddd\bar{d}uu\bar{u}\nu\,,\\
\mathcal{O}_{14, \ (1,1)}^{235}\equiv H\bar{H}ddd\bar{d}QL\bar{e}\nu\,,\\
\mathcal{O}_{14, \ (1,1)}^{236}\equiv H\bar{H}dd\bar{d}QQe\bar{e}\nu\,,\\
\mathcal{O}_{14, \ (1,1)}^{237}\equiv H\bar{H}ddd\bar{d}ue\bar{e}\nu\,,\\
\mathcal{O}_{14, \ (1,1)}^{238}\equiv H\bar{H}dddLL\bar{e}\bar{e}\nu\,,\\
\mathcal{O}_{14, \ (1,1)}^{239}\equiv H\bar{H}ddQLe\bar{e}\bar{e}\nu\,,\\
\mathcal{O}_{14, \ (1,1)}^{240}\equiv H\bar{H}dQQee\bar{e}\bar{e}\nu\,,\\
\mathcal{O}_{14, \ (1,1)}^{241}\equiv H\bar{H}dduee\bar{e}\bar{e}\nu\,,\\
\mathcal{O}_{14, \ (1,1)}^{242}\equiv H\bar{H}ddQQ\bar{Q}L\bar{e}\nu\,,\\
\mathcal{O}_{14, \ (1,1)}^{243}\equiv H\bar{H}dQQQ\bar{Q}e\bar{e}\nu\,,\\
\mathcal{O}_{14, \ (1,1)}^{244}\equiv H\bar{H}dddu\bar{Q}L\bar{e}\nu\,,\\
\mathcal{O}_{14, \ (1,1)}^{245}\equiv H\bar{H}dduQ\bar{Q}e\bar{e}\nu\,,\\
\mathcal{O}_{14, \ (1,1)}^{246}\equiv H\bar{H}d\bar{u}QQQL\bar{e}\nu\,,\\
\mathcal{O}_{14, \ (1,1)}^{247}\equiv H\bar{H}\bar{u}QQQQe\bar{e}\nu\,,\\
\mathcal{O}_{14, \ (1,1)}^{248}\equiv H\bar{H}ddu\bar{u}QL\bar{e}\nu\,,\\
\mathcal{O}_{14, \ (1,1)}^{249}\equiv H\bar{H}du\bar{u}QQe\bar{e}\nu\,,\\
\mathcal{O}_{14, \ (1,1)}^{250}\equiv H\bar{H}dduu\bar{u}e\bar{e}\nu\,,\\
\mathcal{O}_{14, \ (1,1)}^{251}\equiv \bar{H}\bar{H}d\bar{d}\bar{d}QQQQ\nu\,,\\
\mathcal{O}_{14, \ (1,1)}^{252}\equiv \bar{H}\bar{H}dd\bar{d}\bar{d}uQQ\nu\,,\\
\mathcal{O}_{14, \ (1,1)}^{253}\equiv \bar{H}\bar{H}\bar{d}u\bar{u}QQQQ\nu\,,\\
\mathcal{O}_{14, \ (1,1)}^{254}\equiv \bar{H}\bar{H}d\bar{d}uu\bar{u}QQ\nu\,,\\
\mathcal{O}_{14, \ (1,1)}^{255}\equiv \bar{H}\bar{H}d\bar{d}QQQL\bar{e}\nu\,,\\
\mathcal{O}_{14, \ (1,1)}^{256}\equiv \bar{H}\bar{H}\bar{d}QQQQe\bar{e}\nu\,,\\
\mathcal{O}_{14, \ (1,1)}^{257}\equiv \bar{H}\bar{H}dd\bar{d}uQL\bar{e}\nu\,,\\
\mathcal{O}_{14, \ (1,1)}^{258}\equiv \bar{H}\bar{H}d\bar{d}uQQe\bar{e}\nu\,,\\
\mathcal{O}_{14, \ (1,1)}^{259}\equiv \bar{H}\bar{H}dQQLL\bar{e}\bar{e}\nu\,,\\
\mathcal{O}_{14, \ (1,1)}^{260}\equiv \bar{H}\bar{H}QQQLe\bar{e}\bar{e}\nu\,,\\
\mathcal{O}_{14, \ (1,1)}^{261}\equiv \bar{H}\bar{H}dduLL\bar{e}\bar{e}\nu\,,\\
\mathcal{O}_{14, \ (1,1)}^{262}\equiv \bar{H}\bar{H}duQLe\bar{e}\bar{e}\nu\,,\\
\mathcal{O}_{14, \ (1,1)}^{263}\equiv \bar{H}\bar{H}uQQee\bar{e}\bar{e}\nu\,,\\
\mathcal{O}_{14, \ (1,1)}^{264}\equiv \bar{H}\bar{H}QQQQ\bar{Q}L\bar{e}\nu\,,\\
\mathcal{O}_{14, \ (1,1)}^{265}\equiv \bar{H}\bar{H}duQQ\bar{Q}L\bar{e}\nu\,,\\
\mathcal{O}_{14, \ (1,1)}^{266}\equiv \bar{H}\bar{H}uQQQ\bar{Q}e\bar{e}\nu\,,\\
\mathcal{O}_{14, \ (1,1)}^{267}\equiv \bar{H}\bar{H}dduu\bar{Q}L\bar{e}\nu\,,\\
\mathcal{O}_{14, \ (1,1)}^{268}\equiv \bar{H}\bar{H}duuQ\bar{Q}e\bar{e}\nu\,,\\
\mathcal{O}_{14, \ (1,1)}^{269}\equiv \bar{H}\bar{H}u\bar{u}QQQL\bar{e}\nu\,,\\
\mathcal{O}_{14, \ (1,1)}^{270}\equiv \bar{H}\bar{H}duu\bar{u}QL\bar{e}\nu\,,\\
\mathcal{O}_{14, \ (1,1)}^{271}\equiv \bar{H}\bar{H}uu\bar{u}QQe\bar{e}\nu\,,\\
\mathcal{O}_{14, \ (1,1)}^{272}\equiv \bar{H}\bar{H}\bar{d}QQQQL\bar{L}\nu\,,\\
\mathcal{O}_{14, \ (1,1)}^{273}\equiv \bar{H}\bar{H}d\bar{d}uQQL\bar{L}\nu\,,\\
\mathcal{O}_{14, \ (1,1)}^{274}\equiv \bar{H}\bar{H}\bar{d}uQQQ\bar{L}e\nu\,,\\
\mathcal{O}_{14, \ (1,1)}^{275}\equiv \bar{H}\bar{H}dd\bar{d}uuL\bar{L}\nu\,,\\
\mathcal{O}_{14, \ (1,1)}^{276}\equiv \bar{H}\bar{H}d\bar{d}uuQ\bar{L}e\nu\,,\\
\mathcal{O}_{14, \ (1,1)}^{277}\equiv \bar{H}\bar{H}QQQLL\bar{L}\bar{e}\nu\,,\\
\mathcal{O}_{14, \ (1,1)}^{278}\equiv \bar{H}\bar{H}duQLL\bar{L}\bar{e}\nu\,,\\
\mathcal{O}_{14, \ (1,1)}^{279}\equiv \bar{H}\bar{H}uQQL\bar{L}e\bar{e}\nu\,,\\
\mathcal{O}_{14, \ (1,1)}^{280}\equiv \bar{H}\bar{H}duuL\bar{L}e\bar{e}\nu\,,\\
\mathcal{O}_{14, \ (1,1)}^{281}\equiv \bar{H}\bar{H}uuQ\bar{L}ee\bar{e}\nu\,,\\
\mathcal{O}_{14, \ (1,1)}^{282}\equiv \bar{H}\bar{H}uQQLL\bar{L}\bar{L}\nu\,,\\
\mathcal{O}_{14, \ (1,1)}^{283}\equiv \bar{H}\bar{H}duuLL\bar{L}\bar{L}\nu\,,\\
\mathcal{O}_{14, \ (1,1)}^{284}\equiv \bar{H}\bar{H}uuQL\bar{L}\bar{L}e\nu\,,\\
\mathcal{O}_{14, \ (1,1)}^{285}\equiv \bar{H}\bar{H}uuu\bar{L}\bar{L}ee\nu\,,\\
\mathcal{O}_{14, \ (1,1)}^{286}\equiv \bar{H}\bar{H}uQQQ\bar{Q}L\bar{L}\nu\,,\\
\mathcal{O}_{14, \ (1,1)}^{287}\equiv \bar{H}\bar{H}duuQ\bar{Q}L\bar{L}\nu\,,\\
\mathcal{O}_{14, \ (1,1)}^{288}\equiv \bar{H}\bar{H}uuQQ\bar{Q}\bar{L}e\nu\,,\\
\mathcal{O}_{14, \ (1,1)}^{289}\equiv \bar{H}\bar{H}duuu\bar{Q}\bar{L}e\nu\,,\\
\mathcal{O}_{14, \ (1,1)}^{290}\equiv \bar{H}\bar{H}uu\bar{u}QQL\bar{L}\nu\,,\\
\mathcal{O}_{14, \ (1,1)}^{291}\equiv \bar{H}\bar{H}duuu\bar{u}L\bar{L}\nu\,,\\
\mathcal{O}_{14, \ (1,1)}^{292}\equiv \bar{H}\bar{H}uuu\bar{u}Q\bar{L}e\nu\,,\\
\mathcal{O}_{14, \ (1,1)}^{293}\equiv \bar{H}\bar{H}\bar{d}QQQQQ\bar{Q}\nu\,,\\
\mathcal{O}_{14, \ (1,1)}^{294}\equiv \bar{H}\bar{H}d\bar{d}uQQQ\bar{Q}\nu\,,\\
\mathcal{O}_{14, \ (1,1)}^{295}\equiv \bar{H}\bar{H}dd\bar{d}uuQ\bar{Q}\nu\,,\\
\mathcal{O}_{14, \ (1,1)}^{296}\equiv \bar{H}\bar{H}uQQQQ\bar{Q}\bar{Q}\nu\,,\\
\mathcal{O}_{14, \ (1,1)}^{297}\equiv \bar{H}\bar{H}duuQQ\bar{Q}\bar{Q}\nu\,,\\
\mathcal{O}_{14, \ (1,1)}^{298}\equiv \bar{H}\bar{H}dduuu\bar{Q}\bar{Q}\nu\,,\\
\mathcal{O}_{14, \ (1,1)}^{299}\equiv \bar{H}\bar{H}uu\bar{u}QQQ\bar{Q}\nu\,,\\
\mathcal{O}_{14, \ (1,1)}^{300}\equiv \bar{H}\bar{H}duuu\bar{u}Q\bar{Q}\nu\,,\\
\mathcal{O}_{14, \ (1,1)}^{301}\equiv \bar{H}\bar{H}uuu\bar{u}\bar{u}QQ\nu\,,\\
\mathcal{O}_{14, \ (1,1)}^{302}\equiv H\bar{H}dd\bar{d}QQL\bar{L}\nu\,,\\
\mathcal{O}_{14, \ (1,1)}^{303}\equiv H\bar{H}d\bar{d}QQQ\bar{L}e\nu\,,\\
\mathcal{O}_{14, \ (1,1)}^{304}\equiv H\bar{H}ddd\bar{d}uL\bar{L}\nu\,,\\
\mathcal{O}_{14, \ (1,1)}^{305}\equiv H\bar{H}dd\bar{d}uQ\bar{L}e\nu\,,\\
\mathcal{O}_{14, \ (1,1)}^{306}\equiv H\bar{H}ddQLL\bar{L}\bar{e}\nu\,,\\
\mathcal{O}_{14, \ (1,1)}^{307}\equiv H\bar{H}dQQL\bar{L}e\bar{e}\nu\,,\\
\mathcal{O}_{14, \ (1,1)}^{308}\equiv H\bar{H}QQQ\bar{L}ee\bar{e}\nu\,,\\
\mathcal{O}_{14, \ (1,1)}^{309}\equiv H\bar{H}dduL\bar{L}e\bar{e}\nu\,,\\
\mathcal{O}_{14, \ (1,1)}^{310}\equiv H\bar{H}duQ\bar{L}ee\bar{e}\nu\,,\\
\mathcal{O}_{14, \ (1,1)}^{311}\equiv H\bar{H}dQQLL\bar{L}\bar{L}\nu\,,\\
\mathcal{O}_{14, \ (1,1)}^{312}\equiv H\bar{H}QQQL\bar{L}\bar{L}e\nu\,,\\
\mathcal{O}_{14, \ (1,1)}^{313}\equiv H\bar{H}dduLL\bar{L}\bar{L}\nu\,,\\
\mathcal{O}_{14, \ (1,1)}^{314}\equiv H\bar{H}duQL\bar{L}\bar{L}e\nu\,,\\
\mathcal{O}_{14, \ (1,1)}^{315}\equiv H\bar{H}uQQ\bar{L}\bar{L}ee\nu\,,\\
\mathcal{O}_{14, \ (1,1)}^{316}\equiv H\bar{H}duu\bar{L}\bar{L}ee\nu\,,\\
\mathcal{O}_{14, \ (1,1)}^{317}\equiv H\bar{H}dQQQ\bar{Q}L\bar{L}\nu\,,\\
\mathcal{O}_{14, \ (1,1)}^{318}\equiv H\bar{H}QQQQ\bar{Q}\bar{L}e\nu\,,\\
\mathcal{O}_{14, \ (1,1)}^{319}\equiv H\bar{H}dduQ\bar{Q}L\bar{L}\nu\,,\\
\mathcal{O}_{14, \ (1,1)}^{320}\equiv H\bar{H}duQQ\bar{Q}\bar{L}e\nu\,,\\
\mathcal{O}_{14, \ (1,1)}^{321}\equiv H\bar{H}dduu\bar{Q}\bar{L}e\nu\,,\\
\mathcal{O}_{14, \ (1,1)}^{322}\equiv H\bar{H}\bar{u}QQQQL\bar{L}\nu\,,\\
\mathcal{O}_{14, \ (1,1)}^{323}\equiv H\bar{H}du\bar{u}QQL\bar{L}\nu\,,\\
\mathcal{O}_{14, \ (1,1)}^{324}\equiv H\bar{H}u\bar{u}QQQ\bar{L}e\nu\,,\\
\mathcal{O}_{14, \ (1,1)}^{325}\equiv H\bar{H}dduu\bar{u}L\bar{L}\nu\,,\\
\mathcal{O}_{14, \ (1,1)}^{326}\equiv H\bar{H}duu\bar{u}Q\bar{L}e\nu\,,\\
\mathcal{O}_{14, \ (1,1)}^{327}\equiv H\bar{H}dd\bar{d}QQQ\bar{Q}\nu\,,\\
\mathcal{O}_{14, \ (1,1)}^{328}\equiv H\bar{H}ddd\bar{d}uQ\bar{Q}\nu\,,\\
\mathcal{O}_{14, \ (1,1)}^{329}\equiv H\bar{H}dQQQQ\bar{Q}\bar{Q}\nu\,,\\
\mathcal{O}_{14, \ (1,1)}^{330}\equiv H\bar{H}dduQQ\bar{Q}\bar{Q}\nu\,,\\
\mathcal{O}_{14, \ (1,1)}^{331}\equiv H\bar{H}ddduu\bar{Q}\bar{Q}\nu\,,\\
\mathcal{O}_{14, \ (1,1)}^{332}\equiv H\bar{H}\bar{u}QQQQQ\bar{Q}\nu\,,\\
\mathcal{O}_{14, \ (1,1)}^{333}\equiv H\bar{H}du\bar{u}QQQ\bar{Q}\nu\,,\\
\mathcal{O}_{14, \ (1,1)}^{334}\equiv H\bar{H}dduu\bar{u}Q\bar{Q}\nu\,,\\
\mathcal{O}_{14, \ (1,1)}^{335}\equiv H\bar{H}u\bar{u}\bar{u}QQQQ\nu\,,\\
\mathcal{O}_{14, \ (1,1)}^{336}\equiv H\bar{H}duu\bar{u}\bar{u}QQ\nu\,,\\
\mathcal{O}_{14, \ (1,1)}^{337}\equiv H\bar{H}dduuu\bar{u}\bar{u}\nu\,,\\
\mathcal{O}_{14, \ (1,1)}^{338}\equiv HHdddd\bar{d}L\bar{L}\nu\,,\\
\mathcal{O}_{14, \ (1,1)}^{339}\equiv HHddd\bar{d}Q\bar{L}e\nu\,,\\
\mathcal{O}_{14, \ (1,1)}^{340}\equiv HHdddL\bar{L}e\bar{e}\nu\,,\\
\mathcal{O}_{14, \ (1,1)}^{341}\equiv HHddQ\bar{L}ee\bar{e}\nu\,,\\
\mathcal{O}_{14, \ (1,1)}^{342}\equiv HHdddLL\bar{L}\bar{L}\nu\,,\\
\mathcal{O}_{14, \ (1,1)}^{343}\equiv HHddQL\bar{L}\bar{L}e\nu\,,\\
\mathcal{O}_{14, \ (1,1)}^{344}\equiv HHdQQ\bar{L}\bar{L}ee\nu\,,\\
\mathcal{O}_{14, \ (1,1)}^{345}\equiv HHddu\bar{L}\bar{L}ee\nu\,,\\
\mathcal{O}_{14, \ (1,1)}^{346}\equiv HHdddQ\bar{Q}L\bar{L}\nu\,,\\
\mathcal{O}_{14, \ (1,1)}^{347}\equiv HHddQQ\bar{Q}\bar{L}e\nu\,,\\
\mathcal{O}_{14, \ (1,1)}^{348}\equiv HHdddu\bar{Q}\bar{L}e\nu\,,\\
\mathcal{O}_{14, \ (1,1)}^{349}\equiv HHdd\bar{u}QQL\bar{L}\nu\,,\\
\mathcal{O}_{14, \ (1,1)}^{350}\equiv HHd\bar{u}QQQ\bar{L}e\nu\,,\\
\mathcal{O}_{14, \ (1,1)}^{351}\equiv HHdddu\bar{u}L\bar{L}\nu\,,\\
\mathcal{O}_{14, \ (1,1)}^{352}\equiv HHddu\bar{u}Q\bar{L}e\nu\,,\\
\mathcal{O}_{14, \ (1,1)}^{353}\equiv HHdd\bar{d}QQLL\bar{\nu}\,,\\
\mathcal{O}_{14, \ (1,1)}^{354}\equiv HHd\bar{d}QQQLe\bar{\nu}\,,\\
\mathcal{O}_{14, \ (1,1)}^{355}\equiv HH\bar{d}QQQQee\bar{\nu}\,,\\
\mathcal{O}_{14, \ (1,1)}^{356}\equiv HHddd\bar{d}uLL\bar{\nu}\,,\\
\mathcal{O}_{14, \ (1,1)}^{357}\equiv HHdd\bar{d}uQLe\bar{\nu}\,,\\
\mathcal{O}_{14, \ (1,1)}^{358}\equiv HHd\bar{d}uQQee\bar{\nu}\,,\\
\mathcal{O}_{14, \ (1,1)}^{359}\equiv HHddQLLL\bar{e}\bar{\nu}\,,\\
\mathcal{O}_{14, \ (1,1)}^{360}\equiv HHdQQLLe\bar{e}\bar{\nu}\,,\\
\mathcal{O}_{14, \ (1,1)}^{361}\equiv HHQQQLee\bar{e}\bar{\nu}\,,\\
\mathcal{O}_{14, \ (1,1)}^{362}\equiv HHdduLLe\bar{e}\bar{\nu}\,,\\
\mathcal{O}_{14, \ (1,1)}^{363}\equiv HHduQLee\bar{e}\bar{\nu}\,,\\
\mathcal{O}_{14, \ (1,1)}^{364}\equiv HHuQQeee\bar{e}\bar{\nu}\,,\\
\mathcal{O}_{14, \ (1,1)}^{365}\equiv H\bar{H}\bar{d}QQQQLL\bar{\nu}\,,\\
\mathcal{O}_{14, \ (1,1)}^{366}\equiv H\bar{H}d\bar{d}uQQLL\bar{\nu}\,,\\
\mathcal{O}_{14, \ (1,1)}^{367}\equiv H\bar{H}\bar{d}uQQQLe\bar{\nu}\,,\\
\mathcal{O}_{14, \ (1,1)}^{368}\equiv H\bar{H}dd\bar{d}uuLL\bar{\nu}\,,\\
\mathcal{O}_{14, \ (1,1)}^{369}\equiv H\bar{H}d\bar{d}uuQLe\bar{\nu}\,,\\
\mathcal{O}_{14, \ (1,1)}^{370}\equiv H\bar{H}\bar{d}uuQQee\bar{\nu}\,,\\
\mathcal{O}_{14, \ (1,1)}^{371}\equiv H\bar{H}d\bar{d}uuuee\bar{\nu}\,,\\
\mathcal{O}_{14, \ (1,1)}^{372}\equiv H\bar{H}QQQLLL\bar{e}\bar{\nu}\,,\\
\mathcal{O}_{14, \ (1,1)}^{373}\equiv H\bar{H}duQLLL\bar{e}\bar{\nu}\,,\\
\mathcal{O}_{14, \ (1,1)}^{374}\equiv H\bar{H}uQQLLe\bar{e}\bar{\nu}\,,\\
\mathcal{O}_{14, \ (1,1)}^{375}\equiv H\bar{H}duuLLe\bar{e}\bar{\nu}\,,\\
\mathcal{O}_{14, \ (1,1)}^{376}\equiv H\bar{H}uuQLee\bar{e}\bar{\nu}\,,\\
\mathcal{O}_{14, \ (1,1)}^{377}\equiv H\bar{H}uuueee\bar{e}\bar{\nu}\,,\\
\mathcal{O}_{14, \ (1,1)}^{378}\equiv \bar{H}\bar{H}\bar{d}uuQQLL\bar{\nu}\,,\\
\mathcal{O}_{14, \ (1,1)}^{379}\equiv \bar{H}\bar{H}d\bar{d}uuuLL\bar{\nu}\,,\\
\mathcal{O}_{14, \ (1,1)}^{380}\equiv \bar{H}\bar{H}\bar{d}uuuQLe\bar{\nu}\,,\\
\mathcal{O}_{14, \ (1,1)}^{381}\equiv \bar{H}\bar{H}uuQLLL\bar{e}\bar{\nu}\,,\\
\mathcal{O}_{14, \ (1,1)}^{382}\equiv \bar{H}\bar{H}uuuLLe\bar{e}\bar{\nu}\,,\\
\mathcal{O}_{14, \ (1,1)}^{383}\equiv \bar{H}\bar{H}uuuLLL\bar{L}\bar{\nu}\,,\\
\mathcal{O}_{14, \ (1,1)}^{384}\equiv \bar{H}\bar{H}uuuQ\bar{Q}LL\bar{\nu}\,,\\
\mathcal{O}_{14, \ (1,1)}^{385}\equiv \bar{H}\bar{H}uuuu\bar{Q}Le\bar{\nu}\,,\\
\mathcal{O}_{14, \ (1,1)}^{386}\equiv \bar{H}\bar{H}uuuu\bar{u}LL\bar{\nu}\,,\\
\mathcal{O}_{14, \ (1,1)}^{387}\equiv H\bar{H}uQQLLL\bar{L}\bar{\nu}\,,\\
\mathcal{O}_{14, \ (1,1)}^{388}\equiv H\bar{H}duuLLL\bar{L}\bar{\nu}\,,\\
\mathcal{O}_{14, \ (1,1)}^{389}\equiv H\bar{H}uuQLL\bar{L}e\bar{\nu}\,,\\
\mathcal{O}_{14, \ (1,1)}^{390}\equiv H\bar{H}uuuL\bar{L}ee\bar{\nu}\,,\\
\mathcal{O}_{14, \ (1,1)}^{391}\equiv H\bar{H}uQQQ\bar{Q}LL\bar{\nu}\,,\\
\mathcal{O}_{14, \ (1,1)}^{392}\equiv H\bar{H}duuQ\bar{Q}LL\bar{\nu}\,,\\
\mathcal{O}_{14, \ (1,1)}^{393}\equiv H\bar{H}uuQQ\bar{Q}Le\bar{\nu}\,,\\
\mathcal{O}_{14, \ (1,1)}^{394}\equiv H\bar{H}duuu\bar{Q}Le\bar{\nu}\,,\\
\mathcal{O}_{14, \ (1,1)}^{395}\equiv H\bar{H}uuuQ\bar{Q}ee\bar{\nu}\,,\\
\mathcal{O}_{14, \ (1,1)}^{396}\equiv H\bar{H}uu\bar{u}QQLL\bar{\nu}\,,\\
\mathcal{O}_{14, \ (1,1)}^{397}\equiv H\bar{H}duuu\bar{u}LL\bar{\nu}\,,\\
\mathcal{O}_{14, \ (1,1)}^{398}\equiv H\bar{H}uuu\bar{u}QLe\bar{\nu}\,,\\
\mathcal{O}_{14, \ (1,1)}^{399}\equiv H\bar{H}uuuu\bar{u}ee\bar{\nu}\,,\\
\mathcal{O}_{14, \ (1,1)}^{400}\equiv HHdQQLLL\bar{L}\bar{\nu}\,,\\
\mathcal{O}_{14, \ (1,1)}^{401}\equiv HHQQQLL\bar{L}e\bar{\nu}\,,\\
\mathcal{O}_{14, \ (1,1)}^{402}\equiv HHdduLLL\bar{L}\bar{\nu}\,,\\
\mathcal{O}_{14, \ (1,1)}^{403}\equiv HHduQLL\bar{L}e\bar{\nu}\,,\\
\mathcal{O}_{14, \ (1,1)}^{404}\equiv HHuQQL\bar{L}ee\bar{\nu}\,,\\
\mathcal{O}_{14, \ (1,1)}^{405}\equiv HHduuL\bar{L}ee\bar{\nu}\,,\\
\mathcal{O}_{14, \ (1,1)}^{406}\equiv HHuuQ\bar{L}eee\bar{\nu}\,,\\
\mathcal{O}_{14, \ (1,1)}^{407}\equiv HHdQQQ\bar{Q}LL\bar{\nu}\,,\\
\mathcal{O}_{14, \ (1,1)}^{408}\equiv HHQQQQ\bar{Q}Le\bar{\nu}\,,\\
\mathcal{O}_{14, \ (1,1)}^{409}\equiv HHdduQ\bar{Q}LL\bar{\nu}\,,\\
\mathcal{O}_{14, \ (1,1)}^{410}\equiv HHduQQ\bar{Q}Le\bar{\nu}\,,\\
\mathcal{O}_{14, \ (1,1)}^{411}\equiv HHuQQQ\bar{Q}ee\bar{\nu}\,,\\
\mathcal{O}_{14, \ (1,1)}^{412}\equiv HHdduu\bar{Q}Le\bar{\nu}\,,\\
\mathcal{O}_{14, \ (1,1)}^{413}\equiv HHduuQ\bar{Q}ee\bar{\nu}\,,\\
\mathcal{O}_{14, \ (1,1)}^{414}\equiv HH\bar{u}QQQQLL\bar{\nu}\,,\\
\mathcal{O}_{14, \ (1,1)}^{415}\equiv HHdu\bar{u}QQLL\bar{\nu}\,,\\
\mathcal{O}_{14, \ (1,1)}^{416}\equiv HHu\bar{u}QQQLe\bar{\nu}\,,\\
\mathcal{O}_{14, \ (1,1)}^{417}\equiv HHdduu\bar{u}LL\bar{\nu}\,,\\
\mathcal{O}_{14, \ (1,1)}^{418}\equiv HHduu\bar{u}QLe\bar{\nu}\,,\\
\mathcal{O}_{14, \ (1,1)}^{419}\equiv HHuu\bar{u}QQee\bar{\nu}\,,\\
\mathcal{O}_{14, \ (1,1)}^{420}\equiv HHdddd\bar{d}Q\bar{Q}\nu\,,\\
\mathcal{O}_{14, \ (1,1)}^{421}\equiv HHdddQQ\bar{Q}\bar{Q}\nu\,,\\
\mathcal{O}_{14, \ (1,1)}^{422}\equiv HHddddu\bar{Q}\bar{Q}\nu\,,\\
\mathcal{O}_{14, \ (1,1)}^{423}\equiv HHdd\bar{u}QQQ\bar{Q}\nu\,,\\
\mathcal{O}_{14, \ (1,1)}^{424}\equiv HHdddu\bar{u}Q\bar{Q}\nu\,,\\
\mathcal{O}_{14, \ (1,1)}^{425}\equiv HHd\bar{u}\bar{u}QQQQ\nu\,,\\
\mathcal{O}_{14, \ (1,1)}^{426}\equiv HHddu\bar{u}\bar{u}QQ\nu\,,\\
\mathcal{O}_{14, \ (1,1)}^{427}\equiv H\bar{H}dddd\bar{d}\bar{e}\nu\nu\,,\\
\mathcal{O}_{14, \ (1,1)}^{428}\equiv H\bar{H}ddde\bar{e}\bar{e}\nu\nu\,,\\
\mathcal{O}_{14, \ (1,1)}^{429}\equiv H\bar{H}dddQ\bar{Q}\bar{e}\nu\nu\,,\\
\mathcal{O}_{14, \ (1,1)}^{430}\equiv H\bar{H}dd\bar{u}QQ\bar{e}\nu\nu\,,\\
\mathcal{O}_{14, \ (1,1)}^{431}\equiv H\bar{H}dddu\bar{u}\bar{e}\nu\nu\,,\\
\mathcal{O}_{14, \ (1,1)}^{432}\equiv \bar{H}\bar{H}dd\bar{d}QQ\bar{e}\nu\nu\,,\\
\mathcal{O}_{14, \ (1,1)}^{433}\equiv \bar{H}\bar{H}ddQL\bar{e}\bar{e}\nu\nu\,,\\
\mathcal{O}_{14, \ (1,1)}^{434}\equiv \bar{H}\bar{H}dQQe\bar{e}\bar{e}\nu\nu\,,\\
\mathcal{O}_{14, \ (1,1)}^{435}\equiv \bar{H}\bar{H}dQQQ\bar{Q}\bar{e}\nu\nu\,,\\
\mathcal{O}_{14, \ (1,1)}^{436}\equiv \bar{H}\bar{H}dduQ\bar{Q}\bar{e}\nu\nu\,,\\
\mathcal{O}_{14, \ (1,1)}^{437}\equiv \bar{H}\bar{H}\bar{u}QQQQ\bar{e}\nu\nu\,,\\
\mathcal{O}_{14, \ (1,1)}^{438}\equiv \bar{H}\bar{H}du\bar{u}QQ\bar{e}\nu\nu\,,\\
\mathcal{O}_{14, \ (1,1)}^{439}\equiv \bar{H}\bar{H}d\bar{d}QQQ\bar{L}\nu\nu\,,\\
\mathcal{O}_{14, \ (1,1)}^{440}\equiv \bar{H}\bar{H}dd\bar{d}uQ\bar{L}\nu\nu\,,\\
\mathcal{O}_{14, \ (1,1)}^{441}\equiv \bar{H}\bar{H}dQQL\bar{L}\bar{e}\nu\nu\,,\\
\mathcal{O}_{14, \ (1,1)}^{442}\equiv \bar{H}\bar{H}QQQ\bar{L}e\bar{e}\nu\nu\,,\\
\mathcal{O}_{14, \ (1,1)}^{443}\equiv \bar{H}\bar{H}dduL\bar{L}\bar{e}\nu\nu\,,\\
\mathcal{O}_{14, \ (1,1)}^{444}\equiv \bar{H}\bar{H}duQ\bar{L}e\bar{e}\nu\nu\,,\\
\mathcal{O}_{14, \ (1,1)}^{445}\equiv \bar{H}\bar{H}QQQL\bar{L}\bar{L}\nu\nu\,,\\
\mathcal{O}_{14, \ (1,1)}^{446}\equiv \bar{H}\bar{H}duQL\bar{L}\bar{L}\nu\nu\,,\\
\mathcal{O}_{14, \ (1,1)}^{447}\equiv \bar{H}\bar{H}uQQ\bar{L}\bar{L}e\nu\nu\,,\\
\mathcal{O}_{14, \ (1,1)}^{448}\equiv \bar{H}\bar{H}duu\bar{L}\bar{L}e\nu\nu\,,\\
\mathcal{O}_{14, \ (1,1)}^{449}\equiv \bar{H}\bar{H}QQQQ\bar{Q}\bar{L}\nu\nu\,,\\
\mathcal{O}_{14, \ (1,1)}^{450}\equiv \bar{H}\bar{H}duQQ\bar{Q}\bar{L}\nu\nu\,,\\
\mathcal{O}_{14, \ (1,1)}^{451}\equiv \bar{H}\bar{H}dduu\bar{Q}\bar{L}\nu\nu\,,\\
\mathcal{O}_{14, \ (1,1)}^{452}\equiv \bar{H}\bar{H}u\bar{u}QQQ\bar{L}\nu\nu\,,\\
\mathcal{O}_{14, \ (1,1)}^{453}\equiv \bar{H}\bar{H}duu\bar{u}Q\bar{L}\nu\nu\,,\\
\mathcal{O}_{14, \ (1,1)}^{454}\equiv H\bar{H}ddd\bar{d}Q\bar{L}\nu\nu\,,\\
\mathcal{O}_{14, \ (1,1)}^{455}\equiv H\bar{H}dddL\bar{L}\bar{e}\nu\nu\,,\\
\mathcal{O}_{14, \ (1,1)}^{456}\equiv H\bar{H}ddQ\bar{L}e\bar{e}\nu\nu\,,\\
\mathcal{O}_{14, \ (1,1)}^{457}\equiv H\bar{H}ddQL\bar{L}\bar{L}\nu\nu\,,\\
\mathcal{O}_{14, \ (1,1)}^{458}\equiv H\bar{H}dQQ\bar{L}\bar{L}e\nu\nu\,,\\
\mathcal{O}_{14, \ (1,1)}^{459}\equiv H\bar{H}ddu\bar{L}\bar{L}e\nu\nu\,,\\
\mathcal{O}_{14, \ (1,1)}^{460}\equiv H\bar{H}ddQQ\bar{Q}\bar{L}\nu\nu\,,\\
\mathcal{O}_{14, \ (1,1)}^{461}\equiv H\bar{H}dddu\bar{Q}\bar{L}\nu\nu\,,\\
\mathcal{O}_{14, \ (1,1)}^{462}\equiv H\bar{H}d\bar{u}QQQ\bar{L}\nu\nu\,,\\
\mathcal{O}_{14, \ (1,1)}^{463}\equiv H\bar{H}ddu\bar{u}Q\bar{L}\nu\nu\,,\\
\mathcal{O}_{14, \ (1,1)}^{464}\equiv HHddd\bar{L}\bar{L}e\nu\nu\,,\\
\mathcal{O}_{14, \ (1,1)}^{465}\equiv HHdddd\bar{Q}\bar{L}\nu\nu\,,\\
\mathcal{O}_{14, \ (1,1)}^{466}\equiv HHddd\bar{u}Q\bar{L}\nu\nu\,,\\
\mathcal{O}_{14, \ (1,1)}^{467}\equiv HHddd\bar{d}QL\nu\bar{\nu}\,,\\
\mathcal{O}_{14, \ (1,1)}^{468}\equiv HHdd\bar{d}QQe\nu\bar{\nu}\,,\\
\mathcal{O}_{14, \ (1,1)}^{469}\equiv HHdddLL\bar{e}\nu\bar{\nu}\,,\\
\mathcal{O}_{14, \ (1,1)}^{470}\equiv HHddQLe\bar{e}\nu\bar{\nu}\,,\\
\mathcal{O}_{14, \ (1,1)}^{471}\equiv HHdQQee\bar{e}\nu\bar{\nu}\,,\\
\mathcal{O}_{14, \ (1,1)}^{472}\equiv H\bar{H}d\bar{d}QQQL\nu\bar{\nu}\,,\\
\mathcal{O}_{14, \ (1,1)}^{473}\equiv H\bar{H}\bar{d}QQQQe\nu\bar{\nu}\,,\\
\mathcal{O}_{14, \ (1,1)}^{474}\equiv H\bar{H}dd\bar{d}uQL\nu\bar{\nu}\,,\\
\mathcal{O}_{14, \ (1,1)}^{475}\equiv H\bar{H}d\bar{d}uQQe\nu\bar{\nu}\,,\\
\mathcal{O}_{14, \ (1,1)}^{476}\equiv H\bar{H}dd\bar{d}uue\nu\bar{\nu}\,,\\
\mathcal{O}_{14, \ (1,1)}^{477}\equiv H\bar{H}dQQLL\bar{e}\nu\bar{\nu}\,,\\
\mathcal{O}_{14, \ (1,1)}^{478}\equiv H\bar{H}QQQLe\bar{e}\nu\bar{\nu}\,,\\
\mathcal{O}_{14, \ (1,1)}^{479}\equiv H\bar{H}dduLL\bar{e}\nu\bar{\nu}\,,\\
\mathcal{O}_{14, \ (1,1)}^{480}\equiv H\bar{H}duQLe\bar{e}\nu\bar{\nu}\,,\\
\mathcal{O}_{14, \ (1,1)}^{481}\equiv H\bar{H}uQQee\bar{e}\nu\bar{\nu}\,,\\
\mathcal{O}_{14, \ (1,1)}^{482}\equiv H\bar{H}duuee\bar{e}\nu\bar{\nu}\,,\\
\mathcal{O}_{14, \ (1,1)}^{483}\equiv \bar{H}\bar{H}\bar{d}uQQQL\nu\bar{\nu}\,,\\
\mathcal{O}_{14, \ (1,1)}^{484}\equiv \bar{H}\bar{H}d\bar{d}uuQL\nu\bar{\nu}\,,\\
\mathcal{O}_{14, \ (1,1)}^{485}\equiv \bar{H}\bar{H}\bar{d}uuQQe\nu\bar{\nu}\,,\\
\mathcal{O}_{14, \ (1,1)}^{486}\equiv \bar{H}\bar{H}uQQLL\bar{e}\nu\bar{\nu}\,,\\
\mathcal{O}_{14, \ (1,1)}^{487}\equiv \bar{H}\bar{H}duuLL\bar{e}\nu\bar{\nu}\,,\\
\mathcal{O}_{14, \ (1,1)}^{488}\equiv \bar{H}\bar{H}uuQLe\bar{e}\nu\bar{\nu}\,,\\
\mathcal{O}_{14, \ (1,1)}^{489}\equiv \bar{H}\bar{H}uuQLL\bar{L}\nu\bar{\nu}\,,\\
\mathcal{O}_{14, \ (1,1)}^{490}\equiv \bar{H}\bar{H}uuuL\bar{L}e\nu\bar{\nu}\,,\\
\mathcal{O}_{14, \ (1,1)}^{491}\equiv \bar{H}\bar{H}uuQQ\bar{Q}L\nu\bar{\nu}\,,\\
\mathcal{O}_{14, \ (1,1)}^{492}\equiv \bar{H}\bar{H}duuu\bar{Q}L\nu\bar{\nu}\,,\\
\mathcal{O}_{14, \ (1,1)}^{493}\equiv \bar{H}\bar{H}uuuQ\bar{Q}e\nu\bar{\nu}\,,\\
\mathcal{O}_{14, \ (1,1)}^{494}\equiv \bar{H}\bar{H}uuu\bar{u}QL\nu\bar{\nu}\,,\\
\mathcal{O}_{14, \ (1,1)}^{495}\equiv H\bar{H}QQQLL\bar{L}\nu\bar{\nu}\,,\\
\mathcal{O}_{14, \ (1,1)}^{496}\equiv H\bar{H}duQLL\bar{L}\nu\bar{\nu}\,,\\
\mathcal{O}_{14, \ (1,1)}^{497}\equiv H\bar{H}uQQL\bar{L}e\nu\bar{\nu}\,,\\
\mathcal{O}_{14, \ (1,1)}^{498}\equiv H\bar{H}duuL\bar{L}e\nu\bar{\nu}\,,\\
\mathcal{O}_{14, \ (1,1)}^{499}\equiv H\bar{H}uuQ\bar{L}ee\nu\bar{\nu}\,,\\
\mathcal{O}_{14, \ (1,1)}^{500}\equiv H\bar{H}QQQQ\bar{Q}L\nu\bar{\nu}\,,\\
\mathcal{O}_{14, \ (1,1)}^{501}\equiv H\bar{H}duQQ\bar{Q}L\nu\bar{\nu}\,,\\
\mathcal{O}_{14, \ (1,1)}^{502}\equiv H\bar{H}uQQQ\bar{Q}e\nu\bar{\nu}\,,\\
\mathcal{O}_{14, \ (1,1)}^{503}\equiv H\bar{H}dduu\bar{Q}L\nu\bar{\nu}\,,\\
\mathcal{O}_{14, \ (1,1)}^{504}\equiv H\bar{H}duuQ\bar{Q}e\nu\bar{\nu}\,,\\
\mathcal{O}_{14, \ (1,1)}^{505}\equiv H\bar{H}u\bar{u}QQQL\nu\bar{\nu}\,,\\
\mathcal{O}_{14, \ (1,1)}^{506}\equiv H\bar{H}duu\bar{u}QL\nu\bar{\nu}\,,\\
\mathcal{O}_{14, \ (1,1)}^{507}\equiv H\bar{H}uu\bar{u}QQe\nu\bar{\nu}\,,\\
\mathcal{O}_{14, \ (1,1)}^{508}\equiv H\bar{H}duuu\bar{u}e\nu\bar{\nu}\,,\\
\mathcal{O}_{14, \ (1,1)}^{509}\equiv HHddQLL\bar{L}\nu\bar{\nu}\,,\\
\mathcal{O}_{14, \ (1,1)}^{510}\equiv HHdQQL\bar{L}e\nu\bar{\nu}\,,\\
\mathcal{O}_{14, \ (1,1)}^{511}\equiv HHQQQ\bar{L}ee\nu\bar{\nu}\,,\\
\mathcal{O}_{14, \ (1,1)}^{512}\equiv HHdduL\bar{L}e\nu\bar{\nu}\,,\\
\mathcal{O}_{14, \ (1,1)}^{513}\equiv HHduQ\bar{L}ee\nu\bar{\nu}\,,\\
\mathcal{O}_{14, \ (1,1)}^{514}\equiv H\bar{H}uuQLLL\bar{\nu}\bar{\nu}\,,\\
\mathcal{O}_{14, \ (1,1)}^{515}\equiv H\bar{H}uuuLLe\bar{\nu}\bar{\nu}\,,\\
\mathcal{O}_{14, \ (1,1)}^{516}\equiv HHQQQLLL\bar{\nu}\bar{\nu}\,,\\
\mathcal{O}_{14, \ (1,1)}^{517}\equiv HHduQLLL\bar{\nu}\bar{\nu}\,,\\
\mathcal{O}_{14, \ (1,1)}^{518}\equiv HHuQQLLe\bar{\nu}\bar{\nu}\,,\\
\mathcal{O}_{14, \ (1,1)}^{519}\equiv HHduuLLe\bar{\nu}\bar{\nu}\,,\\
\mathcal{O}_{14, \ (1,1)}^{520}\equiv HHuuQLee\bar{\nu}\bar{\nu}\,,\\
\mathcal{O}_{14, \ (1,1)}^{521}\equiv HHddQQ\bar{Q}L\nu\bar{\nu}\,,\\
\mathcal{O}_{14, \ (1,1)}^{522}\equiv HHdQQQ\bar{Q}e\nu\bar{\nu}\,,\\
\mathcal{O}_{14, \ (1,1)}^{523}\equiv HHdddu\bar{Q}L\nu\bar{\nu}\,,\\
\mathcal{O}_{14, \ (1,1)}^{524}\equiv HHdduQ\bar{Q}e\nu\bar{\nu}\,,\\
\mathcal{O}_{14, \ (1,1)}^{525}\equiv HHd\bar{u}QQQL\nu\bar{\nu}\,,\\
\mathcal{O}_{14, \ (1,1)}^{526}\equiv HH\bar{u}QQQQe\nu\bar{\nu}\,,\\
\mathcal{O}_{14, \ (1,1)}^{527}\equiv HHddu\bar{u}QL\nu\bar{\nu}\,,\\
\mathcal{O}_{14, \ (1,1)}^{528}\equiv HHdu\bar{u}QQe\nu\bar{\nu}\,,\\
\mathcal{O}_{14, \ (1,1)}^{529}\equiv \bar{H}\bar{H}ddQ\bar{L}\bar{e}\nu\nu\nu\,,\\
\mathcal{O}_{14, \ (1,1)}^{530}\equiv \bar{H}\bar{H}dQQ\bar{L}\bar{L}\nu\nu\nu\,,\\
\mathcal{O}_{14, \ (1,1)}^{531}\equiv \bar{H}\bar{H}ddu\bar{L}\bar{L}\nu\nu\nu\,,\\
\mathcal{O}_{14, \ (1,1)}^{532}\equiv H\bar{H}ddd\bar{L}\bar{L}\nu\nu\nu\,,\\
\mathcal{O}_{14, \ (1,1)}^{533}\equiv H\bar{H}dd\bar{d}QQ\nu\nu\bar{\nu}\,,\\
\mathcal{O}_{14, \ (1,1)}^{534}\equiv H\bar{H}ddd\bar{d}u\nu\nu\bar{\nu}\,,\\
\mathcal{O}_{14, \ (1,1)}^{535}\equiv H\bar{H}ddQL\bar{e}\nu\nu\bar{\nu}\,,\\
\mathcal{O}_{14, \ (1,1)}^{536}\equiv H\bar{H}dQQe\bar{e}\nu\nu\bar{\nu}\,,\\
\mathcal{O}_{14, \ (1,1)}^{537}\equiv H\bar{H}ddue\bar{e}\nu\nu\bar{\nu}\,,\\
\mathcal{O}_{14, \ (1,1)}^{538}\equiv \bar{H}\bar{H}\bar{d}QQQQ\nu\nu\bar{\nu}\,,\\
\mathcal{O}_{14, \ (1,1)}^{539}\equiv \bar{H}\bar{H}d\bar{d}uQQ\nu\nu\bar{\nu}\,,\\
\mathcal{O}_{14, \ (1,1)}^{540}\equiv \bar{H}\bar{H}QQQL\bar{e}\nu\nu\bar{\nu}\,,\\
\mathcal{O}_{14, \ (1,1)}^{541}\equiv \bar{H}\bar{H}duQL\bar{e}\nu\nu\bar{\nu}\,,\\
\mathcal{O}_{14, \ (1,1)}^{542}\equiv \bar{H}\bar{H}uQQe\bar{e}\nu\nu\bar{\nu}\,,\\
\mathcal{O}_{14, \ (1,1)}^{543}\equiv \bar{H}\bar{H}uQQL\bar{L}\nu\nu\bar{\nu}\,,\\
\mathcal{O}_{14, \ (1,1)}^{544}\equiv \bar{H}\bar{H}duuL\bar{L}\nu\nu\bar{\nu}\,,\\
\mathcal{O}_{14, \ (1,1)}^{545}\equiv \bar{H}\bar{H}uuQ\bar{L}e\nu\nu\bar{\nu}\,,\\
\mathcal{O}_{14, \ (1,1)}^{546}\equiv \bar{H}\bar{H}uQQQ\bar{Q}\nu\nu\bar{\nu}\,,\\
\mathcal{O}_{14, \ (1,1)}^{547}\equiv \bar{H}\bar{H}duuQ\bar{Q}\nu\nu\bar{\nu}\,,\\
\mathcal{O}_{14, \ (1,1)}^{548}\equiv \bar{H}\bar{H}uu\bar{u}QQ\nu\nu\bar{\nu}\,,\\
\mathcal{O}_{14, \ (1,1)}^{549}\equiv H\bar{H}dQQL\bar{L}\nu\nu\bar{\nu}\,,\\
\mathcal{O}_{14, \ (1,1)}^{550}\equiv H\bar{H}QQQ\bar{L}e\nu\nu\bar{\nu}\,,\\
\mathcal{O}_{14, \ (1,1)}^{551}\equiv H\bar{H}dduL\bar{L}\nu\nu\bar{\nu}\,,\\
\mathcal{O}_{14, \ (1,1)}^{552}\equiv H\bar{H}duQ\bar{L}e\nu\nu\bar{\nu}\,,\\
\mathcal{O}_{14, \ (1,1)}^{553}\equiv H\bar{H}dQQQ\bar{Q}\nu\nu\bar{\nu}\,,\\
\mathcal{O}_{14, \ (1,1)}^{554}\equiv H\bar{H}dduQ\bar{Q}\nu\nu\bar{\nu}\,,\\
\mathcal{O}_{14, \ (1,1)}^{555}\equiv H\bar{H}\bar{u}QQQQ\nu\nu\bar{\nu}\,,\\
\mathcal{O}_{14, \ (1,1)}^{556}\equiv H\bar{H}du\bar{u}QQ\nu\nu\bar{\nu}\,,\\
\mathcal{O}_{14, \ (1,1)}^{557}\equiv H\bar{H}dduu\bar{u}\nu\nu\bar{\nu}\,,\\
\mathcal{O}_{14, \ (1,1)}^{558}\equiv HHdddL\bar{L}\nu\nu\bar{\nu}\,,\\
\mathcal{O}_{14, \ (1,1)}^{559}\equiv HHddQ\bar{L}e\nu\nu\bar{\nu}\,,\\
\mathcal{O}_{14, \ (1,1)}^{560}\equiv HHdQQLL\nu\bar{\nu}\bar{\nu}\,,\\
\mathcal{O}_{14, \ (1,1)}^{561}\equiv \bar{H}\bar{H}uuuLL\nu\bar{\nu}\bar{\nu}\,,\\
\mathcal{O}_{14, \ (1,1)}^{562}\equiv H\bar{H}uQQLL\nu\bar{\nu}\bar{\nu}\,,\\
\mathcal{O}_{14, \ (1,1)}^{563}\equiv H\bar{H}duuLL\nu\bar{\nu}\bar{\nu}\,,\\
\mathcal{O}_{14, \ (1,1)}^{564}\equiv H\bar{H}uuQLe\nu\bar{\nu}\bar{\nu}\,,\\
\mathcal{O}_{14, \ (1,1)}^{565}\equiv H\bar{H}uuuee\nu\bar{\nu}\bar{\nu}\,,\\
\mathcal{O}_{14, \ (1,1)}^{566}\equiv HHQQQLe\nu\bar{\nu}\bar{\nu}\,,\\
\mathcal{O}_{14, \ (1,1)}^{567}\equiv HHdduLL\nu\bar{\nu}\bar{\nu}\,,\\
\mathcal{O}_{14, \ (1,1)}^{568}\equiv HHduQLe\nu\bar{\nu}\bar{\nu}\,,\\
\mathcal{O}_{14, \ (1,1)}^{569}\equiv HHuQQee\nu\bar{\nu}\bar{\nu}\,,\\
\mathcal{O}_{14, \ (1,1)}^{570}\equiv HHdddQ\bar{Q}\nu\nu\bar{\nu}\,,\\
\mathcal{O}_{14, \ (1,1)}^{571}\equiv HHdd\bar{u}QQ\nu\nu\bar{\nu}\,,\\
\mathcal{O}_{14, \ (1,1)}^{572}\equiv H\bar{H}ddd\bar{e}\nu\nu\nu\bar{\nu}\,,\\
\mathcal{O}_{14, \ (1,1)}^{573}\equiv \bar{H}\bar{H}dQQ\bar{e}\nu\nu\nu\bar{\nu}\,,\\
\mathcal{O}_{14, \ (1,1)}^{574}\equiv \bar{H}\bar{H}QQQ\bar{L}\nu\nu\nu\bar{\nu}\,,\\
\mathcal{O}_{14, \ (1,1)}^{575}\equiv \bar{H}\bar{H}duQ\bar{L}\nu\nu\nu\bar{\nu}\,,\\
\mathcal{O}_{14, \ (1,1)}^{576}\equiv H\bar{H}ddQ\bar{L}\nu\nu\nu\bar{\nu}\,,\\
\mathcal{O}_{14, \ (1,1)}^{577}\equiv HHddQL\nu\nu\bar{\nu}\bar{\nu}\,,\\
\mathcal{O}_{14, \ (1,1)}^{578}\equiv HHdQQe\nu\nu\bar{\nu}\bar{\nu}\,,\\
\mathcal{O}_{14, \ (1,1)}^{579}\equiv \bar{H}\bar{H}uuQL\nu\nu\bar{\nu}\bar{\nu}\,,\\
\mathcal{O}_{14, \ (1,1)}^{580}\equiv H\bar{H}QQQL\nu\nu\bar{\nu}\bar{\nu}\,,\\
\mathcal{O}_{14, \ (1,1)}^{581}\equiv H\bar{H}duQL\nu\nu\bar{\nu}\bar{\nu}\,,\\
\mathcal{O}_{14, \ (1,1)}^{582}\equiv H\bar{H}uQQe\nu\nu\bar{\nu}\bar{\nu}\,,\\
\mathcal{O}_{14, \ (1,1)}^{583}\equiv H\bar{H}duue\nu\nu\bar{\nu}\bar{\nu}\,,\\
\mathcal{O}_{14, \ (1,1)}^{584}\equiv H\bar{H}dQQ\nu\nu\nu\bar{\nu}\bar{\nu}\,,\\
\mathcal{O}_{14, \ (1,1)}^{585}\equiv \bar{H}\bar{H}uQQ\nu\nu\nu\bar{\nu}\bar{\nu}\,,\\
\mathcal{O}_{14, \ (1,1)}^{586}\equiv H\bar{H}ddu\nu\nu\nu\bar{\nu}\bar{\nu}\,,\\
\mathcal{O}_{14, \ (1,1)}^{587}\equiv HHH\bar{H}\bar{H}dd\bar{d}QQL\,,\\
\mathcal{O}_{14, \ (1,1)}^{588}\equiv HHH\bar{H}\bar{H}d\bar{d}QQQe\,,\\
\mathcal{O}_{14, \ (1,1)}^{589}\equiv HHH\bar{H}\bar{H}ddd\bar{d}uL\,,\\
\mathcal{O}_{14, \ (1,1)}^{590}\equiv HHH\bar{H}\bar{H}dd\bar{d}uQe\,,\\
\mathcal{O}_{14, \ (1,1)}^{591}\equiv HHH\bar{H}\bar{H}ddQLL\bar{e}\,,\\
\mathcal{O}_{14, \ (1,1)}^{592}\equiv HHH\bar{H}\bar{H}dQQLe\bar{e}\,,\\
\mathcal{O}_{14, \ (1,1)}^{593}\equiv HHH\bar{H}\bar{H}QQQee\bar{e}\,,\\
\mathcal{O}_{14, \ (1,1)}^{594}\equiv HHH\bar{H}\bar{H}dduLe\bar{e}\,,\\
\mathcal{O}_{14, \ (1,1)}^{595}\equiv HHH\bar{H}\bar{H}duQee\bar{e}\,,\\
\mathcal{O}_{14, \ (1,1)}^{596}\equiv HH\bar{H}\bar{H}\bar{H}\bar{d}QQQQL\,,\\
\mathcal{O}_{14, \ (1,1)}^{597}\equiv HH\bar{H}\bar{H}\bar{H}d\bar{d}uQQL\,,\\
\mathcal{O}_{14, \ (1,1)}^{598}\equiv HH\bar{H}\bar{H}\bar{H}\bar{d}uQQQe\,,\\
\mathcal{O}_{14, \ (1,1)}^{599}\equiv HH\bar{H}\bar{H}\bar{H}dd\bar{d}uuL\,,\\
\mathcal{O}_{14, \ (1,1)}^{600}\equiv HH\bar{H}\bar{H}\bar{H}d\bar{d}uuQe\,,\\
\mathcal{O}_{14, \ (1,1)}^{601}\equiv HH\bar{H}\bar{H}\bar{H}QQQLL\bar{e}\,,\\
\mathcal{O}_{14, \ (1,1)}^{602}\equiv HH\bar{H}\bar{H}\bar{H}duQLL\bar{e}\,,\\
\mathcal{O}_{14, \ (1,1)}^{603}\equiv HH\bar{H}\bar{H}\bar{H}uQQLe\bar{e}\,,\\
\mathcal{O}_{14, \ (1,1)}^{604}\equiv HH\bar{H}\bar{H}\bar{H}duuLe\bar{e}\,,\\
\mathcal{O}_{14, \ (1,1)}^{605}\equiv HH\bar{H}\bar{H}\bar{H}uuQee\bar{e}\,,\\
\mathcal{O}_{14, \ (1,1)}^{606}\equiv H\bar{H}\bar{H}\bar{H}\bar{H}\bar{d}uuQQL\,,\\
\mathcal{O}_{14, \ (1,1)}^{607}\equiv H\bar{H}\bar{H}\bar{H}\bar{H}uuQLL\bar{e}\,,\\
\mathcal{O}_{14, \ (1,1)}^{608}\equiv H\bar{H}\bar{H}\bar{H}\bar{H}uuuLL\bar{L}\,,\\
\mathcal{O}_{14, \ (1,1)}^{609}\equiv H\bar{H}\bar{H}\bar{H}\bar{H}uuuQ\bar{Q}L\,,\\
\mathcal{O}_{14, \ (1,1)}^{610}\equiv HH\bar{H}\bar{H}\bar{H}uQQLL\bar{L}\,,\\
\mathcal{O}_{14, \ (1,1)}^{611}\equiv HH\bar{H}\bar{H}\bar{H}duuLL\bar{L}\,,\\
\mathcal{O}_{14, \ (1,1)}^{612}\equiv HH\bar{H}\bar{H}\bar{H}uuQL\bar{L}e\,,\\
\mathcal{O}_{14, \ (1,1)}^{613}\equiv HH\bar{H}\bar{H}\bar{H}uuu\bar{L}ee\,,\\
\mathcal{O}_{14, \ (1,1)}^{614}\equiv HH\bar{H}\bar{H}\bar{H}uQQQ\bar{Q}L\,,\\
\mathcal{O}_{14, \ (1,1)}^{615}\equiv HH\bar{H}\bar{H}\bar{H}duuQ\bar{Q}L\,,\\
\mathcal{O}_{14, \ (1,1)}^{616}\equiv HH\bar{H}\bar{H}\bar{H}uuQQ\bar{Q}e\,,\\
\mathcal{O}_{14, \ (1,1)}^{617}\equiv HH\bar{H}\bar{H}\bar{H}duuu\bar{Q}e\,,\\
\mathcal{O}_{14, \ (1,1)}^{618}\equiv HH\bar{H}\bar{H}\bar{H}uu\bar{u}QQL\,,\\
\mathcal{O}_{14, \ (1,1)}^{619}\equiv HH\bar{H}\bar{H}\bar{H}duuu\bar{u}L\,,\\
\mathcal{O}_{14, \ (1,1)}^{620}\equiv HH\bar{H}\bar{H}\bar{H}uuu\bar{u}Qe\,,\\
\mathcal{O}_{14, \ (1,1)}^{621}\equiv HHH\bar{H}\bar{H}dQQLL\bar{L}\,,\\
\mathcal{O}_{14, \ (1,1)}^{622}\equiv HHH\bar{H}\bar{H}QQQL\bar{L}e\,,\\
\mathcal{O}_{14, \ (1,1)}^{623}\equiv HHH\bar{H}\bar{H}dduLL\bar{L}\,,\\
\mathcal{O}_{14, \ (1,1)}^{624}\equiv HHH\bar{H}\bar{H}duQL\bar{L}e\,,\\
\mathcal{O}_{14, \ (1,1)}^{625}\equiv HHH\bar{H}\bar{H}uQQ\bar{L}ee\,,\\
\mathcal{O}_{14, \ (1,1)}^{626}\equiv HHH\bar{H}\bar{H}duu\bar{L}ee\,,\\
\mathcal{O}_{14, \ (1,1)}^{627}\equiv HHH\bar{H}\bar{H}dQQQ\bar{Q}L\,,\\
\mathcal{O}_{14, \ (1,1)}^{628}\equiv HHH\bar{H}\bar{H}QQQQ\bar{Q}e\,,\\
\mathcal{O}_{14, \ (1,1)}^{629}\equiv HHH\bar{H}\bar{H}dduQ\bar{Q}L\,,\\
\mathcal{O}_{14, \ (1,1)}^{630}\equiv HHH\bar{H}\bar{H}duQQ\bar{Q}e\,,\\
\mathcal{O}_{14, \ (1,1)}^{631}\equiv HHH\bar{H}\bar{H}dduu\bar{Q}e\,,\\
\mathcal{O}_{14, \ (1,1)}^{632}\equiv HHH\bar{H}\bar{H}\bar{u}QQQQL\,,\\
\mathcal{O}_{14, \ (1,1)}^{633}\equiv HHH\bar{H}\bar{H}du\bar{u}QQL\,,\\
\mathcal{O}_{14, \ (1,1)}^{634}\equiv HHH\bar{H}\bar{H}u\bar{u}QQQe\,,\\
\mathcal{O}_{14, \ (1,1)}^{635}\equiv HHH\bar{H}\bar{H}dduu\bar{u}L\,,\\
\mathcal{O}_{14, \ (1,1)}^{636}\equiv HHH\bar{H}\bar{H}duu\bar{u}Qe\,,\\
\mathcal{O}_{14, \ (1,1)}^{637}\equiv HHHH\bar{H}dddLL\bar{L}\,,\\
\mathcal{O}_{14, \ (1,1)}^{638}\equiv HHHH\bar{H}ddQL\bar{L}e\,,\\
\mathcal{O}_{14, \ (1,1)}^{639}\equiv HHHH\bar{H}dQQ\bar{L}ee\,,\\
\mathcal{O}_{14, \ (1,1)}^{640}\equiv HHHH\bar{H}dddQ\bar{Q}L\,,\\
\mathcal{O}_{14, \ (1,1)}^{641}\equiv HHHH\bar{H}ddQQ\bar{Q}e\,,\\
\mathcal{O}_{14, \ (1,1)}^{642}\equiv HHHH\bar{H}dd\bar{u}QQL\,,\\
\mathcal{O}_{14, \ (1,1)}^{643}\equiv HHHH\bar{H}d\bar{u}QQQe\,,\\
\mathcal{O}_{14, \ (1,1)}^{644}\equiv HHH\bar{H}\bar{H}ddd\bar{d}Q\nu\,,\\
\mathcal{O}_{14, \ (1,1)}^{645}\equiv HHH\bar{H}\bar{H}dddL\bar{e}\nu\,,\\
\mathcal{O}_{14, \ (1,1)}^{646}\equiv HHH\bar{H}\bar{H}ddQe\bar{e}\nu\,,\\
\mathcal{O}_{14, \ (1,1)}^{647}\equiv HH\bar{H}\bar{H}\bar{H}d\bar{d}QQQ\nu\,,\\
\mathcal{O}_{14, \ (1,1)}^{648}\equiv HH\bar{H}\bar{H}\bar{H}dd\bar{d}uQ\nu\,,\\
\mathcal{O}_{14, \ (1,1)}^{649}\equiv HH\bar{H}\bar{H}\bar{H}dQQL\bar{e}\nu\,,\\
\mathcal{O}_{14, \ (1,1)}^{650}\equiv HH\bar{H}\bar{H}\bar{H}QQQe\bar{e}\nu\,,\\
\mathcal{O}_{14, \ (1,1)}^{651}\equiv HH\bar{H}\bar{H}\bar{H}dduL\bar{e}\nu\,,\\
\mathcal{O}_{14, \ (1,1)}^{652}\equiv HH\bar{H}\bar{H}\bar{H}duQe\bar{e}\nu\,,\\
\mathcal{O}_{14, \ (1,1)}^{653}\equiv H\bar{H}\bar{H}\bar{H}\bar{H}\bar{d}uQQQ\nu\,,\\
\mathcal{O}_{14, \ (1,1)}^{654}\equiv H\bar{H}\bar{H}\bar{H}\bar{H}uQQL\bar{e}\nu\,,\\
\mathcal{O}_{14, \ (1,1)}^{655}\equiv H\bar{H}\bar{H}\bar{H}\bar{H}uuQL\bar{L}\nu\,,\\
\mathcal{O}_{14, \ (1,1)}^{656}\equiv H\bar{H}\bar{H}\bar{H}\bar{H}uuQQ\bar{Q}\nu\,,\\
\mathcal{O}_{14, \ (1,1)}^{657}\equiv HH\bar{H}\bar{H}\bar{H}QQQL\bar{L}\nu\,,\\
\mathcal{O}_{14, \ (1,1)}^{658}\equiv HH\bar{H}\bar{H}\bar{H}duQL\bar{L}\nu\,,\\
\mathcal{O}_{14, \ (1,1)}^{659}\equiv HH\bar{H}\bar{H}\bar{H}uQQ\bar{L}e\nu\,,\\
\mathcal{O}_{14, \ (1,1)}^{660}\equiv HH\bar{H}\bar{H}\bar{H}duu\bar{L}e\nu\,,\\
\mathcal{O}_{14, \ (1,1)}^{661}\equiv HH\bar{H}\bar{H}\bar{H}QQQQ\bar{Q}\nu\,,\\
\mathcal{O}_{14, \ (1,1)}^{662}\equiv HH\bar{H}\bar{H}\bar{H}duQQ\bar{Q}\nu\,,\\
\mathcal{O}_{14, \ (1,1)}^{663}\equiv HH\bar{H}\bar{H}\bar{H}dduu\bar{Q}\nu\,,\\
\mathcal{O}_{14, \ (1,1)}^{664}\equiv HH\bar{H}\bar{H}\bar{H}u\bar{u}QQQ\nu\,,\\
\mathcal{O}_{14, \ (1,1)}^{665}\equiv HH\bar{H}\bar{H}\bar{H}duu\bar{u}Q\nu\,,\\
\mathcal{O}_{14, \ (1,1)}^{666}\equiv HHH\bar{H}\bar{H}ddQL\bar{L}\nu\,,\\
\mathcal{O}_{14, \ (1,1)}^{667}\equiv HHH\bar{H}\bar{H}dQQ\bar{L}e\nu\,,\\
\mathcal{O}_{14, \ (1,1)}^{668}\equiv HHH\bar{H}\bar{H}ddu\bar{L}e\nu\,,\\
\mathcal{O}_{14, \ (1,1)}^{669}\equiv HHH\bar{H}\bar{H}ddQQ\bar{Q}\nu\,,\\
\mathcal{O}_{14, \ (1,1)}^{670}\equiv HHH\bar{H}\bar{H}dddu\bar{Q}\nu\,,\\
\mathcal{O}_{14, \ (1,1)}^{671}\equiv HHH\bar{H}\bar{H}d\bar{u}QQQ\nu\,,\\
\mathcal{O}_{14, \ (1,1)}^{672}\equiv HHH\bar{H}\bar{H}ddu\bar{u}Q\nu\,,\\
\mathcal{O}_{14, \ (1,1)}^{673}\equiv HHHH\bar{H}ddQLL\bar{\nu}\,,\\
\mathcal{O}_{14, \ (1,1)}^{674}\equiv HHHH\bar{H}dQQLe\bar{\nu}\,,\\
\mathcal{O}_{14, \ (1,1)}^{675}\equiv HH\bar{H}\bar{H}\bar{H}uuQLL\bar{\nu}\,,\\
\mathcal{O}_{14, \ (1,1)}^{676}\equiv HH\bar{H}\bar{H}\bar{H}uuuLe\bar{\nu}\,,\\
\mathcal{O}_{14, \ (1,1)}^{677}\equiv HHH\bar{H}\bar{H}QQQLL\bar{\nu}\,,\\
\mathcal{O}_{14, \ (1,1)}^{678}\equiv HHH\bar{H}\bar{H}duQLL\bar{\nu}\,,\\
\mathcal{O}_{14, \ (1,1)}^{679}\equiv HHH\bar{H}\bar{H}uQQLe\bar{\nu}\,,\\
\mathcal{O}_{14, \ (1,1)}^{680}\equiv HHH\bar{H}\bar{H}duuLe\bar{\nu}\,,\\
\mathcal{O}_{14, \ (1,1)}^{681}\equiv HHH\bar{H}\bar{H}uuQee\bar{\nu}\,,\\
\mathcal{O}_{14, \ (1,1)}^{682}\equiv HHHH\bar{H}QQQee\bar{\nu}\,,\\
\mathcal{O}_{14, \ (1,1)}^{683}\equiv HH\bar{H}\bar{H}\bar{H}ddQ\bar{e}\nu\nu\,,\\
\mathcal{O}_{14, \ (1,1)}^{684}\equiv H\bar{H}\bar{H}\bar{H}\bar{H}QQQ\bar{e}\nu\nu\,,\\
\mathcal{O}_{14, \ (1,1)}^{685}\equiv H\bar{H}\bar{H}\bar{H}\bar{H}uQQ\bar{L}\nu\nu\,,\\
\mathcal{O}_{14, \ (1,1)}^{686}\equiv HH\bar{H}\bar{H}\bar{H}dQQ\bar{L}\nu\nu\,,\\
\mathcal{O}_{14, \ (1,1)}^{687}\equiv HH\bar{H}\bar{H}\bar{H}ddu\bar{L}\nu\nu\,,\\
\mathcal{O}_{14, \ (1,1)}^{688}\equiv HHH\bar{H}\bar{H}ddd\bar{L}\nu\nu\,,\\
\mathcal{O}_{14, \ (1,1)}^{689}\equiv HHH\bar{H}\bar{H}dQQL\nu\bar{\nu}\,,\\
\mathcal{O}_{14, \ (1,1)}^{690}\equiv HH\bar{H}\bar{H}\bar{H}uQQL\nu\bar{\nu}\,,\\
\mathcal{O}_{14, \ (1,1)}^{691}\equiv HH\bar{H}\bar{H}\bar{H}duuL\nu\bar{\nu}\,,\\
\mathcal{O}_{14, \ (1,1)}^{692}\equiv HH\bar{H}\bar{H}\bar{H}uuQe\nu\bar{\nu}\,,\\
\mathcal{O}_{14, \ (1,1)}^{693}\equiv HHH\bar{H}\bar{H}QQQe\nu\bar{\nu}\,,\\
\mathcal{O}_{14, \ (1,1)}^{694}\equiv HHH\bar{H}\bar{H}dduL\nu\bar{\nu}\,,\\
\mathcal{O}_{14, \ (1,1)}^{695}\equiv HHH\bar{H}\bar{H}duQe\nu\bar{\nu}\,,\\
\mathcal{O}_{14, \ (1,1)}^{696}\equiv HHH\bar{H}\bar{H}ddQ\nu\nu\bar{\nu}\,,\\
\mathcal{O}_{14, \ (1,1)}^{697}\equiv HH\bar{H}\bar{H}\bar{H}QQQ\nu\nu\bar{\nu}\,,\\
\mathcal{O}_{14, \ (1,1)}^{698}\equiv HH\bar{H}\bar{H}\bar{H}duQ\nu\nu\bar{\nu}\,,\\
\mathcal{O}_{14, \ (1,1)}^{699}\equiv HHHHH\bar{H}\bar{H}\bar{H}ddQL\,,\\
\mathcal{O}_{14, \ (1,1)}^{700}\equiv HHHHH\bar{H}\bar{H}\bar{H}dQQe\,,\\
\mathcal{O}_{14, \ (1,1)}^{701}\equiv HHH\bar{H}\bar{H}\bar{H}\bar{H}\bar{H}uuQL\,,\\
\mathcal{O}_{14, \ (1,1)}^{702}\equiv HHHH\bar{H}\bar{H}\bar{H}\bar{H}QQQL\,,\\
\mathcal{O}_{14, \ (1,1)}^{703}\equiv HHHH\bar{H}\bar{H}\bar{H}\bar{H}duQL\,,\\
\mathcal{O}_{14, \ (1,1)}^{704}\equiv HHHH\bar{H}\bar{H}\bar{H}\bar{H}uQQe\,,\\
\mathcal{O}_{14, \ (1,1)}^{705}\equiv HHHH\bar{H}\bar{H}\bar{H}\bar{H}duue\,,\\
\mathcal{O}_{14, \ (1,1)}^{706}\equiv HHHH\bar{H}\bar{H}\bar{H}\bar{H}dQQ\nu\,,\\
\mathcal{O}_{14, \ (1,1)}^{707}\equiv HHH\bar{H}\bar{H}\bar{H}\bar{H}\bar{H}uQQ\nu\,,\\
\mathcal{O}_{14, \ (1,1)}^{708}\equiv HHHH\bar{H}\bar{H}\bar{H}\bar{H}ddu\nu\,.
 \label{eq:dequal14basisB1L1}
 \end{math}
 \end{spacing}
 \end{multicols}

 \subsection{\texorpdfstring{$\Delta B=1, \Delta L=5$}{Delta B=1,Delta L=5}}
 \begin{multicols}{3}
 \begin{spacing}{1.5}
 \noindent
 \begin{math}
\mathcal{O}_{14, \ (1,5)}^{1}\equiv HHuuQLLLLL\,,\\
\mathcal{O}_{14, \ (1,5)}^{2}\equiv HHuuuLLLLe\,,\\
\mathcal{O}_{14, \ (1,5)}^{3}\equiv H\bar{H}uuuLLLL\nu\,,\\
\mathcal{O}_{14, \ (1,5)}^{4}\equiv HHuQQLLLL\nu\,,\\
\mathcal{O}_{14, \ (1,5)}^{5}\equiv HHduuLLLL\nu\,,\\
\mathcal{O}_{14, \ (1,5)}^{6}\equiv HHuuQLLLe\nu\,,\\
\mathcal{O}_{14, \ (1,5)}^{7}\equiv HHuuuLLee\nu\,,\\
\mathcal{O}_{14, \ (1,5)}^{8}\equiv H\bar{H}uuQLLL\nu\nu\,,\\
\mathcal{O}_{14, \ (1,5)}^{9}\equiv H\bar{H}uuuLLe\nu\nu\,,\\
\mathcal{O}_{14, \ (1,5)}^{10}\equiv HHQQQLLL\nu\nu\,,\\
\mathcal{O}_{14, \ (1,5)}^{11}\equiv HHduQLLL\nu\nu\,,\\
\mathcal{O}_{14, \ (1,5)}^{12}\equiv HHuQQLLe\nu\nu\,,\\
\mathcal{O}_{14, \ (1,5)}^{13}\equiv HHduuLLe\nu\nu\,,\\
\mathcal{O}_{14, \ (1,5)}^{14}\equiv HHuuQLee\nu\nu\,,\\
\mathcal{O}_{14, \ (1,5)}^{15}\equiv HHdQQLL\nu\nu\nu\,,\\
\mathcal{O}_{14, \ (1,5)}^{16}\equiv \bar{H}\bar{H}uuuLL\nu\nu\nu\,,\\
\mathcal{O}_{14, \ (1,5)}^{17}\equiv H\bar{H}uQQLL\nu\nu\nu\,,\\
\mathcal{O}_{14, \ (1,5)}^{18}\equiv H\bar{H}duuLL\nu\nu\nu\,,\\
\mathcal{O}_{14, \ (1,5)}^{19}\equiv H\bar{H}uuQLe\nu\nu\nu\,,\\
\mathcal{O}_{14, \ (1,5)}^{20}\equiv H\bar{H}uuuee\nu\nu\nu\,,\\
\mathcal{O}_{14, \ (1,5)}^{21}\equiv HHQQQLe\nu\nu\nu\,,\\
\mathcal{O}_{14, \ (1,5)}^{22}\equiv HHdduLL\nu\nu\nu\,,\\
\mathcal{O}_{14, \ (1,5)}^{23}\equiv HHduQLe\nu\nu\nu\,,\\
\mathcal{O}_{14, \ (1,5)}^{24}\equiv HHuQQee\nu\nu\nu\,,\\
\mathcal{O}_{14, \ (1,5)}^{25}\equiv HHddQL\nu\nu\nu\nu\,,\\
\mathcal{O}_{14, \ (1,5)}^{26}\equiv HHdQQe\nu\nu\nu\nu\,,\\
\mathcal{O}_{14, \ (1,5)}^{27}\equiv \bar{H}\bar{H}uuQL\nu\nu\nu\nu\,,\\
\mathcal{O}_{14, \ (1,5)}^{28}\equiv H\bar{H}QQQL\nu\nu\nu\nu\,,\\
\mathcal{O}_{14, \ (1,5)}^{29}\equiv H\bar{H}duQL\nu\nu\nu\nu\,,\\
\mathcal{O}_{14, \ (1,5)}^{30}\equiv H\bar{H}uQQe\nu\nu\nu\nu\,,\\
\mathcal{O}_{14, \ (1,5)}^{31}\equiv H\bar{H}duue\nu\nu\nu\nu\,,\\
\mathcal{O}_{14, \ (1,5)}^{32}\equiv H\bar{H}dQQ\nu\nu\nu\nu\nu\,,\\
\mathcal{O}_{14, \ (1,5)}^{33}\equiv \bar{H}\bar{H}uQQ\nu\nu\nu\nu\nu\,,\\
\mathcal{O}_{14, \ (1,5)}^{34}\equiv H\bar{H}ddu\nu\nu\nu\nu\nu\,.
 \label{eq:dequal14basisB1L5}
 \end{math}
 \end{spacing}
 \end{multicols}

 \subsection{\texorpdfstring{$\Delta B=2, \Delta L=-2$}{Delta B=2,Delta L=-2}}
 \begin{multicols}{3}
 \begin{spacing}{1.5}
 \noindent
 \begin{math}
\mathcal{O}_{14, \ (2,-2)}^{1}\equiv H\bar{H}dddddd\bar{e}\bar{e}\,,\\
\mathcal{O}_{14, \ (2,-2)}^{2}\equiv \bar{H}\bar{H}ddddQQ\bar{e}\bar{e}\,,\\
\mathcal{O}_{14, \ (2,-2)}^{3}\equiv \bar{H}\bar{H}dddQQQ\bar{L}\bar{e}\,,\\
\mathcal{O}_{14, \ (2,-2)}^{4}\equiv \bar{H}\bar{H}dddduQ\bar{L}\bar{e}\,,\\
\mathcal{O}_{14, \ (2,-2)}^{5}\equiv \bar{H}\bar{H}ddQQQQ\bar{L}\bar{L}\,,\\
\mathcal{O}_{14, \ (2,-2)}^{6}\equiv \bar{H}\bar{H}ddduQQ\bar{L}\bar{L}\,,\\
\mathcal{O}_{14, \ (2,-2)}^{7}\equiv \bar{H}\bar{H}dddduu\bar{L}\bar{L}\,,\\
\mathcal{O}_{14, \ (2,-2)}^{8}\equiv H\bar{H}dddddQ\bar{L}\bar{e}\,,\\
\mathcal{O}_{14, \ (2,-2)}^{9}\equiv H\bar{H}ddddQQ\bar{L}\bar{L}\,,\\
\mathcal{O}_{14, \ (2,-2)}^{10}\equiv H\bar{H}dddddu\bar{L}\bar{L}\,,\\
\mathcal{O}_{14, \ (2,-2)}^{11}\equiv HHdddddd\bar{L}\bar{L}\,,\\
\mathcal{O}_{14, \ (2,-2)}^{12}\equiv H\bar{H}ddddQQ\bar{e}\bar{\nu}\,,\\
\mathcal{O}_{14, \ (2,-2)}^{13}\equiv H\bar{H}dddddu\bar{e}\bar{\nu}\,,\\
\mathcal{O}_{14, \ (2,-2)}^{14}\equiv \bar{H}\bar{H}ddQQQQ\bar{e}\bar{\nu}\,,\\
\mathcal{O}_{14, \ (2,-2)}^{15}\equiv \bar{H}\bar{H}ddduQQ\bar{e}\bar{\nu}\,,\\
\mathcal{O}_{14, \ (2,-2)}^{16}\equiv \bar{H}\bar{H}dQQQQQ\bar{L}\bar{\nu}\,,\\
\mathcal{O}_{14, \ (2,-2)}^{17}\equiv \bar{H}\bar{H}dduQQQ\bar{L}\bar{\nu}\,,\\
\mathcal{O}_{14, \ (2,-2)}^{18}\equiv \bar{H}\bar{H}ddduuQ\bar{L}\bar{\nu}\,,\\
\mathcal{O}_{14, \ (2,-2)}^{19}\equiv H\bar{H}dddQQQ\bar{L}\bar{\nu}\,,\\
\mathcal{O}_{14, \ (2,-2)}^{20}\equiv H\bar{H}dddduQ\bar{L}\bar{\nu}\,,\\
\mathcal{O}_{14, \ (2,-2)}^{21}\equiv HHdddddQ\bar{L}\bar{\nu}\,,\\
\mathcal{O}_{14, \ (2,-2)}^{22}\equiv HHddddQQ\bar{\nu}\bar{\nu}\,,\\
\mathcal{O}_{14, \ (2,-2)}^{23}\equiv H\bar{H}ddQQQQ\bar{\nu}\bar{\nu}\,,\\
\mathcal{O}_{14, \ (2,-2)}^{24}\equiv \bar{H}\bar{H}QQQQQQ\bar{\nu}\bar{\nu}\,,\\
\mathcal{O}_{14, \ (2,-2)}^{25}\equiv \bar{H}\bar{H}duQQQQ\bar{\nu}\bar{\nu}\,,\\
\mathcal{O}_{14, \ (2,-2)}^{26}\equiv \bar{H}\bar{H}dduuQQ\bar{\nu}\bar{\nu}\,,\\
\mathcal{O}_{14, \ (2,-2)}^{27}\equiv H\bar{H}ddduQQ\bar{\nu}\bar{\nu}\,,\\
\mathcal{O}_{14, \ (2,-2)}^{28}\equiv H\bar{H}dddduu\bar{\nu}\bar{\nu}\,.
 \label{eq:dequal14basisB2L-2}
 \end{math}
 \end{spacing}
 \end{multicols}

 \subsection{\texorpdfstring{$\Delta B=2, \Delta L=2$}{Delta B=2,Delta L=2}}
 \begin{multicols}{3}
 \begin{spacing}{1.5}
 \noindent
 \begin{math}
\mathcal{O}_{14, \ (2,2)}^{1}\equiv HHddQQQQLL\,,\\
\mathcal{O}_{14, \ (2,2)}^{2}\equiv HHdQQQQQLe\,,\\
\mathcal{O}_{14, \ (2,2)}^{3}\equiv \bar{H}\bar{H}uuQQQQLL\,,\\
\mathcal{O}_{14, \ (2,2)}^{4}\equiv \bar{H}\bar{H}duuuQQLL\,,\\
\mathcal{O}_{14, \ (2,2)}^{5}\equiv \bar{H}\bar{H}uuuQQQLe\,,\\
\mathcal{O}_{14, \ (2,2)}^{6}\equiv \bar{H}\bar{H}dduuuuLL\,,\\
\mathcal{O}_{14, \ (2,2)}^{7}\equiv \bar{H}\bar{H}duuuuQLe\,,\\
\mathcal{O}_{14, \ (2,2)}^{8}\equiv \bar{H}\bar{H}uuuuQQee\,,\\
\mathcal{O}_{14, \ (2,2)}^{9}\equiv H\bar{H}QQQQQQLL\,,\\
\mathcal{O}_{14, \ (2,2)}^{10}\equiv H\bar{H}duQQQQLL\,,\\
\mathcal{O}_{14, \ (2,2)}^{11}\equiv H\bar{H}uQQQQQLe\,,\\
\mathcal{O}_{14, \ (2,2)}^{12}\equiv H\bar{H}dduuQQLL\,,\\
\mathcal{O}_{14, \ (2,2)}^{13}\equiv H\bar{H}duuQQQLe\,,\\
\mathcal{O}_{14, \ (2,2)}^{14}\equiv H\bar{H}uuQQQQee\,,\\
\mathcal{O}_{14, \ (2,2)}^{15}\equiv H\bar{H}ddduuuLL\,,\\
\mathcal{O}_{14, \ (2,2)}^{16}\equiv H\bar{H}dduuuQLe\,,\\
\mathcal{O}_{14, \ (2,2)}^{17}\equiv H\bar{H}duuuQQee\,,\\
\mathcal{O}_{14, \ (2,2)}^{18}\equiv H\bar{H}dduuuuee\,,\\
\mathcal{O}_{14, \ (2,2)}^{19}\equiv HHQQQQQQee\,,\\
\mathcal{O}_{14, \ (2,2)}^{20}\equiv HHddduQQLL\,,\\
\mathcal{O}_{14, \ (2,2)}^{21}\equiv HHdduQQQLe\,,\\
\mathcal{O}_{14, \ (2,2)}^{22}\equiv HHduQQQQee\,,\\
\mathcal{O}_{14, \ (2,2)}^{23}\equiv HHdddduuLL\,,\\
\mathcal{O}_{14, \ (2,2)}^{24}\equiv HHddduuQLe\,,\\
\mathcal{O}_{14, \ (2,2)}^{25}\equiv HHdduuQQee\,,\\
\mathcal{O}_{14, \ (2,2)}^{26}\equiv HHdddQQQL\nu\,,\\
\mathcal{O}_{14, \ (2,2)}^{27}\equiv HHddQQQQe\nu\,,\\
\mathcal{O}_{14, \ (2,2)}^{28}\equiv H\bar{H}dQQQQQL\nu\,,\\
\mathcal{O}_{14, \ (2,2)}^{29}\equiv \bar{H}\bar{H}uQQQQQL\nu\,,\\
\mathcal{O}_{14, \ (2,2)}^{30}\equiv \bar{H}\bar{H}duuQQQL\nu\,,\\
\mathcal{O}_{14, \ (2,2)}^{31}\equiv \bar{H}\bar{H}uuQQQQe\nu\,,\\
\mathcal{O}_{14, \ (2,2)}^{32}\equiv \bar{H}\bar{H}dduuuQL\nu\,,\\
\mathcal{O}_{14, \ (2,2)}^{33}\equiv \bar{H}\bar{H}duuuQQe\nu\,,\\
\mathcal{O}_{14, \ (2,2)}^{34}\equiv H\bar{H}QQQQQQe\nu\,,\\
\mathcal{O}_{14, \ (2,2)}^{35}\equiv H\bar{H}dduQQQL\nu\,,\\
\mathcal{O}_{14, \ (2,2)}^{36}\equiv H\bar{H}duQQQQe\nu\,,\\
\mathcal{O}_{14, \ (2,2)}^{37}\equiv H\bar{H}ddduuQL\nu\,,\\
\mathcal{O}_{14, \ (2,2)}^{38}\equiv H\bar{H}dduuQQe\nu\,,\\
\mathcal{O}_{14, \ (2,2)}^{39}\equiv H\bar{H}ddduuue\nu\,,\\
\mathcal{O}_{14, \ (2,2)}^{40}\equiv HHdddduQL\nu\,,\\
\mathcal{O}_{14, \ (2,2)}^{41}\equiv HHddduQQe\nu\,,\\
\mathcal{O}_{14, \ (2,2)}^{42}\equiv HHddddQQ\nu\nu\,,\\
\mathcal{O}_{14, \ (2,2)}^{43}\equiv H\bar{H}ddQQQQ\nu\nu\,,\\
\mathcal{O}_{14, \ (2,2)}^{44}\equiv \bar{H}\bar{H}QQQQQQ\nu\nu\,,\\
\mathcal{O}_{14, \ (2,2)}^{45}\equiv \bar{H}\bar{H}duQQQQ\nu\nu\,,\\
\mathcal{O}_{14, \ (2,2)}^{46}\equiv \bar{H}\bar{H}dduuQQ\nu\nu\,,\\
\mathcal{O}_{14, \ (2,2)}^{47}\equiv H\bar{H}ddduQQ\nu\nu\,,\\
\mathcal{O}_{14, \ (2,2)}^{48}\equiv H\bar{H}dddduu\nu\nu\,.
 \label{eq:dequal14basisB2L2}
 \end{math}
 \end{spacing}
 \end{multicols}

\clearpage
 

\section{Dimension 15}
\label{sec:d15}
 
\begin{figure}
    \centering
    \includegraphics[width=1\linewidth]{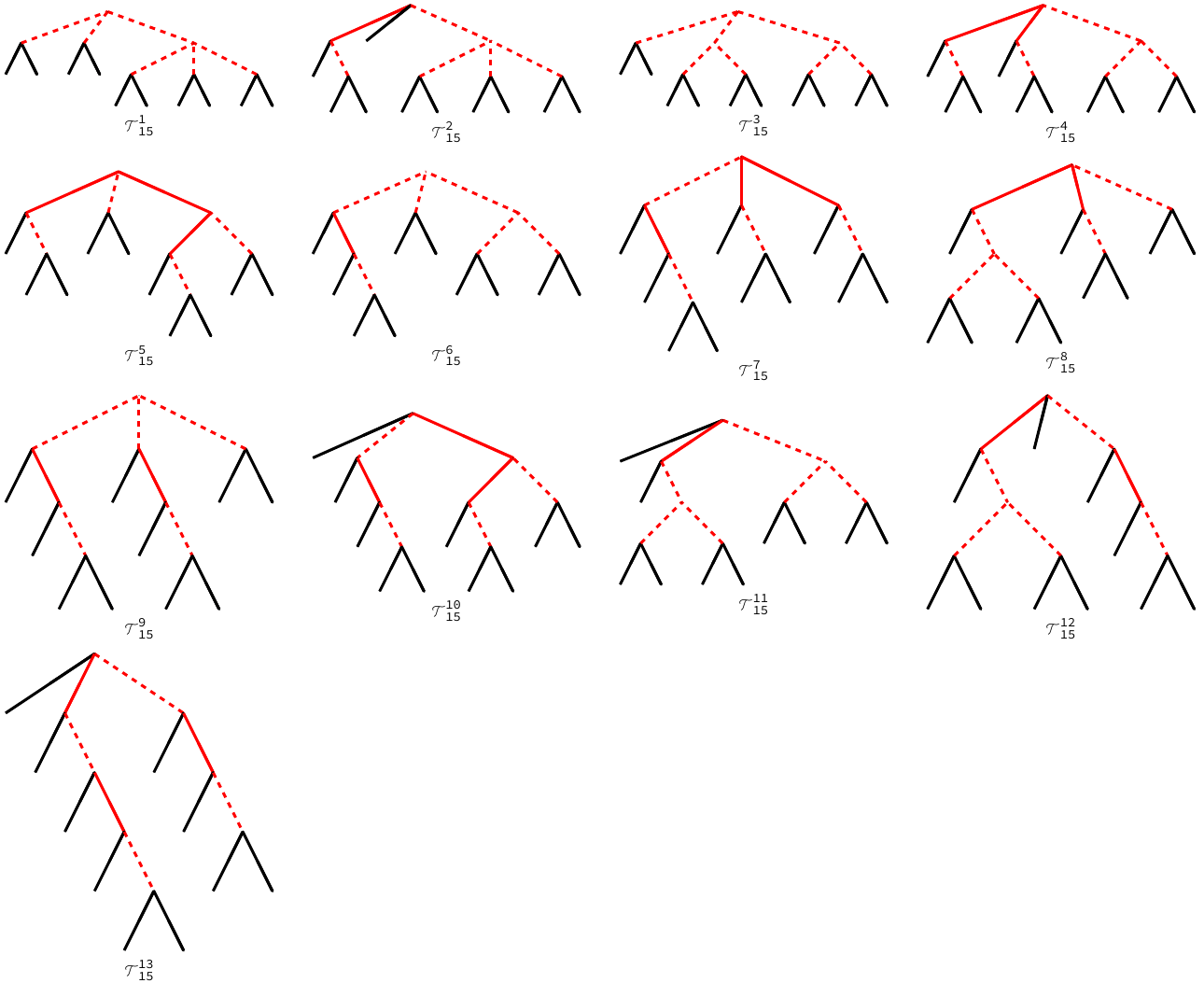}
    \caption{Feynman-diagram topologies for the subset of $d=15$ BNV operators discussed in this article, all with 10 external fermions.}
    \label{fig:d=15_topologies}
\end{figure}

BNV operators at $d=15$ are quite interesting because they give rise to four new breaking patterns: $(\Delta B ,\Delta L) =(1,7)$, $(2,-4)$, $(2,+4)$, and $(3,1)$, each of which could be made dominant through a $U(1)_{B + a L}$ symmetry~\cite{Weinberg:1980bf}. The different topologies for these cases are shown in Fig.~\ref{fig:d=15_topologies}. 
 For the sake of article length, we are not showing the other $(\Delta B ,\Delta L)$ operators, $(1,-1)$, $(1,-5)$, $(2,0)$ and $(1,3)$, which can not be symmetry protected using $B+a L$, nor are we UV-completing them.

  \subsection{\texorpdfstring{$\Delta B=1, \Delta L=7$}{Delta B=1,Delta L=7}}

The $(\Delta B ,\Delta L) =(1,7)$ operators require between three and six right-handed neutrinos and thus do not show up in the SMEFT, or at least not at $d=15$.

 \begin{multicols}{3}
 \begin{spacing}{1.5}
 \noindent
 \begin{math}
\mathcal{O}_{15, \, (1,7)}^{1}\equiv uuuLLLL\nu\nu\nu\,,\\
\mathcal{O}_{15, \, (1,7)}^{2}\equiv uuQLLL\nu\nu\nu\nu\,,\\
\mathcal{O}_{15, \, (1,7)}^{3}\equiv uuuLLe\nu\nu\nu\nu\,,\\
\mathcal{O}_{15, \, (1,7)}^{4}\equiv uQQLL\nu\nu\nu\nu\nu\,,\\
\mathcal{O}_{15, \, (1,7)}^{5}\equiv duuLL\nu\nu\nu\nu\nu\,,\\
\mathcal{O}_{15, \, (1,7)}^{6}\equiv uuQLe\nu\nu\nu\nu\nu\,,\\
\mathcal{O}_{15, \, (1,7)}^{7}\equiv uuuee\nu\nu\nu\nu\nu\,,\\
\mathcal{O}_{15, \, (1,7)}^{8}\equiv QQQL\nu\nu\nu\nu\nu\nu\,,\\
\mathcal{O}_{15, \, (1,7)}^{9}\equiv duQL\nu\nu\nu\nu\nu\nu\,,\\
\mathcal{O}_{15, \, (1,7)}^{10}\equiv uQQe\nu\nu\nu\nu\nu\nu\,,\\
\mathcal{O}_{15, \, (1,7)}^{11}\equiv duue\nu\nu\nu\nu\nu\nu\,.
 \label{eq:dequal15basisB1L7}
 \end{math}
 \end{spacing}
 \end{multicols}

 \subsection{\texorpdfstring{$\Delta B=2, \Delta L=-4$}{Delta B=2,Delta L=-4}}
 \begin{multicols}{3}
 \begin{spacing}{1.5}
 \noindent
 \begin{math}
\mathcal{O}_{15, \, (2,-4)}^{1}\equiv dddddd\bar{L}\bar{L}\bar{L}\bar{L}\,,\\
\mathcal{O}_{15, \, (2,-4)}^{2}\equiv dddddd\bar{L}\bar{L}\bar{e}\bar{\nu}\,,\\
\mathcal{O}_{15, \, (2,-4)}^{3}\equiv dddddQ\bar{L}\bar{L}\bar{L}\bar{\nu}\,,\\
\mathcal{O}_{15, \, (2,-4)}^{4}\equiv dddddd\bar{e}\bar{e}\bar{\nu}\bar{\nu}\,,\\
\mathcal{O}_{15, \, (2,-4)}^{5}\equiv dddddQ\bar{L}\bar{e}\bar{\nu}\bar{\nu}\,,\\
\mathcal{O}_{15, \, (2,-4)}^{6}\equiv ddddQQ\bar{L}\bar{L}\bar{\nu}\bar{\nu}\,,\\
\mathcal{O}_{15, \, (2,-4)}^{7}\equiv dddddu\bar{L}\bar{L}\bar{\nu}\bar{\nu}\,,\\
\mathcal{O}_{15, \, (2,-4)}^{8}\equiv ddddQQ\bar{e}\bar{\nu}\bar{\nu}\bar{\nu}\,,\\
\mathcal{O}_{15, \, (2,-4)}^{9}\equiv dddddu\bar{e}\bar{\nu}\bar{\nu}\bar{\nu}\,,\\
\mathcal{O}_{15, \, (2,-4)}^{10}\equiv dddQQQ\bar{L}\bar{\nu}\bar{\nu}\bar{\nu}\,,\\
\mathcal{O}_{15, \, (2,-4)}^{11}\equiv dddduQ\bar{L}\bar{\nu}\bar{\nu}\bar{\nu}\,,\\
\mathcal{O}_{15, \, (2,-4)}^{12}\equiv ddQQQQ\bar{\nu}\bar{\nu}\bar{\nu}\bar{\nu}\,,\\
\mathcal{O}_{15, \, (2,-4)}^{13}\equiv ddduQQ\bar{\nu}\bar{\nu}\bar{\nu}\bar{\nu}\,,\\
\mathcal{O}_{15, \, (2,-4)}^{14}\equiv dddduu\bar{\nu}\bar{\nu}\bar{\nu}\bar{\nu}\,.
 \label{eq:dequal15basisB2L-4}
 \end{math}
 \end{spacing}
 \end{multicols}

 \subsection{\texorpdfstring{$\Delta B=2, \Delta L=4$}{Delta B=2,Delta L=4}}
 \begin{multicols}{3}
 \begin{spacing}{1.5}
 \noindent
 \begin{math}
\mathcal{O}_{15, \, (2,4)}^{1}\equiv uuQQQQLLLL\,,\\
\mathcal{O}_{15, \, (2,4)}^{2}\equiv duuuQQLLLL\,,\\
\mathcal{O}_{15, \, (2,4)}^{3}\equiv uuuQQQLLLe\,,\\
\mathcal{O}_{15, \, (2,4)}^{4}\equiv dduuuuLLLL\,,\\
\mathcal{O}_{15, \, (2,4)}^{5}\equiv duuuuQLLLe\,,\\
\mathcal{O}_{15, \, (2,4)}^{6}\equiv uuuuQQLLee\,,\\
\mathcal{O}_{15, \, (2,4)}^{7}\equiv duuuuuLLee\,,\\
\mathcal{O}_{15, \, (2,4)}^{8}\equiv uuuuuQLeee\,,\\
\mathcal{O}_{15, \, (2,4)}^{9}\equiv uuuuuueeee\,,\\
\mathcal{O}_{15, \, (2,4)}^{10}\equiv uQQQQQLLL\nu\,,\\
\mathcal{O}_{15, \, (2,4)}^{11}\equiv duuQQQLLL\nu\,,\\
\mathcal{O}_{15, \, (2,4)}^{12}\equiv uuQQQQLLe\nu\,,\\
\mathcal{O}_{15, \, (2,4)}^{13}\equiv dduuuQLLL\nu\,,\\
\mathcal{O}_{15, \, (2,4)}^{14}\equiv duuuQQLLe\nu\,,\\
\mathcal{O}_{15, \, (2,4)}^{15}\equiv uuuQQQLee\nu\,,\\
\mathcal{O}_{15, \, (2,4)}^{16}\equiv dduuuuLLe\nu\,,\\
\mathcal{O}_{15, \, (2,4)}^{17}\equiv duuuuQLee\nu\,,\\
\mathcal{O}_{15, \, (2,4)}^{18}\equiv uuuuQQeee\nu\,,\\
\mathcal{O}_{15, \, (2,4)}^{19}\equiv duuuuueee\nu\,,\\
\mathcal{O}_{15, \, (2,4)}^{20}\equiv QQQQQQLL\nu\nu\,,\\
\mathcal{O}_{15, \, (2,4)}^{21}\equiv duQQQQLL\nu\nu\,,\\
\mathcal{O}_{15, \, (2,4)}^{22}\equiv uQQQQQLe\nu\nu\,,\\
\mathcal{O}_{15, \, (2,4)}^{23}\equiv dduuQQLL\nu\nu\,,\\
\mathcal{O}_{15, \, (2,4)}^{24}\equiv duuQQQLe\nu\nu\,,\\
\mathcal{O}_{15, \, (2,4)}^{25}\equiv uuQQQQee\nu\nu\,,\\
\mathcal{O}_{15, \, (2,4)}^{26}\equiv ddduuuLL\nu\nu\,,\\
\mathcal{O}_{15, \, (2,4)}^{27}\equiv dduuuQLe\nu\nu\,,\\
\mathcal{O}_{15, \, (2,4)}^{28}\equiv duuuQQee\nu\nu\,,\\
\mathcal{O}_{15, \, (2,4)}^{29}\equiv dduuuuee\nu\nu\,,\\
\mathcal{O}_{15, \, (2,4)}^{30}\equiv dQQQQQL\nu\nu\nu\,,\\
\mathcal{O}_{15, \, (2,4)}^{31}\equiv QQQQQQe\nu\nu\nu\,,\\
\mathcal{O}_{15, \, (2,4)}^{32}\equiv dduQQQL\nu\nu\nu\,,\\
\mathcal{O}_{15, \, (2,4)}^{33}\equiv duQQQQe\nu\nu\nu\,,\\
\mathcal{O}_{15, \, (2,4)}^{34}\equiv ddduuQL\nu\nu\nu\,,\\
\mathcal{O}_{15, \, (2,4)}^{35}\equiv dduuQQe\nu\nu\nu\,,\\
\mathcal{O}_{15, \, (2,4)}^{36}\equiv ddduuue\nu\nu\nu\,,\\
\mathcal{O}_{15, \, (2,4)}^{37}\equiv ddQQQQ\nu\nu\nu\nu\,,\\
\mathcal{O}_{15, \, (2,4)}^{38}\equiv ddduQQ\nu\nu\nu\nu\,,\\
\mathcal{O}_{15, \, (2,4)}^{39}\equiv dddduu\nu\nu\nu\nu\,.
 \label{eq:dequal15basisB2L4}
 \end{math}
 \end{spacing}
 \end{multicols}

 \subsection{\texorpdfstring{$\Delta B=3, \Delta L=1$}{Delta B=3,Delta L=1}}

  The $(\Delta B ,\Delta L) =(3,1)$ SMEFT operators (plus one $\nu$SMEFT operator) have been discussed in Ref.~\cite{Babu:2003qh}, although six of theirs are not Lorentz singlets due to an odd number of left-handed or right-handed fields;
 some of the induced tri-nucleon decays have been searched for experimentally in neutrinoless double-beta decay experiments~\cite{EXO-200:2017hwz,Majorana:2018pdo,GERDA:2023uuw,Majorana:2024rmx}.

 \begin{multicols}{3}
 \begin{spacing}{1.5}
 \noindent
 \begin{math}
\mathcal{O}_{15, \, (3,1)}^{1}\equiv ddQQQQQQQL\,,\\
\mathcal{O}_{15, \, (3,1)}^{2}\equiv dQQQQQQQQe\,,\\
\mathcal{O}_{15, \, (3,1)}^{3}\equiv ddduQQQQQL\,,\\
\mathcal{O}_{15, \, (3,1)}^{4}\equiv dduQQQQQQe\,,\\
\mathcal{O}_{15, \, (3,1)}^{5}\equiv dddduuQQQL\,,\\
\mathcal{O}_{15, \, (3,1)}^{6}\equiv ddduuQQQQe\,,\\
\mathcal{O}_{15, \, (3,1)}^{7}\equiv ddddduuuQL\,,\\
\mathcal{O}_{15, \, (3,1)}^{8}\equiv dddduuuQQe\,,\\
\mathcal{O}_{15, \, (3,1)}^{9}\equiv ddddduuuue\,,\\
\mathcal{O}_{15, \, (3,1)}^{10}\equiv dddQQQQQQ\nu\,,\\
\mathcal{O}_{15, \, (3,1)}^{11}\equiv dddduQQQQ\nu\,,\\
\mathcal{O}_{15, \, (3,1)}^{12}\equiv ddddduuQQ\nu\,,\\
\mathcal{O}_{15, \, (3,1)}^{13}\equiv dddddduuu\nu\,.
 \label{eq:dequal15basisB3L1}
 \end{math}
 \end{spacing}
 \end{multicols}

\pagebreak


 \bibliographystyle{JHEP}
 \bibliography{biblio.bib}

\end{document}

%% file: Tables_dim6/Table_dim6_Top1_B1_L1.tex
 \begin{center}
 \begin{SplitTable}[2.0cm] 
  \SplitHeader{$\mathcal{O}^j{(\mathcal{T}_ {6} ^{ 1 })}$}{New Particles} 
   \vspace{0.1cm} 

 \SplitRow{ $\mathcal{O}_{6, \; (1,1)}^{1}$} {{$S_{2}$, $V_{1}$, $V_{2}$}} 
 \SplitRow{ $\mathcal{O}_{6, \; (1,1)}^{2}$} {{$V_{1}$, $S_{2}$}} 
 \SplitRow{ $\mathcal{O}_{6, \; (1,1)}^{3}$} {{$S_{4}$, $S_{2}$}} 
 \SplitRow{ $\mathcal{O}_{6, \; (1,1)}^{4}$} {{$S_{1}$, $S_{2}$}} 
 \SplitRow{ $\mathcal{O}_{6, \; (1,1)}^{5}$} {{$V_{2}$, $S_{2}$}} 
 \SplitRow{ $\mathcal{O}_{6, \; (1,1)}^{6}$} {{$S_{2}$, $S_{3}$}} 
 \end{SplitTable}
   \vspace{0.4cm} 
 \captionof{table}{List of new particles that generate the dimension-6 operators of Eq.~\eqref{eq:dequal6basis_simple}, corresponding to  topology 1 (see Fig.~\ref{fig:d=6_topology}).} 
 \label{tab:T6Top1B1L1}
 \end{center}

%% file: Tables_dim7/Table_dim7_Top1_B1_L-1.tex
 \begin{center}
 \begin{SplitTable}[2.0cm] 
  \SplitHeader{$\mathcal{O}^j{(\mathcal{T}_ {7} ^{ 1 })}$}{New Particles} 
   \vspace{0.1cm} 
 \SplitRow{ $\mathcal{O}_{7, \; (1,-1)}^{1}$} {{$V_{4}V_{2}$, $S_{3}S_{6}$}} 
 \SplitRow{ $\mathcal{O}_{7, \; (1,-1)}^{2}$} {{$S_{5}S_{4}$, $S_{2}S_{5}$, $V_{2}V_{5}$, $V_{4}V_{2}$}} 
 \SplitRow{ $\mathcal{O}_{7, \; (1,-1)}^{3}$} {{$S_{3}S_{6}$, $S_{2}S_{5}$}} 
 \SplitRow{ $\mathcal{O}_{7, \; (1,-1)}^{4}$} {{$S_{3}S_{5}$}} 
 \SplitRow{ $\mathcal{O}_{7, \; (1,-1)}^{5}$} {{$V_{3}V_{2}$, $S_{3}S_{5}$}} 
 \SplitRow{ $\mathcal{O}_{7, \; (1,-1)}^{6}$} {{$S_{5}S_{4}$, $S_{2}S_{5}$}} 
 \SplitRow{ $\mathcal{O}_{7, \; (1,-1)}^{7}$} {{$V_{4}V_{2}$, $V_{3}V_{1}$, $S_{2}S_{5}$}} 
 \end{SplitTable}
   \vspace{0.4cm} 
 \captionof{table}{Dimension 7 $(\Delta B, \Delta L)$ = (1, -1) UV completions for topology 1 (see Fig.~\ref{fig:d=7_topologies}).} 
 \label{tab:T7Top1B1L-1}
 \end{center}

%% file: Tables_dim7/Table_dim7_Top2_B1_L-1.tex
 \begin{center}
 \begin{SplitTable}[2.0cm] 
  \SplitHeader{$\mathcal{O}^j{(\mathcal{T}_ {7} ^{ 2 })}$}{New Particles} 
   \vspace{0.1cm} 
 \SplitRow{ $\mathcal{O}_{7, \; (1,-1)}^{1}$} {{$F_{7}V_{2}$, $F_{5}S_{3}$, $F_{3}V_{2}$, $F_{7}S_{6}$, $F_{5}V_{4}$, $F_{3}S_{3}$, $F_{7}V_{4}$}} 
 \SplitRow{ $\mathcal{O}_{7, \; (1,-1)}^{2}$} {{$F_{7}S_{4}$, $F_{7}S_{2}$, $F_{9}V_{2}$, $F_{5}V_{2}$, $F_{4}S_{4}$, $F_{1}S_{2}$, $F_{9}V_{5}$, $F_{5}V_{4}$, $F_{4}V_{2}$, $F_{1}V_{2}$, $F_{7}V_{5}$, $F_{7}V_{4}$, $F_{9}S_{5}$, $F_{5}S_{5}$}} 
 \SplitRow{ $\mathcal{O}_{7, \; (1,-1)}^{3}$} {{$F_{8}S_{3}$, $F_{7}S_{2}$, $F_{1}S_{3}$, $F_{7}S_{5}$, $F_{1}S_{2}$, $F_{8}S_{5}$, $F_{7}S_{6}$}} 
 \SplitRow{ $\mathcal{O}_{7, \; (1,-1)}^{4}$} {{$F_{8}S_{3}$, $F_{2}S_{3}$, $F_{8}S_{5}$}} 
 \SplitRow{ $\mathcal{O}_{7, \; (1,-1)}^{5}$} {{$F_{8}V_{2}$, $F_{6}S_{3}$, $F_{3}V_{2}$, $F_{8}S_{5}$, $F_{6}V_{3}$, $F_{3}S_{3}$, $F_{8}V_{3}$}} 
 \SplitRow{ $\mathcal{O}_{7, \; (1,-1)}^{6}$} {{$F_{9}S_{4}$, $F_{5}S_{2}$, $F_{3}S_{4}$, $F_{3}S_{2}$, $F_{9}S_{5}$, $F_{5}S_{5}$}} 
 \SplitRow{ $\mathcal{O}_{7, \; (1,-1)}^{7}$} {{$F_{8}V_{2}$, $F_{7}V_{1}$, $F_{5}S_{2}$, $F_{3}V_{2}$, $F_{7}S_{5}$, $F_{5}V_{3}$, $F_{3}V_{1}$, $F_{8}S_{5}$, $F_{5}V_{4}$, $F_{3}S_{2}$, $F_{8}V_{3}$, $F_{7}V_{4}$}} 
 \end{SplitTable}
   \vspace{0.4cm} 
 \captionof{table}{Dimension 7 $(\Delta B, \Delta L)$ = (1, -1) UV completions for topology 2 (see Fig.~\ref{fig:d=7_topologies}).} 
 \label{tab:T7Top2B1L-1}
 \end{center}

%% file: Tables_dim8/Table_dim8_Top1_B1_L1.tex
 \begin{center}
 \begin{SplitTable}[2.0cm] 
  \SplitHeader{$\mathcal{O}^j{(\mathcal{T}_ {8} ^{ 1 })}$}{New Particles} 
   \vspace{0.1cm} 
 \SplitRow{ $\mathcal{O}_{8, \; (1,1)}^{1}$} {{$S_{9}S_{3}S_{4}$, $S_{9}V_{1}V_{2}$}} 
 \SplitRow{ $\mathcal{O}_{8, \; (1,1)}^{2}$} {{$S_{9}V_{1}V_{2}$, $S_{9}S_{1}S_{4}$}} 
 \SplitRow{ $\mathcal{O}_{8, \; (1,1)}^{3}$} {{$S_{9}S_{1}S_{4}$, $S_{9}V_{1}V_{2}$}} 
 \SplitRow{ $\mathcal{O}_{8, \; (1,1)}^{4}$} {{$S_{8}S_{4}S_{4}$, $S_{8}S_{2}S_{4}$, $S_{7}S_{4}S_{4}$, $S_{7}S_{2}S_{2}$}} 
 \SplitRow{ $\mathcal{O}_{8, \; (1,1)}^{5}$} {{$S_{8}S_{2}S_{4}$, $S_{7}S_{2}S_{2}$, $S_{8}V_{1}V_{1}$, $S_{7}V_{1}V_{1}$, $S_{8}V_{2}V_{2}$, $S_{7}V_{2}V_{2}$}} 
 \SplitRow{ $\mathcal{O}_{8, \; (1,1)}^{6}$} {{$S_{8}V_{1}V_{1}$, $S_{7}V_{1}V_{1}$, $S_{8}S_{2}S_{4}$, $S_{7}S_{2}S_{2}$}} 
 \SplitRow{ $\mathcal{O}_{8, \; (1,1)}^{7}$} {{$S_{7}S_{1}S_{1}$, $S_{7}S_{2}S_{2}$}} 
 \SplitRow{ $\mathcal{O}_{8, \; (1,1)}^{8}$} {{$S_{8}V_{2}V_{2}$, $S_{7}V_{2}V_{2}$, $S_{8}S_{2}S_{4}$, $S_{7}S_{2}S_{2}$}} 
 \SplitRow{ $\mathcal{O}_{8, \; (1,1)}^{9}$} {{$S_{9}V_{1}V_{2}$, $S_{9}S_{3}S_{4}$}} 
 \SplitRow{ $\mathcal{O}_{8, \; (1,1)}^{10}$} {{$S_{7}S_{2}S_{2}$, $S_{7}S_{3}S_{3}$}} 
 \end{SplitTable}
   \vspace{0.4cm} 
 \captionof{table}{Dimension 8 $(\Delta B, \Delta L)$ = (1, 1) UV completions for topology 1 (see Fig.~\ref{fig:d=8_topologies}).} 
 \label{tab:T8Top1B1L1}
 \end{center}

%% file: Tables_dim8/Table_dim8_Top2_B1_L1.tex
 \begin{center}
 \begin{SplitTable}[2.0cm] 
  \SplitHeader{$\mathcal{O}^j{(\mathcal{T}_ {8} ^{ 2 })}$}{New Particles} 
   \vspace{0.1cm} 
 \SplitRow{ $\mathcal{O}_{8, \; (1,1)}^{1}$} {{$S_{3}S_{5}S_{4}$, $V_{1}V_{2}V_{8}$, $V_{3}V_{1}V_{2}$}} 
 \SplitRow{ $\mathcal{O}_{8, \; (1,1)}^{2}$} {{$V_{1}V_{2}V_{8}$, $V_{3}V_{1}V_{2}$, $S_{1}S_{11}S_{4}$}} 
 \SplitRow{ $\mathcal{O}_{8, \; (1,1)}^{3}$} {{$S_{1}S_{11}S_{4}$, $V_{1}V_{2}V_{8}$, $V_{3}V_{1}V_{2}$}} 
 \SplitRow{ $\mathcal{O}_{8, \; (1,1)}^{4}$} {{$S_{4}S_{4}S_{13}$, $S_{5}S_{4}S_{4}$, $S_{2}S_{5}S_{4}$, $S_{2}S_{2}S_{5}$, $S_{4}S_{4}S_{12}$, $S_{11}S_{4}S_{4}$, $S_{2}S_{11}S_{4}$, $S_{2}S_{2}S_{11}$}} 
 \SplitRow{ $\mathcal{O}_{8, \; (1,1)}^{5}$} {{$S_{2}S_{5}S_{4}$, $S_{2}S_{2}S_{5}$, $V_{1}V_{1}V_{8}$, $V_{3}V_{1}V_{1}$, $V_{2}V_{2}V_{8}$, $V_{3}V_{2}V_{2}$, $V_{2}V_{2}V_{5}$, $V_{4}V_{2}V_{2}$, $V_{1}V_{1}V_{7}$, $V_{6}V_{1}V_{1}$, $S_{2}S_{11}S_{4}$, $S_{2}S_{2}S_{11}$}} 
 \SplitRow{ $\mathcal{O}_{8, \; (1,1)}^{6}$} {{$V_{1}V_{1}V_{8}$, $V_{3}V_{1}V_{1}$, $S_{2}S_{11}S_{4}$, $S_{2}S_{2}S_{11}$, $S_{2}S_{5}S_{4}$, $S_{2}S_{2}S_{5}$, $V_{1}V_{1}V_{7}$, $V_{6}V_{1}V_{1}$}} 
 \SplitRow{ $\mathcal{O}_{8, \; (1,1)}^{7}$} {{$S_{1}S_{1}S_{11}$, $S_{2}S_{2}S_{5}$, $S_{2}S_{2}S_{11}$, $S_{1}S_{1}S_{10}$}} 
 \SplitRow{ $\mathcal{O}_{8, \; (1,1)}^{8}$} {{$V_{2}V_{2}V_{5}$, $V_{4}V_{2}V_{2}$, $S_{2}S_{11}S_{4}$, $S_{2}S_{2}S_{11}$, $S_{2}S_{5}S_{4}$, $S_{2}S_{2}S_{5}$, $V_{2}V_{2}V_{8}$, $V_{3}V_{2}V_{2}$}} 
 \SplitRow{ $\mathcal{O}_{8, \; (1,1)}^{9}$} {{$V_{1}V_{2}V_{8}$, $V_{3}V_{1}V_{2}$, $S_{3}S_{5}S_{4}$}} 
 \SplitRow{ $\mathcal{O}_{8, \; (1,1)}^{10}$} {{$S_{2}S_{2}S_{5}$, $S_{3}S_{3}S_{5}$, $S_{3}S_{3}S_{6}$, $S_{2}S_{2}S_{11}$}} 
 \end{SplitTable}
   \vspace{0.4cm} 
 \captionof{table}{Dimension 8 $(\Delta B, \Delta L)$ = (1, 1) UV completions for topology 2 (see Fig.~\ref{fig:d=8_topologies}).} 
 \label{tab:T8Top2B1L1}
 \end{center}

%% file: Tables_dim8/Table_dim8_Top3_B1_L1.tex
 \begin{center}
 \begin{SplitTable}[2.0cm] 
  \SplitHeader{$\mathcal{O}^j{(\mathcal{T}_ {8} ^{ 3 })}$}{New Particles} 
   \vspace{0.1cm} 
 \SplitRow{ $\mathcal{O}_{8, \; (1,1)}^{1}$} {{$F_{8}F_{8}S_{4}$, $F_{8}F_{14}V_{1}$, $F_{6}F_{8}V_{1}$, $F_{4}F_{8}V_{2}$, $F_{1}F_{8}V_{2}$, $F_{4}F_{14}S_{3}$, $F_{1}F_{6}S_{3}$}} 
 \SplitRow{ $\mathcal{O}_{8, \; (1,1)}^{2}$} {{$F_{8}F_{14}V_{1}$, $F_{6}F_{8}V_{1}$, $F_{3}F_{8}S_{4}$, $F_{14}F_{14}S_{1}$, $F_{6}F_{6}S_{1}$, $F_{3}F_{14}V_{2}$, $F_{3}F_{6}V_{2}$}} 
 \SplitRow{ $\mathcal{O}_{8, \; (1,1)}^{3}$} {{$F_{8}F_{8}S_{4}$, $F_{8}F_{9}V_{2}$, $F_{5}F_{8}V_{2}$, $F_{11}F_{8}V_{1}$, $F_{2}F_{8}V_{1}$, $F_{11}F_{9}S_{1}$, $F_{2}F_{5}S_{1}$}} 
 \SplitRow{ $\mathcal{O}_{8, \; (1,1)}^{4}$} {{$F_{9}F_{14}S_{4}$, $F_{9}F_{14}S_{2}$, $F_{6}F_{9}S_{4}$, $F_{5}F_{14}S_{4}$, $F_{5}F_{6}S_{2}$, $F_{4}F_{9}S_{4}$, $F_{4}F_{9}S_{2}$, $F_{1}F_{9}S_{4}$, $F_{4}F_{5}S_{4}$, $F_{1}F_{5}S_{2}$, $F_{11}F_{14}S_{4}$, $F_{11}F_{14}S_{2}$, $F_{11}F_{6}S_{4}$, $F_{2}F_{14}S_{4}$, $F_{2}F_{6}S_{2}$}} 
 \SplitRow{ $\mathcal{O}_{8, \; (1,1)}^{5}$} {{$F_{8}F_{8}S_{4}$, $F_{8}F_{8}S_{2}$, $F_{8}F_{14}V_{1}$, $F_{6}F_{8}V_{1}$, $F_{4}F_{8}V_{2}$, $F_{1}F_{8}V_{2}$, $F_{7}F_{13}S_{4}$, $F_{7}F_{13}S_{2}$, $F_{7}F_{14}V_{2}$, $F_{6}F_{7}V_{2}$, $F_{4}F_{7}V_{1}$, $F_{1}F_{7}V_{1}$, $F_{13}F_{9}V_{1}$, $F_{5}F_{13}V_{1}$, $F_{8}F_{9}V_{2}$, $F_{5}F_{8}V_{2}$, $F_{4}F_{9}S_{2}$, $F_{1}F_{5}S_{2}$, $F_{11}F_{13}V_{2}$, $F_{2}F_{13}V_{2}$, $F_{11}F_{8}V_{1}$, $F_{2}F_{8}V_{1}$, $F_{11}F_{14}S_{2}$, $F_{2}F_{6}S_{2}$}} 
 \SplitRow{ $\mathcal{O}_{8, \; (1,1)}^{6}$} {{$F_{8}F_{14}V_{1}$, $F_{6}F_{8}V_{1}$, $F_{3}F_{8}S_{4}$, $F_{3}F_{8}S_{2}$, $F_{13}F_{9}V_{1}$, $F_{5}F_{13}V_{1}$, $F_{9}F_{14}S_{2}$, $F_{5}F_{6}S_{2}$, $F_{3}F_{9}V_{1}$, $F_{3}F_{5}V_{1}$, $F_{10}F_{13}S_{4}$, $F_{10}F_{13}S_{2}$, $F_{10}F_{14}V_{1}$, $F_{10}F_{6}V_{1}$}} 
 \SplitRow{ $\mathcal{O}_{8, \; (1,1)}^{7}$} {{$F_{8}F_{13}S_{1}$, $F_{8}F_{8}S_{2}$, $F_{3}F_{8}S_{2}$, $F_{7}F_{13}S_{2}$, $F_{3}F_{7}S_{1}$, $F_{10}F_{13}S_{2}$, $F_{10}F_{8}S_{1}$}} 
 \SplitRow{ $\mathcal{O}_{8, \; (1,1)}^{8}$} {{$F_{7}F_{14}V_{2}$, $F_{6}F_{7}V_{2}$, $F_{3}F_{7}S_{4}$, $F_{3}F_{7}S_{2}$, $F_{8}F_{9}V_{2}$, $F_{5}F_{8}V_{2}$, $F_{9}F_{14}S_{2}$, $F_{5}F_{6}S_{2}$, $F_{3}F_{9}V_{2}$, $F_{3}F_{5}V_{2}$, $F_{3}F_{8}S_{4}$, $F_{3}F_{8}S_{2}$, $F_{3}F_{14}V_{2}$, $F_{3}F_{6}V_{2}$}} 
 \SplitRow{ $\mathcal{O}_{8, \; (1,1)}^{9}$} {{$F_{8}F_{9}V_{2}$, $F_{5}F_{8}V_{2}$, $F_{3}F_{8}S_{4}$, $F_{9}F_{9}S_{3}$, $F_{5}F_{5}S_{3}$, $F_{3}F_{9}V_{1}$, $F_{3}F_{5}V_{1}$}} 
 \SplitRow{ $\mathcal{O}_{8, \; (1,1)}^{10}$} {{$F_{8}F_{8}S_{2}$, $F_{3}F_{8}S_{3}$, $F_{7}F_{13}S_{2}$, $F_{7}F_{8}S_{3}$, $F_{3}F_{7}S_{2}$, $F_{3}F_{13}S_{3}$, $F_{3}F_{8}S_{2}$}} 
 \end{SplitTable}
   \vspace{0.4cm} 
 \captionof{table}{Dimension 8 $(\Delta B, \Delta L)$ = (1, 1) UV completions for topology 3 (see Fig.~\ref{fig:d=8_topologies}).} 
 \label{tab:T8Top3B1L1}
 \end{center}

%% file: Tables_dim8/Table_dim8_Top4_B1_L1.tex
 \begin{center}
 \begin{SplitTable}[2.0cm] 
  \SplitHeader{$\mathcal{O}^j{(\mathcal{T}_ {8} ^{ 4 })}$}{New Particles} 
   \vspace{0.1cm} 
 \SplitRow{ $\mathcal{O}_{8, \; (1,1)}^{1}$} {{$F_{8}F_{14}S_{4}$, $F_{13}F_{14}V_{1}$, $F_{6}F_{13}V_{1}$, $F_{3}F_{4}V_{2}$, $F_{1}F_{3}V_{2}$, $F_{8}F_{14}V_{1}$, $F_{3}F_{4}S_{3}$, $F_{1}F_{3}S_{3}$, $F_{8}F_{14}V_{2}$, $F_{13}F_{14}S_{3}$, $F_{6}F_{13}S_{3}$}} 
 \SplitRow{ $\mathcal{O}_{8, \; (1,1)}^{2}$} {{$F_{13}F_{14}V_{1}$, $F_{6}F_{13}V_{1}$, $F_{3}F_{4}S_{4}$, $F_{8}F_{14}V_{1}$, $F_{13}F_{14}S_{1}$, $F_{6}F_{13}S_{1}$, $F_{3}F_{4}V_{2}$, $F_{8}F_{14}S_{4}$, $F_{13}F_{14}V_{2}$, $F_{6}F_{13}V_{2}$}} 
 \SplitRow{ $\mathcal{O}_{8, \; (1,1)}^{3}$} {{$F_{8}F_{9}S_{4}$, $F_{7}F_{9}V_{2}$, $F_{5}F_{7}V_{2}$, $F_{10}F_{11}V_{1}$, $F_{2}F_{10}V_{1}$, $F_{8}F_{9}V_{2}$, $F_{10}F_{11}S_{1}$, $F_{2}F_{10}S_{1}$, $F_{8}F_{9}V_{1}$, $F_{7}F_{9}S_{1}$, $F_{5}F_{7}S_{1}$}} 
 \SplitRow{ $\mathcal{O}_{8, \; (1,1)}^{4}$} {{$F_{14}F_{15}S_{4}$, $F_{8}F_{14}S_{4}$, $F_{6}F_{8}S_{4}$, $F_{8}F_{14}S_{2}$, $F_{6}F_{8}S_{2}$, $F_{4}F_{12}S_{4}$, $F_{3}F_{4}S_{4}$, $F_{1}F_{3}S_{4}$, $F_{3}F_{4}S_{2}$, $F_{1}F_{3}S_{2}$, $F_{9}F_{15}S_{4}$, $F_{8}F_{9}S_{4}$, $F_{5}F_{8}S_{4}$, $F_{8}F_{9}S_{2}$, $F_{5}F_{8}S_{2}$, $F_{11}F_{12}S_{4}$, $F_{3}F_{11}S_{4}$, $F_{2}F_{3}S_{4}$, $F_{3}F_{11}S_{2}$, $F_{2}F_{3}S_{2}$}} 
 \SplitRow{ $\mathcal{O}_{8, \; (1,1)}^{5}$} {{$F_{8}F_{9}S_{4}$, $F_{5}F_{8}S_{2}$, $F_{8}F_{14}V_{1}$, $F_{6}F_{8}V_{1}$, $F_{3}F_{4}V_{2}$, $F_{1}F_{3}V_{2}$, $F_{13}F_{14}S_{4}$, $F_{6}F_{13}S_{2}$, $F_{8}F_{14}V_{2}$, $F_{6}F_{8}V_{2}$, $F_{3}F_{4}V_{1}$, $F_{1}F_{3}V_{1}$, $F_{13}F_{14}V_{1}$, $F_{6}F_{13}V_{1}$, $F_{8}F_{9}V_{2}$, $F_{5}F_{8}V_{2}$, $F_{3}F_{4}S_{2}$, $F_{1}F_{3}S_{2}$, $F_{13}F_{14}V_{2}$, $F_{6}F_{13}V_{2}$, $F_{8}F_{9}V_{1}$, $F_{5}F_{8}V_{1}$, $F_{8}F_{14}S_{2}$, $F_{6}F_{8}S_{2}$, $F_{7}F_{9}S_{4}$, $F_{5}F_{7}S_{2}$, $F_{3}F_{11}V_{2}$, $F_{2}F_{3}V_{2}$, $F_{8}F_{14}S_{4}$, $F_{3}F_{11}V_{1}$, $F_{2}F_{3}V_{1}$, $F_{7}F_{9}V_{2}$, $F_{5}F_{7}V_{2}$, $F_{3}F_{11}S_{2}$, $F_{2}F_{3}S_{2}$, $F_{7}F_{9}V_{1}$, $F_{5}F_{7}V_{1}$, $F_{8}F_{9}S_{2}$}} 
 \SplitRow{ $\mathcal{O}_{8, \; (1,1)}^{6}$} {{$F_{8}F_{14}V_{1}$, $F_{6}F_{8}V_{1}$, $F_{3}F_{11}S_{4}$, $F_{2}F_{3}S_{2}$, $F_{13}F_{14}V_{1}$, $F_{6}F_{13}V_{1}$, $F_{8}F_{14}S_{2}$, $F_{6}F_{8}S_{2}$, $F_{3}F_{11}V_{1}$, $F_{2}F_{3}V_{1}$, $F_{13}F_{14}S_{4}$, $F_{6}F_{13}S_{2}$, $F_{8}F_{9}V_{1}$, $F_{5}F_{8}V_{1}$, $F_{10}F_{11}S_{4}$, $F_{2}F_{10}S_{2}$, $F_{8}F_{9}S_{2}$, $F_{5}F_{8}S_{2}$, $F_{10}F_{11}V_{1}$, $F_{2}F_{10}V_{1}$, $F_{8}F_{14}S_{4}$}} 
 \SplitRow{ $\mathcal{O}_{8, \; (1,1)}^{7}$} {{$F_{6}F_{13}S_{1}$, $F_{5}F_{8}S_{2}$, $F_{2}F_{3}S_{2}$, $F_{6}F_{13}S_{2}$, $F_{2}F_{3}S_{1}$, $F_{5}F_{8}S_{1}$, $F_{6}F_{8}S_{1}$, $F_{5}F_{7}S_{2}$, $F_{2}F_{10}S_{2}$, $F_{6}F_{8}S_{2}$, $F_{2}F_{10}S_{1}$, $F_{5}F_{7}S_{1}$}} 
 \SplitRow{ $\mathcal{O}_{8, \; (1,1)}^{8}$} {{$F_{8}F_{14}V_{2}$, $F_{6}F_{8}V_{2}$, $F_{3}F_{4}S_{4}$, $F_{1}F_{3}S_{2}$, $F_{8}F_{9}V_{2}$, $F_{5}F_{8}V_{2}$, $F_{8}F_{14}S_{2}$, $F_{6}F_{8}S_{2}$, $F_{3}F_{4}V_{2}$, $F_{1}F_{3}V_{2}$, $F_{8}F_{9}S_{4}$, $F_{5}F_{8}S_{2}$, $F_{7}F_{9}V_{2}$, $F_{5}F_{7}V_{2}$, $F_{8}F_{9}S_{2}$, $F_{7}F_{9}S_{4}$, $F_{5}F_{7}S_{2}$}} 
 \SplitRow{ $\mathcal{O}_{8, \; (1,1)}^{9}$} {{$F_{7}F_{9}V_{2}$, $F_{5}F_{7}V_{2}$, $F_{3}F_{11}S_{4}$, $F_{8}F_{9}V_{2}$, $F_{7}F_{9}S_{3}$, $F_{5}F_{7}S_{3}$, $F_{3}F_{11}V_{1}$, $F_{8}F_{9}S_{4}$, $F_{7}F_{9}V_{1}$, $F_{5}F_{7}V_{1}$}} 
 \SplitRow{ $\mathcal{O}_{8, \; (1,1)}^{10}$} {{$F_{5}F_{8}S_{2}$, $F_{1}F_{3}S_{3}$, $F_{6}F_{13}S_{2}$, $F_{5}F_{8}S_{3}$, $F_{1}F_{3}S_{2}$, $F_{6}F_{13}S_{3}$, $F_{5}F_{7}S_{2}$, $F_{6}F_{8}S_{2}$, $F_{5}F_{7}S_{3}$, $F_{6}F_{8}S_{3}$}} 
 \end{SplitTable}
   \vspace{0.4cm} 
 \captionof{table}{Dimension 8 $(\Delta B, \Delta L)$ = (1, 1) UV completions for topology 4 (see Fig.~\ref{fig:d=8_topologies}).} 
 \label{tab:T8Top4B1L1}
 \end{center}

%% file: Tables_dim8/Table_dim8_Top5_B1_L1.tex
 \begin{center}
 \begin{SplitTable}[2.0cm] 
  \SplitHeader{$\mathcal{O}^j{(\mathcal{T}_ {8} ^{ 5 })}$}{New Particles} 
   \vspace{0.1cm} 
 \SplitRow{ $\mathcal{O}_{8, \; (1,1)}^{1}$} {{$F_{4}V_{2}V_{8}$, $F_{1}V_{3}V_{2}$, $F_{14}V_{1}V_{8}$, $F_{6}V_{3}V_{1}$, $F_{8}S_{5}S_{4}$, $F_{4}S_{3}S_{5}$, $F_{1}S_{3}S_{5}$, $F_{8}V_{1}V_{8}$, $F_{8}V_{3}V_{1}$, $F_{14}S_{3}S_{5}$, $F_{6}S_{3}S_{5}$, $F_{8}V_{2}V_{8}$, $F_{8}V_{3}V_{2}$}} 
 \SplitRow{ $\mathcal{O}_{8, \; (1,1)}^{2}$} {{$F_{3}S_{11}S_{4}$, $F_{14}V_{1}V_{8}$, $F_{6}V_{3}V_{1}$, $F_{3}V_{2}V_{8}$, $F_{3}V_{3}V_{2}$, $F_{14}S_{1}S_{11}$, $F_{6}S_{1}S_{11}$, $F_{8}V_{1}V_{8}$, $F_{8}V_{3}V_{1}$, $F_{14}V_{2}V_{8}$, $F_{6}V_{3}V_{2}$, $F_{8}S_{11}S_{4}$}} 
 \SplitRow{ $\mathcal{O}_{8, \; (1,1)}^{3}$} {{$F_{11}V_{1}V_{8}$, $F_{2}V_{3}V_{1}$, $F_{9}V_{2}V_{8}$, $F_{5}V_{3}V_{2}$, $F_{8}S_{11}S_{4}$, $F_{11}S_{1}S_{11}$, $F_{2}S_{1}S_{11}$, $F_{8}V_{2}V_{8}$, $F_{8}V_{3}V_{2}$, $F_{9}S_{1}S_{11}$, $F_{5}S_{1}S_{11}$, $F_{8}V_{1}V_{8}$, $F_{8}V_{3}V_{1}$}} 
 \SplitRow{ $\mathcal{O}_{8, \; (1,1)}^{4}$} {{$F_{4}S_{4}S_{13}$, $F_{4}S_{5}S_{4}$, $F_{4}S_{2}S_{5}$, $F_{1}S_{5}S_{4}$, $F_{1}S_{2}S_{5}$, $F_{14}S_{4}S_{12}$, $F_{14}S_{11}S_{4}$, $F_{14}S_{2}S_{11}$, $F_{6}S_{11}S_{4}$, $F_{6}S_{2}S_{11}$, $F_{14}S_{4}S_{13}$, $F_{14}S_{5}S_{4}$, $F_{14}S_{2}S_{5}$, $F_{6}S_{5}S_{4}$, $F_{6}S_{2}S_{5}$, $F_{11}S_{4}S_{12}$, $F_{11}S_{11}S_{4}$, $F_{11}S_{2}S_{11}$, $F_{2}S_{11}S_{4}$, $F_{2}S_{2}S_{11}$, $F_{9}S_{4}S_{13}$, $F_{9}S_{5}S_{4}$, $F_{9}S_{2}S_{5}$, $F_{5}S_{5}S_{4}$, $F_{5}S_{2}S_{5}$, $F_{9}S_{4}S_{12}$, $F_{9}S_{11}S_{4}$, $F_{9}S_{2}S_{11}$, $F_{5}S_{11}S_{4}$, $F_{5}S_{2}S_{11}$}} 
 \SplitRow{ $\mathcal{O}_{8, \; (1,1)}^{5}$} {{$F_{4}V_{2}V_{5}$, $F_{1}V_{4}V_{2}$, $F_{14}V_{1}V_{7}$, $F_{6}V_{6}V_{1}$, $F_{8}S_{11}S_{4}$, $F_{8}S_{2}S_{11}$, $F_{4}V_{1}V_{8}$, $F_{1}V_{3}V_{1}$, $F_{14}V_{2}V_{8}$, $F_{6}V_{3}V_{2}$, $F_{13}S_{11}S_{4}$, $F_{13}S_{2}S_{11}$, $F_{4}S_{2}S_{5}$, $F_{1}S_{2}S_{5}$, $F_{8}V_{2}V_{8}$, $F_{8}V_{3}V_{2}$, $F_{13}V_{1}V_{7}$, $F_{13}V_{6}V_{1}$, $F_{14}S_{2}S_{5}$, $F_{6}S_{2}S_{5}$, $F_{8}V_{1}V_{8}$, $F_{8}V_{3}V_{1}$, $F_{13}V_{2}V_{5}$, $F_{13}V_{4}V_{2}$, $F_{11}V_{2}V_{8}$, $F_{2}V_{3}V_{2}$, $F_{9}V_{1}V_{8}$, $F_{5}V_{3}V_{1}$, $F_{7}S_{5}S_{4}$, $F_{7}S_{2}S_{5}$, $F_{11}V_{1}V_{7}$, $F_{2}V_{6}V_{1}$, $F_{9}V_{2}V_{5}$, $F_{5}V_{4}V_{2}$, $F_{8}S_{5}S_{4}$, $F_{8}S_{2}S_{5}$, $F_{11}S_{2}S_{11}$, $F_{2}S_{2}S_{11}$, $F_{7}V_{2}V_{5}$, $F_{7}V_{4}V_{2}$, $F_{9}S_{2}S_{11}$, $F_{5}S_{2}S_{11}$, $F_{7}V_{1}V_{7}$, $F_{7}V_{6}V_{1}$}} 
 \SplitRow{ $\mathcal{O}_{8, \; (1,1)}^{6}$} {{$F_{3}S_{5}S_{4}$, $F_{3}S_{2}S_{5}$, $F_{14}V_{1}V_{7}$, $F_{6}V_{6}V_{1}$, $F_{3}V_{1}V_{8}$, $F_{3}V_{3}V_{1}$, $F_{14}S_{2}S_{11}$, $F_{6}S_{2}S_{11}$, $F_{13}V_{1}V_{7}$, $F_{13}V_{6}V_{1}$, $F_{14}V_{1}V_{8}$, $F_{6}V_{3}V_{1}$, $F_{13}S_{5}S_{4}$, $F_{13}S_{2}S_{5}$, $F_{10}S_{11}S_{4}$, $F_{10}S_{2}S_{11}$, $F_{9}V_{1}V_{8}$, $F_{5}V_{3}V_{1}$, $F_{10}V_{1}V_{7}$, $F_{10}V_{6}V_{1}$, $F_{9}S_{2}S_{5}$, $F_{5}S_{2}S_{5}$, $F_{8}V_{1}V_{8}$, $F_{8}V_{3}V_{1}$, $F_{9}V_{1}V_{7}$, $F_{5}V_{6}V_{1}$, $F_{8}S_{11}S_{4}$, $F_{8}S_{2}S_{11}$}} 
 \SplitRow{ $\mathcal{O}_{8, \; (1,1)}^{7}$} {{$F_{3}S_{2}S_{5}$, $F_{8}S_{2}S_{11}$, $F_{13}S_{1}S_{10}$, $F_{3}S_{1}S_{11}$, $F_{13}S_{2}S_{11}$, $F_{8}S_{1}S_{11}$, $F_{13}S_{2}S_{5}$, $F_{10}S_{2}S_{11}$, $F_{7}S_{2}S_{5}$, $F_{10}S_{1}S_{10}$, $F_{8}S_{2}S_{5}$, $F_{7}S_{1}S_{10}$}} 
 \SplitRow{ $\mathcal{O}_{8, \; (1,1)}^{8}$} {{$F_{3}S_{5}S_{4}$, $F_{3}S_{2}S_{5}$, $F_{14}V_{2}V_{8}$, $F_{6}V_{3}V_{2}$, $F_{3}V_{2}V_{5}$, $F_{3}V_{4}V_{2}$, $F_{14}S_{2}S_{11}$, $F_{6}S_{2}S_{11}$, $F_{8}V_{2}V_{8}$, $F_{8}V_{3}V_{2}$, $F_{14}V_{2}V_{5}$, $F_{6}V_{4}V_{2}$, $F_{8}S_{5}S_{4}$, $F_{8}S_{2}S_{5}$, $F_{3}S_{11}S_{4}$, $F_{3}S_{2}S_{11}$, $F_{9}V_{2}V_{5}$, $F_{5}V_{4}V_{2}$, $F_{3}V_{2}V_{8}$, $F_{3}V_{3}V_{2}$, $F_{9}S_{2}S_{5}$, $F_{5}S_{2}S_{5}$, $F_{7}V_{2}V_{5}$, $F_{7}V_{4}V_{2}$, $F_{9}V_{2}V_{8}$, $F_{5}V_{3}V_{2}$, $F_{7}S_{11}S_{4}$, $F_{7}S_{2}S_{11}$}} 
 \SplitRow{ $\mathcal{O}_{8, \; (1,1)}^{9}$} {{$F_{3}S_{5}S_{4}$, $F_{9}V_{2}V_{8}$, $F_{5}V_{3}V_{2}$, $F_{3}V_{1}V_{8}$, $F_{3}V_{3}V_{1}$, $F_{9}S_{3}S_{5}$, $F_{5}S_{3}S_{5}$, $F_{8}V_{2}V_{8}$, $F_{8}V_{3}V_{2}$, $F_{9}V_{1}V_{8}$, $F_{5}V_{3}V_{1}$, $F_{8}S_{5}S_{4}$}} 
 \SplitRow{ $\mathcal{O}_{8, \; (1,1)}^{10}$} {{$F_{3}S_{3}S_{6}$, $F_{8}S_{2}S_{11}$, $F_{3}S_{2}S_{5}$, $F_{8}S_{3}S_{5}$, $F_{13}S_{2}S_{11}$, $F_{8}S_{2}S_{5}$, $F_{13}S_{3}S_{6}$, $F_{3}S_{3}S_{5}$, $F_{7}S_{2}S_{5}$, $F_{3}S_{2}S_{11}$, $F_{7}S_{3}S_{6}$, $F_{7}S_{2}S_{11}$}} 
 \end{SplitTable}
   \vspace{0.4cm} 
 \captionof{table}{Dimension 8 $(\Delta B, \Delta L)$ = (1, 1) UV completions for topology 5 (see Fig.~\ref{fig:d=8_topologies}).} 
 \label{tab:T8Top5B1L1}
 \end{center}

%% file: Tables_dim8/Table_dim8_Top6_B1_L1.tex
 \begin{center}
 \begin{SplitTable}[2.0cm] 
  \SplitHeader{$\mathcal{O}^j{(\mathcal{T}_ {8} ^{ 6 })}$}{New Particles} 
   \vspace{0.1cm} 
 \SplitRow{ $\mathcal{O}_{8, \; (1,1)}^{1}$} {{$S_{3}S_{4}$, $V_{1}V_{2}$}} 
 \SplitRow{ $\mathcal{O}_{8, \; (1,1)}^{2}$} {{$V_{1}V_{2}$, $S_{1}S_{4}$}} 
 \SplitRow{ $\mathcal{O}_{8, \; (1,1)}^{3}$} {{$S_{1}S_{4}$, $V_{1}V_{2}$}} 
 \SplitRow{ $\mathcal{O}_{8, \; (1,1)}^{4}$} {{$S_{4}S_{4}$, $S_{2}S_{4}$, $S_{2}S_{2}$}} 
 \SplitRow{ $\mathcal{O}_{8, \; (1,1)}^{5}$} {{$S_{2}S_{4}$, $S_{2}S_{2}$, $V_{1}V_{1}$, $V_{2}V_{2}$}} 
 \SplitRow{ $\mathcal{O}_{8, \; (1,1)}^{6}$} {{$V_{1}V_{1}$, $S_{2}S_{4}$, $S_{2}S_{2}$}} 
 \SplitRow{ $\mathcal{O}_{8, \; (1,1)}^{7}$} {{$S_{1}S_{1}$, $S_{2}S_{2}$}} 
 \SplitRow{ $\mathcal{O}_{8, \; (1,1)}^{8}$} {{$V_{2}V_{2}$, $S_{2}S_{4}$, $S_{2}S_{2}$}} 
 \SplitRow{ $\mathcal{O}_{8, \; (1,1)}^{9}$} {{$V_{1}V_{2}$, $S_{3}S_{4}$}} 
 \SplitRow{ $\mathcal{O}_{8, \; (1,1)}^{10}$} {{$S_{2}S_{2}$, $S_{3}S_{3}$}} 
 \end{SplitTable}
   \vspace{0.4cm} 
 \captionof{table}{Dimension 8 $(\Delta B, \Delta L)$ = (1, 1) UV completions for topology 6 (see Fig.~\ref{fig:d=8_topologies}).} 
 \label{tab:T8Top6B1L1}
 \end{center}

%% file: Tables_dim8/Table_dim8_Top7_B1_L1.tex
 \begin{center}
 \begin{SplitTable}[2.0cm] 
  \SplitHeader{$\mathcal{O}^j{(\mathcal{T}_ {8} ^{ 7 })}$}{New Particles} 
   \vspace{0.1cm} 
 \SplitRow{ $\mathcal{O}_{8, \; (1,1)}^{1}$} {{$F_{14}S_{9}S_{4}$, $F_{14}S_{9}V_{1}$, $F_{14}S_{9}V_{2}$, $F_{13}S_{9}V_{1}$, $F_{13}S_{9}S_{3}$, $F_{3}S_{9}V_{2}$, $F_{3}S_{9}S_{3}$}} 
 \SplitRow{ $\mathcal{O}_{8, \; (1,1)}^{2}$} {{$F_{14}S_{9}V_{1}$, $F_{14}S_{9}S_{4}$, $F_{13}S_{9}V_{1}$, $F_{13}S_{9}S_{1}$, $F_{13}S_{9}V_{2}$, $F_{4}S_{9}S_{4}$, $F_{4}S_{9}V_{2}$}} 
 \SplitRow{ $\mathcal{O}_{8, \; (1,1)}^{3}$} {{$F_{9}S_{9}S_{4}$, $F_{9}S_{9}V_{2}$, $F_{9}S_{9}V_{1}$, $F_{7}S_{9}V_{2}$, $F_{7}S_{9}S_{1}$, $F_{10}S_{9}V_{1}$, $F_{10}S_{9}S_{1}$}} 
 \SplitRow{ $\mathcal{O}_{8, \; (1,1)}^{4}$} {{$F_{15}S_{8}S_{4}$, $F_{8}S_{8}S_{4}$, $F_{8}S_{7}S_{4}$, $F_{8}S_{8}S_{2}$, $F_{8}S_{7}S_{2}$, $F_{12}S_{8}S_{4}$, $F_{3}S_{8}S_{4}$, $F_{3}S_{7}S_{4}$, $F_{3}S_{8}S_{2}$, $F_{3}S_{7}S_{2}$}} 
 \SplitRow{ $\mathcal{O}_{8, \; (1,1)}^{5}$} {{$F_{14}S_{8}S_{4}$, $F_{6}S_{7}S_{2}$, $F_{14}S_{8}V_{1}$, $F_{6}S_{7}V_{1}$, $F_{14}S_{8}V_{2}$, $F_{6}S_{7}V_{2}$, $F_{9}S_{8}S_{4}$, $F_{5}S_{7}S_{2}$, $F_{9}S_{8}V_{2}$, $F_{5}S_{7}V_{2}$, $F_{9}S_{8}V_{1}$, $F_{5}S_{7}V_{1}$, $F_{8}S_{8}V_{1}$, $F_{8}S_{7}V_{1}$, $F_{8}S_{8}V_{2}$, $F_{8}S_{7}V_{2}$, $F_{8}S_{8}S_{2}$, $F_{8}S_{7}S_{2}$, $F_{3}S_{8}V_{2}$, $F_{3}S_{7}V_{2}$, $F_{3}S_{8}V_{1}$, $F_{3}S_{7}V_{1}$, $F_{3}S_{8}S_{2}$, $F_{3}S_{7}S_{2}$}} 
 \SplitRow{ $\mathcal{O}_{8, \; (1,1)}^{6}$} {{$F_{14}S_{8}V_{1}$, $F_{6}S_{7}V_{1}$, $F_{14}S_{8}S_{4}$, $F_{6}S_{7}S_{2}$, $F_{8}S_{8}V_{1}$, $F_{8}S_{7}V_{1}$, $F_{8}S_{8}S_{2}$, $F_{8}S_{7}S_{2}$, $F_{11}S_{8}S_{4}$, $F_{2}S_{7}S_{2}$, $F_{11}S_{8}V_{1}$, $F_{2}S_{7}V_{1}$}} 
 \SplitRow{ $\mathcal{O}_{8, \; (1,1)}^{7}$} {{$F_{6}S_{7}S_{1}$, $F_{6}S_{7}S_{2}$, $F_{5}S_{7}S_{2}$, $F_{5}S_{7}S_{1}$, $F_{2}S_{7}S_{2}$, $F_{2}S_{7}S_{1}$}} 
 \SplitRow{ $\mathcal{O}_{8, \; (1,1)}^{8}$} {{$F_{9}S_{8}V_{2}$, $F_{5}S_{7}V_{2}$, $F_{9}S_{8}S_{4}$, $F_{5}S_{7}S_{2}$, $F_{8}S_{8}V_{2}$, $F_{8}S_{7}V_{2}$, $F_{8}S_{8}S_{2}$, $F_{8}S_{7}S_{2}$, $F_{4}S_{8}S_{4}$, $F_{1}S_{7}S_{2}$, $F_{4}S_{8}V_{2}$, $F_{1}S_{7}V_{2}$}} 
 \SplitRow{ $\mathcal{O}_{8, \; (1,1)}^{9}$} {{$F_{9}S_{9}V_{2}$, $F_{9}S_{9}S_{4}$, $F_{7}S_{9}V_{2}$, $F_{7}S_{9}S_{3}$, $F_{7}S_{9}V_{1}$, $F_{11}S_{9}S_{4}$, $F_{11}S_{9}V_{1}$}} 
 \SplitRow{ $\mathcal{O}_{8, \; (1,1)}^{10}$} {{$F_{6}S_{7}S_{2}$, $F_{6}S_{7}S_{3}$, $F_{5}S_{7}S_{2}$, $F_{5}S_{7}S_{3}$, $F_{1}S_{7}S_{3}$, $F_{1}S_{7}S_{2}$}} 
 \end{SplitTable}
   \vspace{0.4cm} 
 \captionof{table}{Dimension 8 $(\Delta B, \Delta L)$ = (1, 1) UV completions for topology 7 (see Fig.~\ref{fig:d=8_topologies}).} 
 \label{tab:T8Top7B1L1}
 \end{center}

%% file: Tables_dim8/Table_dim8_Top8_B1_L1.tex
 \begin{center}
 \begin{SplitTable}[2.0cm] 
  \SplitHeader{$\mathcal{O}^j{(\mathcal{T}_ {8} ^{ 8 })}$}{New Particles} 
   \vspace{0.1cm} 
 \SplitRow{ $\mathcal{O}_{8, \; (1,1)}^{1}$} {{$F_{4}F_{14}V_{8}$, $F_{1}F_{6}V_{3}$, $F_{4}F_{8}S_{5}$, $F_{1}F_{8}S_{5}$, $F_{4}F_{8}V_{8}$, $F_{1}F_{8}V_{3}$, $F_{8}F_{14}S_{5}$, $F_{6}F_{8}S_{5}$, $F_{8}F_{14}V_{8}$, $F_{6}F_{8}V_{3}$, $F_{8}F_{8}V_{8}$, $F_{8}F_{8}V_{3}$}} 
 \SplitRow{ $\mathcal{O}_{8, \; (1,1)}^{2}$} {{$F_{3}F_{14}V_{8}$, $F_{3}F_{6}V_{3}$, $F_{3}F_{14}S_{11}$, $F_{3}F_{6}S_{11}$, $F_{14}F_{14}V_{8}$, $F_{6}F_{6}V_{3}$, $F_{3}F_{8}V_{8}$, $F_{3}F_{8}V_{3}$, $F_{8}F_{14}V_{8}$, $F_{6}F_{8}V_{3}$, $F_{8}F_{14}S_{11}$, $F_{6}F_{8}S_{11}$}} 
 \SplitRow{ $\mathcal{O}_{8, \; (1,1)}^{3}$} {{$F_{11}F_{9}V_{8}$, $F_{2}F_{5}V_{3}$, $F_{11}F_{8}S_{11}$, $F_{2}F_{8}S_{11}$, $F_{11}F_{8}V_{8}$, $F_{2}F_{8}V_{3}$, $F_{8}F_{9}S_{11}$, $F_{5}F_{8}S_{11}$, $F_{8}F_{9}V_{8}$, $F_{5}F_{8}V_{3}$, $F_{8}F_{8}V_{8}$, $F_{8}F_{8}V_{3}$}} 
 \SplitRow{ $\mathcal{O}_{8, \; (1,1)}^{4}$} {{$F_{4}F_{9}S_{13}$, $F_{4}F_{9}S_{5}$, $F_{4}F_{5}S_{5}$, $F_{1}F_{9}S_{5}$, $F_{1}F_{5}S_{5}$, $F_{11}F_{14}S_{12}$, $F_{11}F_{14}S_{11}$, $F_{2}F_{14}S_{11}$, $F_{11}F_{6}S_{11}$, $F_{2}F_{6}S_{11}$, $F_{9}F_{14}S_{13}$, $F_{9}F_{14}S_{5}$, $F_{5}F_{14}S_{5}$, $F_{6}F_{9}S_{5}$, $F_{5}F_{6}S_{5}$, $F_{9}F_{14}S_{12}$, $F_{9}F_{14}S_{11}$, $F_{5}F_{14}S_{11}$, $F_{6}F_{9}S_{11}$, $F_{5}F_{6}S_{11}$}} 
 \SplitRow{ $\mathcal{O}_{8, \; (1,1)}^{5}$} {{$F_{4}F_{9}V_{8}$, $F_{1}F_{5}V_{3}$, $F_{11}F_{14}V_{8}$, $F_{2}F_{6}V_{3}$, $F_{4}F_{7}S_{5}$, $F_{1}F_{7}S_{5}$, $F_{11}F_{8}V_{8}$, $F_{2}F_{8}V_{3}$, $F_{7}F_{14}S_{5}$, $F_{6}F_{7}S_{5}$, $F_{8}F_{9}V_{8}$, $F_{5}F_{8}V_{3}$, $F_{4}F_{9}V_{5}$, $F_{1}F_{5}V_{4}$, $F_{11}F_{14}V_{7}$, $F_{2}F_{6}V_{6}$, $F_{4}F_{8}S_{5}$, $F_{1}F_{8}S_{5}$, $F_{11}F_{13}V_{7}$, $F_{2}F_{13}V_{6}$, $F_{8}F_{14}S_{5}$, $F_{6}F_{8}S_{5}$, $F_{13}F_{9}V_{5}$, $F_{5}F_{13}V_{4}$, $F_{4}F_{7}V_{5}$, $F_{1}F_{7}V_{4}$, $F_{11}F_{8}S_{11}$, $F_{2}F_{8}S_{11}$, $F_{4}F_{8}V_{8}$, $F_{1}F_{8}V_{3}$, $F_{11}F_{13}S_{11}$, $F_{2}F_{13}S_{11}$, $F_{8}F_{8}V_{8}$, $F_{8}F_{8}V_{3}$, $F_{7}F_{13}V_{5}$, $F_{7}F_{13}V_{4}$, $F_{7}F_{14}V_{7}$, $F_{6}F_{7}V_{6}$, $F_{8}F_{9}S_{11}$, $F_{5}F_{8}S_{11}$, $F_{8}F_{14}V_{8}$, $F_{6}F_{8}V_{3}$, $F_{13}F_{9}S_{11}$, $F_{5}F_{13}S_{11}$, $F_{7}F_{13}V_{7}$, $F_{7}F_{13}V_{6}$}} 
 \SplitRow{ $\mathcal{O}_{8, \; (1,1)}^{6}$} {{$F_{3}F_{9}V_{8}$, $F_{3}F_{5}V_{3}$, $F_{10}F_{14}S_{11}$, $F_{10}F_{6}S_{11}$, $F_{9}F_{14}V_{8}$, $F_{5}F_{6}V_{3}$, $F_{3}F_{9}S_{5}$, $F_{3}F_{5}S_{5}$, $F_{10}F_{14}V_{7}$, $F_{10}F_{6}V_{6}$, $F_{3}F_{8}V_{8}$, $F_{3}F_{8}V_{3}$, $F_{10}F_{13}V_{7}$, $F_{10}F_{13}V_{6}$, $F_{8}F_{14}V_{8}$, $F_{6}F_{8}V_{3}$, $F_{13}F_{9}S_{5}$, $F_{5}F_{13}S_{5}$, $F_{9}F_{14}V_{7}$, $F_{5}F_{6}V_{6}$, $F_{8}F_{14}S_{11}$, $F_{6}F_{8}S_{11}$, $F_{13}F_{9}V_{7}$, $F_{5}F_{13}V_{6}$}} 
 \SplitRow{ $\mathcal{O}_{8, \; (1,1)}^{7}$} {{$F_{3}F_{7}S_{5}$, $F_{10}F_{8}S_{11}$, $F_{3}F_{8}S_{11}$, $F_{10}F_{13}S_{11}$, $F_{8}F_{8}S_{11}$, $F_{7}F_{13}S_{5}$, $F_{3}F_{8}S_{5}$, $F_{10}F_{13}S_{10}$, $F_{8}F_{13}S_{5}$, $F_{7}F_{13}S_{10}$, $F_{8}F_{13}S_{11}$}} 
 \SplitRow{ $\mathcal{O}_{8, \; (1,1)}^{8}$} {{$F_{3}F_{9}V_{5}$, $F_{3}F_{5}V_{4}$, $F_{3}F_{14}S_{11}$, $F_{3}F_{6}S_{11}$, $F_{9}F_{14}V_{5}$, $F_{5}F_{6}V_{4}$, $F_{3}F_{9}S_{5}$, $F_{3}F_{5}S_{5}$, $F_{3}F_{14}V_{8}$, $F_{3}F_{6}V_{3}$, $F_{3}F_{7}V_{5}$, $F_{3}F_{7}V_{4}$, $F_{3}F_{8}V_{8}$, $F_{3}F_{8}V_{3}$, $F_{7}F_{14}V_{5}$, $F_{6}F_{7}V_{4}$, $F_{8}F_{9}S_{5}$, $F_{5}F_{8}S_{5}$, $F_{9}F_{14}V_{8}$, $F_{5}F_{6}V_{3}$, $F_{7}F_{14}S_{11}$, $F_{6}F_{7}S_{11}$, $F_{8}F_{9}V_{8}$, $F_{5}F_{8}V_{3}$}} 
 \SplitRow{ $\mathcal{O}_{8, \; (1,1)}^{9}$} {{$F_{3}F_{9}V_{8}$, $F_{3}F_{5}V_{3}$, $F_{3}F_{9}S_{5}$, $F_{3}F_{5}S_{5}$, $F_{9}F_{9}V_{8}$, $F_{5}F_{5}V_{3}$, $F_{3}F_{8}V_{8}$, $F_{3}F_{8}V_{3}$, $F_{8}F_{9}V_{8}$, $F_{5}F_{8}V_{3}$, $F_{8}F_{9}S_{5}$, $F_{5}F_{8}S_{5}$}} 
 \SplitRow{ $\mathcal{O}_{8, \; (1,1)}^{10}$} {{$F_{3}F_{7}S_{5}$, $F_{3}F_{8}S_{5}$, $F_{7}F_{8}S_{5}$, $F_{3}F_{7}S_{6}$, $F_{3}F_{8}S_{11}$, $F_{3}F_{13}S_{11}$, $F_{8}F_{8}S_{5}$, $F_{7}F_{13}S_{6}$, $F_{7}F_{8}S_{11}$, $F_{7}F_{13}S_{11}$}} 
 \end{SplitTable}
   \vspace{0.4cm} 
 \captionof{table}{Dimension 8 $(\Delta B, \Delta L)$ = (1, 1) UV completions for topology 8 (see Fig.~\ref{fig:d=8_topologies}).} 
 \label{tab:T8Top8B1L1}
 \end{center}

%% file: Tables_dim9/Table_dim9_Top1_B2_L0.tex
 \begin{center}
 \begin{SplitTable}[2.0cm] 
  \SplitHeader{$\mathcal{O}^j{(\mathcal{T}_ {9} ^{ 1 })}$}{New Particles} 
   \vspace{0.1cm} 
 \SplitRow{ $\mathcal{O}_{9, \; (2,0)}^{1}$} {{$S_{23}S_{24}S_{24}$, $S_{22}S_{22}S_{23}$, $S_{23}S_{4}S_{4}$, $S_{23}S_{2}S_{2}$, $S_{24}S_{3}S_{4}$, $S_{22}S_{2}S_{3}$, $S_{3}S_{4}S_{4}$, $S_{2}S_{2}S_{3}$, $S_{24}V_{19}V_{19}$, $S_{22}V_{19}V_{19}$, $S_{4}V_{19}V_{2}$, $S_{2}V_{19}V_{2}$, $S_{24}V_{2}V_{2}$, $S_{22}V_{2}V_{2}$, $S_{4}V_{2}V_{2}$, $S_{2}V_{2}V_{2}$}} 
 \SplitRow{ $\mathcal{O}_{9, \; (2,0)}^{2}$} {{$S_{22}S_{22}S_{23}$, $S_{22}S_{2}S_{3}$, $S_{23}S_{2}S_{2}$, $S_{2}S_{2}S_{3}$, $S_{22}V_{19}V_{19}$, $S_{22}V_{2}V_{2}$, $S_{2}V_{19}V_{2}$, $S_{2}V_{2}V_{2}$, $S_{23}V_{18}V_{19}$, $S_{3}V_{18}V_{2}$, $S_{23}V_{1}V_{2}$, $S_{3}V_{19}V_{1}$, $S_{3}V_{1}V_{2}$}} 
 \SplitRow{ $\mathcal{O}_{9, \; (2,0)}^{3}$} {{$S_{21}S_{23}S_{23}$, $S_{21}S_{3}S_{3}$, $S_{23}S_{1}S_{3}$, $S_{1}S_{3}S_{3}$, $S_{22}S_{22}S_{23}$, $S_{22}S_{2}S_{3}$, $S_{23}S_{2}S_{2}$, $S_{2}S_{2}S_{3}$}} 
 \end{SplitTable}
   \vspace{0.4cm} 
 \captionof{table}{Dimension 9 $(\Delta B, \Delta L)$ = (2, 0) UV completions for topology 1 (see Fig.~\ref{fig:d=9_topologies}).} 
 \label{tab:T9Top1B2L0}
 \end{center}

%% file: Tables_dim9/Table_dim9_Top2_B2_L0.tex
 \begin{center}
 \begin{SplitTable}[2.0cm] 
  \SplitHeader{$\mathcal{O}^j{(\mathcal{T}_ {9} ^{ 2 })}$}{New Particles} 
   \vspace{0.1cm} 
 \SplitRow{ $\mathcal{O}_{9, \; (2,0)}^{1}$} {{$F_{30}S_{24}S_{24}$, $F_{30}S_{24}S_{4}$, $F_{27}S_{22}S_{22}$, $F_{27}S_{22}S_{2}$, $F_{30}S_{4}S_{4}$, $F_{4}S_{4}S_{4}$, $F_{27}S_{2}S_{2}$, $F_{1}S_{2}S_{2}$, $F_{29}S_{24}V_{19}$, $F_{29}S_{24}V_{2}$, $F_{29}S_{22}V_{19}$, $F_{29}S_{22}V_{2}$, $F_{29}S_{4}V_{19}$, $F_{29}S_{4}V_{2}$, $F_{3}S_{4}V_{2}$, $F_{29}S_{2}V_{19}$, $F_{29}S_{2}V_{2}$, $F_{3}S_{2}V_{2}$, $F_{30}S_{24}V_{19}$, $F_{30}S_{4}V_{19}$, $F_{27}S_{22}V_{19}$, $F_{27}S_{2}V_{19}$, $F_{30}S_{24}V_{2}$, $F_{30}S_{4}V_{2}$, $F_{27}S_{22}V_{2}$, $F_{27}S_{2}V_{2}$, $F_{4}S_{4}V_{2}$, $F_{1}S_{2}V_{2}$, $F_{29}S_{23}S_{24}$, $F_{29}S_{24}S_{3}$, $F_{29}S_{22}S_{23}$, $F_{29}S_{22}S_{3}$, $F_{29}S_{23}S_{4}$, $F_{29}S_{3}S_{4}$, $F_{3}S_{3}S_{4}$, $F_{29}S_{23}S_{2}$, $F_{29}S_{2}S_{3}$, $F_{3}S_{2}S_{3}$, $F_{30}V_{19}V_{19}$, $F_{30}V_{19}V_{2}$, $F_{27}V_{19}V_{19}$, $F_{27}V_{19}V_{2}$, $F_{30}V_{2}V_{2}$, $F_{27}V_{2}V_{2}$, $F_{4}V_{2}V_{2}$, $F_{1}V_{2}V_{2}$}} 
 \SplitRow{ $\mathcal{O}_{9, \; (2,0)}^{2}$} {{$F_{27}S_{22}S_{23}$, $F_{27}S_{22}S_{3}$, $F_{27}S_{23}S_{2}$, $F_{27}S_{2}S_{3}$, $F_{1}S_{2}S_{3}$, $F_{29}V_{19}V_{19}$, $F_{29}V_{19}V_{2}$, $F_{29}V_{2}V_{2}$, $F_{3}V_{2}V_{2}$, $F_{28}S_{22}S_{23}$, $F_{28}S_{23}S_{2}$, $F_{28}S_{22}S_{3}$, $F_{28}S_{2}S_{3}$, $F_{2}S_{2}S_{3}$, $F_{27}S_{23}V_{19}$, $F_{27}S_{3}V_{19}$, $F_{27}S_{23}V_{2}$, $F_{27}S_{3}V_{2}$, $F_{1}S_{3}V_{2}$, $F_{29}S_{23}V_{19}$, $F_{29}S_{23}V_{2}$, $F_{29}S_{3}V_{19}$, $F_{29}S_{3}V_{2}$, $F_{3}S_{3}V_{2}$, $F_{27}S_{22}S_{22}$, $F_{27}S_{22}S_{2}$, $F_{27}S_{2}S_{2}$, $F_{1}S_{2}S_{2}$, $F_{29}V_{18}V_{19}$, $F_{29}V_{19}V_{1}$, $F_{29}V_{18}V_{2}$, $F_{29}V_{1}V_{2}$, $F_{3}V_{1}V_{2}$, $F_{27}S_{22}V_{19}$, $F_{27}S_{2}V_{19}$, $F_{27}S_{22}V_{2}$, $F_{27}S_{2}V_{2}$, $F_{1}S_{2}V_{2}$, $F_{29}S_{23}V_{18}$, $F_{29}S_{23}V_{1}$, $F_{29}S_{3}V_{18}$, $F_{29}S_{3}V_{1}$, $F_{3}S_{3}V_{1}$, $F_{28}S_{23}V_{18}$, $F_{28}S_{3}V_{18}$, $F_{28}S_{23}V_{1}$, $F_{28}S_{3}V_{1}$, $F_{2}S_{3}V_{1}$, $F_{29}S_{22}V_{19}$, $F_{29}S_{22}V_{2}$, $F_{29}S_{2}V_{19}$, $F_{29}S_{2}V_{2}$, $F_{3}S_{2}V_{2}$, $F_{29}S_{22}S_{23}$, $F_{29}S_{23}S_{2}$, $F_{29}S_{22}S_{3}$, $F_{29}S_{2}S_{3}$, $F_{3}S_{2}S_{3}$}} 
 \SplitRow{ $\mathcal{O}_{9, \; (2,0)}^{3}$} {{$F_{27}S_{23}S_{23}$, $F_{27}S_{23}S_{3}$, $F_{27}S_{3}S_{3}$, $F_{1}S_{3}S_{3}$, $F_{28}S_{22}S_{23}$, $F_{28}S_{23}S_{2}$, $F_{28}S_{22}S_{3}$, $F_{28}S_{2}S_{3}$, $F_{2}S_{2}S_{3}$, $F_{27}S_{22}S_{23}$, $F_{27}S_{22}S_{3}$, $F_{27}S_{23}S_{2}$, $F_{27}S_{2}S_{3}$, $F_{1}S_{2}S_{3}$, $F_{28}S_{21}S_{23}$, $F_{28}S_{23}S_{1}$, $F_{28}S_{21}S_{3}$, $F_{28}S_{1}S_{3}$, $F_{2}S_{1}S_{3}$, $F_{27}S_{22}S_{22}$, $F_{27}S_{22}S_{2}$, $F_{27}S_{2}S_{2}$, $F_{1}S_{2}S_{2}$}} 
 \end{SplitTable}
   \vspace{0.4cm} 
 \captionof{table}{Dimension 9 $(\Delta B, \Delta L)$ = (2, 0) UV completions for topology 2 (see Fig.~\ref{fig:d=9_topologies}).} 
 \label{tab:T9Top2B2L0}
 \end{center}

%% file: Tables_dim9/Table_dim9_Top1_B1_L3.tex
 \begin{center}
 \begin{SplitTable}[2.0cm] 
  \SplitHeader{$\mathcal{O}^j{(\mathcal{T}_ {9} ^{ 1 })}$}{New Particles} 
   \vspace{0.1cm} 
 \SplitRow{ $\mathcal{O}_{9, \; (1,3)}^{1}$} {{$S_{9}S_{1}S_{4}$, $S_{14}S_{1}S_{2}$, $S_{9}V_{1}V_{2}$, $S_{14}V_{1}V_{2}$, $S_{4}V_{2}V_{2}$, $S_{2}V_{2}V_{2}$}} 
 \SplitRow{ $\mathcal{O}_{9, \; (1,3)}^{2}$} {{$S_{14}S_{1}S_{2}$, $S_{1}V_{12}V_{2}$, $S_{2}V_{2}V_{2}$}} 
 \SplitRow{ $\mathcal{O}_{9, \; (1,3)}^{3}$} {{$S_{4}V_{11}V_{1}$, $S_{2}V_{11}V_{1}$, $S_{9}V_{1}V_{2}$, $S_{14}V_{1}V_{2}$, $S_{4}V_{11}V_{2}$, $S_{2}V_{11}V_{2}$, $S_{4}V_{2}V_{2}$, $S_{2}V_{2}V_{2}$, $S_{9}S_{3}S_{4}$, $S_{14}S_{2}S_{3}$, $S_{3}S_{4}S_{4}$, $S_{2}S_{2}S_{3}$}} 
 \SplitRow{ $\mathcal{O}_{9, \; (1,3)}^{4}$} {{$S_{1}V_{11}V_{1}$, $S_{14}S_{1}S_{2}$, $S_{2}V_{11}V_{2}$, $S_{14}S_{2}S_{3}$, $S_{2}V_{2}V_{2}$, $S_{3}V_{1}V_{2}$}} 
 \SplitRow{ $\mathcal{O}_{9, \; (1,3)}^{5}$} {{$S_{1}V_{11}V_{1}$, $S_{14}S_{1}S_{2}$, $S_{1}V_{12}V_{2}$, $S_{2}V_{11}V_{1}$, $S_{14}V_{1}V_{2}$, $S_{3}V_{12}V_{1}$, $S_{2}V_{2}V_{2}$, $S_{2}S_{2}S_{3}$, $S_{3}V_{1}V_{2}$}} 
 \SplitRow{ $\mathcal{O}_{9, \; (1,3)}^{6}$} {{$S_{14}S_{1}S_{2}$, $S_{15}S_{1}S_{3}$, $S_{2}S_{2}S_{3}$}} 
 \SplitRow{ $\mathcal{O}_{9, \; (1,3)}^{7}$} {{$S_{7}S_{4}S_{4}$, $S_{7}S_{2}S_{2}$, $S_{4}V_{11}V_{2}$, $S_{2}V_{11}V_{2}$, $S_{4}V_{2}V_{2}$, $S_{2}V_{2}V_{2}$}} 
 \SplitRow{ $\mathcal{O}_{9, \; (1,3)}^{8}$} {{$S_{7}S_{2}S_{2}$, $S_{2}V_{11}V_{2}$, $S_{7}V_{1}V_{1}$, $S_{2}V_{11}V_{1}$, $S_{7}V_{2}V_{2}$, $S_{2}V_{2}V_{2}$, $S_{3}V_{11}V_{2}$, $S_{3}V_{1}V_{2}$, $S_{2}S_{2}S_{3}$}} 
 \SplitRow{ $\mathcal{O}_{9, \; (1,3)}^{9}$} {{$S_{7}V_{1}V_{1}$, $S_{14}V_{1}V_{2}$, $S_{7}S_{2}S_{2}$, $S_{2}V_{2}V_{2}$, $S_{14}S_{2}S_{3}$, $S_{3}V_{1}V_{2}$}} 
 \SplitRow{ $\mathcal{O}_{9, \; (1,3)}^{10}$} {{$S_{7}S_{1}S_{1}$, $S_{14}S_{1}S_{2}$, $S_{7}S_{2}S_{2}$, $S_{14}S_{2}S_{3}$, $S_{2}S_{2}S_{3}$, $S_{1}S_{3}S_{3}$}} 
 \SplitRow{ $\mathcal{O}_{9, \; (1,3)}^{11}$} {{$S_{7}V_{2}V_{2}$, $S_{7}S_{2}S_{2}$, $S_{2}V_{2}V_{2}$}} 
 \SplitRow{ $\mathcal{O}_{9, \; (1,3)}^{12}$} {{$S_{7}S_{2}S_{2}$, $S_{7}S_{3}S_{3}$, $S_{2}S_{2}S_{3}$}} 
 \end{SplitTable}
   \vspace{0.4cm} 
 \captionof{table}{Dimension 9 $(\Delta B, \Delta L)$ = (1, 3) UV completions for topology 1 (see Fig.~\ref{fig:d=9_topologies}).} 
 \label{tab:T9Top1B1L3}
 \end{center}

%% file: Tables_dim9/Table_dim9_Top2_B1_L3.tex
 \begin{center}
 \begin{SplitTable}[2.0cm] 
  \SplitHeader{$\mathcal{O}^j{(\mathcal{T}_ {9} ^{ 2 })}$}{New Particles} 
   \vspace{0.1cm} 
 \SplitRow{ $\mathcal{O}_{9, \; (1,3)}^{1}$} {{$F_{9}S_{9}S_{4}$, $F_{5}S_{14}S_{2}$, $F_{7}S_{9}V_{2}$, $F_{7}S_{14}V_{2}$, $F_{9}S_{9}V_{2}$, $F_{5}S_{14}V_{2}$, $F_{10}S_{9}V_{1}$, $F_{10}S_{14}V_{1}$, $F_{7}S_{4}V_{2}$, $F_{7}S_{2}V_{2}$, $F_{9}S_{4}V_{2}$, $F_{5}S_{2}V_{2}$, $F_{9}S_{9}V_{1}$, $F_{5}S_{14}V_{1}$, $F_{10}S_{9}S_{1}$, $F_{10}S_{14}S_{1}$, $F_{9}V_{2}V_{2}$, $F_{5}V_{2}V_{2}$, $F_{7}S_{9}S_{1}$, $F_{7}S_{14}S_{1}$, $F_{7}S_{1}S_{4}$, $F_{7}S_{1}S_{2}$, $F_{9}V_{1}V_{2}$, $F_{5}V_{1}V_{2}$}} 
 \SplitRow{ $\mathcal{O}_{9, \; (1,3)}^{2}$} {{$F_{5}S_{14}S_{2}$, $F_{7}V_{12}V_{2}$, $F_{16}S_{14}S_{1}$, $F_{7}V_{2}V_{2}$, $F_{5}S_{14}S_{1}$, $F_{16}S_{1}V_{12}$, $F_{5}S_{2}V_{2}$, $F_{7}S_{2}V_{2}$, $F_{7}S_{1}V_{12}$, $F_{5}S_{1}V_{2}$, $F_{7}S_{1}V_{2}$, $F_{7}S_{1}S_{2}$}} 
 \SplitRow{ $\mathcal{O}_{9, \; (1,3)}^{3}$} {{$F_{9}S_{4}V_{11}$, $F_{5}S_{2}V_{11}$, $F_{7}S_{9}V_{2}$, $F_{7}S_{14}V_{2}$, $F_{9}S_{9}V_{2}$, $F_{5}S_{14}V_{2}$, $F_{8}S_{4}V_{11}$, $F_{8}S_{2}V_{11}$, $F_{11}S_{4}V_{11}$, $F_{2}S_{2}V_{11}$, $F_{9}S_{4}V_{2}$, $F_{5}S_{2}V_{2}$, $F_{7}S_{4}V_{2}$, $F_{7}S_{2}V_{2}$, $F_{11}S_{9}S_{4}$, $F_{2}S_{14}S_{2}$, $F_{9}S_{4}S_{4}$, $F_{5}S_{2}S_{2}$, $F_{9}S_{9}S_{4}$, $F_{5}S_{14}S_{2}$, $F_{9}V_{11}V_{2}$, $F_{5}V_{11}V_{2}$, $F_{7}S_{9}S_{3}$, $F_{7}S_{14}S_{3}$, $F_{11}V_{11}V_{1}$, $F_{2}V_{11}V_{1}$, $F_{9}V_{2}V_{2}$, $F_{5}V_{2}V_{2}$, $F_{7}S_{3}S_{4}$, $F_{7}S_{2}S_{3}$, $F_{8}S_{3}S_{4}$, $F_{8}S_{2}S_{3}$, $F_{9}V_{11}V_{1}$, $F_{5}V_{11}V_{1}$, $F_{11}S_{9}V_{1}$, $F_{2}S_{14}V_{1}$, $F_{8}S_{4}V_{2}$, $F_{8}S_{2}V_{2}$, $F_{7}S_{9}V_{1}$, $F_{7}S_{14}V_{1}$, $F_{9}V_{1}V_{2}$, $F_{5}V_{1}V_{2}$, $F_{9}S_{4}V_{1}$, $F_{5}S_{2}V_{1}$, $F_{7}S_{4}V_{1}$, $F_{7}S_{2}V_{1}$}} 
 \SplitRow{ $\mathcal{O}_{9, \; (1,3)}^{4}$} {{$F_{8}V_{11}V_{1}$, $F_{5}S_{14}S_{2}$, $F_{7}V_{11}V_{2}$, $F_{17}S_{14}S_{3}$, $F_{5}S_{14}S_{3}$, $F_{8}V_{11}V_{2}$, $F_{2}S_{2}V_{11}$, $F_{7}S_{2}V_{2}$, $F_{17}S_{3}V_{1}$, $F_{8}S_{3}V_{1}$, $F_{5}S_{2}V_{2}$, $F_{8}S_{2}V_{11}$, $F_{2}S_{14}S_{2}$, $F_{7}V_{1}V_{2}$, $F_{8}V_{1}V_{2}$, $F_{2}S_{1}V_{11}$, $F_{8}S_{3}V_{2}$, $F_{5}S_{3}V_{2}$, $F_{7}S_{1}V_{11}$, $F_{2}S_{14}S_{1}$, $F_{8}V_{2}V_{2}$, $F_{17}S_{14}S_{1}$, $F_{7}S_{1}S_{2}$, $F_{8}S_{2}S_{3}$, $F_{7}S_{1}V_{1}$, $F_{8}S_{2}V_{2}$, $F_{17}S_{1}V_{1}$}} 
 \SplitRow{ $\mathcal{O}_{9, \; (1,3)}^{5}$} {{$F_{8}V_{11}V_{1}$, $F_{5}S_{14}S_{2}$, $F_{7}V_{12}V_{2}$, $F_{5}S_{2}V_{11}$, $F_{7}S_{14}V_{2}$, $F_{17}S_{3}V_{12}$, $F_{7}S_{3}V_{12}$, $F_{5}S_{14}V_{2}$, $F_{8}S_{2}V_{11}$, $F_{10}V_{11}V_{1}$, $F_{7}V_{2}V_{2}$, $F_{17}S_{2}S_{3}$, $F_{5}S_{2}S_{3}$, $F_{10}S_{14}V_{1}$, $F_{5}S_{2}V_{2}$, $F_{17}S_{3}V_{1}$, $F_{8}S_{3}V_{1}$, $F_{7}S_{2}V_{2}$, $F_{5}S_{14}V_{1}$, $F_{10}V_{12}V_{1}$, $F_{5}S_{2}S_{2}$, $F_{7}V_{1}V_{2}$, $F_{8}V_{1}V_{2}$, $F_{7}V_{12}V_{1}$, $F_{10}S_{1}V_{11}$, $F_{5}S_{3}V_{2}$, $F_{7}S_{3}V_{2}$, $F_{5}S_{1}V_{11}$, $F_{10}S_{14}S_{1}$, $F_{8}S_{2}S_{3}$, $F_{7}S_{2}S_{3}$, $F_{7}S_{14}S_{1}$, $F_{10}S_{1}V_{12}$, $F_{8}S_{2}V_{2}$, $F_{17}S_{1}V_{12}$, $F_{5}S_{1}V_{2}$, $F_{5}S_{3}V_{1}$, $F_{7}S_{1}V_{2}$, $F_{5}S_{1}S_{2}$, $F_{17}S_{1}S_{2}$, $F_{7}S_{1}V_{1}$, $F_{5}S_{2}V_{1}$, $F_{7}S_{2}V_{1}$, $F_{17}S_{1}V_{1}$}} 
 \SplitRow{ $\mathcal{O}_{9, \; (1,3)}^{6}$} {{$F_{5}S_{14}S_{2}$, $F_{17}S_{15}S_{3}$, $F_{16}S_{14}S_{1}$, $F_{5}S_{2}S_{3}$, $F_{17}S_{2}S_{3}$, $F_{5}S_{14}S_{1}$, $F_{16}S_{15}S_{1}$, $F_{5}S_{2}S_{2}$, $F_{17}S_{15}S_{1}$, $F_{5}S_{1}S_{3}$, $F_{5}S_{1}S_{2}$, $F_{17}S_{1}S_{2}$}} 
 \SplitRow{ $\mathcal{O}_{9, \; (1,3)}^{7}$} {{$F_{8}S_{7}S_{4}$, $F_{8}S_{7}S_{2}$, $F_{9}V_{11}V_{2}$, $F_{5}V_{11}V_{2}$, $F_{3}S_{7}S_{4}$, $F_{3}S_{7}S_{2}$, $F_{9}V_{2}V_{2}$, $F_{5}V_{2}V_{2}$, $F_{3}S_{4}V_{11}$, $F_{3}S_{2}V_{11}$, $F_{8}S_{4}V_{2}$, $F_{8}S_{2}V_{2}$, $F_{9}S_{4}V_{2}$, $F_{5}S_{2}V_{2}$, $F_{9}S_{4}V_{11}$, $F_{5}S_{2}V_{11}$, $F_{9}S_{4}S_{4}$, $F_{5}S_{2}S_{2}$}} 
 \SplitRow{ $\mathcal{O}_{9, \; (1,3)}^{8}$} {{$F_{5}S_{7}S_{2}$, $F_{7}V_{11}V_{2}$, $F_{8}V_{11}V_{2}$, $F_{6}S_{7}S_{2}$, $F_{8}S_{7}V_{1}$, $F_{5}S_{2}V_{11}$, $F_{8}S_{2}V_{11}$, $F_{6}S_{7}V_{1}$, $F_{3}S_{7}V_{2}$, $F_{5}S_{2}V_{2}$, $F_{7}S_{2}V_{2}$, $F_{6}S_{7}V_{2}$, $F_{3}V_{11}V_{2}$, $F_{8}V_{1}V_{2}$, $F_{5}S_{2}S_{2}$, $F_{7}V_{1}V_{2}$, $F_{8}S_{7}V_{2}$, $F_{5}S_{3}V_{11}$, $F_{7}S_{3}V_{11}$, $F_{5}S_{7}V_{2}$, $F_{3}S_{7}V_{1}$, $F_{5}S_{3}V_{2}$, $F_{8}S_{3}V_{2}$, $F_{5}S_{7}V_{1}$, $F_{3}V_{11}V_{1}$, $F_{8}V_{2}V_{2}$, $F_{5}S_{2}S_{3}$, $F_{6}S_{2}S_{3}$, $F_{7}V_{11}V_{1}$, $F_{3}S_{7}S_{2}$, $F_{7}S_{2}S_{3}$, $F_{8}S_{2}S_{3}$, $F_{8}S_{7}S_{2}$, $F_{3}S_{2}V_{11}$, $F_{7}S_{3}V_{1}$, $F_{6}S_{3}V_{1}$, $F_{8}S_{2}V_{2}$, $F_{5}S_{2}V_{1}$, $F_{6}S_{3}V_{2}$, $F_{7}S_{2}V_{1}$, $F_{7}V_{1}V_{1}$}} 
 \SplitRow{ $\mathcal{O}_{9, \; (1,3)}^{9}$} {{$F_{8}S_{7}V_{1}$, $F_{7}S_{14}V_{2}$, $F_{5}S_{14}V_{2}$, $F_{6}S_{7}V_{1}$, $F_{2}S_{7}S_{2}$, $F_{7}V_{2}V_{2}$, $F_{6}S_{7}S_{2}$, $F_{2}S_{14}S_{2}$, $F_{8}V_{1}V_{2}$, $F_{7}V_{1}V_{2}$, $F_{5}S_{14}S_{2}$, $F_{8}S_{7}S_{2}$, $F_{7}S_{14}S_{3}$, $F_{2}S_{7}V_{1}$, $F_{7}S_{3}V_{2}$, $F_{5}S_{3}V_{2}$, $F_{2}S_{14}V_{1}$, $F_{8}S_{2}V_{2}$, $F_{7}S_{3}V_{1}$, $F_{6}S_{3}V_{1}$, $F_{5}S_{2}V_{2}$, $F_{7}S_{14}V_{1}$, $F_{5}S_{2}S_{3}$, $F_{6}S_{2}S_{3}$, $F_{7}V_{1}V_{1}$, $F_{5}S_{2}S_{2}$}} 
 \SplitRow{ $\mathcal{O}_{9, \; (1,3)}^{10}$} {{$F_{6}S_{7}S_{1}$, $F_{5}S_{14}S_{2}$, $F_{5}S_{7}S_{2}$, $F_{17}S_{14}S_{3}$, $F_{5}S_{14}S_{3}$, $F_{6}S_{7}S_{2}$, $F_{2}S_{7}S_{2}$, $F_{17}S_{2}S_{3}$, $F_{5}S_{2}S_{3}$, $F_{2}S_{14}S_{2}$, $F_{5}S_{2}S_{2}$, $F_{17}S_{1}S_{3}$, $F_{6}S_{1}S_{3}$, $F_{2}S_{7}S_{1}$, $F_{5}S_{3}S_{3}$, $F_{5}S_{7}S_{1}$, $F_{2}S_{14}S_{1}$, $F_{6}S_{2}S_{3}$, $F_{17}S_{14}S_{1}$, $F_{5}S_{1}S_{2}$, $F_{17}S_{1}S_{2}$, $F_{17}S_{1}S_{1}$}} 
 \SplitRow{ $\mathcal{O}_{9, \; (1,3)}^{11}$} {{$F_{8}S_{7}V_{2}$, $F_{5}S_{7}V_{2}$, $F_{1}S_{7}S_{2}$, $F_{8}V_{2}V_{2}$, $F_{5}S_{7}S_{2}$, $F_{8}S_{7}S_{2}$, $F_{1}S_{7}V_{2}$, $F_{8}S_{2}V_{2}$, $F_{5}S_{2}V_{2}$, $F_{5}S_{2}S_{2}$}} 
 \SplitRow{ $\mathcal{O}_{9, \; (1,3)}^{12}$} {{$F_{5}S_{7}S_{2}$, $F_{6}S_{7}S_{2}$, $F_{1}S_{7}S_{3}$, $F_{5}S_{2}S_{2}$, $F_{6}S_{7}S_{3}$, $F_{5}S_{7}S_{3}$, $F_{1}S_{7}S_{2}$, $F_{5}S_{2}S_{3}$, $F_{6}S_{2}S_{3}$, $F_{6}S_{3}S_{3}$}} 
 \end{SplitTable}
   \vspace{0.4cm} 
 \captionof{table}{Dimension 9 $(\Delta B, \Delta L)$ = (1, 3) UV completions for topology 2 (see Fig.~\ref{fig:d=9_topologies}).} 
 \label{tab:T9Top2B1L3}
 \end{center}

%% file: Tables_dim9/Table_dim9_Top1_B1_L-1.tex
 \begin{center}
 \begin{SplitTable}[2.0cm] 
  \SplitHeader{$\mathcal{O}^j{(\mathcal{T}_ {9} ^{ 1 })}$}{New Particles} 
   \vspace{0.1cm} 
 \SplitRow{ $\mathcal{O}_{9, \; (1,-1)}^{1}$} {{$S_{23}S_{1}S_{3}$, $S_{1}S_{3}S_{3}$, $S_{23}V_{4}V_{20}$, $S_{3}V_{4}V_{20}$, $S_{3}V_{9}V_{4}$}} 
 \SplitRow{ $\mathcal{O}_{9, \; (1,-1)}^{2}$} {{$S_{15}S_{1}S_{3}$, $S_{1}V_{4}V_{4}$, $S_{3}V_{9}V_{4}$}} 
 \SplitRow{ $\mathcal{O}_{9, \; (1,-1)}^{3}$} {{$S_{23}V_{1}V_{2}$, $S_{3}V_{19}V_{1}$, $S_{3}V_{1}V_{2}$, $S_{25}V_{19}V_{4}$, $S_{25}V_{4}V_{2}$, $S_{16}V_{4}V_{2}$, $S_{23}V_{4}V_{20}$, $S_{3}V_{4}V_{20}$, $S_{3}V_{9}V_{4}$, $S_{23}S_{6}S_{25}$, $S_{3}S_{6}S_{25}$, $S_{16}S_{3}S_{6}$}} 
 \SplitRow{ $\mathcal{O}_{9, \; (1,-1)}^{4}$} {{$S_{23}S_{2}S_{2}$, $S_{22}S_{2}S_{3}$, $S_{2}S_{2}S_{3}$, $S_{2}V_{19}V_{2}$, $S_{2}V_{2}V_{2}$, $S_{22}V_{4}V_{21}$, $S_{2}V_{4}V_{21}$, $S_{2}V_{10}V_{4}$, $S_{25}V_{19}V_{4}$, $S_{25}V_{4}V_{2}$, $S_{16}V_{4}V_{2}$, $S_{6}V_{19}V_{21}$, $S_{6}V_{2}V_{21}$, $S_{6}V_{10}V_{2}$, $S_{23}S_{6}S_{25}$, $S_{3}S_{6}S_{25}$, $S_{16}S_{3}S_{6}$}} 
 \SplitRow{ $\mathcal{O}_{9, \; (1,-1)}^{5}$} {{$S_{22}S_{2}S_{3}$, $S_{23}S_{2}S_{2}$, $S_{2}S_{2}S_{3}$, $S_{23}V_{14}V_{21}$, $S_{3}V_{14}V_{21}$, $S_{3}V_{10}V_{14}$, $S_{23}V_{4}V_{20}$, $S_{3}V_{4}V_{20}$, $S_{3}V_{9}V_{4}$, $S_{22}V_{4}V_{21}$, $S_{2}V_{4}V_{21}$, $S_{2}V_{10}V_{4}$}} 
 \SplitRow{ $\mathcal{O}_{9, \; (1,-1)}^{6}$} {{$S_{23}V_{1}V_{2}$, $S_{3}V_{19}V_{1}$, $S_{3}V_{1}V_{2}$, $S_{5}V_{19}V_{20}$, $S_{5}V_{2}V_{20}$, $S_{5}V_{9}V_{2}$, $S_{23}S_{5}S_{25}$, $S_{3}S_{5}S_{25}$, $S_{16}S_{3}S_{5}$, $S_{23}V_{4}V_{20}$, $S_{3}V_{4}V_{20}$, $S_{3}V_{9}V_{4}$}} 
 \SplitRow{ $\mathcal{O}_{9, \; (1,-1)}^{7}$} {{$S_{3}V_{12}V_{1}$, $S_{5}V_{4}V_{1}$, $S_{16}S_{3}S_{5}$, $S_{3}V_{9}V_{4}$}} 
 \SplitRow{ $\mathcal{O}_{9, \; (1,-1)}^{8}$} {{$S_{3}V_{12}V_{1}$, $S_{1}V_{12}V_{2}$, $S_{5}V_{4}V_{1}$, $S_{1}S_{5}S_{6}$, $S_{5}V_{9}V_{2}$, $S_{1}V_{4}V_{4}$, $S_{3}V_{9}V_{4}$, $S_{16}V_{4}V_{2}$, $S_{16}S_{3}S_{6}$}} 
 \SplitRow{ $\mathcal{O}_{9, \; (1,-1)}^{9}$} {{$S_{9}S_{3}S_{4}$, $S_{14}S_{2}S_{3}$, $S_{9}V_{1}V_{2}$, $S_{14}V_{1}V_{2}$, $S_{5}S_{5}S_{4}$, $S_{2}S_{5}S_{5}$, $S_{5}V_{1}V_{5}$, $S_{5}V_{4}V_{1}$, $S_{5}V_{13}V_{2}$, $S_{5}V_{9}V_{2}$, $S_{3}V_{13}V_{5}$, $S_{3}V_{9}V_{4}$}} 
 \SplitRow{ $\mathcal{O}_{9, \; (1,-1)}^{10}$} {{$S_{9}V_{1}V_{2}$, $S_{14}V_{1}V_{2}$, $S_{9}S_{1}S_{4}$, $S_{14}S_{1}S_{2}$, $S_{5}V_{1}V_{5}$, $S_{5}V_{4}V_{1}$, $S_{16}S_{5}S_{4}$, $S_{16}S_{2}S_{5}$, $S_{1}V_{5}V_{5}$, $S_{1}V_{4}V_{4}$, $S_{16}V_{2}V_{5}$, $S_{16}V_{4}V_{2}$}} 
 \SplitRow{ $\mathcal{O}_{9, \; (1,-1)}^{11}$} {{$S_{14}S_{1}S_{2}$, $S_{14}S_{2}S_{3}$, $S_{1}S_{5}S_{6}$, $S_{16}S_{3}S_{6}$, $S_{2}S_{5}S_{5}$, $S_{16}S_{2}S_{5}$}} 
 \SplitRow{ $\mathcal{O}_{9, \; (1,-1)}^{12}$} {{$S_{23}S_{4}S_{4}$, $S_{24}S_{3}S_{4}$, $S_{3}S_{4}S_{4}$, $S_{23}S_{2}S_{2}$, $S_{22}S_{2}S_{3}$, $S_{2}S_{2}S_{3}$, $S_{4}V_{19}V_{2}$, $S_{4}V_{2}V_{2}$, $S_{2}V_{19}V_{2}$, $S_{2}V_{2}V_{2}$, $S_{24}S_{5}S_{25}$, $S_{22}S_{5}S_{25}$, $S_{5}S_{4}S_{25}$, $S_{2}S_{5}S_{25}$, $S_{16}S_{5}S_{4}$, $S_{16}S_{2}S_{5}$, $S_{5}V_{19}V_{22}$, $S_{5}V_{2}V_{22}$, $S_{5}V_{19}V_{20}$, $S_{5}V_{2}V_{20}$, $S_{5}V_{13}V_{2}$, $S_{5}V_{9}V_{2}$, $S_{25}V_{19}V_{5}$, $S_{25}V_{2}V_{5}$, $S_{16}V_{2}V_{5}$, $S_{25}V_{19}V_{4}$, $S_{25}V_{4}V_{2}$, $S_{16}V_{4}V_{2}$, $S_{23}V_{5}V_{22}$, $S_{3}V_{5}V_{22}$, $S_{3}V_{13}V_{5}$, $S_{23}V_{4}V_{20}$, $S_{3}V_{4}V_{20}$, $S_{3}V_{9}V_{4}$}} 
 \SplitRow{ $\mathcal{O}_{9, \; (1,-1)}^{13}$} {{$S_{22}S_{2}S_{3}$, $S_{23}S_{2}S_{2}$, $S_{2}S_{2}S_{3}$, $S_{23}S_{6}S_{25}$, $S_{3}S_{6}S_{25}$, $S_{16}S_{3}S_{6}$, $S_{23}S_{5}S_{25}$, $S_{3}S_{5}S_{25}$, $S_{16}S_{3}S_{5}$, $S_{22}S_{5}S_{25}$, $S_{2}S_{5}S_{25}$, $S_{16}S_{2}S_{5}$}} 
 \SplitRow{ $\mathcal{O}_{9, \; (1,-1)}^{14}$} {{$S_{4}V_{19}V_{2}$, $S_{2}V_{19}V_{2}$, $S_{24}V_{2}V_{2}$, $S_{22}V_{2}V_{2}$, $S_{4}V_{2}V_{2}$, $S_{2}V_{2}V_{2}$, $S_{24}S_{5}S_{25}$, $S_{22}S_{5}S_{25}$, $S_{5}S_{4}S_{25}$, $S_{2}S_{5}S_{25}$, $S_{16}S_{5}S_{4}$, $S_{16}S_{2}S_{5}$, $S_{24}V_{5}V_{21}$, $S_{4}V_{5}V_{21}$, $S_{4}V_{10}V_{5}$, $S_{22}V_{4}V_{21}$, $S_{2}V_{4}V_{21}$, $S_{2}V_{10}V_{4}$, $S_{25}V_{19}V_{5}$, $S_{25}V_{2}V_{5}$, $S_{16}V_{2}V_{5}$, $S_{25}V_{19}V_{4}$, $S_{25}V_{4}V_{2}$, $S_{16}V_{4}V_{2}$}} 
 \SplitRow{ $\mathcal{O}_{9, \; (1,-1)}^{15}$} {{$S_{22}V_{2}V_{2}$, $S_{2}V_{19}V_{2}$, $S_{2}V_{2}V_{2}$, $S_{3}V_{18}V_{2}$, $S_{23}V_{1}V_{2}$, $S_{3}V_{1}V_{2}$, $S_{6}V_{19}V_{21}$, $S_{6}V_{2}V_{21}$, $S_{6}V_{10}V_{2}$, $S_{23}S_{6}S_{25}$, $S_{3}S_{6}S_{25}$, $S_{16}S_{3}S_{6}$, $S_{5}V_{19}V_{20}$, $S_{5}V_{2}V_{20}$, $S_{5}V_{9}V_{2}$, $S_{5}V_{18}V_{21}$, $S_{5}V_{1}V_{21}$, $S_{5}V_{10}V_{1}$, $S_{22}S_{5}S_{25}$, $S_{2}S_{5}S_{25}$, $S_{16}S_{2}S_{5}$, $S_{23}V_{4}V_{20}$, $S_{3}V_{4}V_{20}$, $S_{3}V_{9}V_{4}$, $S_{22}V_{4}V_{21}$, $S_{2}V_{4}V_{21}$, $S_{2}V_{10}V_{4}$}} 
 \SplitRow{ $\mathcal{O}_{9, \; (1,-1)}^{16}$} {{$S_{14}S_{2}S_{3}$, $S_{2}S_{5}S_{5}$, $S_{16}S_{3}S_{5}$}} 
 \SplitRow{ $\mathcal{O}_{9, \; (1,-1)}^{17}$} {{$S_{23}S_{2}S_{2}$, $S_{22}S_{2}S_{3}$, $S_{2}S_{2}S_{3}$, $S_{2}V_{19}V_{2}$, $S_{2}V_{2}V_{2}$, $S_{22}V_{3}V_{20}$, $S_{2}V_{3}V_{20}$, $S_{2}V_{9}V_{3}$, $S_{25}V_{19}V_{3}$, $S_{25}V_{3}V_{2}$, $S_{16}V_{3}V_{2}$, $S_{5}V_{19}V_{20}$, $S_{5}V_{2}V_{20}$, $S_{5}V_{9}V_{2}$, $S_{23}S_{5}S_{25}$, $S_{3}S_{5}S_{25}$, $S_{16}S_{3}S_{5}$}} 
 \SplitRow{ $\mathcal{O}_{9, \; (1,-1)}^{18}$} {{$S_{22}S_{2}S_{3}$, $S_{23}S_{2}S_{2}$, $S_{2}S_{2}S_{3}$, $S_{23}V_{4}V_{20}$, $S_{3}V_{4}V_{20}$, $S_{3}V_{9}V_{4}$, $S_{23}V_{3}V_{21}$, $S_{3}V_{3}V_{21}$, $S_{3}V_{10}V_{3}$, $S_{22}V_{3}V_{20}$, $S_{2}V_{3}V_{20}$, $S_{2}V_{9}V_{3}$}} 
 \SplitRow{ $\mathcal{O}_{9, \; (1,-1)}^{19}$} {{$S_{14}S_{2}S_{3}$, $S_{14}V_{1}V_{2}$, $S_{2}V_{3}V_{4}$, $S_{6}V_{3}V_{1}$, $S_{16}V_{3}V_{2}$, $S_{5}V_{4}V_{1}$, $S_{16}S_{3}S_{5}$, $S_{16}V_{4}V_{2}$, $S_{16}S_{3}S_{6}$}} 
 \SplitRow{ $\mathcal{O}_{9, \; (1,-1)}^{20}$} {{$S_{14}V_{1}V_{2}$, $S_{14}S_{1}S_{2}$, $S_{6}V_{3}V_{1}$, $S_{2}V_{9}V_{3}$, $S_{5}V_{4}V_{1}$, $S_{1}S_{5}S_{6}$, $S_{5}V_{9}V_{2}$, $S_{2}V_{10}V_{4}$, $S_{6}V_{10}V_{2}$}} 
 \SplitRow{ $\mathcal{O}_{9, \; (1,-1)}^{21}$} {{$S_{14}S_{1}S_{2}$, $S_{14}S_{2}S_{3}$, $S_{1}V_{4}V_{4}$, $S_{3}V_{9}V_{4}$, $S_{1}V_{3}V_{14}$, $S_{2}V_{3}V_{4}$, $S_{2}V_{9}V_{3}$, $S_{3}V_{10}V_{14}$, $S_{2}V_{10}V_{4}$}} 
 \SplitRow{ $\mathcal{O}_{9, \; (1,-1)}^{22}$} {{$S_{4}V_{11}V_{2}$, $S_{2}V_{11}V_{2}$, $S_{4}V_{11}V_{1}$, $S_{2}V_{11}V_{1}$, $S_{4}V_{3}V_{5}$, $S_{2}V_{3}V_{4}$, $S_{4}V_{13}V_{3}$, $S_{2}V_{9}V_{3}$, $S_{5}S_{5}S_{4}$, $S_{2}S_{5}S_{5}$, $S_{5}V_{1}V_{5}$, $S_{5}V_{4}V_{1}$, $S_{5}V_{13}V_{2}$, $S_{5}V_{9}V_{2}$, $S_{16}S_{5}S_{4}$, $S_{16}S_{2}S_{5}$, $S_{16}V_{2}V_{5}$, $S_{16}V_{4}V_{2}$}} 
 \SplitRow{ $\mathcal{O}_{9, \; (1,-1)}^{23}$} {{$S_{4}V_{11}V_{1}$, $S_{2}V_{11}V_{1}$, $S_{5}V_{1}V_{5}$, $S_{5}V_{4}V_{1}$, $S_{16}S_{5}S_{4}$, $S_{16}S_{2}S_{5}$, $S_{4}V_{10}V_{5}$, $S_{2}V_{10}V_{4}$}} 
 \SplitRow{ $\mathcal{O}_{9, \; (1,-1)}^{24}$} {{$S_{2}V_{11}V_{1}$, $S_{3}V_{11}V_{2}$, $S_{5}V_{4}V_{1}$, $S_{3}V_{9}V_{4}$, $S_{6}V_{3}V_{1}$, $S_{5}V_{3}V_{2}$, $S_{2}V_{9}V_{3}$, $S_{16}S_{3}S_{6}$, $S_{16}S_{2}S_{5}$}} 
 \SplitRow{ $\mathcal{O}_{9, \; (1,-1)}^{25}$} {{$S_{2}V_{11}V_{1}$, $S_{1}V_{11}V_{1}$, $S_{2}V_{11}V_{2}$, $S_{5}V_{4}V_{1}$, $S_{1}V_{4}V_{4}$, $S_{16}V_{4}V_{2}$, $S_{6}V_{3}V_{1}$, $S_{2}V_{3}V_{4}$, $S_{16}V_{3}V_{1}$, $S_{1}S_{5}S_{6}$, $S_{2}S_{5}S_{5}$, $S_{16}S_{2}S_{5}$, $S_{6}V_{10}V_{2}$, $S_{5}V_{10}V_{1}$, $S_{2}V_{10}V_{4}$}} 
 \SplitRow{ $\mathcal{O}_{9, \; (1,-1)}^{26}$} {{$S_{4}V_{19}V_{2}$, $S_{2}V_{19}V_{2}$, $S_{24}V_{2}V_{2}$, $S_{22}V_{2}V_{2}$, $S_{4}V_{2}V_{2}$, $S_{2}V_{2}V_{2}$, $S_{24}V_{3}V_{22}$, $S_{4}V_{3}V_{22}$, $S_{22}V_{3}V_{20}$, $S_{2}V_{3}V_{20}$, $S_{4}V_{13}V_{3}$, $S_{2}V_{9}V_{3}$, $S_{24}S_{5}S_{25}$, $S_{22}S_{5}S_{25}$, $S_{5}S_{4}S_{25}$, $S_{2}S_{5}S_{25}$, $S_{16}S_{5}S_{4}$, $S_{16}S_{2}S_{5}$, $S_{5}V_{19}V_{22}$, $S_{5}V_{2}V_{22}$, $S_{5}V_{19}V_{20}$, $S_{5}V_{2}V_{20}$, $S_{5}V_{13}V_{2}$, $S_{5}V_{9}V_{2}$}} 
 \SplitRow{ $\mathcal{O}_{9, \; (1,-1)}^{27}$} {{$S_{22}V_{2}V_{2}$, $S_{2}V_{19}V_{2}$, $S_{2}V_{2}V_{2}$, $S_{3}V_{18}V_{2}$, $S_{23}V_{1}V_{2}$, $S_{3}V_{1}V_{2}$, $S_{25}V_{19}V_{4}$, $S_{25}V_{4}V_{2}$, $S_{16}V_{4}V_{2}$, $S_{23}V_{4}V_{20}$, $S_{3}V_{4}V_{20}$, $S_{3}V_{9}V_{4}$, $S_{25}V_{19}V_{3}$, $S_{25}V_{3}V_{2}$, $S_{16}V_{3}V_{2}$, $S_{25}V_{18}V_{3}$, $S_{25}V_{3}V_{1}$, $S_{16}V_{3}V_{1}$, $S_{22}V_{3}V_{20}$, $S_{2}V_{3}V_{20}$, $S_{2}V_{9}V_{3}$, $S_{23}S_{5}S_{25}$, $S_{3}S_{5}S_{25}$, $S_{16}S_{3}S_{5}$, $S_{22}S_{5}S_{25}$, $S_{2}S_{5}S_{25}$, $S_{16}S_{2}S_{5}$}} 
 \SplitRow{ $\mathcal{O}_{9, \; (1,-1)}^{28}$} {{$S_{24}S_{3}S_{4}$, $S_{22}S_{2}S_{3}$, $S_{3}S_{4}S_{4}$, $S_{2}S_{2}S_{3}$, $S_{24}S_{5}S_{25}$, $S_{22}S_{5}S_{25}$, $S_{5}S_{4}S_{25}$, $S_{2}S_{5}S_{25}$, $S_{16}S_{5}S_{4}$, $S_{16}S_{2}S_{5}$}} 
 \SplitRow{ $\mathcal{O}_{9, \; (1,-1)}^{29}$} {{$S_{22}S_{2}S_{3}$, $S_{2}S_{2}S_{3}$, $S_{3}V_{18}V_{2}$, $S_{3}V_{19}V_{1}$, $S_{3}V_{1}V_{2}$, $S_{22}V_{4}V_{21}$, $S_{2}V_{4}V_{21}$, $S_{2}V_{10}V_{4}$, $S_{25}V_{19}V_{4}$, $S_{25}V_{4}V_{2}$, $S_{16}V_{4}V_{2}$, $S_{22}V_{3}V_{20}$, $S_{2}V_{3}V_{20}$, $S_{2}V_{9}V_{3}$, $S_{25}V_{18}V_{3}$, $S_{25}V_{3}V_{1}$, $S_{16}V_{3}V_{1}$, $S_{5}V_{19}V_{20}$, $S_{5}V_{2}V_{20}$, $S_{5}V_{9}V_{2}$, $S_{5}V_{18}V_{21}$, $S_{5}V_{1}V_{21}$, $S_{5}V_{10}V_{1}$, $S_{22}S_{5}S_{25}$, $S_{2}S_{5}S_{25}$, $S_{16}S_{2}S_{5}$}} 
 \SplitRow{ $\mathcal{O}_{9, \; (1,-1)}^{30}$} {{$S_{21}S_{3}S_{3}$, $S_{23}S_{1}S_{3}$, $S_{1}S_{3}S_{3}$, $S_{22}S_{2}S_{3}$, $S_{2}S_{2}S_{3}$, $S_{23}V_{4}V_{20}$, $S_{3}V_{4}V_{20}$, $S_{3}V_{9}V_{4}$, $S_{22}V_{4}V_{21}$, $S_{2}V_{4}V_{21}$, $S_{2}V_{10}V_{4}$, $S_{22}V_{3}V_{20}$, $S_{2}V_{3}V_{20}$, $S_{2}V_{9}V_{3}$, $S_{21}V_{3}V_{21}$, $S_{1}V_{3}V_{21}$, $S_{1}V_{10}V_{3}$}} 
 \SplitRow{ $\mathcal{O}_{9, \; (1,-1)}^{31}$} {{$S_{14}S_{2}S_{3}$, $S_{2}V_{3}V_{4}$, $S_{3}V_{10}V_{3}$, $S_{3}V_{9}V_{4}$}} 
 \SplitRow{ $\mathcal{O}_{9, \; (1,-1)}^{32}$} {{$S_{3}V_{11}V_{2}$, $S_{2}V_{11}V_{2}$, $S_{5}V_{3}V_{2}$, $S_{2}V_{3}V_{4}$, $S_{16}V_{3}V_{2}$, $S_{2}S_{5}S_{5}$, $S_{16}S_{3}S_{5}$, $S_{5}V_{9}V_{2}$, $S_{3}V_{9}V_{4}$}} 
 \SplitRow{ $\mathcal{O}_{9, \; (1,-1)}^{33}$} {{$S_{7}S_{4}S_{4}$, $S_{7}S_{2}S_{2}$, $S_{5}S_{5}S_{4}$, $S_{2}S_{5}S_{5}$, $S_{16}S_{5}S_{4}$, $S_{16}S_{2}S_{5}$}} 
 \SplitRow{ $\mathcal{O}_{9, \; (1,-1)}^{34}$} {{$S_{7}S_{2}S_{2}$, $S_{7}V_{1}V_{1}$, $S_{7}V_{2}V_{2}$, $S_{2}V_{3}V_{4}$, $S_{5}V_{4}V_{1}$, $S_{16}V_{4}V_{2}$, $S_{5}V_{3}V_{2}$, $S_{16}V_{3}V_{1}$, $S_{16}S_{2}S_{5}$}} 
 \SplitRow{ $\mathcal{O}_{9, \; (1,-1)}^{35}$} {{$S_{7}V_{1}V_{1}$, $S_{7}S_{2}S_{2}$, $S_{5}V_{4}V_{1}$, $S_{2}V_{10}V_{4}$, $S_{2}S_{5}S_{5}$, $S_{5}V_{10}V_{1}$}} 
 \SplitRow{ $\mathcal{O}_{9, \; (1,-1)}^{36}$} {{$S_{7}S_{1}S_{1}$, $S_{7}S_{2}S_{2}$, $S_{1}V_{4}V_{4}$, $S_{2}V_{3}V_{4}$, $S_{2}V_{10}V_{4}$, $S_{1}V_{10}V_{3}$}} 
 \SplitRow{ $\mathcal{O}_{9, \; (1,-1)}^{37}$} {{$S_{7}V_{2}V_{2}$, $S_{7}S_{2}S_{2}$, $S_{5}V_{3}V_{2}$, $S_{2}V_{9}V_{3}$, $S_{2}S_{5}S_{5}$, $S_{5}V_{9}V_{2}$}} 
 \SplitRow{ $\mathcal{O}_{9, \; (1,-1)}^{38}$} {{$S_{7}S_{2}S_{2}$, $S_{7}S_{3}S_{3}$, $S_{2}V_{3}V_{4}$, $S_{3}V_{9}V_{4}$, $S_{3}V_{3}V_{3}$, $S_{2}V_{9}V_{3}$}} 
 \end{SplitTable}
   \vspace{0.4cm} 
 \captionof{table}{Dimension 9 $(\Delta B, \Delta L)$ = (1, -1) UV completions for topology 1 (see Fig.~\ref{fig:d=9_topologies}).} 
 \label{tab:T9Top1B1L-1}
 \end{center}

%% file: Tables_dim9/Table_dim9_Top2_B1_L-1.tex
 \begin{center}
 \begin{SplitTable}[2.0cm] 
  \SplitHeader{$\mathcal{O}^j{(\mathcal{T}_ {9} ^{ 2 })}$}{New Particles} 
   \vspace{0.1cm} 
 \SplitRow{ $\mathcal{O}_{9, \; (1,-1)}^{1}$} {{$F_{5}S_{23}S_{3}$, $F_{22}S_{23}S_{3}$, $F_{5}S_{3}S_{3}$, $F_{28}S_{23}V_{20}$, $F_{28}S_{3}V_{20}$, $F_{2}S_{3}V_{9}$, $F_{22}S_{23}V_{20}$, $F_{5}S_{3}V_{20}$, $F_{5}S_{3}V_{9}$, $F_{28}S_{23}V_{4}$, $F_{28}S_{3}V_{4}$, $F_{2}S_{3}V_{4}$, $F_{22}S_{3}V_{4}$, $F_{5}S_{23}V_{4}$, $F_{5}S_{3}V_{4}$, $F_{28}S_{23}S_{1}$, $F_{28}S_{1}S_{3}$, $F_{2}S_{1}S_{3}$, $F_{22}V_{4}V_{20}$, $F_{5}V_{4}V_{20}$, $F_{5}V_{9}V_{4}$}} 
 \SplitRow{ $\mathcal{O}_{9, \; (1,-1)}^{2}$} {{$F_{5}S_{1}S_{3}$, $F_{18}S_{1}V_{4}$, $F_{2}S_{3}V_{9}$, $F_{5}S_{3}V_{9}$, $F_{5}S_{1}V_{4}$, $F_{18}S_{3}V_{4}$, $F_{5}S_{3}V_{4}$, $F_{18}S_{15}S_{1}$, $F_{5}V_{9}V_{4}$, $F_{18}S_{15}S_{3}$, $F_{5}V_{4}V_{4}$, $F_{2}S_{15}S_{3}$}} 
 \SplitRow{ $\mathcal{O}_{9, \; (1,-1)}^{3}$} {{$F_{5}S_{3}V_{19}$, $F_{22}S_{23}V_{2}$, $F_{5}S_{3}V_{2}$, $F_{7}S_{23}V_{2}$, $F_{24}S_{3}V_{19}$, $F_{7}S_{3}V_{2}$, $F_{29}S_{25}V_{19}$, $F_{29}S_{25}V_{2}$, $F_{3}S_{16}V_{2}$, $F_{28}S_{23}V_{20}$, $F_{28}S_{3}V_{20}$, $F_{2}S_{3}V_{9}$, $F_{22}S_{23}V_{20}$, $F_{5}S_{3}V_{20}$, $F_{5}S_{3}V_{9}$, $F_{24}S_{25}V_{19}$, $F_{7}S_{25}V_{2}$, $F_{7}S_{16}V_{2}$, $F_{29}S_{23}S_{25}$, $F_{29}S_{3}S_{25}$, $F_{3}S_{16}S_{3}$, $F_{22}S_{23}S_{25}$, $F_{5}S_{3}S_{25}$, $F_{5}S_{16}S_{3}$, $F_{29}V_{19}V_{4}$, $F_{29}V_{4}V_{2}$, $F_{3}V_{4}V_{2}$, $F_{28}S_{23}S_{6}$, $F_{28}S_{3}S_{6}$, $F_{2}S_{3}S_{6}$, $F_{24}S_{3}S_{6}$, $F_{7}S_{23}S_{6}$, $F_{7}S_{3}S_{6}$, $F_{22}V_{4}V_{2}$, $F_{5}V_{19}V_{4}$, $F_{5}V_{4}V_{2}$, $F_{29}S_{23}V_{4}$, $F_{29}S_{3}V_{4}$, $F_{3}S_{3}V_{4}$, $F_{24}S_{3}V_{4}$, $F_{7}S_{23}V_{4}$, $F_{7}S_{3}V_{4}$, $F_{29}V_{19}V_{1}$, $F_{29}V_{1}V_{2}$, $F_{3}V_{1}V_{2}$, $F_{22}V_{4}V_{20}$, $F_{5}V_{4}V_{20}$, $F_{5}V_{9}V_{4}$, $F_{24}S_{6}S_{25}$, $F_{7}S_{6}S_{25}$, $F_{7}S_{16}S_{6}$, $F_{29}S_{23}V_{1}$, $F_{29}S_{3}V_{1}$, $F_{3}S_{3}V_{1}$, $F_{22}S_{25}V_{4}$, $F_{5}S_{25}V_{4}$, $F_{5}S_{16}V_{4}$, $F_{24}S_{25}V_{4}$, $F_{7}S_{25}V_{4}$, $F_{7}S_{16}V_{4}$, $F_{28}S_{23}V_{1}$, $F_{28}S_{3}V_{1}$, $F_{2}S_{3}V_{1}$}} 
 \SplitRow{ $\mathcal{O}_{9, \; (1,-1)}^{4}$} {{$F_{5}S_{22}S_{3}$, $F_{22}S_{23}S_{2}$, $F_{5}S_{2}S_{3}$, $F_{7}V_{19}V_{2}$, $F_{24}V_{19}V_{2}$, $F_{7}V_{2}V_{2}$, $F_{17}S_{23}S_{2}$, $F_{21}S_{22}S_{3}$, $F_{17}S_{2}S_{3}$, $F_{27}S_{22}V_{21}$, $F_{27}S_{2}V_{21}$, $F_{1}S_{2}V_{10}$, $F_{29}S_{25}V_{19}$, $F_{29}S_{25}V_{2}$, $F_{3}S_{16}V_{2}$, $F_{24}S_{25}V_{19}$, $F_{7}S_{25}V_{2}$, $F_{7}S_{16}V_{2}$, $F_{21}S_{22}V_{21}$, $F_{17}S_{2}V_{21}$, $F_{17}S_{2}V_{10}$, $F_{27}V_{19}V_{21}$, $F_{27}V_{2}V_{21}$, $F_{1}V_{10}V_{2}$, $F_{29}S_{23}S_{25}$, $F_{29}S_{3}S_{25}$, $F_{3}S_{16}S_{3}$, $F_{22}S_{23}S_{25}$, $F_{5}S_{3}S_{25}$, $F_{5}S_{16}S_{3}$, $F_{24}V_{19}V_{21}$, $F_{7}V_{2}V_{21}$, $F_{7}V_{10}V_{2}$, $F_{27}S_{22}V_{4}$, $F_{27}S_{2}V_{4}$, $F_{1}S_{2}V_{4}$, $F_{29}S_{6}V_{19}$, $F_{29}S_{6}V_{2}$, $F_{3}S_{6}V_{2}$, $F_{24}S_{6}V_{2}$, $F_{7}S_{6}V_{19}$, $F_{7}S_{6}V_{2}$, $F_{22}S_{2}V_{4}$, $F_{5}S_{22}V_{4}$, $F_{5}S_{2}V_{4}$, $F_{27}V_{19}V_{4}$, $F_{27}V_{4}V_{2}$, $F_{1}V_{4}V_{2}$, $F_{29}S_{23}S_{6}$, $F_{29}S_{3}S_{6}$, $F_{3}S_{3}S_{6}$, $F_{21}S_{3}S_{6}$, $F_{17}S_{23}S_{6}$, $F_{17}S_{3}S_{6}$, $F_{24}V_{4}V_{2}$, $F_{7}V_{19}V_{4}$, $F_{7}V_{4}V_{2}$, $F_{27}S_{22}S_{2}$, $F_{27}S_{2}S_{2}$, $F_{1}S_{2}S_{2}$, $F_{24}S_{6}S_{25}$, $F_{7}S_{6}S_{25}$, $F_{7}S_{16}S_{6}$, $F_{27}S_{2}V_{19}$, $F_{27}S_{2}V_{2}$, $F_{1}S_{2}V_{2}$, $F_{22}S_{25}V_{4}$, $F_{5}S_{25}V_{4}$, $F_{5}S_{16}V_{4}$, $F_{24}S_{6}V_{21}$, $F_{7}S_{6}V_{21}$, $F_{7}S_{6}V_{10}$, $F_{21}S_{6}V_{21}$, $F_{17}S_{6}V_{21}$, $F_{17}S_{6}V_{10}$, $F_{24}S_{25}V_{4}$, $F_{7}S_{25}V_{4}$, $F_{7}S_{16}V_{4}$, $F_{29}S_{2}V_{19}$, $F_{29}S_{2}V_{2}$, $F_{3}S_{2}V_{2}$, $F_{29}S_{23}S_{2}$, $F_{29}S_{2}S_{3}$, $F_{3}S_{2}S_{3}$, $F_{24}V_{4}V_{21}$, $F_{7}V_{4}V_{21}$, $F_{7}V_{10}V_{4}$}} 
 \SplitRow{ $\mathcal{O}_{9, \; (1,-1)}^{5}$} {{$F_{17}S_{23}S_{2}$, $F_{21}S_{22}S_{3}$, $F_{17}S_{2}S_{3}$, $F_{5}S_{22}S_{3}$, $F_{22}S_{23}S_{2}$, $F_{5}S_{2}S_{3}$, $F_{27}S_{23}V_{21}$, $F_{27}S_{3}V_{21}$, $F_{1}S_{3}V_{10}$, $F_{22}S_{23}V_{21}$, $F_{5}S_{3}V_{21}$, $F_{5}S_{3}V_{10}$, $F_{28}S_{23}V_{20}$, $F_{28}S_{3}V_{20}$, $F_{2}S_{3}V_{9}$, $F_{27}S_{22}V_{21}$, $F_{27}S_{2}V_{21}$, $F_{1}S_{2}V_{10}$, $F_{21}S_{22}V_{21}$, $F_{17}S_{2}V_{21}$, $F_{17}S_{2}V_{10}$, $F_{22}S_{23}V_{20}$, $F_{5}S_{3}V_{20}$, $F_{5}S_{3}V_{9}$, $F_{27}S_{23}V_{4}$, $F_{27}S_{3}V_{4}$, $F_{1}S_{3}V_{4}$, $F_{21}S_{3}V_{4}$, $F_{17}S_{23}V_{4}$, $F_{17}S_{3}V_{4}$, $F_{28}S_{23}V_{14}$, $F_{28}S_{3}V_{14}$, $F_{2}S_{3}V_{14}$, $F_{27}S_{22}V_{4}$, $F_{27}S_{2}V_{4}$, $F_{1}S_{2}V_{4}$, $F_{22}S_{2}V_{4}$, $F_{5}S_{22}V_{4}$, $F_{5}S_{2}V_{4}$, $F_{21}S_{3}V_{14}$, $F_{17}S_{23}V_{14}$, $F_{17}S_{3}V_{14}$, $F_{28}S_{23}S_{2}$, $F_{28}S_{2}S_{3}$, $F_{2}S_{2}S_{3}$, $F_{21}V_{4}V_{21}$, $F_{17}V_{4}V_{21}$, $F_{17}V_{10}V_{4}$, $F_{22}V_{4}V_{21}$, $F_{5}V_{4}V_{21}$, $F_{5}V_{10}V_{4}$, $F_{27}S_{23}S_{2}$, $F_{27}S_{2}S_{3}$, $F_{1}S_{2}S_{3}$, $F_{27}S_{22}S_{2}$, $F_{27}S_{2}S_{2}$, $F_{1}S_{2}S_{2}$, $F_{21}V_{14}V_{21}$, $F_{17}V_{14}V_{21}$, $F_{17}V_{10}V_{14}$, $F_{22}V_{4}V_{20}$, $F_{5}V_{4}V_{20}$, $F_{5}V_{9}V_{4}$}} 
 \SplitRow{ $\mathcal{O}_{9, \; (1,-1)}^{6}$} {{$F_{8}S_{3}V_{19}$, $F_{25}S_{23}V_{2}$, $F_{8}S_{3}V_{2}$, $F_{5}S_{23}V_{2}$, $F_{22}S_{3}V_{19}$, $F_{5}S_{3}V_{2}$, $F_{29}V_{19}V_{20}$, $F_{29}V_{2}V_{20}$, $F_{3}V_{9}V_{2}$, $F_{28}S_{23}S_{25}$, $F_{28}S_{3}S_{25}$, $F_{2}S_{16}S_{3}$, $F_{25}S_{23}S_{25}$, $F_{8}S_{3}S_{25}$, $F_{8}S_{16}S_{3}$, $F_{22}V_{19}V_{20}$, $F_{5}V_{2}V_{20}$, $F_{5}V_{9}V_{2}$, $F_{29}S_{23}V_{20}$, $F_{29}S_{3}V_{20}$, $F_{3}S_{3}V_{9}$, $F_{25}S_{23}V_{20}$, $F_{8}S_{3}V_{20}$, $F_{8}S_{3}V_{9}$, $F_{29}S_{5}V_{19}$, $F_{29}S_{5}V_{2}$, $F_{3}S_{5}V_{2}$, $F_{28}S_{23}V_{4}$, $F_{28}S_{3}V_{4}$, $F_{2}S_{3}V_{4}$, $F_{22}S_{3}V_{4}$, $F_{5}S_{23}V_{4}$, $F_{5}S_{3}V_{4}$, $F_{25}S_{5}V_{2}$, $F_{8}S_{5}V_{19}$, $F_{8}S_{5}V_{2}$, $F_{29}S_{23}S_{5}$, $F_{29}S_{3}S_{5}$, $F_{3}S_{3}S_{5}$, $F_{22}S_{3}S_{5}$, $F_{5}S_{23}S_{5}$, $F_{5}S_{3}S_{5}$, $F_{29}V_{19}V_{1}$, $F_{29}V_{1}V_{2}$, $F_{3}V_{1}V_{2}$, $F_{25}S_{5}S_{25}$, $F_{8}S_{5}S_{25}$, $F_{8}S_{16}S_{5}$, $F_{22}V_{4}V_{20}$, $F_{5}V_{4}V_{20}$, $F_{5}V_{9}V_{4}$, $F_{29}S_{23}V_{1}$, $F_{29}S_{3}V_{1}$, $F_{3}S_{3}V_{1}$, $F_{25}S_{5}V_{20}$, $F_{8}S_{5}V_{20}$, $F_{8}S_{5}V_{9}$, $F_{22}S_{5}V_{20}$, $F_{5}S_{5}V_{20}$, $F_{5}S_{5}V_{9}$, $F_{28}S_{23}V_{1}$, $F_{28}S_{3}V_{1}$, $F_{2}S_{3}V_{1}$}} 
 \SplitRow{ $\mathcal{O}_{9, \; (1,-1)}^{7}$} {{$F_{8}S_{3}V_{1}$, $F_{5}S_{3}V_{1}$, $F_{13}V_{4}V_{1}$, $F_{2}S_{16}S_{3}$, $F_{8}S_{16}S_{3}$, $F_{5}V_{4}V_{1}$, $F_{13}S_{3}V_{4}$, $F_{8}S_{3}V_{4}$, $F_{13}S_{5}V_{1}$, $F_{2}S_{3}V_{9}$, $F_{5}S_{3}V_{9}$, $F_{8}S_{5}V_{1}$, $F_{13}S_{3}S_{5}$, $F_{5}S_{3}S_{5}$, $F_{13}V_{12}V_{1}$, $F_{8}S_{16}S_{5}$, $F_{5}V_{9}V_{4}$, $F_{13}S_{3}V_{12}$, $F_{8}S_{5}V_{4}$, $F_{5}S_{5}V_{4}$, $F_{2}S_{3}V_{12}$}} 
 \SplitRow{ $\mathcal{O}_{9, \; (1,-1)}^{8}$} {{$F_{8}S_{3}V_{1}$, $F_{5}S_{1}V_{2}$, $F_{7}S_{1}V_{2}$, $F_{5}S_{3}V_{1}$, $F_{13}V_{4}V_{1}$, $F_{18}S_{1}S_{6}$, $F_{3}V_{9}V_{2}$, $F_{5}V_{9}V_{2}$, $F_{7}S_{1}S_{6}$, $F_{5}V_{4}V_{1}$, $F_{13}S_{1}V_{4}$, $F_{3}S_{3}V_{9}$, $F_{8}S_{3}V_{9}$, $F_{7}S_{1}V_{4}$, $F_{13}V_{4}V_{2}$, $F_{18}S_{3}S_{6}$, $F_{8}S_{3}S_{6}$, $F_{5}V_{4}V_{2}$, $F_{13}S_{5}V_{1}$, $F_{18}S_{1}V_{4}$, $F_{3}S_{16}V_{2}$, $F_{7}S_{16}V_{2}$, $F_{5}S_{1}V_{4}$, $F_{8}S_{5}V_{1}$, $F_{13}S_{1}S_{5}$, $F_{3}S_{16}S_{3}$, $F_{5}S_{16}S_{3}$, $F_{5}S_{1}S_{5}$, $F_{13}S_{5}V_{2}$, $F_{18}S_{3}V_{4}$, $F_{5}S_{3}V_{4}$, $F_{7}S_{5}V_{2}$, $F_{13}V_{12}V_{1}$, $F_{5}V_{9}V_{4}$, $F_{7}S_{16}S_{6}$, $F_{13}S_{1}V_{12}$, $F_{8}S_{5}V_{9}$, $F_{7}S_{16}V_{4}$, $F_{5}S_{16}V_{4}$, $F_{5}S_{5}V_{9}$, $F_{18}S_{1}V_{12}$, $F_{13}V_{12}V_{2}$, $F_{8}S_{5}S_{6}$, $F_{5}V_{4}V_{4}$, $F_{7}S_{5}S_{6}$, $F_{3}V_{12}V_{2}$, $F_{18}S_{3}V_{12}$, $F_{5}S_{5}V_{4}$, $F_{7}S_{5}V_{4}$, $F_{3}S_{3}V_{12}$}} 
 \SplitRow{ $\mathcal{O}_{9, \; (1,-1)}^{9}$} {{$F_{8}S_{3}S_{4}$, $F_{8}S_{2}S_{3}$, $F_{9}V_{1}V_{2}$, $F_{5}V_{1}V_{2}$, $F_{14}S_{5}S_{4}$, $F_{6}S_{2}S_{5}$, $F_{13}V_{1}V_{5}$, $F_{13}V_{4}V_{1}$, $F_{3}V_{13}V_{2}$, $F_{3}V_{9}V_{2}$, $F_{9}V_{13}V_{2}$, $F_{5}V_{9}V_{2}$, $F_{9}V_{1}V_{5}$, $F_{5}V_{4}V_{1}$, $F_{8}S_{5}S_{4}$, $F_{8}S_{2}S_{5}$, $F_{14}S_{5}V_{1}$, $F_{6}S_{5}V_{1}$, $F_{3}S_{3}V_{13}$, $F_{3}S_{3}V_{9}$, $F_{8}S_{3}V_{13}$, $F_{8}S_{3}V_{9}$, $F_{9}S_{5}V_{1}$, $F_{5}S_{5}V_{1}$, $F_{14}S_{5}V_{2}$, $F_{6}S_{5}V_{2}$, $F_{13}S_{3}V_{5}$, $F_{13}S_{3}V_{4}$, $F_{8}S_{3}V_{5}$, $F_{8}S_{3}V_{4}$, $F_{9}S_{5}V_{2}$, $F_{5}S_{5}V_{2}$, $F_{14}S_{9}S_{4}$, $F_{6}S_{14}S_{2}$, $F_{9}V_{13}V_{5}$, $F_{5}V_{9}V_{4}$, $F_{14}S_{9}V_{1}$, $F_{6}S_{14}V_{1}$, $F_{8}S_{5}V_{13}$, $F_{8}S_{5}V_{9}$, $F_{9}S_{5}V_{13}$, $F_{5}S_{5}V_{9}$, $F_{13}S_{9}V_{1}$, $F_{13}S_{14}V_{1}$, $F_{14}S_{9}V_{2}$, $F_{6}S_{14}V_{2}$, $F_{8}S_{5}V_{5}$, $F_{8}S_{5}V_{4}$, $F_{9}S_{5}V_{5}$, $F_{5}S_{5}V_{4}$, $F_{3}S_{9}V_{2}$, $F_{3}S_{14}V_{2}$, $F_{13}S_{9}S_{3}$, $F_{13}S_{14}S_{3}$, $F_{9}S_{5}S_{5}$, $F_{5}S_{5}S_{5}$, $F_{3}S_{9}S_{3}$, $F_{3}S_{14}S_{3}$}} 
 \SplitRow{ $\mathcal{O}_{9, \; (1,-1)}^{10}$} {{$F_{9}V_{1}V_{2}$, $F_{5}V_{1}V_{2}$, $F_{7}S_{1}S_{4}$, $F_{7}S_{1}S_{2}$, $F_{13}V_{1}V_{5}$, $F_{13}V_{4}V_{1}$, $F_{4}S_{16}S_{4}$, $F_{1}S_{16}S_{2}$, $F_{7}S_{16}S_{4}$, $F_{7}S_{16}S_{2}$, $F_{9}V_{1}V_{5}$, $F_{5}V_{4}V_{1}$, $F_{14}S_{5}V_{1}$, $F_{6}S_{5}V_{1}$, $F_{13}S_{1}V_{5}$, $F_{13}S_{1}V_{4}$, $F_{4}S_{16}V_{2}$, $F_{1}S_{16}V_{2}$, $F_{9}S_{16}V_{2}$, $F_{5}S_{16}V_{2}$, $F_{7}S_{1}V_{5}$, $F_{7}S_{1}V_{4}$, $F_{9}S_{5}V_{1}$, $F_{5}S_{5}V_{1}$, $F_{14}S_{5}S_{4}$, $F_{6}S_{2}S_{5}$, $F_{13}V_{2}V_{5}$, $F_{13}V_{4}V_{2}$, $F_{9}V_{2}V_{5}$, $F_{5}V_{4}V_{2}$, $F_{7}S_{5}S_{4}$, $F_{7}S_{2}S_{5}$, $F_{14}S_{9}V_{1}$, $F_{6}S_{14}V_{1}$, $F_{9}S_{16}V_{5}$, $F_{5}S_{16}V_{4}$, $F_{7}S_{16}V_{5}$, $F_{7}S_{16}V_{4}$, $F_{13}S_{9}V_{1}$, $F_{13}S_{14}V_{1}$, $F_{14}S_{9}S_{4}$, $F_{6}S_{14}S_{2}$, $F_{9}V_{5}V_{5}$, $F_{5}V_{4}V_{4}$, $F_{4}S_{9}S_{4}$, $F_{1}S_{14}S_{2}$, $F_{13}S_{9}S_{1}$, $F_{13}S_{14}S_{1}$, $F_{9}S_{16}S_{5}$, $F_{5}S_{16}S_{5}$, $F_{13}S_{9}V_{2}$, $F_{13}S_{14}V_{2}$, $F_{9}S_{5}V_{5}$, $F_{5}S_{5}V_{4}$, $F_{7}S_{5}V_{5}$, $F_{7}S_{5}V_{4}$, $F_{4}S_{9}V_{2}$, $F_{1}S_{14}V_{2}$}} 
 \SplitRow{ $\mathcal{O}_{9, \; (1,-1)}^{11}$} {{$F_{7}S_{1}S_{2}$, $F_{8}S_{2}S_{3}$, $F_{6}S_{1}S_{5}$, $F_{1}S_{16}S_{3}$, $F_{8}S_{16}S_{3}$, $F_{7}S_{1}S_{5}$, $F_{18}S_{1}S_{6}$, $F_{6}S_{2}S_{5}$, $F_{1}S_{16}S_{2}$, $F_{7}S_{16}S_{2}$, $F_{8}S_{2}S_{5}$, $F_{7}S_{1}S_{6}$, $F_{18}S_{3}S_{6}$, $F_{7}S_{2}S_{5}$, $F_{8}S_{3}S_{6}$, $F_{18}S_{14}S_{1}$, $F_{7}S_{16}S_{5}$, $F_{8}S_{16}S_{5}$, $F_{6}S_{14}S_{1}$, $F_{18}S_{14}S_{3}$, $F_{7}S_{5}S_{5}$, $F_{1}S_{14}S_{3}$, $F_{6}S_{14}S_{2}$, $F_{7}S_{16}S_{6}$, $F_{7}S_{5}S_{6}$, $F_{8}S_{5}S_{6}$, $F_{1}S_{14}S_{2}$}} 
 \SplitRow{ $\mathcal{O}_{9, \; (1,-1)}^{12}$} {{$F_{8}S_{24}S_{3}$, $F_{8}S_{22}S_{3}$, $F_{25}S_{23}S_{4}$, $F_{8}S_{3}S_{4}$, $F_{25}S_{23}S_{2}$, $F_{8}S_{2}S_{3}$, $F_{9}V_{19}V_{2}$, $F_{5}V_{19}V_{2}$, $F_{26}V_{19}V_{2}$, $F_{22}V_{19}V_{2}$, $F_{9}V_{2}V_{2}$, $F_{5}V_{2}V_{2}$, $F_{7}S_{23}S_{4}$, $F_{7}S_{23}S_{2}$, $F_{24}S_{24}S_{3}$, $F_{24}S_{22}S_{3}$, $F_{7}S_{3}S_{4}$, $F_{7}S_{2}S_{3}$, $F_{30}S_{24}S_{25}$, $F_{27}S_{22}S_{25}$, $F_{30}S_{4}S_{25}$, $F_{4}S_{16}S_{4}$, $F_{27}S_{2}S_{25}$, $F_{1}S_{16}S_{2}$, $F_{29}V_{19}V_{22}$, $F_{29}V_{19}V_{20}$, $F_{29}V_{2}V_{22}$, $F_{29}V_{2}V_{20}$, $F_{3}V_{13}V_{2}$, $F_{3}V_{9}V_{2}$, $F_{26}V_{19}V_{22}$, $F_{9}V_{2}V_{22}$, $F_{22}V_{19}V_{20}$, $F_{5}V_{2}V_{20}$, $F_{9}V_{13}V_{2}$, $F_{5}V_{9}V_{2}$, $F_{24}S_{24}S_{25}$, $F_{24}S_{22}S_{25}$, $F_{7}S_{4}S_{25}$, $F_{7}S_{2}S_{25}$, $F_{7}S_{16}S_{4}$, $F_{7}S_{16}S_{2}$, $F_{30}S_{25}V_{19}$, $F_{27}S_{25}V_{19}$, $F_{30}S_{25}V_{2}$, $F_{27}S_{25}V_{2}$, $F_{4}S_{16}V_{2}$, $F_{1}S_{16}V_{2}$, $F_{29}S_{23}V_{22}$, $F_{29}S_{23}V_{20}$, $F_{29}S_{3}V_{22}$, $F_{29}S_{3}V_{20}$, $F_{3}S_{3}V_{13}$, $F_{3}S_{3}V_{9}$, $F_{25}S_{23}V_{22}$, $F_{8}S_{3}V_{22}$, $F_{25}S_{23}V_{20}$, $F_{8}S_{3}V_{20}$, $F_{8}S_{3}V_{13}$, $F_{8}S_{3}V_{9}$, $F_{26}S_{25}V_{19}$, $F_{22}S_{25}V_{19}$, $F_{9}S_{25}V_{2}$, $F_{5}S_{25}V_{2}$, $F_{9}S_{16}V_{2}$, $F_{5}S_{16}V_{2}$, $F_{30}S_{24}S_{5}$, $F_{27}S_{22}S_{5}$, $F_{30}S_{5}S_{4}$, $F_{4}S_{5}S_{4}$, $F_{27}S_{2}S_{5}$, $F_{1}S_{2}S_{5}$, $F_{29}V_{19}V_{5}$, $F_{29}V_{19}V_{4}$, $F_{29}V_{2}V_{5}$, $F_{29}V_{4}V_{2}$, $F_{3}V_{2}V_{5}$, $F_{3}V_{4}V_{2}$, $F_{26}V_{2}V_{5}$, $F_{9}V_{19}V_{5}$, $F_{9}V_{2}V_{5}$, $F_{22}V_{4}V_{2}$, $F_{5}V_{19}V_{4}$, $F_{5}V_{4}V_{2}$, $F_{25}S_{5}S_{4}$, $F_{25}S_{2}S_{5}$, $F_{8}S_{24}S_{5}$, $F_{8}S_{22}S_{5}$, $F_{8}S_{5}S_{4}$, $F_{8}S_{2}S_{5}$, $F_{30}S_{5}V_{19}$, $F_{27}S_{5}V_{19}$, $F_{30}S_{5}V_{2}$, $F_{27}S_{5}V_{2}$, $F_{4}S_{5}V_{2}$, $F_{1}S_{5}V_{2}$, $F_{29}S_{23}V_{5}$, $F_{29}S_{23}V_{4}$, $F_{29}S_{3}V_{5}$, $F_{29}S_{3}V_{4}$, $F_{3}S_{3}V_{5}$, $F_{3}S_{3}V_{4}$, $F_{24}S_{3}V_{5}$, $F_{7}S_{23}V_{5}$, $F_{7}S_{3}V_{5}$, $F_{24}S_{3}V_{4}$, $F_{7}S_{23}V_{4}$, $F_{7}S_{3}V_{4}$, $F_{26}S_{5}V_{2}$, $F_{22}S_{5}V_{2}$, $F_{9}S_{5}V_{19}$, $F_{9}S_{5}V_{2}$, $F_{5}S_{5}V_{19}$, $F_{5}S_{5}V_{2}$, $F_{30}S_{24}S_{4}$, $F_{27}S_{22}S_{2}$, $F_{30}S_{4}S_{4}$, $F_{4}S_{4}S_{4}$, $F_{27}S_{2}S_{2}$, $F_{1}S_{2}S_{2}$, $F_{26}V_{5}V_{22}$, $F_{9}V_{5}V_{22}$, $F_{22}V_{4}V_{20}$, $F_{5}V_{4}V_{20}$, $F_{9}V_{13}V_{5}$, $F_{5}V_{9}V_{4}$, $F_{30}S_{4}V_{19}$, $F_{27}S_{2}V_{19}$, $F_{30}S_{4}V_{2}$, $F_{27}S_{2}V_{2}$, $F_{4}S_{4}V_{2}$, $F_{1}S_{2}V_{2}$, $F_{25}S_{5}V_{22}$, $F_{8}S_{5}V_{22}$, $F_{25}S_{5}V_{20}$, $F_{8}S_{5}V_{20}$, $F_{8}S_{5}V_{13}$, $F_{8}S_{5}V_{9}$, $F_{26}S_{25}V_{5}$, $F_{22}S_{25}V_{4}$, $F_{9}S_{25}V_{5}$, $F_{5}S_{25}V_{4}$, $F_{9}S_{16}V_{5}$, $F_{5}S_{16}V_{4}$, $F_{24}S_{25}V_{5}$, $F_{7}S_{25}V_{5}$, $F_{7}S_{16}V_{5}$, $F_{24}S_{25}V_{4}$, $F_{7}S_{25}V_{4}$, $F_{7}S_{16}V_{4}$, $F_{26}S_{5}V_{22}$, $F_{22}S_{5}V_{20}$, $F_{9}S_{5}V_{22}$, $F_{9}S_{5}V_{13}$, $F_{5}S_{5}V_{20}$, $F_{5}S_{5}V_{9}$, $F_{29}S_{4}V_{19}$, $F_{29}S_{4}V_{2}$, $F_{3}S_{4}V_{2}$, $F_{29}S_{2}V_{19}$, $F_{29}S_{2}V_{2}$, $F_{3}S_{2}V_{2}$, $F_{29}S_{23}S_{4}$, $F_{29}S_{23}S_{2}$, $F_{29}S_{3}S_{4}$, $F_{29}S_{2}S_{3}$, $F_{3}S_{3}S_{4}$, $F_{3}S_{2}S_{3}$, $F_{26}S_{5}S_{25}$, $F_{22}S_{5}S_{25}$, $F_{9}S_{5}S_{25}$, $F_{5}S_{5}S_{25}$, $F_{9}S_{16}S_{5}$, $F_{5}S_{16}S_{5}$}} 
 \SplitRow{ $\mathcal{O}_{9, \; (1,-1)}^{13}$} {{$F_{7}S_{23}S_{2}$, $F_{24}S_{22}S_{3}$, $F_{7}S_{2}S_{3}$, $F_{8}S_{22}S_{3}$, $F_{25}S_{23}S_{2}$, $F_{8}S_{2}S_{3}$, $F_{27}S_{23}S_{25}$, $F_{27}S_{3}S_{25}$, $F_{1}S_{16}S_{3}$, $F_{25}S_{23}S_{25}$, $F_{8}S_{3}S_{25}$, $F_{8}S_{16}S_{3}$, $F_{28}S_{23}S_{25}$, $F_{28}S_{3}S_{25}$, $F_{2}S_{16}S_{3}$, $F_{27}S_{22}S_{25}$, $F_{27}S_{2}S_{25}$, $F_{1}S_{16}S_{2}$, $F_{24}S_{22}S_{25}$, $F_{7}S_{2}S_{25}$, $F_{7}S_{16}S_{2}$, $F_{27}S_{23}S_{5}$, $F_{27}S_{3}S_{5}$, $F_{1}S_{3}S_{5}$, $F_{24}S_{3}S_{5}$, $F_{7}S_{23}S_{5}$, $F_{7}S_{3}S_{5}$, $F_{28}S_{23}S_{6}$, $F_{28}S_{3}S_{6}$, $F_{2}S_{3}S_{6}$, $F_{27}S_{22}S_{5}$, $F_{27}S_{2}S_{5}$, $F_{1}S_{2}S_{5}$, $F_{25}S_{2}S_{5}$, $F_{8}S_{22}S_{5}$, $F_{8}S_{2}S_{5}$, $F_{24}S_{3}S_{6}$, $F_{7}S_{23}S_{6}$, $F_{7}S_{3}S_{6}$, $F_{28}S_{23}S_{2}$, $F_{28}S_{2}S_{3}$, $F_{2}S_{2}S_{3}$, $F_{24}S_{5}S_{25}$, $F_{7}S_{5}S_{25}$, $F_{7}S_{16}S_{5}$, $F_{25}S_{5}S_{25}$, $F_{8}S_{5}S_{25}$, $F_{8}S_{16}S_{5}$, $F_{27}S_{23}S_{2}$, $F_{27}S_{2}S_{3}$, $F_{1}S_{2}S_{3}$, $F_{27}S_{22}S_{2}$, $F_{27}S_{2}S_{2}$, $F_{1}S_{2}S_{2}$, $F_{24}S_{6}S_{25}$, $F_{7}S_{6}S_{25}$, $F_{7}S_{16}S_{6}$}} 
 \SplitRow{ $\mathcal{O}_{9, \; (1,-1)}^{14}$} {{$F_{9}S_{24}V_{2}$, $F_{5}S_{22}V_{2}$, $F_{26}S_{4}V_{19}$, $F_{9}S_{4}V_{2}$, $F_{22}S_{2}V_{19}$, $F_{5}S_{2}V_{2}$, $F_{7}S_{4}V_{19}$, $F_{7}S_{2}V_{19}$, $F_{24}S_{24}V_{2}$, $F_{24}S_{22}V_{2}$, $F_{7}S_{4}V_{2}$, $F_{7}S_{2}V_{2}$, $F_{30}S_{24}S_{25}$, $F_{27}S_{22}S_{25}$, $F_{30}S_{4}S_{25}$, $F_{4}S_{16}S_{4}$, $F_{27}S_{2}S_{25}$, $F_{1}S_{16}S_{2}$, $F_{24}S_{24}S_{25}$, $F_{24}S_{22}S_{25}$, $F_{7}S_{4}S_{25}$, $F_{7}S_{2}S_{25}$, $F_{7}S_{16}S_{4}$, $F_{7}S_{16}S_{2}$, $F_{29}S_{24}V_{21}$, $F_{29}S_{22}V_{21}$, $F_{29}S_{4}V_{21}$, $F_{3}S_{4}V_{10}$, $F_{29}S_{2}V_{21}$, $F_{3}S_{2}V_{10}$, $F_{30}S_{25}V_{19}$, $F_{27}S_{25}V_{19}$, $F_{30}S_{25}V_{2}$, $F_{27}S_{25}V_{2}$, $F_{4}S_{16}V_{2}$, $F_{1}S_{16}V_{2}$, $F_{26}S_{25}V_{19}$, $F_{22}S_{25}V_{19}$, $F_{9}S_{25}V_{2}$, $F_{5}S_{25}V_{2}$, $F_{9}S_{16}V_{2}$, $F_{5}S_{16}V_{2}$, $F_{24}S_{24}V_{21}$, $F_{24}S_{22}V_{21}$, $F_{7}S_{4}V_{21}$, $F_{7}S_{2}V_{21}$, $F_{7}S_{4}V_{10}$, $F_{7}S_{2}V_{10}$, $F_{30}S_{24}V_{5}$, $F_{27}S_{22}V_{4}$, $F_{30}S_{4}V_{5}$, $F_{4}S_{4}V_{5}$, $F_{27}S_{2}V_{4}$, $F_{1}S_{2}V_{4}$, $F_{26}S_{4}V_{5}$, $F_{9}S_{24}V_{5}$, $F_{9}S_{4}V_{5}$, $F_{22}S_{2}V_{4}$, $F_{5}S_{22}V_{4}$, $F_{5}S_{2}V_{4}$, $F_{29}S_{24}S_{5}$, $F_{29}S_{22}S_{5}$, $F_{29}S_{5}S_{4}$, $F_{3}S_{5}S_{4}$, $F_{29}S_{2}S_{5}$, $F_{3}S_{2}S_{5}$, $F_{30}V_{19}V_{5}$, $F_{27}V_{19}V_{4}$, $F_{30}V_{2}V_{5}$, $F_{27}V_{4}V_{2}$, $F_{4}V_{2}V_{5}$, $F_{1}V_{4}V_{2}$, $F_{24}V_{2}V_{5}$, $F_{7}V_{19}V_{5}$, $F_{7}V_{2}V_{5}$, $F_{24}V_{4}V_{2}$, $F_{7}V_{19}V_{4}$, $F_{7}V_{4}V_{2}$, $F_{26}S_{5}S_{4}$, $F_{22}S_{2}S_{5}$, $F_{9}S_{24}S_{5}$, $F_{9}S_{5}S_{4}$, $F_{5}S_{22}S_{5}$, $F_{5}S_{2}S_{5}$, $F_{29}S_{24}V_{2}$, $F_{29}S_{22}V_{2}$, $F_{29}S_{4}V_{2}$, $F_{3}S_{4}V_{2}$, $F_{29}S_{2}V_{2}$, $F_{3}S_{2}V_{2}$, $F_{26}S_{25}V_{5}$, $F_{22}S_{25}V_{4}$, $F_{9}S_{25}V_{5}$, $F_{5}S_{25}V_{4}$, $F_{9}S_{16}V_{5}$, $F_{5}S_{16}V_{4}$, $F_{24}S_{25}V_{5}$, $F_{7}S_{25}V_{5}$, $F_{7}S_{16}V_{5}$, $F_{24}S_{25}V_{4}$, $F_{7}S_{25}V_{4}$, $F_{7}S_{16}V_{4}$, $F_{30}S_{24}V_{2}$, $F_{30}S_{4}V_{2}$, $F_{27}S_{22}V_{2}$, $F_{27}S_{2}V_{2}$, $F_{4}S_{4}V_{2}$, $F_{1}S_{2}V_{2}$, $F_{30}V_{19}V_{2}$, $F_{27}V_{19}V_{2}$, $F_{30}V_{2}V_{2}$, $F_{27}V_{2}V_{2}$, $F_{4}V_{2}V_{2}$, $F_{1}V_{2}V_{2}$, $F_{26}S_{5}S_{25}$, $F_{22}S_{5}S_{25}$, $F_{9}S_{5}S_{25}$, $F_{5}S_{5}S_{25}$, $F_{9}S_{16}S_{5}$, $F_{5}S_{16}S_{5}$, $F_{24}V_{5}V_{21}$, $F_{24}V_{4}V_{21}$, $F_{7}V_{5}V_{21}$, $F_{7}V_{4}V_{21}$, $F_{7}V_{10}V_{5}$, $F_{7}V_{10}V_{4}$}} 
 \SplitRow{ $\mathcal{O}_{9, \; (1,-1)}^{15}$} {{$F_{7}S_{2}V_{19}$, $F_{24}S_{22}V_{2}$, $F_{7}S_{2}V_{2}$, $F_{17}S_{23}V_{1}$, $F_{21}S_{3}V_{18}$, $F_{17}S_{3}V_{1}$, $F_{8}S_{3}V_{18}$, $F_{25}S_{23}V_{1}$, $F_{8}S_{3}V_{1}$, $F_{5}S_{22}V_{2}$, $F_{22}S_{2}V_{19}$, $F_{5}S_{2}V_{2}$, $F_{29}V_{19}V_{21}$, $F_{29}V_{2}V_{21}$, $F_{3}V_{10}V_{2}$, $F_{27}S_{23}S_{25}$, $F_{27}S_{3}S_{25}$, $F_{1}S_{16}S_{3}$, $F_{25}S_{23}S_{25}$, $F_{8}S_{3}S_{25}$, $F_{8}S_{16}S_{3}$, $F_{22}V_{19}V_{21}$, $F_{5}V_{2}V_{21}$, $F_{5}V_{10}V_{2}$, $F_{29}V_{19}V_{20}$, $F_{29}V_{2}V_{20}$, $F_{3}V_{9}V_{2}$, $F_{29}V_{18}V_{21}$, $F_{29}V_{1}V_{21}$, $F_{3}V_{10}V_{1}$, $F_{27}S_{22}S_{25}$, $F_{27}S_{2}S_{25}$, $F_{1}S_{16}S_{2}$, $F_{24}S_{22}S_{25}$, $F_{7}S_{2}S_{25}$, $F_{7}S_{16}S_{2}$, $F_{21}V_{18}V_{21}$, $F_{17}V_{1}V_{21}$, $F_{17}V_{10}V_{1}$, $F_{22}V_{19}V_{20}$, $F_{5}V_{2}V_{20}$, $F_{5}V_{9}V_{2}$, $F_{29}S_{23}V_{20}$, $F_{29}S_{3}V_{20}$, $F_{3}S_{3}V_{9}$, $F_{29}S_{22}V_{21}$, $F_{29}S_{2}V_{21}$, $F_{3}S_{2}V_{10}$, $F_{24}S_{22}V_{21}$, $F_{7}S_{2}V_{21}$, $F_{7}S_{2}V_{10}$, $F_{25}S_{23}V_{20}$, $F_{8}S_{3}V_{20}$, $F_{8}S_{3}V_{9}$, $F_{29}S_{5}V_{19}$, $F_{29}S_{5}V_{2}$, $F_{3}S_{5}V_{2}$, $F_{27}S_{23}V_{4}$, $F_{27}S_{3}V_{4}$, $F_{1}S_{3}V_{4}$, $F_{21}S_{3}V_{4}$, $F_{17}S_{23}V_{4}$, $F_{17}S_{3}V_{4}$, $F_{24}S_{5}V_{2}$, $F_{7}S_{5}V_{19}$, $F_{7}S_{5}V_{2}$, $F_{29}S_{6}V_{19}$, $F_{29}S_{6}V_{2}$, $F_{3}S_{6}V_{2}$, $F_{29}S_{5}V_{18}$, $F_{29}S_{5}V_{1}$, $F_{3}S_{5}V_{1}$, $F_{27}S_{22}V_{4}$, $F_{27}S_{2}V_{4}$, $F_{1}S_{2}V_{4}$, $F_{22}S_{2}V_{4}$, $F_{5}S_{22}V_{4}$, $F_{5}S_{2}V_{4}$, $F_{25}S_{5}V_{1}$, $F_{8}S_{5}V_{18}$, $F_{8}S_{5}V_{1}$, $F_{24}S_{6}V_{2}$, $F_{7}S_{6}V_{19}$, $F_{7}S_{6}V_{2}$, $F_{29}S_{23}S_{6}$, $F_{29}S_{3}S_{6}$, $F_{3}S_{3}S_{6}$, $F_{29}S_{22}S_{5}$, $F_{29}S_{2}S_{5}$, $F_{3}S_{2}S_{5}$, $F_{22}S_{2}S_{5}$, $F_{5}S_{22}S_{5}$, $F_{5}S_{2}S_{5}$, $F_{21}S_{3}S_{6}$, $F_{17}S_{23}S_{6}$, $F_{17}S_{3}S_{6}$, $F_{29}V_{19}V_{2}$, $F_{29}V_{2}V_{2}$, $F_{3}V_{2}V_{2}$, $F_{24}S_{5}S_{25}$, $F_{7}S_{5}S_{25}$, $F_{7}S_{16}S_{5}$, $F_{21}V_{4}V_{21}$, $F_{17}V_{4}V_{21}$, $F_{17}V_{10}V_{4}$, $F_{22}V_{4}V_{21}$, $F_{5}V_{4}V_{21}$, $F_{5}V_{10}V_{4}$, $F_{25}S_{5}S_{25}$, $F_{8}S_{5}S_{25}$, $F_{8}S_{16}S_{5}$, $F_{29}S_{23}V_{2}$, $F_{29}S_{3}V_{2}$, $F_{3}S_{3}V_{2}$, $F_{24}S_{5}V_{21}$, $F_{7}S_{5}V_{21}$, $F_{7}S_{5}V_{10}$, $F_{22}S_{5}V_{21}$, $F_{5}S_{5}V_{21}$, $F_{5}S_{5}V_{10}$, $F_{27}S_{23}V_{2}$, $F_{27}S_{3}V_{2}$, $F_{1}S_{3}V_{2}$, $F_{29}V_{18}V_{2}$, $F_{29}V_{1}V_{2}$, $F_{3}V_{1}V_{2}$, $F_{24}S_{6}S_{25}$, $F_{7}S_{6}S_{25}$, $F_{7}S_{16}S_{6}$, $F_{22}V_{4}V_{20}$, $F_{5}V_{4}V_{20}$, $F_{5}V_{9}V_{4}$, $F_{29}S_{22}V_{2}$, $F_{29}S_{2}V_{2}$, $F_{3}S_{2}V_{2}$, $F_{24}S_{6}V_{21}$, $F_{7}S_{6}V_{21}$, $F_{7}S_{6}V_{10}$, $F_{25}S_{5}V_{20}$, $F_{8}S_{5}V_{20}$, $F_{8}S_{5}V_{9}$, $F_{22}S_{5}V_{20}$, $F_{5}S_{5}V_{20}$, $F_{5}S_{5}V_{9}$, $F_{21}S_{6}V_{21}$, $F_{17}S_{6}V_{21}$, $F_{17}S_{6}V_{10}$, $F_{27}S_{22}V_{2}$, $F_{27}S_{2}V_{2}$, $F_{1}S_{2}V_{2}$}} 
 \SplitRow{ $\mathcal{O}_{9, \; (1,-1)}^{16}$} {{$F_{8}S_{2}S_{3}$, $F_{6}S_{2}S_{5}$, $F_{2}S_{16}S_{3}$, $F_{8}S_{16}S_{3}$, $F_{8}S_{2}S_{5}$, $F_{6}S_{3}S_{5}$, $F_{8}S_{3}S_{5}$, $F_{6}S_{14}S_{2}$, $F_{8}S_{16}S_{5}$, $F_{6}S_{14}S_{3}$, $F_{8}S_{5}S_{5}$, $F_{2}S_{14}S_{3}$}} 
 \SplitRow{ $\mathcal{O}_{9, \; (1,-1)}^{17}$} {{$F_{6}S_{22}S_{3}$, $F_{23}S_{23}S_{2}$, $F_{6}S_{2}S_{3}$, $F_{8}V_{19}V_{2}$, $F_{25}V_{19}V_{2}$, $F_{8}V_{2}V_{2}$, $F_{5}S_{23}S_{2}$, $F_{22}S_{22}S_{3}$, $F_{5}S_{2}S_{3}$, $F_{27}S_{22}V_{20}$, $F_{27}S_{2}V_{20}$, $F_{1}S_{2}V_{9}$, $F_{29}S_{25}V_{19}$, $F_{29}S_{25}V_{2}$, $F_{3}S_{16}V_{2}$, $F_{25}S_{25}V_{19}$, $F_{8}S_{25}V_{2}$, $F_{8}S_{16}V_{2}$, $F_{22}S_{22}V_{20}$, $F_{5}S_{2}V_{20}$, $F_{5}S_{2}V_{9}$, $F_{27}V_{19}V_{20}$, $F_{27}V_{2}V_{20}$, $F_{1}V_{9}V_{2}$, $F_{29}S_{23}S_{25}$, $F_{29}S_{3}S_{25}$, $F_{3}S_{16}S_{3}$, $F_{23}S_{23}S_{25}$, $F_{6}S_{3}S_{25}$, $F_{6}S_{16}S_{3}$, $F_{25}V_{19}V_{20}$, $F_{8}V_{2}V_{20}$, $F_{8}V_{9}V_{2}$, $F_{27}S_{22}V_{3}$, $F_{27}S_{2}V_{3}$, $F_{1}S_{2}V_{3}$, $F_{29}S_{5}V_{19}$, $F_{29}S_{5}V_{2}$, $F_{3}S_{5}V_{2}$, $F_{25}S_{5}V_{2}$, $F_{8}S_{5}V_{19}$, $F_{8}S_{5}V_{2}$, $F_{23}S_{2}V_{3}$, $F_{6}S_{22}V_{3}$, $F_{6}S_{2}V_{3}$, $F_{27}V_{19}V_{3}$, $F_{27}V_{3}V_{2}$, $F_{1}V_{3}V_{2}$, $F_{29}S_{23}S_{5}$, $F_{29}S_{3}S_{5}$, $F_{3}S_{3}S_{5}$, $F_{22}S_{3}S_{5}$, $F_{5}S_{23}S_{5}$, $F_{5}S_{3}S_{5}$, $F_{25}V_{3}V_{2}$, $F_{8}V_{19}V_{3}$, $F_{8}V_{3}V_{2}$, $F_{27}S_{22}S_{2}$, $F_{27}S_{2}S_{2}$, $F_{1}S_{2}S_{2}$, $F_{25}S_{5}S_{25}$, $F_{8}S_{5}S_{25}$, $F_{8}S_{16}S_{5}$, $F_{27}S_{2}V_{19}$, $F_{27}S_{2}V_{2}$, $F_{1}S_{2}V_{2}$, $F_{23}S_{25}V_{3}$, $F_{6}S_{25}V_{3}$, $F_{6}S_{16}V_{3}$, $F_{25}S_{5}V_{20}$, $F_{8}S_{5}V_{20}$, $F_{8}S_{5}V_{9}$, $F_{22}S_{5}V_{20}$, $F_{5}S_{5}V_{20}$, $F_{5}S_{5}V_{9}$, $F_{25}S_{25}V_{3}$, $F_{8}S_{25}V_{3}$, $F_{8}S_{16}V_{3}$, $F_{29}S_{2}V_{19}$, $F_{29}S_{2}V_{2}$, $F_{3}S_{2}V_{2}$, $F_{29}S_{23}S_{2}$, $F_{29}S_{2}S_{3}$, $F_{3}S_{2}S_{3}$, $F_{25}V_{3}V_{20}$, $F_{8}V_{3}V_{20}$, $F_{8}V_{9}V_{3}$}} 
 \SplitRow{ $\mathcal{O}_{9, \; (1,-1)}^{18}$} {{$F_{5}S_{23}S_{2}$, $F_{22}S_{22}S_{3}$, $F_{5}S_{2}S_{3}$, $F_{6}S_{22}S_{3}$, $F_{23}S_{23}S_{2}$, $F_{6}S_{2}S_{3}$, $F_{27}S_{23}V_{20}$, $F_{27}S_{3}V_{20}$, $F_{1}S_{3}V_{9}$, $F_{23}S_{23}V_{20}$, $F_{6}S_{3}V_{20}$, $F_{6}S_{3}V_{9}$, $F_{28}S_{23}V_{21}$, $F_{28}S_{3}V_{21}$, $F_{2}S_{3}V_{10}$, $F_{27}S_{22}V_{20}$, $F_{27}S_{2}V_{20}$, $F_{1}S_{2}V_{9}$, $F_{22}S_{22}V_{20}$, $F_{5}S_{2}V_{20}$, $F_{5}S_{2}V_{9}$, $F_{23}S_{23}V_{21}$, $F_{6}S_{3}V_{21}$, $F_{6}S_{3}V_{10}$, $F_{27}S_{23}V_{3}$, $F_{27}S_{3}V_{3}$, $F_{1}S_{3}V_{3}$, $F_{22}S_{3}V_{3}$, $F_{5}S_{23}V_{3}$, $F_{5}S_{3}V_{3}$, $F_{28}S_{23}V_{4}$, $F_{28}S_{3}V_{4}$, $F_{2}S_{3}V_{4}$, $F_{27}S_{22}V_{3}$, $F_{27}S_{2}V_{3}$, $F_{1}S_{2}V_{3}$, $F_{23}S_{2}V_{3}$, $F_{6}S_{22}V_{3}$, $F_{6}S_{2}V_{3}$, $F_{22}S_{3}V_{4}$, $F_{5}S_{23}V_{4}$, $F_{5}S_{3}V_{4}$, $F_{28}S_{23}S_{2}$, $F_{28}S_{2}S_{3}$, $F_{2}S_{2}S_{3}$, $F_{22}V_{3}V_{20}$, $F_{5}V_{3}V_{20}$, $F_{5}V_{9}V_{3}$, $F_{23}V_{3}V_{20}$, $F_{6}V_{3}V_{20}$, $F_{6}V_{9}V_{3}$, $F_{27}S_{23}S_{2}$, $F_{27}S_{2}S_{3}$, $F_{1}S_{2}S_{3}$, $F_{27}S_{22}S_{2}$, $F_{27}S_{2}S_{2}$, $F_{1}S_{2}S_{2}$, $F_{22}V_{4}V_{20}$, $F_{5}V_{4}V_{20}$, $F_{5}V_{9}V_{4}$, $F_{23}V_{3}V_{21}$, $F_{6}V_{3}V_{21}$, $F_{6}V_{10}V_{3}$}} 
 \SplitRow{ $\mathcal{O}_{9, \; (1,-1)}^{19}$} {{$F_{6}S_{2}S_{3}$, $F_{8}V_{1}V_{2}$, $F_{7}V_{1}V_{2}$, $F_{5}S_{2}S_{3}$, $F_{6}S_{2}V_{4}$, $F_{13}S_{6}V_{1}$, $F_{3}S_{16}V_{2}$, $F_{8}S_{16}V_{2}$, $F_{7}S_{6}V_{1}$, $F_{5}S_{2}V_{4}$, $F_{6}V_{4}V_{1}$, $F_{3}S_{16}S_{3}$, $F_{6}S_{16}S_{3}$, $F_{7}V_{4}V_{1}$, $F_{6}V_{4}V_{2}$, $F_{13}S_{3}S_{6}$, $F_{6}S_{3}S_{6}$, $F_{8}V_{4}V_{2}$, $F_{6}S_{2}V_{3}$, $F_{13}S_{5}V_{1}$, $F_{7}S_{16}V_{2}$, $F_{8}S_{5}V_{1}$, $F_{6}V_{3}V_{1}$, $F_{5}S_{16}S_{3}$, $F_{8}V_{3}V_{1}$, $F_{6}V_{3}V_{2}$, $F_{13}S_{3}S_{5}$, $F_{5}S_{3}S_{5}$, $F_{7}V_{3}V_{2}$, $F_{6}S_{14}S_{2}$, $F_{8}S_{16}S_{5}$, $F_{7}S_{16}S_{6}$, $F_{6}S_{14}V_{1}$, $F_{6}S_{16}V_{3}$, $F_{7}S_{16}V_{4}$, $F_{5}S_{16}V_{4}$, $F_{8}S_{16}V_{3}$, $F_{13}S_{14}V_{1}$, $F_{6}S_{14}V_{2}$, $F_{6}S_{6}V_{3}$, $F_{8}S_{5}V_{4}$, $F_{5}S_{5}V_{4}$, $F_{7}S_{6}V_{3}$, $F_{3}S_{14}V_{2}$, $F_{13}S_{14}S_{3}$, $F_{8}V_{3}V_{4}$, $F_{7}V_{3}V_{4}$, $F_{3}S_{14}S_{3}$}} 
 \SplitRow{ $\mathcal{O}_{9, \; (1,-1)}^{20}$} {{$F_{8}V_{1}V_{2}$, $F_{17}S_{1}S_{2}$, $F_{5}S_{1}S_{2}$, $F_{7}V_{1}V_{2}$, $F_{13}S_{6}V_{1}$, $F_{1}S_{2}V_{9}$, $F_{5}S_{2}V_{9}$, $F_{7}S_{6}V_{1}$, $F_{6}V_{4}V_{1}$, $F_{13}S_{1}S_{6}$, $F_{1}V_{9}V_{2}$, $F_{8}V_{9}V_{2}$, $F_{17}S_{1}S_{6}$, $F_{7}V_{4}V_{1}$, $F_{6}S_{2}V_{4}$, $F_{13}S_{6}V_{2}$, $F_{8}S_{6}V_{2}$, $F_{5}S_{2}V_{4}$, $F_{13}S_{5}V_{1}$, $F_{1}S_{2}V_{10}$, $F_{17}S_{2}V_{10}$, $F_{8}S_{5}V_{1}$, $F_{6}V_{3}V_{1}$, $F_{13}S_{1}S_{5}$, $F_{1}V_{10}V_{2}$, $F_{7}V_{10}V_{2}$, $F_{5}S_{1}S_{5}$, $F_{8}V_{3}V_{1}$, $F_{6}S_{2}V_{3}$, $F_{13}S_{5}V_{2}$, $F_{7}S_{5}V_{2}$, $F_{17}S_{2}V_{3}$, $F_{6}S_{14}V_{1}$, $F_{8}S_{5}V_{9}$, $F_{17}S_{6}V_{10}$, $F_{7}S_{6}V_{10}$, $F_{5}S_{5}V_{9}$, $F_{13}S_{14}V_{1}$, $F_{6}S_{14}S_{2}$, $F_{8}S_{5}S_{6}$, $F_{7}S_{5}S_{6}$, $F_{1}S_{14}S_{2}$, $F_{13}S_{14}S_{1}$, $F_{8}V_{9}V_{3}$, $F_{7}V_{10}V_{4}$, $F_{13}S_{14}V_{2}$, $F_{8}S_{6}V_{3}$, $F_{5}S_{5}V_{4}$, $F_{7}S_{5}V_{4}$, $F_{17}S_{6}V_{3}$, $F_{1}S_{14}V_{2}$}} 
 \SplitRow{ $\mathcal{O}_{9, \; (1,-1)}^{21}$} {{$F_{5}S_{1}S_{2}$, $F_{5}S_{2}S_{3}$, $F_{6}S_{2}S_{3}$, $F_{17}S_{1}S_{2}$, $F_{6}S_{1}V_{4}$, $F_{1}S_{3}V_{9}$, $F_{6}S_{3}V_{9}$, $F_{17}S_{1}V_{4}$, $F_{18}S_{1}V_{14}$, $F_{6}S_{2}V_{4}$, $F_{1}S_{2}V_{9}$, $F_{5}S_{2}V_{9}$, $F_{5}S_{2}V_{4}$, $F_{17}S_{1}V_{14}$, $F_{18}S_{3}V_{14}$, $F_{6}S_{3}V_{14}$, $F_{6}S_{1}V_{3}$, $F_{1}S_{3}V_{10}$, $F_{5}S_{3}V_{10}$, $F_{5}S_{1}V_{3}$, $F_{18}S_{1}V_{4}$, $F_{6}S_{2}V_{3}$, $F_{1}S_{2}V_{10}$, $F_{17}S_{2}V_{10}$, $F_{5}S_{1}V_{4}$, $F_{18}S_{3}V_{4}$, $F_{17}S_{2}V_{3}$, $F_{5}S_{3}V_{4}$, $F_{18}S_{14}S_{1}$, $F_{5}V_{9}V_{3}$, $F_{5}V_{10}V_{4}$, $F_{17}V_{10}V_{4}$, $F_{6}V_{9}V_{3}$, $F_{6}S_{14}S_{1}$, $F_{18}S_{14}S_{3}$, $F_{5}V_{3}V_{4}$, $F_{17}V_{3}V_{4}$, $F_{1}S_{14}S_{3}$, $F_{6}S_{14}S_{2}$, $F_{5}V_{9}V_{4}$, $F_{17}V_{10}V_{14}$, $F_{5}V_{4}V_{4}$, $F_{6}V_{3}V_{14}$, $F_{17}V_{3}V_{14}$, $F_{1}S_{14}S_{2}$}} 
 \SplitRow{ $\mathcal{O}_{9, \; (1,-1)}^{22}$} {{$F_{8}S_{4}V_{2}$, $F_{8}S_{2}V_{2}$, $F_{7}S_{4}V_{1}$, $F_{7}S_{2}V_{1}$, $F_{9}S_{4}V_{1}$, $F_{5}S_{2}V_{1}$, $F_{9}S_{4}V_{2}$, $F_{5}S_{2}V_{2}$, $F_{14}S_{4}V_{5}$, $F_{6}S_{2}V_{4}$, $F_{4}S_{4}V_{13}$, $F_{1}S_{2}V_{9}$, $F_{9}S_{4}V_{13}$, $F_{5}S_{2}V_{9}$, $F_{9}S_{4}V_{5}$, $F_{5}S_{2}V_{4}$, $F_{8}S_{5}S_{4}$, $F_{8}S_{2}S_{5}$, $F_{14}V_{1}V_{5}$, $F_{6}V_{4}V_{1}$, $F_{4}V_{13}V_{2}$, $F_{1}V_{9}V_{2}$, $F_{8}V_{13}V_{2}$, $F_{8}V_{9}V_{2}$, $F_{7}V_{1}V_{5}$, $F_{7}V_{4}V_{1}$, $F_{9}S_{5}S_{4}$, $F_{5}S_{2}S_{5}$, $F_{14}V_{2}V_{5}$, $F_{6}V_{4}V_{2}$, $F_{8}V_{2}V_{5}$, $F_{8}V_{4}V_{2}$, $F_{14}S_{5}S_{4}$, $F_{6}S_{2}S_{5}$, $F_{4}S_{16}S_{4}$, $F_{1}S_{16}S_{2}$, $F_{7}S_{16}S_{4}$, $F_{7}S_{16}S_{2}$, $F_{8}S_{4}V_{3}$, $F_{8}S_{2}V_{3}$, $F_{14}S_{5}V_{1}$, $F_{6}S_{5}V_{1}$, $F_{4}S_{16}V_{2}$, $F_{1}S_{16}V_{2}$, $F_{9}S_{16}V_{2}$, $F_{5}S_{16}V_{2}$, $F_{9}S_{5}V_{1}$, $F_{5}S_{5}V_{1}$, $F_{14}S_{5}V_{2}$, $F_{6}S_{5}V_{2}$, $F_{9}S_{5}V_{2}$, $F_{5}S_{5}V_{2}$, $F_{7}S_{4}V_{3}$, $F_{7}S_{2}V_{3}$, $F_{8}S_{4}V_{11}$, $F_{8}S_{2}V_{11}$, $F_{8}S_{5}V_{13}$, $F_{8}S_{5}V_{9}$, $F_{7}S_{16}V_{5}$, $F_{7}S_{16}V_{4}$, $F_{9}S_{16}V_{5}$, $F_{5}S_{16}V_{4}$, $F_{9}S_{5}V_{13}$, $F_{5}S_{5}V_{9}$, $F_{14}S_{4}V_{11}$, $F_{6}S_{2}V_{11}$, $F_{8}S_{5}V_{5}$, $F_{8}S_{5}V_{4}$, $F_{9}S_{5}V_{5}$, $F_{5}S_{5}V_{4}$, $F_{4}S_{4}V_{11}$, $F_{1}S_{2}V_{11}$, $F_{14}V_{11}V_{1}$, $F_{6}V_{11}V_{1}$, $F_{8}V_{13}V_{3}$, $F_{8}V_{9}V_{3}$, $F_{9}S_{16}S_{5}$, $F_{5}S_{16}S_{5}$, $F_{14}V_{11}V_{2}$, $F_{6}V_{11}V_{2}$, $F_{8}V_{3}V_{5}$, $F_{8}V_{3}V_{4}$, $F_{9}S_{5}S_{5}$, $F_{5}S_{5}S_{5}$, $F_{7}V_{3}V_{5}$, $F_{7}V_{3}V_{4}$, $F_{4}V_{11}V_{2}$, $F_{1}V_{11}V_{2}$}} 
 \SplitRow{ $\mathcal{O}_{9, \; (1,-1)}^{23}$} {{$F_{9}S_{4}V_{1}$, $F_{5}S_{2}V_{1}$, $F_{7}S_{4}V_{1}$, $F_{7}S_{2}V_{1}$, $F_{14}V_{1}V_{5}$, $F_{6}V_{4}V_{1}$, $F_{3}S_{16}S_{4}$, $F_{3}S_{16}S_{2}$, $F_{9}S_{16}S_{4}$, $F_{5}S_{16}S_{2}$, $F_{7}V_{1}V_{5}$, $F_{7}V_{4}V_{1}$, $F_{14}S_{4}V_{5}$, $F_{6}S_{2}V_{4}$, $F_{9}S_{4}V_{5}$, $F_{5}S_{2}V_{4}$, $F_{14}S_{5}V_{1}$, $F_{6}S_{5}V_{1}$, $F_{3}S_{4}V_{10}$, $F_{3}S_{2}V_{10}$, $F_{7}S_{4}V_{10}$, $F_{7}S_{2}V_{10}$, $F_{9}S_{5}V_{1}$, $F_{5}S_{5}V_{1}$, $F_{14}S_{5}S_{4}$, $F_{6}S_{2}S_{5}$, $F_{7}S_{5}S_{4}$, $F_{7}S_{2}S_{5}$, $F_{14}V_{11}V_{1}$, $F_{6}V_{11}V_{1}$, $F_{9}S_{16}S_{5}$, $F_{5}S_{16}S_{5}$, $F_{7}V_{10}V_{5}$, $F_{7}V_{10}V_{4}$, $F_{14}S_{4}V_{11}$, $F_{6}S_{2}V_{11}$, $F_{9}S_{5}V_{5}$, $F_{5}S_{5}V_{4}$, $F_{7}S_{5}V_{5}$, $F_{7}S_{5}V_{4}$, $F_{3}S_{4}V_{11}$, $F_{3}S_{2}V_{11}$}} 
 \SplitRow{ $\mathcal{O}_{9, \; (1,-1)}^{24}$} {{$F_{5}S_{2}V_{1}$, $F_{8}S_{3}V_{2}$, $F_{6}S_{3}V_{2}$, $F_{7}S_{2}V_{1}$, $F_{8}S_{5}V_{1}$, $F_{1}S_{3}V_{9}$, $F_{6}S_{3}V_{9}$, $F_{7}S_{5}V_{1}$, $F_{13}S_{6}V_{1}$, $F_{8}S_{5}V_{2}$, $F_{1}S_{2}V_{9}$, $F_{5}S_{2}V_{9}$, $F_{7}S_{6}V_{1}$, $F_{13}S_{3}S_{6}$, $F_{8}S_{2}S_{5}$, $F_{5}S_{2}S_{5}$, $F_{6}S_{3}S_{6}$, $F_{8}V_{3}V_{1}$, $F_{1}S_{16}S_{3}$, $F_{8}S_{16}S_{3}$, $F_{5}V_{3}V_{1}$, $F_{13}V_{4}V_{1}$, $F_{8}V_{3}V_{2}$, $F_{1}S_{16}S_{2}$, $F_{7}S_{16}S_{2}$, $F_{6}V_{3}V_{2}$, $F_{5}V_{4}V_{1}$, $F_{13}S_{3}V_{4}$, $F_{8}S_{2}V_{3}$, $F_{7}S_{2}V_{3}$, $F_{8}S_{3}V_{4}$, $F_{13}V_{11}V_{1}$, $F_{5}V_{9}V_{3}$, $F_{8}S_{16}S_{5}$, $F_{7}S_{16}S_{5}$, $F_{6}V_{9}V_{3}$, $F_{8}V_{11}V_{1}$, $F_{13}S_{3}V_{11}$, $F_{5}S_{5}V_{3}$, $F_{7}S_{5}V_{3}$, $F_{1}S_{3}V_{11}$, $F_{8}V_{11}V_{2}$, $F_{5}V_{9}V_{4}$, $F_{7}S_{16}S_{6}$, $F_{8}S_{2}V_{11}$, $F_{5}S_{5}V_{4}$, $F_{6}S_{6}V_{3}$, $F_{7}S_{6}V_{3}$, $F_{8}S_{5}V_{4}$, $F_{1}S_{2}V_{11}$}} 
 \SplitRow{ $\mathcal{O}_{9, \; (1,-1)}^{25}$} {{$F_{5}S_{2}V_{1}$, $F_{7}S_{1}V_{1}$, $F_{5}S_{2}V_{2}$, $F_{8}S_{2}V_{2}$, $F_{17}S_{1}V_{1}$, $F_{7}S_{2}V_{1}$, $F_{8}S_{5}V_{1}$, $F_{6}S_{1}V_{4}$, $F_{3}S_{16}V_{2}$, $F_{8}S_{16}V_{2}$, $F_{17}S_{1}V_{4}$, $F_{7}S_{5}V_{1}$, $F_{13}S_{6}V_{1}$, $F_{6}S_{2}V_{4}$, $F_{3}S_{16}V_{1}$, $F_{7}S_{16}V_{1}$, $F_{5}S_{2}V_{4}$, $F_{7}S_{6}V_{1}$, $F_{13}S_{1}S_{6}$, $F_{8}S_{2}S_{5}$, $F_{3}S_{16}S_{2}$, $F_{5}S_{16}S_{2}$, $F_{5}S_{2}S_{5}$, $F_{17}S_{1}S_{6}$, $F_{13}S_{6}V_{2}$, $F_{8}S_{6}V_{2}$, $F_{8}V_{3}V_{1}$, $F_{6}S_{1}S_{5}$, $F_{3}V_{10}V_{2}$, $F_{5}V_{10}V_{2}$, $F_{7}S_{1}S_{5}$, $F_{5}V_{3}V_{1}$, $F_{13}V_{4}V_{1}$, $F_{6}S_{2}S_{5}$, $F_{3}V_{10}V_{1}$, $F_{17}V_{10}V_{1}$, $F_{5}V_{4}V_{1}$, $F_{13}S_{1}V_{4}$, $F_{8}S_{2}V_{3}$, $F_{3}S_{2}V_{10}$, $F_{7}S_{2}V_{10}$, $F_{7}S_{1}V_{4}$, $F_{13}V_{4}V_{2}$, $F_{7}S_{2}S_{5}$, $F_{17}V_{3}V_{1}$, $F_{5}V_{4}V_{2}$, $F_{13}V_{11}V_{1}$, $F_{7}S_{16}S_{5}$, $F_{5}V_{10}V_{4}$, $F_{17}V_{10}V_{4}$, $F_{8}S_{16}S_{5}$, $F_{8}V_{11}V_{1}$, $F_{13}S_{1}V_{11}$, $F_{5}S_{16}V_{3}$, $F_{5}S_{5}V_{10}$, $F_{7}S_{5}V_{10}$, $F_{8}S_{16}V_{3}$, $F_{6}S_{1}V_{11}$, $F_{13}V_{11}V_{2}$, $F_{5}V_{3}V_{4}$, $F_{7}S_{5}S_{5}$, $F_{17}V_{3}V_{4}$, $F_{3}V_{11}V_{2}$, $F_{8}S_{2}V_{11}$, $F_{5}S_{16}V_{4}$, $F_{17}S_{6}V_{10}$, $F_{7}S_{6}V_{10}$, $F_{7}S_{16}V_{4}$, $F_{6}S_{2}V_{11}$, $F_{5}V_{4}V_{4}$, $F_{8}S_{5}S_{6}$, $F_{7}S_{5}S_{6}$, $F_{3}V_{11}V_{1}$, $F_{7}S_{5}V_{4}$, $F_{8}S_{6}V_{3}$, $F_{17}S_{6}V_{3}$, $F_{5}S_{5}V_{4}$, $F_{3}S_{2}V_{11}$}} 
 \SplitRow{ $\mathcal{O}_{9, \; (1,-1)}^{26}$} {{$F_{8}S_{24}V_{2}$, $F_{8}S_{22}V_{2}$, $F_{25}S_{4}V_{19}$, $F_{8}S_{4}V_{2}$, $F_{25}S_{2}V_{19}$, $F_{8}S_{2}V_{2}$, $F_{9}S_{4}V_{19}$, $F_{5}S_{2}V_{19}$, $F_{26}S_{24}V_{2}$, $F_{22}S_{22}V_{2}$, $F_{9}S_{4}V_{2}$, $F_{5}S_{2}V_{2}$, $F_{30}S_{24}V_{22}$, $F_{27}S_{22}V_{20}$, $F_{30}S_{4}V_{22}$, $F_{4}S_{4}V_{13}$, $F_{27}S_{2}V_{20}$, $F_{1}S_{2}V_{9}$, $F_{26}S_{24}V_{22}$, $F_{9}S_{4}V_{22}$, $F_{22}S_{22}V_{20}$, $F_{5}S_{2}V_{20}$, $F_{9}S_{4}V_{13}$, $F_{5}S_{2}V_{9}$, $F_{29}S_{24}S_{25}$, $F_{29}S_{22}S_{25}$, $F_{29}S_{4}S_{25}$, $F_{3}S_{16}S_{4}$, $F_{29}S_{2}S_{25}$, $F_{3}S_{16}S_{2}$, $F_{30}V_{19}V_{22}$, $F_{27}V_{19}V_{20}$, $F_{30}V_{2}V_{22}$, $F_{27}V_{2}V_{20}$, $F_{4}V_{13}V_{2}$, $F_{1}V_{9}V_{2}$, $F_{25}V_{19}V_{22}$, $F_{8}V_{2}V_{22}$, $F_{25}V_{19}V_{20}$, $F_{8}V_{2}V_{20}$, $F_{8}V_{13}V_{2}$, $F_{8}V_{9}V_{2}$, $F_{26}S_{24}S_{25}$, $F_{22}S_{22}S_{25}$, $F_{9}S_{4}S_{25}$, $F_{5}S_{2}S_{25}$, $F_{9}S_{16}S_{4}$, $F_{5}S_{16}S_{2}$, $F_{30}S_{24}S_{5}$, $F_{27}S_{22}S_{5}$, $F_{30}S_{5}S_{4}$, $F_{4}S_{5}S_{4}$, $F_{27}S_{2}S_{5}$, $F_{1}S_{2}S_{5}$, $F_{25}S_{5}S_{4}$, $F_{25}S_{2}S_{5}$, $F_{8}S_{24}S_{5}$, $F_{8}S_{22}S_{5}$, $F_{8}S_{5}S_{4}$, $F_{8}S_{2}S_{5}$, $F_{29}S_{24}V_{3}$, $F_{29}S_{22}V_{3}$, $F_{29}S_{4}V_{3}$, $F_{3}S_{4}V_{3}$, $F_{29}S_{2}V_{3}$, $F_{3}S_{2}V_{3}$, $F_{30}S_{5}V_{19}$, $F_{27}S_{5}V_{19}$, $F_{30}S_{5}V_{2}$, $F_{27}S_{5}V_{2}$, $F_{4}S_{5}V_{2}$, $F_{1}S_{5}V_{2}$, $F_{26}S_{5}V_{2}$, $F_{22}S_{5}V_{2}$, $F_{9}S_{5}V_{19}$, $F_{9}S_{5}V_{2}$, $F_{5}S_{5}V_{19}$, $F_{5}S_{5}V_{2}$, $F_{25}S_{4}V_{3}$, $F_{25}S_{2}V_{3}$, $F_{8}S_{24}V_{3}$, $F_{8}S_{22}V_{3}$, $F_{8}S_{4}V_{3}$, $F_{8}S_{2}V_{3}$, $F_{29}S_{24}V_{2}$, $F_{29}S_{22}V_{2}$, $F_{29}S_{4}V_{2}$, $F_{3}S_{4}V_{2}$, $F_{29}S_{2}V_{2}$, $F_{3}S_{2}V_{2}$, $F_{25}S_{5}V_{22}$, $F_{8}S_{5}V_{22}$, $F_{25}S_{5}V_{20}$, $F_{8}S_{5}V_{20}$, $F_{8}S_{5}V_{13}$, $F_{8}S_{5}V_{9}$, $F_{26}S_{5}V_{22}$, $F_{22}S_{5}V_{20}$, $F_{9}S_{5}V_{22}$, $F_{9}S_{5}V_{13}$, $F_{5}S_{5}V_{20}$, $F_{5}S_{5}V_{9}$, $F_{30}S_{24}V_{2}$, $F_{30}S_{4}V_{2}$, $F_{27}S_{22}V_{2}$, $F_{27}S_{2}V_{2}$, $F_{4}S_{4}V_{2}$, $F_{1}S_{2}V_{2}$, $F_{30}V_{19}V_{2}$, $F_{27}V_{19}V_{2}$, $F_{30}V_{2}V_{2}$, $F_{27}V_{2}V_{2}$, $F_{4}V_{2}V_{2}$, $F_{1}V_{2}V_{2}$, $F_{25}V_{3}V_{22}$, $F_{8}V_{3}V_{22}$, $F_{25}V_{3}V_{20}$, $F_{8}V_{3}V_{20}$, $F_{8}V_{13}V_{3}$, $F_{8}V_{9}V_{3}$, $F_{26}S_{5}S_{25}$, $F_{22}S_{5}S_{25}$, $F_{9}S_{5}S_{25}$, $F_{5}S_{5}S_{25}$, $F_{9}S_{16}S_{5}$, $F_{5}S_{16}S_{5}$}} 
 \SplitRow{ $\mathcal{O}_{9, \; (1,-1)}^{27}$} {{$F_{5}S_{2}V_{19}$, $F_{22}S_{22}V_{2}$, $F_{5}S_{2}V_{2}$, $F_{7}S_{23}V_{1}$, $F_{24}S_{3}V_{18}$, $F_{7}S_{3}V_{1}$, $F_{6}S_{3}V_{18}$, $F_{23}S_{23}V_{1}$, $F_{6}S_{3}V_{1}$, $F_{8}S_{22}V_{2}$, $F_{25}S_{2}V_{19}$, $F_{8}S_{2}V_{2}$, $F_{29}S_{25}V_{19}$, $F_{29}S_{25}V_{2}$, $F_{3}S_{16}V_{2}$, $F_{27}S_{23}V_{20}$, $F_{27}S_{3}V_{20}$, $F_{1}S_{3}V_{9}$, $F_{23}S_{23}V_{20}$, $F_{6}S_{3}V_{20}$, $F_{6}S_{3}V_{9}$, $F_{25}S_{25}V_{19}$, $F_{8}S_{25}V_{2}$, $F_{8}S_{16}V_{2}$, $F_{29}S_{25}V_{18}$, $F_{29}S_{25}V_{1}$, $F_{3}S_{16}V_{1}$, $F_{27}S_{22}V_{20}$, $F_{27}S_{2}V_{20}$, $F_{1}S_{2}V_{9}$, $F_{22}S_{22}V_{20}$, $F_{5}S_{2}V_{20}$, $F_{5}S_{2}V_{9}$, $F_{24}S_{25}V_{18}$, $F_{7}S_{25}V_{1}$, $F_{7}S_{16}V_{1}$, $F_{29}S_{23}S_{25}$, $F_{29}S_{3}S_{25}$, $F_{3}S_{16}S_{3}$, $F_{29}S_{22}S_{25}$, $F_{29}S_{2}S_{25}$, $F_{3}S_{16}S_{2}$, $F_{22}S_{22}S_{25}$, $F_{5}S_{2}S_{25}$, $F_{5}S_{16}S_{2}$, $F_{23}S_{23}S_{25}$, $F_{6}S_{3}S_{25}$, $F_{6}S_{16}S_{3}$, $F_{29}V_{19}V_{3}$, $F_{29}V_{3}V_{2}$, $F_{3}V_{3}V_{2}$, $F_{27}S_{23}S_{5}$, $F_{27}S_{3}S_{5}$, $F_{1}S_{3}S_{5}$, $F_{24}S_{3}S_{5}$, $F_{7}S_{23}S_{5}$, $F_{7}S_{3}S_{5}$, $F_{22}V_{3}V_{2}$, $F_{5}V_{19}V_{3}$, $F_{5}V_{3}V_{2}$, $F_{29}V_{19}V_{4}$, $F_{29}V_{4}V_{2}$, $F_{3}V_{4}V_{2}$, $F_{29}V_{18}V_{3}$, $F_{29}V_{3}V_{1}$, $F_{3}V_{3}V_{1}$, $F_{27}S_{22}S_{5}$, $F_{27}S_{2}S_{5}$, $F_{1}S_{2}S_{5}$, $F_{25}S_{2}S_{5}$, $F_{8}S_{22}S_{5}$, $F_{8}S_{2}S_{5}$, $F_{23}V_{3}V_{1}$, $F_{6}V_{18}V_{3}$, $F_{6}V_{3}V_{1}$, $F_{22}V_{4}V_{2}$, $F_{5}V_{19}V_{4}$, $F_{5}V_{4}V_{2}$, $F_{29}S_{23}V_{4}$, $F_{29}S_{3}V_{4}$, $F_{3}S_{3}V_{4}$, $F_{29}S_{22}V_{3}$, $F_{29}S_{2}V_{3}$, $F_{3}S_{2}V_{3}$, $F_{25}S_{2}V_{3}$, $F_{8}S_{22}V_{3}$, $F_{8}S_{2}V_{3}$, $F_{24}S_{3}V_{4}$, $F_{7}S_{23}V_{4}$, $F_{7}S_{3}V_{4}$, $F_{29}V_{19}V_{2}$, $F_{29}V_{2}V_{2}$, $F_{3}V_{2}V_{2}$, $F_{22}V_{3}V_{20}$, $F_{5}V_{3}V_{20}$, $F_{5}V_{9}V_{3}$, $F_{24}S_{5}S_{25}$, $F_{7}S_{5}S_{25}$, $F_{7}S_{16}S_{5}$, $F_{25}S_{5}S_{25}$, $F_{8}S_{5}S_{25}$, $F_{8}S_{16}S_{5}$, $F_{23}V_{3}V_{20}$, $F_{6}V_{3}V_{20}$, $F_{6}V_{9}V_{3}$, $F_{29}S_{23}V_{2}$, $F_{29}S_{3}V_{2}$, $F_{3}S_{3}V_{2}$, $F_{22}S_{25}V_{3}$, $F_{5}S_{25}V_{3}$, $F_{5}S_{16}V_{3}$, $F_{25}S_{25}V_{3}$, $F_{8}S_{25}V_{3}$, $F_{8}S_{16}V_{3}$, $F_{27}S_{23}V_{2}$, $F_{27}S_{3}V_{2}$, $F_{1}S_{3}V_{2}$, $F_{29}V_{18}V_{2}$, $F_{29}V_{1}V_{2}$, $F_{3}V_{1}V_{2}$, $F_{22}V_{4}V_{20}$, $F_{5}V_{4}V_{20}$, $F_{5}V_{9}V_{4}$, $F_{29}S_{22}V_{2}$, $F_{29}S_{2}V_{2}$, $F_{3}S_{2}V_{2}$, $F_{22}S_{25}V_{4}$, $F_{5}S_{25}V_{4}$, $F_{5}S_{16}V_{4}$, $F_{23}S_{25}V_{3}$, $F_{6}S_{25}V_{3}$, $F_{6}S_{16}V_{3}$, $F_{24}S_{25}V_{4}$, $F_{7}S_{25}V_{4}$, $F_{7}S_{16}V_{4}$, $F_{27}S_{22}V_{2}$, $F_{27}S_{2}V_{2}$, $F_{1}S_{2}V_{2}$}} 
 \SplitRow{ $\mathcal{O}_{9, \; (1,-1)}^{28}$} {{$F_{9}S_{24}S_{4}$, $F_{5}S_{22}S_{2}$, $F_{26}S_{24}S_{4}$, $F_{9}S_{4}S_{4}$, $F_{22}S_{22}S_{2}$, $F_{5}S_{2}S_{2}$, $F_{29}S_{24}S_{25}$, $F_{29}S_{22}S_{25}$, $F_{29}S_{4}S_{25}$, $F_{3}S_{16}S_{4}$, $F_{29}S_{2}S_{25}$, $F_{3}S_{16}S_{2}$, $F_{26}S_{24}S_{25}$, $F_{22}S_{22}S_{25}$, $F_{9}S_{4}S_{25}$, $F_{5}S_{2}S_{25}$, $F_{9}S_{16}S_{4}$, $F_{5}S_{16}S_{2}$, $F_{29}S_{24}S_{5}$, $F_{29}S_{22}S_{5}$, $F_{29}S_{5}S_{4}$, $F_{3}S_{5}S_{4}$, $F_{29}S_{2}S_{5}$, $F_{3}S_{2}S_{5}$, $F_{26}S_{5}S_{4}$, $F_{22}S_{2}S_{5}$, $F_{9}S_{24}S_{5}$, $F_{9}S_{5}S_{4}$, $F_{5}S_{22}S_{5}$, $F_{5}S_{2}S_{5}$, $F_{29}S_{24}S_{3}$, $F_{29}S_{22}S_{3}$, $F_{29}S_{3}S_{4}$, $F_{3}S_{3}S_{4}$, $F_{29}S_{2}S_{3}$, $F_{3}S_{2}S_{3}$, $F_{26}S_{5}S_{25}$, $F_{22}S_{5}S_{25}$, $F_{9}S_{5}S_{25}$, $F_{5}S_{5}S_{25}$, $F_{9}S_{16}S_{5}$, $F_{5}S_{16}S_{5}$}} 
 \SplitRow{ $\mathcal{O}_{9, \; (1,-1)}^{29}$} {{$F_{5}S_{22}S_{2}$, $F_{22}S_{22}S_{2}$, $F_{5}S_{2}S_{2}$, $F_{7}V_{19}V_{1}$, $F_{24}V_{18}V_{2}$, $F_{7}V_{1}V_{2}$, $F_{8}V_{18}V_{2}$, $F_{25}V_{19}V_{1}$, $F_{8}V_{1}V_{2}$, $F_{28}S_{22}V_{21}$, $F_{28}S_{2}V_{21}$, $F_{2}S_{2}V_{10}$, $F_{29}S_{25}V_{19}$, $F_{29}S_{25}V_{2}$, $F_{3}S_{16}V_{2}$, $F_{25}S_{25}V_{19}$, $F_{8}S_{25}V_{2}$, $F_{8}S_{16}V_{2}$, $F_{22}S_{22}V_{21}$, $F_{5}S_{2}V_{21}$, $F_{5}S_{2}V_{10}$, $F_{27}S_{22}V_{20}$, $F_{27}S_{2}V_{20}$, $F_{1}S_{2}V_{9}$, $F_{29}S_{25}V_{18}$, $F_{29}S_{25}V_{1}$, $F_{3}S_{16}V_{1}$, $F_{24}S_{25}V_{18}$, $F_{7}S_{25}V_{1}$, $F_{7}S_{16}V_{1}$, $F_{22}S_{22}V_{20}$, $F_{5}S_{2}V_{20}$, $F_{5}S_{2}V_{9}$, $F_{27}V_{19}V_{20}$, $F_{27}V_{2}V_{20}$, $F_{1}V_{9}V_{2}$, $F_{28}V_{18}V_{21}$, $F_{28}V_{1}V_{21}$, $F_{2}V_{10}V_{1}$, $F_{29}S_{22}S_{25}$, $F_{29}S_{2}S_{25}$, $F_{3}S_{16}S_{2}$, $F_{22}S_{22}S_{25}$, $F_{5}S_{2}S_{25}$, $F_{5}S_{16}S_{2}$, $F_{24}V_{18}V_{21}$, $F_{7}V_{1}V_{21}$, $F_{7}V_{10}V_{1}$, $F_{25}V_{19}V_{20}$, $F_{8}V_{2}V_{20}$, $F_{8}V_{9}V_{2}$, $F_{28}S_{22}V_{3}$, $F_{28}S_{2}V_{3}$, $F_{2}S_{2}V_{3}$, $F_{29}S_{5}V_{19}$, $F_{29}S_{5}V_{2}$, $F_{3}S_{5}V_{2}$, $F_{24}S_{5}V_{2}$, $F_{7}S_{5}V_{19}$, $F_{7}S_{5}V_{2}$, $F_{22}S_{2}V_{3}$, $F_{5}S_{22}V_{3}$, $F_{5}S_{2}V_{3}$, $F_{27}S_{22}V_{4}$, $F_{27}S_{2}V_{4}$, $F_{1}S_{2}V_{4}$, $F_{29}S_{5}V_{18}$, $F_{29}S_{5}V_{1}$, $F_{3}S_{5}V_{1}$, $F_{25}S_{5}V_{1}$, $F_{8}S_{5}V_{18}$, $F_{8}S_{5}V_{1}$, $F_{22}S_{2}V_{4}$, $F_{5}S_{22}V_{4}$, $F_{5}S_{2}V_{4}$, $F_{27}V_{19}V_{4}$, $F_{27}V_{4}V_{2}$, $F_{1}V_{4}V_{2}$, $F_{28}V_{18}V_{3}$, $F_{28}V_{3}V_{1}$, $F_{2}V_{3}V_{1}$, $F_{29}S_{22}S_{5}$, $F_{29}S_{2}S_{5}$, $F_{3}S_{2}S_{5}$, $F_{22}S_{2}S_{5}$, $F_{5}S_{22}S_{5}$, $F_{5}S_{2}S_{5}$, $F_{25}V_{3}V_{1}$, $F_{8}V_{18}V_{3}$, $F_{8}V_{3}V_{1}$, $F_{24}V_{4}V_{2}$, $F_{7}V_{19}V_{4}$, $F_{7}V_{4}V_{2}$, $F_{27}S_{22}S_{3}$, $F_{27}S_{2}S_{3}$, $F_{1}S_{2}S_{3}$, $F_{24}S_{5}S_{25}$, $F_{7}S_{5}S_{25}$, $F_{7}S_{16}S_{5}$, $F_{25}S_{5}S_{25}$, $F_{8}S_{5}S_{25}$, $F_{8}S_{16}S_{5}$, $F_{28}S_{22}S_{3}$, $F_{28}S_{2}S_{3}$, $F_{2}S_{2}S_{3}$, $F_{27}S_{3}V_{19}$, $F_{27}S_{3}V_{2}$, $F_{1}S_{3}V_{2}$, $F_{22}S_{25}V_{3}$, $F_{5}S_{25}V_{3}$, $F_{5}S_{16}V_{3}$, $F_{24}S_{5}V_{21}$, $F_{7}S_{5}V_{21}$, $F_{7}S_{5}V_{10}$, $F_{22}S_{5}V_{21}$, $F_{5}S_{5}V_{21}$, $F_{5}S_{5}V_{10}$, $F_{25}S_{25}V_{3}$, $F_{8}S_{25}V_{3}$, $F_{8}S_{16}V_{3}$, $F_{29}S_{3}V_{19}$, $F_{29}S_{3}V_{2}$, $F_{3}S_{3}V_{2}$, $F_{28}S_{3}V_{18}$, $F_{28}S_{3}V_{1}$, $F_{2}S_{3}V_{1}$, $F_{22}S_{25}V_{4}$, $F_{5}S_{25}V_{4}$, $F_{5}S_{16}V_{4}$, $F_{25}S_{5}V_{20}$, $F_{8}S_{5}V_{20}$, $F_{8}S_{5}V_{9}$, $F_{22}S_{5}V_{20}$, $F_{5}S_{5}V_{20}$, $F_{5}S_{5}V_{9}$, $F_{24}S_{25}V_{4}$, $F_{7}S_{25}V_{4}$, $F_{7}S_{16}V_{4}$, $F_{29}S_{3}V_{18}$, $F_{29}S_{3}V_{1}$, $F_{3}S_{3}V_{1}$, $F_{29}S_{22}S_{3}$, $F_{29}S_{2}S_{3}$, $F_{3}S_{2}S_{3}$, $F_{24}V_{4}V_{21}$, $F_{7}V_{4}V_{21}$, $F_{7}V_{10}V_{4}$, $F_{25}V_{3}V_{20}$, $F_{8}V_{3}V_{20}$, $F_{8}V_{9}V_{3}$}} 
 \SplitRow{ $\mathcal{O}_{9, \; (1,-1)}^{30}$} {{$F_{17}S_{23}S_{1}$, $F_{21}S_{21}S_{3}$, $F_{17}S_{1}S_{3}$, $F_{5}S_{22}S_{2}$, $F_{22}S_{22}S_{2}$, $F_{5}S_{2}S_{2}$, $F_{6}S_{21}S_{3}$, $F_{23}S_{23}S_{1}$, $F_{6}S_{1}S_{3}$, $F_{27}S_{23}V_{20}$, $F_{27}S_{3}V_{20}$, $F_{1}S_{3}V_{9}$, $F_{28}S_{22}V_{21}$, $F_{28}S_{2}V_{21}$, $F_{2}S_{2}V_{10}$, $F_{22}S_{22}V_{21}$, $F_{5}S_{2}V_{21}$, $F_{5}S_{2}V_{10}$, $F_{23}S_{23}V_{20}$, $F_{6}S_{3}V_{20}$, $F_{6}S_{3}V_{9}$, $F_{27}S_{22}V_{20}$, $F_{27}S_{2}V_{20}$, $F_{1}S_{2}V_{9}$, $F_{28}S_{21}V_{21}$, $F_{28}S_{1}V_{21}$, $F_{2}S_{1}V_{10}$, $F_{21}S_{21}V_{21}$, $F_{17}S_{1}V_{21}$, $F_{17}S_{1}V_{10}$, $F_{22}S_{22}V_{20}$, $F_{5}S_{2}V_{20}$, $F_{5}S_{2}V_{9}$, $F_{27}S_{23}V_{4}$, $F_{27}S_{3}V_{4}$, $F_{1}S_{3}V_{4}$, $F_{28}S_{22}V_{3}$, $F_{28}S_{2}V_{3}$, $F_{2}S_{2}V_{3}$, $F_{22}S_{2}V_{3}$, $F_{5}S_{22}V_{3}$, $F_{5}S_{2}V_{3}$, $F_{21}S_{3}V_{4}$, $F_{17}S_{23}V_{4}$, $F_{17}S_{3}V_{4}$, $F_{27}S_{22}V_{4}$, $F_{27}S_{2}V_{4}$, $F_{1}S_{2}V_{4}$, $F_{28}S_{21}V_{3}$, $F_{28}S_{1}V_{3}$, $F_{2}S_{1}V_{3}$, $F_{23}S_{1}V_{3}$, $F_{6}S_{21}V_{3}$, $F_{6}S_{1}V_{3}$, $F_{22}S_{2}V_{4}$, $F_{5}S_{22}V_{4}$, $F_{5}S_{2}V_{4}$, $F_{27}S_{23}S_{3}$, $F_{27}S_{3}S_{3}$, $F_{1}S_{3}S_{3}$, $F_{22}V_{3}V_{21}$, $F_{5}V_{3}V_{21}$, $F_{5}V_{10}V_{3}$, $F_{27}S_{22}S_{3}$, $F_{27}S_{2}S_{3}$, $F_{1}S_{2}S_{3}$, $F_{21}V_{4}V_{21}$, $F_{17}V_{4}V_{21}$, $F_{17}V_{10}V_{4}$, $F_{22}V_{3}V_{20}$, $F_{5}V_{3}V_{20}$, $F_{5}V_{9}V_{3}$, $F_{23}V_{3}V_{20}$, $F_{6}V_{3}V_{20}$, $F_{6}V_{9}V_{3}$, $F_{22}V_{4}V_{21}$, $F_{5}V_{4}V_{21}$, $F_{5}V_{10}V_{4}$, $F_{28}S_{22}S_{3}$, $F_{28}S_{2}S_{3}$, $F_{2}S_{2}S_{3}$, $F_{28}S_{21}S_{3}$, $F_{28}S_{1}S_{3}$, $F_{2}S_{1}S_{3}$, $F_{22}V_{4}V_{20}$, $F_{5}V_{4}V_{20}$, $F_{5}V_{9}V_{4}$}} 
 \SplitRow{ $\mathcal{O}_{9, \; (1,-1)}^{31}$} {{$F_{6}S_{2}S_{3}$, $F_{5}S_{2}S_{3}$, $F_{6}S_{2}V_{4}$, $F_{2}S_{3}V_{10}$, $F_{6}S_{3}V_{10}$, $F_{5}S_{2}V_{4}$, $F_{6}S_{3}V_{4}$, $F_{6}S_{2}V_{3}$, $F_{2}S_{3}V_{9}$, $F_{5}S_{3}V_{9}$, $F_{6}S_{3}V_{3}$, $F_{5}S_{3}V_{3}$, $F_{6}S_{14}S_{2}$, $F_{6}V_{10}V_{3}$, $F_{5}V_{9}V_{4}$, $F_{6}S_{14}S_{3}$, $F_{6}V_{3}V_{4}$, $F_{5}V_{3}V_{4}$, $F_{2}S_{14}S_{3}$}} 
 \SplitRow{ $\mathcal{O}_{9, \; (1,-1)}^{32}$} {{$F_{6}S_{3}V_{2}$, $F_{8}S_{2}V_{2}$, $F_{5}S_{2}V_{2}$, $F_{8}S_{3}V_{2}$, $F_{8}S_{5}V_{2}$, $F_{6}S_{2}V_{4}$, $F_{3}S_{16}V_{2}$, $F_{8}S_{16}V_{2}$, $F_{5}S_{2}V_{4}$, $F_{8}S_{2}S_{5}$, $F_{3}S_{16}S_{3}$, $F_{6}S_{16}S_{3}$, $F_{5}S_{2}S_{5}$, $F_{6}S_{3}V_{4}$, $F_{8}V_{3}V_{2}$, $F_{6}S_{2}S_{5}$, $F_{3}V_{9}V_{2}$, $F_{5}V_{9}V_{2}$, $F_{6}V_{3}V_{2}$, $F_{8}S_{2}V_{3}$, $F_{3}S_{3}V_{9}$, $F_{8}S_{3}V_{9}$, $F_{6}S_{3}S_{5}$, $F_{8}S_{3}S_{5}$, $F_{5}V_{3}V_{2}$, $F_{8}V_{11}V_{2}$, $F_{8}S_{16}S_{5}$, $F_{5}V_{9}V_{4}$, $F_{8}S_{2}V_{11}$, $F_{6}S_{16}V_{3}$, $F_{5}S_{5}V_{9}$, $F_{8}S_{5}V_{9}$, $F_{8}S_{16}V_{3}$, $F_{6}S_{2}V_{11}$, $F_{6}V_{3}V_{4}$, $F_{8}S_{5}S_{5}$, $F_{5}V_{3}V_{4}$, $F_{3}V_{11}V_{2}$, $F_{6}S_{3}V_{11}$, $F_{8}S_{5}V_{3}$, $F_{5}S_{5}V_{3}$, $F_{3}S_{3}V_{11}$}} 
 \SplitRow{ $\mathcal{O}_{9, \; (1,-1)}^{33}$} {{$F_{9}S_{4}S_{4}$, $F_{5}S_{2}S_{2}$, $F_{8}S_{5}S_{4}$, $F_{8}S_{2}S_{5}$, $F_{3}S_{16}S_{4}$, $F_{3}S_{16}S_{2}$, $F_{9}S_{16}S_{4}$, $F_{5}S_{16}S_{2}$, $F_{9}S_{5}S_{4}$, $F_{5}S_{2}S_{5}$, $F_{8}S_{7}S_{4}$, $F_{8}S_{7}S_{2}$, $F_{9}S_{16}S_{5}$, $F_{5}S_{16}S_{5}$, $F_{9}S_{5}S_{5}$, $F_{5}S_{5}S_{5}$, $F_{3}S_{7}S_{4}$, $F_{3}S_{7}S_{2}$}} 
 \SplitRow{ $\mathcal{O}_{9, \; (1,-1)}^{34}$} {{$F_{5}S_{2}S_{2}$, $F_{7}V_{1}V_{1}$, $F_{8}V_{2}V_{2}$, $F_{5}S_{2}V_{3}$, $F_{8}S_{5}V_{1}$, $F_{3}S_{16}V_{2}$, $F_{8}S_{16}V_{2}$, $F_{7}S_{5}V_{1}$, $F_{6}S_{2}V_{4}$, $F_{8}S_{5}V_{2}$, $F_{3}S_{16}V_{1}$, $F_{7}S_{16}V_{1}$, $F_{5}S_{2}V_{4}$, $F_{6}V_{4}V_{1}$, $F_{5}V_{3}V_{2}$, $F_{3}S_{16}S_{2}$, $F_{5}S_{16}S_{2}$, $F_{8}V_{3}V_{2}$, $F_{7}V_{4}V_{1}$, $F_{6}V_{4}V_{2}$, $F_{5}V_{3}V_{1}$, $F_{8}S_{2}S_{5}$, $F_{5}S_{2}S_{5}$, $F_{7}V_{3}V_{1}$, $F_{8}V_{4}V_{2}$, $F_{6}S_{7}S_{2}$, $F_{7}S_{16}S_{5}$, $F_{8}S_{16}S_{5}$, $F_{5}S_{7}S_{2}$, $F_{6}S_{7}V_{1}$, $F_{5}S_{16}V_{3}$, $F_{8}S_{16}V_{3}$, $F_{8}S_{7}V_{1}$, $F_{6}S_{7}V_{2}$, $F_{5}S_{5}V_{3}$, $F_{7}S_{5}V_{3}$, $F_{3}S_{7}V_{2}$, $F_{5}S_{7}V_{2}$, $F_{5}S_{16}V_{4}$, $F_{7}S_{16}V_{4}$, $F_{8}S_{7}V_{2}$, $F_{5}S_{7}V_{1}$, $F_{5}S_{5}V_{4}$, $F_{8}S_{5}V_{4}$, $F_{3}S_{7}V_{1}$, $F_{8}S_{7}S_{2}$, $F_{7}V_{3}V_{4}$, $F_{8}V_{3}V_{4}$, $F_{3}S_{7}S_{2}$}} 
 \SplitRow{ $\mathcal{O}_{9, \; (1,-1)}^{35}$} {{$F_{7}V_{1}V_{1}$, $F_{5}S_{2}S_{2}$, $F_{8}S_{5}V_{1}$, $F_{2}S_{2}V_{10}$, $F_{5}S_{2}V_{10}$, $F_{7}S_{5}V_{1}$, $F_{6}V_{4}V_{1}$, $F_{8}S_{2}S_{5}$, $F_{2}V_{10}V_{1}$, $F_{7}V_{10}V_{1}$, $F_{5}S_{2}S_{5}$, $F_{7}V_{4}V_{1}$, $F_{6}S_{2}V_{4}$, $F_{5}S_{2}V_{4}$, $F_{6}S_{7}V_{1}$, $F_{7}S_{5}V_{10}$, $F_{5}S_{5}V_{10}$, $F_{8}S_{7}V_{1}$, $F_{6}S_{7}S_{2}$, $F_{7}S_{5}S_{5}$, $F_{2}S_{7}S_{2}$, $F_{8}S_{7}S_{2}$, $F_{7}V_{10}V_{4}$, $F_{7}S_{5}V_{4}$, $F_{5}S_{5}V_{4}$, $F_{2}S_{7}V_{1}$}} 
 \SplitRow{ $\mathcal{O}_{9, \; (1,-1)}^{36}$} {{$F_{17}S_{1}S_{1}$, $F_{5}S_{2}S_{2}$, $F_{6}S_{1}V_{4}$, $F_{5}S_{2}V_{3}$, $F_{2}S_{2}V_{10}$, $F_{5}S_{2}V_{10}$, $F_{17}S_{1}V_{4}$, $F_{6}S_{2}V_{4}$, $F_{2}S_{1}V_{10}$, $F_{17}S_{1}V_{10}$, $F_{5}S_{2}V_{4}$, $F_{5}S_{1}V_{3}$, $F_{17}S_{1}V_{3}$, $F_{6}S_{7}S_{1}$, $F_{5}V_{10}V_{3}$, $F_{6}S_{7}S_{2}$, $F_{17}V_{10}V_{4}$, $F_{5}V_{10}V_{4}$, $F_{5}S_{7}S_{2}$, $F_{17}V_{3}V_{4}$, $F_{5}V_{3}V_{4}$, $F_{2}S_{7}S_{2}$, $F_{5}S_{7}S_{1}$, $F_{5}V_{4}V_{4}$, $F_{2}S_{7}S_{1}$}} 
 \SplitRow{ $\mathcal{O}_{9, \; (1,-1)}^{37}$} {{$F_{8}V_{2}V_{2}$, $F_{5}S_{2}S_{2}$, $F_{8}S_{5}V_{2}$, $F_{1}S_{2}V_{9}$, $F_{5}S_{2}V_{9}$, $F_{5}V_{3}V_{2}$, $F_{8}S_{2}S_{5}$, $F_{1}V_{9}V_{2}$, $F_{8}V_{9}V_{2}$, $F_{5}S_{2}S_{5}$, $F_{8}V_{3}V_{2}$, $F_{5}S_{2}V_{3}$, $F_{5}S_{7}V_{2}$, $F_{8}S_{5}V_{9}$, $F_{5}S_{5}V_{9}$, $F_{8}S_{7}V_{2}$, $F_{5}S_{7}S_{2}$, $F_{8}S_{5}S_{5}$, $F_{1}S_{7}S_{2}$, $F_{8}S_{7}S_{2}$, $F_{8}V_{9}V_{3}$, $F_{8}S_{5}V_{3}$, $F_{5}S_{5}V_{3}$, $F_{1}S_{7}V_{2}$}} 
 \SplitRow{ $\mathcal{O}_{9, \; (1,-1)}^{38}$} {{$F_{5}S_{2}S_{2}$, $F_{6}S_{3}S_{3}$, $F_{5}S_{2}V_{3}$, $F_{1}S_{3}V_{9}$, $F_{6}S_{3}V_{9}$, $F_{6}S_{2}V_{4}$, $F_{5}S_{3}V_{3}$, $F_{1}S_{2}V_{9}$, $F_{5}S_{2}V_{9}$, $F_{6}S_{3}V_{3}$, $F_{5}S_{2}V_{4}$, $F_{6}S_{3}V_{4}$, $F_{6}S_{7}S_{2}$, $F_{5}V_{9}V_{3}$, $F_{6}V_{9}V_{3}$, $F_{5}S_{7}S_{2}$, $F_{6}S_{7}S_{3}$, $F_{5}V_{3}V_{3}$, $F_{1}S_{7}S_{3}$, $F_{5}S_{7}S_{3}$, $F_{5}V_{9}V_{4}$, $F_{5}V_{3}V_{4}$, $F_{6}V_{3}V_{4}$, $F_{1}S_{7}S_{2}$}} 
 \end{SplitTable}
   \vspace{0.4cm} 
 \captionof{table}{Dimension 9 $(\Delta B, \Delta L)$ = (1, -1) UV completions for topology 2 (see Fig.~\ref{fig:d=9_topologies}).} 
 \label{tab:T9Top2B1L-1}
 \end{center}

%% file: Tables_dim9/Table_dim9_Top3_B1_L-1.tex
 \begin{center}
 \begin{SplitTable}[2.0cm] 
  \SplitHeader{$\mathcal{O}^j{(\mathcal{T}_ {9} ^{ 3 })}$}{New Particles} 
   \vspace{0.1cm} 
 \SplitRow{ $\mathcal{O}_{9, \; (1,-1)}^{39}$} {{$S_{9}V_{4}V_{2}$, $S_{9}S_{3}S_{6}$, $S_{8}V_{4}V_{2}$, $S_{7}V_{4}V_{2}$, $S_{8}S_{3}S_{6}$, $S_{7}S_{3}S_{6}$}} 
 \SplitRow{ $\mathcal{O}_{9, \; (1,-1)}^{40}$} {{$S_{9}S_{5}S_{4}$, $S_{9}S_{2}S_{5}$, $S_{9}V_{2}V_{5}$, $S_{9}V_{4}V_{2}$, $S_{8}S_{5}S_{4}$, $S_{8}S_{2}S_{5}$, $S_{7}S_{5}S_{4}$, $S_{7}S_{2}S_{5}$, $S_{8}V_{2}V_{5}$, $S_{8}V_{4}V_{2}$, $S_{7}V_{2}V_{5}$, $S_{7}V_{4}V_{2}$}} 
 \SplitRow{ $\mathcal{O}_{9, \; (1,-1)}^{41}$} {{$S_{9}S_{3}S_{6}$, $S_{9}S_{2}S_{5}$, $S_{8}S_{3}S_{6}$, $S_{7}S_{3}S_{6}$, $S_{8}S_{2}S_{5}$, $S_{7}S_{2}S_{5}$}} 
 \SplitRow{ $\mathcal{O}_{9, \; (1,-1)}^{42}$} {{$S_{8}S_{3}S_{5}$, $S_{7}S_{3}S_{5}$, $S_{9}S_{3}S_{5}$}} 
 \SplitRow{ $\mathcal{O}_{9, \; (1,-1)}^{43}$} {{$S_{9}S_{6}S_{4}$}} 
 \SplitRow{ $\mathcal{O}_{9, \; (1,-1)}^{44}$} {{$S_{9}S_{6}S_{4}$, $S_{9}V_{1}V_{5}$}} 
 \SplitRow{ $\mathcal{O}_{9, \; (1,-1)}^{45}$} {{$S_{8}V_{3}V_{2}$, $S_{7}V_{3}V_{2}$, $S_{8}S_{3}S_{5}$, $S_{7}S_{3}S_{5}$, $S_{9}V_{3}V_{2}$, $S_{9}S_{3}S_{5}$}} 
 \SplitRow{ $\mathcal{O}_{9, \; (1,-1)}^{46}$} {{$S_{9}S_{5}S_{4}$, $S_{9}S_{2}S_{5}$, $S_{8}S_{5}S_{4}$, $S_{8}S_{2}S_{5}$, $S_{7}S_{5}S_{4}$, $S_{7}S_{2}S_{5}$}} 
 \SplitRow{ $\mathcal{O}_{9, \; (1,-1)}^{47}$} {{$S_{9}V_{4}V_{2}$, $S_{9}V_{3}V_{1}$, $S_{9}S_{2}S_{5}$, $S_{8}V_{4}V_{2}$, $S_{7}V_{4}V_{2}$, $S_{8}V_{3}V_{1}$, $S_{7}V_{3}V_{1}$, $S_{8}S_{2}S_{5}$, $S_{7}S_{2}S_{5}$}} 
 \end{SplitTable}
   \vspace{0.4cm} 
 \captionof{table}{Dimension 9 $(\Delta B, \Delta L)$ = (1, -1) UV completions for topology 3 (see Fig.~\ref{fig:d=9_topologies}).} 
 \label{tab:T9Top3B1L-1}
 \end{center}

%% file: Tables_dim9/Table_dim9_Top4_B1_L-1.tex
 \begin{center}
 \begin{SplitTable}[2.0cm] 
  \SplitHeader{$\mathcal{O}^j{(\mathcal{T}_ {9} ^{ 4 })}$}{New Particles} 
   \vspace{0.1cm} 
 \SplitRow{ $\mathcal{O}_{9, \; (1,-1)}^{39}$} {{$S_{8}V_{4}V_{2}V_{5}$, $S_{7}V_{4}V_{4}V_{2}$, $S_{8}S_{3}S_{6}S_{6}$, $S_{7}S_{3}S_{6}S_{6}$, $S_{8}S_{3}S_{6}S_{19}$, $S_{7}S_{3}S_{3}S_{6}$, $S_{8}V_{4}V_{2}V_{2}$, $S_{7}V_{4}V_{2}V_{2}$, $S_{9}V_{4}V_{2}V_{8}$, $S_{9}S_{3}S_{5}S_{6}$, $S_{9}S_{3}S_{6}S_{20}$, $S_{9}V_{4}V_{2}V_{15}$}} 
 \SplitRow{ $\mathcal{O}_{9, \; (1,-1)}^{40}$} {{$S_{8}S_{5}S_{4}S_{13}$, $S_{8}S_{5}S_{5}S_{4}$, $S_{7}S_{5}S_{5}S_{4}$, $S_{8}S_{2}S_{5}S_{5}$, $S_{7}S_{2}S_{5}S_{5}$, $S_{8}V_{2}V_{5}V_{5}$, $S_{8}V_{4}V_{2}V_{5}$, $S_{7}V_{2}V_{5}V_{5}$, $S_{7}V_{4}V_{4}V_{2}$, $S_{8}V_{2}V_{5}V_{16}$, $S_{8}V_{2}V_{2}V_{5}$, $S_{7}V_{2}V_{2}V_{5}$, $S_{8}V_{4}V_{2}V_{2}$, $S_{7}V_{4}V_{2}V_{2}$, $S_{8}S_{5}S_{4}S_{4}$, $S_{8}S_{2}S_{5}S_{4}$, $S_{7}S_{5}S_{4}S_{4}$, $S_{7}S_{2}S_{2}S_{5}$, $S_{9}S_{5}S_{4}S_{12}$, $S_{9}S_{11}S_{5}S_{4}$, $S_{9}S_{2}S_{11}S_{5}$, $S_{9}V_{2}V_{8}V_{5}$, $S_{9}V_{4}V_{2}V_{8}$, $S_{9}V_{3}V_{2}V_{5}$, $S_{9}V_{2}V_{5}V_{17}$, $S_{9}V_{2}V_{15}V_{5}$, $S_{9}V_{4}V_{2}V_{15}$, $S_{9}S_{5}S_{4}S_{19}$, $S_{9}S_{2}S_{5}S_{19}$, $S_{9}S_{3}S_{5}S_{4}$}} 
 \SplitRow{ $\mathcal{O}_{9, \; (1,-1)}^{41}$} {{$S_{8}S_{3}S_{6}S_{6}$, $S_{7}S_{3}S_{6}S_{6}$, $S_{8}S_{2}S_{5}S_{5}$, $S_{7}S_{2}S_{5}S_{5}$, $S_{8}S_{2}S_{5}S_{4}$, $S_{7}S_{2}S_{2}S_{5}$, $S_{8}S_{3}S_{6}S_{19}$, $S_{7}S_{3}S_{3}S_{6}$, $S_{9}S_{3}S_{5}S_{6}$, $S_{9}S_{2}S_{11}S_{5}$, $S_{9}S_{2}S_{5}S_{19}$, $S_{9}S_{3}S_{6}S_{20}$}} 
 \SplitRow{ $\mathcal{O}_{9, \; (1,-1)}^{42}$} {{$S_{9}S_{3}S_{5}S_{6}$, $S_{9}S_{3}S_{5}S_{4}$, $S_{8}S_{3}S_{5}S_{5}$, $S_{7}S_{3}S_{5}S_{5}$, $S_{8}S_{3}S_{5}S_{19}$, $S_{7}S_{3}S_{3}S_{5}$}} 
 \SplitRow{ $\mathcal{O}_{9, \; (1,-1)}^{43}$} {{$S_{9}S_{6}S_{4}S_{19}$, $S_{9}S_{3}S_{6}S_{4}$, $S_{9}S_{6}S_{4}S_{13}$, $S_{9}S_{5}S_{6}S_{4}$}} 
 \SplitRow{ $\mathcal{O}_{9, \; (1,-1)}^{44}$} {{$S_{9}S_{6}S_{4}S_{19}$, $S_{9}S_{3}S_{6}S_{4}$, $S_{9}V_{1}V_{5}V_{16}$, $S_{9}V_{1}V_{2}V_{5}$, $S_{9}V_{1}V_{8}V_{5}$, $S_{9}V_{3}V_{1}V_{5}$, $S_{9}S_{6}S_{4}S_{13}$, $S_{9}S_{5}S_{6}S_{4}$}} 
 \SplitRow{ $\mathcal{O}_{9, \; (1,-1)}^{45}$} {{$S_{9}V_{3}V_{2}V_{5}$, $S_{9}S_{3}S_{5}S_{6}$, $S_{9}S_{3}S_{5}S_{4}$, $S_{9}V_{3}V_{1}V_{2}$, $S_{8}V_{3}V_{2}V_{8}$, $S_{7}V_{3}V_{3}V_{2}$, $S_{8}S_{3}S_{5}S_{5}$, $S_{7}S_{3}S_{5}S_{5}$, $S_{8}S_{3}S_{5}S_{19}$, $S_{7}S_{3}S_{3}S_{5}$, $S_{8}V_{3}V_{2}V_{2}$, $S_{7}V_{3}V_{2}V_{2}$}} 
 \SplitRow{ $\mathcal{O}_{9, \; (1,-1)}^{46}$} {{$S_{8}S_{5}S_{4}S_{13}$, $S_{8}S_{5}S_{5}S_{4}$, $S_{7}S_{5}S_{5}S_{4}$, $S_{8}S_{2}S_{5}S_{5}$, $S_{7}S_{2}S_{5}S_{5}$, $S_{8}S_{5}S_{4}S_{4}$, $S_{8}S_{2}S_{5}S_{4}$, $S_{7}S_{5}S_{4}S_{4}$, $S_{7}S_{2}S_{2}S_{5}$, $S_{9}S_{5}S_{4}S_{12}$, $S_{9}S_{11}S_{5}S_{4}$, $S_{9}S_{2}S_{11}S_{5}$, $S_{9}S_{5}S_{4}S_{19}$, $S_{9}S_{2}S_{5}S_{19}$, $S_{9}S_{3}S_{5}S_{4}$}} 
 \SplitRow{ $\mathcal{O}_{9, \; (1,-1)}^{47}$} {{$S_{8}V_{4}V_{2}V_{5}$, $S_{7}V_{4}V_{4}V_{2}$, $S_{8}V_{3}V_{1}V_{8}$, $S_{7}V_{3}V_{3}V_{1}$, $S_{8}S_{2}S_{5}S_{5}$, $S_{7}S_{2}S_{5}S_{5}$, $S_{8}S_{2}S_{5}S_{4}$, $S_{7}S_{2}S_{2}S_{5}$, $S_{8}V_{3}V_{1}V_{1}$, $S_{7}V_{3}V_{1}V_{1}$, $S_{8}V_{4}V_{2}V_{2}$, $S_{7}V_{4}V_{2}V_{2}$, $S_{9}V_{4}V_{2}V_{8}$, $S_{9}V_{3}V_{1}V_{7}$, $S_{9}S_{2}S_{11}S_{5}$, $S_{9}S_{2}S_{5}S_{19}$, $S_{9}V_{3}V_{1}V_{2}$, $S_{9}V_{4}V_{2}V_{15}$}} 
 \end{SplitTable}
   \vspace{0.4cm} 
 \captionof{table}{Dimension 9 $(\Delta B, \Delta L)$ = (1, -1) UV completions for topology 4 (see Fig.~\ref{fig:d=9_topologies}).} 
 \label{tab:T9Top4B1L-1}
 \end{center}

%% file: Tables_dim9/Table_dim9_Top5_B1_L-1.tex
 \begin{center}
 \begin{SplitTable}[2.0cm] 
  \SplitHeader{$\mathcal{O}^j{(\mathcal{T}_ {9} ^{ 5 })}$}{New Particles} 
   \vspace{0.1cm} 
 \SplitRow{ $\mathcal{O}_{9, \; (1,-1)}^{39}$} {{$V_{4}V_{2}V_{2}V_{8}$, $S_{3}S_{5}S_{6}S_{19}$, $S_{3}S_{3}S_{5}S_{6}$, $S_{3}S_{6}S_{6}S_{20}$, $V_{4}V_{2}V_{15}V_{5}$, $V_{4}V_{4}V_{2}V_{15}$, $V_{4}V_{2}V_{2}V_{5}$, $V_{4}V_{4}V_{2}V_{2}$, $S_{3}S_{6}S_{6}S_{19}$, $S_{3}S_{3}S_{6}S_{6}$}} 
 \SplitRow{ $\mathcal{O}_{9, \; (1,-1)}^{40}$} {{$S_{5}S_{4}S_{4}S_{12}$, $S_{11}S_{5}S_{4}S_{4}$, $S_{2}S_{11}S_{5}S_{4}$, $S_{2}S_{2}S_{11}S_{5}$, $V_{2}V_{8}V_{5}V_{16}$, $V_{2}V_{2}V_{8}V_{5}$, $V_{3}V_{2}V_{2}V_{5}$, $V_{4}V_{2}V_{2}V_{8}$, $V_{2}V_{5}V_{5}V_{17}$, $V_{2}V_{15}V_{5}V_{5}$, $V_{4}V_{2}V_{15}V_{5}$, $V_{4}V_{4}V_{2}V_{15}$, $S_{5}S_{4}S_{19}S_{13}$, $S_{5}S_{5}S_{4}S_{19}$, $S_{3}S_{5}S_{5}S_{4}$, $S_{2}S_{5}S_{5}S_{19}$, $S_{5}S_{4}S_{4}S_{13}$, $S_{5}S_{5}S_{4}S_{4}$, $S_{2}S_{5}S_{5}S_{4}$, $S_{2}S_{2}S_{5}S_{5}$, $V_{2}V_{5}V_{5}V_{16}$, $V_{2}V_{2}V_{5}V_{5}$, $V_{4}V_{2}V_{2}V_{5}$, $V_{4}V_{4}V_{2}V_{2}$}} 
 \SplitRow{ $\mathcal{O}_{9, \; (1,-1)}^{41}$} {{$S_{3}S_{5}S_{6}S_{19}$, $S_{3}S_{3}S_{5}S_{6}$, $S_{2}S_{11}S_{5}S_{4}$, $S_{2}S_{2}S_{11}S_{5}$, $S_{2}S_{5}S_{5}S_{19}$, $S_{3}S_{6}S_{6}S_{20}$, $S_{3}S_{6}S_{6}S_{19}$, $S_{3}S_{3}S_{6}S_{6}$, $S_{2}S_{5}S_{5}S_{4}$, $S_{2}S_{2}S_{5}S_{5}$}} 
 \SplitRow{ $\mathcal{O}_{9, \; (1,-1)}^{42}$} {{$S_{3}S_{5}S_{5}S_{19}$, $S_{3}S_{3}S_{5}S_{5}$, $S_{3}S_{5}S_{5}S_{4}$, $S_{3}S_{5}S_{6}S_{19}$, $S_{3}S_{3}S_{5}S_{6}$}} 
 \SplitRow{ $\mathcal{O}_{9, \; (1,-1)}^{43}$} {{$S_{6}S_{4}S_{19}S_{13}$, $S_{5}S_{6}S_{4}S_{19}$, $S_{3}S_{5}S_{6}S_{4}$}} 
 \SplitRow{ $\mathcal{O}_{9, \; (1,-1)}^{44}$} {{$S_{6}S_{4}S_{19}S_{13}$, $S_{5}S_{6}S_{4}S_{19}$, $S_{3}S_{5}S_{6}S_{4}$, $V_{1}V_{8}V_{5}V_{16}$, $V_{1}V_{2}V_{8}V_{5}$, $V_{3}V_{1}V_{2}V_{5}$}} 
 \SplitRow{ $\mathcal{O}_{9, \; (1,-1)}^{45}$} {{$V_{3}V_{2}V_{2}V_{8}$, $V_{3}V_{3}V_{2}V_{2}$, $S_{3}S_{5}S_{5}S_{19}$, $S_{3}S_{3}S_{5}S_{5}$, $V_{3}V_{1}V_{2}V_{8}$, $V_{3}V_{3}V_{1}V_{2}$, $S_{3}S_{5}S_{5}S_{4}$, $S_{3}S_{5}S_{6}S_{19}$, $S_{3}S_{3}S_{5}S_{6}$, $V_{3}V_{2}V_{2}V_{5}$}} 
 \SplitRow{ $\mathcal{O}_{9, \; (1,-1)}^{46}$} {{$S_{5}S_{4}S_{4}S_{12}$, $S_{11}S_{5}S_{4}S_{4}$, $S_{2}S_{11}S_{5}S_{4}$, $S_{2}S_{2}S_{11}S_{5}$, $S_{5}S_{4}S_{19}S_{13}$, $S_{5}S_{5}S_{4}S_{19}$, $S_{3}S_{5}S_{5}S_{4}$, $S_{2}S_{5}S_{5}S_{19}$, $S_{5}S_{4}S_{4}S_{13}$, $S_{5}S_{5}S_{4}S_{4}$, $S_{2}S_{5}S_{5}S_{4}$, $S_{2}S_{2}S_{5}S_{5}$}} 
 \SplitRow{ $\mathcal{O}_{9, \; (1,-1)}^{47}$} {{$V_{4}V_{2}V_{2}V_{8}$, $V_{3}V_{1}V_{1}V_{7}$, $S_{2}S_{11}S_{5}S_{4}$, $S_{2}S_{2}S_{11}S_{5}$, $S_{2}S_{5}S_{5}S_{19}$, $V_{3}V_{1}V_{2}V_{8}$, $V_{3}V_{3}V_{1}V_{2}$, $V_{4}V_{2}V_{15}V_{5}$, $V_{4}V_{4}V_{2}V_{15}$, $V_{4}V_{2}V_{2}V_{5}$, $V_{4}V_{4}V_{2}V_{2}$, $V_{3}V_{1}V_{1}V_{8}$, $V_{3}V_{3}V_{1}V_{1}$, $S_{2}S_{5}S_{5}S_{4}$, $S_{2}S_{2}S_{5}S_{5}$}} 
 \end{SplitTable}
   \vspace{0.4cm} 
 \captionof{table}{Dimension 9 $(\Delta B, \Delta L)$ = (1, -1) UV completions for topology 5 (see Fig.~\ref{fig:d=9_topologies}).} 
 \label{tab:T9Top5B1L-1}
 \end{center}

%% file: Tables_dim9/Table_dim9_Top6_B1_L-1.tex
 \begin{center}
 \begin{SplitTable}[2.0cm] 
  \SplitHeader{$\mathcal{O}^j{(\mathcal{T}_ {9} ^{ 6 })}$}{New Particles} 
   \vspace{0.1cm} 
 \SplitRow{ $\mathcal{O}_{9, \; (1,-1)}^{39}$} {{$F_{3}F_{8}V_{2}V_{5}$, $F_{3}F_{8}V_{4}V_{2}$, $F_{3}F_{14}S_{3}S_{6}$, $F_{10}F_{7}V_{2}V_{5}$, $F_{10}F_{7}V_{4}V_{2}$, $F_{7}F_{8}S_{6}S_{19}$, $F_{7}F_{8}S_{3}S_{6}$, $F_{7}F_{14}V_{4}V_{2}$, $F_{10}F_{9}S_{3}S_{6}$, $F_{10}F_{5}S_{3}S_{6}$, $F_{8}F_{9}V_{4}V_{2}$, $F_{5}F_{8}V_{4}V_{2}$, $F_{3}F_{7}V_{2}V_{8}$, $F_{3}F_{9}S_{3}S_{5}$, $F_{3}F_{5}S_{3}S_{5}$, $F_{7}F_{7}S_{6}S_{20}$, $F_{7}F_{9}V_{4}V_{15}$, $F_{5}F_{7}V_{4}V_{15}$}} 
 \SplitRow{ $\mathcal{O}_{9, \; (1,-1)}^{40}$} {{$F_{4}F_{8}S_{4}S_{13}$, $F_{4}F_{8}S_{5}S_{4}$, $F_{1}F_{8}S_{5}S_{4}$, $F_{4}F_{8}S_{2}S_{5}$, $F_{1}F_{8}S_{2}S_{5}$, $F_{4}F_{14}V_{2}V_{5}$, $F_{4}F_{6}V_{2}V_{5}$, $F_{1}F_{14}V_{2}V_{5}$, $F_{4}F_{14}V_{4}V_{2}$, $F_{1}F_{6}V_{4}V_{2}$, $F_{11}F_{7}S_{4}S_{13}$, $F_{11}F_{7}S_{5}S_{4}$, $F_{2}F_{7}S_{5}S_{4}$, $F_{11}F_{7}S_{2}S_{5}$, $F_{7}F_{14}V_{5}V_{16}$, $F_{7}F_{14}V_{2}V_{5}$, $F_{6}F_{7}V_{2}V_{5}$, $F_{7}F_{14}V_{4}V_{2}$, $F_{11}F_{9}V_{2}V_{5}$, $F_{2}F_{9}V_{2}V_{5}$, $F_{11}F_{5}V_{2}V_{5}$, $F_{11}F_{9}V_{4}V_{2}$, $F_{2}F_{5}V_{4}V_{2}$, $F_{8}F_{9}V_{5}V_{16}$, $F_{8}F_{9}V_{2}V_{5}$, $F_{5}F_{8}V_{2}V_{5}$, $F_{8}F_{9}V_{4}V_{2}$, $F_{5}F_{8}V_{4}V_{2}$, $F_{9}F_{14}S_{5}S_{4}$, $F_{6}F_{9}S_{5}S_{4}$, $F_{5}F_{14}S_{5}S_{4}$, $F_{9}F_{14}S_{2}S_{5}$, $F_{5}F_{6}S_{2}S_{5}$, $F_{4}F_{7}S_{4}S_{12}$, $F_{4}F_{7}S_{11}S_{4}$, $F_{1}F_{7}S_{11}S_{4}$, $F_{4}F_{7}S_{2}S_{11}$, $F_{1}F_{7}S_{2}S_{11}$, $F_{4}F_{9}V_{2}V_{8}$, $F_{4}F_{5}V_{2}V_{8}$, $F_{1}F_{9}V_{2}V_{8}$, $F_{4}F_{9}V_{3}V_{2}$, $F_{1}F_{5}V_{3}V_{2}$, $F_{7}F_{9}V_{5}V_{17}$, $F_{7}F_{9}V_{15}V_{5}$, $F_{5}F_{7}V_{15}V_{5}$, $F_{7}F_{9}V_{4}V_{15}$, $F_{5}F_{7}V_{4}V_{15}$, $F_{9}F_{9}S_{5}S_{19}$, $F_{5}F_{9}S_{5}S_{19}$, $F_{9}F_{9}S_{3}S_{5}$, $F_{5}F_{5}S_{3}S_{5}$}} 
 \SplitRow{ $\mathcal{O}_{9, \; (1,-1)}^{41}$} {{$F_{4}F_{13}S_{3}S_{6}$, $F_{1}F_{13}S_{3}S_{6}$, $F_{4}F_{8}S_{2}S_{5}$, $F_{1}F_{8}S_{2}S_{5}$, $F_{11}F_{8}S_{3}S_{6}$, $F_{8}F_{8}S_{5}S_{4}$, $F_{8}F_{8}S_{2}S_{5}$, $F_{11}F_{7}S_{2}S_{5}$, $F_{7}F_{13}S_{5}S_{4}$, $F_{7}F_{13}S_{2}S_{5}$, $F_{7}F_{8}S_{6}S_{19}$, $F_{7}F_{8}S_{3}S_{6}$, $F_{4}F_{8}S_{3}S_{5}$, $F_{1}F_{8}S_{3}S_{5}$, $F_{4}F_{7}S_{2}S_{11}$, $F_{1}F_{7}S_{2}S_{11}$, $F_{7}F_{8}S_{5}S_{19}$, $F_{7}F_{7}S_{6}S_{20}$}} 
 \SplitRow{ $\mathcal{O}_{9, \; (1,-1)}^{42}$} {{$F_{11}F_{8}S_{3}S_{6}$, $F_{2}F_{8}S_{3}S_{6}$, $F_{8}F_{8}S_{5}S_{4}$, $F_{4}F_{8}S_{3}S_{5}$, $F_{11}F_{7}S_{3}S_{5}$, $F_{2}F_{7}S_{3}S_{5}$, $F_{7}F_{8}S_{5}S_{19}$, $F_{7}F_{8}S_{3}S_{5}$}} 
 \SplitRow{ $\mathcal{O}_{9, \; (1,-1)}^{43}$} {{$F_{5}F_{9}S_{6}S_{19}$, $F_{9}F_{9}S_{3}S_{6}$, $F_{5}F_{5}S_{3}S_{6}$, $F_{3}F_{9}S_{4}S_{13}$, $F_{3}F_{9}S_{5}S_{4}$, $F_{3}F_{5}S_{5}S_{4}$}} 
 \SplitRow{ $\mathcal{O}_{9, \; (1,-1)}^{44}$} {{$F_{5}F_{9}S_{6}S_{19}$, $F_{9}F_{9}S_{3}S_{6}$, $F_{5}F_{5}S_{3}S_{6}$, $F_{8}F_{9}V_{5}V_{16}$, $F_{8}F_{9}V_{2}V_{5}$, $F_{5}F_{8}V_{2}V_{5}$, $F_{1}F_{9}V_{1}V_{8}$, $F_{4}F_{5}V_{1}V_{8}$, $F_{4}F_{9}V_{3}V_{1}$, $F_{1}F_{5}V_{3}V_{1}$, $F_{4}F_{8}S_{4}S_{13}$, $F_{4}F_{8}S_{5}S_{4}$, $F_{1}F_{8}S_{5}S_{4}$}} 
 \SplitRow{ $\mathcal{O}_{9, \; (1,-1)}^{45}$} {{$F_{3}F_{8}V_{2}V_{5}$, $F_{3}F_{14}S_{3}S_{6}$, $F_{3}F_{6}S_{3}S_{6}$, $F_{8}F_{8}S_{5}S_{4}$, $F_{8}F_{14}V_{3}V_{1}$, $F_{6}F_{8}V_{3}V_{1}$, $F_{3}F_{8}V_{2}V_{8}$, $F_{3}F_{8}V_{3}V_{2}$, $F_{3}F_{14}S_{3}S_{5}$, $F_{3}F_{6}S_{3}S_{5}$, $F_{3}F_{7}V_{2}V_{8}$, $F_{3}F_{7}V_{3}V_{2}$, $F_{7}F_{8}S_{5}S_{19}$, $F_{7}F_{8}S_{3}S_{5}$, $F_{7}F_{14}V_{3}V_{2}$, $F_{6}F_{7}V_{3}V_{2}$, $F_{3}F_{9}S_{3}S_{5}$, $F_{8}F_{9}V_{3}V_{2}$}} 
 \SplitRow{ $\mathcal{O}_{9, \; (1,-1)}^{46}$} {{$F_{3}F_{14}S_{4}S_{13}$, $F_{3}F_{14}S_{5}S_{4}$, $F_{3}F_{6}S_{5}S_{4}$, $F_{3}F_{14}S_{2}S_{5}$, $F_{3}F_{9}S_{4}S_{13}$, $F_{3}F_{9}S_{5}S_{4}$, $F_{3}F_{5}S_{5}S_{4}$, $F_{3}F_{9}S_{2}S_{5}$, $F_{3}F_{5}S_{2}S_{5}$, $F_{9}F_{14}S_{5}S_{4}$, $F_{6}F_{9}S_{5}S_{4}$, $F_{5}F_{14}S_{5}S_{4}$, $F_{9}F_{14}S_{2}S_{5}$, $F_{5}F_{6}S_{2}S_{5}$, $F_{3}F_{9}S_{4}S_{12}$, $F_{3}F_{9}S_{11}S_{4}$, $F_{3}F_{5}S_{11}S_{4}$, $F_{3}F_{9}S_{2}S_{11}$, $F_{3}F_{5}S_{2}S_{11}$, $F_{9}F_{9}S_{5}S_{19}$, $F_{5}F_{9}S_{5}S_{19}$, $F_{9}F_{9}S_{3}S_{5}$, $F_{5}F_{5}S_{3}S_{5}$}} 
 \SplitRow{ $\mathcal{O}_{9, \; (1,-1)}^{47}$} {{$F_{3}F_{13}V_{2}V_{5}$, $F_{3}F_{13}V_{4}V_{2}$, $F_{3}F_{8}V_{1}V_{8}$, $F_{3}F_{8}V_{3}V_{1}$, $F_{3}F_{14}S_{2}S_{5}$, $F_{3}F_{8}V_{2}V_{5}$, $F_{3}F_{8}V_{4}V_{2}$, $F_{8}F_{8}S_{5}S_{4}$, $F_{8}F_{8}S_{2}S_{5}$, $F_{8}F_{14}V_{3}V_{1}$, $F_{3}F_{7}V_{1}V_{8}$, $F_{3}F_{7}V_{3}V_{1}$, $F_{7}F_{13}S_{5}S_{4}$, $F_{7}F_{13}S_{2}S_{5}$, $F_{7}F_{14}V_{4}V_{2}$, $F_{3}F_{9}S_{2}S_{5}$, $F_{3}F_{5}S_{2}S_{5}$, $F_{13}F_{9}V_{3}V_{1}$, $F_{5}F_{13}V_{3}V_{1}$, $F_{8}F_{9}V_{4}V_{2}$, $F_{5}F_{8}V_{4}V_{2}$, $F_{3}F_{8}V_{2}V_{8}$, $F_{3}F_{7}V_{1}V_{7}$, $F_{3}F_{9}S_{2}S_{11}$, $F_{3}F_{5}S_{2}S_{11}$, $F_{7}F_{8}S_{5}S_{19}$, $F_{8}F_{9}V_{3}V_{2}$, $F_{5}F_{8}V_{3}V_{2}$, $F_{7}F_{9}V_{4}V_{15}$, $F_{5}F_{7}V_{4}V_{15}$}} 
 \end{SplitTable}
   \vspace{0.4cm} 
 \captionof{table}{Dimension 9 $(\Delta B, \Delta L)$ = (1, -1) UV completions for topology 6 (see Fig.~\ref{fig:d=9_topologies}).} 
 \label{tab:T9Top6B1L-1}
 \end{center}

%% file: Tables_dim9/Table_dim9_Top7_B1_L-1.tex
 \begin{center}
 \begin{SplitTable}[2.0cm] 
  \SplitHeader{$\mathcal{O}^j{(\mathcal{T}_ {9} ^{ 7 })}$}{New Particles} 
   \vspace{0.1cm} 
 \SplitRow{ $\mathcal{O}_{9, \; (1,-1)}^{39}$} {{$F_{7}F_{8}F_{9}V_{2}$, $F_{5}F_{7}F_{8}V_{2}$, $F_{5}F_{8}F_{14}S_{3}$, $F_{3}F_{10}F_{11}V_{2}$, $F_{2}F_{3}F_{10}V_{2}$, $F_{7}F_{8}F_{9}S_{6}$, $F_{5}F_{7}F_{8}S_{6}$, $F_{5}F_{8}F_{14}V_{4}$, $F_{3}F_{10}F_{11}S_{3}$, $F_{2}F_{3}F_{10}S_{3}$, $F_{7}F_{8}F_{9}V_{4}$, $F_{5}F_{7}F_{8}V_{4}$, $F_{7}F_{7}F_{9}V_{2}$, $F_{5}F_{7}F_{7}V_{2}$, $F_{5}F_{8}F_{9}S_{3}$, $F_{5}F_{5}F_{8}S_{3}$, $F_{3}F_{3}F_{11}V_{2}$, $F_{2}F_{3}F_{3}V_{2}$, $F_{7}F_{7}F_{9}S_{6}$, $F_{5}F_{7}F_{7}S_{6}$, $F_{5}F_{8}F_{9}V_{4}$, $F_{5}F_{5}F_{8}V_{4}$, $F_{3}F_{3}F_{11}S_{3}$, $F_{2}F_{3}F_{3}S_{3}$, $F_{7}F_{7}F_{9}V_{4}$, $F_{5}F_{7}F_{7}V_{4}$, $F_{7}F_{7}F_{19}V_{2}$, $F_{5}F_{7}F_{9}S_{3}$, $F_{5}F_{5}F_{7}S_{3}$, $F_{3}F_{3}F_{4}V_{2}$, $F_{7}F_{7}F_{19}S_{6}$, $F_{5}F_{7}F_{9}V_{4}$, $F_{5}F_{5}F_{7}V_{4}$, $F_{3}F_{3}F_{4}S_{3}$, $F_{7}F_{7}F_{19}V_{4}$}} 
 \SplitRow{ $\mathcal{O}_{9, \; (1,-1)}^{40}$} {{$F_{8}F_{9}F_{20}S_{4}$, $F_{7}F_{8}F_{9}S_{4}$, $F_{7}F_{8}F_{9}S_{2}$, $F_{5}F_{7}F_{8}S_{4}$, $F_{5}F_{7}F_{8}S_{2}$, $F_{9}F_{14}F_{15}V_{2}$, $F_{8}F_{9}F_{14}V_{2}$, $F_{5}F_{8}F_{14}V_{2}$, $F_{6}F_{8}F_{9}V_{2}$, $F_{4}F_{11}F_{12}S_{4}$, $F_{3}F_{4}F_{11}S_{4}$, $F_{1}F_{3}F_{11}S_{2}$, $F_{2}F_{3}F_{4}S_{4}$, $F_{9}F_{14}F_{15}V_{5}$, $F_{8}F_{9}F_{14}V_{5}$, $F_{5}F_{8}F_{14}V_{4}$, $F_{6}F_{8}F_{9}V_{5}$, $F_{4}F_{11}F_{12}V_{2}$, $F_{3}F_{4}F_{11}V_{2}$, $F_{1}F_{3}F_{11}V_{2}$, $F_{2}F_{3}F_{4}V_{2}$, $F_{8}F_{9}F_{20}V_{5}$, $F_{7}F_{8}F_{9}V_{5}$, $F_{7}F_{8}F_{9}V_{4}$, $F_{5}F_{7}F_{8}V_{5}$, $F_{5}F_{7}F_{8}V_{4}$, $F_{9}F_{14}F_{15}S_{5}$, $F_{8}F_{9}F_{14}S_{5}$, $F_{5}F_{8}F_{14}S_{5}$, $F_{6}F_{8}F_{9}S_{5}$, $F_{7}F_{9}F_{20}S_{4}$, $F_{7}F_{7}F_{9}S_{4}$, $F_{7}F_{7}F_{9}S_{2}$, $F_{5}F_{7}F_{7}S_{4}$, $F_{5}F_{7}F_{7}S_{2}$, $F_{9}F_{9}F_{15}V_{2}$, $F_{8}F_{9}F_{9}V_{2}$, $F_{5}F_{8}F_{9}V_{2}$, $F_{5}F_{5}F_{8}V_{2}$, $F_{4}F_{4}F_{12}S_{4}$, $F_{3}F_{4}F_{4}S_{4}$, $F_{1}F_{3}F_{4}S_{2}$, $F_{1}F_{3}F_{4}S_{4}$, $F_{1}F_{1}F_{3}S_{2}$, $F_{9}F_{9}F_{15}V_{5}$, $F_{8}F_{9}F_{9}V_{5}$, $F_{5}F_{8}F_{9}V_{4}$, $F_{5}F_{8}F_{9}V_{5}$, $F_{5}F_{5}F_{8}V_{4}$, $F_{4}F_{4}F_{12}V_{2}$, $F_{3}F_{4}F_{4}V_{2}$, $F_{1}F_{3}F_{4}V_{2}$, $F_{1}F_{1}F_{3}V_{2}$, $F_{7}F_{9}F_{20}V_{5}$, $F_{7}F_{7}F_{9}V_{5}$, $F_{7}F_{7}F_{9}V_{4}$, $F_{5}F_{7}F_{7}V_{5}$, $F_{5}F_{7}F_{7}V_{4}$, $F_{9}F_{9}F_{15}S_{5}$, $F_{8}F_{9}F_{9}S_{5}$, $F_{5}F_{8}F_{9}S_{5}$, $F_{5}F_{5}F_{8}S_{5}$, $F_{7}F_{19}F_{20}S_{4}$, $F_{7}F_{7}F_{19}S_{4}$, $F_{7}F_{7}F_{19}S_{2}$, $F_{9}F_{9}F_{20}V_{2}$, $F_{7}F_{9}F_{9}V_{2}$, $F_{5}F_{7}F_{9}V_{2}$, $F_{5}F_{5}F_{7}V_{2}$, $F_{9}F_{9}F_{20}V_{5}$, $F_{7}F_{9}F_{9}V_{5}$, $F_{5}F_{7}F_{9}V_{4}$, $F_{5}F_{7}F_{9}V_{5}$, $F_{5}F_{5}F_{7}V_{4}$, $F_{7}F_{19}F_{20}V_{5}$, $F_{7}F_{7}F_{19}V_{5}$, $F_{7}F_{7}F_{19}V_{4}$, $F_{9}F_{9}F_{20}S_{5}$, $F_{7}F_{9}F_{9}S_{5}$, $F_{5}F_{7}F_{9}S_{5}$, $F_{5}F_{5}F_{7}S_{5}$}} 
 \SplitRow{ $\mathcal{O}_{9, \; (1,-1)}^{41}$} {{$F_{8}F_{13}F_{14}S_{3}$, $F_{6}F_{8}F_{13}S_{3}$, $F_{7}F_{8}F_{9}S_{2}$, $F_{5}F_{7}F_{8}S_{2}$, $F_{1}F_{3}F_{11}S_{3}$, $F_{7}F_{8}F_{9}S_{5}$, $F_{5}F_{7}F_{8}S_{5}$, $F_{1}F_{3}F_{11}S_{2}$, $F_{8}F_{13}F_{14}S_{5}$, $F_{6}F_{8}F_{13}S_{5}$, $F_{7}F_{8}F_{9}S_{6}$, $F_{5}F_{7}F_{8}S_{6}$, $F_{8}F_{8}F_{14}S_{3}$, $F_{6}F_{8}F_{8}S_{3}$, $F_{7}F_{7}F_{9}S_{2}$, $F_{5}F_{7}F_{7}S_{2}$, $F_{1}F_{3}F_{4}S_{3}$, $F_{1}F_{1}F_{3}S_{3}$, $F_{7}F_{7}F_{9}S_{5}$, $F_{5}F_{7}F_{7}S_{5}$, $F_{1}F_{3}F_{4}S_{2}$, $F_{1}F_{1}F_{3}S_{2}$, $F_{8}F_{8}F_{14}S_{5}$, $F_{6}F_{8}F_{8}S_{5}$, $F_{7}F_{7}F_{9}S_{6}$, $F_{5}F_{7}F_{7}S_{6}$, $F_{8}F_{8}F_{9}S_{3}$, $F_{7}F_{7}F_{19}S_{2}$, $F_{7}F_{7}F_{19}S_{5}$, $F_{8}F_{8}F_{9}S_{5}$, $F_{7}F_{7}F_{19}S_{6}$}} 
 \SplitRow{ $\mathcal{O}_{9, \; (1,-1)}^{42}$} {{$F_{8}F_{8}F_{14}S_{3}$, $F_{2}F_{10}F_{11}S_{3}$, $F_{2}F_{2}F_{10}S_{3}$, $F_{8}F_{8}F_{14}S_{5}$, $F_{8}F_{8}F_{9}S_{3}$, $F_{5}F_{8}F_{8}S_{3}$, $F_{2}F_{3}F_{11}S_{3}$, $F_{2}F_{2}F_{3}S_{3}$, $F_{8}F_{8}F_{9}S_{5}$, $F_{5}F_{8}F_{8}S_{5}$, $F_{7}F_{8}F_{9}S_{3}$, $F_{5}F_{7}F_{8}S_{3}$, $F_{2}F_{3}F_{4}S_{3}$, $F_{7}F_{8}F_{9}S_{5}$, $F_{5}F_{7}F_{8}S_{5}$}} 
 \SplitRow{ $\mathcal{O}_{9, \; (1,-1)}^{43}$} {{$F_{19}F_{9}F_{20}S_{6}$, $F_{7}F_{19}F_{9}S_{6}$, $F_{5}F_{7}F_{19}S_{6}$, $F_{3}F_{4}F_{12}S_{4}$, $F_{19}F_{9}F_{20}S_{4}$, $F_{7}F_{19}F_{9}S_{4}$, $F_{5}F_{7}F_{19}S_{4}$}} 
 \SplitRow{ $\mathcal{O}_{9, \; (1,-1)}^{44}$} {{$F_{19}F_{9}F_{20}S_{6}$, $F_{7}F_{19}F_{9}S_{6}$, $F_{5}F_{7}F_{19}S_{6}$, $F_{8}F_{9}F_{20}V_{5}$, $F_{4}F_{11}F_{12}V_{1}$, $F_{3}F_{4}F_{11}V_{1}$, $F_{1}F_{3}F_{11}V_{1}$, $F_{19}F_{9}F_{20}V_{5}$, $F_{7}F_{19}F_{9}V_{5}$, $F_{5}F_{7}F_{19}V_{5}$, $F_{4}F_{11}F_{12}S_{4}$, $F_{3}F_{4}F_{11}S_{4}$, $F_{1}F_{3}F_{11}S_{4}$, $F_{19}F_{9}F_{20}V_{1}$, $F_{7}F_{19}F_{9}V_{1}$, $F_{5}F_{7}F_{19}V_{1}$, $F_{8}F_{9}F_{20}S_{4}$}} 
 \SplitRow{ $\mathcal{O}_{9, \; (1,-1)}^{45}$} {{$F_{8}F_{8}F_{14}V_{2}$, $F_{6}F_{13}F_{14}S_{3}$, $F_{6}F_{6}F_{13}S_{3}$, $F_{3}F_{3}F_{11}V_{2}$, $F_{8}F_{8}F_{14}S_{5}$, $F_{6}F_{13}F_{14}V_{3}$, $F_{6}F_{6}F_{13}V_{3}$, $F_{3}F_{3}F_{11}S_{3}$, $F_{8}F_{8}F_{14}V_{3}$, $F_{8}F_{8}F_{9}V_{2}$, $F_{5}F_{8}F_{8}V_{2}$, $F_{6}F_{8}F_{14}S_{3}$, $F_{6}F_{6}F_{8}S_{3}$, $F_{3}F_{3}F_{4}V_{2}$, $F_{1}F_{3}F_{3}V_{2}$, $F_{8}F_{8}F_{9}S_{5}$, $F_{5}F_{8}F_{8}S_{5}$, $F_{6}F_{8}F_{14}V_{3}$, $F_{6}F_{6}F_{8}V_{3}$, $F_{3}F_{3}F_{4}S_{3}$, $F_{1}F_{3}F_{3}S_{3}$, $F_{8}F_{8}F_{9}V_{3}$, $F_{5}F_{8}F_{8}V_{3}$, $F_{7}F_{8}F_{9}V_{2}$, $F_{5}F_{7}F_{8}V_{2}$, $F_{6}F_{8}F_{9}S_{3}$, $F_{7}F_{8}F_{9}S_{5}$, $F_{5}F_{7}F_{8}S_{5}$, $F_{6}F_{8}F_{9}V_{3}$, $F_{7}F_{8}F_{9}V_{3}$, $F_{5}F_{7}F_{8}V_{3}$}} 
 \SplitRow{ $\mathcal{O}_{9, \; (1,-1)}^{46}$} {{$F_{9}F_{14}F_{15}S_{4}$, $F_{8}F_{9}F_{14}S_{4}$, $F_{5}F_{8}F_{14}S_{2}$, $F_{6}F_{8}F_{9}S_{4}$, $F_{3}F_{4}F_{12}S_{4}$, $F_{3}F_{3}F_{4}S_{4}$, $F_{3}F_{3}F_{4}S_{2}$, $F_{1}F_{3}F_{3}S_{4}$, $F_{1}F_{3}F_{3}S_{2}$, $F_{9}F_{14}F_{15}S_{5}$, $F_{8}F_{9}F_{14}S_{5}$, $F_{5}F_{8}F_{14}S_{5}$, $F_{6}F_{8}F_{9}S_{5}$, $F_{9}F_{9}F_{15}S_{4}$, $F_{8}F_{9}F_{9}S_{4}$, $F_{5}F_{8}F_{9}S_{2}$, $F_{5}F_{8}F_{9}S_{4}$, $F_{5}F_{5}F_{8}S_{2}$, $F_{9}F_{9}F_{15}S_{5}$, $F_{8}F_{9}F_{9}S_{5}$, $F_{5}F_{8}F_{9}S_{5}$, $F_{5}F_{5}F_{8}S_{5}$, $F_{9}F_{9}F_{20}S_{4}$, $F_{7}F_{9}F_{9}S_{4}$, $F_{5}F_{7}F_{9}S_{2}$, $F_{5}F_{7}F_{9}S_{4}$, $F_{5}F_{5}F_{7}S_{2}$, $F_{3}F_{11}F_{12}S_{4}$, $F_{3}F_{3}F_{11}S_{4}$, $F_{3}F_{3}F_{11}S_{2}$, $F_{9}F_{9}F_{20}S_{5}$, $F_{7}F_{9}F_{9}S_{5}$, $F_{5}F_{7}F_{9}S_{5}$, $F_{5}F_{5}F_{7}S_{5}$}} 
 \SplitRow{ $\mathcal{O}_{9, \; (1,-1)}^{47}$} {{$F_{8}F_{13}F_{14}V_{2}$, $F_{6}F_{8}F_{13}V_{2}$, $F_{7}F_{8}F_{9}V_{1}$, $F_{5}F_{7}F_{8}V_{1}$, $F_{5}F_{8}F_{14}S_{2}$, $F_{3}F_{3}F_{4}V_{2}$, $F_{1}F_{3}F_{3}V_{2}$, $F_{7}F_{8}F_{9}S_{5}$, $F_{5}F_{7}F_{8}S_{5}$, $F_{5}F_{8}F_{14}V_{3}$, $F_{3}F_{3}F_{4}V_{1}$, $F_{1}F_{3}F_{3}V_{1}$, $F_{8}F_{13}F_{14}S_{5}$, $F_{6}F_{8}F_{13}S_{5}$, $F_{5}F_{8}F_{14}V_{4}$, $F_{3}F_{3}F_{4}S_{2}$, $F_{1}F_{3}F_{3}S_{2}$, $F_{8}F_{13}F_{14}V_{3}$, $F_{6}F_{8}F_{13}V_{3}$, $F_{7}F_{8}F_{9}V_{4}$, $F_{5}F_{7}F_{8}V_{4}$, $F_{8}F_{8}F_{14}V_{2}$, $F_{6}F_{8}F_{8}V_{2}$, $F_{7}F_{7}F_{9}V_{1}$, $F_{5}F_{7}F_{7}V_{1}$, $F_{5}F_{8}F_{9}S_{2}$, $F_{5}F_{5}F_{8}S_{2}$, $F_{7}F_{7}F_{9}S_{5}$, $F_{5}F_{7}F_{7}S_{5}$, $F_{5}F_{8}F_{9}V_{3}$, $F_{5}F_{5}F_{8}V_{3}$, $F_{8}F_{8}F_{14}S_{5}$, $F_{6}F_{8}F_{8}S_{5}$, $F_{5}F_{8}F_{9}V_{4}$, $F_{5}F_{5}F_{8}V_{4}$, $F_{8}F_{8}F_{14}V_{3}$, $F_{6}F_{8}F_{8}V_{3}$, $F_{7}F_{7}F_{9}V_{4}$, $F_{5}F_{7}F_{7}V_{4}$, $F_{8}F_{8}F_{9}V_{2}$, $F_{7}F_{7}F_{19}V_{1}$, $F_{5}F_{7}F_{9}S_{2}$, $F_{5}F_{5}F_{7}S_{2}$, $F_{3}F_{3}F_{11}V_{2}$, $F_{7}F_{7}F_{19}S_{5}$, $F_{5}F_{7}F_{9}V_{3}$, $F_{5}F_{5}F_{7}V_{3}$, $F_{3}F_{3}F_{11}V_{1}$, $F_{8}F_{8}F_{9}S_{5}$, $F_{5}F_{7}F_{9}V_{4}$, $F_{5}F_{5}F_{7}V_{4}$, $F_{3}F_{3}F_{11}S_{2}$, $F_{8}F_{8}F_{9}V_{3}$, $F_{7}F_{7}F_{19}V_{4}$}} 
 \end{SplitTable}
   \vspace{0.4cm} 
 \captionof{table}{Dimension 9 $(\Delta B, \Delta L)$ = (1, -1) UV completions for topology 7 (see Fig.~\ref{fig:d=9_topologies}).} 
 \label{tab:T9Top7B1L-1}
 \end{center}

%% file: Tables_dim9/Table_dim9_Top8_B1_L-1.tex
 \begin{center}
 \begin{SplitTable}[2.0cm] 
  \SplitHeader{$\mathcal{O}^j{(\mathcal{T}_ {9} ^{ 8 })}$}{New Particles} 
   \vspace{0.1cm} 
 \SplitRow{ $\mathcal{O}_{9, \; (1,-1)}^{39}$} {{$F_{3}F_{8}F_{14}S_{3}$, $F_{3}F_{8}F_{9}V_{2}$, $F_{3}F_{5}F_{8}V_{2}$, $F_{7}F_{8}F_{14}V_{4}$, $F_{7}F_{8}F_{9}S_{6}$, $F_{5}F_{7}F_{8}S_{6}$, $F_{10}F_{11}F_{7}V_{2}$, $F_{2}F_{10}F_{7}V_{2}$, $F_{8}F_{9}F_{9}V_{4}$, $F_{5}F_{5}F_{8}V_{4}$, $F_{10}F_{11}F_{9}S_{3}$, $F_{2}F_{10}F_{5}S_{3}$, $F_{10}F_{7}F_{9}S_{3}$, $F_{10}F_{5}F_{7}S_{3}$, $F_{10}F_{7}F_{19}V_{2}$, $F_{7}F_{8}F_{9}V_{4}$, $F_{5}F_{7}F_{8}V_{4}$, $F_{7}F_{8}F_{19}S_{6}$, $F_{3}F_{4}F_{8}V_{2}$, $F_{7}F_{19}F_{14}V_{4}$, $F_{3}F_{4}F_{14}S_{3}$, $F_{3}F_{8}F_{9}S_{3}$, $F_{3}F_{5}F_{8}S_{3}$, $F_{3}F_{7}F_{9}V_{2}$, $F_{3}F_{5}F_{7}V_{2}$, $F_{7}F_{7}F_{9}S_{6}$, $F_{5}F_{7}F_{7}S_{6}$, $F_{3}F_{11}F_{7}V_{2}$, $F_{2}F_{3}F_{7}V_{2}$, $F_{7}F_{9}F_{9}V_{4}$, $F_{5}F_{5}F_{7}V_{4}$, $F_{3}F_{11}F_{9}S_{3}$, $F_{2}F_{3}F_{5}S_{3}$}} 
 \SplitRow{ $\mathcal{O}_{9, \; (1,-1)}^{40}$} {{$F_{4}F_{14}F_{15}V_{2}$, $F_{4}F_{8}F_{14}V_{2}$, $F_{1}F_{8}F_{14}V_{2}$, $F_{4}F_{6}F_{8}V_{2}$, $F_{1}F_{6}F_{8}V_{2}$, $F_{4}F_{8}F_{9}S_{4}$, $F_{4}F_{8}F_{9}S_{2}$, $F_{1}F_{8}F_{9}S_{4}$, $F_{4}F_{5}F_{8}S_{4}$, $F_{1}F_{5}F_{8}S_{2}$, $F_{7}F_{14}F_{15}V_{5}$, $F_{7}F_{8}F_{14}V_{5}$, $F_{7}F_{8}F_{14}V_{4}$, $F_{6}F_{7}F_{8}V_{5}$, $F_{11}F_{12}F_{7}S_{4}$, $F_{3}F_{11}F_{7}S_{4}$, $F_{3}F_{11}F_{7}S_{2}$, $F_{2}F_{3}F_{7}S_{4}$, $F_{9}F_{14}F_{15}S_{5}$, $F_{8}F_{9}F_{14}S_{5}$, $F_{5}F_{8}F_{14}S_{5}$, $F_{6}F_{8}F_{9}S_{5}$, $F_{5}F_{6}F_{8}S_{5}$, $F_{8}F_{9}F_{9}V_{5}$, $F_{8}F_{9}F_{9}V_{4}$, $F_{5}F_{8}F_{9}V_{5}$, $F_{5}F_{5}F_{8}V_{4}$, $F_{11}F_{12}F_{9}V_{2}$, $F_{3}F_{11}F_{9}V_{2}$, $F_{3}F_{11}F_{5}V_{2}$, $F_{2}F_{3}F_{9}V_{2}$, $F_{2}F_{3}F_{5}V_{2}$, $F_{11}F_{9}F_{20}V_{2}$, $F_{11}F_{7}F_{9}V_{2}$, $F_{2}F_{7}F_{9}V_{2}$, $F_{11}F_{5}F_{7}V_{2}$, $F_{2}F_{5}F_{7}V_{2}$, $F_{11}F_{7}F_{19}S_{4}$, $F_{11}F_{7}F_{19}S_{2}$, $F_{2}F_{7}F_{19}S_{4}$, $F_{8}F_{9}F_{20}V_{5}$, $F_{7}F_{8}F_{9}V_{5}$, $F_{7}F_{8}F_{9}V_{4}$, $F_{5}F_{7}F_{8}V_{5}$, $F_{5}F_{7}F_{8}V_{4}$, $F_{4}F_{12}F_{8}S_{4}$, $F_{3}F_{4}F_{8}S_{4}$, $F_{3}F_{4}F_{8}S_{2}$, $F_{1}F_{3}F_{8}S_{4}$, $F_{1}F_{3}F_{8}S_{2}$, $F_{9}F_{14}F_{20}S_{5}$, $F_{7}F_{9}F_{14}S_{5}$, $F_{6}F_{7}F_{9}S_{5}$, $F_{5}F_{7}F_{14}S_{5}$, $F_{5}F_{6}F_{7}S_{5}$, $F_{7}F_{19}F_{14}V_{5}$, $F_{7}F_{19}F_{14}V_{4}$, $F_{6}F_{7}F_{19}V_{5}$, $F_{4}F_{12}F_{14}V_{2}$, $F_{3}F_{4}F_{14}V_{2}$, $F_{3}F_{4}F_{6}V_{2}$, $F_{1}F_{3}F_{14}V_{2}$, $F_{1}F_{3}F_{6}V_{2}$, $F_{4}F_{9}F_{15}V_{2}$, $F_{4}F_{8}F_{9}V_{2}$, $F_{1}F_{8}F_{9}V_{2}$, $F_{4}F_{5}F_{8}V_{2}$, $F_{1}F_{5}F_{8}V_{2}$, $F_{4}F_{7}F_{9}S_{4}$, $F_{4}F_{7}F_{9}S_{2}$, $F_{1}F_{7}F_{9}S_{4}$, $F_{4}F_{5}F_{7}S_{4}$, $F_{1}F_{5}F_{7}S_{2}$, $F_{7}F_{9}F_{15}V_{5}$, $F_{4}F_{12}F_{7}S_{4}$, $F_{3}F_{4}F_{7}S_{4}$, $F_{3}F_{4}F_{7}S_{2}$, $F_{1}F_{3}F_{7}S_{4}$, $F_{1}F_{3}F_{7}S_{2}$, $F_{9}F_{9}F_{15}S_{5}$, $F_{8}F_{9}F_{9}S_{5}$, $F_{5}F_{8}F_{9}S_{5}$, $F_{5}F_{5}F_{8}S_{5}$, $F_{7}F_{9}F_{9}V_{5}$, $F_{7}F_{9}F_{9}V_{4}$, $F_{5}F_{7}F_{9}V_{5}$, $F_{5}F_{5}F_{7}V_{4}$, $F_{4}F_{12}F_{9}V_{2}$, $F_{3}F_{4}F_{9}V_{2}$, $F_{3}F_{4}F_{5}V_{2}$, $F_{1}F_{3}F_{9}V_{2}$, $F_{1}F_{3}F_{5}V_{2}$}} 
 \SplitRow{ $\mathcal{O}_{9, \; (1,-1)}^{41}$} {{$F_{4}F_{8}F_{9}S_{2}$, $F_{1}F_{5}F_{8}S_{2}$, $F_{4}F_{13}F_{14}S_{3}$, $F_{1}F_{6}F_{13}S_{3}$, $F_{8}F_{8}F_{9}S_{5}$, $F_{5}F_{8}F_{8}S_{5}$, $F_{3}F_{11}F_{8}S_{3}$, $F_{7}F_{8}F_{9}S_{6}$, $F_{5}F_{7}F_{8}S_{6}$, $F_{7}F_{13}F_{14}S_{5}$, $F_{6}F_{7}F_{13}S_{5}$, $F_{3}F_{11}F_{7}S_{2}$, $F_{11}F_{7}F_{19}S_{2}$, $F_{11}F_{8}F_{9}S_{3}$, $F_{7}F_{13}F_{19}S_{5}$, $F_{3}F_{4}F_{13}S_{3}$, $F_{1}F_{3}F_{13}S_{3}$, $F_{7}F_{8}F_{19}S_{6}$, $F_{3}F_{4}F_{8}S_{2}$, $F_{1}F_{3}F_{8}S_{2}$, $F_{4}F_{7}F_{9}S_{2}$, $F_{1}F_{5}F_{7}S_{2}$, $F_{4}F_{8}F_{14}S_{3}$, $F_{1}F_{6}F_{8}S_{3}$, $F_{7}F_{8}F_{9}S_{5}$, $F_{5}F_{7}F_{8}S_{5}$, $F_{3}F_{4}F_{8}S_{3}$, $F_{1}F_{3}F_{8}S_{3}$, $F_{7}F_{7}F_{9}S_{6}$, $F_{5}F_{7}F_{7}S_{6}$, $F_{7}F_{8}F_{14}S_{5}$, $F_{6}F_{7}F_{8}S_{5}$, $F_{3}F_{4}F_{7}S_{2}$, $F_{1}F_{3}F_{7}S_{2}$}} 
 \SplitRow{ $\mathcal{O}_{9, \; (1,-1)}^{42}$} {{$F_{11}F_{8}F_{9}S_{3}$, $F_{2}F_{5}F_{8}S_{3}$, $F_{8}F_{8}F_{9}S_{5}$, $F_{5}F_{8}F_{8}S_{5}$, $F_{3}F_{11}F_{8}S_{3}$, $F_{2}F_{3}F_{8}S_{3}$, $F_{4}F_{8}F_{14}S_{3}$, $F_{7}F_{8}F_{14}S_{5}$, $F_{10}F_{11}F_{7}S_{3}$, $F_{2}F_{10}F_{7}S_{3}$, $F_{11}F_{7}F_{9}S_{3}$, $F_{2}F_{5}F_{7}S_{3}$, $F_{7}F_{8}F_{9}S_{5}$, $F_{5}F_{7}F_{8}S_{5}$, $F_{3}F_{4}F_{8}S_{3}$}} 
 \SplitRow{ $\mathcal{O}_{9, \; (1,-1)}^{43}$} {{$F_{3}F_{4}F_{9}S_{4}$, $F_{3}F_{4}F_{5}S_{4}$, $F_{9}F_{9}F_{20}S_{6}$, $F_{7}F_{9}F_{9}S_{6}$, $F_{5}F_{7}F_{9}S_{6}$, $F_{5}F_{5}F_{7}S_{6}$, $F_{3}F_{9}F_{20}S_{4}$, $F_{3}F_{7}F_{9}S_{4}$, $F_{3}F_{5}F_{7}S_{4}$}} 
 \SplitRow{ $\mathcal{O}_{9, \; (1,-1)}^{44}$} {{$F_{4}F_{12}F_{9}V_{1}$, $F_{3}F_{4}F_{9}V_{1}$, $F_{3}F_{4}F_{5}V_{1}$, $F_{1}F_{3}F_{9}V_{1}$, $F_{1}F_{3}F_{5}V_{1}$, $F_{8}F_{9}F_{9}V_{5}$, $F_{5}F_{8}F_{9}V_{5}$, $F_{9}F_{9}F_{20}S_{6}$, $F_{7}F_{9}F_{9}S_{6}$, $F_{5}F_{7}F_{9}S_{6}$, $F_{5}F_{5}F_{7}S_{6}$, $F_{4}F_{12}F_{8}S_{4}$, $F_{3}F_{4}F_{8}S_{4}$, $F_{1}F_{3}F_{8}S_{4}$, $F_{8}F_{9}F_{20}V_{5}$, $F_{7}F_{8}F_{9}V_{5}$, $F_{5}F_{7}F_{8}V_{5}$, $F_{4}F_{8}F_{9}S_{4}$, $F_{1}F_{8}F_{9}S_{4}$, $F_{4}F_{9}F_{20}V_{1}$, $F_{4}F_{7}F_{9}V_{1}$, $F_{1}F_{7}F_{9}V_{1}$, $F_{4}F_{5}F_{7}V_{1}$, $F_{1}F_{5}F_{7}V_{1}$}} 
 \SplitRow{ $\mathcal{O}_{9, \; (1,-1)}^{45}$} {{$F_{3}F_{8}F_{14}S_{3}$, $F_{3}F_{6}F_{8}S_{3}$, $F_{3}F_{8}F_{9}V_{2}$, $F_{3}F_{5}F_{8}V_{2}$, $F_{8}F_{8}F_{14}V_{3}$, $F_{6}F_{8}F_{8}V_{3}$, $F_{8}F_{8}F_{9}S_{5}$, $F_{5}F_{8}F_{8}S_{5}$, $F_{3}F_{4}F_{8}V_{2}$, $F_{1}F_{3}F_{8}V_{2}$, $F_{8}F_{9}F_{14}V_{3}$, $F_{5}F_{6}F_{8}V_{3}$, $F_{3}F_{4}F_{14}S_{3}$, $F_{1}F_{3}F_{6}S_{3}$, $F_{3}F_{13}F_{14}S_{3}$, $F_{3}F_{6}F_{13}S_{3}$, $F_{3}F_{8}F_{14}V_{2}$, $F_{7}F_{13}F_{14}V_{3}$, $F_{6}F_{7}F_{13}V_{3}$, $F_{7}F_{8}F_{14}S_{5}$, $F_{3}F_{11}F_{7}V_{2}$, $F_{3}F_{11}F_{9}S_{3}$, $F_{3}F_{8}F_{9}S_{3}$, $F_{3}F_{7}F_{9}V_{2}$, $F_{3}F_{5}F_{7}V_{2}$, $F_{8}F_{8}F_{9}V_{3}$, $F_{7}F_{8}F_{9}S_{5}$, $F_{5}F_{7}F_{8}S_{5}$, $F_{7}F_{9}F_{14}V_{3}$, $F_{5}F_{6}F_{7}V_{3}$}} 
 \SplitRow{ $\mathcal{O}_{9, \; (1,-1)}^{46}$} {{$F_{3}F_{14}F_{15}S_{4}$, $F_{3}F_{8}F_{14}S_{4}$, $F_{3}F_{8}F_{14}S_{2}$, $F_{3}F_{6}F_{8}S_{4}$, $F_{9}F_{14}F_{15}S_{5}$, $F_{8}F_{9}F_{14}S_{5}$, $F_{5}F_{8}F_{14}S_{5}$, $F_{6}F_{8}F_{9}S_{5}$, $F_{5}F_{6}F_{8}S_{5}$, $F_{3}F_{4}F_{9}S_{4}$, $F_{3}F_{4}F_{9}S_{2}$, $F_{3}F_{4}F_{5}S_{4}$, $F_{1}F_{3}F_{9}S_{4}$, $F_{1}F_{3}F_{5}S_{2}$, $F_{3}F_{9}F_{20}S_{4}$, $F_{3}F_{7}F_{9}S_{4}$, $F_{3}F_{7}F_{9}S_{2}$, $F_{3}F_{5}F_{7}S_{4}$, $F_{3}F_{5}F_{7}S_{2}$, $F_{9}F_{14}F_{20}S_{5}$, $F_{7}F_{9}F_{14}S_{5}$, $F_{6}F_{7}F_{9}S_{5}$, $F_{5}F_{7}F_{14}S_{5}$, $F_{5}F_{6}F_{7}S_{5}$, $F_{3}F_{11}F_{14}S_{4}$, $F_{3}F_{11}F_{14}S_{2}$, $F_{3}F_{11}F_{6}S_{4}$, $F_{3}F_{9}F_{15}S_{4}$, $F_{3}F_{8}F_{9}S_{4}$, $F_{3}F_{8}F_{9}S_{2}$, $F_{3}F_{5}F_{8}S_{4}$, $F_{3}F_{5}F_{8}S_{2}$, $F_{9}F_{9}F_{15}S_{5}$, $F_{8}F_{9}F_{9}S_{5}$, $F_{5}F_{8}F_{9}S_{5}$, $F_{5}F_{5}F_{8}S_{5}$}} 
 \SplitRow{ $\mathcal{O}_{9, \; (1,-1)}^{47}$} {{$F_{3}F_{8}F_{14}S_{2}$, $F_{3}F_{8}F_{9}V_{1}$, $F_{3}F_{5}F_{8}V_{1}$, $F_{3}F_{13}F_{14}V_{2}$, $F_{3}F_{6}F_{13}V_{2}$, $F_{8}F_{8}F_{14}V_{3}$, $F_{8}F_{8}F_{9}S_{5}$, $F_{5}F_{8}F_{8}S_{5}$, $F_{3}F_{4}F_{8}V_{2}$, $F_{1}F_{3}F_{8}V_{2}$, $F_{7}F_{8}F_{14}V_{4}$, $F_{7}F_{13}F_{14}S_{5}$, $F_{6}F_{7}F_{13}S_{5}$, $F_{3}F_{4}F_{7}V_{1}$, $F_{1}F_{3}F_{7}V_{1}$, $F_{8}F_{9}F_{9}V_{4}$, $F_{5}F_{5}F_{8}V_{4}$, $F_{13}F_{9}F_{14}V_{3}$, $F_{5}F_{6}F_{13}V_{3}$, $F_{3}F_{4}F_{9}S_{2}$, $F_{1}F_{3}F_{5}S_{2}$, $F_{3}F_{7}F_{9}S_{2}$, $F_{3}F_{5}F_{7}S_{2}$, $F_{3}F_{7}F_{19}V_{1}$, $F_{3}F_{8}F_{9}V_{2}$, $F_{7}F_{13}F_{9}V_{3}$, $F_{5}F_{7}F_{13}V_{3}$, $F_{7}F_{13}F_{19}S_{5}$, $F_{3}F_{11}F_{13}V_{2}$, $F_{7}F_{8}F_{9}V_{4}$, $F_{5}F_{7}F_{8}V_{4}$, $F_{3}F_{11}F_{8}V_{1}$, $F_{7}F_{19}F_{14}V_{4}$, $F_{8}F_{9}F_{14}V_{3}$, $F_{3}F_{11}F_{14}S_{2}$, $F_{3}F_{8}F_{9}S_{2}$, $F_{3}F_{5}F_{8}S_{2}$, $F_{3}F_{7}F_{9}V_{1}$, $F_{3}F_{5}F_{7}V_{1}$, $F_{3}F_{8}F_{14}V_{2}$, $F_{3}F_{6}F_{8}V_{2}$, $F_{8}F_{8}F_{9}V_{3}$, $F_{5}F_{8}F_{8}V_{3}$, $F_{7}F_{8}F_{9}S_{5}$, $F_{5}F_{7}F_{8}S_{5}$, $F_{7}F_{8}F_{14}S_{5}$, $F_{6}F_{7}F_{8}S_{5}$, $F_{7}F_{9}F_{9}V_{4}$, $F_{5}F_{5}F_{7}V_{4}$, $F_{5}F_{6}F_{8}V_{3}$}} 
 \end{SplitTable}
   \vspace{0.4cm} 
 \captionof{table}{Dimension 9 $(\Delta B, \Delta L)$ = (1, -1) UV completions for topology 8 (see Fig.~\ref{fig:d=9_topologies}).} 
 \label{tab:T9Top8B1L-1}
 \end{center}

%% file: Tables_dim9/Table_dim9_Top9_B1_L-1.tex
 \begin{center}
 \begin{SplitTable}[2.0cm] 
  \SplitHeader{$\mathcal{O}^j{(\mathcal{T}_ {9} ^{ 9 })}$}{New Particles} 
   \vspace{0.1cm} 
 \SplitRow{ $\mathcal{O}_{9, \; (1,-1)}^{39}$} {{$F_{14}S_{3}S_{6}S_{20}$, $F_{8}V_{2}V_{15}V_{5}$, $F_{8}V_{4}V_{2}V_{15}$, $F_{14}V_{4}V_{2}V_{8}$, $F_{8}S_{5}S_{6}S_{19}$, $F_{8}S_{3}S_{5}S_{6}$, $F_{10}V_{2}V_{15}V_{5}$, $F_{10}V_{4}V_{2}V_{15}$, $F_{8}V_{4}V_{2}V_{8}$, $F_{10}S_{3}S_{6}S_{20}$, $F_{9}S_{3}S_{5}S_{19}$, $F_{5}S_{3}S_{3}S_{5}$, $F_{7}V_{2}V_{2}V_{8}$, $F_{9}V_{4}V_{15}V_{5}$, $F_{5}V_{4}V_{4}V_{15}$, $F_{7}S_{6}S_{6}S_{20}$, $F_{3}V_{2}V_{2}V_{8}$, $F_{7}V_{4}V_{15}V_{5}$, $F_{7}V_{4}V_{4}V_{15}$, $F_{3}S_{3}S_{5}S_{19}$, $F_{3}S_{3}S_{3}S_{5}$, $F_{9}S_{3}S_{6}S_{19}$, $F_{5}S_{3}S_{3}S_{6}$, $F_{7}V_{2}V_{2}V_{5}$, $F_{7}V_{4}V_{2}V_{2}$, $F_{9}V_{4}V_{2}V_{5}$, $F_{5}V_{4}V_{4}V_{2}$, $F_{7}S_{6}S_{6}S_{19}$, $F_{7}S_{3}S_{6}S_{6}$, $F_{3}V_{2}V_{2}V_{5}$, $F_{3}V_{4}V_{2}V_{2}$, $F_{7}V_{4}V_{2}V_{5}$, $F_{7}V_{4}V_{4}V_{2}$, $F_{3}S_{3}S_{6}S_{19}$, $F_{3}S_{3}S_{3}S_{6}$}} 
 \SplitRow{ $\mathcal{O}_{9, \; (1,-1)}^{40}$} {{$F_{14}V_{2}V_{5}V_{17}$, $F_{14}V_{2}V_{15}V_{5}$, $F_{6}V_{2}V_{15}V_{5}$, $F_{14}V_{4}V_{2}V_{15}$, $F_{6}V_{4}V_{2}V_{15}$, $F_{8}S_{4}S_{19}S_{13}$, $F_{8}S_{5}S_{4}S_{19}$, $F_{8}S_{3}S_{5}S_{4}$, $F_{8}S_{2}S_{5}S_{19}$, $F_{14}V_{8}V_{5}V_{16}$, $F_{14}V_{2}V_{8}V_{5}$, $F_{6}V_{3}V_{2}V_{5}$, $F_{14}V_{4}V_{2}V_{8}$, $F_{11}S_{4}S_{19}S_{13}$, $F_{11}S_{5}S_{4}S_{19}$, $F_{2}S_{3}S_{5}S_{4}$, $F_{11}S_{2}S_{5}S_{19}$, $F_{14}S_{5}S_{4}S_{12}$, $F_{14}S_{11}S_{5}S_{4}$, $F_{6}S_{11}S_{5}S_{4}$, $F_{14}S_{2}S_{11}S_{5}$, $F_{6}S_{2}S_{11}S_{5}$, $F_{8}V_{8}V_{5}V_{16}$, $F_{8}V_{2}V_{8}V_{5}$, $F_{8}V_{3}V_{2}V_{5}$, $F_{8}V_{4}V_{2}V_{8}$, $F_{11}V_{2}V_{5}V_{17}$, $F_{11}V_{2}V_{15}V_{5}$, $F_{2}V_{2}V_{15}V_{5}$, $F_{11}V_{4}V_{2}V_{15}$, $F_{2}V_{4}V_{2}V_{15}$, $F_{9}V_{2}V_{8}V_{16}$, $F_{9}V_{2}V_{2}V_{8}$, $F_{5}V_{2}V_{2}V_{8}$, $F_{9}V_{3}V_{2}V_{2}$, $F_{5}V_{3}V_{2}V_{2}$, $F_{7}S_{4}S_{4}S_{12}$, $F_{7}S_{11}S_{4}S_{4}$, $F_{7}S_{2}S_{11}S_{4}$, $F_{7}S_{2}S_{2}S_{11}$, $F_{9}V_{5}V_{5}V_{17}$, $F_{9}V_{15}V_{5}V_{5}$, $F_{5}V_{4}V_{15}V_{5}$, $F_{9}V_{4}V_{15}V_{5}$, $F_{5}V_{4}V_{4}V_{15}$, $F_{4}S_{4}S_{4}S_{12}$, $F_{4}S_{11}S_{4}S_{4}$, $F_{1}S_{2}S_{11}S_{4}$, $F_{4}S_{2}S_{11}S_{4}$, $F_{1}S_{2}S_{2}S_{11}$, $F_{9}S_{5}S_{19}S_{13}$, $F_{9}S_{5}S_{5}S_{19}$, $F_{5}S_{5}S_{5}S_{19}$, $F_{9}S_{3}S_{5}S_{5}$, $F_{5}S_{3}S_{5}S_{5}$, $F_{7}V_{5}V_{5}V_{17}$, $F_{7}V_{15}V_{5}V_{5}$, $F_{7}V_{4}V_{15}V_{5}$, $F_{7}V_{4}V_{4}V_{15}$, $F_{4}V_{2}V_{8}V_{16}$, $F_{4}V_{2}V_{2}V_{8}$, $F_{1}V_{2}V_{2}V_{8}$, $F_{4}V_{3}V_{2}V_{2}$, $F_{1}V_{3}V_{2}V_{2}$, $F_{9}V_{2}V_{5}V_{16}$, $F_{9}V_{2}V_{2}V_{5}$, $F_{5}V_{2}V_{2}V_{5}$, $F_{9}V_{4}V_{2}V_{2}$, $F_{5}V_{4}V_{2}V_{2}$, $F_{7}S_{4}S_{4}S_{13}$, $F_{7}S_{5}S_{4}S_{4}$, $F_{7}S_{2}S_{5}S_{4}$, $F_{7}S_{2}S_{2}S_{5}$, $F_{9}V_{5}V_{5}V_{16}$, $F_{9}V_{2}V_{5}V_{5}$, $F_{5}V_{4}V_{2}V_{5}$, $F_{9}V_{4}V_{2}V_{5}$, $F_{5}V_{4}V_{4}V_{2}$, $F_{4}S_{4}S_{4}S_{13}$, $F_{4}S_{5}S_{4}S_{4}$, $F_{1}S_{2}S_{5}S_{4}$, $F_{4}S_{2}S_{5}S_{4}$, $F_{1}S_{2}S_{2}S_{5}$, $F_{9}S_{5}S_{4}S_{13}$, $F_{9}S_{5}S_{5}S_{4}$, $F_{5}S_{5}S_{5}S_{4}$, $F_{9}S_{2}S_{5}S_{5}$, $F_{5}S_{2}S_{5}S_{5}$, $F_{7}V_{5}V_{5}V_{16}$, $F_{7}V_{2}V_{5}V_{5}$, $F_{7}V_{4}V_{2}V_{5}$, $F_{7}V_{4}V_{4}V_{2}$, $F_{4}V_{2}V_{5}V_{16}$, $F_{4}V_{2}V_{2}V_{5}$, $F_{1}V_{2}V_{2}V_{5}$, $F_{4}V_{4}V_{2}V_{2}$, $F_{1}V_{4}V_{2}V_{2}$}} 
 \SplitRow{ $\mathcal{O}_{9, \; (1,-1)}^{41}$} {{$F_{8}S_{2}S_{5}S_{19}$, $F_{13}S_{3}S_{6}S_{20}$, $F_{8}S_{11}S_{5}S_{4}$, $F_{8}S_{2}S_{11}S_{5}$, $F_{11}S_{3}S_{6}S_{20}$, $F_{8}S_{5}S_{6}S_{19}$, $F_{8}S_{3}S_{5}S_{6}$, $F_{13}S_{11}S_{5}S_{4}$, $F_{13}S_{2}S_{11}S_{5}$, $F_{11}S_{2}S_{5}S_{19}$, $F_{7}S_{2}S_{11}S_{4}$, $F_{7}S_{2}S_{2}S_{11}$, $F_{8}S_{3}S_{5}S_{19}$, $F_{8}S_{3}S_{3}S_{5}$, $F_{7}S_{5}S_{5}S_{19}$, $F_{4}S_{3}S_{5}S_{19}$, $F_{1}S_{3}S_{3}S_{5}$, $F_{7}S_{6}S_{6}S_{20}$, $F_{8}S_{5}S_{5}S_{19}$, $F_{4}S_{2}S_{11}S_{4}$, $F_{1}S_{2}S_{2}S_{11}$, $F_{7}S_{2}S_{5}S_{4}$, $F_{7}S_{2}S_{2}S_{5}$, $F_{8}S_{3}S_{6}S_{19}$, $F_{8}S_{3}S_{3}S_{6}$, $F_{7}S_{5}S_{5}S_{4}$, $F_{7}S_{2}S_{5}S_{5}$, $F_{4}S_{3}S_{6}S_{19}$, $F_{1}S_{3}S_{3}S_{6}$, $F_{7}S_{6}S_{6}S_{19}$, $F_{7}S_{3}S_{6}S_{6}$, $F_{8}S_{5}S_{5}S_{4}$, $F_{8}S_{2}S_{5}S_{5}$, $F_{4}S_{2}S_{5}S_{4}$, $F_{1}S_{2}S_{2}S_{5}$}} 
 \SplitRow{ $\mathcal{O}_{9, \; (1,-1)}^{42}$} {{$F_{8}S_{3}S_{5}S_{19}$, $F_{8}S_{3}S_{3}S_{5}$, $F_{8}S_{5}S_{5}S_{19}$, $F_{8}S_{3}S_{5}S_{5}$, $F_{11}S_{3}S_{5}S_{19}$, $F_{2}S_{3}S_{3}S_{5}$, $F_{8}S_{3}S_{6}S_{19}$, $F_{8}S_{3}S_{3}S_{6}$, $F_{8}S_{5}S_{5}S_{4}$, $F_{11}S_{3}S_{6}S_{19}$, $F_{2}S_{3}S_{3}S_{6}$, $F_{7}S_{3}S_{5}S_{4}$, $F_{7}S_{5}S_{6}S_{19}$, $F_{7}S_{3}S_{5}S_{6}$, $F_{4}S_{3}S_{5}S_{4}$}} 
 \SplitRow{ $\mathcal{O}_{9, \; (1,-1)}^{43}$} {{$F_{3}S_{4}S_{19}S_{13}$, $F_{3}S_{5}S_{4}S_{19}$, $F_{3}S_{3}S_{5}S_{4}$, $F_{9}S_{6}S_{19}S_{13}$, $F_{9}S_{5}S_{6}S_{19}$, $F_{5}S_{5}S_{6}S_{19}$, $F_{9}S_{3}S_{5}S_{6}$, $F_{5}S_{3}S_{5}S_{6}$, $F_{9}S_{4}S_{19}S_{13}$, $F_{9}S_{5}S_{4}S_{19}$, $F_{5}S_{3}S_{5}S_{4}$}} 
 \SplitRow{ $\mathcal{O}_{9, \; (1,-1)}^{44}$} {{$F_{4}V_{1}V_{8}V_{16}$, $F_{4}V_{1}V_{2}V_{8}$, $F_{1}V_{1}V_{2}V_{8}$, $F_{4}V_{3}V_{1}V_{2}$, $F_{1}V_{3}V_{1}V_{2}$, $F_{8}V_{8}V_{5}V_{16}$, $F_{8}V_{2}V_{8}V_{5}$, $F_{8}V_{3}V_{2}V_{5}$, $F_{9}S_{6}S_{19}S_{13}$, $F_{9}S_{5}S_{6}S_{19}$, $F_{5}S_{5}S_{6}S_{19}$, $F_{9}S_{3}S_{5}S_{6}$, $F_{5}S_{3}S_{5}S_{6}$, $F_{4}S_{4}S_{19}S_{13}$, $F_{4}S_{5}S_{4}S_{19}$, $F_{1}S_{3}S_{5}S_{4}$, $F_{9}V_{8}V_{5}V_{16}$, $F_{9}V_{2}V_{8}V_{5}$, $F_{5}V_{3}V_{2}V_{5}$, $F_{8}S_{4}S_{19}S_{13}$, $F_{8}S_{5}S_{4}S_{19}$, $F_{8}S_{3}S_{5}S_{4}$, $F_{9}V_{1}V_{8}V_{16}$, $F_{9}V_{1}V_{2}V_{8}$, $F_{5}V_{1}V_{2}V_{8}$, $F_{9}V_{3}V_{1}V_{2}$, $F_{5}V_{3}V_{1}V_{2}$}} 
 \SplitRow{ $\mathcal{O}_{9, \; (1,-1)}^{45}$} {{$F_{14}S_{3}S_{5}S_{19}$, $F_{6}S_{3}S_{3}S_{5}$, $F_{8}V_{2}V_{2}V_{8}$, $F_{8}V_{3}V_{2}V_{2}$, $F_{14}V_{3}V_{2}V_{8}$, $F_{6}V_{3}V_{3}V_{2}$, $F_{8}S_{5}S_{5}S_{19}$, $F_{8}S_{3}S_{5}S_{5}$, $F_{3}V_{2}V_{2}V_{8}$, $F_{3}V_{3}V_{2}V_{2}$, $F_{8}V_{3}V_{2}V_{8}$, $F_{8}V_{3}V_{3}V_{2}$, $F_{3}S_{3}S_{5}S_{19}$, $F_{3}S_{3}S_{3}S_{5}$, $F_{14}S_{3}S_{6}S_{19}$, $F_{6}S_{3}S_{3}S_{6}$, $F_{8}V_{2}V_{2}V_{5}$, $F_{14}V_{3}V_{1}V_{8}$, $F_{6}V_{3}V_{3}V_{1}$, $F_{8}S_{5}S_{5}S_{4}$, $F_{3}V_{2}V_{2}V_{5}$, $F_{8}V_{3}V_{1}V_{8}$, $F_{8}V_{3}V_{3}V_{1}$, $F_{3}S_{3}S_{6}S_{19}$, $F_{3}S_{3}S_{3}S_{6}$, $F_{9}S_{3}S_{5}S_{4}$, $F_{7}V_{1}V_{2}V_{8}$, $F_{7}V_{3}V_{1}V_{2}$, $F_{9}V_{3}V_{2}V_{5}$, $F_{7}S_{5}S_{6}S_{19}$, $F_{7}S_{3}S_{5}S_{6}$, $F_{3}V_{1}V_{2}V_{8}$, $F_{3}V_{3}V_{1}V_{2}$, $F_{7}V_{3}V_{2}V_{5}$, $F_{3}S_{3}S_{5}S_{4}$}} 
 \SplitRow{ $\mathcal{O}_{9, \; (1,-1)}^{46}$} {{$F_{14}S_{4}S_{19}S_{13}$, $F_{14}S_{5}S_{4}S_{19}$, $F_{6}S_{3}S_{5}S_{4}$, $F_{14}S_{2}S_{5}S_{19}$, $F_{14}S_{5}S_{4}S_{12}$, $F_{14}S_{11}S_{5}S_{4}$, $F_{6}S_{11}S_{5}S_{4}$, $F_{14}S_{2}S_{11}S_{5}$, $F_{6}S_{2}S_{11}S_{5}$, $F_{3}S_{4}S_{19}S_{13}$, $F_{3}S_{5}S_{4}S_{19}$, $F_{3}S_{3}S_{5}S_{4}$, $F_{3}S_{2}S_{5}S_{19}$, $F_{9}S_{4}S_{4}S_{12}$, $F_{9}S_{11}S_{4}S_{4}$, $F_{5}S_{2}S_{11}S_{4}$, $F_{9}S_{2}S_{11}S_{4}$, $F_{5}S_{2}S_{2}S_{11}$, $F_{9}S_{5}S_{19}S_{13}$, $F_{9}S_{5}S_{5}S_{19}$, $F_{5}S_{5}S_{5}S_{19}$, $F_{9}S_{3}S_{5}S_{5}$, $F_{5}S_{3}S_{5}S_{5}$, $F_{3}S_{4}S_{4}S_{12}$, $F_{3}S_{11}S_{4}S_{4}$, $F_{3}S_{2}S_{11}S_{4}$, $F_{3}S_{2}S_{2}S_{11}$, $F_{9}S_{4}S_{4}S_{13}$, $F_{9}S_{5}S_{4}S_{4}$, $F_{5}S_{2}S_{5}S_{4}$, $F_{9}S_{2}S_{5}S_{4}$, $F_{5}S_{2}S_{2}S_{5}$, $F_{9}S_{5}S_{4}S_{13}$, $F_{9}S_{5}S_{5}S_{4}$, $F_{5}S_{5}S_{5}S_{4}$, $F_{9}S_{2}S_{5}S_{5}$, $F_{5}S_{2}S_{5}S_{5}$, $F_{3}S_{4}S_{4}S_{13}$, $F_{3}S_{5}S_{4}S_{4}$, $F_{3}S_{2}S_{5}S_{4}$, $F_{3}S_{2}S_{2}S_{5}$}} 
 \SplitRow{ $\mathcal{O}_{9, \; (1,-1)}^{47}$} {{$F_{14}S_{2}S_{5}S_{19}$, $F_{8}V_{1}V_{2}V_{8}$, $F_{8}V_{3}V_{1}V_{2}$, $F_{13}V_{2}V_{15}V_{5}$, $F_{13}V_{4}V_{2}V_{15}$, $F_{14}V_{3}V_{1}V_{7}$, $F_{8}S_{11}S_{5}S_{4}$, $F_{8}S_{2}S_{11}S_{5}$, $F_{3}V_{2}V_{15}V_{5}$, $F_{3}V_{4}V_{2}V_{15}$, $F_{14}V_{4}V_{2}V_{8}$, $F_{13}S_{11}S_{5}S_{4}$, $F_{13}S_{2}S_{11}S_{5}$, $F_{3}V_{1}V_{2}V_{8}$, $F_{3}V_{3}V_{1}V_{2}$, $F_{8}V_{4}V_{2}V_{8}$, $F_{13}V_{3}V_{1}V_{7}$, $F_{3}S_{2}S_{5}S_{19}$, $F_{9}S_{2}S_{11}S_{4}$, $F_{5}S_{2}S_{2}S_{11}$, $F_{7}V_{1}V_{1}V_{7}$, $F_{8}V_{2}V_{2}V_{8}$, $F_{9}V_{3}V_{2}V_{8}$, $F_{5}V_{3}V_{3}V_{2}$, $F_{7}S_{5}S_{5}S_{19}$, $F_{3}V_{2}V_{2}V_{8}$, $F_{9}V_{4}V_{15}V_{5}$, $F_{5}V_{4}V_{4}V_{15}$, $F_{8}S_{5}S_{5}S_{19}$, $F_{3}V_{1}V_{1}V_{7}$, $F_{7}V_{4}V_{15}V_{5}$, $F_{7}V_{4}V_{4}V_{15}$, $F_{8}V_{3}V_{2}V_{8}$, $F_{8}V_{3}V_{3}V_{2}$, $F_{3}S_{2}S_{11}S_{4}$, $F_{3}S_{2}S_{2}S_{11}$, $F_{9}S_{2}S_{5}S_{4}$, $F_{5}S_{2}S_{2}S_{5}$, $F_{7}V_{1}V_{1}V_{8}$, $F_{7}V_{3}V_{1}V_{1}$, $F_{8}V_{2}V_{2}V_{5}$, $F_{8}V_{4}V_{2}V_{2}$, $F_{9}V_{3}V_{1}V_{8}$, $F_{5}V_{3}V_{3}V_{1}$, $F_{7}S_{5}S_{5}S_{4}$, $F_{7}S_{2}S_{5}S_{5}$, $F_{3}V_{2}V_{2}V_{5}$, $F_{3}V_{4}V_{2}V_{2}$, $F_{9}V_{4}V_{2}V_{5}$, $F_{5}V_{4}V_{4}V_{2}$, $F_{8}S_{5}S_{5}S_{4}$, $F_{8}S_{2}S_{5}S_{5}$, $F_{3}V_{1}V_{1}V_{8}$, $F_{3}V_{3}V_{1}V_{1}$, $F_{7}V_{4}V_{2}V_{5}$, $F_{7}V_{4}V_{4}V_{2}$, $F_{8}V_{3}V_{1}V_{8}$, $F_{8}V_{3}V_{3}V_{1}$, $F_{3}S_{2}S_{5}S_{4}$, $F_{3}S_{2}S_{2}S_{5}$}} 
 \end{SplitTable}
   \vspace{0.4cm} 
 \captionof{table}{Dimension 9 $(\Delta B, \Delta L)$ = (1, -1) UV completions for topology 9 (see Fig.~\ref{fig:d=9_topologies}).} 
 \label{tab:T9Top9B1L-1}
 \end{center}

%% file: Tables_dim9/Table_dim9_Top10_B1_L-1.tex
 \begin{center}
 \begin{SplitTable}[2.0cm] 
  \SplitHeader{$\mathcal{O}^j{(\mathcal{T}_ {9} ^{ 10 })}$}{New Particles} 
   \vspace{0.1cm} 
 \SplitRow{ $\mathcal{O}_{9, \; (1,-1)}^{39}$} {{$V_{4}V_{2}V_{2}$, $S_{3}S_{6}S_{19}$, $S_{3}S_{3}S_{6}$, $S_{3}S_{6}S_{6}$, $V_{4}V_{2}V_{5}$, $V_{4}V_{4}V_{2}$, $V_{4}V_{2}V_{15}$, $S_{3}S_{6}S_{20}$, $S_{3}S_{5}S_{6}$, $V_{4}V_{2}V_{8}$}} 
 \SplitRow{ $\mathcal{O}_{9, \; (1,-1)}^{40}$} {{$S_{5}S_{4}S_{4}$, $S_{2}S_{5}S_{4}$, $S_{2}S_{2}S_{5}$, $V_{2}V_{5}V_{16}$, $V_{2}V_{2}V_{5}$, $V_{4}V_{2}V_{2}$, $V_{2}V_{5}V_{5}$, $V_{4}V_{2}V_{5}$, $V_{4}V_{4}V_{2}$, $S_{5}S_{4}S_{13}$, $S_{5}S_{5}S_{4}$, $S_{2}S_{5}S_{5}$, $S_{5}S_{4}S_{19}$, $S_{2}S_{5}S_{19}$, $S_{3}S_{5}S_{4}$, $V_{2}V_{5}V_{17}$, $V_{2}V_{15}V_{5}$, $V_{4}V_{2}V_{15}$, $V_{2}V_{8}V_{5}$, $V_{4}V_{2}V_{8}$, $V_{3}V_{2}V_{5}$, $S_{5}S_{4}S_{12}$, $S_{11}S_{5}S_{4}$, $S_{2}S_{11}S_{5}$}} 
 \SplitRow{ $\mathcal{O}_{9, \; (1,-1)}^{41}$} {{$S_{3}S_{6}S_{19}$, $S_{3}S_{3}S_{6}$, $S_{2}S_{5}S_{4}$, $S_{2}S_{2}S_{5}$, $S_{2}S_{5}S_{5}$, $S_{3}S_{6}S_{6}$, $S_{3}S_{6}S_{20}$, $S_{2}S_{5}S_{19}$, $S_{2}S_{11}S_{5}$, $S_{3}S_{5}S_{6}$}} 
 \SplitRow{ $\mathcal{O}_{9, \; (1,-1)}^{42}$} {{$S_{3}S_{5}S_{4}$, $S_{3}S_{5}S_{6}$, $S_{3}S_{5}S_{19}$, $S_{3}S_{3}S_{5}$, $S_{3}S_{5}S_{5}$}} 
 \SplitRow{ $\mathcal{O}_{9, \; (1,-1)}^{43}$} {{$S_{6}S_{4}S_{13}$, $S_{5}S_{6}S_{4}$, $S_{6}S_{4}S_{19}$, $S_{3}S_{6}S_{4}$}} 
 \SplitRow{ $\mathcal{O}_{9, \; (1,-1)}^{44}$} {{$S_{6}S_{4}S_{13}$, $S_{5}S_{6}S_{4}$, $V_{1}V_{8}V_{5}$, $V_{3}V_{1}V_{5}$, $V_{1}V_{5}V_{16}$, $V_{1}V_{2}V_{5}$, $S_{6}S_{4}S_{19}$, $S_{3}S_{6}S_{4}$}} 
 \SplitRow{ $\mathcal{O}_{9, \; (1,-1)}^{45}$} {{$V_{3}V_{1}V_{2}$, $S_{3}S_{5}S_{4}$, $S_{3}S_{5}S_{6}$, $V_{3}V_{2}V_{5}$, $V_{3}V_{2}V_{2}$, $S_{3}S_{5}S_{19}$, $S_{3}S_{3}S_{5}$, $S_{3}S_{5}S_{5}$, $V_{3}V_{2}V_{8}$, $V_{3}V_{3}V_{2}$}} 
 \SplitRow{ $\mathcal{O}_{9, \; (1,-1)}^{46}$} {{$S_{5}S_{4}S_{4}$, $S_{2}S_{5}S_{4}$, $S_{2}S_{2}S_{5}$, $S_{5}S_{4}S_{13}$, $S_{5}S_{5}S_{4}$, $S_{2}S_{5}S_{5}$, $S_{5}S_{4}S_{19}$, $S_{2}S_{5}S_{19}$, $S_{3}S_{5}S_{4}$, $S_{5}S_{4}S_{12}$, $S_{11}S_{5}S_{4}$, $S_{2}S_{11}S_{5}$}} 
 \SplitRow{ $\mathcal{O}_{9, \; (1,-1)}^{47}$} {{$V_{4}V_{2}V_{2}$, $V_{3}V_{1}V_{1}$, $S_{2}S_{5}S_{4}$, $S_{2}S_{2}S_{5}$, $S_{2}S_{5}S_{5}$, $V_{3}V_{1}V_{8}$, $V_{3}V_{3}V_{1}$, $V_{4}V_{2}V_{5}$, $V_{4}V_{4}V_{2}$, $V_{4}V_{2}V_{15}$, $V_{3}V_{1}V_{2}$, $S_{2}S_{5}S_{19}$, $S_{2}S_{11}S_{5}$, $V_{3}V_{1}V_{7}$, $V_{4}V_{2}V_{8}$}} 
 \end{SplitTable}
   \vspace{0.4cm} 
 \captionof{table}{Dimension 9 $(\Delta B, \Delta L)$ = (1, -1) UV completions for topology 10 (see Fig.~\ref{fig:d=9_topologies}).} 
 \label{tab:T9Top10B1L-1}
 \end{center}

%% file: Tables_dim9/Table_dim9_Top11_B1_L-1.tex
 \begin{center}
 \begin{SplitTable}[2.0cm] 
  \SplitHeader{$\mathcal{O}^j{(\mathcal{T}_ {9} ^{ 11 })}$}{New Particles} 
   \vspace{0.1cm} 
 \SplitRow{ $\mathcal{O}_{9, \; (1,-1)}^{39}$} {{$F_{3}F_{11}S_{8}V_{2}$, $F_{2}F_{3}S_{7}V_{2}$, $F_{3}F_{11}S_{8}S_{3}$, $F_{2}F_{3}S_{7}S_{3}$, $F_{7}F_{9}S_{8}V_{2}$, $F_{5}F_{7}S_{7}V_{2}$, $F_{7}F_{9}S_{8}S_{6}$, $F_{5}F_{7}S_{7}S_{6}$, $F_{7}F_{9}S_{8}V_{4}$, $F_{5}F_{7}S_{7}V_{4}$, $F_{5}F_{8}S_{8}S_{3}$, $F_{5}F_{8}S_{7}S_{3}$, $F_{5}F_{8}S_{8}V_{4}$, $F_{5}F_{8}S_{7}V_{4}$, $F_{3}F_{4}S_{9}V_{2}$, $F_{3}F_{4}S_{9}S_{3}$, $F_{7}F_{19}S_{9}V_{2}$, $F_{7}F_{19}S_{9}S_{6}$, $F_{7}F_{19}S_{9}V_{4}$, $F_{5}F_{7}S_{9}S_{3}$, $F_{5}F_{7}S_{9}V_{4}$}} 
 \SplitRow{ $\mathcal{O}_{9, \; (1,-1)}^{40}$} {{$F_{4}F_{12}S_{8}S_{4}$, $F_{3}F_{4}S_{8}S_{4}$, $F_{1}F_{3}S_{8}S_{2}$, $F_{3}F_{4}S_{7}S_{4}$, $F_{1}F_{3}S_{7}S_{2}$, $F_{4}F_{12}S_{8}V_{2}$, $F_{3}F_{4}S_{8}V_{2}$, $F_{1}F_{3}S_{8}V_{2}$, $F_{3}F_{4}S_{7}V_{2}$, $F_{1}F_{3}S_{7}V_{2}$, $F_{9}F_{20}S_{8}S_{4}$, $F_{7}F_{9}S_{8}S_{4}$, $F_{7}F_{9}S_{8}S_{2}$, $F_{5}F_{7}S_{7}S_{4}$, $F_{5}F_{7}S_{7}S_{2}$, $F_{9}F_{20}S_{8}V_{5}$, $F_{7}F_{9}S_{8}V_{5}$, $F_{7}F_{9}S_{8}V_{4}$, $F_{5}F_{7}S_{7}V_{5}$, $F_{5}F_{7}S_{7}V_{4}$, $F_{9}F_{15}S_{8}V_{2}$, $F_{8}F_{9}S_{8}V_{2}$, $F_{5}F_{8}S_{8}V_{2}$, $F_{8}F_{9}S_{7}V_{2}$, $F_{5}F_{8}S_{7}V_{2}$, $F_{9}F_{15}S_{8}V_{5}$, $F_{8}F_{9}S_{8}V_{5}$, $F_{5}F_{8}S_{8}V_{4}$, $F_{8}F_{9}S_{7}V_{5}$, $F_{5}F_{8}S_{7}V_{4}$, $F_{9}F_{15}S_{8}S_{5}$, $F_{8}F_{9}S_{8}S_{5}$, $F_{5}F_{8}S_{8}S_{5}$, $F_{8}F_{9}S_{7}S_{5}$, $F_{5}F_{8}S_{7}S_{5}$, $F_{4}F_{12}S_{9}S_{4}$, $F_{3}F_{4}S_{9}S_{4}$, $F_{1}F_{3}S_{9}S_{2}$, $F_{4}F_{12}S_{9}V_{2}$, $F_{3}F_{4}S_{9}V_{2}$, $F_{1}F_{3}S_{9}V_{2}$, $F_{19}F_{20}S_{9}S_{4}$, $F_{7}F_{19}S_{9}S_{4}$, $F_{7}F_{19}S_{9}S_{2}$, $F_{19}F_{20}S_{9}V_{5}$, $F_{7}F_{19}S_{9}V_{5}$, $F_{7}F_{19}S_{9}V_{4}$, $F_{9}F_{20}S_{9}V_{2}$, $F_{7}F_{9}S_{9}V_{2}$, $F_{5}F_{7}S_{9}V_{2}$, $F_{9}F_{20}S_{9}V_{5}$, $F_{7}F_{9}S_{9}V_{5}$, $F_{5}F_{7}S_{9}V_{4}$, $F_{9}F_{20}S_{9}S_{5}$, $F_{7}F_{9}S_{9}S_{5}$, $F_{5}F_{7}S_{9}S_{5}$}} 
 \SplitRow{ $\mathcal{O}_{9, \; (1,-1)}^{41}$} {{$F_{1}F_{3}S_{8}S_{3}$, $F_{1}F_{3}S_{7}S_{3}$, $F_{1}F_{3}S_{8}S_{2}$, $F_{1}F_{3}S_{7}S_{2}$, $F_{8}F_{14}S_{8}S_{3}$, $F_{6}F_{8}S_{7}S_{3}$, $F_{8}F_{14}S_{8}S_{5}$, $F_{6}F_{8}S_{7}S_{5}$, $F_{7}F_{9}S_{8}S_{2}$, $F_{5}F_{7}S_{7}S_{2}$, $F_{7}F_{9}S_{8}S_{5}$, $F_{5}F_{7}S_{7}S_{5}$, $F_{7}F_{9}S_{8}S_{6}$, $F_{5}F_{7}S_{7}S_{6}$, $F_{1}F_{3}S_{9}S_{3}$, $F_{1}F_{3}S_{9}S_{2}$, $F_{8}F_{9}S_{9}S_{3}$, $F_{8}F_{9}S_{9}S_{5}$, $F_{7}F_{19}S_{9}S_{2}$, $F_{7}F_{19}S_{9}S_{5}$, $F_{7}F_{19}S_{9}S_{6}$}} 
 \SplitRow{ $\mathcal{O}_{9, \; (1,-1)}^{42}$} {{$F_{2}F_{10}S_{9}S_{3}$, $F_{8}F_{14}S_{9}S_{3}$, $F_{8}F_{14}S_{9}S_{5}$, $F_{2}F_{3}S_{8}S_{3}$, $F_{2}F_{3}S_{7}S_{3}$, $F_{8}F_{9}S_{8}S_{3}$, $F_{5}F_{8}S_{7}S_{3}$, $F_{8}F_{9}S_{8}S_{5}$, $F_{5}F_{8}S_{7}S_{5}$}} 
 \SplitRow{ $\mathcal{O}_{9, \; (1,-1)}^{43}$} {{$F_{19}F_{20}S_{9}S_{6}$, $F_{7}F_{19}S_{9}S_{6}$, $F_{19}F_{20}S_{9}S_{4}$, $F_{7}F_{19}S_{9}S_{4}$, $F_{4}F_{12}S_{9}S_{4}$}} 
 \SplitRow{ $\mathcal{O}_{9, \; (1,-1)}^{44}$} {{$F_{19}F_{20}S_{9}S_{6}$, $F_{7}F_{19}S_{9}S_{6}$, $F_{19}F_{20}S_{9}V_{5}$, $F_{7}F_{19}S_{9}V_{5}$, $F_{19}F_{20}S_{9}V_{1}$, $F_{7}F_{19}S_{9}V_{1}$, $F_{9}F_{20}S_{9}V_{5}$, $F_{9}F_{20}S_{9}S_{4}$, $F_{11}F_{12}S_{9}V_{1}$, $F_{3}F_{11}S_{9}V_{1}$, $F_{11}F_{12}S_{9}S_{4}$, $F_{3}F_{11}S_{9}S_{4}$}} 
 \SplitRow{ $\mathcal{O}_{9, \; (1,-1)}^{45}$} {{$F_{3}F_{11}S_{9}V_{2}$, $F_{3}F_{11}S_{9}S_{3}$, $F_{8}F_{14}S_{9}V_{2}$, $F_{8}F_{14}S_{9}S_{5}$, $F_{8}F_{14}S_{9}V_{3}$, $F_{6}F_{13}S_{9}S_{3}$, $F_{6}F_{13}S_{9}V_{3}$, $F_{3}F_{4}S_{8}V_{2}$, $F_{1}F_{3}S_{7}V_{2}$, $F_{3}F_{4}S_{8}S_{3}$, $F_{1}F_{3}S_{7}S_{3}$, $F_{8}F_{9}S_{8}V_{2}$, $F_{5}F_{8}S_{7}V_{2}$, $F_{8}F_{9}S_{8}S_{5}$, $F_{5}F_{8}S_{7}S_{5}$, $F_{8}F_{9}S_{8}V_{3}$, $F_{5}F_{8}S_{7}V_{3}$, $F_{6}F_{8}S_{8}S_{3}$, $F_{6}F_{8}S_{7}S_{3}$, $F_{6}F_{8}S_{8}V_{3}$, $F_{6}F_{8}S_{7}V_{3}$}} 
 \SplitRow{ $\mathcal{O}_{9, \; (1,-1)}^{46}$} {{$F_{4}F_{12}S_{8}S_{4}$, $F_{3}F_{4}S_{8}S_{4}$, $F_{3}F_{4}S_{8}S_{2}$, $F_{1}F_{3}S_{7}S_{4}$, $F_{1}F_{3}S_{7}S_{2}$, $F_{9}F_{15}S_{8}S_{4}$, $F_{8}F_{9}S_{8}S_{4}$, $F_{5}F_{8}S_{8}S_{2}$, $F_{8}F_{9}S_{7}S_{4}$, $F_{5}F_{8}S_{7}S_{2}$, $F_{9}F_{15}S_{8}S_{5}$, $F_{8}F_{9}S_{8}S_{5}$, $F_{5}F_{8}S_{8}S_{5}$, $F_{8}F_{9}S_{7}S_{5}$, $F_{5}F_{8}S_{7}S_{5}$, $F_{11}F_{12}S_{9}S_{4}$, $F_{3}F_{11}S_{9}S_{4}$, $F_{3}F_{11}S_{9}S_{2}$, $F_{9}F_{20}S_{9}S_{4}$, $F_{7}F_{9}S_{9}S_{4}$, $F_{5}F_{7}S_{9}S_{2}$, $F_{9}F_{20}S_{9}S_{5}$, $F_{7}F_{9}S_{9}S_{5}$, $F_{5}F_{7}S_{9}S_{5}$}} 
 \SplitRow{ $\mathcal{O}_{9, \; (1,-1)}^{47}$} {{$F_{3}F_{4}S_{8}V_{2}$, $F_{1}F_{3}S_{7}V_{2}$, $F_{3}F_{4}S_{8}V_{1}$, $F_{1}F_{3}S_{7}V_{1}$, $F_{3}F_{4}S_{8}S_{2}$, $F_{1}F_{3}S_{7}S_{2}$, $F_{8}F_{14}S_{8}V_{2}$, $F_{6}F_{8}S_{7}V_{2}$, $F_{8}F_{14}S_{8}S_{5}$, $F_{6}F_{8}S_{7}S_{5}$, $F_{8}F_{14}S_{8}V_{3}$, $F_{6}F_{8}S_{7}V_{3}$, $F_{7}F_{9}S_{8}V_{1}$, $F_{5}F_{7}S_{7}V_{1}$, $F_{7}F_{9}S_{8}S_{5}$, $F_{5}F_{7}S_{7}S_{5}$, $F_{7}F_{9}S_{8}V_{4}$, $F_{5}F_{7}S_{7}V_{4}$, $F_{5}F_{8}S_{8}S_{2}$, $F_{5}F_{8}S_{7}S_{2}$, $F_{5}F_{8}S_{8}V_{3}$, $F_{5}F_{8}S_{7}V_{3}$, $F_{5}F_{8}S_{8}V_{4}$, $F_{5}F_{8}S_{7}V_{4}$, $F_{3}F_{11}S_{9}V_{2}$, $F_{3}F_{11}S_{9}V_{1}$, $F_{3}F_{11}S_{9}S_{2}$, $F_{8}F_{9}S_{9}V_{2}$, $F_{8}F_{9}S_{9}S_{5}$, $F_{8}F_{9}S_{9}V_{3}$, $F_{7}F_{19}S_{9}V_{1}$, $F_{7}F_{19}S_{9}S_{5}$, $F_{7}F_{19}S_{9}V_{4}$, $F_{5}F_{7}S_{9}S_{2}$, $F_{5}F_{7}S_{9}V_{3}$, $F_{5}F_{7}S_{9}V_{4}$}} 
 \end{SplitTable}
   \vspace{0.4cm} 
 \captionof{table}{Dimension 9 $(\Delta B, \Delta L)$ = (1, -1) UV completions for topology 11 (see Fig.~\ref{fig:d=9_topologies}).} 
 \label{tab:T9Top11B1L-1}
 \end{center}

%% file: Tables_dim9/Table_dim9_Top12_B1_L-1.tex
 \begin{center}
 \begin{SplitTable}[2.0cm] 
  \SplitHeader{$\mathcal{O}^j{(\mathcal{T}_ {9} ^{ 12 })}$}{New Particles} 
   \vspace{0.1cm} 
 \SplitRow{ $\mathcal{O}_{9, \; (1,-1)}^{39}$} {{$F_{7}F_{14}V_{2}V_{8}$, $F_{8}F_{9}S_{5}S_{19}$, $F_{5}F_{8}S_{3}S_{5}$, $F_{7}F_{8}V_{2}V_{8}$, $F_{7}F_{14}S_{6}S_{20}$, $F_{8}F_{9}V_{15}V_{5}$, $F_{5}F_{8}V_{4}V_{15}$, $F_{3}F_{14}V_{2}V_{8}$, $F_{10}F_{9}V_{15}V_{5}$, $F_{10}F_{5}V_{4}V_{15}$, $F_{3}F_{8}V_{2}V_{8}$, $F_{10}F_{7}S_{6}S_{20}$, $F_{7}F_{8}V_{15}V_{5}$, $F_{7}F_{8}V_{4}V_{15}$, $F_{3}F_{8}S_{5}S_{19}$, $F_{3}F_{8}S_{3}S_{5}$, $F_{10}F_{7}V_{15}V_{5}$, $F_{10}F_{7}V_{4}V_{15}$, $F_{7}F_{9}V_{2}V_{5}$, $F_{5}F_{7}V_{4}V_{2}$, $F_{7}F_{9}S_{6}S_{19}$, $F_{5}F_{7}S_{3}S_{6}$, $F_{7}F_{7}V_{2}V_{5}$, $F_{7}F_{7}V_{4}V_{2}$, $F_{3}F_{9}V_{2}V_{5}$, $F_{3}F_{5}V_{4}V_{2}$, $F_{3}F_{7}V_{2}V_{5}$, $F_{3}F_{7}V_{4}V_{2}$, $F_{3}F_{7}S_{6}S_{19}$, $F_{3}F_{7}S_{3}S_{6}$}} 
 \SplitRow{ $\mathcal{O}_{9, \; (1,-1)}^{40}$} {{$F_{9}F_{14}V_{8}V_{16}$, $F_{9}F_{14}V_{2}V_{8}$, $F_{6}F_{9}V_{3}V_{2}$, $F_{5}F_{14}V_{2}V_{8}$, $F_{5}F_{6}V_{3}V_{2}$, $F_{7}F_{14}S_{4}S_{12}$, $F_{7}F_{14}S_{11}S_{4}$, $F_{6}F_{7}S_{11}S_{4}$, $F_{7}F_{14}S_{2}S_{11}$, $F_{6}F_{7}S_{2}S_{11}$, $F_{8}F_{9}V_{8}V_{16}$, $F_{8}F_{9}V_{2}V_{8}$, $F_{8}F_{9}V_{3}V_{2}$, $F_{5}F_{8}V_{2}V_{8}$, $F_{5}F_{8}V_{3}V_{2}$, $F_{9}F_{14}V_{5}V_{17}$, $F_{9}F_{14}V_{15}V_{5}$, $F_{6}F_{9}V_{15}V_{5}$, $F_{5}F_{14}V_{4}V_{15}$, $F_{5}F_{6}V_{4}V_{15}$, $F_{4}F_{14}S_{4}S_{12}$, $F_{4}F_{14}S_{11}S_{4}$, $F_{4}F_{6}S_{11}S_{4}$, $F_{1}F_{14}S_{2}S_{11}$, $F_{1}F_{6}S_{2}S_{11}$, $F_{11}F_{9}V_{5}V_{17}$, $F_{11}F_{9}V_{15}V_{5}$, $F_{2}F_{9}V_{15}V_{5}$, $F_{11}F_{5}V_{4}V_{15}$, $F_{2}F_{5}V_{4}V_{15}$, $F_{7}F_{14}V_{5}V_{17}$, $F_{7}F_{14}V_{15}V_{5}$, $F_{6}F_{7}V_{15}V_{5}$, $F_{7}F_{14}V_{4}V_{15}$, $F_{6}F_{7}V_{4}V_{15}$, $F_{8}F_{9}S_{19}S_{13}$, $F_{8}F_{9}S_{5}S_{19}$, $F_{8}F_{9}S_{3}S_{5}$, $F_{5}F_{8}S_{5}S_{19}$, $F_{5}F_{8}S_{3}S_{5}$, $F_{4}F_{14}V_{8}V_{16}$, $F_{4}F_{14}V_{2}V_{8}$, $F_{4}F_{6}V_{3}V_{2}$, $F_{1}F_{14}V_{2}V_{8}$, $F_{1}F_{6}V_{3}V_{2}$, $F_{11}F_{9}S_{19}S_{13}$, $F_{11}F_{9}S_{5}S_{19}$, $F_{2}F_{9}S_{3}S_{5}$, $F_{11}F_{5}S_{5}S_{19}$, $F_{2}F_{5}S_{3}S_{5}$, $F_{4}F_{8}V_{8}V_{16}$, $F_{4}F_{8}V_{2}V_{8}$, $F_{4}F_{8}V_{3}V_{2}$, $F_{1}F_{8}V_{2}V_{8}$, $F_{1}F_{8}V_{3}V_{2}$, $F_{11}F_{7}V_{5}V_{17}$, $F_{11}F_{7}V_{15}V_{5}$, $F_{2}F_{7}V_{15}V_{5}$, $F_{11}F_{7}V_{4}V_{15}$, $F_{2}F_{7}V_{4}V_{15}$, $F_{9}F_{9}V_{5}V_{16}$, $F_{9}F_{9}V_{2}V_{5}$, $F_{5}F_{9}V_{4}V_{2}$, $F_{5}F_{9}V_{2}V_{5}$, $F_{5}F_{5}V_{4}V_{2}$, $F_{7}F_{9}S_{4}S_{13}$, $F_{7}F_{9}S_{5}S_{4}$, $F_{5}F_{7}S_{5}S_{4}$, $F_{7}F_{9}S_{2}S_{5}$, $F_{5}F_{7}S_{2}S_{5}$, $F_{7}F_{9}V_{5}V_{16}$, $F_{7}F_{9}V_{2}V_{5}$, $F_{7}F_{9}V_{4}V_{2}$, $F_{5}F_{7}V_{2}V_{5}$, $F_{5}F_{7}V_{4}V_{2}$, $F_{4}F_{9}S_{4}S_{13}$, $F_{4}F_{9}S_{5}S_{4}$, $F_{4}F_{5}S_{5}S_{4}$, $F_{1}F_{9}S_{2}S_{5}$, $F_{1}F_{5}S_{2}S_{5}$, $F_{4}F_{9}V_{5}V_{16}$, $F_{4}F_{9}V_{2}V_{5}$, $F_{1}F_{9}V_{2}V_{5}$, $F_{4}F_{5}V_{4}V_{2}$, $F_{1}F_{5}V_{4}V_{2}$, $F_{4}F_{7}V_{5}V_{16}$, $F_{4}F_{7}V_{2}V_{5}$, $F_{4}F_{7}V_{4}V_{2}$, $F_{1}F_{7}V_{2}V_{5}$, $F_{1}F_{7}V_{4}V_{2}$}} 
 \SplitRow{ $\mathcal{O}_{9, \; (1,-1)}^{41}$} {{$F_{7}F_{8}S_{11}S_{4}$, $F_{7}F_{8}S_{2}S_{11}$, $F_{8}F_{8}S_{5}S_{19}$, $F_{8}F_{8}S_{3}S_{5}$, $F_{7}F_{13}S_{11}S_{4}$, $F_{7}F_{13}S_{2}S_{11}$, $F_{7}F_{8}S_{5}S_{19}$, $F_{4}F_{8}S_{5}S_{19}$, $F_{1}F_{8}S_{3}S_{5}$, $F_{11}F_{7}S_{5}S_{19}$, $F_{7}F_{13}S_{6}S_{20}$, $F_{4}F_{8}S_{11}S_{4}$, $F_{1}F_{8}S_{2}S_{11}$, $F_{11}F_{7}S_{6}S_{20}$, $F_{4}F_{13}S_{11}S_{4}$, $F_{1}F_{13}S_{2}S_{11}$, $F_{11}F_{8}S_{5}S_{19}$, $F_{7}F_{7}S_{5}S_{4}$, $F_{7}F_{7}S_{2}S_{5}$, $F_{7}F_{8}S_{6}S_{19}$, $F_{7}F_{8}S_{3}S_{6}$, $F_{7}F_{8}S_{5}S_{4}$, $F_{7}F_{8}S_{2}S_{5}$, $F_{4}F_{7}S_{6}S_{19}$, $F_{1}F_{7}S_{3}S_{6}$, $F_{4}F_{7}S_{5}S_{4}$, $F_{1}F_{7}S_{2}S_{5}$, $F_{4}F_{8}S_{5}S_{4}$, $F_{1}F_{8}S_{2}S_{5}$}} 
 \SplitRow{ $\mathcal{O}_{9, \; (1,-1)}^{42}$} {{$F_{8}F_{8}S_{5}S_{19}$, $F_{8}F_{8}S_{3}S_{5}$, $F_{11}F_{8}S_{5}S_{19}$, $F_{2}F_{8}S_{3}S_{5}$, $F_{7}F_{8}S_{5}S_{4}$, $F_{7}F_{8}S_{6}S_{19}$, $F_{7}F_{8}S_{3}S_{6}$, $F_{4}F_{8}S_{5}S_{4}$, $F_{11}F_{7}S_{6}S_{19}$, $F_{2}F_{7}S_{3}S_{6}$}} 
 \SplitRow{ $\mathcal{O}_{9, \; (1,-1)}^{43}$} {{$F_{3}F_{9}S_{19}S_{13}$, $F_{3}F_{9}S_{5}S_{19}$, $F_{3}F_{9}S_{3}S_{5}$, $F_{3}F_{5}S_{5}S_{19}$, $F_{3}F_{5}S_{3}S_{5}$, $F_{9}F_{9}S_{19}S_{13}$, $F_{9}F_{9}S_{5}S_{19}$, $F_{5}F_{9}S_{3}S_{5}$, $F_{5}F_{9}S_{5}S_{19}$, $F_{5}F_{5}S_{3}S_{5}$}} 
 \SplitRow{ $\mathcal{O}_{9, \; (1,-1)}^{44}$} {{$F_{4}F_{8}V_{8}V_{16}$, $F_{4}F_{8}V_{2}V_{8}$, $F_{1}F_{8}V_{2}V_{8}$, $F_{4}F_{8}V_{3}V_{2}$, $F_{1}F_{8}V_{3}V_{2}$, $F_{4}F_{9}S_{19}S_{13}$, $F_{4}F_{9}S_{5}S_{19}$, $F_{1}F_{9}S_{3}S_{5}$, $F_{4}F_{5}S_{5}S_{19}$, $F_{1}F_{5}S_{3}S_{5}$, $F_{4}F_{9}V_{8}V_{16}$, $F_{4}F_{9}V_{2}V_{8}$, $F_{4}F_{5}V_{3}V_{2}$, $F_{1}F_{9}V_{2}V_{8}$, $F_{1}F_{5}V_{3}V_{2}$, $F_{8}F_{9}S_{19}S_{13}$, $F_{8}F_{9}S_{5}S_{19}$, $F_{8}F_{9}S_{3}S_{5}$, $F_{5}F_{8}S_{5}S_{19}$, $F_{5}F_{8}S_{3}S_{5}$, $F_{8}F_{9}V_{8}V_{16}$, $F_{8}F_{9}V_{2}V_{8}$, $F_{5}F_{8}V_{2}V_{8}$, $F_{8}F_{9}V_{3}V_{2}$, $F_{5}F_{8}V_{3}V_{2}$, $F_{9}F_{9}V_{8}V_{16}$, $F_{9}F_{9}V_{2}V_{8}$, $F_{5}F_{9}V_{2}V_{8}$, $F_{5}F_{9}V_{3}V_{2}$, $F_{5}F_{5}V_{3}V_{2}$}} 
 \SplitRow{ $\mathcal{O}_{9, \; (1,-1)}^{45}$} {{$F_{8}F_{14}V_{2}V_{8}$, $F_{6}F_{8}V_{3}V_{2}$, $F_{8}F_{14}S_{5}S_{19}$, $F_{6}F_{8}S_{3}S_{5}$, $F_{8}F_{8}V_{2}V_{8}$, $F_{8}F_{8}V_{3}V_{2}$, $F_{3}F_{14}V_{2}V_{8}$, $F_{3}F_{6}V_{3}V_{2}$, $F_{3}F_{8}V_{2}V_{8}$, $F_{3}F_{8}V_{3}V_{2}$, $F_{3}F_{8}S_{5}S_{19}$, $F_{3}F_{8}S_{3}S_{5}$, $F_{7}F_{14}V_{1}V_{8}$, $F_{6}F_{7}V_{3}V_{1}$, $F_{8}F_{9}S_{5}S_{4}$, $F_{7}F_{8}V_{1}V_{8}$, $F_{7}F_{8}V_{3}V_{1}$, $F_{7}F_{14}S_{6}S_{19}$, $F_{6}F_{7}S_{3}S_{6}$, $F_{8}F_{9}V_{2}V_{5}$, $F_{3}F_{14}V_{1}V_{8}$, $F_{3}F_{6}V_{3}V_{1}$, $F_{3}F_{9}V_{2}V_{5}$, $F_{3}F_{8}V_{1}V_{8}$, $F_{3}F_{8}V_{3}V_{1}$, $F_{3}F_{7}S_{6}S_{19}$, $F_{3}F_{7}S_{3}S_{6}$, $F_{7}F_{8}V_{2}V_{5}$, $F_{3}F_{8}S_{5}S_{4}$, $F_{3}F_{7}V_{2}V_{5}$}} 
 \SplitRow{ $\mathcal{O}_{9, \; (1,-1)}^{46}$} {{$F_{9}F_{14}S_{4}S_{12}$, $F_{9}F_{14}S_{11}S_{4}$, $F_{6}F_{9}S_{11}S_{4}$, $F_{5}F_{14}S_{2}S_{11}$, $F_{5}F_{6}S_{2}S_{11}$, $F_{9}F_{14}S_{19}S_{13}$, $F_{9}F_{14}S_{5}S_{19}$, $F_{6}F_{9}S_{3}S_{5}$, $F_{5}F_{14}S_{5}S_{19}$, $F_{5}F_{6}S_{3}S_{5}$, $F_{3}F_{14}S_{4}S_{12}$, $F_{3}F_{14}S_{11}S_{4}$, $F_{3}F_{6}S_{11}S_{4}$, $F_{3}F_{14}S_{2}S_{11}$, $F_{3}F_{6}S_{2}S_{11}$, $F_{3}F_{9}S_{19}S_{13}$, $F_{3}F_{9}S_{5}S_{19}$, $F_{3}F_{9}S_{3}S_{5}$, $F_{3}F_{5}S_{5}S_{19}$, $F_{3}F_{5}S_{3}S_{5}$, $F_{9}F_{9}S_{4}S_{13}$, $F_{9}F_{9}S_{5}S_{4}$, $F_{5}F_{9}S_{5}S_{4}$, $F_{5}F_{9}S_{2}S_{5}$, $F_{5}F_{5}S_{2}S_{5}$, $F_{3}F_{9}S_{4}S_{13}$, $F_{3}F_{9}S_{5}S_{4}$, $F_{3}F_{5}S_{5}S_{4}$, $F_{3}F_{9}S_{2}S_{5}$, $F_{3}F_{5}S_{2}S_{5}$}} 
 \SplitRow{ $\mathcal{O}_{9, \; (1,-1)}^{47}$} {{$F_{7}F_{14}V_{1}V_{7}$, $F_{8}F_{9}S_{11}S_{4}$, $F_{5}F_{8}S_{2}S_{11}$, $F_{8}F_{14}V_{2}V_{8}$, $F_{13}F_{9}S_{11}S_{4}$, $F_{5}F_{13}S_{2}S_{11}$, $F_{8}F_{8}V_{2}V_{8}$, $F_{7}F_{13}V_{1}V_{7}$, $F_{7}F_{14}S_{5}S_{19}$, $F_{8}F_{9}V_{2}V_{8}$, $F_{5}F_{8}V_{3}V_{2}$, $F_{3}F_{14}V_{2}V_{8}$, $F_{3}F_{9}V_{2}V_{8}$, $F_{3}F_{5}V_{3}V_{2}$, $F_{3}F_{8}V_{2}V_{8}$, $F_{3}F_{7}S_{5}S_{19}$, $F_{8}F_{14}S_{5}S_{19}$, $F_{13}F_{9}V_{15}V_{5}$, $F_{5}F_{13}V_{4}V_{15}$, $F_{3}F_{14}V_{1}V_{7}$, $F_{3}F_{9}V_{15}V_{5}$, $F_{3}F_{5}V_{4}V_{15}$, $F_{3}F_{13}V_{1}V_{7}$, $F_{3}F_{8}S_{5}S_{19}$, $F_{8}F_{8}V_{3}V_{2}$, $F_{7}F_{13}V_{15}V_{5}$, $F_{7}F_{13}V_{4}V_{15}$, $F_{3}F_{8}S_{11}S_{4}$, $F_{3}F_{8}S_{2}S_{11}$, $F_{3}F_{7}V_{15}V_{5}$, $F_{3}F_{7}V_{4}V_{15}$, $F_{3}F_{13}S_{11}S_{4}$, $F_{3}F_{13}S_{2}S_{11}$, $F_{3}F_{8}V_{3}V_{2}$, $F_{7}F_{9}V_{1}V_{8}$, $F_{5}F_{7}V_{3}V_{1}$, $F_{7}F_{9}S_{5}S_{4}$, $F_{5}F_{7}S_{2}S_{5}$, $F_{8}F_{9}V_{2}V_{5}$, $F_{5}F_{8}V_{4}V_{2}$, $F_{8}F_{9}S_{5}S_{4}$, $F_{5}F_{8}S_{2}S_{5}$, $F_{7}F_{8}V_{2}V_{5}$, $F_{7}F_{8}V_{4}V_{2}$, $F_{7}F_{8}V_{1}V_{8}$, $F_{7}F_{8}V_{3}V_{1}$, $F_{3}F_{9}V_{2}V_{5}$, $F_{3}F_{5}V_{4}V_{2}$, $F_{3}F_{9}V_{1}V_{8}$, $F_{3}F_{5}V_{3}V_{1}$, $F_{3}F_{7}V_{2}V_{5}$, $F_{3}F_{7}V_{4}V_{2}$, $F_{3}F_{7}S_{5}S_{4}$, $F_{3}F_{7}S_{2}S_{5}$, $F_{3}F_{8}V_{1}V_{8}$, $F_{3}F_{8}V_{3}V_{1}$, $F_{3}F_{8}S_{5}S_{4}$, $F_{3}F_{8}S_{2}S_{5}$}} 
 \end{SplitTable}
   \vspace{0.4cm} 
 \captionof{table}{Dimension 9 $(\Delta B, \Delta L)$ = (1, -1) UV completions for topology 12 (see Fig.~\ref{fig:d=9_topologies}).} 
 \label{tab:T9Top12B1L-1}
 \end{center}

%% file: Tables_dim9/Table_dim9_Top13_B1_L-1.tex
 \begin{center}
 \begin{SplitTable}[2.0cm] 
  \SplitHeader{$\mathcal{O}^j{(\mathcal{T}_ {9} ^{ 13 })}$}{New Particles} 
   \vspace{0.1cm} 
 \SplitRow{ $\mathcal{O}_{9, \; (1,-1)}^{39}$} {{$S_{16}S_{8}V_{4}V_{2}$, $S_{7}S_{16}V_{4}V_{2}$, $S_{16}S_{8}S_{3}S_{6}$, $S_{7}S_{16}S_{3}S_{6}$, $S_{16}S_{9}V_{4}V_{2}$, $S_{16}S_{9}S_{3}S_{6}$}} 
 \SplitRow{ $\mathcal{O}_{9, \; (1,-1)}^{40}$} {{$S_{8}S_{17}S_{5}S_{4}$, $S_{16}S_{8}S_{5}S_{4}$, $S_{16}S_{8}S_{2}S_{5}$, $S_{7}S_{16}S_{5}S_{4}$, $S_{7}S_{16}S_{2}S_{5}$, $S_{8}S_{17}V_{2}V_{5}$, $S_{16}S_{8}V_{2}V_{5}$, $S_{16}S_{8}V_{4}V_{2}$, $S_{7}S_{16}V_{2}V_{5}$, $S_{7}S_{16}V_{4}V_{2}$, $S_{9}S_{17}S_{5}S_{4}$, $S_{16}S_{9}S_{5}S_{4}$, $S_{16}S_{9}S_{2}S_{5}$, $S_{9}S_{17}V_{2}V_{5}$, $S_{16}S_{9}V_{2}V_{5}$, $S_{16}S_{9}V_{4}V_{2}$}} 
 \SplitRow{ $\mathcal{O}_{9, \; (1,-1)}^{41}$} {{$S_{16}S_{8}S_{3}S_{6}$, $S_{7}S_{16}S_{3}S_{6}$, $S_{16}S_{8}S_{2}S_{5}$, $S_{7}S_{16}S_{2}S_{5}$, $S_{16}S_{9}S_{3}S_{6}$, $S_{16}S_{9}S_{2}S_{5}$}} 
 \SplitRow{ $\mathcal{O}_{9, \; (1,-1)}^{42}$} {{$S_{16}S_{9}S_{3}S_{5}$, $S_{16}S_{8}S_{3}S_{5}$, $S_{7}S_{16}S_{3}S_{5}$}} 
 \SplitRow{ $\mathcal{O}_{9, \; (1,-1)}^{43}$} {{$S_{9}S_{18}S_{6}S_{4}$}} 
 \SplitRow{ $\mathcal{O}_{9, \; (1,-1)}^{44}$} {{$S_{9}S_{18}S_{6}S_{4}$, $S_{9}S_{18}V_{1}V_{5}$}} 
 \SplitRow{ $\mathcal{O}_{9, \; (1,-1)}^{45}$} {{$S_{16}S_{9}V_{3}V_{2}$, $S_{16}S_{9}S_{3}S_{5}$, $S_{16}S_{8}V_{3}V_{2}$, $S_{7}S_{16}V_{3}V_{2}$, $S_{16}S_{8}S_{3}S_{5}$, $S_{7}S_{16}S_{3}S_{5}$}} 
 \SplitRow{ $\mathcal{O}_{9, \; (1,-1)}^{46}$} {{$S_{8}S_{17}S_{5}S_{4}$, $S_{16}S_{8}S_{5}S_{4}$, $S_{16}S_{8}S_{2}S_{5}$, $S_{7}S_{16}S_{5}S_{4}$, $S_{7}S_{16}S_{2}S_{5}$, $S_{9}S_{17}S_{5}S_{4}$, $S_{16}S_{9}S_{5}S_{4}$, $S_{16}S_{9}S_{2}S_{5}$}} 
 \SplitRow{ $\mathcal{O}_{9, \; (1,-1)}^{47}$} {{$S_{16}S_{8}V_{4}V_{2}$, $S_{7}S_{16}V_{4}V_{2}$, $S_{16}S_{8}V_{3}V_{1}$, $S_{7}S_{16}V_{3}V_{1}$, $S_{16}S_{8}S_{2}S_{5}$, $S_{7}S_{16}S_{2}S_{5}$, $S_{16}S_{9}V_{4}V_{2}$, $S_{16}S_{9}V_{3}V_{1}$, $S_{16}S_{9}S_{2}S_{5}$}} 
 \end{SplitTable}
   \vspace{0.4cm} 
 \captionof{table}{Dimension 9 $(\Delta B, \Delta L)$ = (1, -1) UV completions for topology 13 (see Fig.~\ref{fig:d=9_topologies}).} 
 \label{tab:T9Top13B1L-1}
 \end{center}

%% file: Tables_dim9/Table_dim9_Top14_B1_L-1.tex
 \begin{center}
 \begin{SplitTable}[2.0cm] 
  \SplitHeader{$\mathcal{O}^j{(\mathcal{T}_ {9} ^{ 14 })}$}{New Particles} 
   \vspace{0.1cm} 
 \SplitRow{ $\mathcal{O}_{9, \; (1,-1)}^{39}$} {{$F_{7}F_{8}S_{9}V_{2}$, $F_{5}F_{14}S_{9}S_{3}$, $F_{3}F_{10}S_{9}V_{2}$, $F_{7}F_{8}S_{9}S_{6}$, $F_{5}F_{14}S_{9}V_{4}$, $F_{3}F_{10}S_{9}S_{3}$, $F_{7}F_{8}S_{9}V_{4}$, $F_{7}F_{7}S_{8}V_{2}$, $F_{7}F_{7}S_{7}V_{2}$, $F_{5}F_{9}S_{8}S_{3}$, $F_{5}F_{5}S_{7}S_{3}$, $F_{3}F_{3}S_{8}V_{2}$, $F_{3}F_{3}S_{7}V_{2}$, $F_{7}F_{7}S_{8}S_{6}$, $F_{7}F_{7}S_{7}S_{6}$, $F_{5}F_{9}S_{8}V_{4}$, $F_{5}F_{5}S_{7}V_{4}$, $F_{3}F_{3}S_{8}S_{3}$, $F_{3}F_{3}S_{7}S_{3}$, $F_{7}F_{7}S_{8}V_{4}$, $F_{7}F_{7}S_{7}V_{4}$}} 
 \SplitRow{ $\mathcal{O}_{9, \; (1,-1)}^{40}$} {{$F_{8}F_{20}S_{9}S_{4}$, $F_{7}F_{8}S_{9}S_{4}$, $F_{7}F_{8}S_{9}S_{2}$, $F_{9}F_{14}S_{9}V_{2}$, $F_{6}F_{9}S_{9}V_{2}$, $F_{5}F_{14}S_{9}V_{2}$, $F_{4}F_{11}S_{9}S_{4}$, $F_{2}F_{4}S_{9}S_{4}$, $F_{1}F_{11}S_{9}S_{2}$, $F_{9}F_{14}S_{9}V_{5}$, $F_{6}F_{9}S_{9}V_{5}$, $F_{5}F_{14}S_{9}V_{4}$, $F_{4}F_{11}S_{9}V_{2}$, $F_{2}F_{4}S_{9}V_{2}$, $F_{1}F_{11}S_{9}V_{2}$, $F_{8}F_{20}S_{9}V_{5}$, $F_{7}F_{8}S_{9}V_{5}$, $F_{7}F_{8}S_{9}V_{4}$, $F_{9}F_{14}S_{9}S_{5}$, $F_{6}F_{9}S_{9}S_{5}$, $F_{5}F_{14}S_{9}S_{5}$, $F_{7}F_{20}S_{8}S_{4}$, $F_{7}F_{7}S_{8}S_{4}$, $F_{7}F_{7}S_{7}S_{4}$, $F_{7}F_{7}S_{8}S_{2}$, $F_{7}F_{7}S_{7}S_{2}$, $F_{9}F_{9}S_{8}V_{2}$, $F_{5}F_{9}S_{8}V_{2}$, $F_{9}F_{9}S_{7}V_{2}$, $F_{5}F_{5}S_{7}V_{2}$, $F_{4}F_{4}S_{8}S_{4}$, $F_{1}F_{4}S_{8}S_{4}$, $F_{4}F_{4}S_{7}S_{4}$, $F_{1}F_{4}S_{8}S_{2}$, $F_{1}F_{1}S_{7}S_{2}$, $F_{9}F_{9}S_{8}V_{5}$, $F_{5}F_{9}S_{8}V_{5}$, $F_{9}F_{9}S_{7}V_{5}$, $F_{5}F_{9}S_{8}V_{4}$, $F_{5}F_{5}S_{7}V_{4}$, $F_{4}F_{4}S_{8}V_{2}$, $F_{1}F_{4}S_{8}V_{2}$, $F_{4}F_{4}S_{7}V_{2}$, $F_{1}F_{1}S_{7}V_{2}$, $F_{7}F_{20}S_{8}V_{5}$, $F_{7}F_{7}S_{8}V_{5}$, $F_{7}F_{7}S_{7}V_{5}$, $F_{7}F_{7}S_{8}V_{4}$, $F_{7}F_{7}S_{7}V_{4}$, $F_{9}F_{9}S_{8}S_{5}$, $F_{5}F_{9}S_{8}S_{5}$, $F_{9}F_{9}S_{7}S_{5}$, $F_{5}F_{5}S_{7}S_{5}$}} 
 \SplitRow{ $\mathcal{O}_{9, \; (1,-1)}^{41}$} {{$F_{8}F_{13}S_{9}S_{3}$, $F_{7}F_{8}S_{9}S_{2}$, $F_{1}F_{11}S_{9}S_{3}$, $F_{7}F_{8}S_{9}S_{5}$, $F_{1}F_{11}S_{9}S_{2}$, $F_{8}F_{13}S_{9}S_{5}$, $F_{7}F_{8}S_{9}S_{6}$, $F_{8}F_{8}S_{8}S_{3}$, $F_{8}F_{8}S_{7}S_{3}$, $F_{7}F_{7}S_{8}S_{2}$, $F_{7}F_{7}S_{7}S_{2}$, $F_{1}F_{4}S_{8}S_{3}$, $F_{1}F_{1}S_{7}S_{3}$, $F_{7}F_{7}S_{8}S_{5}$, $F_{7}F_{7}S_{7}S_{5}$, $F_{1}F_{4}S_{8}S_{2}$, $F_{1}F_{1}S_{7}S_{2}$, $F_{8}F_{8}S_{8}S_{5}$, $F_{8}F_{8}S_{7}S_{5}$, $F_{7}F_{7}S_{8}S_{6}$, $F_{7}F_{7}S_{7}S_{6}$}} 
 \SplitRow{ $\mathcal{O}_{9, \; (1,-1)}^{42}$} {{$F_{8}F_{8}S_{8}S_{3}$, $F_{8}F_{8}S_{7}S_{3}$, $F_{2}F_{11}S_{8}S_{3}$, $F_{2}F_{2}S_{7}S_{3}$, $F_{8}F_{8}S_{8}S_{5}$, $F_{8}F_{8}S_{7}S_{5}$, $F_{7}F_{8}S_{9}S_{3}$, $F_{2}F_{4}S_{9}S_{3}$, $F_{7}F_{8}S_{9}S_{5}$}} 
 \SplitRow{ $\mathcal{O}_{9, \; (1,-1)}^{43}$} {{$F_{19}F_{9}S_{9}S_{6}$, $F_{5}F_{19}S_{9}S_{6}$, $F_{3}F_{12}S_{9}S_{4}$, $F_{19}F_{9}S_{9}S_{4}$, $F_{5}F_{19}S_{9}S_{4}$}} 
 \SplitRow{ $\mathcal{O}_{9, \; (1,-1)}^{44}$} {{$F_{19}F_{9}S_{9}S_{6}$, $F_{5}F_{19}S_{9}S_{6}$, $F_{8}F_{20}S_{9}V_{5}$, $F_{4}F_{11}S_{9}V_{1}$, $F_{1}F_{11}S_{9}V_{1}$, $F_{19}F_{9}S_{9}V_{5}$, $F_{5}F_{19}S_{9}V_{5}$, $F_{4}F_{11}S_{9}S_{4}$, $F_{1}F_{11}S_{9}S_{4}$, $F_{19}F_{9}S_{9}V_{1}$, $F_{5}F_{19}S_{9}V_{1}$, $F_{8}F_{20}S_{9}S_{4}$}} 
 \SplitRow{ $\mathcal{O}_{9, \; (1,-1)}^{45}$} {{$F_{8}F_{8}S_{8}V_{2}$, $F_{8}F_{8}S_{7}V_{2}$, $F_{6}F_{14}S_{8}S_{3}$, $F_{6}F_{6}S_{7}S_{3}$, $F_{3}F_{3}S_{8}V_{2}$, $F_{3}F_{3}S_{7}V_{2}$, $F_{8}F_{8}S_{8}S_{5}$, $F_{8}F_{8}S_{7}S_{5}$, $F_{6}F_{14}S_{8}V_{3}$, $F_{6}F_{6}S_{7}V_{3}$, $F_{3}F_{3}S_{8}S_{3}$, $F_{3}F_{3}S_{7}S_{3}$, $F_{8}F_{8}S_{8}V_{3}$, $F_{8}F_{8}S_{7}V_{3}$, $F_{7}F_{8}S_{9}V_{2}$, $F_{6}F_{9}S_{9}S_{3}$, $F_{3}F_{3}S_{9}V_{2}$, $F_{7}F_{8}S_{9}S_{5}$, $F_{6}F_{9}S_{9}V_{3}$, $F_{3}F_{3}S_{9}S_{3}$, $F_{7}F_{8}S_{9}V_{3}$}} 
 \SplitRow{ $\mathcal{O}_{9, \; (1,-1)}^{46}$} {{$F_{9}F_{14}S_{9}S_{4}$, $F_{6}F_{9}S_{9}S_{4}$, $F_{5}F_{14}S_{9}S_{2}$, $F_{3}F_{12}S_{9}S_{4}$, $F_{3}F_{3}S_{9}S_{4}$, $F_{3}F_{3}S_{9}S_{2}$, $F_{9}F_{14}S_{9}S_{5}$, $F_{6}F_{9}S_{9}S_{5}$, $F_{5}F_{14}S_{9}S_{5}$, $F_{9}F_{9}S_{8}S_{4}$, $F_{5}F_{9}S_{8}S_{4}$, $F_{9}F_{9}S_{7}S_{4}$, $F_{5}F_{9}S_{8}S_{2}$, $F_{5}F_{5}S_{7}S_{2}$, $F_{3}F_{12}S_{8}S_{4}$, $F_{3}F_{3}S_{8}S_{4}$, $F_{3}F_{3}S_{7}S_{4}$, $F_{3}F_{3}S_{8}S_{2}$, $F_{3}F_{3}S_{7}S_{2}$, $F_{9}F_{9}S_{8}S_{5}$, $F_{5}F_{9}S_{8}S_{5}$, $F_{9}F_{9}S_{7}S_{5}$, $F_{5}F_{5}S_{7}S_{5}$}} 
 \SplitRow{ $\mathcal{O}_{9, \; (1,-1)}^{47}$} {{$F_{8}F_{13}S_{9}V_{2}$, $F_{7}F_{8}S_{9}V_{1}$, $F_{5}F_{14}S_{9}S_{2}$, $F_{3}F_{3}S_{9}V_{2}$, $F_{7}F_{8}S_{9}S_{5}$, $F_{5}F_{14}S_{9}V_{3}$, $F_{3}F_{3}S_{9}V_{1}$, $F_{8}F_{13}S_{9}S_{5}$, $F_{5}F_{14}S_{9}V_{4}$, $F_{3}F_{3}S_{9}S_{2}$, $F_{8}F_{13}S_{9}V_{3}$, $F_{7}F_{8}S_{9}V_{4}$, $F_{8}F_{8}S_{8}V_{2}$, $F_{8}F_{8}S_{7}V_{2}$, $F_{7}F_{7}S_{8}V_{1}$, $F_{7}F_{7}S_{7}V_{1}$, $F_{5}F_{9}S_{8}S_{2}$, $F_{5}F_{5}S_{7}S_{2}$, $F_{3}F_{3}S_{8}V_{2}$, $F_{3}F_{3}S_{7}V_{2}$, $F_{7}F_{7}S_{8}S_{5}$, $F_{7}F_{7}S_{7}S_{5}$, $F_{5}F_{9}S_{8}V_{3}$, $F_{5}F_{5}S_{7}V_{3}$, $F_{3}F_{3}S_{8}V_{1}$, $F_{3}F_{3}S_{7}V_{1}$, $F_{8}F_{8}S_{8}S_{5}$, $F_{8}F_{8}S_{7}S_{5}$, $F_{5}F_{9}S_{8}V_{4}$, $F_{5}F_{5}S_{7}V_{4}$, $F_{3}F_{3}S_{8}S_{2}$, $F_{3}F_{3}S_{7}S_{2}$, $F_{8}F_{8}S_{8}V_{3}$, $F_{8}F_{8}S_{7}V_{3}$, $F_{7}F_{7}S_{8}V_{4}$, $F_{7}F_{7}S_{7}V_{4}$}} 
 \end{SplitTable}
   \vspace{0.4cm} 
 \captionof{table}{Dimension 9 $(\Delta B, \Delta L)$ = (1, -1) UV completions for topology 14 (see Fig.~\ref{fig:d=9_topologies}).} 
 \label{tab:T9Top14B1L-1}
 \end{center}

%% file: Tables_dim9/Table_dim9_Top15_B1_L-1.tex
 \begin{center}
 \begin{SplitTable}[2.0cm] 
  \SplitHeader{$\mathcal{O}^j{(\mathcal{T}_ {9} ^{ 15 })}$}{New Particles} 
   \vspace{0.1cm} 
 \SplitRow{ $\mathcal{O}_{9, \; (1,-1)}^{39}$} {{$F_{8}F_{9}V_{2}V_{5}$, $F_{5}F_{8}V_{4}V_{2}$, $F_{8}F_{14}S_{3}S_{6}$, $F_{10}F_{11}V_{2}V_{5}$, $F_{2}F_{10}V_{4}V_{2}$, $F_{8}F_{9}S_{6}S_{19}$, $F_{5}F_{8}S_{3}S_{6}$, $F_{8}F_{14}V_{4}V_{2}$, $F_{10}F_{11}S_{3}S_{6}$, $F_{2}F_{10}S_{3}S_{6}$, $F_{8}F_{9}V_{4}V_{2}$, $F_{7}F_{19}V_{2}V_{8}$, $F_{7}F_{9}S_{3}S_{5}$, $F_{5}F_{7}S_{3}S_{5}$, $F_{3}F_{4}V_{2}V_{8}$, $F_{7}F_{19}S_{6}S_{20}$, $F_{7}F_{9}V_{4}V_{15}$, $F_{5}F_{7}V_{4}V_{15}$, $F_{3}F_{4}S_{3}S_{5}$, $F_{7}F_{19}V_{4}V_{15}$, $F_{7}F_{9}V_{2}V_{5}$, $F_{5}F_{7}V_{4}V_{2}$, $F_{8}F_{9}S_{3}S_{6}$, $F_{3}F_{11}V_{2}V_{5}$, $F_{2}F_{3}V_{4}V_{2}$, $F_{7}F_{9}S_{6}S_{19}$, $F_{5}F_{7}S_{3}S_{6}$, $F_{3}F_{11}S_{3}S_{6}$, $F_{2}F_{3}S_{3}S_{6}$, $F_{7}F_{9}V_{4}V_{2}$}} 
 \SplitRow{ $\mathcal{O}_{9, \; (1,-1)}^{40}$} {{$F_{8}F_{9}S_{4}S_{13}$, $F_{8}F_{9}S_{5}S_{4}$, $F_{8}F_{9}S_{2}S_{5}$, $F_{5}F_{8}S_{5}S_{4}$, $F_{5}F_{8}S_{2}S_{5}$, $F_{14}F_{15}V_{2}V_{5}$, $F_{8}F_{14}V_{2}V_{5}$, $F_{8}F_{14}V_{4}V_{2}$, $F_{6}F_{8}V_{2}V_{5}$, $F_{6}F_{8}V_{4}V_{2}$, $F_{11}F_{12}S_{4}S_{13}$, $F_{3}F_{11}S_{5}S_{4}$, $F_{3}F_{11}S_{2}S_{5}$, $F_{2}F_{3}S_{5}S_{4}$, $F_{14}F_{15}V_{5}V_{16}$, $F_{11}F_{12}V_{2}V_{5}$, $F_{3}F_{11}V_{2}V_{5}$, $F_{3}F_{11}V_{4}V_{2}$, $F_{2}F_{3}V_{2}V_{5}$, $F_{2}F_{3}V_{4}V_{2}$, $F_{8}F_{9}V_{5}V_{16}$, $F_{8}F_{9}V_{2}V_{5}$, $F_{8}F_{9}V_{4}V_{2}$, $F_{5}F_{8}V_{2}V_{5}$, $F_{5}F_{8}V_{4}V_{2}$, $F_{14}F_{15}S_{5}S_{4}$, $F_{8}F_{14}S_{5}S_{4}$, $F_{8}F_{14}S_{2}S_{5}$, $F_{6}F_{8}S_{5}S_{4}$, $F_{6}F_{8}S_{2}S_{5}$, $F_{7}F_{19}S_{4}S_{12}$, $F_{7}F_{19}S_{11}S_{4}$, $F_{7}F_{19}S_{2}S_{11}$, $F_{9}F_{20}V_{2}V_{8}$, $F_{7}F_{9}V_{2}V_{8}$, $F_{7}F_{9}V_{3}V_{2}$, $F_{5}F_{7}V_{2}V_{8}$, $F_{5}F_{7}V_{3}V_{2}$, $F_{4}F_{12}S_{4}S_{12}$, $F_{3}F_{4}S_{11}S_{4}$, $F_{3}F_{4}S_{2}S_{11}$, $F_{1}F_{3}S_{11}S_{4}$, $F_{1}F_{3}S_{2}S_{11}$, $F_{9}F_{20}V_{5}V_{17}$, $F_{7}F_{9}V_{15}V_{5}$, $F_{7}F_{9}V_{4}V_{15}$, $F_{5}F_{7}V_{15}V_{5}$, $F_{5}F_{7}V_{4}V_{15}$, $F_{4}F_{12}V_{2}V_{8}$, $F_{3}F_{4}V_{2}V_{8}$, $F_{3}F_{4}V_{3}V_{2}$, $F_{1}F_{3}V_{2}V_{8}$, $F_{1}F_{3}V_{3}V_{2}$, $F_{7}F_{19}V_{5}V_{17}$, $F_{7}F_{19}V_{15}V_{5}$, $F_{7}F_{19}V_{4}V_{15}$, $F_{9}F_{20}S_{5}S_{19}$, $F_{7}F_{9}S_{5}S_{19}$, $F_{7}F_{9}S_{3}S_{5}$, $F_{5}F_{7}S_{5}S_{19}$, $F_{5}F_{7}S_{3}S_{5}$, $F_{7}F_{9}S_{4}S_{13}$, $F_{7}F_{9}S_{5}S_{4}$, $F_{7}F_{9}S_{2}S_{5}$, $F_{5}F_{7}S_{5}S_{4}$, $F_{5}F_{7}S_{2}S_{5}$, $F_{9}F_{15}V_{2}V_{5}$, $F_{4}F_{12}S_{4}S_{13}$, $F_{3}F_{4}S_{5}S_{4}$, $F_{3}F_{4}S_{2}S_{5}$, $F_{1}F_{3}S_{5}S_{4}$, $F_{1}F_{3}S_{2}S_{5}$, $F_{9}F_{15}V_{5}V_{16}$, $F_{4}F_{12}V_{2}V_{5}$, $F_{3}F_{4}V_{2}V_{5}$, $F_{3}F_{4}V_{4}V_{2}$, $F_{1}F_{3}V_{2}V_{5}$, $F_{1}F_{3}V_{4}V_{2}$, $F_{7}F_{9}V_{5}V_{16}$, $F_{7}F_{9}V_{2}V_{5}$, $F_{7}F_{9}V_{4}V_{2}$, $F_{5}F_{7}V_{2}V_{5}$, $F_{5}F_{7}V_{4}V_{2}$, $F_{9}F_{15}S_{5}S_{4}$}} 
 \SplitRow{ $\mathcal{O}_{9, \; (1,-1)}^{41}$} {{$F_{13}F_{14}S_{3}S_{6}$, $F_{6}F_{13}S_{3}S_{6}$, $F_{8}F_{9}S_{2}S_{5}$, $F_{5}F_{8}S_{2}S_{5}$, $F_{3}F_{11}S_{3}S_{6}$, $F_{8}F_{9}S_{5}S_{4}$, $F_{3}F_{11}S_{2}S_{5}$, $F_{13}F_{14}S_{5}S_{4}$, $F_{6}F_{13}S_{2}S_{5}$, $F_{8}F_{9}S_{6}S_{19}$, $F_{5}F_{8}S_{3}S_{6}$, $F_{8}F_{9}S_{3}S_{5}$, $F_{7}F_{19}S_{2}S_{11}$, $F_{3}F_{4}S_{3}S_{5}$, $F_{1}F_{3}S_{3}S_{5}$, $F_{7}F_{19}S_{5}S_{19}$, $F_{3}F_{4}S_{2}S_{11}$, $F_{1}F_{3}S_{2}S_{11}$, $F_{8}F_{9}S_{5}S_{19}$, $F_{7}F_{19}S_{6}S_{20}$, $F_{8}F_{14}S_{3}S_{6}$, $F_{6}F_{8}S_{3}S_{6}$, $F_{7}F_{9}S_{2}S_{5}$, $F_{5}F_{7}S_{2}S_{5}$, $F_{3}F_{4}S_{3}S_{6}$, $F_{1}F_{3}S_{3}S_{6}$, $F_{7}F_{9}S_{5}S_{4}$, $F_{3}F_{4}S_{2}S_{5}$, $F_{1}F_{3}S_{2}S_{5}$, $F_{8}F_{14}S_{5}S_{4}$, $F_{6}F_{8}S_{2}S_{5}$, $F_{7}F_{9}S_{6}S_{19}$, $F_{5}F_{7}S_{3}S_{6}$}} 
 \SplitRow{ $\mathcal{O}_{9, \; (1,-1)}^{42}$} {{$F_{8}F_{9}S_{3}S_{5}$, $F_{5}F_{8}S_{3}S_{5}$, $F_{3}F_{11}S_{3}S_{5}$, $F_{2}F_{3}S_{3}S_{5}$, $F_{8}F_{9}S_{5}S_{19}$, $F_{8}F_{14}S_{3}S_{6}$, $F_{10}F_{11}S_{3}S_{6}$, $F_{2}F_{10}S_{3}S_{6}$, $F_{8}F_{14}S_{5}S_{4}$, $F_{7}F_{9}S_{3}S_{5}$, $F_{5}F_{7}S_{3}S_{5}$, $F_{3}F_{4}S_{3}S_{5}$, $F_{7}F_{9}S_{5}S_{19}$}} 
 \SplitRow{ $\mathcal{O}_{9, \; (1,-1)}^{43}$} {{$F_{9}F_{20}S_{6}S_{19}$, $F_{7}F_{9}S_{6}S_{19}$, $F_{7}F_{9}S_{3}S_{6}$, $F_{5}F_{7}S_{6}S_{19}$, $F_{5}F_{7}S_{3}S_{6}$, $F_{3}F_{4}S_{4}S_{13}$, $F_{3}F_{4}S_{5}S_{4}$, $F_{9}F_{20}S_{4}S_{13}$, $F_{7}F_{9}S_{5}S_{4}$, $F_{5}F_{7}S_{5}S_{4}$}} 
 \SplitRow{ $\mathcal{O}_{9, \; (1,-1)}^{44}$} {{$F_{9}F_{20}S_{6}S_{19}$, $F_{7}F_{9}S_{6}S_{19}$, $F_{7}F_{9}S_{3}S_{6}$, $F_{5}F_{7}S_{6}S_{19}$, $F_{5}F_{7}S_{3}S_{6}$, $F_{8}F_{9}V_{5}V_{16}$, $F_{8}F_{9}V_{2}V_{5}$, $F_{4}F_{12}V_{1}V_{8}$, $F_{3}F_{4}V_{1}V_{8}$, $F_{3}F_{4}V_{3}V_{1}$, $F_{1}F_{3}V_{1}V_{8}$, $F_{1}F_{3}V_{3}V_{1}$, $F_{9}F_{20}V_{5}V_{16}$, $F_{7}F_{9}V_{2}V_{5}$, $F_{5}F_{7}V_{2}V_{5}$, $F_{4}F_{12}S_{4}S_{13}$, $F_{3}F_{4}S_{5}S_{4}$, $F_{1}F_{3}S_{5}S_{4}$, $F_{9}F_{20}V_{1}V_{8}$, $F_{7}F_{9}V_{1}V_{8}$, $F_{7}F_{9}V_{3}V_{1}$, $F_{5}F_{7}V_{1}V_{8}$, $F_{5}F_{7}V_{3}V_{1}$, $F_{8}F_{9}S_{4}S_{13}$, $F_{8}F_{9}S_{5}S_{4}$}} 
 \SplitRow{ $\mathcal{O}_{9, \; (1,-1)}^{45}$} {{$F_{8}F_{9}V_{2}V_{8}$, $F_{5}F_{8}V_{3}V_{2}$, $F_{8}F_{14}S_{3}S_{5}$, $F_{6}F_{8}S_{3}S_{5}$, $F_{3}F_{4}V_{2}V_{8}$, $F_{1}F_{3}V_{3}V_{2}$, $F_{8}F_{9}S_{5}S_{19}$, $F_{5}F_{8}S_{3}S_{5}$, $F_{8}F_{14}V_{3}V_{2}$, $F_{6}F_{8}V_{3}V_{2}$, $F_{3}F_{4}S_{3}S_{5}$, $F_{1}F_{3}S_{3}S_{5}$, $F_{8}F_{9}V_{3}V_{2}$, $F_{8}F_{14}V_{2}V_{5}$, $F_{13}F_{14}S_{3}S_{6}$, $F_{6}F_{13}S_{3}S_{6}$, $F_{3}F_{11}V_{2}V_{5}$, $F_{8}F_{14}S_{5}S_{4}$, $F_{13}F_{14}V_{3}V_{1}$, $F_{6}F_{13}V_{3}V_{1}$, $F_{3}F_{11}S_{3}S_{6}$, $F_{8}F_{14}V_{3}V_{1}$, $F_{7}F_{9}V_{2}V_{8}$, $F_{5}F_{7}V_{3}V_{2}$, $F_{8}F_{9}S_{3}S_{5}$, $F_{7}F_{9}S_{5}S_{19}$, $F_{5}F_{7}S_{3}S_{5}$, $F_{7}F_{9}V_{3}V_{2}$}} 
 \SplitRow{ $\mathcal{O}_{9, \; (1,-1)}^{46}$} {{$F_{14}F_{15}S_{4}S_{13}$, $F_{8}F_{14}S_{5}S_{4}$, $F_{8}F_{14}S_{2}S_{5}$, $F_{6}F_{8}S_{5}S_{4}$, $F_{3}F_{4}S_{4}S_{13}$, $F_{3}F_{4}S_{5}S_{4}$, $F_{3}F_{4}S_{2}S_{5}$, $F_{1}F_{3}S_{5}S_{4}$, $F_{1}F_{3}S_{2}S_{5}$, $F_{14}F_{15}S_{5}S_{4}$, $F_{6}F_{8}S_{2}S_{5}$, $F_{9}F_{20}S_{4}S_{12}$, $F_{7}F_{9}S_{11}S_{4}$, $F_{7}F_{9}S_{2}S_{11}$, $F_{5}F_{7}S_{11}S_{4}$, $F_{5}F_{7}S_{2}S_{11}$, $F_{3}F_{11}S_{4}S_{12}$, $F_{3}F_{11}S_{11}S_{4}$, $F_{3}F_{11}S_{2}S_{11}$, $F_{9}F_{20}S_{5}S_{19}$, $F_{7}F_{9}S_{5}S_{19}$, $F_{7}F_{9}S_{3}S_{5}$, $F_{5}F_{7}S_{5}S_{19}$, $F_{5}F_{7}S_{3}S_{5}$, $F_{9}F_{15}S_{4}S_{13}$, $F_{8}F_{9}S_{5}S_{4}$, $F_{8}F_{9}S_{2}S_{5}$, $F_{5}F_{8}S_{5}S_{4}$, $F_{5}F_{8}S_{2}S_{5}$, $F_{9}F_{15}S_{5}S_{4}$}} 
 \SplitRow{ $\mathcal{O}_{9, \; (1,-1)}^{47}$} {{$F_{13}F_{14}V_{2}V_{5}$, $F_{6}F_{13}V_{4}V_{2}$, $F_{8}F_{9}V_{1}V_{8}$, $F_{5}F_{8}V_{3}V_{1}$, $F_{8}F_{14}S_{2}S_{5}$, $F_{3}F_{4}V_{2}V_{5}$, $F_{1}F_{3}V_{4}V_{2}$, $F_{8}F_{9}S_{5}S_{4}$, $F_{5}F_{8}S_{2}S_{5}$, $F_{8}F_{14}V_{3}V_{1}$, $F_{3}F_{4}V_{1}V_{8}$, $F_{1}F_{3}V_{3}V_{1}$, $F_{13}F_{14}S_{5}S_{4}$, $F_{6}F_{13}S_{2}S_{5}$, $F_{8}F_{14}V_{4}V_{2}$, $F_{3}F_{4}S_{2}S_{5}$, $F_{1}F_{3}S_{2}S_{5}$, $F_{13}F_{14}V_{3}V_{1}$, $F_{6}F_{13}V_{3}V_{1}$, $F_{8}F_{9}V_{4}V_{2}$, $F_{5}F_{8}V_{4}V_{2}$, $F_{8}F_{9}V_{2}V_{8}$, $F_{7}F_{19}V_{1}V_{7}$, $F_{7}F_{9}S_{2}S_{11}$, $F_{5}F_{7}S_{2}S_{11}$, $F_{3}F_{11}V_{2}V_{8}$, $F_{7}F_{19}S_{5}S_{19}$, $F_{7}F_{9}V_{3}V_{2}$, $F_{5}F_{7}V_{3}V_{2}$, $F_{3}F_{11}V_{1}V_{7}$, $F_{8}F_{9}S_{5}S_{19}$, $F_{7}F_{9}V_{4}V_{15}$, $F_{5}F_{7}V_{4}V_{15}$, $F_{3}F_{11}S_{2}S_{11}$, $F_{8}F_{9}V_{3}V_{2}$, $F_{7}F_{19}V_{4}V_{15}$, $F_{8}F_{14}V_{2}V_{5}$, $F_{6}F_{8}V_{4}V_{2}$, $F_{7}F_{9}V_{1}V_{8}$, $F_{5}F_{7}V_{3}V_{1}$, $F_{8}F_{9}S_{2}S_{5}$, $F_{7}F_{9}S_{5}S_{4}$, $F_{5}F_{7}S_{2}S_{5}$, $F_{8}F_{9}V_{3}V_{1}$, $F_{8}F_{14}S_{5}S_{4}$, $F_{6}F_{8}S_{2}S_{5}$, $F_{6}F_{8}V_{3}V_{1}$, $F_{7}F_{9}V_{4}V_{2}$, $F_{5}F_{7}V_{4}V_{2}$}} 
 \end{SplitTable}
   \vspace{0.4cm} 
 \captionof{table}{Dimension 9 $(\Delta B, \Delta L)$ = (1, -1) UV completions for topology 15 (see Fig.~\ref{fig:d=9_topologies}).} 
 \label{tab:T9Top15B1L-1}
 \end{center}

%% file: Tables_dim9/Table_dim9_Top16_B1_L-1.tex
 \begin{center}
 \begin{SplitTable}[2.0cm] 
  \SplitHeader{$\mathcal{O}^j{(\mathcal{T}_ {9} ^{ 16 })}$}{New Particles} 
   \vspace{0.1cm} 
 \SplitRow{ $\mathcal{O}_{9, \; (1,-1)}^{39}$} {{$F_{14}S_{9}S_{3}S_{20}$, $F_{8}S_{9}V_{2}V_{15}$, $F_{14}S_{9}V_{4}V_{8}$, $F_{8}S_{9}S_{5}S_{6}$, $F_{10}S_{9}V_{2}V_{15}$, $F_{8}S_{9}V_{4}V_{8}$, $F_{10}S_{9}S_{3}S_{20}$, $F_{9}S_{8}S_{3}S_{19}$, $F_{5}S_{7}S_{3}S_{3}$, $F_{7}S_{8}V_{2}V_{2}$, $F_{7}S_{7}V_{2}V_{2}$, $F_{9}S_{8}V_{4}V_{5}$, $F_{5}S_{7}V_{4}V_{4}$, $F_{7}S_{8}S_{6}S_{6}$, $F_{7}S_{7}S_{6}S_{6}$, $F_{3}S_{8}V_{2}V_{2}$, $F_{3}S_{7}V_{2}V_{2}$, $F_{7}S_{8}V_{4}V_{5}$, $F_{7}S_{7}V_{4}V_{4}$, $F_{3}S_{8}S_{3}S_{19}$, $F_{3}S_{7}S_{3}S_{3}$}} 
 \SplitRow{ $\mathcal{O}_{9, \; (1,-1)}^{40}$} {{$F_{14}S_{9}V_{2}V_{17}$, $F_{14}S_{9}V_{2}V_{15}$, $F_{6}S_{9}V_{2}V_{15}$, $F_{8}S_{9}S_{4}S_{19}$, $F_{8}S_{9}S_{2}S_{19}$, $F_{8}S_{9}S_{3}S_{4}$, $F_{14}S_{9}V_{8}V_{5}$, $F_{14}S_{9}V_{4}V_{8}$, $F_{6}S_{9}V_{3}V_{5}$, $F_{11}S_{9}S_{4}S_{19}$, $F_{11}S_{9}S_{2}S_{19}$, $F_{2}S_{9}S_{3}S_{4}$, $F_{14}S_{9}S_{5}S_{12}$, $F_{14}S_{9}S_{11}S_{5}$, $F_{6}S_{9}S_{11}S_{5}$, $F_{8}S_{9}V_{8}V_{5}$, $F_{8}S_{9}V_{4}V_{8}$, $F_{8}S_{9}V_{3}V_{5}$, $F_{11}S_{9}V_{2}V_{17}$, $F_{11}S_{9}V_{2}V_{15}$, $F_{2}S_{9}V_{2}V_{15}$, $F_{9}S_{8}V_{2}V_{16}$, $F_{9}S_{8}V_{2}V_{2}$, $F_{9}S_{7}V_{2}V_{2}$, $F_{5}S_{8}V_{2}V_{2}$, $F_{5}S_{7}V_{2}V_{2}$, $F_{7}S_{8}S_{4}S_{4}$, $F_{7}S_{8}S_{2}S_{4}$, $F_{7}S_{7}S_{4}S_{4}$, $F_{7}S_{7}S_{2}S_{2}$, $F_{9}S_{8}V_{5}V_{5}$, $F_{9}S_{8}V_{4}V_{5}$, $F_{9}S_{7}V_{5}V_{5}$, $F_{5}S_{8}V_{4}V_{5}$, $F_{5}S_{7}V_{4}V_{4}$, $F_{4}S_{8}S_{4}S_{4}$, $F_{4}S_{8}S_{2}S_{4}$, $F_{4}S_{7}S_{4}S_{4}$, $F_{1}S_{8}S_{2}S_{4}$, $F_{1}S_{7}S_{2}S_{2}$, $F_{9}S_{8}S_{5}S_{13}$, $F_{9}S_{8}S_{5}S_{5}$, $F_{9}S_{7}S_{5}S_{5}$, $F_{5}S_{8}S_{5}S_{5}$, $F_{5}S_{7}S_{5}S_{5}$, $F_{7}S_{8}V_{5}V_{5}$, $F_{7}S_{8}V_{4}V_{5}$, $F_{7}S_{7}V_{5}V_{5}$, $F_{7}S_{7}V_{4}V_{4}$, $F_{4}S_{8}V_{2}V_{16}$, $F_{4}S_{8}V_{2}V_{2}$, $F_{4}S_{7}V_{2}V_{2}$, $F_{1}S_{8}V_{2}V_{2}$, $F_{1}S_{7}V_{2}V_{2}$}} 
 \SplitRow{ $\mathcal{O}_{9, \; (1,-1)}^{41}$} {{$F_{8}S_{9}S_{2}S_{19}$, $F_{13}S_{9}S_{3}S_{20}$, $F_{8}S_{9}S_{11}S_{5}$, $F_{11}S_{9}S_{3}S_{20}$, $F_{8}S_{9}S_{5}S_{6}$, $F_{13}S_{9}S_{11}S_{5}$, $F_{11}S_{9}S_{2}S_{19}$, $F_{7}S_{8}S_{2}S_{4}$, $F_{7}S_{7}S_{2}S_{2}$, $F_{8}S_{8}S_{3}S_{19}$, $F_{8}S_{7}S_{3}S_{3}$, $F_{7}S_{8}S_{5}S_{5}$, $F_{7}S_{7}S_{5}S_{5}$, $F_{4}S_{8}S_{3}S_{19}$, $F_{1}S_{7}S_{3}S_{3}$, $F_{7}S_{8}S_{6}S_{6}$, $F_{7}S_{7}S_{6}S_{6}$, $F_{8}S_{8}S_{5}S_{5}$, $F_{8}S_{7}S_{5}S_{5}$, $F_{4}S_{8}S_{2}S_{4}$, $F_{1}S_{7}S_{2}S_{2}$}} 
 \SplitRow{ $\mathcal{O}_{9, \; (1,-1)}^{42}$} {{$F_{8}S_{8}S_{3}S_{19}$, $F_{8}S_{7}S_{3}S_{3}$, $F_{8}S_{8}S_{5}S_{5}$, $F_{8}S_{7}S_{5}S_{5}$, $F_{11}S_{8}S_{3}S_{19}$, $F_{2}S_{7}S_{3}S_{3}$, $F_{7}S_{9}S_{3}S_{4}$, $F_{7}S_{9}S_{5}S_{6}$, $F_{4}S_{9}S_{3}S_{4}$}} 
 \SplitRow{ $\mathcal{O}_{9, \; (1,-1)}^{43}$} {{$F_{3}S_{9}S_{4}S_{19}$, $F_{3}S_{9}S_{3}S_{4}$, $F_{9}S_{9}S_{6}S_{13}$, $F_{9}S_{9}S_{5}S_{6}$, $F_{5}S_{9}S_{5}S_{6}$, $F_{9}S_{9}S_{4}S_{19}$, $F_{5}S_{9}S_{3}S_{4}$}} 
 \SplitRow{ $\mathcal{O}_{9, \; (1,-1)}^{44}$} {{$F_{4}S_{9}V_{1}V_{16}$, $F_{4}S_{9}V_{1}V_{2}$, $F_{1}S_{9}V_{1}V_{2}$, $F_{8}S_{9}V_{8}V_{5}$, $F_{8}S_{9}V_{3}V_{5}$, $F_{9}S_{9}S_{6}S_{13}$, $F_{9}S_{9}S_{5}S_{6}$, $F_{5}S_{9}S_{5}S_{6}$, $F_{4}S_{9}S_{4}S_{19}$, $F_{1}S_{9}S_{3}S_{4}$, $F_{9}S_{9}V_{8}V_{5}$, $F_{5}S_{9}V_{3}V_{5}$, $F_{8}S_{9}S_{4}S_{19}$, $F_{8}S_{9}S_{3}S_{4}$, $F_{9}S_{9}V_{1}V_{16}$, $F_{9}S_{9}V_{1}V_{2}$, $F_{5}S_{9}V_{1}V_{2}$}} 
 \SplitRow{ $\mathcal{O}_{9, \; (1,-1)}^{45}$} {{$F_{14}S_{8}S_{3}S_{19}$, $F_{6}S_{7}S_{3}S_{3}$, $F_{8}S_{8}V_{2}V_{2}$, $F_{8}S_{7}V_{2}V_{2}$, $F_{14}S_{8}V_{3}V_{8}$, $F_{6}S_{7}V_{3}V_{3}$, $F_{8}S_{8}S_{5}S_{5}$, $F_{8}S_{7}S_{5}S_{5}$, $F_{3}S_{8}V_{2}V_{2}$, $F_{3}S_{7}V_{2}V_{2}$, $F_{8}S_{8}V_{3}V_{8}$, $F_{8}S_{7}V_{3}V_{3}$, $F_{3}S_{8}S_{3}S_{19}$, $F_{3}S_{7}S_{3}S_{3}$, $F_{9}S_{9}S_{3}S_{4}$, $F_{7}S_{9}V_{1}V_{2}$, $F_{9}S_{9}V_{3}V_{5}$, $F_{7}S_{9}S_{5}S_{6}$, $F_{3}S_{9}V_{1}V_{2}$, $F_{7}S_{9}V_{3}V_{5}$, $F_{3}S_{9}S_{3}S_{4}$}} 
 \SplitRow{ $\mathcal{O}_{9, \; (1,-1)}^{46}$} {{$F_{14}S_{9}S_{4}S_{19}$, $F_{14}S_{9}S_{2}S_{19}$, $F_{6}S_{9}S_{3}S_{4}$, $F_{14}S_{9}S_{5}S_{12}$, $F_{14}S_{9}S_{11}S_{5}$, $F_{6}S_{9}S_{11}S_{5}$, $F_{3}S_{9}S_{4}S_{19}$, $F_{3}S_{9}S_{2}S_{19}$, $F_{3}S_{9}S_{3}S_{4}$, $F_{9}S_{8}S_{4}S_{4}$, $F_{9}S_{8}S_{2}S_{4}$, $F_{9}S_{7}S_{4}S_{4}$, $F_{5}S_{8}S_{2}S_{4}$, $F_{5}S_{7}S_{2}S_{2}$, $F_{9}S_{8}S_{5}S_{13}$, $F_{9}S_{8}S_{5}S_{5}$, $F_{9}S_{7}S_{5}S_{5}$, $F_{5}S_{8}S_{5}S_{5}$, $F_{5}S_{7}S_{5}S_{5}$, $F_{3}S_{8}S_{4}S_{4}$, $F_{3}S_{8}S_{2}S_{4}$, $F_{3}S_{7}S_{4}S_{4}$, $F_{3}S_{7}S_{2}S_{2}$}} 
 \SplitRow{ $\mathcal{O}_{9, \; (1,-1)}^{47}$} {{$F_{14}S_{9}S_{2}S_{19}$, $F_{8}S_{9}V_{1}V_{2}$, $F_{13}S_{9}V_{2}V_{15}$, $F_{14}S_{9}V_{3}V_{7}$, $F_{8}S_{9}S_{11}S_{5}$, $F_{3}S_{9}V_{2}V_{15}$, $F_{14}S_{9}V_{4}V_{8}$, $F_{13}S_{9}S_{11}S_{5}$, $F_{3}S_{9}V_{1}V_{2}$, $F_{8}S_{9}V_{4}V_{8}$, $F_{13}S_{9}V_{3}V_{7}$, $F_{3}S_{9}S_{2}S_{19}$, $F_{9}S_{8}S_{2}S_{4}$, $F_{5}S_{7}S_{2}S_{2}$, $F_{7}S_{8}V_{1}V_{1}$, $F_{7}S_{7}V_{1}V_{1}$, $F_{8}S_{8}V_{2}V_{2}$, $F_{8}S_{7}V_{2}V_{2}$, $F_{9}S_{8}V_{3}V_{8}$, $F_{5}S_{7}V_{3}V_{3}$, $F_{7}S_{8}S_{5}S_{5}$, $F_{7}S_{7}S_{5}S_{5}$, $F_{3}S_{8}V_{2}V_{2}$, $F_{3}S_{7}V_{2}V_{2}$, $F_{9}S_{8}V_{4}V_{5}$, $F_{5}S_{7}V_{4}V_{4}$, $F_{8}S_{8}S_{5}S_{5}$, $F_{8}S_{7}S_{5}S_{5}$, $F_{3}S_{8}V_{1}V_{1}$, $F_{3}S_{7}V_{1}V_{1}$, $F_{7}S_{8}V_{4}V_{5}$, $F_{7}S_{7}V_{4}V_{4}$, $F_{8}S_{8}V_{3}V_{8}$, $F_{8}S_{7}V_{3}V_{3}$, $F_{3}S_{8}S_{2}S_{4}$, $F_{3}S_{7}S_{2}S_{2}$}} 
 \end{SplitTable}
   \vspace{0.4cm} 
 \captionof{table}{Dimension 9 $(\Delta B, \Delta L)$ = (1, -1) UV completions for topology 16 (see Fig.~\ref{fig:d=9_topologies}).} 
 \label{tab:T9Top16B1L-1}
 \end{center}

%% file: Tables_dim9/Table_dim9_Top17_B1_L-1.tex
 \begin{center}
 \begin{SplitTable}[2.0cm] 
  \SplitHeader{$\mathcal{O}^j{(\mathcal{T}_ {9} ^{ 17 })}$}{New Particles} 
   \vspace{0.1cm} 
 \SplitRow{ $\mathcal{O}_{9, \; (1,-1)}^{39}$} {{$F_{7}V_{2}V_{2}$, $F_{9}S_{3}S_{19}$, $F_{5}S_{3}S_{3}$, $F_{3}V_{2}V_{2}$, $F_{7}S_{6}S_{6}$, $F_{9}V_{4}V_{5}$, $F_{5}V_{4}V_{4}$, $F_{3}S_{3}S_{19}$, $F_{3}S_{3}S_{3}$, $F_{7}V_{4}V_{5}$, $F_{7}V_{4}V_{4}$, $F_{8}V_{2}V_{15}$, $F_{14}S_{3}S_{20}$, $F_{10}V_{2}V_{15}$, $F_{8}S_{5}S_{6}$, $F_{14}V_{4}V_{8}$, $F_{10}S_{3}S_{20}$, $F_{8}V_{4}V_{8}$}} 
 \SplitRow{ $\mathcal{O}_{9, \; (1,-1)}^{40}$} {{$F_{7}S_{4}S_{4}$, $F_{7}S_{2}S_{4}$, $F_{7}S_{2}S_{2}$, $F_{9}V_{2}V_{16}$, $F_{9}V_{2}V_{2}$, $F_{5}V_{2}V_{2}$, $F_{4}S_{4}S_{4}$, $F_{4}S_{2}S_{4}$, $F_{1}S_{2}S_{4}$, $F_{1}S_{2}S_{2}$, $F_{9}V_{5}V_{5}$, $F_{9}V_{4}V_{5}$, $F_{5}V_{4}V_{5}$, $F_{5}V_{4}V_{4}$, $F_{4}V_{2}V_{16}$, $F_{4}V_{2}V_{2}$, $F_{1}V_{2}V_{2}$, $F_{7}V_{5}V_{5}$, $F_{7}V_{4}V_{5}$, $F_{7}V_{4}V_{4}$, $F_{9}S_{5}S_{13}$, $F_{9}S_{5}S_{5}$, $F_{5}S_{5}S_{5}$, $F_{8}S_{4}S_{19}$, $F_{8}S_{2}S_{19}$, $F_{8}S_{3}S_{4}$, $F_{14}V_{2}V_{17}$, $F_{14}V_{2}V_{15}$, $F_{6}V_{2}V_{15}$, $F_{11}S_{4}S_{19}$, $F_{11}S_{2}S_{19}$, $F_{2}S_{3}S_{4}$, $F_{14}V_{8}V_{5}$, $F_{14}V_{4}V_{8}$, $F_{6}V_{3}V_{5}$, $F_{11}V_{2}V_{17}$, $F_{11}V_{2}V_{15}$, $F_{2}V_{2}V_{15}$, $F_{8}V_{8}V_{5}$, $F_{8}V_{4}V_{8}$, $F_{8}V_{3}V_{5}$, $F_{14}S_{5}S_{12}$, $F_{14}S_{11}S_{5}$, $F_{6}S_{11}S_{5}$}} 
 \SplitRow{ $\mathcal{O}_{9, \; (1,-1)}^{41}$} {{$F_{8}S_{3}S_{19}$, $F_{8}S_{3}S_{3}$, $F_{7}S_{2}S_{4}$, $F_{7}S_{2}S_{2}$, $F_{4}S_{3}S_{19}$, $F_{1}S_{3}S_{3}$, $F_{7}S_{5}S_{5}$, $F_{4}S_{2}S_{4}$, $F_{1}S_{2}S_{2}$, $F_{8}S_{5}S_{5}$, $F_{7}S_{6}S_{6}$, $F_{13}S_{3}S_{20}$, $F_{8}S_{2}S_{19}$, $F_{11}S_{3}S_{20}$, $F_{8}S_{11}S_{5}$, $F_{11}S_{2}S_{19}$, $F_{13}S_{11}S_{5}$, $F_{8}S_{5}S_{6}$}} 
 \SplitRow{ $\mathcal{O}_{9, \; (1,-1)}^{42}$} {{$F_{7}S_{3}S_{4}$, $F_{4}S_{3}S_{4}$, $F_{7}S_{5}S_{6}$, $F_{8}S_{3}S_{19}$, $F_{8}S_{3}S_{3}$, $F_{11}S_{3}S_{19}$, $F_{2}S_{3}S_{3}$, $F_{8}S_{5}S_{5}$}} 
 \SplitRow{ $\mathcal{O}_{9, \; (1,-1)}^{43}$} {{$F_{9}S_{6}S_{13}$, $F_{9}S_{5}S_{6}$, $F_{5}S_{5}S_{6}$, $F_{3}S_{4}S_{19}$, $F_{3}S_{3}S_{4}$, $F_{9}S_{4}S_{19}$, $F_{5}S_{3}S_{4}$}} 
 \SplitRow{ $\mathcal{O}_{9, \; (1,-1)}^{44}$} {{$F_{9}S_{6}S_{13}$, $F_{9}S_{5}S_{6}$, $F_{5}S_{5}S_{6}$, $F_{8}V_{8}V_{5}$, $F_{8}V_{3}V_{5}$, $F_{4}V_{1}V_{16}$, $F_{4}V_{1}V_{2}$, $F_{1}V_{1}V_{2}$, $F_{9}V_{8}V_{5}$, $F_{5}V_{3}V_{5}$, $F_{4}S_{4}S_{19}$, $F_{1}S_{3}S_{4}$, $F_{9}V_{1}V_{16}$, $F_{9}V_{1}V_{2}$, $F_{5}V_{1}V_{2}$, $F_{8}S_{4}S_{19}$, $F_{8}S_{3}S_{4}$}} 
 \SplitRow{ $\mathcal{O}_{9, \; (1,-1)}^{45}$} {{$F_{7}V_{1}V_{2}$, $F_{9}S_{3}S_{4}$, $F_{3}V_{1}V_{2}$, $F_{7}S_{5}S_{6}$, $F_{9}V_{3}V_{5}$, $F_{3}S_{3}S_{4}$, $F_{7}V_{3}V_{5}$, $F_{8}V_{2}V_{2}$, $F_{14}S_{3}S_{19}$, $F_{6}S_{3}S_{3}$, $F_{3}V_{2}V_{2}$, $F_{8}S_{5}S_{5}$, $F_{14}V_{3}V_{8}$, $F_{6}V_{3}V_{3}$, $F_{3}S_{3}S_{19}$, $F_{3}S_{3}S_{3}$, $F_{8}V_{3}V_{8}$, $F_{8}V_{3}V_{3}$}} 
 \SplitRow{ $\mathcal{O}_{9, \; (1,-1)}^{46}$} {{$F_{9}S_{4}S_{4}$, $F_{9}S_{2}S_{4}$, $F_{5}S_{2}S_{4}$, $F_{5}S_{2}S_{2}$, $F_{3}S_{4}S_{4}$, $F_{3}S_{2}S_{4}$, $F_{3}S_{2}S_{2}$, $F_{9}S_{5}S_{13}$, $F_{9}S_{5}S_{5}$, $F_{5}S_{5}S_{5}$, $F_{14}S_{4}S_{19}$, $F_{14}S_{2}S_{19}$, $F_{6}S_{3}S_{4}$, $F_{3}S_{4}S_{19}$, $F_{3}S_{2}S_{19}$, $F_{3}S_{3}S_{4}$, $F_{14}S_{5}S_{12}$, $F_{14}S_{11}S_{5}$, $F_{6}S_{11}S_{5}$}} 
 \SplitRow{ $\mathcal{O}_{9, \; (1,-1)}^{47}$} {{$F_{8}V_{2}V_{2}$, $F_{7}V_{1}V_{1}$, $F_{9}S_{2}S_{4}$, $F_{5}S_{2}S_{2}$, $F_{3}V_{2}V_{2}$, $F_{7}S_{5}S_{5}$, $F_{9}V_{3}V_{8}$, $F_{5}V_{3}V_{3}$, $F_{3}V_{1}V_{1}$, $F_{8}S_{5}S_{5}$, $F_{9}V_{4}V_{5}$, $F_{5}V_{4}V_{4}$, $F_{3}S_{2}S_{4}$, $F_{3}S_{2}S_{2}$, $F_{8}V_{3}V_{8}$, $F_{8}V_{3}V_{3}$, $F_{7}V_{4}V_{5}$, $F_{7}V_{4}V_{4}$, $F_{13}V_{2}V_{15}$, $F_{8}V_{1}V_{2}$, $F_{14}S_{2}S_{19}$, $F_{3}V_{2}V_{15}$, $F_{8}S_{11}S_{5}$, $F_{14}V_{3}V_{7}$, $F_{3}V_{1}V_{2}$, $F_{13}S_{11}S_{5}$, $F_{14}V_{4}V_{8}$, $F_{3}S_{2}S_{19}$, $F_{13}V_{3}V_{7}$, $F_{8}V_{4}V_{8}$}} 
 \end{SplitTable}
   \vspace{0.4cm} 
 \captionof{table}{Dimension 9 $(\Delta B, \Delta L)$ = (1, -1) UV completions for topology 17 (see Fig.~\ref{fig:d=9_topologies}).} 
 \label{tab:T9Top17B1L-1}
 \end{center}

%% file: Tables_dim9/Table_dim9_Top18_B1_L-1.tex
 \begin{center}
 \begin{SplitTable}[2.0cm] 
  \SplitHeader{$\mathcal{O}^j{(\mathcal{T}_ {9} ^{ 18 })}$}{New Particles} 
   \vspace{0.1cm} 
 \SplitRow{ $\mathcal{O}_{9, \; (1,-1)}^{39}$} {{$F_{3}S_{16}S_{8}V_{2}$, $F_{3}S_{7}S_{16}V_{2}$, $F_{3}S_{16}S_{8}S_{3}$, $F_{3}S_{7}S_{16}S_{3}$, $F_{7}S_{16}S_{8}V_{2}$, $F_{7}S_{7}S_{16}V_{2}$, $F_{7}S_{16}S_{8}S_{6}$, $F_{7}S_{7}S_{16}S_{6}$, $F_{7}S_{16}S_{8}V_{4}$, $F_{7}S_{7}S_{16}V_{4}$, $F_{5}S_{16}S_{8}S_{3}$, $F_{5}S_{7}S_{16}S_{3}$, $F_{5}S_{16}S_{8}V_{4}$, $F_{5}S_{7}S_{16}V_{4}$, $F_{3}S_{16}S_{9}V_{2}$, $F_{3}S_{16}S_{9}S_{3}$, $F_{7}S_{16}S_{9}V_{2}$, $F_{7}S_{16}S_{9}S_{6}$, $F_{7}S_{16}S_{9}V_{4}$, $F_{5}S_{16}S_{9}S_{3}$, $F_{5}S_{16}S_{9}V_{4}$}} 
 \SplitRow{ $\mathcal{O}_{9, \; (1,-1)}^{40}$} {{$F_{4}S_{8}S_{17}S_{4}$, $F_{4}S_{16}S_{8}S_{4}$, $F_{1}S_{16}S_{8}S_{2}$, $F_{4}S_{7}S_{16}S_{4}$, $F_{1}S_{7}S_{16}S_{2}$, $F_{4}S_{8}S_{17}V_{2}$, $F_{4}S_{16}S_{8}V_{2}$, $F_{1}S_{16}S_{8}V_{2}$, $F_{4}S_{7}S_{16}V_{2}$, $F_{1}S_{7}S_{16}V_{2}$, $F_{20}S_{8}S_{17}S_{4}$, $F_{7}S_{16}S_{8}S_{4}$, $F_{7}S_{16}S_{8}S_{2}$, $F_{7}S_{7}S_{16}S_{4}$, $F_{7}S_{7}S_{16}S_{2}$, $F_{20}S_{8}S_{17}V_{5}$, $F_{7}S_{16}S_{8}V_{5}$, $F_{7}S_{16}S_{8}V_{4}$, $F_{7}S_{7}S_{16}V_{5}$, $F_{7}S_{7}S_{16}V_{4}$, $F_{9}S_{8}S_{17}V_{2}$, $F_{9}S_{16}S_{8}V_{2}$, $F_{5}S_{16}S_{8}V_{2}$, $F_{9}S_{7}S_{16}V_{2}$, $F_{5}S_{7}S_{16}V_{2}$, $F_{9}S_{8}S_{17}V_{5}$, $F_{9}S_{16}S_{8}V_{5}$, $F_{5}S_{16}S_{8}V_{4}$, $F_{9}S_{7}S_{16}V_{5}$, $F_{5}S_{7}S_{16}V_{4}$, $F_{9}S_{8}S_{17}S_{5}$, $F_{9}S_{16}S_{8}S_{5}$, $F_{5}S_{16}S_{8}S_{5}$, $F_{9}S_{7}S_{16}S_{5}$, $F_{5}S_{7}S_{16}S_{5}$, $F_{4}S_{9}S_{17}S_{4}$, $F_{4}S_{16}S_{9}S_{4}$, $F_{1}S_{16}S_{9}S_{2}$, $F_{4}S_{9}S_{17}V_{2}$, $F_{4}S_{16}S_{9}V_{2}$, $F_{1}S_{16}S_{9}V_{2}$, $F_{20}S_{9}S_{17}S_{4}$, $F_{7}S_{16}S_{9}S_{4}$, $F_{7}S_{16}S_{9}S_{2}$, $F_{20}S_{9}S_{17}V_{5}$, $F_{7}S_{16}S_{9}V_{5}$, $F_{7}S_{16}S_{9}V_{4}$, $F_{9}S_{9}S_{17}V_{2}$, $F_{9}S_{16}S_{9}V_{2}$, $F_{5}S_{16}S_{9}V_{2}$, $F_{9}S_{9}S_{17}V_{5}$, $F_{9}S_{16}S_{9}V_{5}$, $F_{5}S_{16}S_{9}V_{4}$, $F_{9}S_{9}S_{17}S_{5}$, $F_{9}S_{16}S_{9}S_{5}$, $F_{5}S_{16}S_{9}S_{5}$}} 
 \SplitRow{ $\mathcal{O}_{9, \; (1,-1)}^{41}$} {{$F_{1}S_{16}S_{8}S_{3}$, $F_{1}S_{7}S_{16}S_{3}$, $F_{1}S_{16}S_{8}S_{2}$, $F_{1}S_{7}S_{16}S_{2}$, $F_{8}S_{16}S_{8}S_{3}$, $F_{8}S_{7}S_{16}S_{3}$, $F_{8}S_{16}S_{8}S_{5}$, $F_{8}S_{7}S_{16}S_{5}$, $F_{7}S_{16}S_{8}S_{2}$, $F_{7}S_{7}S_{16}S_{2}$, $F_{7}S_{16}S_{8}S_{5}$, $F_{7}S_{7}S_{16}S_{5}$, $F_{7}S_{16}S_{8}S_{6}$, $F_{7}S_{7}S_{16}S_{6}$, $F_{1}S_{16}S_{9}S_{3}$, $F_{1}S_{16}S_{9}S_{2}$, $F_{8}S_{16}S_{9}S_{3}$, $F_{8}S_{16}S_{9}S_{5}$, $F_{7}S_{16}S_{9}S_{2}$, $F_{7}S_{16}S_{9}S_{5}$, $F_{7}S_{16}S_{9}S_{6}$}} 
 \SplitRow{ $\mathcal{O}_{9, \; (1,-1)}^{42}$} {{$F_{2}S_{16}S_{9}S_{3}$, $F_{8}S_{16}S_{9}S_{3}$, $F_{8}S_{16}S_{9}S_{5}$, $F_{2}S_{16}S_{8}S_{3}$, $F_{2}S_{7}S_{16}S_{3}$, $F_{8}S_{16}S_{8}S_{3}$, $F_{8}S_{7}S_{16}S_{3}$, $F_{8}S_{16}S_{8}S_{5}$, $F_{8}S_{7}S_{16}S_{5}$}} 
 \SplitRow{ $\mathcal{O}_{9, \; (1,-1)}^{43}$} {{$F_{19}S_{9}S_{18}S_{6}$, $F_{19}S_{9}S_{18}S_{4}$, $F_{12}S_{9}S_{18}S_{4}$}} 
 \SplitRow{ $\mathcal{O}_{9, \; (1,-1)}^{44}$} {{$F_{19}S_{9}S_{18}S_{6}$, $F_{19}S_{9}S_{18}V_{5}$, $F_{19}S_{9}S_{18}V_{1}$, $F_{20}S_{9}S_{18}V_{5}$, $F_{20}S_{9}S_{18}S_{4}$, $F_{11}S_{9}S_{18}V_{1}$, $F_{11}S_{9}S_{18}S_{4}$}} 
 \SplitRow{ $\mathcal{O}_{9, \; (1,-1)}^{45}$} {{$F_{3}S_{16}S_{9}V_{2}$, $F_{3}S_{16}S_{9}S_{3}$, $F_{8}S_{16}S_{9}V_{2}$, $F_{8}S_{16}S_{9}S_{5}$, $F_{8}S_{16}S_{9}V_{3}$, $F_{6}S_{16}S_{9}S_{3}$, $F_{6}S_{16}S_{9}V_{3}$, $F_{3}S_{16}S_{8}V_{2}$, $F_{3}S_{7}S_{16}V_{2}$, $F_{3}S_{16}S_{8}S_{3}$, $F_{3}S_{7}S_{16}S_{3}$, $F_{8}S_{16}S_{8}V_{2}$, $F_{8}S_{7}S_{16}V_{2}$, $F_{8}S_{16}S_{8}S_{5}$, $F_{8}S_{7}S_{16}S_{5}$, $F_{8}S_{16}S_{8}V_{3}$, $F_{8}S_{7}S_{16}V_{3}$, $F_{6}S_{16}S_{8}S_{3}$, $F_{6}S_{7}S_{16}S_{3}$, $F_{6}S_{16}S_{8}V_{3}$, $F_{6}S_{7}S_{16}V_{3}$}} 
 \SplitRow{ $\mathcal{O}_{9, \; (1,-1)}^{46}$} {{$F_{12}S_{8}S_{17}S_{4}$, $F_{3}S_{16}S_{8}S_{4}$, $F_{3}S_{16}S_{8}S_{2}$, $F_{3}S_{7}S_{16}S_{4}$, $F_{3}S_{7}S_{16}S_{2}$, $F_{9}S_{8}S_{17}S_{4}$, $F_{9}S_{16}S_{8}S_{4}$, $F_{5}S_{16}S_{8}S_{2}$, $F_{9}S_{7}S_{16}S_{4}$, $F_{5}S_{7}S_{16}S_{2}$, $F_{9}S_{8}S_{17}S_{5}$, $F_{9}S_{16}S_{8}S_{5}$, $F_{5}S_{16}S_{8}S_{5}$, $F_{9}S_{7}S_{16}S_{5}$, $F_{5}S_{7}S_{16}S_{5}$, $F_{12}S_{9}S_{17}S_{4}$, $F_{3}S_{16}S_{9}S_{4}$, $F_{3}S_{16}S_{9}S_{2}$, $F_{9}S_{9}S_{17}S_{4}$, $F_{9}S_{16}S_{9}S_{4}$, $F_{5}S_{16}S_{9}S_{2}$, $F_{9}S_{9}S_{17}S_{5}$, $F_{9}S_{16}S_{9}S_{5}$, $F_{5}S_{16}S_{9}S_{5}$}} 
 \SplitRow{ $\mathcal{O}_{9, \; (1,-1)}^{47}$} {{$F_{3}S_{16}S_{8}V_{2}$, $F_{3}S_{7}S_{16}V_{2}$, $F_{3}S_{16}S_{8}V_{1}$, $F_{3}S_{7}S_{16}V_{1}$, $F_{3}S_{16}S_{8}S_{2}$, $F_{3}S_{7}S_{16}S_{2}$, $F_{8}S_{16}S_{8}V_{2}$, $F_{8}S_{7}S_{16}V_{2}$, $F_{8}S_{16}S_{8}S_{5}$, $F_{8}S_{7}S_{16}S_{5}$, $F_{8}S_{16}S_{8}V_{3}$, $F_{8}S_{7}S_{16}V_{3}$, $F_{7}S_{16}S_{8}V_{1}$, $F_{7}S_{7}S_{16}V_{1}$, $F_{7}S_{16}S_{8}S_{5}$, $F_{7}S_{7}S_{16}S_{5}$, $F_{7}S_{16}S_{8}V_{4}$, $F_{7}S_{7}S_{16}V_{4}$, $F_{5}S_{16}S_{8}S_{2}$, $F_{5}S_{7}S_{16}S_{2}$, $F_{5}S_{16}S_{8}V_{3}$, $F_{5}S_{7}S_{16}V_{3}$, $F_{5}S_{16}S_{8}V_{4}$, $F_{5}S_{7}S_{16}V_{4}$, $F_{3}S_{16}S_{9}V_{2}$, $F_{3}S_{16}S_{9}V_{1}$, $F_{3}S_{16}S_{9}S_{2}$, $F_{8}S_{16}S_{9}V_{2}$, $F_{8}S_{16}S_{9}S_{5}$, $F_{8}S_{16}S_{9}V_{3}$, $F_{7}S_{16}S_{9}V_{1}$, $F_{7}S_{16}S_{9}S_{5}$, $F_{7}S_{16}S_{9}V_{4}$, $F_{5}S_{16}S_{9}S_{2}$, $F_{5}S_{16}S_{9}V_{3}$, $F_{5}S_{16}S_{9}V_{4}$}} 
 \end{SplitTable}
   \vspace{0.4cm} 
 \captionof{table}{Dimension 9 $(\Delta B, \Delta L)$ = (1, -1) UV completions for topology 18 (see Fig.~\ref{fig:d=9_topologies}).} 
 \label{tab:T9Top18B1L-1}
 \end{center}

%% file: Tables_dim9/Table_dim9_Top19_B1_L-1.tex
 \begin{center}
 \begin{SplitTable}[2.0cm] 
  \SplitHeader{$\mathcal{O}^j{(\mathcal{T}_ {9} ^{ 19 })}$}{New Particles} 
   \vspace{0.1cm} 
 \SplitRow{ $\mathcal{O}_{9, \; (1,-1)}^{39}$} {{$F_{7}F_{7}F_{14}S_{20}$, $F_{7}F_{8}F_{9}V_{15}$, $F_{5}F_{7}F_{8}V_{15}$, $F_{3}F_{7}F_{14}V_{8}$, $F_{3}F_{8}F_{9}S_{5}$, $F_{3}F_{5}F_{8}S_{5}$, $F_{10}F_{7}F_{9}V_{15}$, $F_{10}F_{5}F_{7}V_{15}$, $F_{3}F_{7}F_{8}V_{8}$, $F_{10}F_{7}F_{7}S_{20}$, $F_{7}F_{8}F_{9}S_{19}$, $F_{5}F_{7}F_{8}S_{3}$, $F_{7}F_{7}F_{14}V_{2}$, $F_{7}F_{8}F_{9}V_{2}$, $F_{5}F_{7}F_{8}V_{2}$, $F_{3}F_{8}F_{9}V_{5}$, $F_{3}F_{5}F_{8}V_{4}$, $F_{3}F_{7}F_{14}S_{6}$, $F_{10}F_{7}F_{9}V_{5}$, $F_{10}F_{5}F_{7}V_{4}$, $F_{3}F_{7}F_{14}V_{2}$, $F_{10}F_{7}F_{9}S_{6}$, $F_{10}F_{5}F_{7}S_{6}$, $F_{3}F_{8}F_{9}V_{2}$, $F_{3}F_{5}F_{8}V_{2}$, $F_{3}F_{7}F_{8}V_{5}$, $F_{3}F_{7}F_{8}V_{4}$, $F_{10}F_{7}F_{7}V_{5}$, $F_{10}F_{7}F_{7}V_{4}$, $F_{3}F_{7}F_{8}S_{19}$, $F_{3}F_{7}F_{8}S_{3}$}} 
 \SplitRow{ $\mathcal{O}_{9, \; (1,-1)}^{40}$} {{$F_{7}F_{9}F_{14}V_{17}$, $F_{7}F_{9}F_{14}V_{15}$, $F_{5}F_{7}F_{14}V_{15}$, $F_{6}F_{7}F_{9}V_{15}$, $F_{5}F_{6}F_{7}V_{15}$, $F_{8}F_{9}F_{9}S_{19}$, $F_{5}F_{8}F_{9}S_{19}$, $F_{8}F_{9}F_{9}S_{3}$, $F_{5}F_{5}F_{8}S_{3}$, $F_{4}F_{9}F_{14}V_{8}$, $F_{4}F_{5}F_{14}V_{8}$, $F_{1}F_{9}F_{14}V_{8}$, $F_{4}F_{6}F_{9}V_{3}$, $F_{1}F_{5}F_{6}V_{3}$, $F_{11}F_{9}F_{9}S_{19}$, $F_{11}F_{5}F_{9}S_{19}$, $F_{2}F_{9}F_{9}S_{3}$, $F_{2}F_{5}F_{5}S_{3}$, $F_{4}F_{7}F_{14}S_{12}$, $F_{4}F_{7}F_{14}S_{11}$, $F_{1}F_{7}F_{14}S_{11}$, $F_{4}F_{6}F_{7}S_{11}$, $F_{1}F_{6}F_{7}S_{11}$, $F_{4}F_{8}F_{9}V_{8}$, $F_{4}F_{5}F_{8}V_{8}$, $F_{1}F_{8}F_{9}V_{8}$, $F_{4}F_{8}F_{9}V_{3}$, $F_{1}F_{5}F_{8}V_{3}$, $F_{11}F_{7}F_{9}V_{17}$, $F_{11}F_{7}F_{9}V_{15}$, $F_{11}F_{5}F_{7}V_{15}$, $F_{2}F_{7}F_{9}V_{15}$, $F_{2}F_{5}F_{7}V_{15}$, $F_{7}F_{9}F_{14}V_{16}$, $F_{7}F_{9}F_{14}V_{2}$, $F_{6}F_{7}F_{9}V_{2}$, $F_{5}F_{7}F_{14}V_{2}$, $F_{5}F_{6}F_{7}V_{2}$, $F_{8}F_{9}F_{9}V_{16}$, $F_{8}F_{9}F_{9}V_{2}$, $F_{5}F_{8}F_{9}V_{2}$, $F_{5}F_{5}F_{8}V_{2}$, $F_{7}F_{9}F_{14}S_{4}$, $F_{6}F_{7}F_{9}S_{4}$, $F_{5}F_{7}F_{14}S_{4}$, $F_{7}F_{9}F_{14}S_{2}$, $F_{5}F_{6}F_{7}S_{2}$, $F_{4}F_{9}F_{14}V_{5}$, $F_{4}F_{6}F_{9}V_{5}$, $F_{1}F_{9}F_{14}V_{5}$, $F_{4}F_{5}F_{14}V_{4}$, $F_{1}F_{5}F_{6}V_{4}$, $F_{11}F_{9}F_{9}V_{5}$, $F_{2}F_{9}F_{9}V_{5}$, $F_{11}F_{5}F_{9}V_{5}$, $F_{11}F_{5}F_{9}V_{4}$, $F_{2}F_{5}F_{5}V_{4}$, $F_{4}F_{9}F_{14}S_{4}$, $F_{4}F_{6}F_{9}S_{4}$, $F_{4}F_{5}F_{14}S_{4}$, $F_{1}F_{9}F_{14}S_{2}$, $F_{1}F_{5}F_{6}S_{2}$, $F_{4}F_{8}F_{9}S_{13}$, $F_{4}F_{8}F_{9}S_{5}$, $F_{1}F_{8}F_{9}S_{5}$, $F_{4}F_{5}F_{8}S_{5}$, $F_{1}F_{5}F_{8}S_{5}$, $F_{4}F_{7}F_{14}V_{5}$, $F_{4}F_{6}F_{7}V_{5}$, $F_{1}F_{7}F_{14}V_{5}$, $F_{4}F_{7}F_{14}V_{4}$, $F_{1}F_{6}F_{7}V_{4}$, $F_{11}F_{7}F_{9}S_{13}$, $F_{11}F_{7}F_{9}S_{5}$, $F_{2}F_{7}F_{9}S_{5}$, $F_{11}F_{5}F_{7}S_{5}$, $F_{2}F_{5}F_{7}S_{5}$, $F_{4}F_{7}F_{14}V_{16}$, $F_{4}F_{7}F_{14}V_{2}$, $F_{4}F_{6}F_{7}V_{2}$, $F_{1}F_{7}F_{14}V_{2}$, $F_{1}F_{6}F_{7}V_{2}$, $F_{11}F_{7}F_{9}V_{5}$, $F_{2}F_{7}F_{9}V_{5}$, $F_{11}F_{5}F_{7}V_{5}$, $F_{11}F_{7}F_{9}V_{4}$, $F_{2}F_{5}F_{7}V_{4}$, $F_{4}F_{8}F_{9}V_{16}$, $F_{4}F_{8}F_{9}V_{2}$, $F_{4}F_{5}F_{8}V_{2}$, $F_{1}F_{8}F_{9}V_{2}$, $F_{1}F_{5}F_{8}V_{2}$}} 
 \SplitRow{ $\mathcal{O}_{9, \; (1,-1)}^{41}$} {{$F_{7}F_{8}F_{8}S_{19}$, $F_{7}F_{7}F_{13}S_{20}$, $F_{4}F_{7}F_{8}S_{11}$, $F_{1}F_{7}F_{8}S_{11}$, $F_{11}F_{7}F_{7}S_{20}$, $F_{4}F_{8}F_{8}S_{5}$, $F_{1}F_{8}F_{8}S_{5}$, $F_{4}F_{7}F_{13}S_{11}$, $F_{1}F_{7}F_{13}S_{11}$, $F_{11}F_{7}F_{8}S_{19}$, $F_{7}F_{8}F_{8}S_{4}$, $F_{7}F_{8}F_{8}S_{2}$, $F_{7}F_{7}F_{13}S_{4}$, $F_{7}F_{7}F_{13}S_{2}$, $F_{7}F_{8}F_{8}S_{3}$, $F_{4}F_{7}F_{8}S_{5}$, $F_{1}F_{7}F_{8}S_{5}$, $F_{11}F_{7}F_{7}S_{5}$, $F_{4}F_{7}F_{8}S_{19}$, $F_{1}F_{7}F_{8}S_{3}$, $F_{4}F_{7}F_{13}S_{6}$, $F_{1}F_{7}F_{13}S_{6}$, $F_{11}F_{7}F_{8}S_{6}$, $F_{4}F_{8}F_{8}S_{4}$, $F_{1}F_{8}F_{8}S_{2}$, $F_{11}F_{7}F_{8}S_{5}$, $F_{4}F_{7}F_{13}S_{4}$, $F_{1}F_{7}F_{13}S_{2}$}} 
 \SplitRow{ $\mathcal{O}_{9, \; (1,-1)}^{42}$} {{$F_{7}F_{8}F_{8}S_{19}$, $F_{7}F_{8}F_{8}S_{3}$, $F_{4}F_{8}F_{8}S_{5}$, $F_{11}F_{7}F_{8}S_{5}$, $F_{2}F_{7}F_{8}S_{5}$, $F_{11}F_{7}F_{8}S_{19}$, $F_{2}F_{7}F_{8}S_{3}$, $F_{7}F_{8}F_{8}S_{4}$, $F_{11}F_{7}F_{8}S_{6}$, $F_{2}F_{7}F_{8}S_{6}$, $F_{4}F_{8}F_{8}S_{4}$}} 
 \SplitRow{ $\mathcal{O}_{9, \; (1,-1)}^{43}$} {{$F_{3}F_{5}F_{9}S_{19}$, $F_{3}F_{9}F_{9}S_{3}$, $F_{3}F_{5}F_{5}S_{3}$, $F_{3}F_{9}F_{9}S_{13}$, $F_{3}F_{9}F_{9}S_{5}$, $F_{3}F_{5}F_{9}S_{5}$, $F_{3}F_{5}F_{5}S_{5}$, $F_{5}F_{9}F_{9}S_{19}$, $F_{5}F_{9}F_{9}S_{3}$, $F_{5}F_{5}F_{5}S_{3}$}} 
 \SplitRow{ $\mathcal{O}_{9, \; (1,-1)}^{44}$} {{$F_{4}F_{8}F_{9}V_{16}$, $F_{4}F_{8}F_{9}V_{2}$, $F_{4}F_{5}F_{8}V_{2}$, $F_{1}F_{8}F_{9}V_{2}$, $F_{1}F_{5}F_{8}V_{2}$, $F_{1}F_{8}F_{9}V_{8}$, $F_{4}F_{5}F_{8}V_{8}$, $F_{4}F_{8}F_{9}V_{3}$, $F_{1}F_{5}F_{8}V_{3}$, $F_{4}F_{8}F_{9}S_{13}$, $F_{4}F_{8}F_{9}S_{5}$, $F_{1}F_{8}F_{9}S_{5}$, $F_{4}F_{5}F_{8}S_{5}$, $F_{1}F_{5}F_{8}S_{5}$, $F_{4}F_{5}F_{9}S_{19}$, $F_{1}F_{9}F_{9}S_{3}$, $F_{1}F_{5}F_{5}S_{3}$, $F_{1}F_{9}F_{9}V_{8}$, $F_{4}F_{5}F_{9}V_{8}$, $F_{4}F_{5}F_{9}V_{3}$, $F_{1}F_{5}F_{5}V_{3}$, $F_{5}F_{8}F_{9}S_{19}$, $F_{8}F_{9}F_{9}S_{3}$, $F_{5}F_{5}F_{8}S_{3}$, $F_{8}F_{9}F_{9}V_{16}$, $F_{8}F_{9}F_{9}V_{2}$, $F_{5}F_{8}F_{9}V_{2}$, $F_{5}F_{5}F_{8}V_{2}$}} 
 \SplitRow{ $\mathcal{O}_{9, \; (1,-1)}^{45}$} {{$F_{7}F_{8}F_{14}S_{19}$, $F_{6}F_{7}F_{8}S_{3}$, $F_{7}F_{8}F_{14}V_{2}$, $F_{6}F_{7}F_{8}V_{2}$, $F_{8}F_{8}F_{9}V_{2}$, $F_{3}F_{8}F_{14}V_{8}$, $F_{3}F_{6}F_{8}V_{3}$, $F_{3}F_{8}F_{14}S_{5}$, $F_{3}F_{6}F_{8}S_{5}$, $F_{3}F_{7}F_{14}V_{8}$, $F_{3}F_{6}F_{7}V_{3}$, $F_{3}F_{7}F_{14}V_{2}$, $F_{3}F_{6}F_{7}V_{2}$, $F_{3}F_{8}F_{9}S_{5}$, $F_{3}F_{8}F_{9}V_{2}$, $F_{3}F_{8}F_{8}V_{8}$, $F_{3}F_{8}F_{8}V_{3}$, $F_{3}F_{7}F_{8}V_{8}$, $F_{3}F_{7}F_{8}V_{3}$, $F_{3}F_{7}F_{8}S_{19}$, $F_{3}F_{7}F_{8}S_{3}$, $F_{8}F_{8}F_{9}S_{4}$, $F_{7}F_{8}F_{14}V_{1}$, $F_{6}F_{7}F_{8}V_{1}$, $F_{3}F_{8}F_{9}V_{5}$, $F_{3}F_{7}F_{14}S_{6}$, $F_{3}F_{6}F_{7}S_{6}$, $F_{3}F_{8}F_{14}V_{1}$, $F_{3}F_{6}F_{8}V_{1}$, $F_{3}F_{7}F_{8}V_{5}$, $F_{3}F_{8}F_{8}S_{4}$}} 
 \SplitRow{ $\mathcal{O}_{9, \; (1,-1)}^{46}$} {{$F_{9}F_{9}F_{14}S_{19}$, $F_{5}F_{9}F_{14}S_{19}$, $F_{6}F_{9}F_{9}S_{3}$, $F_{5}F_{5}F_{6}S_{3}$, $F_{3}F_{9}F_{14}S_{12}$, $F_{3}F_{9}F_{14}S_{11}$, $F_{3}F_{5}F_{14}S_{11}$, $F_{3}F_{6}F_{9}S_{11}$, $F_{3}F_{5}F_{6}S_{11}$, $F_{3}F_{9}F_{9}S_{19}$, $F_{3}F_{5}F_{9}S_{19}$, $F_{3}F_{9}F_{9}S_{3}$, $F_{3}F_{5}F_{5}S_{3}$, $F_{9}F_{9}F_{14}S_{4}$, $F_{6}F_{9}F_{9}S_{4}$, $F_{5}F_{9}F_{14}S_{4}$, $F_{5}F_{9}F_{14}S_{2}$, $F_{5}F_{5}F_{6}S_{2}$, $F_{3}F_{9}F_{14}S_{13}$, $F_{3}F_{9}F_{14}S_{5}$, $F_{3}F_{6}F_{9}S_{5}$, $F_{3}F_{5}F_{14}S_{5}$, $F_{3}F_{5}F_{6}S_{5}$, $F_{3}F_{9}F_{9}S_{13}$, $F_{3}F_{9}F_{9}S_{5}$, $F_{3}F_{5}F_{9}S_{5}$, $F_{3}F_{5}F_{5}S_{5}$, $F_{3}F_{9}F_{14}S_{4}$, $F_{3}F_{6}F_{9}S_{4}$, $F_{3}F_{5}F_{14}S_{4}$, $F_{3}F_{9}F_{14}S_{2}$, $F_{3}F_{5}F_{6}S_{2}$}} 
 \SplitRow{ $\mathcal{O}_{9, \; (1,-1)}^{47}$} {{$F_{7}F_{8}F_{14}S_{19}$, $F_{8}F_{8}F_{9}V_{2}$, $F_{5}F_{8}F_{8}V_{2}$, $F_{7}F_{13}F_{9}V_{15}$, $F_{5}F_{7}F_{13}V_{15}$, $F_{3}F_{7}F_{14}V_{7}$, $F_{3}F_{8}F_{9}S_{11}$, $F_{3}F_{5}F_{8}S_{11}$, $F_{3}F_{7}F_{9}V_{15}$, $F_{3}F_{5}F_{7}V_{15}$, $F_{3}F_{8}F_{14}V_{8}$, $F_{3}F_{13}F_{9}S_{11}$, $F_{3}F_{5}F_{13}S_{11}$, $F_{3}F_{8}F_{9}V_{2}$, $F_{3}F_{5}F_{8}V_{2}$, $F_{3}F_{8}F_{8}V_{8}$, $F_{3}F_{7}F_{13}V_{7}$, $F_{3}F_{7}F_{8}S_{19}$, $F_{8}F_{8}F_{9}S_{4}$, $F_{5}F_{8}F_{8}S_{2}$, $F_{7}F_{8}F_{14}V_{1}$, $F_{7}F_{13}F_{9}S_{4}$, $F_{5}F_{7}F_{13}S_{2}$, $F_{7}F_{8}F_{14}V_{2}$, $F_{7}F_{13}F_{9}V_{1}$, $F_{5}F_{7}F_{13}V_{1}$, $F_{3}F_{8}F_{9}V_{8}$, $F_{3}F_{5}F_{8}V_{3}$, $F_{3}F_{7}F_{14}S_{5}$, $F_{3}F_{7}F_{9}V_{8}$, $F_{3}F_{5}F_{7}V_{3}$, $F_{3}F_{7}F_{14}V_{2}$, $F_{3}F_{7}F_{9}S_{5}$, $F_{3}F_{5}F_{7}S_{5}$, $F_{3}F_{13}F_{9}V_{5}$, $F_{3}F_{5}F_{13}V_{4}$, $F_{3}F_{8}F_{14}S_{5}$, $F_{3}F_{8}F_{9}V_{5}$, $F_{3}F_{5}F_{8}V_{4}$, $F_{3}F_{8}F_{14}V_{1}$, $F_{3}F_{8}F_{9}S_{5}$, $F_{3}F_{5}F_{8}S_{5}$, $F_{3}F_{13}F_{9}V_{1}$, $F_{3}F_{5}F_{13}V_{1}$, $F_{3}F_{7}F_{13}V_{5}$, $F_{3}F_{7}F_{13}V_{4}$, $F_{3}F_{8}F_{8}V_{3}$, $F_{3}F_{7}F_{8}V_{5}$, $F_{3}F_{7}F_{8}V_{4}$, $F_{3}F_{8}F_{8}S_{4}$, $F_{3}F_{8}F_{8}S_{2}$, $F_{3}F_{7}F_{8}V_{8}$, $F_{3}F_{7}F_{8}V_{3}$, $F_{3}F_{7}F_{13}S_{4}$, $F_{3}F_{7}F_{13}S_{2}$}} 
 \end{SplitTable}
   \vspace{0.4cm} 
 \captionof{table}{Dimension 9 $(\Delta B, \Delta L)$ = (1, -1) UV completions for topology 19 (see Fig.~\ref{fig:d=9_topologies}).} 
 \label{tab:T9Top19B1L-1}
 \end{center}

%% file: Tables_dim9/Table_dim9_Top20_B1_L-1.tex
 \begin{center}
 \begin{SplitTable}[2.0cm] 
  \SplitHeader{$\mathcal{O}^j{(\mathcal{T}_ {9} ^{ 20 })}$}{New Particles} 
   \vspace{0.1cm} 
 \SplitRow{ $\mathcal{O}_{9, \; (1,-1)}^{39}$} {{$F_{7}F_{19}F_{14}V_{8}$, $F_{7}F_{8}F_{9}S_{5}$, $F_{5}F_{7}F_{8}S_{5}$, $F_{7}F_{8}F_{19}V_{8}$, $F_{7}F_{19}F_{14}S_{20}$, $F_{7}F_{8}F_{9}V_{15}$, $F_{5}F_{7}F_{8}V_{15}$, $F_{3}F_{4}F_{14}V_{8}$, $F_{10}F_{7}F_{9}V_{15}$, $F_{10}F_{5}F_{7}V_{15}$, $F_{3}F_{4}F_{8}V_{8}$, $F_{10}F_{7}F_{19}S_{20}$, $F_{7}F_{8}F_{19}V_{15}$, $F_{3}F_{4}F_{8}S_{5}$, $F_{10}F_{7}F_{19}V_{15}$, $F_{8}F_{9}F_{9}V_{5}$, $F_{5}F_{5}F_{8}V_{4}$, $F_{7}F_{8}F_{14}S_{6}$, $F_{7}F_{8}F_{9}V_{5}$, $F_{5}F_{7}F_{8}V_{4}$, $F_{8}F_{9}F_{9}S_{19}$, $F_{5}F_{5}F_{8}S_{3}$, $F_{7}F_{8}F_{14}V_{2}$, $F_{10}F_{11}F_{9}V_{5}$, $F_{2}F_{10}F_{5}V_{4}$, $F_{3}F_{8}F_{14}V_{2}$, $F_{10}F_{11}F_{7}V_{5}$, $F_{2}F_{10}F_{7}V_{4}$, $F_{3}F_{8}F_{9}S_{19}$, $F_{3}F_{5}F_{8}S_{3}$, $F_{7}F_{8}F_{9}V_{2}$, $F_{5}F_{7}F_{8}V_{2}$, $F_{10}F_{11}F_{7}S_{6}$, $F_{2}F_{10}F_{7}S_{6}$, $F_{3}F_{8}F_{9}V_{2}$, $F_{3}F_{5}F_{8}V_{2}$, $F_{7}F_{9}F_{9}V_{5}$, $F_{5}F_{5}F_{7}V_{4}$, $F_{7}F_{8}F_{9}S_{6}$, $F_{5}F_{7}F_{8}S_{6}$, $F_{7}F_{7}F_{9}V_{5}$, $F_{5}F_{7}F_{7}V_{4}$, $F_{7}F_{9}F_{9}S_{19}$, $F_{5}F_{5}F_{7}S_{3}$, $F_{3}F_{11}F_{9}V_{5}$, $F_{2}F_{3}F_{5}V_{4}$, $F_{3}F_{11}F_{7}V_{5}$, $F_{2}F_{3}F_{7}V_{4}$, $F_{3}F_{7}F_{9}S_{19}$, $F_{3}F_{5}F_{7}S_{3}$, $F_{7}F_{7}F_{9}V_{2}$, $F_{5}F_{7}F_{7}V_{2}$, $F_{3}F_{11}F_{7}S_{6}$, $F_{2}F_{3}F_{7}S_{6}$, $F_{3}F_{7}F_{9}V_{2}$, $F_{3}F_{5}F_{7}V_{2}$}} 
 \SplitRow{ $\mathcal{O}_{9, \; (1,-1)}^{40}$} {{$F_{9}F_{14}F_{20}V_{8}$, $F_{7}F_{9}F_{14}V_{8}$, $F_{6}F_{7}F_{9}V_{3}$, $F_{5}F_{7}F_{14}V_{8}$, $F_{5}F_{6}F_{7}V_{3}$, $F_{7}F_{19}F_{14}S_{12}$, $F_{7}F_{19}F_{14}S_{11}$, $F_{6}F_{7}F_{19}S_{11}$, $F_{8}F_{9}F_{20}V_{8}$, $F_{7}F_{8}F_{9}V_{8}$, $F_{7}F_{8}F_{9}V_{3}$, $F_{5}F_{7}F_{8}V_{8}$, $F_{5}F_{7}F_{8}V_{3}$, $F_{9}F_{14}F_{20}V_{17}$, $F_{7}F_{9}F_{14}V_{15}$, $F_{6}F_{7}F_{9}V_{15}$, $F_{5}F_{7}F_{14}V_{15}$, $F_{5}F_{6}F_{7}V_{15}$, $F_{4}F_{12}F_{14}S_{12}$, $F_{3}F_{4}F_{14}S_{11}$, $F_{3}F_{4}F_{6}S_{11}$, $F_{1}F_{3}F_{14}S_{11}$, $F_{1}F_{3}F_{6}S_{11}$, $F_{11}F_{9}F_{20}V_{17}$, $F_{11}F_{7}F_{9}V_{15}$, $F_{2}F_{7}F_{9}V_{15}$, $F_{11}F_{5}F_{7}V_{15}$, $F_{2}F_{5}F_{7}V_{15}$, $F_{7}F_{19}F_{14}V_{17}$, $F_{7}F_{19}F_{14}V_{15}$, $F_{6}F_{7}F_{19}V_{15}$, $F_{8}F_{9}F_{20}S_{19}$, $F_{7}F_{8}F_{9}S_{19}$, $F_{7}F_{8}F_{9}S_{3}$, $F_{5}F_{7}F_{8}S_{19}$, $F_{5}F_{7}F_{8}S_{3}$, $F_{4}F_{12}F_{14}V_{8}$, $F_{3}F_{4}F_{14}V_{8}$, $F_{3}F_{4}F_{6}V_{3}$, $F_{1}F_{3}F_{14}V_{8}$, $F_{1}F_{3}F_{6}V_{3}$, $F_{11}F_{9}F_{20}S_{19}$, $F_{11}F_{7}F_{9}S_{19}$, $F_{2}F_{7}F_{9}S_{3}$, $F_{11}F_{5}F_{7}S_{19}$, $F_{2}F_{5}F_{7}S_{3}$, $F_{4}F_{12}F_{8}V_{8}$, $F_{3}F_{4}F_{8}V_{8}$, $F_{3}F_{4}F_{8}V_{3}$, $F_{1}F_{3}F_{8}V_{8}$, $F_{1}F_{3}F_{8}V_{3}$, $F_{11}F_{7}F_{19}V_{17}$, $F_{11}F_{7}F_{19}V_{15}$, $F_{2}F_{7}F_{19}V_{15}$, $F_{9}F_{14}F_{15}V_{5}$, $F_{8}F_{9}F_{14}V_{5}$, $F_{5}F_{8}F_{14}V_{4}$, $F_{6}F_{8}F_{9}V_{5}$, $F_{5}F_{6}F_{8}V_{4}$, $F_{8}F_{9}F_{9}S_{13}$, $F_{8}F_{9}F_{9}S_{5}$, $F_{5}F_{8}F_{9}S_{5}$, $F_{5}F_{5}F_{8}S_{5}$, $F_{7}F_{14}F_{15}V_{5}$, $F_{7}F_{8}F_{14}V_{5}$, $F_{7}F_{8}F_{14}V_{4}$, $F_{6}F_{7}F_{8}V_{5}$, $F_{6}F_{7}F_{8}V_{4}$, $F_{9}F_{14}F_{15}V_{16}$, $F_{8}F_{9}F_{14}V_{2}$, $F_{5}F_{8}F_{14}V_{2}$, $F_{6}F_{8}F_{9}V_{2}$, $F_{5}F_{6}F_{8}V_{2}$, $F_{11}F_{12}F_{9}S_{13}$, $F_{3}F_{11}F_{9}S_{5}$, $F_{3}F_{11}F_{5}S_{5}$, $F_{2}F_{3}F_{9}S_{5}$, $F_{2}F_{3}F_{5}S_{5}$, $F_{4}F_{14}F_{15}V_{16}$, $F_{4}F_{8}F_{14}V_{2}$, $F_{1}F_{8}F_{14}V_{2}$, $F_{4}F_{6}F_{8}V_{2}$, $F_{1}F_{6}F_{8}V_{2}$, $F_{8}F_{9}F_{9}V_{16}$, $F_{8}F_{9}F_{9}V_{2}$, $F_{5}F_{8}F_{9}V_{2}$, $F_{5}F_{5}F_{8}V_{2}$, $F_{7}F_{14}F_{15}S_{4}$, $F_{7}F_{8}F_{14}S_{4}$, $F_{7}F_{8}F_{14}S_{2}$, $F_{6}F_{7}F_{8}S_{4}$, $F_{6}F_{7}F_{8}S_{2}$, $F_{11}F_{12}F_{9}V_{5}$, $F_{3}F_{11}F_{9}V_{5}$, $F_{3}F_{11}F_{5}V_{4}$, $F_{2}F_{3}F_{9}V_{5}$, $F_{2}F_{3}F_{5}V_{4}$, $F_{4}F_{14}F_{15}S_{4}$, $F_{4}F_{8}F_{14}S_{4}$, $F_{1}F_{8}F_{14}S_{2}$, $F_{4}F_{6}F_{8}S_{4}$, $F_{1}F_{6}F_{8}S_{2}$, $F_{11}F_{12}F_{7}V_{5}$, $F_{3}F_{11}F_{7}V_{5}$, $F_{3}F_{11}F_{7}V_{4}$, $F_{2}F_{3}F_{7}V_{5}$, $F_{2}F_{3}F_{7}V_{4}$, $F_{4}F_{8}F_{9}V_{16}$, $F_{4}F_{8}F_{9}V_{2}$, $F_{1}F_{8}F_{9}V_{2}$, $F_{4}F_{5}F_{8}V_{2}$, $F_{1}F_{5}F_{8}V_{2}$, $F_{9}F_{9}F_{15}V_{5}$, $F_{8}F_{9}F_{9}V_{5}$, $F_{5}F_{8}F_{9}V_{4}$, $F_{5}F_{8}F_{9}V_{5}$, $F_{5}F_{5}F_{8}V_{4}$, $F_{7}F_{9}F_{9}S_{13}$, $F_{7}F_{9}F_{9}S_{5}$, $F_{5}F_{7}F_{9}S_{5}$, $F_{5}F_{5}F_{7}S_{5}$, $F_{7}F_{9}F_{15}V_{5}$, $F_{7}F_{8}F_{9}V_{5}$, $F_{7}F_{8}F_{9}V_{4}$, $F_{5}F_{7}F_{8}V_{5}$, $F_{5}F_{7}F_{8}V_{4}$, $F_{9}F_{9}F_{15}V_{16}$, $F_{4}F_{12}F_{9}S_{13}$, $F_{3}F_{4}F_{9}S_{5}$, $F_{3}F_{4}F_{5}S_{5}$, $F_{1}F_{3}F_{9}S_{5}$, $F_{1}F_{3}F_{5}S_{5}$, $F_{4}F_{9}F_{15}V_{16}$, $F_{7}F_{9}F_{9}V_{16}$, $F_{7}F_{9}F_{9}V_{2}$, $F_{5}F_{7}F_{9}V_{2}$, $F_{5}F_{5}F_{7}V_{2}$, $F_{7}F_{9}F_{15}S_{4}$, $F_{7}F_{8}F_{9}S_{4}$, $F_{7}F_{8}F_{9}S_{2}$, $F_{5}F_{7}F_{8}S_{4}$, $F_{5}F_{7}F_{8}S_{2}$, $F_{4}F_{12}F_{9}V_{5}$, $F_{3}F_{4}F_{9}V_{5}$, $F_{3}F_{4}F_{5}V_{4}$, $F_{1}F_{3}F_{9}V_{5}$, $F_{1}F_{3}F_{5}V_{4}$, $F_{4}F_{9}F_{15}S_{4}$, $F_{4}F_{8}F_{9}S_{4}$, $F_{1}F_{8}F_{9}S_{2}$, $F_{4}F_{5}F_{8}S_{4}$, $F_{1}F_{5}F_{8}S_{2}$, $F_{4}F_{12}F_{7}V_{5}$, $F_{3}F_{4}F_{7}V_{5}$, $F_{3}F_{4}F_{7}V_{4}$, $F_{1}F_{3}F_{7}V_{5}$, $F_{1}F_{3}F_{7}V_{4}$, $F_{4}F_{7}F_{9}V_{16}$, $F_{4}F_{7}F_{9}V_{2}$, $F_{1}F_{7}F_{9}V_{2}$, $F_{4}F_{5}F_{7}V_{2}$, $F_{1}F_{5}F_{7}V_{2}$}} 
 \SplitRow{ $\mathcal{O}_{9, \; (1,-1)}^{41}$} {{$F_{7}F_{8}F_{19}S_{11}$, $F_{8}F_{8}F_{9}S_{5}$, $F_{7}F_{13}F_{19}S_{11}$, $F_{7}F_{8}F_{19}S_{19}$, $F_{3}F_{4}F_{8}S_{5}$, $F_{1}F_{3}F_{8}S_{5}$, $F_{11}F_{7}F_{19}S_{19}$, $F_{8}F_{8}F_{9}S_{19}$, $F_{7}F_{13}F_{19}S_{20}$, $F_{3}F_{4}F_{8}S_{11}$, $F_{1}F_{3}F_{8}S_{11}$, $F_{11}F_{7}F_{19}S_{20}$, $F_{3}F_{4}F_{13}S_{11}$, $F_{1}F_{3}F_{13}S_{11}$, $F_{11}F_{8}F_{9}S_{19}$, $F_{7}F_{8}F_{9}S_{5}$, $F_{5}F_{7}F_{8}S_{5}$, $F_{7}F_{13}F_{14}S_{6}$, $F_{6}F_{7}F_{13}S_{6}$, $F_{5}F_{8}F_{8}S_{5}$, $F_{7}F_{8}F_{9}S_{4}$, $F_{5}F_{7}F_{8}S_{2}$, $F_{3}F_{11}F_{7}S_{6}$, $F_{4}F_{8}F_{9}S_{4}$, $F_{1}F_{5}F_{8}S_{2}$, $F_{7}F_{13}F_{14}S_{4}$, $F_{6}F_{7}F_{13}S_{2}$, $F_{5}F_{8}F_{8}S_{3}$, $F_{3}F_{11}F_{7}S_{5}$, $F_{4}F_{8}F_{9}S_{19}$, $F_{1}F_{5}F_{8}S_{3}$, $F_{3}F_{11}F_{8}S_{5}$, $F_{4}F_{13}F_{14}S_{4}$, $F_{1}F_{6}F_{13}S_{2}$, $F_{7}F_{7}F_{9}S_{5}$, $F_{5}F_{7}F_{7}S_{5}$, $F_{7}F_{8}F_{14}S_{6}$, $F_{6}F_{7}F_{8}S_{6}$, $F_{7}F_{7}F_{9}S_{4}$, $F_{5}F_{7}F_{7}S_{2}$, $F_{3}F_{4}F_{7}S_{6}$, $F_{1}F_{3}F_{7}S_{6}$, $F_{4}F_{7}F_{9}S_{4}$, $F_{1}F_{5}F_{7}S_{2}$, $F_{7}F_{8}F_{14}S_{4}$, $F_{6}F_{7}F_{8}S_{2}$, $F_{7}F_{8}F_{9}S_{19}$, $F_{5}F_{7}F_{8}S_{3}$, $F_{3}F_{4}F_{7}S_{5}$, $F_{1}F_{3}F_{7}S_{5}$, $F_{4}F_{7}F_{9}S_{19}$, $F_{1}F_{5}F_{7}S_{3}$, $F_{4}F_{8}F_{14}S_{4}$, $F_{1}F_{6}F_{8}S_{2}$}} 
 \SplitRow{ $\mathcal{O}_{9, \; (1,-1)}^{42}$} {{$F_{8}F_{8}F_{9}S_{5}$, $F_{5}F_{8}F_{8}S_{5}$, $F_{8}F_{8}F_{9}S_{19}$, $F_{5}F_{8}F_{8}S_{3}$, $F_{3}F_{11}F_{8}S_{5}$, $F_{2}F_{3}F_{8}S_{5}$, $F_{11}F_{8}F_{9}S_{19}$, $F_{2}F_{5}F_{8}S_{3}$, $F_{7}F_{8}F_{9}S_{5}$, $F_{5}F_{7}F_{8}S_{5}$, $F_{7}F_{8}F_{9}S_{19}$, $F_{5}F_{7}F_{8}S_{3}$, $F_{3}F_{4}F_{8}S_{5}$, $F_{11}F_{7}F_{9}S_{19}$, $F_{2}F_{5}F_{7}S_{3}$, $F_{7}F_{8}F_{14}S_{6}$, $F_{7}F_{8}F_{14}S_{4}$, $F_{10}F_{11}F_{7}S_{6}$, $F_{2}F_{10}F_{7}S_{6}$, $F_{4}F_{8}F_{14}S_{4}$}} 
 \SplitRow{ $\mathcal{O}_{9, \; (1,-1)}^{43}$} {{$F_{3}F_{9}F_{20}S_{19}$, $F_{3}F_{7}F_{9}S_{19}$, $F_{3}F_{7}F_{9}S_{3}$, $F_{3}F_{5}F_{7}S_{19}$, $F_{3}F_{5}F_{7}S_{3}$, $F_{3}F_{4}F_{9}S_{13}$, $F_{3}F_{4}F_{9}S_{5}$, $F_{3}F_{4}F_{5}S_{5}$, $F_{9}F_{9}F_{20}S_{19}$, $F_{7}F_{9}F_{9}S_{19}$, $F_{5}F_{7}F_{9}S_{3}$, $F_{5}F_{7}F_{9}S_{19}$, $F_{5}F_{5}F_{7}S_{3}$, $F_{9}F_{9}F_{20}S_{13}$, $F_{7}F_{9}F_{9}S_{5}$, $F_{5}F_{7}F_{9}S_{5}$, $F_{5}F_{5}F_{7}S_{5}$}} 
 \SplitRow{ $\mathcal{O}_{9, \; (1,-1)}^{44}$} {{$F_{4}F_{8}F_{9}V_{16}$, $F_{4}F_{8}F_{9}V_{2}$, $F_{1}F_{8}F_{9}V_{2}$, $F_{4}F_{12}F_{8}V_{8}$, $F_{3}F_{4}F_{8}V_{8}$, $F_{3}F_{4}F_{8}V_{3}$, $F_{1}F_{3}F_{8}V_{8}$, $F_{1}F_{3}F_{8}V_{3}$, $F_{4}F_{9}F_{20}S_{19}$, $F_{4}F_{7}F_{9}S_{19}$, $F_{1}F_{7}F_{9}S_{3}$, $F_{4}F_{5}F_{7}S_{19}$, $F_{1}F_{5}F_{7}S_{3}$, $F_{4}F_{12}F_{9}V_{8}$, $F_{3}F_{4}F_{9}V_{8}$, $F_{3}F_{4}F_{5}V_{3}$, $F_{1}F_{3}F_{9}V_{8}$, $F_{1}F_{3}F_{5}V_{3}$, $F_{8}F_{9}F_{20}S_{19}$, $F_{7}F_{8}F_{9}S_{19}$, $F_{7}F_{8}F_{9}S_{3}$, $F_{5}F_{7}F_{8}S_{19}$, $F_{5}F_{7}F_{8}S_{3}$, $F_{8}F_{9}F_{9}V_{16}$, $F_{8}F_{9}F_{9}V_{2}$, $F_{5}F_{8}F_{9}V_{2}$, $F_{4}F_{9}F_{20}V_{16}$, $F_{4}F_{7}F_{9}V_{2}$, $F_{1}F_{7}F_{9}V_{2}$, $F_{4}F_{5}F_{7}V_{2}$, $F_{1}F_{5}F_{7}V_{2}$, $F_{4}F_{12}F_{9}S_{13}$, $F_{3}F_{4}F_{9}S_{5}$, $F_{3}F_{4}F_{5}S_{5}$, $F_{1}F_{3}F_{9}S_{5}$, $F_{1}F_{3}F_{5}S_{5}$, $F_{9}F_{9}F_{20}V_{16}$, $F_{7}F_{9}F_{9}V_{2}$, $F_{5}F_{7}F_{9}V_{2}$, $F_{5}F_{5}F_{7}V_{2}$, $F_{8}F_{9}F_{20}V_{8}$, $F_{7}F_{8}F_{9}V_{8}$, $F_{7}F_{8}F_{9}V_{3}$, $F_{5}F_{7}F_{8}V_{8}$, $F_{5}F_{7}F_{8}V_{3}$, $F_{8}F_{9}F_{9}S_{13}$, $F_{8}F_{9}F_{9}S_{5}$, $F_{5}F_{8}F_{9}S_{5}$, $F_{9}F_{9}F_{20}V_{8}$, $F_{7}F_{9}F_{9}V_{8}$, $F_{5}F_{7}F_{9}V_{3}$, $F_{5}F_{7}F_{9}V_{8}$, $F_{5}F_{5}F_{7}V_{3}$}} 
 \SplitRow{ $\mathcal{O}_{9, \; (1,-1)}^{45}$} {{$F_{8}F_{9}F_{14}V_{8}$, $F_{5}F_{6}F_{8}V_{3}$, $F_{8}F_{8}F_{14}S_{5}$, $F_{6}F_{8}F_{8}S_{5}$, $F_{8}F_{8}F_{9}V_{8}$, $F_{5}F_{8}F_{8}V_{3}$, $F_{8}F_{9}F_{14}S_{19}$, $F_{5}F_{6}F_{8}S_{3}$, $F_{8}F_{8}F_{14}V_{2}$, $F_{6}F_{8}F_{8}V_{2}$, $F_{3}F_{4}F_{14}V_{8}$, $F_{1}F_{3}F_{6}V_{3}$, $F_{3}F_{8}F_{14}V_{2}$, $F_{3}F_{6}F_{8}V_{2}$, $F_{3}F_{4}F_{8}V_{8}$, $F_{1}F_{3}F_{8}V_{3}$, $F_{3}F_{8}F_{9}S_{19}$, $F_{3}F_{5}F_{8}S_{3}$, $F_{8}F_{8}F_{9}V_{2}$, $F_{5}F_{8}F_{8}V_{2}$, $F_{3}F_{4}F_{8}S_{5}$, $F_{1}F_{3}F_{8}S_{5}$, $F_{3}F_{8}F_{9}V_{2}$, $F_{3}F_{5}F_{8}V_{2}$, $F_{7}F_{9}F_{14}V_{8}$, $F_{5}F_{6}F_{7}V_{3}$, $F_{8}F_{8}F_{9}S_{5}$, $F_{7}F_{8}F_{9}V_{8}$, $F_{5}F_{7}F_{8}V_{3}$, $F_{7}F_{9}F_{14}S_{19}$, $F_{5}F_{6}F_{7}S_{3}$, $F_{3}F_{7}F_{9}S_{19}$, $F_{3}F_{5}F_{7}S_{3}$, $F_{7}F_{8}F_{9}V_{2}$, $F_{5}F_{7}F_{8}V_{2}$, $F_{3}F_{7}F_{9}V_{2}$, $F_{3}F_{5}F_{7}V_{2}$, $F_{8}F_{9}F_{14}V_{5}$, $F_{7}F_{13}F_{14}S_{6}$, $F_{6}F_{7}F_{13}S_{6}$, $F_{7}F_{8}F_{14}V_{5}$, $F_{8}F_{9}F_{14}S_{4}$, $F_{7}F_{13}F_{14}V_{1}$, $F_{6}F_{7}F_{13}V_{1}$, $F_{3}F_{11}F_{9}V_{5}$, $F_{3}F_{13}F_{14}V_{1}$, $F_{3}F_{6}F_{13}V_{1}$, $F_{3}F_{11}F_{7}V_{5}$, $F_{3}F_{8}F_{14}S_{4}$, $F_{7}F_{8}F_{14}V_{1}$, $F_{3}F_{11}F_{7}S_{6}$, $F_{3}F_{8}F_{14}V_{1}$}} 
 \SplitRow{ $\mathcal{O}_{9, \; (1,-1)}^{46}$} {{$F_{9}F_{14}F_{20}S_{12}$, $F_{7}F_{9}F_{14}S_{11}$, $F_{6}F_{7}F_{9}S_{11}$, $F_{5}F_{7}F_{14}S_{11}$, $F_{5}F_{6}F_{7}S_{11}$, $F_{9}F_{14}F_{20}S_{19}$, $F_{7}F_{9}F_{14}S_{19}$, $F_{6}F_{7}F_{9}S_{3}$, $F_{5}F_{7}F_{14}S_{19}$, $F_{5}F_{6}F_{7}S_{3}$, $F_{3}F_{11}F_{14}S_{12}$, $F_{3}F_{11}F_{14}S_{11}$, $F_{3}F_{11}F_{6}S_{11}$, $F_{3}F_{9}F_{20}S_{19}$, $F_{3}F_{7}F_{9}S_{19}$, $F_{3}F_{7}F_{9}S_{3}$, $F_{3}F_{5}F_{7}S_{19}$, $F_{3}F_{5}F_{7}S_{3}$, $F_{9}F_{14}F_{15}S_{13}$, $F_{8}F_{9}F_{14}S_{5}$, $F_{5}F_{8}F_{14}S_{5}$, $F_{6}F_{8}F_{9}S_{5}$, $F_{5}F_{6}F_{8}S_{5}$, $F_{9}F_{14}F_{15}S_{4}$, $F_{8}F_{9}F_{14}S_{4}$, $F_{5}F_{8}F_{14}S_{2}$, $F_{6}F_{8}F_{9}S_{4}$, $F_{5}F_{6}F_{8}S_{2}$, $F_{3}F_{4}F_{9}S_{13}$, $F_{3}F_{4}F_{9}S_{5}$, $F_{3}F_{4}F_{5}S_{5}$, $F_{1}F_{3}F_{9}S_{5}$, $F_{1}F_{3}F_{5}S_{5}$, $F_{3}F_{14}F_{15}S_{4}$, $F_{3}F_{8}F_{14}S_{4}$, $F_{3}F_{8}F_{14}S_{2}$, $F_{3}F_{6}F_{8}S_{4}$, $F_{3}F_{6}F_{8}S_{2}$, $F_{9}F_{9}F_{15}S_{13}$, $F_{8}F_{9}F_{9}S_{5}$, $F_{5}F_{8}F_{9}S_{5}$, $F_{5}F_{5}F_{8}S_{5}$, $F_{9}F_{9}F_{15}S_{4}$, $F_{8}F_{9}F_{9}S_{4}$, $F_{5}F_{8}F_{9}S_{2}$, $F_{5}F_{8}F_{9}S_{4}$, $F_{5}F_{5}F_{8}S_{2}$, $F_{3}F_{9}F_{15}S_{4}$, $F_{3}F_{8}F_{9}S_{4}$, $F_{3}F_{8}F_{9}S_{2}$, $F_{3}F_{5}F_{8}S_{4}$, $F_{3}F_{5}F_{8}S_{2}$}} 
 \SplitRow{ $\mathcal{O}_{9, \; (1,-1)}^{47}$} {{$F_{7}F_{19}F_{14}V_{7}$, $F_{7}F_{8}F_{9}S_{11}$, $F_{5}F_{7}F_{8}S_{11}$, $F_{8}F_{9}F_{14}V_{8}$, $F_{7}F_{13}F_{9}S_{11}$, $F_{5}F_{7}F_{13}S_{11}$, $F_{8}F_{8}F_{9}V_{8}$, $F_{7}F_{13}F_{19}V_{7}$, $F_{7}F_{19}F_{14}S_{19}$, $F_{7}F_{8}F_{9}V_{2}$, $F_{5}F_{7}F_{8}V_{2}$, $F_{3}F_{11}F_{14}V_{8}$, $F_{3}F_{7}F_{9}V_{2}$, $F_{3}F_{5}F_{7}V_{2}$, $F_{3}F_{11}F_{8}V_{8}$, $F_{3}F_{7}F_{19}S_{19}$, $F_{8}F_{9}F_{14}S_{19}$, $F_{7}F_{13}F_{9}V_{15}$, $F_{5}F_{7}F_{13}V_{15}$, $F_{3}F_{11}F_{14}V_{7}$, $F_{3}F_{7}F_{9}V_{15}$, $F_{3}F_{5}F_{7}V_{15}$, $F_{3}F_{11}F_{13}V_{7}$, $F_{3}F_{8}F_{9}S_{19}$, $F_{8}F_{8}F_{9}V_{2}$, $F_{7}F_{13}F_{19}V_{15}$, $F_{3}F_{11}F_{8}S_{11}$, $F_{3}F_{7}F_{19}V_{15}$, $F_{3}F_{11}F_{13}S_{11}$, $F_{3}F_{8}F_{9}V_{2}$, $F_{8}F_{9}F_{9}V_{8}$, $F_{5}F_{5}F_{8}V_{3}$, $F_{7}F_{8}F_{14}S_{5}$, $F_{13}F_{9}F_{14}V_{5}$, $F_{5}F_{6}F_{13}V_{4}$, $F_{8}F_{8}F_{14}S_{5}$, $F_{7}F_{13}F_{14}V_{5}$, $F_{6}F_{7}F_{13}V_{4}$, $F_{5}F_{8}F_{8}V_{3}$, $F_{8}F_{9}F_{9}S_{4}$, $F_{5}F_{5}F_{8}S_{2}$, $F_{7}F_{8}F_{14}V_{1}$, $F_{3}F_{4}F_{9}V_{5}$, $F_{1}F_{3}F_{5}V_{4}$, $F_{3}F_{8}F_{14}V_{1}$, $F_{3}F_{4}F_{7}V_{5}$, $F_{1}F_{3}F_{7}V_{4}$, $F_{3}F_{8}F_{9}S_{4}$, $F_{3}F_{5}F_{8}S_{2}$, $F_{13}F_{9}F_{14}S_{4}$, $F_{5}F_{6}F_{13}S_{2}$, $F_{8}F_{8}F_{14}V_{2}$, $F_{3}F_{4}F_{9}V_{8}$, $F_{1}F_{3}F_{5}V_{3}$, $F_{3}F_{8}F_{14}V_{2}$, $F_{3}F_{4}F_{8}V_{8}$, $F_{1}F_{3}F_{8}V_{3}$, $F_{3}F_{13}F_{14}S_{4}$, $F_{3}F_{6}F_{13}S_{2}$, $F_{7}F_{13}F_{14}V_{1}$, $F_{6}F_{7}F_{13}V_{1}$, $F_{5}F_{8}F_{8}V_{2}$, $F_{3}F_{4}F_{7}S_{5}$, $F_{1}F_{3}F_{7}S_{5}$, $F_{3}F_{5}F_{8}V_{2}$, $F_{3}F_{4}F_{8}S_{5}$, $F_{1}F_{3}F_{8}S_{5}$, $F_{3}F_{13}F_{14}V_{1}$, $F_{3}F_{6}F_{13}V_{1}$, $F_{7}F_{9}F_{9}V_{8}$, $F_{5}F_{5}F_{7}V_{3}$, $F_{7}F_{8}F_{9}S_{5}$, $F_{5}F_{7}F_{8}S_{5}$, $F_{8}F_{9}F_{14}V_{5}$, $F_{5}F_{6}F_{8}V_{4}$, $F_{8}F_{8}F_{9}S_{5}$, $F_{5}F_{8}F_{8}S_{5}$, $F_{7}F_{8}F_{14}V_{5}$, $F_{6}F_{7}F_{8}V_{4}$, $F_{7}F_{8}F_{9}V_{8}$, $F_{5}F_{7}F_{8}V_{3}$, $F_{7}F_{9}F_{9}S_{4}$, $F_{5}F_{5}F_{7}S_{2}$, $F_{7}F_{8}F_{9}V_{1}$, $F_{5}F_{7}F_{8}V_{1}$, $F_{3}F_{8}F_{9}V_{1}$, $F_{3}F_{5}F_{8}V_{1}$, $F_{3}F_{7}F_{9}S_{4}$, $F_{3}F_{5}F_{7}S_{2}$, $F_{8}F_{9}F_{14}S_{4}$, $F_{5}F_{6}F_{8}S_{2}$, $F_{3}F_{8}F_{14}S_{4}$, $F_{3}F_{6}F_{8}S_{2}$, $F_{6}F_{7}F_{8}V_{1}$, $F_{3}F_{6}F_{8}V_{1}$}} 
 \end{SplitTable}
   \vspace{0.4cm} 
 \captionof{table}{Dimension 9 $(\Delta B, \Delta L)$ = (1, -1) UV completions for topology 20 (see Fig.~\ref{fig:d=9_topologies}).} 
 \label{tab:T9Top20B1L-1}
 \end{center}

%% file: Tables_dim9/Table_dim9_Top21_B1_L-1.tex
 \begin{center}
 \begin{SplitTable}[2.0cm] 
  \SplitHeader{$\mathcal{O}^j{(\mathcal{T}_ {9} ^{ 21 })}$}{New Particles} 
   \vspace{0.1cm} 
 \SplitRow{ $\mathcal{O}_{9, \; (1,-1)}^{39}$} {{$F_{11}F_{7}S_{8}V_{2}$, $F_{2}F_{7}S_{7}V_{2}$, $F_{11}F_{9}S_{8}S_{3}$, $F_{2}F_{5}S_{7}S_{3}$, $F_{3}F_{9}S_{8}V_{2}$, $F_{3}F_{5}S_{7}V_{2}$, $F_{7}F_{9}S_{8}S_{6}$, $F_{5}F_{7}S_{7}S_{6}$, $F_{9}F_{9}S_{8}V_{4}$, $F_{5}F_{5}S_{7}V_{4}$, $F_{3}F_{8}S_{8}S_{3}$, $F_{3}F_{8}S_{7}S_{3}$, $F_{7}F_{8}S_{8}V_{4}$, $F_{7}F_{8}S_{7}V_{4}$, $F_{4}F_{8}S_{9}V_{2}$, $F_{4}F_{14}S_{9}S_{3}$, $F_{10}F_{19}S_{9}V_{2}$, $F_{8}F_{19}S_{9}S_{6}$, $F_{19}F_{14}S_{9}V_{4}$, $F_{10}F_{7}S_{9}S_{3}$, $F_{7}F_{8}S_{9}V_{4}$}} 
 \SplitRow{ $\mathcal{O}_{9, \; (1,-1)}^{40}$} {{$F_{12}F_{7}S_{8}S_{4}$, $F_{3}F_{7}S_{8}S_{4}$, $F_{3}F_{7}S_{8}S_{2}$, $F_{3}F_{7}S_{7}S_{4}$, $F_{3}F_{7}S_{7}S_{2}$, $F_{12}F_{9}S_{8}V_{2}$, $F_{3}F_{9}S_{8}V_{2}$, $F_{3}F_{5}S_{8}V_{2}$, $F_{3}F_{9}S_{7}V_{2}$, $F_{3}F_{5}S_{7}V_{2}$, $F_{4}F_{9}S_{8}S_{4}$, $F_{4}F_{9}S_{8}S_{2}$, $F_{1}F_{9}S_{8}S_{4}$, $F_{4}F_{5}S_{7}S_{4}$, $F_{1}F_{5}S_{7}S_{2}$, $F_{9}F_{9}S_{8}V_{5}$, $F_{9}F_{9}S_{8}V_{4}$, $F_{5}F_{9}S_{8}V_{5}$, $F_{5}F_{9}S_{7}V_{5}$, $F_{5}F_{5}S_{7}V_{4}$, $F_{4}F_{15}S_{8}V_{2}$, $F_{4}F_{8}S_{8}V_{2}$, $F_{1}F_{8}S_{8}V_{2}$, $F_{4}F_{8}S_{7}V_{2}$, $F_{1}F_{8}S_{7}V_{2}$, $F_{7}F_{15}S_{8}V_{5}$, $F_{7}F_{8}S_{8}V_{5}$, $F_{7}F_{8}S_{8}V_{4}$, $F_{7}F_{8}S_{7}V_{5}$, $F_{7}F_{8}S_{7}V_{4}$, $F_{9}F_{15}S_{8}S_{5}$, $F_{8}F_{9}S_{8}S_{5}$, $F_{5}F_{8}S_{8}S_{5}$, $F_{8}F_{9}S_{7}S_{5}$, $F_{5}F_{8}S_{7}S_{5}$, $F_{12}F_{8}S_{9}S_{4}$, $F_{3}F_{8}S_{9}S_{4}$, $F_{3}F_{8}S_{9}S_{2}$, $F_{12}F_{14}S_{9}V_{2}$, $F_{3}F_{14}S_{9}V_{2}$, $F_{3}F_{6}S_{9}V_{2}$, $F_{11}F_{19}S_{9}S_{4}$, $F_{11}F_{19}S_{9}S_{2}$, $F_{2}F_{19}S_{9}S_{4}$, $F_{19}F_{14}S_{9}V_{5}$, $F_{19}F_{14}S_{9}V_{4}$, $F_{6}F_{19}S_{9}V_{5}$, $F_{11}F_{20}S_{9}V_{2}$, $F_{11}F_{7}S_{9}V_{2}$, $F_{2}F_{7}S_{9}V_{2}$, $F_{8}F_{20}S_{9}V_{5}$, $F_{7}F_{8}S_{9}V_{5}$, $F_{7}F_{8}S_{9}V_{4}$, $F_{14}F_{20}S_{9}S_{5}$, $F_{7}F_{14}S_{9}S_{5}$, $F_{6}F_{7}S_{9}S_{5}$}} 
 \SplitRow{ $\mathcal{O}_{9, \; (1,-1)}^{41}$} {{$F_{3}F_{8}S_{8}S_{3}$, $F_{3}F_{8}S_{7}S_{3}$, $F_{3}F_{7}S_{8}S_{2}$, $F_{3}F_{7}S_{7}S_{2}$, $F_{4}F_{14}S_{8}S_{3}$, $F_{1}F_{6}S_{7}S_{3}$, $F_{7}F_{14}S_{8}S_{5}$, $F_{6}F_{7}S_{7}S_{5}$, $F_{4}F_{9}S_{8}S_{2}$, $F_{1}F_{5}S_{7}S_{2}$, $F_{8}F_{9}S_{8}S_{5}$, $F_{5}F_{8}S_{7}S_{5}$, $F_{7}F_{9}S_{8}S_{6}$, $F_{5}F_{7}S_{7}S_{6}$, $F_{3}F_{13}S_{9}S_{3}$, $F_{3}F_{8}S_{9}S_{2}$, $F_{11}F_{9}S_{9}S_{3}$, $F_{8}F_{9}S_{9}S_{5}$, $F_{11}F_{19}S_{9}S_{2}$, $F_{13}F_{19}S_{9}S_{5}$, $F_{8}F_{19}S_{9}S_{6}$}} 
 \SplitRow{ $\mathcal{O}_{9, \; (1,-1)}^{42}$} {{$F_{10}F_{7}S_{9}S_{3}$, $F_{4}F_{14}S_{9}S_{3}$, $F_{7}F_{14}S_{9}S_{5}$, $F_{3}F_{8}S_{8}S_{3}$, $F_{3}F_{8}S_{7}S_{3}$, $F_{11}F_{9}S_{8}S_{3}$, $F_{2}F_{5}S_{7}S_{3}$, $F_{8}F_{9}S_{8}S_{5}$, $F_{5}F_{8}S_{7}S_{5}$}} 
 \SplitRow{ $\mathcal{O}_{9, \; (1,-1)}^{43}$} {{$F_{9}F_{20}S_{9}S_{6}$, $F_{7}F_{9}S_{9}S_{6}$, $F_{5}F_{7}S_{9}S_{6}$, $F_{3}F_{20}S_{9}S_{4}$, $F_{3}F_{7}S_{9}S_{4}$, $F_{4}F_{9}S_{9}S_{4}$, $F_{4}F_{5}S_{9}S_{4}$}} 
 \SplitRow{ $\mathcal{O}_{9, \; (1,-1)}^{44}$} {{$F_{9}F_{20}S_{9}S_{6}$, $F_{7}F_{9}S_{9}S_{6}$, $F_{5}F_{7}S_{9}S_{6}$, $F_{8}F_{20}S_{9}V_{5}$, $F_{7}F_{8}S_{9}V_{5}$, $F_{4}F_{20}S_{9}V_{1}$, $F_{4}F_{7}S_{9}V_{1}$, $F_{1}F_{7}S_{9}V_{1}$, $F_{9}F_{9}S_{9}V_{5}$, $F_{5}F_{9}S_{9}V_{5}$, $F_{4}F_{9}S_{9}S_{4}$, $F_{1}F_{9}S_{9}S_{4}$, $F_{12}F_{9}S_{9}V_{1}$, $F_{3}F_{9}S_{9}V_{1}$, $F_{3}F_{5}S_{9}V_{1}$, $F_{12}F_{8}S_{9}S_{4}$, $F_{3}F_{8}S_{9}S_{4}$}} 
 \SplitRow{ $\mathcal{O}_{9, \; (1,-1)}^{45}$} {{$F_{11}F_{7}S_{9}V_{2}$, $F_{11}F_{9}S_{9}S_{3}$, $F_{3}F_{14}S_{9}V_{2}$, $F_{7}F_{14}S_{9}S_{5}$, $F_{9}F_{14}S_{9}V_{3}$, $F_{3}F_{13}S_{9}S_{3}$, $F_{7}F_{13}S_{9}V_{3}$, $F_{4}F_{8}S_{8}V_{2}$, $F_{1}F_{8}S_{7}V_{2}$, $F_{4}F_{14}S_{8}S_{3}$, $F_{1}F_{6}S_{7}S_{3}$, $F_{3}F_{9}S_{8}V_{2}$, $F_{3}F_{5}S_{7}V_{2}$, $F_{8}F_{9}S_{8}S_{5}$, $F_{5}F_{8}S_{7}S_{5}$, $F_{9}F_{14}S_{8}V_{3}$, $F_{5}F_{6}S_{7}V_{3}$, $F_{3}F_{8}S_{8}S_{3}$, $F_{3}F_{8}S_{7}S_{3}$, $F_{8}F_{8}S_{8}V_{3}$, $F_{8}F_{8}S_{7}V_{3}$}} 
 \SplitRow{ $\mathcal{O}_{9, \; (1,-1)}^{46}$} {{$F_{4}F_{9}S_{8}S_{4}$, $F_{4}F_{9}S_{8}S_{2}$, $F_{4}F_{5}S_{8}S_{4}$, $F_{1}F_{9}S_{7}S_{4}$, $F_{1}F_{5}S_{7}S_{2}$, $F_{3}F_{15}S_{8}S_{4}$, $F_{3}F_{8}S_{8}S_{4}$, $F_{3}F_{8}S_{8}S_{2}$, $F_{3}F_{8}S_{7}S_{4}$, $F_{3}F_{8}S_{7}S_{2}$, $F_{9}F_{15}S_{8}S_{5}$, $F_{8}F_{9}S_{8}S_{5}$, $F_{5}F_{8}S_{8}S_{5}$, $F_{8}F_{9}S_{7}S_{5}$, $F_{5}F_{8}S_{7}S_{5}$, $F_{11}F_{14}S_{9}S_{4}$, $F_{11}F_{14}S_{9}S_{2}$, $F_{11}F_{6}S_{9}S_{4}$, $F_{3}F_{20}S_{9}S_{4}$, $F_{3}F_{7}S_{9}S_{4}$, $F_{3}F_{7}S_{9}S_{2}$, $F_{14}F_{20}S_{9}S_{5}$, $F_{7}F_{14}S_{9}S_{5}$, $F_{6}F_{7}S_{9}S_{5}$}} 
 \SplitRow{ $\mathcal{O}_{9, \; (1,-1)}^{47}$} {{$F_{4}F_{8}S_{8}V_{2}$, $F_{1}F_{8}S_{7}V_{2}$, $F_{4}F_{7}S_{8}V_{1}$, $F_{1}F_{7}S_{7}V_{1}$, $F_{4}F_{9}S_{8}S_{2}$, $F_{1}F_{5}S_{7}S_{2}$, $F_{3}F_{14}S_{8}V_{2}$, $F_{3}F_{6}S_{7}V_{2}$, $F_{7}F_{14}S_{8}S_{5}$, $F_{6}F_{7}S_{7}S_{5}$, $F_{9}F_{14}S_{8}V_{3}$, $F_{5}F_{6}S_{7}V_{3}$, $F_{3}F_{9}S_{8}V_{1}$, $F_{3}F_{5}S_{7}V_{1}$, $F_{8}F_{9}S_{8}S_{5}$, $F_{5}F_{8}S_{7}S_{5}$, $F_{9}F_{9}S_{8}V_{4}$, $F_{5}F_{5}S_{7}V_{4}$, $F_{3}F_{8}S_{8}S_{2}$, $F_{3}F_{8}S_{7}S_{2}$, $F_{8}F_{8}S_{8}V_{3}$, $F_{8}F_{8}S_{7}V_{3}$, $F_{7}F_{8}S_{8}V_{4}$, $F_{7}F_{8}S_{7}V_{4}$, $F_{11}F_{13}S_{9}V_{2}$, $F_{11}F_{8}S_{9}V_{1}$, $F_{11}F_{14}S_{9}S_{2}$, $F_{3}F_{9}S_{9}V_{2}$, $F_{8}F_{9}S_{9}S_{5}$, $F_{9}F_{14}S_{9}V_{3}$, $F_{3}F_{19}S_{9}V_{1}$, $F_{13}F_{19}S_{9}S_{5}$, $F_{19}F_{14}S_{9}V_{4}$, $F_{3}F_{7}S_{9}S_{2}$, $F_{7}F_{13}S_{9}V_{3}$, $F_{7}F_{8}S_{9}V_{4}$}} 
 \end{SplitTable}
   \vspace{0.4cm} 
 \captionof{table}{Dimension 9 $(\Delta B, \Delta L)$ = (1, -1) UV completions for topology 21 (see Fig.~\ref{fig:d=9_topologies}).} 
 \label{tab:T9Top21B1L-1}
 \end{center}

%% file: Tables_dim9/Table_dim9_Top22_B1_L-1.tex
 \begin{center}
 \begin{SplitTable}[2.0cm] 
  \SplitHeader{$\mathcal{O}^j{(\mathcal{T}_ {9} ^{ 22 })}$}{New Particles} 
   \vspace{0.1cm} 
 \SplitRow{ $\mathcal{O}_{9, \; (1,-1)}^{39}$} {{$F_{9}S_{8}V_{2}V_{5}$, $F_{5}S_{7}V_{4}V_{2}$, $F_{8}S_{8}S_{3}S_{6}$, $F_{8}S_{7}S_{3}S_{6}$, $F_{11}S_{8}V_{2}V_{5}$, $F_{2}S_{7}V_{4}V_{2}$, $F_{9}S_{8}S_{6}S_{19}$, $F_{5}S_{7}S_{3}S_{6}$, $F_{8}S_{8}V_{4}V_{2}$, $F_{8}S_{7}V_{4}V_{2}$, $F_{11}S_{8}S_{3}S_{6}$, $F_{2}S_{7}S_{3}S_{6}$, $F_{9}S_{8}V_{4}V_{2}$, $F_{19}S_{9}V_{2}V_{8}$, $F_{7}S_{9}S_{3}S_{5}$, $F_{4}S_{9}V_{2}V_{8}$, $F_{19}S_{9}S_{6}S_{20}$, $F_{7}S_{9}V_{4}V_{15}$, $F_{4}S_{9}S_{3}S_{5}$, $F_{19}S_{9}V_{4}V_{15}$}} 
 \SplitRow{ $\mathcal{O}_{9, \; (1,-1)}^{40}$} {{$F_{9}S_{8}S_{4}S_{13}$, $F_{9}S_{8}S_{5}S_{4}$, $F_{5}S_{7}S_{5}S_{4}$, $F_{9}S_{8}S_{2}S_{5}$, $F_{5}S_{7}S_{2}S_{5}$, $F_{15}S_{8}V_{2}V_{5}$, $F_{8}S_{8}V_{2}V_{5}$, $F_{8}S_{7}V_{2}V_{5}$, $F_{8}S_{8}V_{4}V_{2}$, $F_{8}S_{7}V_{4}V_{2}$, $F_{12}S_{8}S_{4}S_{13}$, $F_{3}S_{8}S_{5}S_{4}$, $F_{3}S_{7}S_{5}S_{4}$, $F_{3}S_{8}S_{2}S_{5}$, $F_{3}S_{7}S_{2}S_{5}$, $F_{15}S_{8}V_{5}V_{16}$, $F_{12}S_{8}V_{2}V_{5}$, $F_{3}S_{8}V_{2}V_{5}$, $F_{3}S_{7}V_{2}V_{5}$, $F_{3}S_{8}V_{4}V_{2}$, $F_{3}S_{7}V_{4}V_{2}$, $F_{9}S_{8}V_{5}V_{16}$, $F_{9}S_{8}V_{2}V_{5}$, $F_{5}S_{7}V_{2}V_{5}$, $F_{9}S_{8}V_{4}V_{2}$, $F_{5}S_{7}V_{4}V_{2}$, $F_{15}S_{8}S_{5}S_{4}$, $F_{8}S_{8}S_{5}S_{4}$, $F_{8}S_{7}S_{5}S_{4}$, $F_{8}S_{8}S_{2}S_{5}$, $F_{8}S_{7}S_{2}S_{5}$, $F_{19}S_{9}S_{4}S_{12}$, $F_{19}S_{9}S_{11}S_{4}$, $F_{19}S_{9}S_{2}S_{11}$, $F_{20}S_{9}V_{2}V_{8}$, $F_{7}S_{9}V_{2}V_{8}$, $F_{7}S_{9}V_{3}V_{2}$, $F_{12}S_{9}S_{4}S_{12}$, $F_{3}S_{9}S_{11}S_{4}$, $F_{3}S_{9}S_{2}S_{11}$, $F_{20}S_{9}V_{5}V_{17}$, $F_{7}S_{9}V_{15}V_{5}$, $F_{7}S_{9}V_{4}V_{15}$, $F_{12}S_{9}V_{2}V_{8}$, $F_{3}S_{9}V_{2}V_{8}$, $F_{3}S_{9}V_{3}V_{2}$, $F_{19}S_{9}V_{5}V_{17}$, $F_{19}S_{9}V_{15}V_{5}$, $F_{19}S_{9}V_{4}V_{15}$, $F_{20}S_{9}S_{5}S_{19}$, $F_{7}S_{9}S_{5}S_{19}$, $F_{7}S_{9}S_{3}S_{5}$}} 
 \SplitRow{ $\mathcal{O}_{9, \; (1,-1)}^{41}$} {{$F_{14}S_{8}S_{3}S_{6}$, $F_{6}S_{7}S_{3}S_{6}$, $F_{9}S_{8}S_{2}S_{5}$, $F_{5}S_{7}S_{2}S_{5}$, $F_{3}S_{8}S_{3}S_{6}$, $F_{3}S_{7}S_{3}S_{6}$, $F_{9}S_{8}S_{5}S_{4}$, $F_{3}S_{8}S_{2}S_{5}$, $F_{3}S_{7}S_{2}S_{5}$, $F_{14}S_{8}S_{5}S_{4}$, $F_{6}S_{7}S_{2}S_{5}$, $F_{9}S_{8}S_{6}S_{19}$, $F_{5}S_{7}S_{3}S_{6}$, $F_{9}S_{9}S_{3}S_{5}$, $F_{19}S_{9}S_{2}S_{11}$, $F_{3}S_{9}S_{3}S_{5}$, $F_{19}S_{9}S_{5}S_{19}$, $F_{3}S_{9}S_{2}S_{11}$, $F_{9}S_{9}S_{5}S_{19}$, $F_{19}S_{9}S_{6}S_{20}$}} 
 \SplitRow{ $\mathcal{O}_{9, \; (1,-1)}^{42}$} {{$F_{14}S_{9}S_{3}S_{6}$, $F_{10}S_{9}S_{3}S_{6}$, $F_{14}S_{9}S_{5}S_{4}$, $F_{9}S_{8}S_{3}S_{5}$, $F_{5}S_{7}S_{3}S_{5}$, $F_{3}S_{8}S_{3}S_{5}$, $F_{3}S_{7}S_{3}S_{5}$, $F_{9}S_{8}S_{5}S_{19}$}} 
 \SplitRow{ $\mathcal{O}_{9, \; (1,-1)}^{43}$} {{$F_{20}S_{9}S_{6}S_{19}$, $F_{7}S_{9}S_{6}S_{19}$, $F_{7}S_{9}S_{3}S_{6}$, $F_{4}S_{9}S_{4}S_{13}$, $F_{4}S_{9}S_{5}S_{4}$, $F_{20}S_{9}S_{4}S_{13}$, $F_{7}S_{9}S_{5}S_{4}$}} 
 \SplitRow{ $\mathcal{O}_{9, \; (1,-1)}^{44}$} {{$F_{20}S_{9}S_{6}S_{19}$, $F_{7}S_{9}S_{6}S_{19}$, $F_{7}S_{9}S_{3}S_{6}$, $F_{9}S_{9}V_{5}V_{16}$, $F_{9}S_{9}V_{2}V_{5}$, $F_{12}S_{9}V_{1}V_{8}$, $F_{3}S_{9}V_{1}V_{8}$, $F_{3}S_{9}V_{3}V_{1}$, $F_{20}S_{9}V_{5}V_{16}$, $F_{7}S_{9}V_{2}V_{5}$, $F_{12}S_{9}S_{4}S_{13}$, $F_{3}S_{9}S_{5}S_{4}$, $F_{20}S_{9}V_{1}V_{8}$, $F_{7}S_{9}V_{1}V_{8}$, $F_{7}S_{9}V_{3}V_{1}$, $F_{9}S_{9}S_{4}S_{13}$, $F_{9}S_{9}S_{5}S_{4}$}} 
 \SplitRow{ $\mathcal{O}_{9, \; (1,-1)}^{45}$} {{$F_{14}S_{9}V_{2}V_{5}$, $F_{13}S_{9}S_{3}S_{6}$, $F_{11}S_{9}V_{2}V_{5}$, $F_{14}S_{9}S_{5}S_{4}$, $F_{13}S_{9}V_{3}V_{1}$, $F_{11}S_{9}S_{3}S_{6}$, $F_{14}S_{9}V_{3}V_{1}$, $F_{9}S_{8}V_{2}V_{8}$, $F_{5}S_{7}V_{3}V_{2}$, $F_{8}S_{8}S_{3}S_{5}$, $F_{8}S_{7}S_{3}S_{5}$, $F_{4}S_{8}V_{2}V_{8}$, $F_{1}S_{7}V_{3}V_{2}$, $F_{9}S_{8}S_{5}S_{19}$, $F_{5}S_{7}S_{3}S_{5}$, $F_{8}S_{8}V_{3}V_{2}$, $F_{8}S_{7}V_{3}V_{2}$, $F_{4}S_{8}S_{3}S_{5}$, $F_{1}S_{7}S_{3}S_{5}$, $F_{9}S_{8}V_{3}V_{2}$}} 
 \SplitRow{ $\mathcal{O}_{9, \; (1,-1)}^{46}$} {{$F_{15}S_{8}S_{4}S_{13}$, $F_{8}S_{8}S_{5}S_{4}$, $F_{8}S_{7}S_{5}S_{4}$, $F_{8}S_{8}S_{2}S_{5}$, $F_{8}S_{7}S_{2}S_{5}$, $F_{4}S_{8}S_{4}S_{13}$, $F_{4}S_{8}S_{5}S_{4}$, $F_{1}S_{7}S_{5}S_{4}$, $F_{4}S_{8}S_{2}S_{5}$, $F_{1}S_{7}S_{2}S_{5}$, $F_{15}S_{8}S_{5}S_{4}$, $F_{20}S_{9}S_{4}S_{12}$, $F_{7}S_{9}S_{11}S_{4}$, $F_{7}S_{9}S_{2}S_{11}$, $F_{11}S_{9}S_{4}S_{12}$, $F_{11}S_{9}S_{11}S_{4}$, $F_{11}S_{9}S_{2}S_{11}$, $F_{20}S_{9}S_{5}S_{19}$, $F_{7}S_{9}S_{5}S_{19}$, $F_{7}S_{9}S_{3}S_{5}$}} 
 \SplitRow{ $\mathcal{O}_{9, \; (1,-1)}^{47}$} {{$F_{14}S_{8}V_{2}V_{5}$, $F_{6}S_{7}V_{4}V_{2}$, $F_{9}S_{8}V_{1}V_{8}$, $F_{5}S_{7}V_{3}V_{1}$, $F_{8}S_{8}S_{2}S_{5}$, $F_{8}S_{7}S_{2}S_{5}$, $F_{4}S_{8}V_{2}V_{5}$, $F_{1}S_{7}V_{4}V_{2}$, $F_{9}S_{8}S_{5}S_{4}$, $F_{5}S_{7}S_{2}S_{5}$, $F_{8}S_{8}V_{3}V_{1}$, $F_{8}S_{7}V_{3}V_{1}$, $F_{4}S_{8}V_{1}V_{8}$, $F_{1}S_{7}V_{3}V_{1}$, $F_{14}S_{8}S_{5}S_{4}$, $F_{6}S_{7}S_{2}S_{5}$, $F_{8}S_{8}V_{4}V_{2}$, $F_{8}S_{7}V_{4}V_{2}$, $F_{4}S_{8}S_{2}S_{5}$, $F_{1}S_{7}S_{2}S_{5}$, $F_{14}S_{8}V_{3}V_{1}$, $F_{6}S_{7}V_{3}V_{1}$, $F_{9}S_{8}V_{4}V_{2}$, $F_{5}S_{7}V_{4}V_{2}$, $F_{9}S_{9}V_{2}V_{8}$, $F_{19}S_{9}V_{1}V_{7}$, $F_{7}S_{9}S_{2}S_{11}$, $F_{11}S_{9}V_{2}V_{8}$, $F_{19}S_{9}S_{5}S_{19}$, $F_{7}S_{9}V_{3}V_{2}$, $F_{11}S_{9}V_{1}V_{7}$, $F_{9}S_{9}S_{5}S_{19}$, $F_{7}S_{9}V_{4}V_{15}$, $F_{11}S_{9}S_{2}S_{11}$, $F_{9}S_{9}V_{3}V_{2}$, $F_{19}S_{9}V_{4}V_{15}$}} 
 \end{SplitTable}
   \vspace{0.4cm} 
 \captionof{table}{Dimension 9 $(\Delta B, \Delta L)$ = (1, -1) UV completions for topology 22 (see Fig.~\ref{fig:d=9_topologies}).} 
 \label{tab:T9Top22B1L-1}
 \end{center}

%% file: Tables_dim9/Table_dim9_Top23_B1_L-1.tex
 \begin{center}
 \begin{SplitTable}[2.0cm] 
  \SplitHeader{$\mathcal{O}^j{(\mathcal{T}_ {9} ^{ 23 })}$}{New Particles} 
   \vspace{0.1cm} 
 \SplitRow{ $\mathcal{O}_{9, \; (1,-1)}^{39}$} {{$F_{19}F_{14}S_{9}V_{8}$, $F_{8}F_{19}S_{9}V_{8}$, $F_{7}F_{8}S_{9}S_{5}$, $F_{4}F_{14}S_{9}V_{8}$, $F_{4}F_{8}S_{9}V_{8}$, $F_{19}F_{14}S_{9}S_{20}$, $F_{10}F_{19}S_{9}S_{20}$, $F_{7}F_{8}S_{9}V_{15}$, $F_{10}F_{7}S_{9}V_{15}$, $F_{4}F_{8}S_{9}S_{5}$, $F_{8}F_{19}S_{9}V_{15}$, $F_{10}F_{19}S_{9}V_{15}$, $F_{9}F_{9}S_{8}V_{5}$, $F_{5}F_{5}S_{7}V_{4}$, $F_{7}F_{9}S_{8}V_{5}$, $F_{5}F_{7}S_{7}V_{4}$, $F_{7}F_{8}S_{8}S_{6}$, $F_{7}F_{8}S_{7}S_{6}$, $F_{11}F_{9}S_{8}V_{5}$, $F_{2}F_{5}S_{7}V_{4}$, $F_{11}F_{7}S_{8}V_{5}$, $F_{2}F_{7}S_{7}V_{4}$, $F_{9}F_{9}S_{8}S_{19}$, $F_{5}F_{5}S_{7}S_{3}$, $F_{3}F_{9}S_{8}S_{19}$, $F_{3}F_{5}S_{7}S_{3}$, $F_{7}F_{8}S_{8}V_{2}$, $F_{7}F_{8}S_{7}V_{2}$, $F_{3}F_{8}S_{8}V_{2}$, $F_{3}F_{8}S_{7}V_{2}$, $F_{11}F_{7}S_{8}S_{6}$, $F_{2}F_{7}S_{7}S_{6}$, $F_{7}F_{9}S_{8}V_{2}$, $F_{5}F_{7}S_{7}V_{2}$, $F_{3}F_{9}S_{8}V_{2}$, $F_{3}F_{5}S_{7}V_{2}$}} 
 \SplitRow{ $\mathcal{O}_{9, \; (1,-1)}^{40}$} {{$F_{19}F_{14}S_{9}S_{12}$, $F_{19}F_{14}S_{9}S_{11}$, $F_{6}F_{19}S_{9}S_{11}$, $F_{14}F_{20}S_{9}V_{8}$, $F_{7}F_{14}S_{9}V_{8}$, $F_{6}F_{7}S_{9}V_{3}$, $F_{8}F_{20}S_{9}V_{8}$, $F_{7}F_{8}S_{9}V_{8}$, $F_{7}F_{8}S_{9}V_{3}$, $F_{12}F_{14}S_{9}S_{12}$, $F_{3}F_{14}S_{9}S_{11}$, $F_{3}F_{6}S_{9}S_{11}$, $F_{14}F_{20}S_{9}V_{17}$, $F_{7}F_{14}S_{9}V_{15}$, $F_{6}F_{7}S_{9}V_{15}$, $F_{11}F_{20}S_{9}V_{17}$, $F_{11}F_{7}S_{9}V_{15}$, $F_{2}F_{7}S_{9}V_{15}$, $F_{12}F_{14}S_{9}V_{8}$, $F_{3}F_{14}S_{9}V_{8}$, $F_{3}F_{6}S_{9}V_{3}$, $F_{12}F_{8}S_{9}V_{8}$, $F_{3}F_{8}S_{9}V_{8}$, $F_{3}F_{8}S_{9}V_{3}$, $F_{19}F_{14}S_{9}V_{17}$, $F_{19}F_{14}S_{9}V_{15}$, $F_{6}F_{19}S_{9}V_{15}$, $F_{11}F_{19}S_{9}V_{17}$, $F_{11}F_{19}S_{9}V_{15}$, $F_{2}F_{19}S_{9}V_{15}$, $F_{8}F_{20}S_{9}S_{19}$, $F_{7}F_{8}S_{9}S_{19}$, $F_{7}F_{8}S_{9}S_{3}$, $F_{11}F_{20}S_{9}S_{19}$, $F_{11}F_{7}S_{9}S_{19}$, $F_{2}F_{7}S_{9}S_{3}$, $F_{9}F_{9}S_{8}S_{13}$, $F_{9}F_{9}S_{8}S_{5}$, $F_{5}F_{9}S_{8}S_{5}$, $F_{5}F_{9}S_{7}S_{5}$, $F_{5}F_{5}S_{7}S_{5}$, $F_{9}F_{15}S_{8}V_{5}$, $F_{8}F_{9}S_{8}V_{5}$, $F_{5}F_{8}S_{8}V_{4}$, $F_{8}F_{9}S_{7}V_{5}$, $F_{5}F_{8}S_{7}V_{4}$, $F_{7}F_{15}S_{8}V_{5}$, $F_{7}F_{8}S_{8}V_{5}$, $F_{7}F_{8}S_{8}V_{4}$, $F_{7}F_{8}S_{7}V_{5}$, $F_{7}F_{8}S_{7}V_{4}$, $F_{12}F_{9}S_{8}S_{13}$, $F_{3}F_{9}S_{8}S_{5}$, $F_{3}F_{5}S_{8}S_{5}$, $F_{3}F_{9}S_{7}S_{5}$, $F_{3}F_{5}S_{7}S_{5}$, $F_{9}F_{15}S_{8}V_{16}$, $F_{8}F_{9}S_{8}V_{2}$, $F_{5}F_{8}S_{8}V_{2}$, $F_{8}F_{9}S_{7}V_{2}$, $F_{5}F_{8}S_{7}V_{2}$, $F_{4}F_{15}S_{8}V_{16}$, $F_{4}F_{8}S_{8}V_{2}$, $F_{1}F_{8}S_{8}V_{2}$, $F_{4}F_{8}S_{7}V_{2}$, $F_{1}F_{8}S_{7}V_{2}$, $F_{12}F_{9}S_{8}V_{5}$, $F_{3}F_{9}S_{8}V_{5}$, $F_{3}F_{5}S_{8}V_{4}$, $F_{3}F_{9}S_{7}V_{5}$, $F_{3}F_{5}S_{7}V_{4}$, $F_{12}F_{7}S_{8}V_{5}$, $F_{3}F_{7}S_{8}V_{5}$, $F_{3}F_{7}S_{8}V_{4}$, $F_{3}F_{7}S_{7}V_{5}$, $F_{3}F_{7}S_{7}V_{4}$, $F_{9}F_{9}S_{8}V_{16}$, $F_{9}F_{9}S_{8}V_{2}$, $F_{5}F_{9}S_{8}V_{2}$, $F_{5}F_{9}S_{7}V_{2}$, $F_{5}F_{5}S_{7}V_{2}$, $F_{4}F_{9}S_{8}V_{16}$, $F_{4}F_{9}S_{8}V_{2}$, $F_{1}F_{9}S_{8}V_{2}$, $F_{4}F_{5}S_{7}V_{2}$, $F_{1}F_{5}S_{7}V_{2}$, $F_{7}F_{15}S_{8}S_{4}$, $F_{7}F_{8}S_{8}S_{4}$, $F_{7}F_{8}S_{8}S_{2}$, $F_{7}F_{8}S_{7}S_{4}$, $F_{7}F_{8}S_{7}S_{2}$, $F_{4}F_{15}S_{8}S_{4}$, $F_{4}F_{8}S_{8}S_{4}$, $F_{1}F_{8}S_{8}S_{2}$, $F_{4}F_{8}S_{7}S_{4}$, $F_{1}F_{8}S_{7}S_{2}$}} 
 \SplitRow{ $\mathcal{O}_{9, \; (1,-1)}^{41}$} {{$F_{8}F_{9}S_{9}S_{5}$, $F_{8}F_{19}S_{9}S_{11}$, $F_{13}F_{19}S_{9}S_{11}$, $F_{3}F_{8}S_{9}S_{5}$, $F_{8}F_{19}S_{9}S_{19}$, $F_{11}F_{19}S_{9}S_{19}$, $F_{3}F_{8}S_{9}S_{11}$, $F_{3}F_{13}S_{9}S_{11}$, $F_{8}F_{9}S_{9}S_{19}$, $F_{11}F_{9}S_{9}S_{19}$, $F_{13}F_{19}S_{9}S_{20}$, $F_{11}F_{19}S_{9}S_{20}$, $F_{7}F_{14}S_{8}S_{6}$, $F_{6}F_{7}S_{7}S_{6}$, $F_{7}F_{9}S_{8}S_{5}$, $F_{5}F_{7}S_{7}S_{5}$, $F_{8}F_{9}S_{8}S_{5}$, $F_{5}F_{8}S_{7}S_{5}$, $F_{3}F_{7}S_{8}S_{6}$, $F_{3}F_{7}S_{7}S_{6}$, $F_{7}F_{9}S_{8}S_{4}$, $F_{5}F_{7}S_{7}S_{2}$, $F_{4}F_{9}S_{8}S_{4}$, $F_{1}F_{5}S_{7}S_{2}$, $F_{3}F_{7}S_{8}S_{5}$, $F_{3}F_{7}S_{7}S_{5}$, $F_{3}F_{8}S_{8}S_{5}$, $F_{3}F_{8}S_{7}S_{5}$, $F_{7}F_{14}S_{8}S_{4}$, $F_{6}F_{7}S_{7}S_{2}$, $F_{4}F_{14}S_{8}S_{4}$, $F_{1}F_{6}S_{7}S_{2}$, $F_{8}F_{9}S_{8}S_{19}$, $F_{5}F_{8}S_{7}S_{3}$, $F_{4}F_{9}S_{8}S_{19}$, $F_{1}F_{5}S_{7}S_{3}$}} 
 \SplitRow{ $\mathcal{O}_{9, \; (1,-1)}^{42}$} {{$F_{8}F_{9}S_{8}S_{5}$, $F_{5}F_{8}S_{7}S_{5}$, $F_{3}F_{8}S_{8}S_{5}$, $F_{3}F_{8}S_{7}S_{5}$, $F_{8}F_{9}S_{8}S_{19}$, $F_{5}F_{8}S_{7}S_{3}$, $F_{11}F_{9}S_{8}S_{19}$, $F_{2}F_{5}S_{7}S_{3}$, $F_{7}F_{14}S_{9}S_{6}$, $F_{10}F_{7}S_{9}S_{6}$, $F_{7}F_{14}S_{9}S_{4}$, $F_{4}F_{14}S_{9}S_{4}$}} 
 \SplitRow{ $\mathcal{O}_{9, \; (1,-1)}^{43}$} {{$F_{3}F_{20}S_{9}S_{19}$, $F_{3}F_{7}S_{9}S_{19}$, $F_{3}F_{7}S_{9}S_{3}$, $F_{9}F_{20}S_{9}S_{19}$, $F_{7}F_{9}S_{9}S_{19}$, $F_{5}F_{7}S_{9}S_{3}$, $F_{4}F_{9}S_{9}S_{13}$, $F_{4}F_{9}S_{9}S_{5}$, $F_{4}F_{5}S_{9}S_{5}$, $F_{9}F_{20}S_{9}S_{13}$, $F_{7}F_{9}S_{9}S_{5}$, $F_{5}F_{7}S_{9}S_{5}$}} 
 \SplitRow{ $\mathcal{O}_{9, \; (1,-1)}^{44}$} {{$F_{4}F_{20}S_{9}S_{19}$, $F_{4}F_{7}S_{9}S_{19}$, $F_{1}F_{7}S_{9}S_{3}$, $F_{8}F_{20}S_{9}S_{19}$, $F_{7}F_{8}S_{9}S_{19}$, $F_{7}F_{8}S_{9}S_{3}$, $F_{4}F_{9}S_{9}V_{16}$, $F_{4}F_{9}S_{9}V_{2}$, $F_{1}F_{9}S_{9}V_{2}$, $F_{9}F_{9}S_{9}V_{16}$, $F_{9}F_{9}S_{9}V_{2}$, $F_{5}F_{9}S_{9}V_{2}$, $F_{12}F_{8}S_{9}V_{8}$, $F_{3}F_{8}S_{9}V_{8}$, $F_{3}F_{8}S_{9}V_{3}$, $F_{12}F_{9}S_{9}V_{8}$, $F_{3}F_{9}S_{9}V_{8}$, $F_{3}F_{5}S_{9}V_{3}$, $F_{4}F_{20}S_{9}V_{16}$, $F_{4}F_{7}S_{9}V_{2}$, $F_{1}F_{7}S_{9}V_{2}$, $F_{9}F_{20}S_{9}V_{16}$, $F_{7}F_{9}S_{9}V_{2}$, $F_{5}F_{7}S_{9}V_{2}$, $F_{12}F_{9}S_{9}S_{13}$, $F_{3}F_{9}S_{9}S_{5}$, $F_{3}F_{5}S_{9}S_{5}$, $F_{8}F_{20}S_{9}V_{8}$, $F_{7}F_{8}S_{9}V_{8}$, $F_{7}F_{8}S_{9}V_{3}$, $F_{9}F_{20}S_{9}V_{8}$, $F_{7}F_{9}S_{9}V_{8}$, $F_{5}F_{7}S_{9}V_{3}$, $F_{9}F_{9}S_{9}S_{13}$, $F_{9}F_{9}S_{9}S_{5}$, $F_{5}F_{9}S_{9}S_{5}$}} 
 \SplitRow{ $\mathcal{O}_{9, \; (1,-1)}^{45}$} {{$F_{9}F_{14}S_{8}V_{8}$, $F_{5}F_{6}S_{7}V_{3}$, $F_{8}F_{9}S_{8}V_{8}$, $F_{5}F_{8}S_{7}V_{3}$, $F_{8}F_{8}S_{8}S_{5}$, $F_{8}F_{8}S_{7}S_{5}$, $F_{4}F_{14}S_{8}V_{8}$, $F_{1}F_{6}S_{7}V_{3}$, $F_{4}F_{8}S_{8}V_{8}$, $F_{1}F_{8}S_{7}V_{3}$, $F_{9}F_{14}S_{8}S_{19}$, $F_{5}F_{6}S_{7}S_{3}$, $F_{3}F_{9}S_{8}S_{19}$, $F_{3}F_{5}S_{7}S_{3}$, $F_{8}F_{8}S_{8}V_{2}$, $F_{8}F_{8}S_{7}V_{2}$, $F_{3}F_{8}S_{8}V_{2}$, $F_{3}F_{8}S_{7}V_{2}$, $F_{4}F_{8}S_{8}S_{5}$, $F_{1}F_{8}S_{7}S_{5}$, $F_{8}F_{9}S_{8}V_{2}$, $F_{5}F_{8}S_{7}V_{2}$, $F_{3}F_{9}S_{8}V_{2}$, $F_{3}F_{5}S_{7}V_{2}$, $F_{9}F_{14}S_{9}V_{5}$, $F_{7}F_{14}S_{9}V_{5}$, $F_{7}F_{13}S_{9}S_{6}$, $F_{11}F_{9}S_{9}V_{5}$, $F_{11}F_{7}S_{9}V_{5}$, $F_{9}F_{14}S_{9}S_{4}$, $F_{3}F_{14}S_{9}S_{4}$, $F_{7}F_{13}S_{9}V_{1}$, $F_{3}F_{13}S_{9}V_{1}$, $F_{11}F_{7}S_{9}S_{6}$, $F_{7}F_{14}S_{9}V_{1}$, $F_{3}F_{14}S_{9}V_{1}$}} 
 \SplitRow{ $\mathcal{O}_{9, \; (1,-1)}^{46}$} {{$F_{14}F_{20}S_{9}S_{12}$, $F_{7}F_{14}S_{9}S_{11}$, $F_{6}F_{7}S_{9}S_{11}$, $F_{11}F_{14}S_{9}S_{12}$, $F_{11}F_{14}S_{9}S_{11}$, $F_{11}F_{6}S_{9}S_{11}$, $F_{14}F_{20}S_{9}S_{19}$, $F_{7}F_{14}S_{9}S_{19}$, $F_{6}F_{7}S_{9}S_{3}$, $F_{3}F_{20}S_{9}S_{19}$, $F_{3}F_{7}S_{9}S_{19}$, $F_{3}F_{7}S_{9}S_{3}$, $F_{9}F_{15}S_{8}S_{13}$, $F_{8}F_{9}S_{8}S_{5}$, $F_{5}F_{8}S_{8}S_{5}$, $F_{8}F_{9}S_{7}S_{5}$, $F_{5}F_{8}S_{7}S_{5}$, $F_{4}F_{9}S_{8}S_{13}$, $F_{4}F_{9}S_{8}S_{5}$, $F_{4}F_{5}S_{8}S_{5}$, $F_{1}F_{9}S_{7}S_{5}$, $F_{1}F_{5}S_{7}S_{5}$, $F_{9}F_{15}S_{8}S_{4}$, $F_{8}F_{9}S_{8}S_{4}$, $F_{5}F_{8}S_{8}S_{2}$, $F_{8}F_{9}S_{7}S_{4}$, $F_{5}F_{8}S_{7}S_{2}$, $F_{3}F_{15}S_{8}S_{4}$, $F_{3}F_{8}S_{8}S_{4}$, $F_{3}F_{8}S_{8}S_{2}$, $F_{3}F_{8}S_{7}S_{4}$, $F_{3}F_{8}S_{7}S_{2}$}} 
 \SplitRow{ $\mathcal{O}_{9, \; (1,-1)}^{47}$} {{$F_{9}F_{14}S_{9}V_{8}$, $F_{8}F_{9}S_{9}V_{8}$, $F_{19}F_{14}S_{9}V_{7}$, $F_{13}F_{19}S_{9}V_{7}$, $F_{7}F_{8}S_{9}S_{11}$, $F_{7}F_{13}S_{9}S_{11}$, $F_{11}F_{14}S_{9}V_{8}$, $F_{11}F_{8}S_{9}V_{8}$, $F_{19}F_{14}S_{9}S_{19}$, $F_{3}F_{19}S_{9}S_{19}$, $F_{7}F_{8}S_{9}V_{2}$, $F_{3}F_{7}S_{9}V_{2}$, $F_{11}F_{14}S_{9}V_{7}$, $F_{11}F_{13}S_{9}V_{7}$, $F_{9}F_{14}S_{9}S_{19}$, $F_{3}F_{9}S_{9}S_{19}$, $F_{7}F_{13}S_{9}V_{15}$, $F_{3}F_{7}S_{9}V_{15}$, $F_{11}F_{8}S_{9}S_{11}$, $F_{11}F_{13}S_{9}S_{11}$, $F_{8}F_{9}S_{9}V_{2}$, $F_{3}F_{9}S_{9}V_{2}$, $F_{13}F_{19}S_{9}V_{15}$, $F_{3}F_{19}S_{9}V_{15}$, $F_{9}F_{14}S_{8}V_{5}$, $F_{5}F_{6}S_{7}V_{4}$, $F_{7}F_{14}S_{8}V_{5}$, $F_{6}F_{7}S_{7}V_{4}$, $F_{9}F_{9}S_{8}V_{8}$, $F_{5}F_{5}S_{7}V_{3}$, $F_{8}F_{9}S_{8}V_{8}$, $F_{5}F_{8}S_{7}V_{3}$, $F_{7}F_{8}S_{8}S_{5}$, $F_{7}F_{8}S_{7}S_{5}$, $F_{8}F_{8}S_{8}S_{5}$, $F_{8}F_{8}S_{7}S_{5}$, $F_{4}F_{9}S_{8}V_{5}$, $F_{1}F_{5}S_{7}V_{4}$, $F_{4}F_{7}S_{8}V_{5}$, $F_{1}F_{7}S_{7}V_{4}$, $F_{9}F_{9}S_{8}S_{4}$, $F_{5}F_{5}S_{7}S_{2}$, $F_{3}F_{9}S_{8}S_{4}$, $F_{3}F_{5}S_{7}S_{2}$, $F_{7}F_{8}S_{8}V_{1}$, $F_{7}F_{8}S_{7}V_{1}$, $F_{3}F_{8}S_{8}V_{1}$, $F_{3}F_{8}S_{7}V_{1}$, $F_{4}F_{9}S_{8}V_{8}$, $F_{1}F_{5}S_{7}V_{3}$, $F_{4}F_{8}S_{8}V_{8}$, $F_{1}F_{8}S_{7}V_{3}$, $F_{9}F_{14}S_{8}S_{4}$, $F_{5}F_{6}S_{7}S_{2}$, $F_{3}F_{14}S_{8}S_{4}$, $F_{3}F_{6}S_{7}S_{2}$, $F_{8}F_{8}S_{8}V_{2}$, $F_{8}F_{8}S_{7}V_{2}$, $F_{3}F_{8}S_{8}V_{2}$, $F_{3}F_{8}S_{7}V_{2}$, $F_{4}F_{7}S_{8}S_{5}$, $F_{1}F_{7}S_{7}S_{5}$, $F_{4}F_{8}S_{8}S_{5}$, $F_{1}F_{8}S_{7}S_{5}$, $F_{7}F_{14}S_{8}V_{1}$, $F_{6}F_{7}S_{7}V_{1}$, $F_{3}F_{14}S_{8}V_{1}$, $F_{3}F_{6}S_{7}V_{1}$, $F_{8}F_{9}S_{8}V_{2}$, $F_{5}F_{8}S_{7}V_{2}$, $F_{3}F_{9}S_{8}V_{2}$, $F_{3}F_{5}S_{7}V_{2}$}} 
 \end{SplitTable}
   \vspace{0.4cm} 
 \captionof{table}{Dimension 9 $(\Delta B, \Delta L)$ = (1, -1) UV completions for topology 23 (see Fig.~\ref{fig:d=9_topologies}).} 
 \label{tab:T9Top23B1L-1}
 \end{center}

%% file: Tables_dim9/Table_dim9_Top24_B1_L-1.tex
 \begin{center}
 \begin{SplitTable}[2.0cm] 
  \SplitHeader{$\mathcal{O}^j{(\mathcal{T}_ {9} ^{ 24 })}$}{New Particles} 
   \vspace{0.1cm} 
 \SplitRow{ $\mathcal{O}_{9, \; (1,-1)}^{39}$} {{$S_{16}V_{4}V_{2}$, $S_{16}S_{3}S_{6}$}} 
 \SplitRow{ $\mathcal{O}_{9, \; (1,-1)}^{40}$} {{$S_{17}S_{5}S_{4}$, $S_{16}S_{5}S_{4}$, $S_{16}S_{2}S_{5}$, $S_{17}V_{2}V_{5}$, $S_{16}V_{2}V_{5}$, $S_{16}V_{4}V_{2}$}} 
 \SplitRow{ $\mathcal{O}_{9, \; (1,-1)}^{41}$} {{$S_{16}S_{3}S_{6}$, $S_{16}S_{2}S_{5}$}} 
 \SplitRow{ $\mathcal{O}_{9, \; (1,-1)}^{42}$} {{$S_{16}S_{3}S_{5}$}} 
 \SplitRow{ $\mathcal{O}_{9, \; (1,-1)}^{43}$} {{$S_{18}S_{6}S_{4}$}} 
 \SplitRow{ $\mathcal{O}_{9, \; (1,-1)}^{44}$} {{$S_{18}S_{6}S_{4}$, $S_{18}V_{1}V_{5}$}} 
 \SplitRow{ $\mathcal{O}_{9, \; (1,-1)}^{45}$} {{$S_{16}V_{3}V_{2}$, $S_{16}S_{3}S_{5}$}} 
 \SplitRow{ $\mathcal{O}_{9, \; (1,-1)}^{46}$} {{$S_{17}S_{5}S_{4}$, $S_{16}S_{5}S_{4}$, $S_{16}S_{2}S_{5}$}} 
 \SplitRow{ $\mathcal{O}_{9, \; (1,-1)}^{47}$} {{$S_{16}V_{4}V_{2}$, $S_{16}V_{3}V_{1}$, $S_{16}S_{2}S_{5}$}} 
 \end{SplitTable}
   \vspace{0.4cm} 
 \captionof{table}{Dimension 9 $(\Delta B, \Delta L)$ = (1, -1) UV completions for topology 24 (see Fig.~\ref{fig:d=9_topologies}).} 
 \label{tab:T9Top24B1L-1}
 \end{center}

%% file: Tables_dim9/Table_dim9_Top25_B1_L-1.tex
 \begin{center}
 \begin{SplitTable}[2.0cm] 
  \SplitHeader{$\mathcal{O}^j{(\mathcal{T}_ {9} ^{ 25 })}$}{New Particles} 
   \vspace{0.1cm} 
 \SplitRow{ $\mathcal{O}_{9, \; (1,-1)}^{39}$} {{$F_{7}S_{16}V_{2}$, $F_{5}S_{16}S_{3}$, $F_{3}S_{16}V_{2}$, $F_{7}S_{16}S_{6}$, $F_{5}S_{16}V_{4}$, $F_{3}S_{16}S_{3}$, $F_{7}S_{16}V_{4}$}} 
 \SplitRow{ $\mathcal{O}_{9, \; (1,-1)}^{40}$} {{$F_{20}S_{17}S_{4}$, $F_{7}S_{16}S_{4}$, $F_{7}S_{16}S_{2}$, $F_{9}S_{17}V_{2}$, $F_{9}S_{16}V_{2}$, $F_{5}S_{16}V_{2}$, $F_{4}S_{17}S_{4}$, $F_{4}S_{16}S_{4}$, $F_{1}S_{16}S_{2}$, $F_{9}S_{17}V_{5}$, $F_{9}S_{16}V_{5}$, $F_{5}S_{16}V_{4}$, $F_{4}S_{17}V_{2}$, $F_{4}S_{16}V_{2}$, $F_{1}S_{16}V_{2}$, $F_{20}S_{17}V_{5}$, $F_{7}S_{16}V_{5}$, $F_{7}S_{16}V_{4}$, $F_{9}S_{17}S_{5}$, $F_{9}S_{16}S_{5}$, $F_{5}S_{16}S_{5}$}} 
 \SplitRow{ $\mathcal{O}_{9, \; (1,-1)}^{41}$} {{$F_{8}S_{16}S_{3}$, $F_{7}S_{16}S_{2}$, $F_{1}S_{16}S_{3}$, $F_{7}S_{16}S_{5}$, $F_{1}S_{16}S_{2}$, $F_{8}S_{16}S_{5}$, $F_{7}S_{16}S_{6}$}} 
 \SplitRow{ $\mathcal{O}_{9, \; (1,-1)}^{42}$} {{$F_{8}S_{16}S_{3}$, $F_{2}S_{16}S_{3}$, $F_{8}S_{16}S_{5}$}} 
 \SplitRow{ $\mathcal{O}_{9, \; (1,-1)}^{43}$} {{$F_{19}S_{18}S_{6}$, $F_{12}S_{18}S_{4}$, $F_{19}S_{18}S_{4}$}} 
 \SplitRow{ $\mathcal{O}_{9, \; (1,-1)}^{44}$} {{$F_{19}S_{18}S_{6}$, $F_{20}S_{18}V_{5}$, $F_{11}S_{18}V_{1}$, $F_{19}S_{18}V_{5}$, $F_{11}S_{18}S_{4}$, $F_{19}S_{18}V_{1}$, $F_{20}S_{18}S_{4}$}} 
 \SplitRow{ $\mathcal{O}_{9, \; (1,-1)}^{45}$} {{$F_{8}S_{16}V_{2}$, $F_{6}S_{16}S_{3}$, $F_{3}S_{16}V_{2}$, $F_{8}S_{16}S_{5}$, $F_{6}S_{16}V_{3}$, $F_{3}S_{16}S_{3}$, $F_{8}S_{16}V_{3}$}} 
 \SplitRow{ $\mathcal{O}_{9, \; (1,-1)}^{46}$} {{$F_{9}S_{17}S_{4}$, $F_{9}S_{16}S_{4}$, $F_{5}S_{16}S_{2}$, $F_{12}S_{17}S_{4}$, $F_{3}S_{16}S_{4}$, $F_{3}S_{16}S_{2}$, $F_{9}S_{17}S_{5}$, $F_{9}S_{16}S_{5}$, $F_{5}S_{16}S_{5}$}} 
 \SplitRow{ $\mathcal{O}_{9, \; (1,-1)}^{47}$} {{$F_{8}S_{16}V_{2}$, $F_{7}S_{16}V_{1}$, $F_{5}S_{16}S_{2}$, $F_{3}S_{16}V_{2}$, $F_{7}S_{16}S_{5}$, $F_{5}S_{16}V_{3}$, $F_{3}S_{16}V_{1}$, $F_{8}S_{16}S_{5}$, $F_{5}S_{16}V_{4}$, $F_{3}S_{16}S_{2}$, $F_{8}S_{16}V_{3}$, $F_{7}S_{16}V_{4}$}} 
 \end{SplitTable}
   \vspace{0.4cm} 
 \captionof{table}{Dimension 9 $(\Delta B, \Delta L)$ = (1, -1) UV completions for topology 25 (see Fig.~\ref{fig:d=9_topologies}).} 
 \label{tab:T9Top25B1L-1}
 \end{center}

%% file: Tables_dim10/Table_dim10_Top1_B1_L-3.tex
 \begin{center}
 \begin{SplitTable}[2.0cm] 
  \SplitHeader{$\mathcal{O}^j{(\mathcal{T}_ {10} ^{ 1 })}$}{New Particles} 
   \vspace{0.1cm} 
 \SplitRow{ $\mathcal{O}_{10, \; (1,-3)}^{1}$} {{$S_{16}S_{9}S_{3}S_{5}$, $S_{14}S_{16}S_{3}S_{5}$, $S_{9}S_{3}S_{5}S_{4}$, $S_{14}S_{2}S_{3}S_{5}$, $S_{5}S_{5}S_{5}S_{4}$, $S_{2}S_{5}S_{5}S_{5}$, $S_{9}S_{3}S_{5}S_{6}$, $S_{14}S_{3}S_{5}S_{6}$}} 
 \SplitRow{ $\mathcal{O}_{10, \; (1,-3)}^{2}$} {{$S_{3}V_{9}V_{11}V_{4}$, $S_{14}S_{16}S_{3}S_{5}$, $S_{3}V_{10}V_{12}V_{3}$, $S_{3}V_{12}V_{3}V_{1}$, $S_{14}S_{2}S_{3}S_{5}$, $S_{3}V_{11}V_{4}V_{2}$, $S_{5}V_{3}V_{4}V_{1}$, $S_{2}S_{5}V_{3}V_{4}$, $S_{3}S_{6}V_{11}V_{4}$, $S_{5}V_{3}V_{4}V_{2}$, $S_{14}S_{3}S_{5}S_{6}$, $S_{3}S_{6}V_{12}V_{3}$}} 
 \SplitRow{ $\mathcal{O}_{10, \; (1,-3)}^{3}$} {{$S_{5}V_{11}V_{13}V_{2}$, $S_{5}V_{9}V_{11}V_{2}$, $S_{16}S_{9}V_{3}V_{2}$, $S_{14}S_{16}V_{3}V_{2}$, $S_{9}V_{3}V_{1}V_{2}$, $S_{14}V_{3}V_{1}V_{2}$, $S_{5}S_{4}V_{11}V_{2}$, $S_{2}S_{5}V_{11}V_{2}$, $S_{3}V_{11}V_{13}V_{5}$, $S_{3}V_{9}V_{11}V_{4}$, $S_{5}V_{3}V_{1}V_{5}$, $S_{5}V_{3}V_{4}V_{1}$, $S_{5}S_{4}V_{3}V_{5}$, $S_{2}S_{5}V_{3}V_{4}$, $S_{3}S_{6}V_{11}V_{5}$, $S_{3}S_{6}V_{11}V_{4}$, $S_{16}S_{9}S_{3}S_{5}$, $S_{14}S_{16}S_{3}S_{5}$, $S_{5}S_{5}S_{5}S_{4}$, $S_{2}S_{5}S_{5}S_{5}$, $S_{9}S_{3}S_{5}S_{6}$, $S_{14}S_{3}S_{5}S_{6}$, $S_{9}S_{3}S_{5}S_{4}$, $S_{14}S_{2}S_{3}S_{5}$, $S_{3}V_{11}V_{2}V_{5}$, $S_{3}V_{11}V_{4}V_{2}$, $S_{5}V_{3}V_{2}V_{5}$, $S_{5}V_{3}V_{4}V_{2}$, $S_{5}V_{11}V_{2}V_{5}$, $S_{5}V_{11}V_{4}V_{2}$, $S_{9}V_{3}V_{2}V_{5}$, $S_{14}V_{3}V_{4}V_{2}$}} 
 \SplitRow{ $\mathcal{O}_{10, \; (1,-3)}^{4}$} {{$S_{7}S_{16}S_{3}S_{6}$, $S_{3}V_{9}V_{11}V_{4}$, $S_{3}V_{11}V_{4}V_{2}$, $S_{7}S_{3}S_{3}S_{6}$, $S_{7}S_{16}S_{2}S_{5}$, $S_{2}V_{9}V_{11}V_{3}$, $S_{2}V_{11}V_{3}V_{1}$, $S_{7}S_{2}S_{2}S_{5}$, $S_{5}V_{3}V_{4}V_{2}$, $S_{5}V_{3}V_{4}V_{1}$, $S_{7}S_{2}S_{5}S_{5}$, $S_{3}S_{6}V_{3}V_{3}$, $S_{6}V_{3}V_{3}V_{1}$, $S_{2}S_{5}V_{3}V_{4}$, $S_{2}S_{5}V_{11}V_{3}$, $S_{7}S_{3}S_{6}S_{6}$, $S_{3}S_{6}V_{11}V_{4}$}} 
 \SplitRow{ $\mathcal{O}_{10, \; (1,-3)}^{5}$} {{$S_{7}S_{3}S_{5}S_{5}$, $S_{7}S_{3}S_{3}S_{5}$, $S_{3}S_{5}V_{11}V_{3}$, $S_{3}V_{11}V_{3}V_{2}$, $S_{3}S_{5}V_{3}V_{3}$, $S_{5}V_{3}V_{3}V_{2}$, $S_{3}V_{10}V_{11}V_{3}$, $S_{7}S_{16}S_{3}S_{5}$}} 
 \SplitRow{ $\mathcal{O}_{10, \; (1,-3)}^{6}$} {{$S_{7}S_{16}V_{4}V_{2}$, $S_{14}S_{16}V_{3}V_{2}$, $S_{14}V_{3}V_{1}V_{2}$, $S_{7}V_{4}V_{2}V_{2}$, $S_{7}S_{16}S_{3}S_{6}$, $S_{6}V_{3}V_{3}V_{1}$, $S_{7}S_{3}S_{6}S_{6}$, $S_{14}S_{16}S_{3}S_{5}$, $S_{5}V_{3}V_{4}V_{1}$, $S_{5}V_{3}V_{4}V_{2}$, $S_{14}S_{3}S_{5}S_{6}$, $S_{14}S_{2}S_{3}S_{5}$, $S_{7}S_{3}S_{3}S_{6}$, $S_{2}S_{5}V_{3}V_{4}$, $S_{7}V_{4}V_{4}V_{2}$, $S_{3}S_{6}V_{3}V_{3}$, $S_{14}V_{3}V_{4}V_{2}$}} 
 \SplitRow{ $\mathcal{O}_{10, \; (1,-3)}^{7}$} {{$S_{7}S_{16}S_{5}S_{4}$, $S_{7}S_{16}S_{2}S_{5}$, $S_{4}V_{11}V_{13}V_{3}$, $S_{2}V_{9}V_{11}V_{3}$, $S_{4}V_{11}V_{3}V_{1}$, $S_{2}V_{11}V_{3}V_{1}$, $S_{7}S_{5}S_{4}S_{4}$, $S_{7}S_{2}S_{2}S_{5}$, $S_{7}S_{16}V_{2}V_{5}$, $S_{7}S_{16}V_{4}V_{2}$, $S_{5}V_{11}V_{13}V_{2}$, $S_{5}V_{9}V_{11}V_{2}$, $S_{5}S_{4}V_{11}V_{2}$, $S_{2}S_{5}V_{11}V_{2}$, $S_{7}V_{2}V_{2}V_{5}$, $S_{7}V_{4}V_{2}V_{2}$, $S_{5}V_{3}V_{1}V_{5}$, $S_{5}V_{3}V_{4}V_{1}$, $S_{5}S_{4}V_{3}V_{5}$, $S_{2}S_{5}V_{3}V_{4}$, $S_{7}V_{2}V_{5}V_{5}$, $S_{7}V_{4}V_{4}V_{2}$, $S_{5}S_{5}S_{5}S_{4}$, $S_{2}S_{5}S_{5}S_{5}$, $S_{5}V_{3}V_{2}V_{5}$, $S_{5}V_{3}V_{4}V_{2}$, $S_{5}V_{11}V_{2}V_{5}$, $S_{5}V_{11}V_{4}V_{2}$, $S_{7}S_{5}S_{5}S_{4}$, $S_{7}S_{2}S_{5}S_{5}$, $S_{5}S_{4}V_{11}V_{3}$, $S_{2}S_{5}V_{11}V_{3}$}} 
 \SplitRow{ $\mathcal{O}_{10, \; (1,-3)}^{8}$} {{$S_{7}V_{3}V_{3}V_{2}$, $S_{7}S_{3}S_{5}S_{5}$, $S_{7}S_{3}S_{3}S_{5}$, $S_{7}V_{3}V_{2}V_{2}$, $S_{3}S_{5}V_{3}V_{3}$, $S_{5}V_{3}V_{3}V_{2}$, $S_{7}S_{16}V_{3}V_{2}$, $S_{7}S_{16}S_{3}S_{5}$}} 
 \SplitRow{ $\mathcal{O}_{10, \; (1,-3)}^{9}$} {{$S_{7}S_{16}V_{4}V_{2}$, $S_{7}V_{4}V_{2}V_{2}$, $S_{7}S_{16}V_{3}V_{1}$, $S_{7}V_{3}V_{1}V_{1}$, $S_{7}S_{16}S_{2}S_{5}$, $S_{5}V_{3}V_{4}V_{2}$, $S_{5}V_{3}V_{4}V_{1}$, $S_{7}S_{2}S_{5}S_{5}$, $S_{7}S_{2}S_{2}S_{5}$, $S_{2}S_{5}V_{3}V_{4}$, $S_{7}V_{3}V_{3}V_{1}$, $S_{7}V_{4}V_{4}V_{2}$}} 
 \SplitRow{ $\mathcal{O}_{10, \; (1,-3)}^{10}$} {{$S_{7}S_{16}S_{5}S_{4}$, $S_{7}S_{16}S_{2}S_{5}$, $S_{7}S_{5}S_{4}S_{4}$, $S_{7}S_{2}S_{2}S_{5}$, $S_{5}S_{5}S_{5}S_{4}$, $S_{2}S_{5}S_{5}S_{5}$, $S_{7}S_{5}S_{5}S_{4}$, $S_{7}S_{2}S_{5}S_{5}$}} 
 \end{SplitTable}
   \vspace{0.4cm} 
 \captionof{table}{Dimension 10 $(\Delta B, \Delta L)$ = (1, -3) UV completions for topology 1 (see Fig.~\ref{fig:d=10_topologies}).} 
 \label{tab:T10Top1B1L-3}
 \end{center}

%% file: Tables_dim10/Table_dim10_Top2_B1_L-3.tex
 \begin{center}
 \begin{SplitTable}[2.0cm] 
  \SplitHeader{$\mathcal{O}^j{(\mathcal{T}_ {10} ^{ 2 })}$}{New Particles} 
   \vspace{0.1cm} 
 \SplitRow{ $\mathcal{O}_{10, \; (1,-3)}^{1}$} {{$F_{8}F_{14}S_{9}S_{5}$, $F_{6}F_{8}S_{14}S_{5}$, $F_{2}F_{10}S_{9}S_{3}$, $F_{2}F_{10}S_{14}S_{3}$, $F_{8}F_{14}S_{5}S_{5}$, $F_{6}F_{8}S_{5}S_{5}$, $F_{8}F_{14}S_{9}S_{3}$, $F_{6}F_{8}S_{14}S_{3}$, $F_{8}F_{14}S_{3}S_{5}$, $F_{6}F_{8}S_{3}S_{5}$}} 
 \SplitRow{ $\mathcal{O}_{10, \; (1,-3)}^{2}$} {{$F_{5}F_{8}V_{11}V_{4}$, $F_{6}F_{8}S_{14}S_{5}$, $F_{6}F_{13}V_{12}V_{3}$, $F_{2}F_{10}S_{3}V_{11}$, $F_{6}F_{8}S_{5}V_{3}$, $F_{6}F_{13}S_{5}V_{3}$, $F_{5}F_{8}S_{3}V_{11}$, $F_{2}F_{10}S_{14}S_{3}$, $F_{5}F_{8}V_{3}V_{4}$, $F_{6}F_{13}V_{3}V_{4}$, $F_{6}F_{8}S_{14}S_{3}$, $F_{2}F_{10}S_{3}V_{12}$, $F_{5}F_{8}S_{5}V_{4}$, $F_{6}F_{8}S_{5}V_{4}$, $F_{6}F_{13}S_{3}V_{12}$, $F_{5}F_{8}S_{3}V_{3}$, $F_{6}F_{8}S_{3}V_{3}$, $F_{5}F_{8}S_{3}S_{5}$, $F_{6}F_{13}S_{3}S_{5}$, $F_{6}F_{8}S_{3}V_{4}$, $F_{6}F_{13}S_{3}V_{4}$}} 
 \SplitRow{ $\mathcal{O}_{10, \; (1,-3)}^{3}$} {{$F_{8}F_{9}S_{5}V_{11}$, $F_{5}F_{8}S_{5}V_{11}$, $F_{8}F_{14}S_{9}V_{3}$, $F_{6}F_{8}S_{14}V_{3}$, $F_{6}F_{13}S_{9}V_{3}$, $F_{6}F_{13}S_{14}V_{3}$, $F_{8}F_{14}S_{5}V_{11}$, $F_{6}F_{8}S_{5}V_{11}$, $F_{3}F_{11}S_{3}V_{11}$, $F_{2}F_{3}S_{3}V_{11}$, $F_{8}F_{9}S_{5}V_{3}$, $F_{5}F_{8}S_{5}V_{3}$, $F_{8}F_{14}S_{5}V_{3}$, $F_{6}F_{8}S_{5}V_{3}$, $F_{8}F_{14}S_{3}V_{11}$, $F_{6}F_{8}S_{3}V_{11}$, $F_{3}F_{11}S_{9}S_{3}$, $F_{2}F_{3}S_{14}S_{3}$, $F_{8}F_{9}S_{5}S_{5}$, $F_{5}F_{8}S_{5}S_{5}$, $F_{6}F_{13}S_{9}S_{3}$, $F_{6}F_{13}S_{14}S_{3}$, $F_{8}F_{9}V_{11}V_{5}$, $F_{5}F_{8}V_{11}V_{4}$, $F_{8}F_{14}S_{9}S_{5}$, $F_{6}F_{8}S_{14}S_{5}$, $F_{3}F_{11}V_{11}V_{2}$, $F_{2}F_{3}V_{11}V_{2}$, $F_{8}F_{9}V_{3}V_{5}$, $F_{5}F_{8}V_{3}V_{4}$, $F_{8}F_{14}S_{5}S_{5}$, $F_{6}F_{8}S_{5}S_{5}$, $F_{6}F_{13}V_{3}V_{5}$, $F_{6}F_{13}V_{3}V_{4}$, $F_{8}F_{9}V_{11}V_{2}$, $F_{5}F_{8}V_{11}V_{2}$, $F_{3}F_{11}S_{9}V_{2}$, $F_{2}F_{3}S_{14}V_{2}$, $F_{8}F_{9}S_{5}V_{5}$, $F_{5}F_{8}S_{5}V_{4}$, $F_{8}F_{14}S_{5}V_{5}$, $F_{6}F_{8}S_{5}V_{4}$, $F_{8}F_{14}S_{9}V_{2}$, $F_{6}F_{8}S_{14}V_{2}$, $F_{8}F_{9}V_{3}V_{2}$, $F_{5}F_{8}V_{3}V_{2}$, $F_{8}F_{14}S_{3}S_{5}$, $F_{6}F_{8}S_{3}S_{5}$, $F_{8}F_{9}S_{5}V_{2}$, $F_{5}F_{8}S_{5}V_{2}$, $F_{6}F_{13}S_{3}V_{5}$, $F_{6}F_{13}S_{3}V_{4}$, $F_{8}F_{14}S_{3}V_{5}$, $F_{6}F_{8}S_{3}V_{4}$, $F_{8}F_{14}S_{5}V_{2}$, $F_{6}F_{8}S_{5}V_{2}$}} 
 \SplitRow{ $\mathcal{O}_{10, \; (1,-3)}^{4}$} {{$F_{5}F_{7}S_{7}S_{6}$, $F_{5}F_{8}V_{11}V_{4}$, $F_{6}F_{8}S_{7}S_{5}$, $F_{6}F_{13}V_{11}V_{3}$, $F_{5}F_{8}V_{11}V_{3}$, $F_{5}F_{7}S_{7}S_{5}$, $F_{1}F_{3}S_{7}S_{2}$, $F_{6}F_{13}V_{3}V_{4}$, $F_{5}F_{8}V_{3}V_{4}$, $F_{5}F_{7}S_{7}S_{2}$, $F_{1}F_{3}S_{2}V_{11}$, $F_{6}F_{8}S_{5}V_{4}$, $F_{6}F_{13}S_{6}V_{3}$, $F_{5}F_{7}S_{6}V_{3}$, $F_{5}F_{8}S_{5}V_{4}$, $F_{5}F_{8}S_{2}V_{11}$, $F_{1}F_{3}S_{7}S_{3}$, $F_{5}F_{8}V_{3}V_{3}$, $F_{6}F_{8}S_{7}S_{3}$, $F_{1}F_{3}S_{3}V_{11}$, $F_{5}F_{7}S_{5}V_{3}$, $F_{5}F_{8}S_{5}V_{3}$, $F_{6}F_{13}S_{3}V_{11}$, $F_{6}F_{8}S_{3}V_{4}$, $F_{5}F_{7}S_{2}V_{3}$, $F_{5}F_{8}S_{2}V_{3}$, $F_{6}F_{13}S_{3}V_{4}$, $F_{6}F_{13}S_{3}S_{6}$, $F_{5}F_{8}S_{2}S_{5}$}} 
 \SplitRow{ $\mathcal{O}_{10, \; (1,-3)}^{5}$} {{$F_{6}F_{8}S_{3}S_{5}$, $F_{5}F_{8}S_{3}V_{3}$, $F_{6}F_{8}S_{3}V_{3}$, $F_{2}F_{3}S_{3}V_{11}$, $F_{5}F_{8}S_{5}V_{3}$, $F_{6}F_{8}S_{5}V_{3}$, $F_{6}F_{8}S_{3}V_{11}$, $F_{2}F_{3}S_{7}S_{3}$, $F_{6}F_{8}V_{3}V_{3}$, $F_{5}F_{8}S_{7}S_{3}$, $F_{5}F_{8}S_{7}S_{5}$, $F_{6}F_{8}V_{11}V_{3}$}} 
 \SplitRow{ $\mathcal{O}_{10, \; (1,-3)}^{6}$} {{$F_{5}F_{7}S_{7}V_{4}$, $F_{6}F_{8}S_{14}V_{3}$, $F_{6}F_{13}S_{14}V_{3}$, $F_{5}F_{8}S_{7}V_{4}$, $F_{2}F_{3}S_{7}S_{3}$, $F_{6}F_{8}V_{3}V_{3}$, $F_{5}F_{8}S_{7}S_{3}$, $F_{2}F_{3}S_{14}S_{3}$, $F_{5}F_{7}V_{3}V_{4}$, $F_{6}F_{8}V_{3}V_{4}$, $F_{6}F_{13}S_{14}S_{3}$, $F_{5}F_{7}S_{7}S_{6}$, $F_{6}F_{8}S_{14}S_{5}$, $F_{2}F_{3}S_{7}V_{2}$, $F_{6}F_{8}S_{5}V_{3}$, $F_{6}F_{13}S_{5}V_{3}$, $F_{5}F_{7}S_{7}V_{2}$, $F_{2}F_{3}S_{14}V_{2}$, $F_{5}F_{7}S_{6}V_{3}$, $F_{6}F_{8}S_{5}V_{4}$, $F_{5}F_{8}S_{5}V_{4}$, $F_{6}F_{13}S_{6}V_{3}$, $F_{6}F_{8}S_{14}V_{2}$, $F_{5}F_{7}V_{3}V_{2}$, $F_{6}F_{13}S_{3}S_{5}$, $F_{5}F_{8}S_{3}S_{5}$, $F_{6}F_{8}V_{3}V_{2}$, $F_{6}F_{8}V_{4}V_{2}$, $F_{6}F_{13}S_{3}S_{6}$}} 
 \SplitRow{ $\mathcal{O}_{10, \; (1,-3)}^{7}$} {{$F_{8}F_{9}S_{7}S_{5}$, $F_{5}F_{8}S_{7}S_{5}$, $F_{8}F_{14}V_{11}V_{3}$, $F_{6}F_{8}V_{11}V_{3}$, $F_{5}F_{7}S_{7}V_{5}$, $F_{5}F_{7}S_{7}V_{4}$, $F_{8}F_{9}S_{5}V_{11}$, $F_{5}F_{8}S_{5}V_{11}$, $F_{8}F_{14}S_{5}V_{11}$, $F_{6}F_{8}S_{5}V_{11}$, $F_{8}F_{9}S_{7}V_{5}$, $F_{5}F_{8}S_{7}V_{4}$, $F_{3}F_{4}S_{7}V_{2}$, $F_{1}F_{3}S_{7}V_{2}$, $F_{8}F_{9}S_{5}V_{3}$, $F_{5}F_{8}S_{5}V_{3}$, $F_{8}F_{14}S_{5}V_{3}$, $F_{6}F_{8}S_{5}V_{3}$, $F_{8}F_{9}S_{7}V_{2}$, $F_{5}F_{8}S_{7}V_{2}$, $F_{3}F_{4}V_{11}V_{2}$, $F_{1}F_{3}V_{11}V_{2}$, $F_{5}F_{7}V_{3}V_{5}$, $F_{5}F_{7}V_{3}V_{4}$, $F_{8}F_{9}S_{5}S_{5}$, $F_{5}F_{8}S_{5}S_{5}$, $F_{8}F_{14}V_{3}V_{5}$, $F_{6}F_{8}V_{3}V_{4}$, $F_{8}F_{14}V_{11}V_{2}$, $F_{6}F_{8}V_{11}V_{2}$, $F_{3}F_{4}S_{7}S_{4}$, $F_{1}F_{3}S_{7}S_{2}$, $F_{8}F_{14}S_{5}S_{5}$, $F_{6}F_{8}S_{5}S_{5}$, $F_{5}F_{7}S_{7}S_{4}$, $F_{5}F_{7}S_{7}S_{2}$, $F_{3}F_{4}S_{4}V_{11}$, $F_{1}F_{3}S_{2}V_{11}$, $F_{8}F_{9}S_{5}V_{5}$, $F_{5}F_{8}S_{5}V_{4}$, $F_{8}F_{14}S_{5}V_{5}$, $F_{6}F_{8}S_{5}V_{4}$, $F_{8}F_{9}S_{4}V_{11}$, $F_{5}F_{8}S_{2}V_{11}$, $F_{5}F_{7}S_{4}V_{3}$, $F_{5}F_{7}S_{2}V_{3}$, $F_{8}F_{9}S_{5}V_{2}$, $F_{5}F_{8}S_{5}V_{2}$, $F_{8}F_{14}S_{5}V_{2}$, $F_{6}F_{8}S_{5}V_{2}$, $F_{8}F_{9}S_{4}V_{3}$, $F_{5}F_{8}S_{2}V_{3}$, $F_{8}F_{9}S_{5}S_{4}$, $F_{5}F_{8}S_{2}S_{5}$, $F_{8}F_{14}V_{2}V_{5}$, $F_{6}F_{8}V_{4}V_{2}$}} 
 \SplitRow{ $\mathcal{O}_{10, \; (1,-3)}^{8}$} {{$F_{5}F_{8}V_{3}V_{2}$, $F_{6}F_{8}S_{3}S_{5}$, $F_{1}F_{3}S_{7}V_{2}$, $F_{5}F_{8}S_{5}V_{3}$, $F_{6}F_{8}S_{5}V_{3}$, $F_{5}F_{8}S_{7}V_{2}$, $F_{1}F_{3}S_{7}S_{3}$, $F_{5}F_{8}V_{3}V_{3}$, $F_{6}F_{8}S_{7}S_{3}$, $F_{5}F_{8}S_{7}S_{5}$, $F_{5}F_{8}S_{7}V_{3}$, $F_{6}F_{8}S_{7}V_{3}$}} 
 \SplitRow{ $\mathcal{O}_{10, \; (1,-3)}^{9}$} {{$F_{5}F_{7}S_{7}V_{4}$, $F_{5}F_{8}S_{7}V_{4}$, $F_{6}F_{8}S_{7}V_{3}$, $F_{5}F_{8}S_{7}V_{3}$, $F_{1}F_{3}S_{7}S_{2}$, $F_{6}F_{8}V_{3}V_{4}$, $F_{5}F_{7}V_{3}V_{4}$, $F_{5}F_{8}S_{7}S_{2}$, $F_{6}F_{8}S_{7}S_{5}$, $F_{5}F_{7}S_{7}S_{5}$, $F_{1}F_{3}S_{7}V_{1}$, $F_{6}F_{8}S_{5}V_{4}$, $F_{5}F_{8}S_{5}V_{4}$, $F_{5}F_{7}S_{7}V_{1}$, $F_{1}F_{3}S_{7}V_{2}$, $F_{5}F_{7}S_{5}V_{3}$, $F_{5}F_{8}S_{5}V_{3}$, $F_{6}F_{8}S_{7}V_{2}$, $F_{6}F_{8}V_{4}V_{2}$, $F_{5}F_{7}V_{3}V_{1}$, $F_{5}F_{8}S_{2}S_{5}$}} 
 \SplitRow{ $\mathcal{O}_{10, \; (1,-3)}^{10}$} {{$F_{8}F_{9}S_{7}S_{5}$, $F_{5}F_{8}S_{7}S_{5}$, $F_{1}F_{3}S_{7}S_{4}$, $F_{1}F_{3}S_{7}S_{2}$, $F_{8}F_{9}S_{5}S_{5}$, $F_{5}F_{8}S_{5}S_{5}$, $F_{8}F_{9}S_{7}S_{4}$, $F_{5}F_{8}S_{7}S_{2}$, $F_{8}F_{9}S_{5}S_{4}$, $F_{5}F_{8}S_{2}S_{5}$}} 
 \end{SplitTable}
   \vspace{0.4cm} 
 \captionof{table}{Dimension 10 $(\Delta B, \Delta L)$ = (1, -3) UV completions for topology 2 (see Fig.~\ref{fig:d=10_topologies}).} 
 \label{tab:T10Top2B1L-3}
 \end{center}

%% file: Tables_dim10/Table_dim10_Top3_B1_L-3.tex
 \begin{center}
 \begin{SplitTable}[2.0cm] 
  \SplitHeader{$\mathcal{O}^j{(\mathcal{T}_ {10} ^{ 3 })}$}{New Particles} 
   \vspace{0.1cm} 
 \SplitRow{ $\mathcal{O}_{10, \; (1,-3)}^{1}$} {{$F_{7}F_{14}S_{9}S_{5}$, $F_{6}F_{7}S_{14}S_{5}$, $F_{7}F_{8}S_{9}S_{5}$, $F_{7}F_{8}S_{14}S_{5}$, $F_{4}F_{14}S_{9}S_{3}$, $F_{1}F_{6}S_{14}S_{3}$, $F_{4}F_{8}S_{5}S_{5}$, $F_{1}F_{8}S_{5}S_{5}$, $F_{2}F_{4}S_{9}S_{3}$, $F_{1}F_{2}S_{14}S_{3}$, $F_{10}F_{7}S_{9}S_{3}$, $F_{10}F_{7}S_{14}S_{3}$, $F_{7}F_{14}S_{5}S_{5}$, $F_{6}F_{7}S_{5}S_{5}$, $F_{7}F_{8}S_{9}S_{3}$, $F_{7}F_{8}S_{14}S_{3}$, $F_{4}F_{14}S_{3}S_{5}$, $F_{1}F_{6}S_{3}S_{5}$, $F_{4}F_{8}S_{3}S_{5}$, $F_{1}F_{8}S_{3}S_{5}$}} 
 \SplitRow{ $\mathcal{O}_{10, \; (1,-3)}^{2}$} {{$F_{7}F_{8}V_{11}V_{4}$, $F_{6}F_{7}S_{14}S_{5}$, $F_{7}F_{13}V_{12}V_{3}$, $F_{6}F_{7}V_{12}V_{3}$, $F_{7}F_{8}S_{14}S_{5}$, $F_{5}F_{7}V_{11}V_{4}$, $F_{3}F_{8}S_{3}V_{11}$, $F_{3}F_{6}S_{5}V_{3}$, $F_{3}F_{8}S_{5}V_{3}$, $F_{2}F_{3}S_{3}V_{11}$, $F_{1}F_{6}S_{14}S_{3}$, $F_{1}F_{6}V_{3}V_{4}$, $F_{1}F_{5}V_{3}V_{4}$, $F_{1}F_{2}S_{14}S_{3}$, $F_{3}F_{13}S_{3}V_{12}$, $F_{3}F_{8}S_{5}V_{4}$, $F_{3}F_{5}S_{5}V_{4}$, $F_{2}F_{3}S_{3}V_{12}$, $F_{10}F_{7}S_{3}V_{11}$, $F_{6}F_{7}S_{5}V_{3}$, $F_{7}F_{13}S_{5}V_{3}$, $F_{5}F_{7}S_{3}V_{11}$, $F_{1}F_{6}S_{3}V_{3}$, $F_{1}F_{5}S_{3}V_{3}$, $F_{3}F_{13}S_{3}S_{5}$, $F_{3}F_{5}S_{3}S_{5}$, $F_{10}F_{7}S_{14}S_{3}$, $F_{7}F_{8}V_{3}V_{4}$, $F_{7}F_{13}V_{3}V_{4}$, $F_{7}F_{8}S_{14}S_{3}$, $F_{3}F_{8}S_{3}V_{3}$, $F_{3}F_{13}S_{3}V_{4}$, $F_{3}F_{8}S_{3}V_{4}$, $F_{10}F_{7}S_{3}V_{12}$, $F_{7}F_{8}S_{5}V_{4}$, $F_{6}F_{7}S_{5}V_{4}$, $F_{6}F_{7}S_{3}V_{12}$, $F_{3}F_{8}S_{3}S_{5}$, $F_{3}F_{6}S_{3}S_{5}$, $F_{1}F_{6}S_{3}V_{4}$}} 
 \SplitRow{ $\mathcal{O}_{10, \; (1,-3)}^{3}$} {{$F_{7}F_{14}S_{5}V_{11}$, $F_{6}F_{7}S_{5}V_{11}$, $F_{7}F_{13}S_{9}V_{3}$, $F_{7}F_{13}S_{14}V_{3}$, $F_{7}F_{8}S_{9}V_{3}$, $F_{7}F_{8}S_{14}V_{3}$, $F_{7}F_{9}S_{5}V_{11}$, $F_{5}F_{7}S_{5}V_{11}$, $F_{4}F_{14}S_{3}V_{11}$, $F_{1}F_{6}S_{3}V_{11}$, $F_{4}F_{8}S_{5}V_{3}$, $F_{1}F_{8}S_{5}V_{3}$, $F_{4}F_{9}S_{5}V_{3}$, $F_{1}F_{5}S_{5}V_{3}$, $F_{3}F_{4}S_{3}V_{11}$, $F_{1}F_{3}S_{3}V_{11}$, $F_{3}F_{13}S_{9}S_{3}$, $F_{3}F_{13}S_{14}S_{3}$, $F_{3}F_{9}S_{5}S_{5}$, $F_{3}F_{5}S_{5}S_{5}$, $F_{3}F_{3}S_{9}S_{3}$, $F_{3}F_{3}S_{14}S_{3}$, $F_{8}F_{9}S_{5}V_{11}$, $F_{5}F_{8}S_{5}V_{11}$, $F_{9}F_{14}S_{9}V_{3}$, $F_{5}F_{6}S_{14}V_{3}$, $F_{6}F_{9}S_{9}V_{3}$, $F_{7}F_{8}V_{11}V_{5}$, $F_{7}F_{8}V_{11}V_{4}$, $F_{7}F_{14}S_{9}S_{5}$, $F_{6}F_{7}S_{14}S_{5}$, $F_{7}F_{8}S_{9}S_{5}$, $F_{7}F_{8}S_{14}S_{5}$, $F_{7}F_{9}V_{11}V_{5}$, $F_{5}F_{7}V_{11}V_{4}$, $F_{4}F_{8}V_{11}V_{2}$, $F_{1}F_{8}V_{11}V_{2}$, $F_{4}F_{6}V_{3}V_{5}$, $F_{1}F_{6}V_{3}V_{4}$, $F_{4}F_{8}S_{5}S_{5}$, $F_{1}F_{8}S_{5}S_{5}$, $F_{4}F_{9}V_{3}V_{5}$, $F_{1}F_{5}V_{3}V_{4}$, $F_{3}F_{4}V_{11}V_{2}$, $F_{1}F_{3}V_{11}V_{2}$, $F_{3}F_{14}S_{9}V_{2}$, $F_{3}F_{6}S_{14}V_{2}$, $F_{3}F_{8}S_{5}V_{5}$, $F_{3}F_{8}S_{5}V_{4}$, $F_{3}F_{9}S_{5}V_{5}$, $F_{3}F_{5}S_{5}V_{4}$, $F_{3}F_{3}S_{9}V_{2}$, $F_{3}F_{3}S_{14}V_{2}$, $F_{11}F_{9}S_{3}V_{11}$, $F_{2}F_{5}S_{3}V_{11}$, $F_{8}F_{9}S_{5}V_{3}$, $F_{5}F_{8}S_{5}V_{3}$, $F_{9}F_{14}S_{5}V_{3}$, $F_{5}F_{6}S_{5}V_{3}$, $F_{8}F_{9}S_{3}V_{11}$, $F_{5}F_{8}S_{3}V_{11}$, $F_{11}F_{7}V_{11}V_{2}$, $F_{2}F_{7}V_{11}V_{2}$, $F_{7}F_{8}V_{3}V_{5}$, $F_{7}F_{8}V_{3}V_{4}$, $F_{7}F_{14}S_{5}S_{5}$, $F_{6}F_{7}S_{5}S_{5}$, $F_{7}F_{13}V_{3}V_{5}$, $F_{7}F_{13}V_{3}V_{4}$, $F_{7}F_{9}V_{11}V_{2}$, $F_{5}F_{7}V_{11}V_{2}$, $F_{4}F_{8}V_{3}V_{2}$, $F_{1}F_{8}V_{3}V_{2}$, $F_{4}F_{8}S_{3}S_{5}$, $F_{1}F_{8}S_{3}S_{5}$, $F_{4}F_{14}S_{3}S_{5}$, $F_{1}F_{6}S_{3}S_{5}$, $F_{4}F_{9}V_{3}V_{2}$, $F_{1}F_{5}V_{3}V_{2}$, $F_{3}F_{14}S_{5}V_{2}$, $F_{3}F_{6}S_{5}V_{2}$, $F_{3}F_{8}S_{3}V_{5}$, $F_{3}F_{8}S_{3}V_{4}$, $F_{3}F_{13}S_{3}V_{5}$, $F_{3}F_{13}S_{3}V_{4}$, $F_{3}F_{9}S_{5}V_{2}$, $F_{3}F_{5}S_{5}V_{2}$, $F_{11}F_{9}S_{9}S_{3}$, $F_{2}F_{5}S_{14}S_{3}$, $F_{8}F_{9}S_{5}S_{5}$, $F_{5}F_{8}S_{5}S_{5}$, $F_{6}F_{9}S_{9}S_{3}$, $F_{5}F_{6}S_{14}S_{3}$, $F_{11}F_{7}S_{9}V_{2}$, $F_{2}F_{7}S_{14}V_{2}$, $F_{7}F_{8}S_{5}V_{5}$, $F_{7}F_{8}S_{5}V_{4}$, $F_{7}F_{14}S_{5}V_{5}$, $F_{6}F_{7}S_{5}V_{4}$, $F_{7}F_{8}S_{9}V_{2}$, $F_{7}F_{8}S_{14}V_{2}$, $F_{4}F_{8}S_{5}V_{2}$, $F_{1}F_{8}S_{5}V_{2}$, $F_{4}F_{6}S_{3}V_{5}$, $F_{1}F_{6}S_{3}V_{4}$, $F_{4}F_{14}S_{3}V_{5}$}} 
 \SplitRow{ $\mathcal{O}_{10, \; (1,-3)}^{4}$} {{$F_{5}F_{7}S_{7}S_{6}$, $F_{7}F_{8}V_{11}V_{4}$, $F_{5}F_{7}V_{11}V_{4}$, $F_{7}F_{7}S_{7}S_{6}$, $F_{5}F_{8}S_{7}S_{5}$, $F_{8}F_{8}V_{11}V_{3}$, $F_{6}F_{8}V_{11}V_{3}$, $F_{8}F_{8}S_{7}S_{5}$, $F_{1}F_{5}S_{7}S_{2}$, $F_{1}F_{5}V_{3}V_{4}$, $F_{1}F_{6}V_{3}V_{4}$, $F_{1}F_{1}S_{7}S_{2}$, $F_{3}F_{8}S_{2}V_{11}$, $F_{3}F_{5}S_{5}V_{4}$, $F_{3}F_{7}S_{6}V_{3}$, $F_{3}F_{6}S_{6}V_{3}$, $F_{3}F_{8}S_{5}V_{4}$, $F_{1}F_{3}S_{2}V_{11}$, $F_{6}F_{7}S_{7}S_{5}$, $F_{7}F_{13}V_{11}V_{3}$, $F_{5}F_{7}V_{11}V_{3}$, $F_{7}F_{7}S_{7}S_{5}$, $F_{1}F_{6}S_{7}S_{3}$, $F_{1}F_{5}V_{3}V_{3}$, $F_{1}F_{1}S_{7}S_{3}$, $F_{3}F_{13}S_{3}V_{11}$, $F_{3}F_{5}S_{5}V_{3}$, $F_{3}F_{7}S_{5}V_{3}$, $F_{1}F_{3}S_{3}V_{11}$, $F_{3}F_{7}S_{7}S_{2}$, $F_{7}F_{13}V_{3}V_{4}$, $F_{7}F_{8}V_{3}V_{4}$, $F_{7}F_{7}S_{7}S_{2}$, $F_{3}F_{8}S_{7}S_{3}$, $F_{8}F_{8}V_{3}V_{3}$, $F_{8}F_{8}S_{7}S_{3}$, $F_{3}F_{13}S_{3}V_{4}$, $F_{3}F_{8}S_{2}V_{3}$, $F_{3}F_{7}S_{2}V_{3}$, $F_{3}F_{8}S_{3}V_{4}$, $F_{3}F_{7}S_{2}V_{11}$, $F_{6}F_{7}S_{5}V_{4}$, $F_{7}F_{13}S_{6}V_{3}$, $F_{5}F_{7}S_{6}V_{3}$, $F_{7}F_{8}S_{5}V_{4}$, $F_{5}F_{7}S_{2}V_{11}$, $F_{3}F_{8}S_{3}V_{11}$, $F_{5}F_{8}S_{5}V_{3}$, $F_{8}F_{8}S_{5}V_{3}$, $F_{6}F_{8}S_{3}V_{11}$, $F_{1}F_{6}S_{3}V_{4}$, $F_{1}F_{5}S_{2}V_{3}$, $F_{3}F_{13}S_{3}S_{6}$, $F_{3}F_{8}S_{2}S_{5}$, $F_{3}F_{5}S_{2}S_{5}$, $F_{3}F_{6}S_{3}S_{6}$}} 
 \SplitRow{ $\mathcal{O}_{10, \; (1,-3)}^{5}$} {{$F_{3}F_{6}S_{3}S_{5}$, $F_{3}F_{8}S_{3}S_{5}$, $F_{2}F_{6}S_{3}V_{3}$, $F_{2}F_{5}S_{3}V_{3}$, $F_{6}F_{8}S_{3}V_{11}$, $F_{8}F_{8}S_{5}V_{3}$, $F_{5}F_{8}S_{5}V_{3}$, $F_{3}F_{8}S_{3}V_{11}$, $F_{3}F_{8}S_{3}V_{3}$, $F_{8}F_{8}S_{7}S_{3}$, $F_{8}F_{8}V_{3}V_{3}$, $F_{3}F_{8}S_{7}S_{3}$, $F_{2}F_{3}S_{3}V_{11}$, $F_{3}F_{8}S_{5}V_{3}$, $F_{3}F_{6}S_{5}V_{3}$, $F_{2}F_{2}S_{7}S_{3}$, $F_{2}F_{6}V_{3}V_{3}$, $F_{2}F_{5}S_{7}S_{3}$, $F_{8}F_{8}S_{7}S_{5}$, $F_{6}F_{8}V_{11}V_{3}$, $F_{8}F_{8}V_{11}V_{3}$, $F_{5}F_{8}S_{7}S_{5}$}} 
 \SplitRow{ $\mathcal{O}_{10, \; (1,-3)}^{6}$} {{$F_{7}F_{8}S_{7}V_{4}$, $F_{7}F_{13}S_{14}V_{3}$, $F_{7}F_{8}S_{14}V_{3}$, $F_{7}F_{7}S_{7}V_{4}$, $F_{3}F_{8}S_{7}S_{3}$, $F_{3}F_{8}V_{3}V_{3}$, $F_{3}F_{3}S_{7}S_{3}$, $F_{3}F_{13}S_{14}S_{3}$, $F_{3}F_{8}V_{3}V_{4}$, $F_{3}F_{7}V_{3}V_{4}$, $F_{3}F_{3}S_{14}S_{3}$, $F_{5}F_{5}S_{7}V_{4}$, $F_{5}F_{6}S_{14}V_{3}$, $F_{5}F_{7}S_{7}S_{6}$, $F_{6}F_{7}S_{14}S_{5}$, $F_{7}F_{8}S_{14}S_{5}$, $F_{7}F_{7}S_{7}S_{6}$, $F_{3}F_{5}S_{7}V_{2}$, $F_{3}F_{6}S_{5}V_{3}$, $F_{3}F_{8}S_{5}V_{3}$, $F_{3}F_{3}S_{7}V_{2}$, $F_{3}F_{6}S_{14}V_{2}$, $F_{3}F_{6}S_{6}V_{3}$, $F_{3}F_{5}S_{5}V_{4}$, $F_{3}F_{8}S_{5}V_{4}$, $F_{3}F_{7}S_{6}V_{3}$, $F_{3}F_{3}S_{14}V_{2}$, $F_{2}F_{5}S_{7}S_{3}$, $F_{5}F_{6}V_{3}V_{3}$, $F_{5}F_{5}S_{7}S_{3}$, $F_{2}F_{7}S_{7}V_{2}$, $F_{6}F_{7}S_{5}V_{3}$, $F_{7}F_{13}S_{5}V_{3}$, $F_{7}F_{7}S_{7}V_{2}$, $F_{3}F_{6}V_{3}V_{2}$, $F_{3}F_{5}S_{3}S_{5}$, $F_{3}F_{13}S_{3}S_{5}$, $F_{3}F_{7}V_{3}V_{2}$, $F_{2}F_{5}S_{14}S_{3}$, $F_{5}F_{5}V_{3}V_{4}$, $F_{5}F_{6}V_{3}V_{4}$, $F_{5}F_{6}S_{14}S_{3}$, $F_{2}F_{7}S_{14}V_{2}$, $F_{5}F_{7}S_{6}V_{3}$, $F_{6}F_{7}S_{5}V_{4}$, $F_{7}F_{8}S_{5}V_{4}$, $F_{7}F_{13}S_{6}V_{3}$, $F_{7}F_{8}S_{14}V_{2}$, $F_{3}F_{5}V_{3}V_{2}$, $F_{3}F_{6}S_{3}S_{5}$, $F_{3}F_{8}S_{3}S_{5}$, $F_{3}F_{8}V_{3}V_{2}$, $F_{3}F_{6}V_{4}V_{2}$, $F_{3}F_{6}S_{3}S_{6}$, $F_{3}F_{13}S_{3}S_{6}$, $F_{3}F_{8}V_{4}V_{2}$}} 
 \SplitRow{ $\mathcal{O}_{10, \; (1,-3)}^{7}$} {{$F_{8}F_{9}S_{7}S_{5}$, $F_{5}F_{8}S_{7}S_{5}$, $F_{9}F_{14}V_{11}V_{3}$, $F_{5}F_{6}V_{11}V_{3}$, $F_{8}F_{9}V_{11}V_{3}$, $F_{5}F_{8}V_{11}V_{3}$, $F_{9}F_{9}S_{7}S_{5}$, $F_{5}F_{5}S_{7}S_{5}$, $F_{7}F_{8}S_{7}V_{5}$, $F_{7}F_{8}S_{7}V_{4}$, $F_{7}F_{14}S_{5}V_{11}$, $F_{6}F_{7}S_{5}V_{11}$, $F_{7}F_{9}S_{5}V_{11}$, $F_{5}F_{7}S_{5}V_{11}$, $F_{7}F_{7}S_{7}V_{5}$, $F_{7}F_{7}S_{7}V_{4}$, $F_{4}F_{8}S_{7}V_{2}$, $F_{1}F_{8}S_{7}V_{2}$, $F_{4}F_{8}S_{5}V_{3}$, $F_{1}F_{8}S_{5}V_{3}$, $F_{4}F_{9}S_{5}V_{3}$, $F_{1}F_{5}S_{5}V_{3}$, $F_{4}F_{4}S_{7}V_{2}$, $F_{1}F_{1}S_{7}V_{2}$, $F_{3}F_{14}V_{11}V_{2}$, $F_{3}F_{6}V_{11}V_{2}$, $F_{3}F_{8}V_{3}V_{5}$, $F_{3}F_{8}V_{3}V_{4}$, $F_{3}F_{9}S_{5}S_{5}$, $F_{3}F_{5}S_{5}S_{5}$, $F_{3}F_{7}V_{3}V_{5}$, $F_{3}F_{7}V_{3}V_{4}$, $F_{3}F_{4}V_{11}V_{2}$, $F_{1}F_{3}V_{11}V_{2}$, $F_{5}F_{9}S_{7}V_{5}$, $F_{5}F_{5}S_{7}V_{4}$, $F_{8}F_{9}S_{5}V_{11}$, $F_{5}F_{8}S_{5}V_{11}$, $F_{9}F_{9}S_{7}V_{5}$, $F_{4}F_{5}S_{7}S_{4}$, $F_{1}F_{5}S_{7}S_{2}$, $F_{4}F_{8}S_{5}S_{5}$, $F_{1}F_{8}S_{5}S_{5}$, $F_{4}F_{4}S_{7}S_{4}$, $F_{1}F_{1}S_{7}S_{2}$, $F_{3}F_{8}S_{4}V_{11}$, $F_{3}F_{8}S_{2}V_{11}$, $F_{3}F_{8}S_{5}V_{5}$, $F_{3}F_{8}S_{5}V_{4}$, $F_{3}F_{9}S_{5}V_{5}$, $F_{3}F_{5}S_{5}V_{4}$, $F_{3}F_{4}S_{4}V_{11}$, $F_{1}F_{3}S_{2}V_{11}$, $F_{3}F_{9}S_{7}V_{2}$, $F_{3}F_{5}S_{7}V_{2}$, $F_{8}F_{9}S_{5}V_{3}$, $F_{5}F_{8}S_{5}V_{3}$, $F_{9}F_{14}S_{5}V_{3}$, $F_{5}F_{6}S_{5}V_{3}$, $F_{9}F_{9}S_{7}V_{2}$, $F_{5}F_{5}S_{7}V_{2}$, $F_{3}F_{7}S_{7}S_{4}$, $F_{3}F_{7}S_{7}S_{2}$, $F_{7}F_{14}S_{5}S_{5}$, $F_{6}F_{7}S_{5}S_{5}$, $F_{7}F_{7}S_{7}S_{4}$, $F_{7}F_{7}S_{7}S_{2}$, $F_{3}F_{8}S_{4}V_{3}$, $F_{3}F_{8}S_{2}V_{3}$, $F_{3}F_{14}S_{5}V_{2}$, $F_{3}F_{6}S_{5}V_{2}$, $F_{3}F_{9}S_{5}V_{2}$, $F_{3}F_{5}S_{5}V_{2}$, $F_{3}F_{7}S_{4}V_{3}$, $F_{3}F_{7}S_{2}V_{3}$, $F_{3}F_{9}V_{11}V_{2}$, $F_{3}F_{5}V_{11}V_{2}$, $F_{5}F_{9}V_{3}V_{5}$, $F_{5}F_{5}V_{3}V_{4}$, $F_{8}F_{9}S_{5}S_{5}$, $F_{5}F_{8}S_{5}S_{5}$, $F_{9}F_{14}V_{3}V_{5}$, $F_{5}F_{6}V_{3}V_{4}$, $F_{8}F_{9}V_{11}V_{2}$, $F_{5}F_{8}V_{11}V_{2}$, $F_{3}F_{7}S_{4}V_{11}$, $F_{3}F_{7}S_{2}V_{11}$, $F_{7}F_{8}S_{5}V_{5}$, $F_{7}F_{8}S_{5}V_{4}$, $F_{7}F_{14}S_{5}V_{5}$, $F_{6}F_{7}S_{5}V_{4}$, $F_{7}F_{9}S_{4}V_{11}$, $F_{5}F_{7}S_{2}V_{11}$, $F_{4}F_{5}S_{4}V_{3}$, $F_{1}F_{5}S_{2}V_{3}$, $F_{4}F_{8}S_{5}V_{2}$, $F_{1}F_{8}S_{5}V_{2}$, $F_{4}F_{9}S_{4}V_{3}$, $F_{3}F_{8}S_{5}S_{4}$, $F_{3}F_{8}S_{2}S_{5}$, $F_{3}F_{14}V_{2}V_{5}$, $F_{3}F_{6}V_{4}V_{2}$, $F_{3}F_{8}V_{2}V_{5}$, $F_{3}F_{8}V_{4}V_{2}$, $F_{3}F_{9}S_{5}S_{4}$, $F_{3}F_{5}S_{2}S_{5}$}} 
 \SplitRow{ $\mathcal{O}_{10, \; (1,-3)}^{8}$} {{$F_{3}F_{8}V_{3}V_{2}$, $F_{3}F_{6}S_{3}S_{5}$, $F_{3}F_{8}S_{3}S_{5}$, $F_{3}F_{5}V_{3}V_{2}$, $F_{8}F_{8}S_{7}V_{2}$, $F_{8}F_{8}S_{5}V_{3}$, $F_{5}F_{8}S_{5}V_{3}$, $F_{1}F_{8}S_{7}V_{2}$, $F_{6}F_{6}S_{7}S_{3}$, $F_{5}F_{6}V_{3}V_{3}$, $F_{1}F_{6}S_{7}S_{3}$, $F_{3}F_{3}S_{7}V_{2}$, $F_{3}F_{8}S_{5}V_{3}$, $F_{3}F_{6}S_{5}V_{3}$, $F_{3}F_{5}S_{7}V_{2}$, $F_{8}F_{8}S_{7}S_{5}$, $F_{5}F_{8}S_{7}S_{5}$, $F_{6}F_{6}S_{7}V_{3}$, $F_{5}F_{6}S_{7}V_{3}$, $F_{3}F_{3}S_{7}S_{3}$, $F_{3}F_{8}V_{3}V_{3}$, $F_{3}F_{8}S_{7}S_{3}$, $F_{8}F_{8}S_{7}V_{3}$}} 
 \SplitRow{ $\mathcal{O}_{10, \; (1,-3)}^{9}$} {{$F_{7}F_{8}S_{7}V_{4}$, $F_{7}F_{7}S_{7}V_{4}$, $F_{8}F_{8}S_{7}V_{3}$, $F_{3}F_{8}S_{7}S_{2}$, $F_{3}F_{7}V_{3}V_{4}$, $F_{3}F_{8}V_{3}V_{4}$, $F_{3}F_{3}S_{7}S_{2}$, $F_{5}F_{5}S_{7}V_{4}$, $F_{5}F_{8}S_{7}S_{5}$, $F_{8}F_{8}S_{7}S_{5}$, $F_{3}F_{5}S_{7}V_{1}$, $F_{3}F_{5}S_{5}V_{4}$, $F_{3}F_{8}S_{5}V_{4}$, $F_{3}F_{3}S_{7}V_{1}$, $F_{5}F_{6}S_{7}V_{3}$, $F_{5}F_{5}S_{7}V_{3}$, $F_{6}F_{7}S_{7}S_{5}$, $F_{7}F_{7}S_{7}S_{5}$, $F_{3}F_{6}S_{7}V_{2}$, $F_{3}F_{5}S_{5}V_{3}$, $F_{3}F_{7}S_{5}V_{3}$, $F_{3}F_{3}S_{7}V_{2}$, $F_{1}F_{5}S_{7}S_{2}$, $F_{5}F_{6}V_{3}V_{4}$, $F_{5}F_{5}V_{3}V_{4}$, $F_{5}F_{5}S_{7}S_{2}$, $F_{1}F_{7}S_{7}V_{1}$, $F_{6}F_{7}S_{5}V_{4}$, $F_{7}F_{8}S_{5}V_{4}$, $F_{7}F_{7}S_{7}V_{1}$, $F_{1}F_{8}S_{7}V_{2}$, $F_{5}F_{8}S_{5}V_{3}$, $F_{8}F_{8}S_{5}V_{3}$, $F_{8}F_{8}S_{7}V_{2}$, $F_{3}F_{6}V_{4}V_{2}$, $F_{3}F_{5}V_{3}V_{1}$, $F_{3}F_{5}S_{2}S_{5}$, $F_{3}F_{8}S_{2}S_{5}$, $F_{3}F_{7}V_{3}V_{1}$, $F_{3}F_{8}V_{4}V_{2}$}} 
 \SplitRow{ $\mathcal{O}_{10, \; (1,-3)}^{10}$} {{$F_{8}F_{9}S_{7}S_{5}$, $F_{5}F_{8}S_{7}S_{5}$, $F_{9}F_{9}S_{7}S_{5}$, $F_{5}F_{5}S_{7}S_{5}$, $F_{3}F_{8}S_{7}S_{4}$, $F_{3}F_{8}S_{7}S_{2}$, $F_{3}F_{9}S_{5}S_{5}$, $F_{3}F_{5}S_{5}S_{5}$, $F_{3}F_{3}S_{7}S_{4}$, $F_{3}F_{3}S_{7}S_{2}$, $F_{1}F_{9}S_{7}S_{4}$, $F_{1}F_{5}S_{7}S_{2}$, $F_{8}F_{9}S_{5}S_{5}$, $F_{5}F_{8}S_{5}S_{5}$, $F_{9}F_{9}S_{7}S_{4}$, $F_{5}F_{5}S_{7}S_{2}$, $F_{3}F_{8}S_{5}S_{4}$, $F_{3}F_{8}S_{2}S_{5}$, $F_{3}F_{9}S_{5}S_{4}$, $F_{3}F_{5}S_{2}S_{5}$}} 
 \end{SplitTable}
   \vspace{0.4cm} 
 \captionof{table}{Dimension 10 $(\Delta B, \Delta L)$ = (1, -3) UV completions for topology 3 (see Fig.~\ref{fig:d=10_topologies}).} 
 \label{tab:T10Top3B1L-3}
 \end{center}

%% file: Tables_dim10/Table_dim10_Top4_B1_L-3.tex
 \begin{center}
 \begin{SplitTable}[2.0cm] 
  \SplitHeader{$\mathcal{O}^j{(\mathcal{T}_ {10} ^{ 4 })}$}{New Particles} 
   \vspace{0.1cm} 
 \SplitRow{ $\mathcal{O}_{10, \; (1,-3)}^{1}$} {{$F_{14}S_{9}S_{5}S_{4}$, $F_{6}S_{14}S_{2}S_{5}$, $F_{8}S_{16}S_{9}S_{5}$, $F_{8}S_{14}S_{16}S_{5}$, $F_{10}S_{9}S_{3}S_{6}$, $F_{10}S_{14}S_{3}S_{6}$, $F_{14}S_{5}S_{5}S_{4}$, $F_{6}S_{2}S_{5}S_{5}$, $F_{8}S_{16}S_{9}S_{3}$, $F_{8}S_{14}S_{16}S_{3}$, $F_{14}S_{9}S_{3}S_{6}$, $F_{6}S_{14}S_{3}S_{6}$, $F_{8}S_{5}S_{5}S_{4}$, $F_{8}S_{2}S_{5}S_{5}$, $F_{2}S_{16}S_{9}S_{3}$, $F_{2}S_{14}S_{16}S_{3}$, $F_{14}S_{3}S_{5}S_{6}$, $F_{6}S_{3}S_{5}S_{6}$, $F_{8}S_{3}S_{5}S_{4}$, $F_{8}S_{2}S_{3}S_{5}$}} 
 \SplitRow{ $\mathcal{O}_{10, \; (1,-3)}^{2}$} {{$F_{8}V_{11}V_{4}V_{2}$, $F_{6}S_{14}S_{2}S_{5}$, $F_{13}V_{12}V_{3}V_{1}$, $F_{6}V_{10}V_{12}V_{3}$, $F_{8}S_{14}S_{16}S_{5}$, $F_{5}V_{9}V_{11}V_{4}$, $F_{10}S_{3}S_{6}V_{11}$, $F_{6}S_{2}S_{5}V_{3}$, $F_{13}S_{5}V_{3}V_{1}$, $F_{5}S_{3}V_{9}V_{11}$, $F_{10}S_{14}S_{3}S_{6}$, $F_{8}V_{3}V_{4}V_{2}$, $F_{13}V_{3}V_{4}V_{1}$, $F_{8}S_{14}S_{16}S_{3}$, $F_{10}S_{3}S_{6}V_{12}$, $F_{8}S_{5}V_{4}V_{2}$, $F_{6}S_{2}S_{5}V_{4}$, $F_{6}S_{3}V_{10}V_{12}$, $F_{8}S_{3}S_{6}V_{11}$, $F_{8}S_{5}V_{3}V_{1}$, $F_{2}S_{3}V_{9}V_{11}$, $F_{8}S_{3}S_{6}V_{3}$, $F_{8}S_{3}V_{3}V_{1}$, $F_{8}S_{3}S_{5}S_{6}$, $F_{6}S_{2}S_{3}S_{5}$, $F_{6}S_{14}S_{3}S_{6}$, $F_{6}V_{3}V_{4}V_{2}$, $F_{5}V_{3}V_{4}V_{1}$, $F_{2}S_{14}S_{16}S_{3}$, $F_{6}S_{3}S_{6}V_{3}$, $F_{5}S_{3}V_{3}V_{1}$, $F_{6}S_{3}S_{6}V_{4}$, $F_{6}S_{3}V_{4}V_{2}$, $F_{13}S_{3}S_{6}V_{12}$, $F_{5}S_{2}S_{5}V_{4}$, $F_{2}S_{3}V_{10}V_{12}$, $F_{13}S_{3}S_{5}S_{6}$, $F_{5}S_{2}S_{3}S_{5}$, $F_{13}S_{3}S_{6}V_{4}$, $F_{8}S_{3}V_{4}V_{2}$}} 
 \SplitRow{ $\mathcal{O}_{10, \; (1,-3)}^{3}$} {{$F_{8}S_{5}S_{4}V_{11}$, $F_{8}S_{2}S_{5}V_{11}$, $F_{14}S_{9}V_{3}V_{1}$, $F_{6}S_{14}V_{3}V_{1}$, $F_{6}S_{16}S_{9}V_{3}$, $F_{6}S_{14}S_{16}V_{3}$, $F_{8}S_{5}V_{11}V_{13}$, $F_{8}S_{5}V_{9}V_{11}$, $F_{11}S_{3}S_{6}V_{11}$, $F_{2}S_{3}S_{6}V_{11}$, $F_{8}S_{5}S_{4}V_{3}$, $F_{8}S_{2}S_{5}V_{3}$, $F_{14}S_{5}V_{3}V_{1}$, $F_{6}S_{5}V_{3}V_{1}$, $F_{8}S_{3}V_{11}V_{13}$, $F_{8}S_{3}V_{9}V_{11}$, $F_{11}S_{9}S_{3}S_{6}$, $F_{2}S_{14}S_{3}S_{6}$, $F_{8}S_{5}S_{5}S_{4}$, $F_{8}S_{2}S_{5}S_{5}$, $F_{6}S_{16}S_{9}S_{3}$, $F_{6}S_{14}S_{16}S_{3}$, $F_{14}S_{5}S_{4}V_{11}$, $F_{6}S_{2}S_{5}V_{11}$, $F_{13}S_{9}V_{3}V_{1}$, $F_{13}S_{14}V_{3}V_{1}$, $F_{8}S_{16}S_{9}V_{3}$, $F_{8}S_{14}S_{16}V_{3}$, $F_{9}S_{5}V_{11}V_{13}$, $F_{5}S_{5}V_{9}V_{11}$, $F_{8}V_{11}V_{2}V_{5}$, $F_{8}V_{11}V_{4}V_{2}$, $F_{14}S_{9}S_{5}S_{4}$, $F_{6}S_{14}S_{2}S_{5}$, $F_{8}S_{16}S_{9}S_{5}$, $F_{8}S_{14}S_{16}S_{5}$, $F_{9}V_{11}V_{13}V_{5}$, $F_{5}V_{9}V_{11}V_{4}$, $F_{11}V_{11}V_{2}V_{5}$, $F_{2}V_{11}V_{4}V_{2}$, $F_{8}V_{3}V_{2}V_{5}$, $F_{8}V_{3}V_{4}V_{2}$, $F_{14}S_{5}S_{5}S_{4}$, $F_{6}S_{2}S_{5}S_{5}$, $F_{13}V_{3}V_{1}V_{5}$, $F_{13}V_{3}V_{4}V_{1}$, $F_{9}V_{11}V_{13}V_{2}$, $F_{5}V_{9}V_{11}V_{2}$, $F_{11}S_{9}V_{2}V_{5}$, $F_{2}S_{14}V_{4}V_{2}$, $F_{8}S_{5}V_{2}V_{5}$, $F_{8}S_{5}V_{4}V_{2}$, $F_{14}S_{5}S_{4}V_{5}$, $F_{6}S_{2}S_{5}V_{4}$, $F_{8}S_{16}S_{9}V_{2}$, $F_{8}S_{14}S_{16}V_{2}$, $F_{14}S_{3}S_{6}V_{11}$, $F_{6}S_{3}S_{6}V_{11}$, $F_{9}S_{5}V_{3}V_{1}$, $F_{5}S_{5}V_{3}V_{1}$, $F_{3}S_{3}V_{11}V_{13}$, $F_{3}S_{3}V_{9}V_{11}$, $F_{6}V_{3}V_{2}V_{5}$, $F_{6}V_{3}V_{4}V_{2}$, $F_{9}V_{3}V_{1}V_{5}$, $F_{5}V_{3}V_{4}V_{1}$, $F_{3}V_{11}V_{13}V_{2}$, $F_{3}V_{9}V_{11}V_{2}$, $F_{8}S_{3}S_{5}S_{4}$, $F_{8}S_{2}S_{3}S_{5}$, $F_{14}S_{3}S_{5}S_{6}$, $F_{6}S_{3}S_{5}S_{6}$, $F_{9}V_{3}V_{1}V_{2}$, $F_{5}V_{3}V_{1}V_{2}$, $F_{6}S_{3}V_{2}V_{5}$, $F_{6}S_{3}V_{4}V_{2}$, $F_{14}S_{3}S_{6}V_{5}$, $F_{6}S_{3}S_{6}V_{4}$, $F_{8}S_{5}S_{4}V_{2}$, $F_{8}S_{2}S_{5}V_{2}$, $F_{13}S_{9}S_{3}S_{6}$, $F_{13}S_{14}S_{3}S_{6}$, $F_{9}S_{5}S_{5}S_{4}$, $F_{5}S_{2}S_{5}S_{5}$, $F_{3}S_{16}S_{9}S_{3}$, $F_{3}S_{14}S_{16}S_{3}$, $F_{14}S_{9}V_{2}V_{5}$, $F_{6}S_{14}V_{4}V_{2}$, $F_{9}S_{5}S_{4}V_{5}$, $F_{5}S_{2}S_{5}V_{4}$, $F_{3}S_{16}S_{9}V_{2}$, $F_{3}S_{14}S_{16}V_{2}$, $F_{14}S_{5}V_{2}V_{5}$, $F_{6}S_{5}V_{4}V_{2}$, $F_{8}S_{3}V_{2}V_{5}$, $F_{8}S_{3}V_{4}V_{2}$, $F_{13}S_{3}S_{6}V_{5}$, $F_{13}S_{3}S_{6}V_{4}$, $F_{9}S_{5}S_{4}V_{2}$, $F_{5}S_{2}S_{5}V_{2}$}} 
 \SplitRow{ $\mathcal{O}_{10, \; (1,-3)}^{4}$} {{$F_{5}S_{7}S_{3}S_{6}$, $F_{8}V_{11}V_{4}V_{2}$, $F_{5}V_{9}V_{11}V_{4}$, $F_{7}S_{7}S_{16}S_{6}$, $F_{6}S_{7}S_{2}S_{5}$, $F_{13}V_{11}V_{3}V_{1}$, $F_{5}V_{9}V_{11}V_{3}$, $F_{7}S_{7}S_{16}S_{5}$, $F_{3}S_{7}S_{2}S_{5}$, $F_{13}V_{3}V_{4}V_{1}$, $F_{8}V_{3}V_{4}V_{2}$, $F_{7}S_{7}S_{16}S_{2}$, $F_{3}S_{2}S_{5}V_{11}$, $F_{6}S_{2}S_{5}V_{4}$, $F_{13}S_{6}V_{3}V_{1}$, $F_{5}S_{3}S_{6}V_{3}$, $F_{8}S_{5}V_{4}V_{2}$, $F_{5}S_{2}V_{9}V_{11}$, $F_{5}S_{7}S_{2}S_{5}$, $F_{8}V_{11}V_{3}V_{1}$, $F_{6}V_{9}V_{11}V_{3}$, $F_{8}S_{7}S_{16}S_{5}$, $F_{3}S_{7}S_{3}S_{6}$, $F_{8}V_{3}V_{3}V_{1}$, $F_{8}S_{7}S_{16}S_{3}$, $F_{3}S_{3}S_{6}V_{11}$, $F_{5}S_{2}S_{5}V_{3}$, $F_{8}S_{5}V_{3}V_{1}$, $F_{6}S_{3}V_{9}V_{11}$, $F_{5}V_{3}V_{4}V_{1}$, $F_{6}V_{3}V_{4}V_{2}$, $F_{1}S_{7}S_{16}S_{2}$, $F_{6}S_{7}S_{3}S_{6}$, $F_{5}V_{3}V_{3}V_{1}$, $F_{1}S_{7}S_{16}S_{3}$, $F_{6}S_{3}S_{6}V_{4}$, $F_{5}S_{2}V_{3}V_{1}$, $F_{6}S_{3}V_{4}V_{2}$, $F_{8}S_{2}S_{5}V_{11}$, $F_{5}S_{2}S_{5}V_{4}$, $F_{7}S_{6}V_{3}V_{1}$, $F_{6}S_{3}S_{6}V_{3}$, $F_{1}S_{2}V_{9}V_{11}$, $F_{13}S_{3}S_{6}V_{11}$, $F_{7}S_{5}V_{3}V_{1}$, $F_{1}S_{3}V_{9}V_{11}$, $F_{13}S_{3}S_{6}V_{4}$, $F_{8}S_{2}S_{5}V_{3}$, $F_{7}S_{2}V_{3}V_{1}$, $F_{8}S_{3}V_{4}V_{2}$, $F_{13}S_{3}S_{6}S_{6}$, $F_{8}S_{2}S_{5}S_{5}$, $F_{5}S_{2}S_{2}S_{5}$, $F_{6}S_{3}S_{3}S_{6}$}} 
 \SplitRow{ $\mathcal{O}_{10, \; (1,-3)}^{5}$} {{$F_{6}S_{3}S_{3}S_{5}$, $F_{8}S_{3}S_{5}S_{5}$, $F_{8}S_{3}V_{3}V_{2}$, $F_{8}S_{3}S_{5}V_{3}$, $F_{2}S_{3}V_{10}V_{11}$, $F_{8}S_{5}V_{3}V_{2}$, $F_{6}S_{3}S_{5}V_{3}$, $F_{8}S_{3}S_{5}V_{11}$, $F_{6}S_{3}V_{3}V_{2}$, $F_{5}S_{3}S_{5}V_{3}$, $F_{2}S_{7}S_{16}S_{3}$, $F_{6}V_{3}V_{3}V_{2}$, $F_{5}S_{7}S_{3}S_{5}$, $F_{6}S_{3}V_{10}V_{11}$, $F_{3}S_{3}S_{5}V_{11}$, $F_{8}S_{7}S_{16}S_{3}$, $F_{8}V_{3}V_{3}V_{2}$, $F_{3}S_{7}S_{3}S_{5}$, $F_{8}S_{7}S_{16}S_{5}$, $F_{6}V_{10}V_{11}V_{3}$, $F_{8}V_{11}V_{3}V_{2}$}} 
 \SplitRow{ $\mathcal{O}_{10, \; (1,-3)}^{6}$} {{$F_{5}S_{7}V_{4}V_{2}$, $F_{6}S_{14}V_{3}V_{1}$, $F_{6}S_{14}S_{16}V_{3}$, $F_{5}S_{7}S_{16}V_{4}$, $F_{2}S_{7}S_{3}S_{6}$, $F_{6}V_{3}V_{3}V_{1}$, $F_{5}S_{7}S_{16}S_{3}$, $F_{2}S_{14}S_{3}S_{6}$, $F_{5}V_{3}V_{4}V_{2}$, $F_{6}V_{3}V_{4}V_{1}$, $F_{6}S_{14}S_{16}S_{3}$, $F_{8}S_{7}V_{4}V_{2}$, $F_{13}S_{14}V_{3}V_{1}$, $F_{8}S_{14}S_{16}V_{3}$, $F_{7}S_{7}S_{16}V_{4}$, $F_{5}S_{7}S_{3}S_{6}$, $F_{6}S_{14}S_{2}S_{5}$, $F_{8}S_{14}S_{16}S_{5}$, $F_{7}S_{7}S_{16}S_{6}$, $F_{2}S_{7}V_{4}V_{2}$, $F_{6}S_{2}S_{5}V_{3}$, $F_{13}S_{5}V_{3}V_{1}$, $F_{7}S_{7}S_{16}V_{2}$, $F_{2}S_{14}V_{4}V_{2}$, $F_{5}S_{3}S_{6}V_{3}$, $F_{6}S_{2}S_{5}V_{4}$, $F_{8}S_{5}V_{4}V_{2}$, $F_{13}S_{6}V_{3}V_{1}$, $F_{8}S_{14}S_{16}V_{2}$, $F_{8}S_{7}S_{3}S_{6}$, $F_{8}V_{3}V_{3}V_{1}$, $F_{3}S_{7}S_{16}S_{3}$, $F_{8}S_{5}V_{3}V_{1}$, $F_{3}S_{7}S_{16}V_{2}$, $F_{6}S_{2}S_{3}S_{5}$, $F_{8}S_{3}S_{5}S_{6}$, $F_{8}V_{3}V_{1}V_{2}$, $F_{13}S_{14}S_{3}S_{6}$, $F_{8}V_{3}V_{4}V_{2}$, $F_{7}V_{3}V_{4}V_{1}$, $F_{3}S_{14}S_{16}S_{3}$, $F_{6}S_{14}V_{4}V_{2}$, $F_{6}S_{3}S_{6}V_{3}$, $F_{5}S_{2}S_{5}V_{4}$, $F_{7}S_{6}V_{3}V_{1}$, $F_{3}S_{14}S_{16}V_{2}$, $F_{6}V_{3}V_{4}V_{2}$, $F_{5}S_{2}S_{3}S_{5}$, $F_{13}S_{3}S_{5}S_{6}$, $F_{7}V_{3}V_{1}V_{2}$, $F_{6}V_{4}V_{4}V_{2}$, $F_{6}S_{3}S_{3}S_{6}$, $F_{13}S_{3}S_{6}S_{6}$, $F_{8}V_{4}V_{2}V_{2}$}} 
 \SplitRow{ $\mathcal{O}_{10, \; (1,-3)}^{7}$} {{$F_{8}S_{7}S_{5}S_{4}$, $F_{8}S_{7}S_{2}S_{5}$, $F_{14}V_{11}V_{3}V_{1}$, $F_{6}V_{11}V_{3}V_{1}$, $F_{8}V_{11}V_{13}V_{3}$, $F_{8}V_{9}V_{11}V_{3}$, $F_{9}S_{7}S_{16}S_{5}$, $F_{5}S_{7}S_{16}S_{5}$, $F_{5}S_{7}V_{2}V_{5}$, $F_{5}S_{7}V_{4}V_{2}$, $F_{8}S_{5}S_{4}V_{11}$, $F_{8}S_{2}S_{5}V_{11}$, $F_{8}S_{5}V_{11}V_{13}$, $F_{8}S_{5}V_{9}V_{11}$, $F_{9}S_{7}S_{16}V_{5}$, $F_{5}S_{7}S_{16}V_{4}$, $F_{3}S_{7}V_{2}V_{5}$, $F_{3}S_{7}V_{4}V_{2}$, $F_{8}S_{5}S_{4}V_{3}$, $F_{8}S_{2}S_{5}V_{3}$, $F_{14}S_{5}V_{3}V_{1}$, $F_{6}S_{5}V_{3}V_{1}$, $F_{9}S_{7}S_{16}V_{2}$, $F_{5}S_{7}S_{16}V_{2}$, $F_{3}V_{11}V_{2}V_{5}$, $F_{3}V_{11}V_{4}V_{2}$, $F_{5}V_{3}V_{2}V_{5}$, $F_{5}V_{3}V_{4}V_{2}$, $F_{8}S_{5}S_{5}S_{4}$, $F_{8}S_{2}S_{5}S_{5}$, $F_{14}V_{3}V_{1}V_{5}$, $F_{6}V_{3}V_{4}V_{1}$, $F_{8}V_{11}V_{13}V_{2}$, $F_{8}V_{9}V_{11}V_{2}$, $F_{8}S_{7}V_{2}V_{5}$, $F_{8}S_{7}V_{4}V_{2}$, $F_{14}S_{5}S_{4}V_{11}$, $F_{6}S_{2}S_{5}V_{11}$, $F_{9}S_{5}V_{11}V_{13}$, $F_{5}S_{5}V_{9}V_{11}$, $F_{7}S_{7}S_{16}V_{5}$, $F_{7}S_{7}S_{16}V_{4}$, $F_{3}S_{7}S_{5}S_{4}$, $F_{3}S_{7}S_{2}S_{5}$, $F_{14}S_{5}S_{5}S_{4}$, $F_{6}S_{2}S_{5}S_{5}$, $F_{7}S_{7}S_{16}S_{4}$, $F_{7}S_{7}S_{16}S_{2}$, $F_{3}S_{5}S_{4}V_{11}$, $F_{3}S_{2}S_{5}V_{11}$, $F_{8}S_{5}V_{2}V_{5}$, $F_{8}S_{5}V_{4}V_{2}$, $F_{14}S_{5}S_{4}V_{5}$, $F_{6}S_{2}S_{5}V_{4}$, $F_{9}S_{4}V_{11}V_{13}$, $F_{5}S_{2}V_{9}V_{11}$, $F_{9}S_{5}V_{3}V_{1}$, $F_{5}S_{5}V_{3}V_{1}$, $F_{4}S_{7}S_{16}V_{2}$, $F_{1}S_{7}S_{16}V_{2}$, $F_{5}S_{7}S_{5}S_{4}$, $F_{5}S_{7}S_{2}S_{5}$, $F_{4}S_{7}S_{16}S_{4}$, $F_{1}S_{7}S_{16}S_{2}$, $F_{5}S_{5}S_{4}V_{3}$, $F_{5}S_{2}S_{5}V_{3}$, $F_{8}S_{5}S_{4}V_{2}$, $F_{8}S_{2}S_{5}V_{2}$, $F_{9}S_{4}V_{3}V_{1}$, $F_{5}S_{2}V_{3}V_{1}$, $F_{14}V_{11}V_{2}V_{5}$, $F_{6}V_{11}V_{4}V_{2}$, $F_{8}V_{3}V_{2}V_{5}$, $F_{8}V_{3}V_{4}V_{2}$, $F_{9}S_{5}S_{5}S_{4}$, $F_{5}S_{2}S_{5}S_{5}$, $F_{7}V_{3}V_{1}V_{5}$, $F_{7}V_{3}V_{4}V_{1}$, $F_{4}V_{11}V_{13}V_{2}$, $F_{1}V_{9}V_{11}V_{2}$, $F_{9}S_{5}S_{4}V_{5}$, $F_{5}S_{2}S_{5}V_{4}$, $F_{4}S_{4}V_{11}V_{13}$, $F_{1}S_{2}V_{9}V_{11}$, $F_{14}S_{5}V_{2}V_{5}$, $F_{6}S_{5}V_{4}V_{2}$, $F_{9}S_{5}S_{4}V_{2}$, $F_{5}S_{2}S_{5}V_{2}$, $F_{7}S_{4}V_{3}V_{1}$, $F_{7}S_{2}V_{3}V_{1}$, $F_{14}V_{2}V_{5}V_{5}$, $F_{6}V_{4}V_{4}V_{2}$, $F_{8}V_{2}V_{2}V_{5}$, $F_{8}V_{4}V_{2}V_{2}$, $F_{9}S_{5}S_{4}S_{4}$, $F_{5}S_{2}S_{2}S_{5}$}} 
 \SplitRow{ $\mathcal{O}_{10, \; (1,-3)}^{8}$} {{$F_{8}V_{3}V_{2}V_{2}$, $F_{6}S_{3}S_{3}S_{5}$, $F_{8}S_{3}S_{5}S_{5}$, $F_{5}V_{3}V_{3}V_{2}$, $F_{3}S_{7}S_{16}V_{2}$, $F_{8}S_{5}V_{3}V_{2}$, $F_{6}S_{3}S_{5}V_{3}$, $F_{5}S_{7}V_{3}V_{2}$, $F_{3}S_{7}S_{16}S_{3}$, $F_{8}V_{3}V_{3}V_{2}$, $F_{8}S_{7}S_{3}S_{5}$, $F_{8}S_{7}S_{16}V_{2}$, $F_{5}S_{3}S_{5}V_{3}$, $F_{1}S_{7}V_{3}V_{2}$, $F_{8}S_{7}S_{16}S_{5}$, $F_{5}S_{7}S_{3}S_{5}$, $F_{8}S_{7}S_{16}V_{3}$, $F_{8}S_{7}V_{3}V_{2}$, $F_{6}S_{7}S_{16}S_{3}$, $F_{1}S_{7}S_{3}S_{5}$, $F_{6}S_{7}S_{16}V_{3}$}} 
 \SplitRow{ $\mathcal{O}_{10, \; (1,-3)}^{9}$} {{$F_{5}S_{7}V_{4}V_{2}$, $F_{5}S_{7}S_{16}V_{4}$, $F_{6}S_{7}V_{3}V_{1}$, $F_{5}S_{7}S_{16}V_{3}$, $F_{1}S_{7}S_{2}S_{5}$, $F_{6}V_{3}V_{4}V_{1}$, $F_{5}V_{3}V_{4}V_{2}$, $F_{5}S_{7}S_{16}S_{2}$, $F_{8}S_{7}V_{4}V_{2}$, $F_{7}S_{7}S_{16}V_{4}$, $F_{6}S_{7}S_{2}S_{5}$, $F_{7}S_{7}S_{16}S_{5}$, $F_{1}S_{7}V_{3}V_{1}$, $F_{6}S_{2}S_{5}V_{4}$, $F_{8}S_{5}V_{4}V_{2}$, $F_{7}S_{7}S_{16}V_{1}$, $F_{8}S_{7}V_{3}V_{1}$, $F_{8}S_{7}S_{16}V_{3}$, $F_{5}S_{7}S_{2}S_{5}$, $F_{8}S_{7}S_{16}S_{5}$, $F_{1}S_{7}V_{4}V_{2}$, $F_{5}S_{2}S_{5}V_{3}$, $F_{8}S_{5}V_{3}V_{1}$, $F_{8}S_{7}S_{16}V_{2}$, $F_{8}S_{7}S_{2}S_{5}$, $F_{7}V_{3}V_{4}V_{1}$, $F_{8}V_{3}V_{4}V_{2}$, $F_{3}S_{7}S_{16}S_{2}$, $F_{5}S_{7}V_{3}V_{1}$, $F_{5}S_{2}S_{5}V_{4}$, $F_{3}S_{7}S_{16}V_{1}$, $F_{6}S_{7}V_{4}V_{2}$, $F_{7}S_{5}V_{3}V_{1}$, $F_{3}S_{7}S_{16}V_{2}$, $F_{6}V_{4}V_{4}V_{2}$, $F_{5}V_{3}V_{3}V_{1}$, $F_{5}S_{2}S_{2}S_{5}$, $F_{8}S_{2}S_{5}S_{5}$, $F_{7}V_{3}V_{1}V_{1}$, $F_{8}V_{4}V_{2}V_{2}$}} 
 \SplitRow{ $\mathcal{O}_{10, \; (1,-3)}^{10}$} {{$F_{8}S_{7}S_{5}S_{4}$, $F_{8}S_{7}S_{2}S_{5}$, $F_{9}S_{7}S_{16}S_{5}$, $F_{5}S_{7}S_{16}S_{5}$, $F_{1}S_{7}S_{5}S_{4}$, $F_{1}S_{7}S_{2}S_{5}$, $F_{8}S_{5}S_{5}S_{4}$, $F_{8}S_{2}S_{5}S_{5}$, $F_{9}S_{7}S_{16}S_{4}$, $F_{5}S_{7}S_{16}S_{2}$, $F_{9}S_{5}S_{5}S_{4}$, $F_{5}S_{2}S_{5}S_{5}$, $F_{3}S_{7}S_{16}S_{4}$, $F_{3}S_{7}S_{16}S_{2}$, $F_{9}S_{5}S_{4}S_{4}$, $F_{5}S_{2}S_{2}S_{5}$}} 
 \end{SplitTable}
   \vspace{0.4cm} 
 \captionof{table}{Dimension 10 $(\Delta B, \Delta L)$ = (1, -3) UV completions for topology 4 (see Fig.~\ref{fig:d=10_topologies}).} 
 \label{tab:T10Top4B1L-3}
 \end{center}

%% file: Tables_dim10/Table_dim10_Top5_B1_L-3.tex
 \begin{center}
 \begin{SplitTable}[2.0cm] 
  \SplitHeader{$\mathcal{O}^j{(\mathcal{T}_ {10} ^{ 5 })}$}{New Particles} 
   \vspace{0.1cm} 
 \SplitRow{ $\mathcal{O}_{10, \; (1,-3)}^{1}$} {{$S_{9}S_{3}S_{5}$, $S_{14}S_{3}S_{5}$, $S_{5}S_{5}S_{5}$}} 
 \SplitRow{ $\mathcal{O}_{10, \; (1,-3)}^{2}$} {{$S_{3}V_{11}V_{4}$, $S_{14}S_{3}S_{5}$, $S_{3}V_{12}V_{3}$, $S_{5}V_{3}V_{4}$}} 
 \SplitRow{ $\mathcal{O}_{10, \; (1,-3)}^{3}$} {{$S_{5}V_{11}V_{2}$, $S_{9}V_{3}V_{2}$, $S_{14}V_{3}V_{2}$, $S_{3}V_{11}V_{5}$, $S_{3}V_{11}V_{4}$, $S_{5}V_{3}V_{5}$, $S_{5}V_{3}V_{4}$, $S_{9}S_{3}S_{5}$, $S_{14}S_{3}S_{5}$, $S_{5}S_{5}S_{5}$}} 
 \SplitRow{ $\mathcal{O}_{10, \; (1,-3)}^{4}$} {{$S_{7}S_{3}S_{6}$, $S_{3}V_{11}V_{4}$, $S_{7}S_{2}S_{5}$, $S_{2}V_{11}V_{3}$, $S_{5}V_{3}V_{4}$, $S_{6}V_{3}V_{3}$}} 
 \SplitRow{ $\mathcal{O}_{10, \; (1,-3)}^{5}$} {{$S_{7}S_{3}S_{5}$, $S_{3}V_{11}V_{3}$, $S_{5}V_{3}V_{3}$}} 
 \SplitRow{ $\mathcal{O}_{10, \; (1,-3)}^{6}$} {{$S_{7}V_{4}V_{2}$, $S_{14}V_{3}V_{2}$, $S_{7}S_{3}S_{6}$, $S_{6}V_{3}V_{3}$, $S_{14}S_{3}S_{5}$, $S_{5}V_{3}V_{4}$}} 
 \SplitRow{ $\mathcal{O}_{10, \; (1,-3)}^{7}$} {{$S_{7}S_{5}S_{4}$, $S_{7}S_{2}S_{5}$, $S_{4}V_{11}V_{3}$, $S_{2}V_{11}V_{3}$, $S_{7}V_{2}V_{5}$, $S_{7}V_{4}V_{2}$, $S_{5}V_{11}V_{2}$, $S_{5}V_{3}V_{5}$, $S_{5}V_{3}V_{4}$, $S_{5}S_{5}S_{5}$}} 
 \SplitRow{ $\mathcal{O}_{10, \; (1,-3)}^{8}$} {{$S_{7}V_{3}V_{2}$, $S_{7}S_{3}S_{5}$, $S_{5}V_{3}V_{3}$}} 
 \SplitRow{ $\mathcal{O}_{10, \; (1,-3)}^{9}$} {{$S_{7}V_{4}V_{2}$, $S_{7}V_{3}V_{1}$, $S_{7}S_{2}S_{5}$, $S_{5}V_{3}V_{4}$}} 
 \SplitRow{ $\mathcal{O}_{10, \; (1,-3)}^{10}$} {{$S_{7}S_{5}S_{4}$, $S_{7}S_{2}S_{5}$, $S_{5}S_{5}S_{5}$}} 
 \end{SplitTable}
   \vspace{0.4cm} 
 \captionof{table}{Dimension 10 $(\Delta B, \Delta L)$ = (1, -3) UV completions for topology 5 (see Fig.~\ref{fig:d=10_topologies}).} 
 \label{tab:T10Top5B1L-3}
 \end{center}

%% file: Tables_dim10/Table_dim10_Top6_B1_L-3.tex
 \begin{center}
 \begin{SplitTable}[2.0cm] 
  \SplitHeader{$\mathcal{O}^j{(\mathcal{T}_ {10} ^{ 6 })}$}{New Particles} 
   \vspace{0.1cm} 
 \SplitRow{ $\mathcal{O}_{10, \; (1,-3)}^{1}$} {{$F_{4}F_{14}S_{9}S_{4}$, $F_{1}F_{6}S_{14}S_{2}$, $F_{10}F_{7}S_{9}S_{6}$, $F_{10}F_{7}S_{14}S_{6}$, $F_{4}F_{14}S_{5}S_{4}$, $F_{1}F_{6}S_{2}S_{5}$, $F_{7}F_{14}S_{9}S_{4}$, $F_{6}F_{7}S_{14}S_{2}$, $F_{4}F_{8}S_{16}S_{5}$, $F_{1}F_{8}S_{16}S_{5}$, $F_{7}F_{14}S_{5}S_{4}$, $F_{6}F_{7}S_{2}S_{5}$, $F_{4}F_{8}S_{16}S_{3}$, $F_{1}F_{8}S_{16}S_{3}$, $F_{7}F_{14}S_{9}S_{6}$, $F_{6}F_{7}S_{14}S_{6}$, $F_{4}F_{8}S_{5}S_{4}$, $F_{1}F_{8}S_{2}S_{5}$, $F_{7}F_{14}S_{5}S_{6}$, $F_{6}F_{7}S_{5}S_{6}$, $F_{4}F_{8}S_{3}S_{4}$, $F_{1}F_{8}S_{2}S_{3}$, $F_{7}F_{8}S_{5}S_{4}$, $F_{7}F_{8}S_{2}S_{5}$, $F_{2}F_{4}S_{16}S_{3}$, $F_{1}F_{2}S_{16}S_{3}$, $F_{7}F_{8}S_{3}S_{4}$, $F_{7}F_{8}S_{2}S_{3}$}} 
 \SplitRow{ $\mathcal{O}_{10, \; (1,-3)}^{2}$} {{$F_{3}F_{8}V_{11}V_{2}$, $F_{1}F_{6}S_{14}S_{2}$, $F_{3}F_{13}V_{12}V_{1}$, $F_{10}F_{7}S_{6}V_{11}$, $F_{1}F_{6}S_{2}V_{3}$, $F_{3}F_{13}S_{5}V_{1}$, $F_{10}F_{7}S_{14}S_{6}$, $F_{3}F_{8}V_{3}V_{2}$, $F_{3}F_{13}V_{4}V_{1}$, $F_{10}F_{7}S_{6}V_{12}$, $F_{3}F_{8}S_{5}V_{2}$, $F_{1}F_{6}S_{2}V_{4}$, $F_{7}F_{8}V_{11}V_{2}$, $F_{1}F_{6}V_{10}V_{3}$, $F_{3}F_{8}S_{16}S_{5}$, $F_{7}F_{8}V_{3}V_{2}$, $F_{3}F_{8}S_{16}S_{3}$, $F_{7}F_{8}S_{5}V_{2}$, $F_{1}F_{6}S_{3}V_{10}$, $F_{6}F_{7}S_{14}S_{2}$, $F_{3}F_{6}V_{10}V_{3}$, $F_{3}F_{5}V_{9}V_{4}$, $F_{6}F_{7}S_{2}V_{3}$, $F_{3}F_{5}S_{3}V_{9}$, $F_{6}F_{7}S_{2}V_{4}$, $F_{3}F_{6}S_{3}V_{10}$, $F_{7}F_{13}V_{12}V_{1}$, $F_{1}F_{5}V_{9}V_{4}$, $F_{7}F_{13}S_{5}V_{1}$, $F_{1}F_{5}S_{3}V_{9}$, $F_{7}F_{13}V_{4}V_{1}$, $F_{7}F_{8}S_{6}V_{11}$, $F_{3}F_{8}S_{5}V_{1}$, $F_{7}F_{8}S_{6}V_{3}$, $F_{3}F_{8}S_{3}V_{1}$, $F_{7}F_{8}S_{5}S_{6}$, $F_{1}F_{6}S_{2}S_{3}$, $F_{2}F_{3}S_{3}V_{9}$, $F_{6}F_{7}S_{2}S_{3}$, $F_{7}F_{8}S_{5}V_{1}$, $F_{1}F_{2}S_{3}V_{9}$, $F_{7}F_{8}S_{3}V_{1}$, $F_{6}F_{7}S_{14}S_{6}$, $F_{3}F_{6}V_{3}V_{2}$, $F_{3}F_{5}V_{4}V_{1}$, $F_{6}F_{7}S_{6}V_{3}$, $F_{3}F_{5}S_{3}V_{1}$, $F_{6}F_{7}S_{6}V_{4}$, $F_{3}F_{6}S_{3}V_{2}$, $F_{6}F_{7}V_{3}V_{2}$, $F_{2}F_{3}S_{16}S_{3}$, $F_{6}F_{7}S_{3}V_{2}$, $F_{5}F_{7}V_{4}V_{1}$, $F_{5}F_{7}S_{3}V_{1}$, $F_{7}F_{13}S_{6}V_{12}$, $F_{1}F_{5}S_{2}V_{4}$, $F_{7}F_{13}S_{5}S_{6}$, $F_{1}F_{5}S_{2}S_{3}$, $F_{7}F_{13}S_{6}V_{4}$, $F_{3}F_{8}S_{3}V_{2}$, $F_{1}F_{2}S_{3}V_{10}$, $F_{7}F_{8}S_{3}V_{2}$, $F_{5}F_{7}S_{2}V_{4}$, $F_{2}F_{3}S_{3}V_{10}$, $F_{5}F_{7}S_{2}S_{3}$}} 
 \SplitRow{ $\mathcal{O}_{10, \; (1,-3)}^{3}$} {{$F_{4}F_{8}S_{4}V_{11}$, $F_{1}F_{8}S_{2}V_{11}$, $F_{3}F_{14}S_{9}V_{1}$, $F_{3}F_{6}S_{14}V_{1}$, $F_{11}F_{7}S_{6}V_{11}$, $F_{2}F_{7}S_{6}V_{11}$, $F_{4}F_{8}S_{4}V_{3}$, $F_{1}F_{8}S_{2}V_{3}$, $F_{3}F_{14}S_{5}V_{1}$, $F_{3}F_{6}S_{5}V_{1}$, $F_{11}F_{7}S_{9}S_{6}$, $F_{2}F_{7}S_{14}S_{6}$, $F_{4}F_{8}S_{5}S_{4}$, $F_{1}F_{8}S_{2}S_{5}$, $F_{7}F_{8}S_{4}V_{11}$, $F_{7}F_{8}S_{2}V_{11}$, $F_{4}F_{6}S_{16}V_{3}$, $F_{1}F_{6}S_{16}V_{3}$, $F_{3}F_{8}S_{5}V_{13}$, $F_{3}F_{8}S_{5}V_{9}$, $F_{7}F_{8}S_{4}V_{3}$, $F_{7}F_{8}S_{2}V_{3}$, $F_{3}F_{8}S_{3}V_{13}$, $F_{3}F_{8}S_{3}V_{9}$, $F_{7}F_{8}S_{5}S_{4}$, $F_{7}F_{8}S_{2}S_{5}$, $F_{4}F_{6}S_{16}S_{3}$, $F_{1}F_{6}S_{16}S_{3}$, $F_{7}F_{14}S_{9}V_{1}$, $F_{6}F_{7}S_{14}V_{1}$, $F_{4}F_{8}S_{5}V_{13}$, $F_{1}F_{8}S_{5}V_{9}$, $F_{7}F_{14}S_{5}V_{1}$, $F_{6}F_{7}S_{5}V_{1}$, $F_{4}F_{8}S_{3}V_{13}$, $F_{1}F_{8}S_{3}V_{9}$, $F_{4}F_{8}V_{11}V_{2}$, $F_{1}F_{8}V_{11}V_{2}$, $F_{3}F_{14}S_{9}S_{4}$, $F_{3}F_{6}S_{14}S_{2}$, $F_{11}F_{7}V_{11}V_{5}$, $F_{2}F_{7}V_{11}V_{4}$, $F_{4}F_{8}V_{3}V_{2}$, $F_{1}F_{8}V_{3}V_{2}$, $F_{3}F_{14}S_{5}S_{4}$, $F_{3}F_{6}S_{2}S_{5}$, $F_{11}F_{7}S_{9}V_{5}$, $F_{2}F_{7}S_{14}V_{4}$, $F_{4}F_{8}S_{5}V_{2}$, $F_{1}F_{8}S_{5}V_{2}$, $F_{4}F_{14}S_{4}V_{11}$, $F_{1}F_{6}S_{2}V_{11}$, $F_{3}F_{13}S_{9}V_{1}$, $F_{3}F_{13}S_{14}V_{1}$, $F_{11}F_{9}V_{11}V_{5}$, $F_{2}F_{5}V_{11}V_{4}$, $F_{4}F_{14}S_{5}S_{4}$, $F_{1}F_{6}S_{2}S_{5}$, $F_{3}F_{13}V_{1}V_{5}$, $F_{3}F_{13}V_{4}V_{1}$, $F_{11}F_{9}S_{9}V_{5}$, $F_{2}F_{5}S_{14}V_{4}$, $F_{4}F_{14}S_{4}V_{5}$, $F_{1}F_{6}S_{2}V_{4}$, $F_{7}F_{14}S_{4}V_{11}$, $F_{6}F_{7}S_{2}V_{11}$, $F_{4}F_{8}S_{16}V_{3}$, $F_{1}F_{8}S_{16}V_{3}$, $F_{3}F_{9}S_{5}V_{13}$, $F_{3}F_{5}S_{5}V_{9}$, $F_{8}F_{9}V_{11}V_{2}$, $F_{5}F_{8}V_{11}V_{2}$, $F_{4}F_{8}S_{16}S_{5}$, $F_{1}F_{8}S_{16}S_{5}$, $F_{3}F_{9}V_{13}V_{5}$, $F_{3}F_{5}V_{9}V_{4}$, $F_{8}F_{9}V_{3}V_{2}$, $F_{5}F_{8}V_{3}V_{2}$, $F_{7}F_{14}S_{5}S_{4}$, $F_{6}F_{7}S_{2}S_{5}$, $F_{3}F_{9}V_{13}V_{2}$, $F_{3}F_{5}V_{9}V_{2}$, $F_{8}F_{9}S_{5}V_{2}$, $F_{5}F_{8}S_{5}V_{2}$, $F_{7}F_{14}S_{4}V_{5}$, $F_{6}F_{7}S_{2}V_{4}$, $F_{4}F_{8}S_{16}V_{2}$, $F_{1}F_{8}S_{16}V_{2}$, $F_{7}F_{13}S_{9}V_{1}$, $F_{7}F_{13}S_{14}V_{1}$, $F_{4}F_{9}S_{5}V_{13}$, $F_{1}F_{5}S_{5}V_{9}$, $F_{9}F_{14}S_{9}S_{4}$, $F_{5}F_{6}S_{14}S_{2}$, $F_{4}F_{9}V_{13}V_{5}$, $F_{1}F_{5}V_{9}V_{4}$, $F_{9}F_{14}S_{5}S_{4}$, $F_{5}F_{6}S_{2}S_{5}$, $F_{7}F_{13}V_{1}V_{5}$, $F_{7}F_{13}V_{4}V_{1}$, $F_{4}F_{9}V_{13}V_{2}$, $F_{1}F_{5}V_{9}V_{2}$, $F_{7}F_{8}V_{11}V_{5}$, $F_{7}F_{8}V_{11}V_{4}$, $F_{4}F_{6}V_{3}V_{2}$, $F_{1}F_{6}V_{3}V_{2}$, $F_{3}F_{8}S_{5}S_{4}$, $F_{3}F_{8}S_{2}S_{5}$, $F_{7}F_{8}V_{3}V_{5}$, $F_{7}F_{8}V_{3}V_{4}$, $F_{3}F_{8}S_{3}S_{4}$, $F_{3}F_{8}S_{2}S_{3}$, $F_{7}F_{8}S_{5}V_{5}$, $F_{7}F_{8}S_{5}V_{4}$, $F_{4}F_{6}S_{3}V_{2}$, $F_{1}F_{6}S_{3}V_{2}$, $F_{7}F_{14}S_{6}V_{11}$, $F_{6}F_{7}S_{6}V_{11}$, $F_{3}F_{9}S_{5}V_{1}$, $F_{3}F_{5}S_{5}V_{1}$, $F_{8}F_{9}V_{11}V_{5}$, $F_{5}F_{8}V_{11}V_{4}$, $F_{3}F_{9}V_{1}V_{5}$, $F_{3}F_{5}V_{4}V_{1}$, $F_{8}F_{9}V_{3}V_{5}$, $F_{5}F_{8}V_{3}V_{4}$, $F_{7}F_{14}S_{5}S_{6}$, $F_{6}F_{7}S_{5}S_{6}$, $F_{3}F_{9}V_{1}V_{2}$, $F_{3}F_{5}V_{1}V_{2}$, $F_{8}F_{9}S_{5}V_{5}$, $F_{5}F_{8}S_{5}V_{4}$, $F_{7}F_{14}S_{6}V_{5}$, $F_{6}F_{7}S_{6}V_{4}$, $F_{4}F_{8}S_{4}V_{2}$, $F_{1}F_{8}S_{2}V_{2}$, $F_{3}F_{3}S_{3}V_{13}$, $F_{3}F_{3}S_{3}V_{9}$, $F_{6}F_{9}V_{3}V_{2}$, $F_{5}F_{6}V_{3}V_{2}$, $F_{3}F_{3}V_{13}V_{2}$, $F_{3}F_{3}V_{9}V_{2}$, $F_{6}F_{9}S_{3}V_{2}$, $F_{5}F_{6}S_{3}V_{2}$, $F_{7}F_{8}S_{4}V_{2}$, $F_{7}F_{8}S_{2}V_{2}$, $F_{7}F_{9}S_{5}V_{1}$, $F_{5}F_{7}S_{5}V_{1}$, $F_{3}F_{4}S_{3}V_{13}$, $F_{1}F_{3}S_{3}V_{9}$, $F_{8}F_{9}S_{5}S_{4}$, $F_{5}F_{8}S_{2}S_{5}$, $F_{7}F_{9}V_{1}V_{5}$, $F_{5}F_{7}V_{4}V_{1}$, $F_{3}F_{4}V_{13}V_{2}$, $F_{1}F_{3}V_{9}V_{2}$, $F_{8}F_{9}S_{3}S_{4}$, $F_{5}F_{8}S_{2}S_{3}$, $F_{7}F_{9}V_{1}V_{2}$, $F_{5}F_{7}V_{1}V_{2}$, $F_{7}F_{14}S_{9}V_{5}$, $F_{6}F_{7}S_{14}V_{4}$, $F_{7}F_{14}S_{5}V_{5}$, $F_{6}F_{7}S_{5}V_{4}$, $F_{4}F_{8}S_{3}V_{2}$, $F_{1}F_{8}S_{3}V_{2}$, $F_{7}F_{13}S_{9}S_{6}$, $F_{7}F_{13}S_{14}S_{6}$, $F_{4}F_{9}S_{5}S_{4}$, $F_{1}F_{5}S_{2}S_{5}$, $F_{9}F_{14}S_{9}V_{5}$, $F_{5}F_{6}S_{14}V_{4}$, $F_{4}F_{9}S_{4}V_{5}$, $F_{1}F_{5}S_{2}V_{4}$, $F_{9}F_{14}S_{5}V_{5}$, $F_{5}F_{6}S_{5}V_{4}$, $F_{7}F_{13}S_{6}V_{5}$, $F_{7}F_{13}S_{6}V_{4}$, $F_{4}F_{9}S_{4}V_{2}$, $F_{1}F_{5}S_{2}V_{2}$, $F_{7}F_{9}S_{5}S_{4}$, $F_{5}F_{7}S_{2}S_{5}$, $F_{3}F_{4}S_{16}S_{3}$, $F_{1}F_{3}S_{16}S_{3}$, $F_{7}F_{9}S_{4}V_{5}$, $F_{5}F_{7}S_{2}V_{4}$, $F_{3}F_{4}S_{16}V_{2}$, $F_{1}F_{3}S_{16}V_{2}$, $F_{8}F_{9}S_{3}V_{2}$, $F_{5}F_{8}S_{3}V_{2}$, $F_{7}F_{9}S_{4}V_{2}$, $F_{5}F_{7}S_{2}V_{2}$}} 
 \SplitRow{ $\mathcal{O}_{10, \; (1,-3)}^{4}$} {{$F_{1}F_{6}S_{7}S_{2}$, $F_{3}F_{13}V_{11}V_{1}$, $F_{3}F_{8}S_{7}S_{5}$, $F_{3}F_{13}V_{4}V_{1}$, $F_{3}F_{8}S_{5}V_{11}$, $F_{1}F_{6}S_{2}V_{4}$, $F_{3}F_{13}S_{6}V_{1}$, $F_{1}F_{5}S_{7}S_{3}$, $F_{3}F_{8}V_{11}V_{2}$, $F_{3}F_{7}S_{7}S_{5}$, $F_{3}F_{8}V_{3}V_{2}$, $F_{3}F_{7}S_{5}V_{11}$, $F_{1}F_{5}S_{3}V_{3}$, $F_{3}F_{8}S_{5}V_{2}$, $F_{5}F_{8}S_{7}S_{3}$, $F_{3}F_{5}V_{9}V_{4}$, $F_{6}F_{7}S_{7}S_{2}$, $F_{3}F_{5}V_{9}V_{3}$, $F_{6}F_{7}S_{2}V_{4}$, $F_{5}F_{8}S_{3}V_{3}$, $F_{3}F_{5}S_{2}V_{9}$, $F_{8}F_{8}V_{11}V_{2}$, $F_{1}F_{5}V_{9}V_{4}$, $F_{3}F_{7}S_{16}S_{6}$, $F_{7}F_{13}V_{11}V_{1}$, $F_{1}F_{5}V_{9}V_{3}$, $F_{3}F_{7}S_{16}S_{5}$, $F_{7}F_{13}V_{4}V_{1}$, $F_{8}F_{8}V_{3}V_{2}$, $F_{3}F_{7}S_{16}S_{2}$, $F_{7}F_{13}S_{6}V_{1}$, $F_{8}F_{8}S_{5}V_{2}$, $F_{1}F_{5}S_{2}V_{9}$, $F_{1}F_{5}S_{7}S_{2}$, $F_{3}F_{8}V_{11}V_{1}$, $F_{3}F_{7}S_{7}S_{6}$, $F_{3}F_{8}V_{3}V_{1}$, $F_{3}F_{7}S_{6}V_{11}$, $F_{1}F_{5}S_{2}V_{3}$, $F_{3}F_{8}S_{5}V_{1}$, $F_{5}F_{7}S_{7}S_{2}$, $F_{3}F_{6}V_{9}V_{3}$, $F_{5}F_{7}S_{2}V_{3}$, $F_{3}F_{6}S_{3}V_{9}$, $F_{7}F_{8}V_{11}V_{1}$, $F_{1}F_{6}V_{9}V_{3}$, $F_{3}F_{8}S_{16}S_{5}$, $F_{7}F_{8}V_{3}V_{1}$, $F_{3}F_{8}S_{16}S_{3}$, $F_{7}F_{8}S_{5}V_{1}$, $F_{1}F_{6}S_{3}V_{9}$, $F_{5}F_{8}S_{7}S_{5}$, $F_{3}F_{5}V_{4}V_{1}$, $F_{6}F_{7}S_{7}S_{6}$, $F_{3}F_{5}V_{3}V_{1}$, $F_{6}F_{7}S_{6}V_{4}$, $F_{5}F_{8}S_{5}V_{3}$, $F_{3}F_{5}S_{2}V_{1}$, $F_{5}F_{7}S_{7}S_{5}$, $F_{3}F_{6}V_{3}V_{2}$, $F_{5}F_{7}S_{5}V_{3}$, $F_{3}F_{6}S_{3}V_{2}$, $F_{5}F_{7}V_{4}V_{1}$, $F_{6}F_{8}V_{3}V_{2}$, $F_{1}F_{3}S_{16}S_{2}$, $F_{5}F_{7}V_{3}V_{1}$, $F_{1}F_{3}S_{16}S_{3}$, $F_{5}F_{7}S_{2}V_{1}$, $F_{6}F_{8}S_{3}V_{2}$, $F_{8}F_{8}S_{5}V_{11}$, $F_{1}F_{5}S_{2}V_{4}$, $F_{3}F_{7}S_{6}V_{1}$, $F_{7}F_{13}S_{6}V_{11}$, $F_{3}F_{7}S_{5}V_{1}$, $F_{7}F_{13}S_{6}V_{4}$, $F_{8}F_{8}S_{5}V_{3}$, $F_{3}F_{7}S_{2}V_{1}$, $F_{7}F_{13}S_{6}S_{6}$, $F_{8}F_{8}S_{5}S_{5}$, $F_{1}F_{5}S_{2}S_{2}$, $F_{7}F_{8}S_{5}V_{11}$, $F_{1}F_{6}S_{3}V_{3}$, $F_{7}F_{8}S_{5}V_{3}$, $F_{3}F_{8}S_{3}V_{2}$, $F_{7}F_{8}S_{5}S_{5}$, $F_{1}F_{6}S_{3}S_{3}$, $F_{5}F_{7}S_{2}V_{4}$, $F_{6}F_{8}S_{3}V_{3}$, $F_{1}F_{3}S_{2}V_{9}$, $F_{1}F_{3}S_{3}V_{9}$, $F_{5}F_{7}S_{2}S_{2}$, $F_{6}F_{8}S_{3}S_{3}$, $F_{7}F_{7}S_{6}V_{1}$, $F_{1}F_{1}S_{2}V_{9}$, $F_{7}F_{7}S_{5}V_{1}$, $F_{1}F_{1}S_{3}V_{9}$, $F_{7}F_{7}S_{2}V_{1}$, $F_{8}F_{8}S_{3}V_{2}$}} 
 \SplitRow{ $\mathcal{O}_{10, \; (1,-3)}^{5}$} {{$F_{8}F_{8}S_{3}V_{2}$, $F_{2}F_{2}S_{3}V_{10}$, $F_{8}F_{8}S_{5}V_{2}$, $F_{6}F_{8}S_{3}S_{3}$, $F_{2}F_{3}S_{3}V_{10}$, $F_{6}F_{8}S_{3}V_{3}$, $F_{2}F_{6}S_{3}S_{3}$, $F_{8}F_{8}S_{5}S_{5}$, $F_{3}F_{8}S_{3}V_{2}$, $F_{8}F_{8}S_{5}V_{3}$, $F_{3}F_{8}S_{5}V_{2}$, $F_{2}F_{6}S_{3}V_{3}$, $F_{8}F_{8}S_{5}V_{11}$, $F_{6}F_{8}S_{3}V_{2}$, $F_{2}F_{3}S_{16}S_{3}$, $F_{6}F_{8}V_{3}V_{2}$, $F_{3}F_{6}S_{3}V_{2}$, $F_{5}F_{8}S_{5}V_{3}$, $F_{3}F_{6}V_{3}V_{2}$, $F_{5}F_{8}S_{7}S_{5}$, $F_{2}F_{6}S_{3}V_{10}$, $F_{3}F_{8}S_{16}S_{3}$, $F_{8}F_{8}V_{3}V_{2}$, $F_{3}F_{8}S_{16}S_{5}$, $F_{2}F_{6}V_{10}V_{3}$, $F_{8}F_{8}V_{11}V_{2}$, $F_{3}F_{6}S_{3}V_{10}$, $F_{5}F_{8}S_{3}V_{3}$, $F_{3}F_{6}V_{10}V_{3}$, $F_{5}F_{8}S_{7}S_{3}$, $F_{2}F_{5}S_{3}V_{3}$, $F_{3}F_{8}S_{5}V_{11}$, $F_{3}F_{8}V_{3}V_{2}$, $F_{3}F_{8}S_{7}S_{5}$, $F_{3}F_{8}V_{11}V_{2}$, $F_{2}F_{5}S_{7}S_{3}$}} 
 \SplitRow{ $\mathcal{O}_{10, \; (1,-3)}^{6}$} {{$F_{3}F_{5}S_{7}V_{2}$, $F_{3}F_{6}S_{14}V_{1}$, $F_{2}F_{7}S_{7}S_{6}$, $F_{3}F_{6}V_{3}V_{1}$, $F_{2}F_{7}S_{14}S_{6}$, $F_{3}F_{5}V_{3}V_{2}$, $F_{3}F_{6}V_{4}V_{1}$, $F_{5}F_{7}S_{7}V_{2}$, $F_{3}F_{6}S_{16}V_{3}$, $F_{5}F_{7}V_{3}V_{2}$, $F_{3}F_{6}S_{16}S_{3}$, $F_{6}F_{7}S_{14}V_{1}$, $F_{3}F_{5}S_{16}V_{4}$, $F_{6}F_{7}V_{3}V_{1}$, $F_{3}F_{5}S_{16}S_{3}$, $F_{6}F_{7}V_{4}V_{1}$, $F_{3}F_{5}S_{7}S_{3}$, $F_{3}F_{6}S_{14}S_{2}$, $F_{2}F_{7}S_{7}V_{4}$, $F_{3}F_{6}S_{2}V_{3}$, $F_{2}F_{7}S_{14}V_{4}$, $F_{3}F_{5}S_{3}V_{3}$, $F_{3}F_{6}S_{2}V_{4}$, $F_{3}F_{8}S_{7}V_{2}$, $F_{3}F_{13}S_{14}V_{1}$, $F_{2}F_{5}S_{7}V_{4}$, $F_{3}F_{13}S_{5}V_{1}$, $F_{2}F_{5}S_{14}V_{4}$, $F_{3}F_{8}S_{5}V_{2}$, $F_{3}F_{13}S_{6}V_{1}$, $F_{7}F_{8}S_{7}V_{2}$, $F_{3}F_{8}S_{16}V_{3}$, $F_{5}F_{5}S_{7}S_{3}$, $F_{3}F_{8}S_{16}S_{5}$, $F_{5}F_{5}S_{3}V_{3}$, $F_{7}F_{8}S_{5}V_{2}$, $F_{3}F_{8}S_{16}V_{2}$, $F_{7}F_{13}S_{14}V_{1}$, $F_{3}F_{7}S_{16}V_{4}$, $F_{5}F_{6}S_{14}S_{2}$, $F_{3}F_{7}S_{16}S_{6}$, $F_{5}F_{6}S_{2}V_{3}$, $F_{7}F_{13}S_{5}V_{1}$, $F_{3}F_{7}S_{16}V_{2}$, $F_{5}F_{6}S_{2}V_{4}$, $F_{7}F_{13}S_{6}V_{1}$, $F_{5}F_{7}S_{7}V_{4}$, $F_{5}F_{7}V_{3}V_{4}$, $F_{3}F_{6}S_{2}S_{3}$, $F_{7}F_{8}S_{7}S_{6}$, $F_{3}F_{8}V_{3}V_{1}$, $F_{5}F_{5}S_{7}V_{4}$, $F_{3}F_{8}S_{5}V_{1}$, $F_{5}F_{5}V_{3}V_{4}$, $F_{7}F_{8}S_{5}S_{6}$, $F_{3}F_{8}V_{1}V_{2}$, $F_{7}F_{8}V_{3}V_{1}$, $F_{3}F_{3}S_{16}S_{3}$, $F_{7}F_{8}S_{5}V_{1}$, $F_{3}F_{3}S_{16}V_{2}$, $F_{5}F_{6}S_{2}S_{3}$, $F_{7}F_{8}V_{1}V_{2}$, $F_{6}F_{7}S_{14}V_{4}$, $F_{3}F_{6}S_{3}V_{3}$, $F_{3}F_{5}S_{2}V_{4}$, $F_{6}F_{7}V_{3}V_{4}$, $F_{3}F_{5}S_{2}S_{3}$, $F_{6}F_{7}V_{4}V_{4}$, $F_{3}F_{6}S_{3}S_{3}$, $F_{7}F_{13}S_{14}S_{6}$, $F_{3}F_{8}V_{3}V_{2}$, $F_{3}F_{7}V_{4}V_{1}$, $F_{5}F_{6}S_{14}V_{4}$, $F_{3}F_{7}S_{6}V_{1}$, $F_{5}F_{6}V_{3}V_{4}$, $F_{7}F_{13}S_{5}S_{6}$, $F_{3}F_{7}V_{1}V_{2}$, $F_{5}F_{6}V_{4}V_{4}$, $F_{7}F_{13}S_{6}S_{6}$, $F_{3}F_{8}V_{2}V_{2}$, $F_{7}F_{8}V_{3}V_{2}$, $F_{5}F_{6}S_{3}V_{3}$, $F_{5}F_{6}S_{3}S_{3}$, $F_{7}F_{8}V_{2}V_{2}$, $F_{7}F_{7}V_{4}V_{1}$, $F_{5}F_{5}S_{2}V_{4}$, $F_{7}F_{7}S_{6}V_{1}$, $F_{5}F_{5}S_{2}S_{3}$, $F_{7}F_{7}V_{1}V_{2}$}} 
 \SplitRow{ $\mathcal{O}_{10, \; (1,-3)}^{7}$} {{$F_{4}F_{5}S_{7}V_{2}$, $F_{1}F_{5}S_{7}V_{2}$, $F_{3}F_{8}S_{4}V_{11}$, $F_{3}F_{8}S_{2}V_{11}$, $F_{3}F_{7}S_{7}V_{5}$, $F_{3}F_{7}S_{7}V_{4}$, $F_{3}F_{8}S_{4}V_{3}$, $F_{3}F_{8}S_{2}V_{3}$, $F_{3}F_{7}V_{11}V_{5}$, $F_{3}F_{7}V_{11}V_{4}$, $F_{4}F_{5}V_{3}V_{2}$, $F_{1}F_{5}V_{3}V_{2}$, $F_{3}F_{8}S_{5}S_{4}$, $F_{3}F_{8}S_{2}S_{5}$, $F_{4}F_{8}S_{7}S_{4}$, $F_{1}F_{8}S_{7}S_{2}$, $F_{3}F_{14}V_{11}V_{1}$, $F_{3}F_{6}V_{11}V_{1}$, $F_{3}F_{9}S_{7}V_{5}$, $F_{3}F_{5}S_{7}V_{4}$, $F_{3}F_{14}S_{5}V_{1}$, $F_{3}F_{6}S_{5}V_{1}$, $F_{3}F_{9}V_{11}V_{5}$, $F_{3}F_{5}V_{11}V_{4}$, $F_{4}F_{8}S_{5}S_{4}$, $F_{1}F_{8}S_{2}S_{5}$, $F_{3}F_{14}V_{1}V_{5}$, $F_{3}F_{6}V_{4}V_{1}$, $F_{7}F_{8}S_{7}S_{4}$, $F_{7}F_{8}S_{7}S_{2}$, $F_{3}F_{8}V_{13}V_{3}$, $F_{3}F_{8}V_{9}V_{3}$, $F_{5}F_{9}S_{7}V_{2}$, $F_{5}F_{5}S_{7}V_{2}$, $F_{3}F_{8}S_{5}V_{13}$, $F_{3}F_{8}S_{5}V_{9}$, $F_{5}F_{9}V_{3}V_{2}$, $F_{5}F_{5}V_{3}V_{2}$, $F_{7}F_{8}S_{5}S_{4}$, $F_{7}F_{8}S_{2}S_{5}$, $F_{3}F_{8}V_{13}V_{2}$, $F_{3}F_{8}V_{9}V_{2}$, $F_{7}F_{14}V_{11}V_{1}$, $F_{6}F_{7}V_{11}V_{1}$, $F_{4}F_{8}V_{13}V_{3}$, $F_{1}F_{8}V_{9}V_{3}$, $F_{3}F_{9}S_{16}S_{5}$, $F_{3}F_{5}S_{16}S_{5}$, $F_{8}F_{9}S_{4}V_{11}$, $F_{5}F_{8}S_{2}V_{11}$, $F_{4}F_{8}S_{5}V_{13}$, $F_{1}F_{8}S_{5}V_{9}$, $F_{3}F_{9}S_{16}V_{5}$, $F_{3}F_{5}S_{16}V_{4}$, $F_{8}F_{9}S_{4}V_{3}$, $F_{5}F_{8}S_{2}V_{3}$, $F_{7}F_{14}S_{5}V_{1}$, $F_{6}F_{7}S_{5}V_{1}$, $F_{3}F_{9}S_{16}V_{2}$, $F_{3}F_{5}S_{16}V_{2}$, $F_{8}F_{9}S_{5}S_{4}$, $F_{5}F_{8}S_{2}S_{5}$, $F_{7}F_{14}V_{1}V_{5}$, $F_{6}F_{7}V_{4}V_{1}$, $F_{4}F_{8}V_{13}V_{2}$, $F_{1}F_{8}V_{9}V_{2}$, $F_{4}F_{8}S_{7}V_{2}$, $F_{1}F_{8}S_{7}V_{2}$, $F_{3}F_{14}S_{4}V_{11}$, $F_{3}F_{6}S_{2}V_{11}$, $F_{3}F_{9}S_{7}S_{5}$, $F_{3}F_{5}S_{7}S_{5}$, $F_{3}F_{14}S_{5}S_{4}$, $F_{3}F_{6}S_{2}S_{5}$, $F_{3}F_{9}S_{5}V_{11}$, $F_{3}F_{5}S_{5}V_{11}$, $F_{4}F_{8}S_{5}V_{2}$, $F_{1}F_{8}S_{5}V_{2}$, $F_{3}F_{14}S_{4}V_{5}$, $F_{3}F_{6}S_{2}V_{4}$, $F_{8}F_{9}S_{7}V_{2}$, $F_{5}F_{8}S_{7}V_{2}$, $F_{3}F_{9}S_{5}V_{13}$, $F_{3}F_{5}S_{5}V_{9}$, $F_{8}F_{9}S_{5}V_{2}$, $F_{5}F_{8}S_{5}V_{2}$, $F_{3}F_{9}S_{4}V_{13}$, $F_{3}F_{5}S_{2}V_{9}$, $F_{9}F_{14}S_{4}V_{11}$, $F_{5}F_{6}S_{2}V_{11}$, $F_{4}F_{9}S_{5}V_{13}$, $F_{1}F_{5}S_{5}V_{9}$, $F_{3}F_{7}S_{16}V_{5}$, $F_{3}F_{7}S_{16}V_{4}$, $F_{9}F_{14}S_{5}S_{4}$, $F_{5}F_{6}S_{2}S_{5}$, $F_{3}F_{7}S_{16}S_{4}$, $F_{3}F_{7}S_{16}S_{2}$, $F_{9}F_{14}S_{4}V_{5}$, $F_{5}F_{6}S_{2}V_{4}$, $F_{4}F_{9}S_{4}V_{13}$, $F_{1}F_{5}S_{2}V_{9}$, $F_{7}F_{8}S_{7}V_{5}$, $F_{7}F_{8}S_{7}V_{4}$, $F_{5}F_{9}S_{7}S_{5}$, $F_{5}F_{5}S_{7}S_{5}$, $F_{5}F_{9}S_{5}V_{3}$, $F_{5}F_{5}S_{5}V_{3}$, $F_{7}F_{8}S_{5}V_{5}$, $F_{7}F_{8}S_{5}V_{4}$, $F_{3}F_{8}S_{4}V_{2}$, $F_{3}F_{8}S_{2}V_{2}$, $F_{8}F_{9}S_{7}V_{5}$, $F_{5}F_{8}S_{7}V_{4}$, $F_{3}F_{9}S_{5}V_{1}$, $F_{3}F_{5}S_{5}V_{1}$, $F_{8}F_{9}S_{5}V_{5}$, $F_{5}F_{8}S_{5}V_{4}$, $F_{3}F_{9}S_{4}V_{1}$, $F_{3}F_{5}S_{2}V_{1}$, $F_{7}F_{9}S_{5}V_{1}$, $F_{5}F_{7}S_{5}V_{1}$, $F_{3}F_{4}S_{16}V_{2}$, $F_{1}F_{3}S_{16}V_{2}$, $F_{3}F_{4}S_{16}S_{4}$, $F_{1}F_{3}S_{16}S_{2}$, $F_{8}F_{9}S_{4}V_{2}$, $F_{5}F_{8}S_{2}V_{2}$, $F_{7}F_{9}S_{4}V_{1}$, $F_{5}F_{7}S_{2}V_{1}$, $F_{7}F_{14}V_{11}V_{5}$, $F_{6}F_{7}V_{11}V_{4}$, $F_{4}F_{8}V_{3}V_{2}$, $F_{1}F_{8}V_{3}V_{2}$, $F_{3}F_{9}S_{5}S_{4}$, $F_{3}F_{5}S_{2}S_{5}$, $F_{8}F_{9}S_{5}V_{11}$, $F_{5}F_{8}S_{5}V_{11}$, $F_{3}F_{9}S_{4}V_{5}$, $F_{3}F_{5}S_{2}V_{4}$, $F_{8}F_{9}S_{5}V_{3}$, $F_{5}F_{8}S_{5}V_{3}$, $F_{7}F_{14}S_{5}V_{5}$, $F_{6}F_{7}S_{5}V_{4}$, $F_{3}F_{9}S_{4}V_{2}$, $F_{3}F_{5}S_{2}V_{2}$, $F_{8}F_{9}S_{5}S_{5}$, $F_{5}F_{8}S_{5}S_{5}$, $F_{7}F_{14}V_{5}V_{5}$, $F_{6}F_{7}V_{4}V_{4}$, $F_{4}F_{8}V_{2}V_{2}$, $F_{1}F_{8}V_{2}V_{2}$, $F_{9}F_{14}V_{11}V_{5}$, $F_{5}F_{6}V_{11}V_{4}$, $F_{4}F_{9}S_{5}S_{4}$, $F_{1}F_{5}S_{2}S_{5}$, $F_{3}F_{7}V_{1}V_{5}$, $F_{3}F_{7}V_{4}V_{1}$, $F_{9}F_{14}S_{5}V_{5}$, $F_{5}F_{6}S_{5}V_{4}$, $F_{3}F_{7}S_{4}V_{1}$, $F_{3}F_{7}S_{2}V_{1}$, $F_{9}F_{14}V_{5}V_{5}$, $F_{5}F_{6}V_{4}V_{4}$, $F_{4}F_{9}S_{4}S_{4}$, $F_{1}F_{5}S_{2}S_{2}$, $F_{8}F_{9}V_{3}V_{2}$, $F_{5}F_{8}V_{3}V_{2}$, $F_{7}F_{9}S_{5}S_{4}$, $F_{5}F_{7}S_{2}S_{5}$, $F_{3}F_{4}V_{13}V_{2}$, $F_{1}F_{3}V_{9}V_{2}$, $F_{3}F_{4}S_{4}V_{13}$, $F_{1}F_{3}S_{2}V_{9}$, $F_{8}F_{9}V_{2}V_{2}$, $F_{5}F_{8}V_{2}V_{2}$, $F_{7}F_{9}S_{4}S_{4}$, $F_{5}F_{7}S_{2}S_{2}$, $F_{9}F_{9}S_{5}S_{4}$, $F_{5}F_{5}S_{2}S_{5}$, $F_{7}F_{7}V_{1}V_{5}$, $F_{7}F_{7}V_{4}V_{1}$, $F_{4}F_{4}V_{13}V_{2}$, $F_{1}F_{1}V_{9}V_{2}$, $F_{9}F_{9}S_{4}V_{5}$, $F_{5}F_{5}S_{2}V_{4}$, $F_{4}F_{4}S_{4}V_{13}$, $F_{1}F_{1}S_{2}V_{9}$, $F_{9}F_{9}S_{4}V_{2}$, $F_{5}F_{5}S_{2}V_{2}$, $F_{7}F_{7}S_{4}V_{1}$, $F_{7}F_{7}S_{2}V_{1}$}} 
 \SplitRow{ $\mathcal{O}_{10, \; (1,-3)}^{8}$} {{$F_{8}F_{8}V_{2}V_{2}$, $F_{6}F_{6}S_{3}S_{3}$, $F_{3}F_{3}S_{16}V_{2}$, $F_{8}F_{8}S_{5}V_{2}$, $F_{6}F_{6}S_{3}V_{3}$, $F_{3}F_{3}S_{16}S_{3}$, $F_{8}F_{8}V_{3}V_{2}$, $F_{3}F_{8}V_{2}V_{2}$, $F_{8}F_{8}S_{5}S_{5}$, $F_{5}F_{6}V_{3}V_{3}$, $F_{3}F_{8}S_{5}V_{2}$, $F_{5}F_{6}S_{7}V_{3}$, $F_{3}F_{8}V_{3}V_{2}$, $F_{8}F_{8}S_{7}S_{5}$, $F_{3}F_{6}S_{3}S_{3}$, $F_{5}F_{8}V_{3}V_{3}$, $F_{3}F_{6}S_{3}V_{3}$, $F_{5}F_{8}S_{7}V_{3}$, $F_{3}F_{8}S_{16}V_{2}$, $F_{5}F_{6}S_{3}V_{3}$, $F_{3}F_{8}S_{16}S_{5}$, $F_{5}F_{6}S_{7}S_{3}$, $F_{3}F_{8}S_{16}V_{3}$, $F_{8}F_{8}S_{7}V_{2}$, $F_{1}F_{6}S_{7}V_{3}$, $F_{3}F_{8}S_{7}V_{2}$, $F_{3}F_{5}S_{3}V_{3}$, $F_{1}F_{8}S_{7}V_{3}$, $F_{3}F_{5}S_{7}S_{3}$, $F_{3}F_{6}S_{16}S_{3}$, $F_{5}F_{8}V_{3}V_{2}$, $F_{3}F_{6}S_{16}V_{3}$, $F_{5}F_{8}S_{7}V_{2}$, $F_{3}F_{5}V_{3}V_{2}$, $F_{1}F_{8}S_{7}S_{5}$, $F_{3}F_{5}S_{7}V_{2}$}} 
 \SplitRow{ $\mathcal{O}_{10, \; (1,-3)}^{9}$} {{$F_{3}F_{6}S_{7}V_{1}$, $F_{1}F_{8}S_{7}S_{5}$, $F_{3}F_{6}V_{4}V_{1}$, $F_{3}F_{5}S_{7}V_{2}$, $F_{1}F_{7}S_{7}S_{5}$, $F_{3}F_{5}V_{3}V_{2}$, $F_{5}F_{8}S_{7}V_{2}$, $F_{3}F_{5}S_{16}V_{4}$, $F_{6}F_{7}S_{7}V_{1}$, $F_{3}F_{5}S_{16}V_{3}$, $F_{6}F_{7}V_{4}V_{1}$, $F_{5}F_{8}V_{3}V_{2}$, $F_{3}F_{5}S_{16}S_{2}$, $F_{3}F_{6}S_{7}S_{2}$, $F_{1}F_{8}S_{7}V_{3}$, $F_{3}F_{6}S_{2}V_{4}$, $F_{3}F_{8}S_{7}V_{2}$, $F_{1}F_{5}S_{7}V_{3}$, $F_{3}F_{8}S_{5}V_{2}$, $F_{8}F_{8}S_{7}V_{2}$, $F_{3}F_{7}S_{16}V_{4}$, $F_{5}F_{6}S_{7}S_{2}$, $F_{3}F_{7}S_{16}S_{5}$, $F_{5}F_{6}S_{2}V_{4}$, $F_{8}F_{8}S_{5}V_{2}$, $F_{3}F_{7}S_{16}V_{1}$, $F_{3}F_{5}S_{7}S_{2}$, $F_{1}F_{7}S_{7}V_{4}$, $F_{3}F_{5}S_{2}V_{3}$, $F_{3}F_{8}S_{7}V_{1}$, $F_{1}F_{5}S_{7}V_{4}$, $F_{3}F_{8}S_{5}V_{1}$, $F_{7}F_{8}S_{7}V_{1}$, $F_{3}F_{8}S_{16}V_{3}$, $F_{5}F_{5}S_{7}S_{2}$, $F_{3}F_{8}S_{16}S_{5}$, $F_{5}F_{5}S_{2}V_{3}$, $F_{7}F_{8}S_{5}V_{1}$, $F_{3}F_{8}S_{16}V_{2}$, $F_{5}F_{8}S_{7}V_{3}$, $F_{3}F_{5}S_{2}V_{4}$, $F_{6}F_{7}S_{7}V_{4}$, $F_{6}F_{7}V_{4}V_{4}$, $F_{5}F_{8}V_{3}V_{3}$, $F_{3}F_{5}S_{2}S_{2}$, $F_{8}F_{8}S_{7}S_{5}$, $F_{3}F_{7}V_{4}V_{1}$, $F_{5}F_{6}S_{7}V_{4}$, $F_{3}F_{7}S_{5}V_{1}$, $F_{5}F_{6}V_{4}V_{4}$, $F_{8}F_{8}S_{5}S_{5}$, $F_{3}F_{7}V_{1}V_{1}$, $F_{7}F_{8}S_{7}S_{5}$, $F_{3}F_{8}V_{3}V_{2}$, $F_{5}F_{5}S_{7}V_{3}$, $F_{5}F_{5}V_{3}V_{3}$, $F_{7}F_{8}S_{5}S_{5}$, $F_{3}F_{8}V_{2}V_{2}$, $F_{7}F_{7}V_{4}V_{1}$, $F_{8}F_{8}V_{3}V_{2}$, $F_{3}F_{3}S_{16}S_{2}$, $F_{5}F_{5}S_{2}V_{4}$, $F_{3}F_{3}S_{16}V_{1}$, $F_{7}F_{7}S_{5}V_{1}$, $F_{3}F_{3}S_{16}V_{2}$, $F_{5}F_{5}S_{2}S_{2}$, $F_{7}F_{7}V_{1}V_{1}$, $F_{8}F_{8}V_{2}V_{2}$}} 
 \SplitRow{ $\mathcal{O}_{10, \; (1,-3)}^{10}$} {{$F_{3}F_{8}S_{7}S_{4}$, $F_{3}F_{8}S_{7}S_{2}$, $F_{1}F_{9}S_{7}S_{5}$, $F_{1}F_{5}S_{7}S_{5}$, $F_{3}F_{8}S_{5}S_{4}$, $F_{3}F_{8}S_{2}S_{5}$, $F_{8}F_{9}S_{7}S_{4}$, $F_{5}F_{8}S_{7}S_{2}$, $F_{3}F_{9}S_{16}S_{5}$, $F_{3}F_{5}S_{16}S_{5}$, $F_{8}F_{9}S_{5}S_{4}$, $F_{5}F_{8}S_{2}S_{5}$, $F_{3}F_{9}S_{16}S_{4}$, $F_{3}F_{5}S_{16}S_{2}$, $F_{8}F_{9}S_{7}S_{5}$, $F_{5}F_{8}S_{7}S_{5}$, $F_{3}F_{9}S_{5}S_{4}$, $F_{3}F_{5}S_{2}S_{5}$, $F_{8}F_{9}S_{5}S_{5}$, $F_{5}F_{8}S_{5}S_{5}$, $F_{3}F_{9}S_{4}S_{4}$, $F_{3}F_{5}S_{2}S_{2}$, $F_{9}F_{9}S_{5}S_{4}$, $F_{5}F_{5}S_{2}S_{5}$, $F_{3}F_{3}S_{16}S_{4}$, $F_{3}F_{3}S_{16}S_{2}$, $F_{9}F_{9}S_{4}S_{4}$, $F_{5}F_{5}S_{2}S_{2}$}} 
 \end{SplitTable}
   \vspace{0.4cm} 
 \captionof{table}{Dimension 10 $(\Delta B, \Delta L)$ = (1, -3) UV completions for topology 6 (see Fig.~\ref{fig:d=10_topologies}).} 
 \label{tab:T10Top6B1L-3}
 \end{center}

%% file: Tables_dim10/Table_dim10_Top7_B1_L-3.tex
 \begin{center}
 \begin{SplitTable}[2.0cm] 
  \SplitHeader{$\mathcal{O}^j{(\mathcal{T}_ {10} ^{ 7 })}$}{New Particles} 
   \vspace{0.1cm} 
 \SplitRow{ $\mathcal{O}_{10, \; (1,-3)}^{1}$} {{$F_{4}S_{9}S_{3}S_{4}$, $F_{1}S_{14}S_{2}S_{3}$, $F_{4}S_{5}S_{5}S_{4}$, $F_{1}S_{2}S_{5}S_{5}$, $F_{7}S_{9}S_{5}S_{6}$, $F_{7}S_{14}S_{5}S_{6}$, $F_{4}S_{16}S_{3}S_{5}$, $F_{1}S_{16}S_{3}S_{5}$, $F_{7}S_{9}S_{3}S_{4}$, $F_{7}S_{14}S_{2}S_{3}$, $F_{7}S_{5}S_{5}S_{4}$, $F_{7}S_{2}S_{5}S_{5}$}} 
 \SplitRow{ $\mathcal{O}_{10, \; (1,-3)}^{2}$} {{$F_{3}S_{3}V_{12}V_{1}$, $F_{1}S_{14}S_{2}S_{3}$, $F_{3}S_{3}V_{11}V_{2}$, $F_{3}S_{5}V_{4}V_{1}$, $F_{1}S_{2}V_{3}V_{4}$, $F_{7}S_{6}V_{11}V_{4}$, $F_{3}S_{5}V_{3}V_{2}$, $F_{7}S_{14}S_{5}S_{6}$, $F_{7}S_{6}V_{12}V_{3}$, $F_{3}S_{16}S_{3}S_{5}$, $F_{1}S_{3}V_{10}V_{3}$, $F_{7}S_{3}V_{11}V_{2}$, $F_{7}S_{5}V_{3}V_{2}$, $F_{3}S_{3}V_{9}V_{4}$, $F_{3}S_{3}V_{10}V_{3}$, $F_{7}S_{14}S_{2}S_{3}$, $F_{7}S_{2}V_{3}V_{4}$, $F_{1}S_{3}V_{9}V_{4}$, $F_{7}S_{3}V_{12}V_{1}$, $F_{7}S_{5}V_{4}V_{1}$}} 
 \SplitRow{ $\mathcal{O}_{10, \; (1,-3)}^{3}$} {{$F_{3}S_{9}S_{3}S_{4}$, $F_{3}S_{14}S_{2}S_{3}$, $F_{4}S_{3}V_{11}V_{2}$, $F_{1}S_{3}V_{11}V_{2}$, $F_{3}S_{5}S_{5}S_{4}$, $F_{3}S_{2}S_{5}S_{5}$, $F_{4}S_{5}V_{3}V_{2}$, $F_{1}S_{5}V_{3}V_{2}$, $F_{7}S_{5}V_{11}V_{5}$, $F_{7}S_{5}V_{11}V_{4}$, $F_{7}S_{9}V_{3}V_{5}$, $F_{7}S_{14}V_{3}V_{4}$, $F_{3}S_{9}V_{1}V_{2}$, $F_{3}S_{14}V_{1}V_{2}$, $F_{4}S_{4}V_{11}V_{2}$, $F_{1}S_{2}V_{11}V_{2}$, $F_{3}S_{5}V_{1}V_{5}$, $F_{3}S_{5}V_{4}V_{1}$, $F_{4}S_{4}V_{3}V_{5}$, $F_{1}S_{2}V_{3}V_{4}$, $F_{7}S_{6}V_{11}V_{5}$, $F_{7}S_{6}V_{11}V_{4}$, $F_{4}S_{5}S_{5}S_{4}$, $F_{1}S_{2}S_{5}S_{5}$, $F_{7}S_{9}S_{5}S_{6}$, $F_{7}S_{14}S_{5}S_{6}$, $F_{9}S_{5}V_{11}V_{5}$, $F_{5}S_{5}V_{11}V_{4}$, $F_{9}S_{9}V_{3}V_{5}$, $F_{5}S_{14}V_{3}V_{4}$, $F_{3}S_{5}V_{13}V_{2}$, $F_{3}S_{5}V_{9}V_{2}$, $F_{4}S_{16}V_{3}V_{2}$, $F_{1}S_{16}V_{3}V_{2}$, $F_{7}S_{4}V_{11}V_{2}$, $F_{7}S_{2}V_{11}V_{2}$, $F_{3}S_{3}V_{13}V_{5}$, $F_{3}S_{3}V_{9}V_{4}$, $F_{7}S_{4}V_{3}V_{5}$, $F_{7}S_{2}V_{3}V_{4}$, $F_{4}S_{16}S_{3}S_{5}$, $F_{1}S_{16}S_{3}S_{5}$, $F_{7}S_{5}S_{5}S_{4}$, $F_{7}S_{2}S_{5}S_{5}$, $F_{9}S_{3}V_{11}V_{2}$, $F_{5}S_{3}V_{11}V_{2}$, $F_{9}S_{5}V_{3}V_{2}$, $F_{5}S_{5}V_{3}V_{2}$, $F_{4}S_{5}V_{13}V_{2}$, $F_{1}S_{5}V_{9}V_{2}$, $F_{7}S_{9}V_{1}V_{2}$, $F_{7}S_{14}V_{1}V_{2}$, $F_{4}S_{3}V_{13}V_{5}$, $F_{1}S_{3}V_{9}V_{4}$, $F_{7}S_{5}V_{1}V_{5}$, $F_{7}S_{5}V_{4}V_{1}$, $F_{9}S_{9}S_{3}S_{4}$, $F_{5}S_{14}S_{2}S_{3}$, $F_{9}S_{5}S_{5}S_{4}$, $F_{5}S_{2}S_{5}S_{5}$}} 
 \SplitRow{ $\mathcal{O}_{10, \; (1,-3)}^{4}$} {{$F_{3}S_{2}V_{11}V_{1}$, $F_{1}S_{7}S_{2}S_{2}$, $F_{3}S_{5}V_{4}V_{1}$, $F_{8}S_{7}S_{5}S_{5}$, $F_{3}S_{6}V_{3}V_{1}$, $F_{1}S_{2}V_{3}V_{4}$, $F_{8}S_{5}V_{11}V_{3}$, $F_{7}S_{7}S_{6}S_{6}$, $F_{7}S_{6}V_{11}V_{4}$, $F_{3}S_{3}V_{11}V_{2}$, $F_{1}S_{7}S_{3}S_{3}$, $F_{3}S_{5}V_{3}V_{2}$, $F_{7}S_{7}S_{5}S_{5}$, $F_{1}S_{3}V_{3}V_{3}$, $F_{7}S_{5}V_{11}V_{3}$, $F_{3}S_{3}V_{9}V_{4}$, $F_{8}S_{7}S_{3}S_{3}$, $F_{3}S_{2}V_{9}V_{3}$, $F_{7}S_{7}S_{2}S_{2}$, $F_{8}S_{3}V_{3}V_{3}$, $F_{7}S_{2}V_{3}V_{4}$, $F_{3}S_{16}S_{3}S_{6}$, $F_{1}S_{3}V_{9}V_{4}$, $F_{8}S_{3}V_{11}V_{2}$, $F_{3}S_{16}S_{2}S_{5}$, $F_{1}S_{2}V_{9}V_{3}$, $F_{7}S_{2}V_{11}V_{1}$, $F_{8}S_{5}V_{3}V_{2}$, $F_{7}S_{5}V_{4}V_{1}$, $F_{7}S_{6}V_{3}V_{1}$}} 
 \SplitRow{ $\mathcal{O}_{10, \; (1,-3)}^{5}$} {{$F_{8}S_{3}V_{11}V_{2}$, $F_{8}S_{5}V_{3}V_{2}$, $F_{2}S_{3}V_{10}V_{3}$, $F_{3}S_{16}S_{3}S_{5}$, $F_{8}S_{7}S_{3}S_{3}$, $F_{8}S_{3}V_{3}V_{3}$, $F_{3}S_{3}V_{10}V_{3}$, $F_{8}S_{7}S_{5}S_{5}$, $F_{2}S_{7}S_{3}S_{3}$, $F_{8}S_{5}V_{11}V_{3}$, $F_{3}S_{3}V_{11}V_{2}$, $F_{2}S_{3}V_{3}V_{3}$, $F_{3}S_{5}V_{3}V_{2}$}} 
 \SplitRow{ $\mathcal{O}_{10, \; (1,-3)}^{6}$} {{$F_{3}S_{14}S_{2}S_{3}$, $F_{3}S_{7}S_{3}S_{3}$, $F_{3}S_{2}V_{3}V_{4}$, $F_{7}S_{7}V_{4}V_{4}$, $F_{3}S_{3}V_{3}V_{3}$, $F_{7}S_{14}V_{3}V_{4}$, $F_{3}S_{14}V_{1}V_{2}$, $F_{3}S_{7}V_{2}V_{2}$, $F_{3}S_{6}V_{3}V_{1}$, $F_{7}S_{7}S_{6}S_{6}$, $F_{3}S_{5}V_{4}V_{1}$, $F_{3}S_{5}V_{3}V_{2}$, $F_{7}S_{14}S_{5}S_{6}$, $F_{5}S_{7}V_{4}V_{4}$, $F_{5}S_{14}V_{3}V_{4}$, $F_{3}S_{16}V_{3}V_{2}$, $F_{7}S_{7}V_{2}V_{2}$, $F_{3}S_{16}S_{3}S_{5}$, $F_{7}S_{5}V_{3}V_{2}$, $F_{5}S_{7}S_{3}S_{3}$, $F_{5}S_{3}V_{3}V_{3}$, $F_{3}S_{16}V_{4}V_{2}$, $F_{7}S_{14}V_{1}V_{2}$, $F_{3}S_{16}S_{3}S_{6}$, $F_{7}S_{6}V_{3}V_{1}$, $F_{7}S_{5}V_{4}V_{1}$, $F_{5}S_{14}S_{2}S_{3}$, $F_{5}S_{2}V_{3}V_{4}$}} 
 \SplitRow{ $\mathcal{O}_{10, \; (1,-3)}^{7}$} {{$F_{3}S_{4}V_{11}V_{2}$, $F_{3}S_{2}V_{11}V_{2}$, $F_{4}S_{7}V_{2}V_{2}$, $F_{1}S_{7}V_{2}V_{2}$, $F_{3}S_{4}V_{3}V_{5}$, $F_{3}S_{2}V_{3}V_{4}$, $F_{7}S_{7}V_{5}V_{5}$, $F_{7}S_{7}V_{4}V_{4}$, $F_{3}S_{5}S_{5}S_{4}$, $F_{3}S_{2}S_{5}S_{5}$, $F_{4}S_{5}V_{3}V_{2}$, $F_{1}S_{5}V_{3}V_{2}$, $F_{7}S_{5}V_{11}V_{5}$, $F_{7}S_{5}V_{11}V_{4}$, $F_{9}S_{7}S_{5}S_{5}$, $F_{5}S_{7}S_{5}S_{5}$, $F_{9}S_{5}V_{11}V_{3}$, $F_{5}S_{5}V_{11}V_{3}$, $F_{3}S_{4}V_{11}V_{1}$, $F_{3}S_{2}V_{11}V_{1}$, $F_{4}S_{7}S_{4}S_{4}$, $F_{1}S_{7}S_{2}S_{2}$, $F_{3}S_{5}V_{1}V_{5}$, $F_{3}S_{5}V_{4}V_{1}$, $F_{9}S_{7}V_{5}V_{5}$, $F_{5}S_{7}V_{4}V_{4}$, $F_{4}S_{5}S_{5}S_{4}$, $F_{1}S_{2}S_{5}S_{5}$, $F_{9}S_{5}V_{11}V_{5}$, $F_{5}S_{5}V_{11}V_{4}$, $F_{3}S_{4}V_{13}V_{3}$, $F_{3}S_{2}V_{9}V_{3}$, $F_{7}S_{7}S_{4}S_{4}$, $F_{7}S_{7}S_{2}S_{2}$, $F_{3}S_{5}V_{13}V_{2}$, $F_{3}S_{5}V_{9}V_{2}$, $F_{9}S_{7}V_{2}V_{2}$, $F_{5}S_{7}V_{2}V_{2}$, $F_{7}S_{5}S_{5}S_{4}$, $F_{7}S_{2}S_{5}S_{5}$, $F_{9}S_{5}V_{3}V_{2}$, $F_{5}S_{5}V_{3}V_{2}$, $F_{3}S_{16}S_{5}S_{4}$, $F_{3}S_{16}S_{2}S_{5}$, $F_{4}S_{4}V_{13}V_{3}$, $F_{1}S_{2}V_{9}V_{3}$, $F_{7}S_{4}V_{11}V_{1}$, $F_{7}S_{2}V_{11}V_{1}$, $F_{3}S_{16}V_{2}V_{5}$, $F_{3}S_{16}V_{4}V_{2}$, $F_{4}S_{5}V_{13}V_{2}$, $F_{1}S_{5}V_{9}V_{2}$, $F_{9}S_{4}V_{11}V_{2}$, $F_{5}S_{2}V_{11}V_{2}$, $F_{7}S_{5}V_{1}V_{5}$, $F_{7}S_{5}V_{4}V_{1}$, $F_{9}S_{4}V_{3}V_{5}$, $F_{5}S_{2}V_{3}V_{4}$, $F_{9}S_{5}S_{5}S_{4}$, $F_{5}S_{2}S_{5}S_{5}$}} 
 \SplitRow{ $\mathcal{O}_{10, \; (1,-3)}^{8}$} {{$F_{6}S_{7}S_{3}S_{3}$, $F_{8}S_{7}V_{2}V_{2}$, $F_{6}S_{3}V_{3}V_{3}$, $F_{8}S_{5}V_{3}V_{2}$, $F_{3}S_{16}V_{3}V_{2}$, $F_{3}S_{16}S_{3}S_{5}$, $F_{6}S_{7}V_{3}V_{3}$, $F_{8}S_{7}S_{5}S_{5}$, $F_{3}S_{7}V_{2}V_{2}$, $F_{3}S_{5}V_{3}V_{2}$, $F_{8}S_{7}V_{3}V_{3}$, $F_{3}S_{7}S_{3}S_{3}$, $F_{3}S_{3}V_{3}V_{3}$}} 
 \SplitRow{ $\mathcal{O}_{10, \; (1,-3)}^{9}$} {{$F_{3}S_{7}S_{2}S_{2}$, $F_{3}S_{2}V_{3}V_{4}$, $F_{8}S_{7}V_{3}V_{3}$, $F_{7}S_{7}V_{4}V_{4}$, $F_{3}S_{7}V_{1}V_{1}$, $F_{3}S_{5}V_{4}V_{1}$, $F_{8}S_{7}S_{5}S_{5}$, $F_{5}S_{7}V_{4}V_{4}$, $F_{3}S_{7}V_{2}V_{2}$, $F_{3}S_{5}V_{3}V_{2}$, $F_{7}S_{7}S_{5}S_{5}$, $F_{5}S_{7}V_{3}V_{3}$, $F_{3}S_{16}V_{4}V_{2}$, $F_{8}S_{7}V_{2}V_{2}$, $F_{3}S_{16}V_{3}V_{1}$, $F_{7}S_{7}V_{1}V_{1}$, $F_{3}S_{16}S_{2}S_{5}$, $F_{8}S_{5}V_{3}V_{2}$, $F_{7}S_{5}V_{4}V_{1}$, $F_{5}S_{7}S_{2}S_{2}$, $F_{5}S_{2}V_{3}V_{4}$}} 
 \SplitRow{ $\mathcal{O}_{10, \; (1,-3)}^{10}$} {{$F_{3}S_{7}S_{4}S_{4}$, $F_{3}S_{7}S_{2}S_{2}$, $F_{3}S_{5}S_{5}S_{4}$, $F_{3}S_{2}S_{5}S_{5}$, $F_{9}S_{7}S_{5}S_{5}$, $F_{5}S_{7}S_{5}S_{5}$, $F_{3}S_{16}S_{5}S_{4}$, $F_{3}S_{16}S_{2}S_{5}$, $F_{9}S_{7}S_{4}S_{4}$, $F_{5}S_{7}S_{2}S_{2}$, $F_{9}S_{5}S_{5}S_{4}$, $F_{5}S_{2}S_{5}S_{5}$}} 
 \end{SplitTable}
   \vspace{0.4cm} 
 \captionof{table}{Dimension 10 $(\Delta B, \Delta L)$ = (1, -3) UV completions for topology 7 (see Fig.~\ref{fig:d=10_topologies}).} 
 \label{tab:T10Top7B1L-3}
 \end{center}

%% file: biblio.bib
@article{Davidson:2022jai,
    author = "Davidson, Sacha and Echenard, Bertrand and Bernstein, Robert H. and Heeck, Julian and Hitlin, David G.",
    title = "{Charged Lepton Flavor Violation}",
    eprint = "2209.00142",
    archivePrefix = "arXiv",
    primaryClass = "hep-ex",
    month = "8",
    year = "2022"
}

@article{Kublbeck:1990xc,
    author = "Kublbeck, J. and Bohm, M. and Denner, Ansgar",
    title = "{Feyn Arts: Computer Algebraic Generation of Feynman Graphs and Amplitudes}",
    reportNumber = "Print-90-0144 (WURZBURG)",
    doi = "10.1016/0010-4655(90)90001-H",
    journal = "Comput. Phys. Commun.",
    volume = "60",
    pages = "165--180",
    year = "1990"
}

@article{Horv_t_2023,
   title="{IGraph/M: graph theory and network analysis for Mathematica}",
   volume={8},
   ISSN={2475-9066},
   url={http://dx.doi.org/10.21105/joss.04899},
   DOI={10.21105/joss.04899},
   number={81},
   journal={Journal of Open Source Software},
   publisher={The Open Journal},
   author={Horvát, Szabolcs and Podkalicki, Jakub and Csárdi, Gábor and Nepusz, Tamás and Traag, Vincent and Zanini, Fabio and Noom, Daniel},
   year={2023},
   month=jan, 
   pages={4899} 
}

@article{Fuentes-Martin:2022jrf,
    author = {Fuentes-Mart\'\i{}n, Javier and K\"onig, Matthias and Pag\`es, Julie and Thomsen, Anders Eller and Wilsch, Felix},
    title = "{A Proof of Concept for Matchete: An Automated Tool for Matching Effective Theories}",
    eprint = "2212.04510",
    archivePrefix = "arXiv",
    primaryClass = "hep-ph",
    reportNumber = "MITP-22-105, TUM-HEP-1443/22, ZU-TH-58/22",
    doi = "10.1140/epjc/s10052-023-11726-1",
    journal = "Eur. Phys. J. C",
    volume = "83",
    number = "7",
    pages = "662",
    year = "2023"
}

@article{FileviezPerez:2022ypk,
    author = "Pavel Fileviez Perez and  Andrea Pocar and  K.S. Babu and  Leah J. Broussard and  Vincenzo Cirigliano and  Julian Heeck and  Susan Gardner and  Ed Kearns and  Andrew J. Long and  Stuart Raby and  Richard Ruiz and  Evelyn Thomson and  Carlos E.M. Wagner and  Mark B. Wise",
    title = "{On Baryon and Lepton Number Violation}",
    eprint = "2208.00010",
    archivePrefix = "arXiv",
    primaryClass = "hep-ph",
    month = "7",
    year = "2022"
}

@article{Heeck:2019kgr,
    author = "Heeck, Julian and Takhistov, Volodymyr",
    title = "{Inclusive Nucleon Decay Searches as a Frontier of Baryon Number Violation}",
    eprint = "1910.07647",
    archivePrefix = "arXiv",
    primaryClass = "hep-ph",
    reportNumber = "UCI-TR-2019-24",
    doi = "10.1103/PhysRevD.101.015005",
    journal = "Phys. Rev. D",
    volume = "101",
    pages = "015005",
    year = "2020"
}

@article{Hambye:2017qix,
    author = "Hambye, Thomas and Heeck, Julian",
    title = "{Proton decay into charged leptons}",
    eprint = "1712.04871",
    archivePrefix = "arXiv",
    primaryClass = "hep-ph",
    reportNumber = "ULB-TH-17-22",
    doi = "10.1103/PhysRevLett.120.171801",
    journal = "Phys. Rev. Lett.",
    volume = "120",
    pages = "171801",
    year = "2018"
}

@article{Weinberg:1979sa,
    author = "Weinberg, Steven",
    title = "{Baryon and Lepton Nonconserving Processes}",
    reportNumber = "HUTP-79-A050",
    doi = "10.1103/PhysRevLett.43.1566",
    journal = "Phys. Rev. Lett.",
    volume = "43",
    pages = "1566--1570",
    year = "1979"
}

@article{Wilczek:1979hc,
    author = "Wilczek, Frank and Zee, A.",
    title = "{Operator Analysis of Nucleon Decay}",
    reportNumber = "Print-79-0709 (PRINCETON)",
    doi = "10.1103/PhysRevLett.43.1571",
    journal = "Phys. Rev. Lett.",
    volume = "43",
    pages = "1571--1573",
    year = "1979"
}

@article{Weinberg:1980bf,
    author = "Weinberg, Steven",
    title = "{Varieties of Baryon and Lepton Nonconservation}",
    reportNumber = "HUTP-80/A023",
    doi = "10.1103/PhysRevD.22.1694",
    journal = "Phys. Rev. D",
    volume = "22",
    pages = "1694",
    year = "1980"
}

@article{Weldon:1980gi,
    author = "Weldon, H. A. and Zee, A.",
    title = "{Operator Analysis of New Physics}",
    reportNumber = "UPR-0153T",
    doi = "10.1016/0550-3213(80)90218-7",
    journal = "Nucl. Phys. B",
    volume = "173",
    pages = "269--290",
    year = "1980"
}

@article{Abbott:1980zj,
    author = "Abbott, L. F. and Wise, Mark B.",
    title = "{The Effective Hamiltonian for Nucleon Decay}",
    reportNumber = "SLAC-PUB-2487",
    doi = "10.1103/PhysRevD.22.2208",
    journal = "Phys. Rev. D",
    volume = "22",
    pages = "2208",
    year = "1980"
}

@article{Dorsner:2012nq,
    author = "Dorsner, Ilja and Fajfer, Svjetlana and Kosnik, Nejc",
    title = "{Heavy and light scalar leptoquarks in proton decay}",
    eprint = "1204.0674",
    archivePrefix = "arXiv",
    primaryClass = "hep-ph",
    reportNumber = "LAL-12-111",
    doi = "10.1103/PhysRevD.86.015013",
    journal = "Phys. Rev. D",
    volume = "86",
    pages = "015013",
    year = "2012"
}

@article{Yoo:2021gql,
    author = "Yoo, Jun-Sik and Aoki, Yasumichi and Boyle, Peter and Izubuchi, Taku and Soni, Amarjit and Syritsyn, Sergey",
    title = "{Proton decay matrix elements on the lattice at physical pion mass}",
    eprint = "2111.01608",
    archivePrefix = "arXiv",
    primaryClass = "hep-lat",
    reportNumber = "RBRC-1333, KEK-CP-0385",
    doi = "10.1103/PhysRevD.105.074501",
    journal = "Phys. Rev. D",
    volume = "105",
    pages = "074501",
    year = "2022"
}

@article{Grzadkowski:2010es,
    author = "Grzadkowski, B. and Iskrzynski, M. and Misiak, M. and Rosiek, J.",
    title = "{Dimension-Six Terms in the Standard Model Lagrangian}",
    eprint = "1008.4884",
    archivePrefix = "arXiv",
    primaryClass = "hep-ph",
    reportNumber = "IFT-9-2010, TTP10-35",
    doi = "10.1007/JHEP10(2010)085",
    journal = "JHEP",
    volume = "10",
    pages = "085",
    year = "2010"
}

@article{Fonseca:2017lem,
    author = "Fonseca, Renato M.",
    editor = "Grzadkowski, Bohdan and Kalinowski, Jan and Krawczyk, Maria",
    title = "{The Sym2Int program: going from symmetries to interactions}",
    eprint = "1703.05221",
    archivePrefix = "arXiv",
    primaryClass = "hep-ph",
    doi = "10.1088/1742-6596/873/1/012045",
    journal = "J. Phys. Conf. Ser.",
    volume = "873",
    number = "1",
    pages = "012045",
    year = "2017"
}

@article{Fonseca:2019yya,
    author = "Fonseca, Renato M.",
    title = "{Enumerating the operators of an effective field theory}",
    eprint = "1907.12584",
    archivePrefix = "arXiv",
    primaryClass = "hep-ph",
    doi = "10.1103/PhysRevD.101.035040",
    journal = "Phys. Rev. D",
    volume = "101",
    pages = "035040",
    year = "2020"
}

@article{Buchmuller:1986zs,
    author = "Buchmuller, W. and Ruckl, R. and Wyler, D.",
    title = "{Leptoquarks in Lepton - Quark Collisions}",
    reportNumber = "DESY-86-150, ITP-UH-14-86",
    doi = "10.1016/0370-2693(87)90637-X",
    journal = "Phys. Lett. B",
    volume = "191",
    pages = "442--448",
    year = "1987",
    note = "[Erratum: Phys.Lett.B 448, 320--320 (1999)]"
}

@article{Dorsner:2016wpm,
    author = "Dor\v{s}ner, I. and Fajfer, S. and Greljo, A. and Kamenik, J. F. and Ko\v{s}nik, N.",
    title = "{Physics of leptoquarks in precision experiments and at particle colliders}",
    eprint = "1603.04993",
    archivePrefix = "arXiv",
    primaryClass = "hep-ph",
    doi = "10.1016/j.physrep.2016.06.001",
    journal = "Phys. Rept.",
    volume = "641",
    pages = "1--68",
    year = "2016"
}

@article{Ma:1998dn,
    author = "Ma, Ernest",
    title = "{Pathways to naturally small neutrino masses}",
    eprint = "hep-ph/9805219",
    archivePrefix = "arXiv",
    reportNumber = "UCRHEP-T-222",
    doi = "10.1103/PhysRevLett.81.1171",
    journal = "Phys. Rev. Lett.",
    volume = "81",
    pages = "1171--1174",
    year = "1998"
}

@article{Farzan:2012ev,
    author = "Farzan, Yasaman and Pascoli, Silvia and Schmidt, Michael A.",
    title = "{Recipes and Ingredients for Neutrino Mass at Loop Level}",
    eprint = "1208.2732",
    archivePrefix = "arXiv",
    primaryClass = "hep-ph",
    reportNumber = "IPM-P-2012-027, IPPP-12-64, DCPT-12-128",
    doi = "10.1007/JHEP03(2013)107",
    journal = "JHEP",
    volume = "03",
    pages = "107",
    year = "2013"
}

@article{Babu:2001ex,
    author = "Babu, K. S. and Leung, Chung Ngoc",
    title = "{Classification of effective neutrino mass operators}",
    eprint = "hep-ph/0106054",
    archivePrefix = "arXiv",
    reportNumber = "OSU-HEP-01-02, UDHEP-02-01",
    doi = "10.1016/S0550-3213(01)00504-1",
    journal = "Nucl. Phys. B",
    volume = "619",
    pages = "667--689",
    year = "2001"
}

@article{deGouvea:2007qla,
    author = "de Gouvea, Andre and Jenkins, James",
    title = "{A Survey of Lepton Number Violation Via Effective Operators}",
    eprint = "0708.1344",
    archivePrefix = "arXiv",
    primaryClass = "hep-ph",
    reportNumber = "NUHEP-TH-07-10",
    doi = "10.1103/PhysRevD.77.013008",
    journal = "Phys. Rev. D",
    volume = "77",
    pages = "013008",
    year = "2008"
}

@article{Angel:2012ug,
    author = "Angel, Paul W. and Rodd, Nicholas L. and Volkas, Raymond R.",
    title = "{Origin of neutrino masses at the LHC: $\Delta L = 2$ effective operators and their ultraviolet completions}",
    eprint = "1212.6111",
    archivePrefix = "arXiv",
    primaryClass = "hep-ph",
    doi = "10.1103/PhysRevD.87.073007",
    journal = "Phys. Rev. D",
    volume = "87",
    pages = "073007",
    year = "2013"
}

@article{Gargalionis:2020xvt,
    author = "Gargalionis, John and Volkas, Raymond R.",
    title = "{Exploding operators for Majorana neutrino masses and beyond}",
    eprint = "2009.13537",
    archivePrefix = "arXiv",
    primaryClass = "hep-ph",
    doi = "10.1007/JHEP01(2021)074",
    journal = "JHEP",
    volume = "01",
    pages = "074",
    year = "2021"
}

@article{terHoeve:2023pvs,
    author = "ter Hoeve, Jaco and Magni, Giacomo and Rojo, Juan and Rossia, Alejo N. and Vryonidou, Eleni",
    title = "{The automation of SMEFT-assisted constraints on UV-complete models}",
    eprint = "2309.04523",
    archivePrefix = "arXiv",
    primaryClass = "hep-ph",
    reportNumber = "Nikhef 2023-011",
    doi = "10.1007/JHEP01(2024)179",
    journal = "JHEP",
    volume = "01",
    pages = "179",
    year = "2024"
}

@article{Beltran:2023ymm,
    author = "Beltr\'an, Rebeca and Cepedello, Ricardo and Hirsch, Martin",
    title = "{Tree-level UV completions for N$_{R}$SMEFT d = 6 and d = 7 operators}",
    eprint = "2306.12578",
    archivePrefix = "arXiv",
    primaryClass = "hep-ph",
    doi = "10.1007/JHEP08(2023)166",
    journal = "JHEP",
    volume = "08",
    pages = "166",
    year = "2023"
}

@article{deBlas:2017xtg,
    author = "de Blas, J. and Criado, J. C. and Perez-Victoria, M. and Santiago, J.",
    title = "{Effective description of general extensions of the Standard Model: the complete tree-level dictionary}",
    eprint = "1711.10391",
    archivePrefix = "arXiv",
    primaryClass = "hep-ph",
    reportNumber = "CERN-TH-2017-251",
    doi = "10.1007/JHEP03(2018)109",
    journal = "JHEP",
    volume = "03",
    pages = "109",
    year = "2018"
}

@article{Li:2023cwy,
    author = "Li, Xu-Xiang and Ren, Zhe and Yub, Jiang-Hao",
    title = "{Complete tree-level dictionary between simplified BSM models and SMEFT $d\leq 7$ operators}",
    eprint = "2307.10380",
    archivePrefix = "arXiv",
    primaryClass = "hep-ph",
    doi = "10.1103/PhysRevD.109.095041",
    journal = "Phys. Rev. D",
    volume = "109",
    pages = "095041",
    year = "2024"
}

@article{Lehman:2014jma,
    author = "Lehman, Landon",
    title = "{Extending the Standard Model Effective Field Theory with the Complete Set of Dimension-7 Operators}",
    eprint = "1410.4193",
    archivePrefix = "arXiv",
    primaryClass = "hep-ph",
    doi = "10.1103/PhysRevD.90.125023",
    journal = "Phys. Rev. D",
    volume = "90",
    pages = "125023",
    year = "2014"
}

@article{Liao:2016hru,
    author = "Liao, Yi and Ma, Xiao-Dong",
    title = "{Renormalization Group Evolution of Dimension-seven Baryon- and Lepton-number-violating Operators}",
    eprint = "1607.07309",
    archivePrefix = "arXiv",
    primaryClass = "hep-ph",
    doi = "10.1007/JHEP11(2016)043",
    journal = "JHEP",
    volume = "11",
    pages = "043",
    year = "2016"
}

@article{Liao:2020jmn,
    author = "Liao, Yi and Ma, Xiao-Dong",
    title = "{An explicit construction of the dimension-9 operator basis in the standard model effective field theory}",
    eprint = "2007.08125",
    archivePrefix = "arXiv",
    primaryClass = "hep-ph",
    doi = "10.1007/JHEP11(2020)152",
    journal = "JHEP",
    volume = "11",
    pages = "152",
    year = "2020"
}

@article{Li:2020xlh,
    author = "Li, Hao-Lin and Ren, Zhe and Xiao, Ming-Lei and Yu, Jiang-Hao and Zheng, Yu-Hui",
    title = "{Complete set of dimension-nine operators in the standard model effective field theory}",
    eprint = "2007.07899",
    archivePrefix = "arXiv",
    primaryClass = "hep-ph",
    doi = "10.1103/PhysRevD.104.015025",
    journal = "Phys. Rev. D",
    volume = "104",
    pages = "015025",
    year = "2021"
}

@article{Murphy:2020rsh,
    author = "Murphy, Christopher W.",
    title = "{Dimension-8 operators in the Standard Model Effective Field Theory}",
    eprint = "2005.00059",
    archivePrefix = "arXiv",
    primaryClass = "hep-ph",
    doi = "10.1007/JHEP10(2020)174",
    journal = "JHEP",
    volume = "10",
    pages = "174",
    year = "2020"
}

@article{Li:2020gnx,
    author = "Li, Hao-Lin and Ren, Zhe and Shu, Jing and Xiao, Ming-Lei and Yu, Jiang-Hao and Zheng, Yu-Hui",
    title = "{Complete set of dimension-eight operators in the standard model effective field theory}",
    eprint = "2005.00008",
    archivePrefix = "arXiv",
    primaryClass = "hep-ph",
    doi = "10.1103/PhysRevD.104.015026",
    journal = "Phys. Rev. D",
    volume = "104",
    pages = "015026",
    year = "2021"
}

@article{Harlander:2023psl,
    author = "Harlander, R. V. and Kempkens, T. and Schaaf, M. C.",
    title = "{Standard model effective field theory up to mass dimension 12}",
    eprint = "2305.06832",
    archivePrefix = "arXiv",
    primaryClass = "hep-ph",
    reportNumber = "TTK-22-47, TTK-22-47; P3H-23-031",
    doi = "10.1103/PhysRevD.108.055020",
    journal = "Phys. Rev. D",
    volume = "108",
    pages = "055020",
    year = "2023"
}

@article{Buchmuller:1985jz,
    author = "Buchmuller, W. and Wyler, D.",
    title = "{Effective Lagrangian Analysis of New Interactions and Flavor Conservation}",
    reportNumber = "CERN-TH-4254/85",
    doi = "10.1016/0550-3213(86)90262-2",
    journal = "Nucl. Phys. B",
    volume = "268",
    pages = "621--653",
    year = "1986"
}

@article{Cepedello:2022pyx,
    author = "Cepedello, Ricardo and Esser, Fabian and Hirsch, Martin and Sanz, Veronica",
    title = "{Mapping the SMEFT to discoverable models}",
    eprint = "2207.13714",
    archivePrefix = "arXiv",
    primaryClass = "hep-ph",
    doi = "10.1007/JHEP09(2022)229",
    journal = "JHEP",
    volume = "09",
    pages = "229",
    year = "2022"
}

@article{Isidori:2023pyp,
    author = "Isidori, Gino and Wilsch, Felix and Wyler, Daniel",
    title = "{The standard model effective field theory at work}",
    eprint = "2303.16922",
    archivePrefix = "arXiv",
    primaryClass = "hep-ph",
    reportNumber = "ZU-TH 14/23",
    doi = "10.1103/RevModPhys.96.015006",
    journal = "Rev. Mod. Phys.",
    volume = "96",
    pages = "015006",
    year = "2024"
}

@article{Li:2022abx,
    author = "Li, Hao-Lin and Ni, Yu-Han and Xiao, Ming-Lei and Yu, Jiang-Hao",
    title = "{The bottom-up EFT: complete UV resonances of the SMEFT operators}",
    eprint = "2204.03660",
    archivePrefix = "arXiv",
    primaryClass = "hep-ph",
    doi = "10.1007/JHEP11(2022)170",
    journal = "JHEP",
    volume = "11",
    pages = "170",
    year = "2022"
}

@article{Li:2021tsq,
    author = "Li, Hao-Lin and Ren, Zhe and Xiao, Ming-Lei and Yu, Jiang-Hao and Zheng, Yu-Hui",
    title = "{Operator bases in effective field theories with sterile neutrinos: d \ensuremath{\leq} 9}",
    eprint = "2105.09329",
    archivePrefix = "arXiv",
    primaryClass = "hep-ph",
    doi = "10.1007/JHEP11(2021)003",
    journal = "JHEP",
    volume = "11",
    pages = "003",
    year = "2021"
}

@article{Bischer:2019ttk,
    author = "Bischer, Ingolf and Rodejohann, Werner",
    title = "{General neutrino interactions from an effective field theory perspective}",
    eprint = "1905.08699",
    archivePrefix = "arXiv",
    primaryClass = "hep-ph",
    doi = "10.1016/j.nuclphysb.2019.114746",
    journal = "Nucl. Phys. B",
    volume = "947",
    pages = "114746",
    year = "2019"
}

@article{Harlander:2023ozs,
    author = "Harlander, Robert V. and Schaaf, Magnus C.",
    title = "{AutoEFT: Automated operator construction for effective field theories}",
    eprint = "2309.15783",
    archivePrefix = "arXiv",
    primaryClass = "hep-ph",
    reportNumber = "TTK-23-25; P3H-23-066",
    doi = "10.1016/j.cpc.2024.109198",
    journal = "Comput. Phys. Commun.",
    volume = "300",
    pages = "109198",
    year = "2024"
}

@article{Li:2023pfw,
    author = "Li, Hao-Lin and Ni, Yu-Han and Xiao, Ming-Lei and Yu, Jiang-Hao",
    title = "{Complete UV resonances of the dimension-8 SMEFT operators}",
    eprint = "2309.15933",
    archivePrefix = "arXiv",
    primaryClass = "hep-ph",
    doi = "10.1007/JHEP05(2024)238",
    journal = "JHEP",
    volume = "05",
    pages = "238",
    year = "2024"
}

@article{Gargalionis:2024nij,
    author = "Gargalionis, John and Herrero-Garc\'\i{}a, Juan and Schmidt, Michael A.",
    title = "{Model-independent estimates for loop-induced baryon-number-violating nucleon decays}",
    eprint = "2401.04768",
    archivePrefix = "arXiv",
    primaryClass = "hep-ph",
    reportNumber = "CPPC-2024-02",
    doi = "10.1007/JHEP06(2024)182",
    journal = "JHEP",
    volume = "06",
    pages = "182",
    year = "2024"
}

@article{Chen:2022gjd,
    author = "Chen, Shao-Long and Xiao, Yu-Qi",
    title = "{The decomposition of neutron-antineutron oscillation operators}",
    eprint = "2211.02813",
    archivePrefix = "arXiv",
    primaryClass = "hep-ph",
    doi = "10.1007/JHEP03(2023)065",
    journal = "JHEP",
    volume = "03",
    pages = "065",
    year = "2023"
}

@misc{oeis,
    Author = {{OEIS Foundation Inc.}},
    howpublished = {Published electronically at \url{http://oeis.org}},
    Title = {The {O}n-{L}ine {E}ncyclopedia of {I}nteger {S}equences},
    Year = 2024
}

@article{Georgi:1974sy,
    author = "Georgi, H. and Glashow, S. L.",
    title = "{Unity of All Elementary Particle Forces}",
    doi = "10.1103/PhysRevLett.32.438",
    journal = "Phys. Rev. Lett.",
    volume = "32",
    pages = "438--441",
    year = "1974"
}

@article{Fritzsch:1974nn,
    author = "Fritzsch, Harald and Minkowski, Peter",
    title = "{Unified Interactions of Leptons and Hadrons}",
    reportNumber = "CALT-68-467",
    doi = "10.1016/0003-4916(75)90211-0",
    journal = "Annals Phys.",
    volume = "93",
    pages = "193--266",
    year = "1975"
}

@article{Dong:2011rh,
    author = "Dong, Zhe and Durieux, Gauthier and Gerard, Jean-Marc and Han, Tao and Maltoni, Fabio",
    title = "{Baryon number violation at the LHC: the top option}",
    eprint = "1107.3805",
    archivePrefix = "arXiv",
    primaryClass = "hep-ph",
    reportNumber = "CP3-11-24, MADPH-11-1573",
    doi = "10.1103/PhysRevD.85.016006",
    journal = "Phys. Rev. D",
    volume = "85",
    pages = "016006",
    year = "2012"
}

@article{Fonseca:2020vke,
    author = "Fonseca, Renato M.",
    title = "{GroupMath: A Mathematica package for group theory calculations}",
    eprint = "2011.01764",
    archivePrefix = "arXiv",
    primaryClass = "hep-th",
    doi = "10.1016/j.cpc.2021.108085",
    journal = "Comput. Phys. Commun.",
    volume = "267",
    pages = "108085",
    year = "2021"
}

@article{Babu:2003qh,
    author = "Babu, K. S. and Gogoladze, Ilia and Wang, Kai",
    title = "{Gauged baryon parity and nucleon stability}",
    eprint = "hep-ph/0306003",
    archivePrefix = "arXiv",
    reportNumber = "OSU-HEP-03-8",
    doi = "10.1016/j.physletb.2003.07.036",
    journal = "Phys. Lett. B",
    volume = "570",
    pages = "32--38",
    year = "2003"
}

@article{Ibanez:1991pr,
    author = "Ibanez, Luis E. and Ross, Graham G.",
    title = "{Discrete gauge symmetries and the origin of baryon and lepton number conservation in supersymmetric versions of the standard model}",
    reportNumber = "CERN-TH-6111-91",
    doi = "10.1016/0550-3213(92)90195-H",
    journal = "Nucl. Phys. B",
    volume = "368",
    pages = "3--37",
    year = "1992"
}

@article{Carone:1995pu,
    author = "Carone, Christopher D. and Murayama, Hitoshi",
    title = "{Realistic models with a light U(1) gauge boson coupled to baryon number}",
    eprint = "hep-ph/9501220",
    archivePrefix = "arXiv",
    reportNumber = "LBL-36598",
    doi = "10.1103/PhysRevD.52.484",
    journal = "Phys. Rev. D",
    volume = "52",
    pages = "484--493",
    year = "1995"
}

@article{Appelquist:2001mj,
    author = "Appelquist, Thomas and Dobrescu, Bogdan A. and Ponton, Eduardo and Yee, Ho-Ung",
    title = "{Proton stability in six-dimensions}",
    eprint = "hep-ph/0107056",
    archivePrefix = "arXiv",
    reportNumber = "YCTP-P5-01",
    doi = "10.1103/PhysRevLett.87.181802",
    journal = "Phys. Rev. Lett.",
    volume = "87",
    pages = "181802",
    year = "2001"
}

@article{Lee:2002mn,
    author = "Lee, Chin-Aik and Shafi, Qaisar and Tavartkiladze, Zurab",
    title = "{Gauge origin of baryon number conservation and suppressed neutrino masses from five-dimensions}",
    eprint = "hep-ph/0206258",
    archivePrefix = "arXiv",
    reportNumber = "BA-02-24",
    doi = "10.1103/PhysRevD.66.055010",
    journal = "Phys. Rev. D",
    volume = "66",
    pages = "055010",
    year = "2002"
}

@article{EXO-200:2017hwz,
    author = "Albert, J. B. and others",
    collaboration = "EXO-200",
    title = "{Search for nucleon decays with EXO-200}",
    eprint = "1710.07670",
    archivePrefix = "arXiv",
    primaryClass = "hep-ex",
    doi = "10.1103/PhysRevD.97.072007",
    journal = "Phys. Rev. D",
    volume = "97",
    pages = "072007",
    year = "2018"
}

@article{Majorana:2018pdo,
    author = "Alvis, S. I. and others",
    collaboration = "Majorana",
    title = "{Search for trinucleon decay in the Majorana Demonstrator}",
    eprint = "1812.01090",
    archivePrefix = "arXiv",
    primaryClass = "hep-ex",
    doi = "10.1103/PhysRevD.99.072004",
    journal = "Phys. Rev. D",
    volume = "99",
    pages = "072004",
    year = "2019"
}

@article{GERDA:2023uuw,
    author = "Agostini, M. and others",
    collaboration = "GERDA",
    title = "{Search for tri-nucleon decays of $^{76}$Ge in GERDA}",
    eprint = "2307.16542",
    archivePrefix = "arXiv",
    primaryClass = "nucl-ex",
    doi = "10.1140/epjc/s10052-023-11862-8",
    journal = "Eur. Phys. J. C",
    volume = "83",
    number = "9",
    pages = "778",
    year = "2023"
}

@article{Mohapatra:2002ug,
    author = "Mohapatra, Rabindra Nath and Perez-Lorenzana, Abdel",
    title = "{Neutrino mass, proton decay and dark matter in TeV scale universal extra dimension models}",
    eprint = "hep-ph/0212254",
    archivePrefix = "arXiv",
    reportNumber = "UMD-PP-03-029",
    doi = "10.1103/PhysRevD.67.075015",
    journal = "Phys. Rev. D",
    volume = "67",
    pages = "075015",
    year = "2003"
}

@article{Fonseca:2018aav,
    author = "Fonseca, Renato M. and Hirsch, Martin",
    title = "{$\Delta L \ge 4$ lepton number violating processes}",
    eprint = "1804.10545",
    archivePrefix = "arXiv",
    primaryClass = "hep-ph",
    reportNumber = "IFIC-18-16, IFIC/18-16",
    doi = "10.1103/PhysRevD.98.015035",
    journal = "Phys. Rev. D",
    volume = "98",
    pages = "015035",
    year = "2018"
}

@article{Fonseca:2018ehk,
    author = "Fonseca, Renato M. and Hirsch, Martin and Srivastava, Rahul",
    title = "{$\Delta L = 3$ processes: Proton decay and the LHC}",
    eprint = "1802.04814",
    archivePrefix = "arXiv",
    primaryClass = "hep-ph",
    reportNumber = "IFIC-18-03",
    doi = "10.1103/PhysRevD.97.075026",
    journal = "Phys. Rev. D",
    volume = "97",
    pages = "075026",
    year = "2018"
}

@article{Durieux:2012gj,
    author = "Durieux, Gauthier and Gerard, Jean-Marc and Maltoni, Fabio and Smith, Christopher",
    title = "{Three-generation baryon and lepton number violation at the LHC}",
    eprint = "1210.6598",
    archivePrefix = "arXiv",
    primaryClass = "hep-ph",
    reportNumber = "CP3-12-44, LPSC12296",
    doi = "10.1016/j.physletb.2013.02.052",
    journal = "Phys. Lett. B",
    volume = "721",
    pages = "82--85",
    year = "2013"
}

@article{Kovalenko:2002eh,
    author = "Kovalenko, Sergey and Schmidt, Ivan",
    title = "{Proton stability in leptoquark models}",
    eprint = "hep-ph/0210187",
    archivePrefix = "arXiv",
    reportNumber = "USM-TH-132",
    doi = "10.1016/S0370-2693(03)00544-6",
    journal = "Phys. Lett. B",
    volume = "562",
    pages = "104--108",
    year = "2003"
}

@article{ODonnell:1993kdg,
    author = "O'Donnell, Patrick J. and Sarkar, Utpal",
    title = "{Three lepton decay mode of the proton}",
    eprint = "hep-ph/9307254",
    archivePrefix = "arXiv",
    reportNumber = "UTPT-93-13",
    doi = "10.1016/0370-2693(93)90667-7",
    journal = "Phys. Lett. B",
    volume = "316",
    pages = "121--126",
    year = "1993"
}

@article{Faroughy:2014tfa,
    author = "Faroughy, Cyrus and Prabhu, Siddharth and Zheng, Bob",
    title = "{Simultaneous B and L Violation: New Signatures from RPV-SUSY}",
    eprint = "1409.5438",
    archivePrefix = "arXiv",
    primaryClass = "hep-ph",
    reportNumber = "MCTP-14-34",
    doi = "10.1007/JHEP06(2015)073",
    journal = "JHEP",
    volume = "06",
    pages = "073",
    year = "2015"
}

@article{Kobach:2016ami,
    author = "Kobach, Andrew",
    title = "{Baryon Number, Lepton Number, and Operator Dimension in the Standard Model}",
    eprint = "1604.05726",
    archivePrefix = "arXiv",
    primaryClass = "hep-ph",
    reportNumber = "PHYS.LETT.-B758-(2016)-455-457",
    doi = "10.1016/j.physletb.2016.05.050",
    journal = "Phys. Lett. B",
    volume = "758",
    pages = "455--457",
    year = "2016"
}

@article{Helset:2019eyc,
    author = "Helset, Andreas and Kobach, Andrew",
    title = "{Baryon Number, Lepton Number, and Operator Dimension in the SMEFT with Flavor Symmetries}",
    eprint = "1909.05853",
    archivePrefix = "arXiv",
    primaryClass = "hep-ph",
    doi = "10.1016/j.physletb.2019.135132",
    journal = "Phys. Lett. B",
    volume = "800",
    pages = "135132",
    year = "2020"
}

@article{Sarkar:1983tm,
    author = "Sarkar, Utpal and Ray, Asim K.",
    title = "{Testing Horizontal Gauge Symmetries From Nucleon Decay Experiments}",
    reportNumber = "Print-83-0571 (VISVA BHARATI)",
    doi = "10.1103/PhysRevD.29.166",
    journal = "Phys. Rev. D",
    volume = "29",
    pages = "166",
    year = "1984"
}

@article{Zee:1981vr,
    author = "Zee, A.",
    title = "{Proton Decay And Horizontal Symmetry}",
    reportNumber = "Print-81-0788 (WASH.U.,SEATTLE), 40048-82-PT3",
    doi = "10.1016/0370-2693(82)90750-X",
    journal = "Phys. Lett. B",
    volume = "109",
    pages = "187--189",
    year = "1982"
}

@article{Babu:2012iv,
    author = "Babu, K. S. and Mohapatra, R. N.",
    title = "{B-L Violating Proton Decay Modes and New Baryogenesis Scenario in SO(10)}",
    eprint = "1207.5771",
    archivePrefix = "arXiv",
    primaryClass = "hep-ph",
    reportNumber = "OSU-HEP-12-08",
    doi = "10.1103/PhysRevLett.109.091803",
    journal = "Phys. Rev. Lett.",
    volume = "109",
    pages = "091803",
    year = "2012"
}

@article{Helo:2019yqp,
    author = "Helo, Juan Carlos and Hirsch, Martin and Ota, Toshihiko",
    title = "{Proton decay at one loop}",
    eprint = "1904.00036",
    archivePrefix = "arXiv",
    primaryClass = "hep-ph",
    doi = "10.1103/PhysRevD.99.095021",
    journal = "Phys. Rev. D",
    volume = "99",
    pages = "095021",
    year = "2019"
}

@article{delAguila:2008ir,
    author = "del Aguila, Francisco and Bar-Shalom, Shaouly and Soni, Amarjit and Wudka, Jose",
    title = "{Heavy Majorana Neutrinos in the Effective Lagrangian Description: Application to Hadron Colliders}",
    eprint = "0806.0876",
    archivePrefix = "arXiv",
    primaryClass = "hep-ph",
    reportNumber = "UG-FT-230-08, CAFPE-100-08, UCI-TR-2008-21, BNL-HET-08-14",
    doi = "10.1016/j.physletb.2008.11.031",
    journal = "Phys. Lett. B",
    volume = "670",
    pages = "399--402",
    year = "2009"
}

@article{Helo:2018bgb,
    author = "Helo, Juan C. and Hirsch, Martin and Ota, Toshihiko",
    title = "{Proton decay and light sterile neutrinos}",
    eprint = "1803.00035",
    archivePrefix = "arXiv",
    primaryClass = "hep-ph",
    doi = "10.1007/JHEP06(2018)047",
    journal = "JHEP",
    volume = "06",
    pages = "047",
    year = "2018"
}

@article{Kleiss:1985gy,
    author = "Kleiss, R. and Stirling, W. James and Ellis, S. D.",
    title = "{A New Monte Carlo Treatment of Multiparticle Phase Space at High-energies}",
    reportNumber = "CERN-TH-4299/85",
    doi = "10.1016/0010-4655(86)90119-0",
    journal = "Comput. Phys. Commun.",
    volume = "40",
    pages = "359",
    year = "1986"
}

@article{JUNO:2024pur,
    author = "Abusleme, Angel and others",
    collaboration = "JUNO",
    title = "{JUNO sensitivity to invisible decay modes of neutrons}",
    eprint = "2405.17792",
    archivePrefix = "arXiv",
    primaryClass = "hep-ex",
    doi = "10.1140/epjc/s10052-024-13638-0",
    journal = "Eur. Phys. J. C",
    volume = "85",
    number = "1",
    pages = "5",
    year = "2025"
}

@article{SNO:2022trz,
    author = "Allega, A. and others",
    collaboration = "SNO+",
    title = "{Improved search for invisible modes of nucleon decay in water with the SNO+detector}",
    eprint = "2205.06400",
    archivePrefix = "arXiv",
    primaryClass = "hep-ex",
    doi = "10.1103/PhysRevD.105.112012",
    journal = "Phys. Rev. D",
    volume = "105",
    pages = "112012",
    year = "2022"
}

@article{Arnold:2012sd,
    author = "Arnold, Jonathan M. and Fornal, Bartosz and Wise, Mark B.",
    title = "{Simplified models with baryon number violation but no proton decay}",
    eprint = "1212.4556",
    archivePrefix = "arXiv",
    primaryClass = "hep-ph",
    reportNumber = "CALT-68-2898",
    doi = "10.1103/PhysRevD.87.075004",
    journal = "Phys. Rev. D",
    volume = "87",
    pages = "075004",
    year = "2013"
}

@article{Klapdor-Kleingrothaus:2002rvk,
    author = "Klapdor-Kleingrothaus, H. V. and Ma, Ernest and Sarkar, Utpal",
    title = "{Baryon and lepton number violation with scalar bilinears}",
    eprint = "hep-ph/0210156",
    archivePrefix = "arXiv",
    reportNumber = "UCRHEP-T345",
    doi = "10.1142/S0217732302008757",
    journal = "Mod. Phys. Lett. A",
    volume = "17",
    pages = "2221",
    year = "2002"
}

@article{Dorsner:2022twk,
    author = "Dor\v{s}ner, Ilja and Fajfer, Svjetlana and Sumensari, Olcyr",
    title = "{Triple-leptoquark interactions for tree- and loop-level proton decays}",
    eprint = "2202.08287",
    archivePrefix = "arXiv",
    primaryClass = "hep-ph",
    doi = "10.1007/JHEP05(2022)183",
    journal = "JHEP",
    volume = "05",
    pages = "183",
    year = "2022"
}

@article{Nussinov:2001rb,
    author = "Nussinov, Shmuel and Shrock, Robert",
    title = "{$n$-$\bar{n}$ oscillations in models with large extra dimensions}",
    eprint = "hep-ph/0112337",
    archivePrefix = "arXiv",
    doi = "10.1103/PhysRevLett.88.171601",
    journal = "Phys. Rev. Lett.",
    volume = "88",
    pages = "171601",
    year = "2002"
}

@article{Girmohanta:2019fsx,
    author = "Girmohanta, Sudhakantha and Shrock, Robert",
    title = "{Baryon-Number-Violating Nucleon and Dinucleon Decays in a Model with Large Extra Dimensions}",
    eprint = "1911.05102",
    archivePrefix = "arXiv",
    primaryClass = "hep-ph",
    reportNumber = "Stony Brook preprint YITP-SB-2019-35",
    doi = "10.1103/PhysRevD.101.015017",
    journal = "Phys. Rev. D",
    volume = "101",
    pages = "015017",
    year = "2020"
}

@article{Arnellos:1982nt,
    author = "Arnellos, Lampros and Marciano, William J.",
    title = "{Hydrogen - Anti-hydrogen Oscillations, Double Proton Decay and Grand Unified Theories}",
    reportNumber = "Print-82-0284 (BROOKHAVEN), BNL-31239",
    doi = "10.1103/PhysRevLett.48.1708",
    journal = "Phys. Rev. Lett.",
    volume = "48",
    pages = "1708",
    year = "1982"
}

@article{Bowes:1996xy,
    author = "Bowes, J. P. and Foot, Robert and Volkas, R. R.",
    title = "{Electric charge quantization from gauge invariance of a Lagrangian: A Catalog of baryon number violating scalar interactions}",
    eprint = "hep-ph/9609290",
    archivePrefix = "arXiv",
    reportNumber = "UM-P-95-115, RCHEP-95-27",
    doi = "10.1103/PhysRevD.54.6936",
    journal = "Phys. Rev. D",
    volume = "54",
    pages = "6936--6943",
    year = "1996"
}

@article{Super-Kamiokande:2020tor,
    author = "Tanaka, M. and others",
    collaboration = "Super-Kamiokande",
    title = "{Search for proton decay into three charged leptons in 0.37 megaton-years exposure of the Super-Kamiokande}",
    eprint = "2001.08011",
    archivePrefix = "arXiv",
    primaryClass = "hep-ex",
    doi = "10.1103/PhysRevD.101.052011",
    journal = "Phys. Rev. D",
    volume = "101",
    pages = "052011",
    year = "2020"
}

@article{Super-Kamiokande:2020wjk,
    author = "Takenaka, A. and others",
    collaboration = "Super-Kamiokande",
    title = "{Search for proton decay via $p\to e^+\pi^0$ and $p\to \mu^+\pi^0$ with an enlarged fiducial volume in Super-Kamiokande I-IV}",
    eprint = "2010.16098",
    archivePrefix = "arXiv",
    primaryClass = "hep-ex",
    doi = "10.1103/PhysRevD.102.112011",
    journal = "Phys. Rev. D",
    volume = "102",
    pages = "112011",
    year = "2020"
}

@article{Super-Kamiokande:2024qbv,
    author = "Taniuchi, N. and others",
    collaboration = "Super-Kamiokande",
    title = "{Search for proton decay via $p\to e^+\eta$ and $p\to\mu^+\eta$ with a 0.37~Mton-year exposure of Super-Kamiokande}",
    eprint = "2409.19633",
    archivePrefix = "arXiv",
    primaryClass = "hep-ex",
    doi = "10.1103/PhysRevD.110.112011",
    journal = "Phys. Rev. D",
    volume = "110",
    pages = "112011",
    year = "2024"
}

@article{Hyper-Kamiokande:2018ofw,
    author = "Abe, K. and others",
    collaboration = "Hyper-Kamiokande",
    title = "{Hyper-Kamiokande Design Report}",
    eprint = "1805.04163",
    archivePrefix = "arXiv",
    primaryClass = "physics.ins-det",
    month = "5",
    year = "2018"
}

@article{DUNE:2020ypp,
    author = "Abi, Babak and others",
    collaboration = "DUNE",
    title = "{Deep Underground Neutrino Experiment (DUNE), Far Detector Technical Design Report, Volume II: DUNE Physics}",
    eprint = "2002.03005",
    archivePrefix = "arXiv",
    primaryClass = "hep-ex",
    reportNumber = "FERMILAB-PUB-20-025-ND, FERMILAB-DESIGN-2020-02",
    month = "2",
    year = "2020"
}

@article{JUNO:2015zny,
    author = "An, Fengpeng and others",
    collaboration = "JUNO",
    title = "{Neutrino Physics with JUNO}",
    eprint = "1507.05613",
    archivePrefix = "arXiv",
    primaryClass = "physics.ins-det",
    doi = "10.1088/0954-3899/43/3/030401",
    journal = "J. Phys. G",
    volume = "43",
    number = "3",
    pages = "030401",
    year = "2016"
}

@article{Minkowski:1977sc,
    author = "Minkowski, Peter",
    title = "{$\mu \to e\gamma$ at a Rate of One Out of $10^{9}$ Muon Decays?}",
    reportNumber = "Print-77-0182 (BERN)",
    doi = "10.1016/0370-2693(77)90435-X",
    journal = "Phys. Lett. B",
    volume = "67",
    pages = "421--428",
    year = "1977"
}

@article{Yanagida:1979as,
    author = "Yanagida, Tsutomu",
    editor = "Sawada, Osamu and Sugamoto, Akio",
    title = "{Horizontal gauge symmetry and masses of neutrinos}",
    reportNumber = "KEK-79-18-95",
    journal = "Conf. Proc. C",
    volume = "7902131",
    pages = "95--99",
    year = "1979"
}

@article{Gell-Mann:1979vob,
    author = "Gell-Mann, Murray and Ramond, Pierre and Slansky, Richard",
    title = "{Complex Spinors and Unified Theories}",
    eprint = "1306.4669",
    archivePrefix = "arXiv",
    primaryClass = "hep-th",
    reportNumber = "PRINT-80-0576",
    journal = "Conf. Proc. C",
    volume = "790927",
    pages = "315--321",
    year = "1979"
}

@article{Mohapatra:1979ia,
    author = "Mohapatra, Rabindra N. and Senjanovic, Goran",
    title = "{Neutrino Mass and Spontaneous Parity Nonconservation}",
    reportNumber = "MDDP-TR-80-060, MDDP-PP-80-105, CCNY-HEP-79-10",
    doi = "10.1103/PhysRevLett.44.912",
    journal = "Phys. Rev. Lett.",
    volume = "44",
    pages = "912",
    year = "1980"
}

@article{Magg:1980ut,
    author = "Magg, M. and Wetterich, C.",
    title = "{Neutrino Mass Problem and Gauge Hierarchy}",
    reportNumber = "CERN-TH-2829",
    doi = "10.1016/0370-2693(80)90825-4",
    journal = "Phys. Lett. B",
    volume = "94",
    pages = "61--64",
    year = "1980"
}

@article{Schechter:1980gr,
    author = "Schechter, J. and Valle, J. W. F.",
    title = "{Neutrino Masses in $SU(2) \times U(1)$ Theories}",
    reportNumber = "SU-4217-167, COO-3533-167",
    doi = "10.1103/PhysRevD.22.2227",
    journal = "Phys. Rev. D",
    volume = "22",
    pages = "2227",
    year = "1980"
}

@article{Wetterich:1981bx,
    author = "Wetterich, C.",
    title = "{Neutrino Masses and the Scale of $B-L$ Violation}",
    reportNumber = "FREIBURG-THEP-81-2",
    doi = "10.1016/0550-3213(81)90279-0",
    journal = "Nucl. Phys. B",
    volume = "187",
    pages = "343--375",
    year = "1981"
}

@article{Lazarides:1980nt,
    author = "Lazarides, George and Shafi, Q. and Wetterich, C.",
    title = "{Proton Lifetime and Fermion Masses in an SO(10) Model}",
    reportNumber = "FREIBURG-THEP-80-2",
    doi = "10.1016/0550-3213(81)90354-0",
    journal = "Nucl. Phys. B",
    volume = "181",
    pages = "287--300",
    year = "1981"
}

@article{Mohapatra:1980yp,
    author = "Mohapatra, Rabindra N. and Senjanovic, Goran",
    title = "{Neutrino Masses and Mixings in Gauge Models with Spontaneous Parity Violation}",
    reportNumber = "FERMILAB-PUB-80-061-THY, FERMILAB-PUB-80-061-T",
    doi = "10.1103/PhysRevD.23.165",
    journal = "Phys. Rev. D",
    volume = "23",
    pages = "165",
    year = "1981"
}

@article{Cheng:1980qt,
    author = "Cheng, T. P. and Li, Ling-Fong",
    title = "{Neutrino Masses, Mixings and Oscillations in $SU(2) \times U(1)$ Models of Electroweak Interactions}",
    reportNumber = "PRINT-80-0511 (CARNEGIE-MELLON), COO-3066-152",
    doi = "10.1103/PhysRevD.22.2860",
    journal = "Phys. Rev. D",
    volume = "22",
    pages = "2860",
    year = "1980"
}

@article{Foot:1988aq,
    author = "Foot, Robert and Lew, H. and He, X. G. and Joshi, Girish C.",
    title = "{Seesaw Neutrino Masses Induced by a Triplet of Leptons}",
    reportNumber = "UM-P-88/89, OZ-P-88/7",
    doi = "10.1007/BF01415558",
    journal = "Z. Phys. C",
    volume = "44",
    pages = "441",
    year = "1989"
}

@article{Rodejohann:2011mu,
    author = "Rodejohann, Werner",
    title = "{Neutrino-less Double Beta Decay and Particle Physics}",
    eprint = "1106.1334",
    archivePrefix = "arXiv",
    primaryClass = "hep-ph",
    doi = "10.1142/S0218301311020186",
    journal = "Int. J. Mod. Phys. E",
    volume = "20",
    pages = "1833--1930",
    year = "2011"
}

@article{Fonseca:2016jbm,
    author = "Fonseca, Renato M. and Hirsch, Martin",
    title = "{Gauge vectors and double beta decay}",
    eprint = "1612.04272",
    archivePrefix = "arXiv",
    primaryClass = "hep-ph",
    reportNumber = "IFIC-16-50",
    doi = "10.1103/PhysRevD.95.035033",
    journal = "Phys. Rev. D",
    volume = "95",
    pages = "035033",
    year = "2017"
}

@article{Heeck:2024jei,
    author = "Heeck, Julian and Watkins, Dima",
    title = "{Baryon number violation involving tau leptons}",
    eprint = "2405.18478",
    archivePrefix = "arXiv",
    primaryClass = "hep-ph",
    doi = "10.1007/JHEP07(2024)170",
    journal = "JHEP",
    volume = "07",
    pages = "170",
    year = "2024"
}

@article{Majorana:2024rmx,
    author = "Arnquist, I. J. and others",
    collaboration = "Majorana",
    title = "{Rare multinucleon decays with the full data sets of the Majorana Demonstrator}",
    eprint = "2412.16047",
    archivePrefix = "arXiv",
    primaryClass = "nucl-ex",
    reportNumber = "LA-UR-24-33090",
    doi = "10.1103/2pgp-hvst",
    journal = "Phys. Rev. C",
    volume = "112",
    pages = "L022501",
    year = "2025"
}

@article{Heeck:2025btc,
    author = "Heeck, Julian and Sokhashvili, Diana",
    title = "{Revisiting the connection of baryon number, lepton number, and operator dimension}",
    eprint = "2505.06172",
    archivePrefix = "arXiv",
    primaryClass = "hep-ph",
    doi = "10.1016/j.physletb.2025.139791",
    journal = "Phys. Lett. B",
    volume = "868",
    pages = "139791",
    year = "2025"
}

@article{Super-Kamiokande:2018apg,
    author = "Sussman, S. and others",
    collaboration = "Super-Kamiokande",
    title = "{Dinucleon and Nucleon Decay to Two-Body Final States with no Hadrons in Super-Kamiokande}",
    eprint = "1811.12430",
    archivePrefix = "arXiv",
    primaryClass = "hep-ex",
    month = "11",
    year = "2018"
}

@article{Theia:2019non,
    author = "Askins, M. and others",
    collaboration = "Theia",
    title = "{THEIA: an advanced optical neutrino detector}",
    eprint = "1911.03501",
    archivePrefix = "arXiv",
    primaryClass = "physics.ins-det",
    doi = "10.1140/epjc/s10052-020-7977-8",
    journal = "Eur. Phys. J. C",
    volume = "80",
    number = "5",
    pages = "416",
    year = "2020"
}

@article{Broussard:2025opd,
    author = "Leah J. Broussard and Andreas Crivellin and Martin Hoferichter and Sergey Syritsyn and Yasumichi Aoki and Joshua L. Barrow and Arnau Bas i Beneito and Zurab Berezhiani and Nicola Fulvio Calabria and Svjetlana Fajfer and Susan Gardner and Julian Heeck and Cailian Jiang and Luca Naterop and Alexey A. Petrov and Robert Shrock and Adrian Thompson and Ubirajara van Kolck and Michael L. Wagman and Linyan Wan and John Womersley and Jun-Sik Yoo",
    title = "{Baryon number violation: from nuclear matrix elements to BSM physics}",
    eprint = "2504.16983",
    archivePrefix = "arXiv",
    primaryClass = "hep-ph",
    reportNumber = "INT-PUB-25-009, LA-UR-25-23551, PSI-PR-25-07, YITP-SB-2025-08, ZU-TH 23/25, FERMILAB-CONF-25-0265-T",
    doi = "10.1088/1361-6471/adf081",
    journal = "J. Phys. G",
    volume = "52",
    number = "8",
    pages = "083001",
    year = "2025"
}

@article{Rao:1983sd,
    author = "Rao, Sumathi and Shrock, Robert E.",
    title = "{Six Fermion ($B-L$) Violating Operators of Arbitrary Generational Structure}",
    reportNumber = "ITP-SB-82-56R",
    doi = "10.1016/0550-3213(84)90365-1",
    journal = "Nucl. Phys. B",
    volume = "232",
    pages = "143--179",
    year = "1984"
}

@article{deGouvea:2014lva,
    author = "de Gouvea, Andre and Herrero-Garcia, Juan and Kobach, Andrew",
    title = "{Neutrino Masses, Grand Unification, and Baryon Number Violation}",
    eprint = "1404.4057",
    archivePrefix = "arXiv",
    primaryClass = "hep-ph",
    reportNumber = "NUHEP-TH-14-04",
    doi = "10.1103/PhysRevD.90.016011",
    journal = "Phys. Rev. D",
    volume = "90",
    pages = "016011",
    year = "2014"
}

@article{Degrande:2012wf,
    author = "Degrande, Celine and Greiner, Nicolas and Kilian, Wolfgang and Mattelaer, Olivier and Mebane, Harrison and Stelzer, Tim and Willenbrock, Scott and Zhang, Cen",
    title = "{Effective Field Theory: A Modern Approach to Anomalous Couplings}",
    eprint = "1205.4231",
    archivePrefix = "arXiv",
    primaryClass = "hep-ph",
    reportNumber = "MPP-2011-149, SI-HEP-2011-17, CP3-12-25",
    doi = "10.1016/j.aop.2013.04.016",
    journal = "Annals Phys.",
    volume = "335",
    pages = "21--32",
    year = "2013"
}

@article{Henning:2015alf,
    author = "Henning, Brian and Lu, Xiaochuan and Melia, Tom and Murayama, Hitoshi",
    title = "{2, 84, 30, 993, 560, 15456, 11962, 261485, ...: Higher dimension operators in the SM EFT}",
    eprint = "1512.03433",
    archivePrefix = "arXiv",
    primaryClass = "hep-ph",
    reportNumber = "UCB-PTH-15-14, IPMU15-0207",
    doi = "10.1007/JHEP08(2017)016",
    journal = "JHEP",
    volume = "08",
    pages = "016",
    year = "2017",
    note = "[Erratum: JHEP 09, 019 (2019)]"
}

@article{Heeck:2025uwh,
    author = "Heeck, Julian and Shoemaker, Ian M.",
    title = "{Nucleon Decays into Light New Particles in Neutrino Detectors}",
    eprint = "2506.08090",
    archivePrefix = "arXiv",
    primaryClass = "hep-ph",
    doi = "10.1103/cxvm-p412",
    journal = "Phys. Rev. Lett.",
    volume = "135",
    pages = "111804",
    year = "2025"
}

@article{Heeck:2020nbq,
    author = "Heeck, Julian",
    title = "{Light particles with baryon and lepton numbers}",
    eprint = "2009.01256",
    archivePrefix = "arXiv",
    primaryClass = "hep-ph",
    doi = "10.1016/j.physletb.2020.136043",
    journal = "Phys. Lett. B",
    volume = "813",
    pages = "136043",
    year = "2021"
}

@article{Li:2025slp,
    author = "Li, Tong and Schmidt, Michael A. and Yao, Chang-Yuan",
    title = "{Baryon-number-violating nucleon decays in sterile neutrino effective field theories}",
    eprint = "2502.14303",
    archivePrefix = "arXiv",
    primaryClass = "hep-ph",
    reportNumber = "CPPC-2025-02",
    doi = "10.1007/JHEP06(2025)077",
    journal = "JHEP",
    volume = "06",
    pages = "077",
    year = "2025"
}

@article{Li:2024liy,
    author = "Li, Tong and Schmidt, Michael A. and Yao, Chang-Yuan",
    title = "{Baryon-number-violating nucleon decays in ALP effective field theories}",
    eprint = "2406.11382",
    archivePrefix = "arXiv",
    primaryClass = "hep-ph",
    reportNumber = "CPPC-2024-05, DESY-24-082",
    doi = "10.1007/JHEP08(2024)221",
    journal = "JHEP",
    volume = "08",
    pages = "221",
    year = "2024"
}

@article{Domingo:2024qoj,
    author = {Domingo, Florian and Dreiner, Herbi K. and K\"ohler, Dominik and Nangia, Saurabh and Shah, Apoorva},
    title = "{A novel proton decay signature at DUNE, JUNO, and Hyper-K}",
    eprint = "2403.18502",
    archivePrefix = "arXiv",
    primaryClass = "hep-ph",
    reportNumber = "BONN-TH-2024-08",
    doi = "10.1007/JHEP05(2024)258",
    journal = "JHEP",
    volume = "05",
    pages = "258",
    year = "2024"
}

@article{Fajfer:2020tqf,
    author = "Fajfer, Svjetlana and Susi\v{c}, David",
    title = "{Colored scalar mediated nucleon decays to an invisible fermion}",
    eprint = "2010.08367",
    archivePrefix = "arXiv",
    primaryClass = "hep-ph",
    doi = "10.1103/PhysRevD.103.055012",
    journal = "Phys. Rev. D",
    volume = "103",
    pages = "055012",
    year = "2021"
}

@article{BaBar:2011ouc,
    author = "Lees, J. P. and others",
    collaboration = "BaBar",
    title = "{Searches for Rare or Forbidden Semileptonic Charm Decays}",
    eprint = "1107.4465",
    archivePrefix = "arXiv",
    primaryClass = "hep-ex",
    reportNumber = "BABAR-PUB-11-008, SLAC-PUB-14482",
    doi = "10.1103/PhysRevD.84.072006",
    journal = "Phys. Rev. D",
    volume = "84",
    pages = "072006",
    year = "2011"
}

@misc{Heeck2025derivative,
    author = "Heeck, Julian and Sokhashvili, Diana",
    title = "{Opening up derivative baryon-number-violating operators}",
    year = "2026",
    howpublished = "to appear"
}

@article{Super-Kamiokande:2015pys,
    author = "Takhistov, V. and others",
    collaboration = "Super-Kamiokande",
    title = "{Search for Nucleon and Dinucleon Decays with an Invisible Particle and a Charged Lepton in the Final State at the Super-Kamiokande Experiment}",
    eprint = "1508.05530",
    archivePrefix = "arXiv",
    primaryClass = "hep-ex",
    doi = "10.1103/PhysRevLett.115.121803",
    journal = "Phys. Rev. Lett.",
    volume = "115",
    pages = "121803",
    year = "2015"
}

@article{McKeen:2020zni,
    author = "McKeen, David and Pospelov, Maxim",
    title = "{How Long Does the Hydrogen Atom Live?}",
    eprint = "2003.02270",
    archivePrefix = "arXiv",
    primaryClass = "hep-ph",
    doi = "10.3390/universe9110473",
    journal = "Universe",
    volume = "9",
    number = "11",
    pages = "473",
    year = "2023"
}

@article{Liao:2025vlj,
    author = "Liao, Yi and Ma, Xiao-Dong and Wang, Hao-Lin",
    title = "{New Chiral Structures for Baryon Number Violating Nucleon Decays}",
    eprint = "2504.14855",
    archivePrefix = "arXiv",
    primaryClass = "hep-ph",
    doi = "10.1103/d8m7-5xxx",
    journal = "Phys. Rev. Lett.",
    volume = "135",
    number = "16",
    pages = "161801",
    year = "2025"
}

@article{Aebischer:2025qhh,
    author = "Aebischer, Jason and Buras, Andrzej J. and Kumar, Jacky",
    title = "{SMEFT ATLAS: The Landscape Beyond the Standard Model}",
    eprint = "2507.05926",
    archivePrefix = "arXiv",
    primaryClass = "hep-ph",
    reportNumber = "AJB-25-1, CERN-TH-2025-129, LA-UR-24-24665",
    month = "7",
    year = "2025"
}

@article{Bhoonah:2025qzd,
    author = "Bhoonah, Amit and Burk, Francis and Liu, Da and Ou, Tong and Sathyan, Deepak",
    title = "{A Baryon and Lepton Number Violation Model Testable at the LHC}",
    eprint = "2508.21064",
    archivePrefix = "arXiv",
    primaryClass = "hep-ph",
    reportNumber = "PITT-PACC-2507, MI-HET-865",
    month = "8",
    year = "2025"
}

@article{Dorsner:2025pwe,
    author = "Dorsner, Ilja and Fajfer, Svjetlana and Saad, Shaikh",
    title = "{Neutrino Mass Induced $n$-$\overline{n}$ Oscillation}",
    eprint = "2510.16103",
    archivePrefix = "arXiv",
    primaryClass = "hep-ph",
    month = "10",
    year = "2025"
}

@article{Ma:2025mjy,
    author = "Ma, Xiao-Dong and Schmidt, Michael A. and Zhang, Weihang",
    title = "{Baryon-number-violating nucleon decays in SMEFT extended with a light scalar}",
    eprint = "2511.02169",
    archivePrefix = "arXiv",
    primaryClass = "hep-ph",
    doi = "10.1007/JHEP02(2026)175",
    journal = "JHEP",
    volume = "02",
    pages = "175",
    year = "2026"
}

@article{Sakharov:1967dj,
    author = "Sakharov, A. D.",
    title = "{Violation of CP Invariance, C asymmetry, and baryon asymmetry of the universe}",
    doi = "10.1070/PU1991v034n05ABEH002497",
    journal = "Pisma Zh. Eksp. Teor. Fiz.",
    volume = "5",
    pages = "32--35",
    year = "1967"
}

@article{Helo:2025kgx,
    author = "Helo, J. C. and Ota, T. and Hirsch, M.",
    title = "{Invisible neutron decay and light BSM particles}",
    eprint = "2510.14940",
    archivePrefix = "arXiv",
    primaryClass = "hep-ph",
    doi = "10.1103/jg6k-239w",
    journal = "Phys. Rev. D",
    volume = "113",
    pages = "035002",
    year = "2026"
}

@article{Takhistov:2026aln,
    author = "Takhistov, Volodymyr",
    title = "{Experimental Tests of Baryon and Lepton Number Conservation}",
    eprint = "2602.09097",
    archivePrefix = "arXiv",
    primaryClass = "hep-ph",
    reportNumber = "KEK-QUP-2026-0002, KEK-TH-2801",
    month = "2",
    year = "2026"
}
